\documentclass[aps,preprint,nofootinbib,superscriptaddress]{revtex4}%
\usepackage{amssymb}
\usepackage{amsmath}
\usepackage{epsfig}
\usepackage{amsfonts}
\usepackage{graphicx}
\usepackage{hyperref}%
\providecommand{\U}[1]{\protect\rule{.1in}{.1in}}
\begin{document}
\title{Review on High energy String Scattering Amplitudes and Symmetries of String Theory}
\author{Jen-Chi Lee}
\email{jcclee@cc.nctu.edu.tw}
\affiliation{Department of Electrophysics, National Chiao-Tung University, Hsinchu, Taiwan, R.O.C.}
\author{Yi Yang}
\email{yiyang@mail.nctu.edu.tw}
\affiliation{Department of Electrophysics, National Chiao-Tung University, Hsinchu, Taiwan, R.O.C.}
\date{\today }

\begin{abstract}
We review high energy symmetries of string theory at both the fixed angle or
Gross regime (GR) and the fixed momentum transfer or Regge regime (RR). We
calculated in details high energy string scattering amplitudes at arbitrary
mass levels for both regimes. We discovered infinite linear relations among
fixed angle string amplitudes conjectured by Gross in 1988 from decoupling of
high energy zero-norm states (ZNS), and infinite recurrence relations among
Regge string amplitudes from Kummer function $U$ and Appell function $F_{1}$.

However, the linear relations we obtained in the GR corrected [27-32] the
saddle point calculations of Gross, Gross and Mende and Gross and Manes [1-5].
Our results were consistent with the decoupling of high energy ZNS or
unitarity of the theory while those of them were not. In addition, for the
case of high energy closed string scatterings, our results [36] differ from
theirs by an oscillating prefactor which was crucial to recover the KLT
relation valid for all energies.

In the GR/RR regime, all high energy string amplitudes can be solved by these
linear/recurrence relations so that all GR/RR string amplitudes can be
expressed in terms of one single GR/RR string amplitude. In addition, we found
an interesting link between string amplitudes of the two regimes, and
discovered that at each mass level the ratios among fixed angle amplitudes can
be extracted from Regge string scattering amplitudes. This result enables us
to argue that the known $SL(5,C)$ dynamical symmetry of the Appell function
$F_{1}$ is crucial to probe high energy spacetime symmetry of string theory.

Keywords: Symmetries of strings, Hard string scattering amplitudes, Regge
string scattering amplitudes, High energy limits, Zero norm states, Linear
relations, Recurrence relations

\end{abstract}
\maketitle
\tableofcontents
%

\setcounter{equation}{0}
\renewcommand{\theequation}{\arabic{section}.\arabic{equation}}%

\setcounter{section}{-1}

\section{Introduction and overview}

One of the fundamental issues of string theory is its spacetime symmetry
structure. It has long been believed that string theory consists of huge
hidden symmetries. This is strongly suggested by the UV finiteness of quantum
string theory, which contains no free parameter and an infinite number of
states. On the other hand, the high energy, fixed angle behavior of string
scattering amplitudes was known to be very soft exponential fall-off, while
that of a local quantum field theory was power law. Presumably, it is these
huge hidden symmetries which soften the UV structure of quantum string theory.
In a local quantum field theory, a symmetry principle was postulated, which
can be used to determine the interaction of the theory. In string theory, on
the contrary, it is the interaction, prescribed by the very tight quantum
consistency conditions due to the extendedness of string, which determines the
form of the symmetry.

Historically, the first key progress to understand symmetry of string theory
was to study, instead of low energy field theory limit, the high energy, fixed
angle behavior of hard string scattering (HSS) amplitudes \cite{GM,GM1, Gross,
Gross1,GrossManes}. This was motivated by the spontaneously broken symmetries
in gauge field theories which were hidden at low energy, but became evident in
the high energy behavior of the theory. There were two main conjectures of
Gross's \cite{Gross,Gross1} pioneer work in 1988 on this subject. The first
one was the existence of an infinite number of linear relations among the
scattering amplitudes of different string states that were valid order by
order in string perturbation theory at high energies, fixed angle regime or
Gross regime (GR). The second was that this symmetry was so powerful as to
determine the scattering amplitudes of all the infinite number of string
states in terms of a single dilaton (tachyon for the case of open string)
scattering amplitudes. However, the symmetry charges of his proposed stringy
symmetries were not understood and the proportionality constants or ratios
among scattering amplitudes of different string states were not calculated.

The second key to uncover the fundamental symmetry of string theory was the
realization of the the importance of zero norm states (ZNS) in the old
covariant first quantized (OCFQ) string spectrum. It was proposed that
\cite{Lee,Lee-Ov,LeePRL} spacetime symmetry charges of string theory originate
from an infinite number of ZNS with arbitrary high spin in the spectrum. In
the late eighties the decoupling of ZNS was also used in the literature to the
measure in the study of multi-string vertices by seveal authors and was
closely related to the so-called group theoretic approach of stringy
scattering amplitudes. This subject will not be covered in this review. For
more details see the review in \cite{West9}. In the context of $\sigma$-model
approach of string theory, one turns on background fields on the worldsheet
energy momentum tensor $T$. Conformal invariance of the worldsheet then
requires, in addition to $D=26$, cancellation of various $q$-number anomalies
and results to equations of motion of the background fields \cite{GSW}. It was
then shown that \cite{Lee} for each \textit{spacetime} ZNS, one can
systematically construct a \textit{worldsheet} $(1,1)$ primary field $\delta
T_{\Phi}$ such that%
\begin{equation}
T_{\Phi}+\delta T_{\Phi}=T_{\Phi+\delta\Phi} \label{BZNS}%
\end{equation}
is satisfied to some order of weak field approximation in the $\sigma$-model
background fields $\beta$ function calculation. In the above equation,
$T_{\Phi}$ is the worldsheet energy momentum tensor with background fields
$\Phi$ and $T_{\Phi+\delta\Phi}$ is the new energy momentum tensor with new
background fields $\Phi+\delta\Phi$. Thus for each ZNS one can construct a
spacetime symmetry transformation for string background fields.

In addition to the positive norm physical propagating states, there are two
types of physical ZNS in the old covariant first quantized open bosonic string
spectrum: \cite{GSW}%
\begin{equation}
\text{Type I}:L_{-1}\left\vert x\right\rangle ,\text{ where }L_{1}\left\vert
x\right\rangle =L_{2}\left\vert x\right\rangle =0,\text{ }L_{0}\left\vert
x\right\rangle =0;
\end{equation}%
\begin{equation}
\text{Type II}:(L_{-2}+\frac{3}{2}L_{-1}^{2})\left\vert \widetilde{x}%
\right\rangle ,\text{ where }L_{1}\left\vert \widetilde{x}\right\rangle
=L_{2}\left\vert \widetilde{x}\right\rangle =0,\text{ }(L_{0}+1)\left\vert
\widetilde{x}\right\rangle =0.
\end{equation}
While type I states have zero-norm at any spacetime dimension, type II states
have zero-norm \emph{only} at $D=26$. For example, among other stringy
symmetries, an inter-particle symmetry transformation for two propagating
states at mass level $M^{2}=4$ of open bosonic string can be generated
\cite{Lee}
\begin{equation}
\delta C_{(\mu\nu\lambda)}=\frac{1}{2}\partial_{(\mu}\partial_{\nu}%
\theta_{\lambda)}^{2}-2\eta_{(\mu\nu}\theta_{\lambda)}^{2},\delta C_{[\mu\nu
]}=9\partial_{\lbrack\mu}\theta_{\nu]}^{2}, \label{01}%
\end{equation}
where $\partial^{\mu}\theta_{\mu}^{2}=0,(\partial^{2}-4)\theta_{\mu}^{2}=0$
which are the on-shell conditions of the $D_{2}$ vector ZNS with polarization
$\theta_{\mu}^{2}$ \cite{Lee}%

\begin{equation}
|D_{2}\rangle=[(\frac{1}{2}k_{\mu}k_{\nu}\theta_{\lambda}^{2}+2\eta_{\mu\nu
}\theta_{\lambda}^{2})\alpha_{-1}^{\mu}\alpha_{-1}^{\nu}\alpha_{-1}^{\lambda
}+9k_{\mu}\theta_{\nu}^{2}\alpha_{-2}^{[\mu}\alpha_{-1}^{\nu]}-6\theta_{\mu
}^{2}\alpha_{-3}^{\mu}]\left\vert 0,k\right\rangle ,\text{ \ }k\cdot\theta
^{2}=0, \label{02}%
\end{equation}
and $C_{(\mu\nu\lambda)}$ and $C_{[\mu\nu]}$ are the background fields of the
symmetric spin-three and antisymmetric spin-two propagating states respectively.

In the even higher mass levels, $M^{2}=6$ for example, a new phenomenon begins
to show up. There are ambiguities in defining positive-norm spin-two and
scalar states due to the existence of ZNS in the same Young representations
\cite{LeePRL}. As a result, the degenerate spin two and scalar positive-norm
states can be gauged to the higher rank fields, the symmetric spin four
$D_{\mu\nu\alpha\beta}$ and mixed-symmetric spin three $D_{\mu\nu\alpha}$ in
the first order weak field approximation. In fact, for instance, it can be
shown \cite{Lee3} that the scattering amplitude involving the positive-norm
spin-two state can be expressed in terms of those of spin-four and
mixed-symmetric spin-three states due to the existence of a
\textit{degenerate} type I and a type II spin-two ZNS. This stringy phenomenon
seems to persist to higher mass levels.

This calculation is consistent with the result in the HSS limit. In fact, it
can be shown that in the HSS limit all the scattering amplitudes of leading
order in energy at each fixed mass level can be expressed in terms of that of
the leading trajectory string state with transverse polarizations on the
scattering plane. See Eq.(\ref{03}), Eq.(\ref{055}) and Eq.(\ref{04}) below.
One can also justify this decoupling by WSFT to be discussed in section I.D.
Finally one expects this decoupling to persist even if one includes the higher
order corrections in weak field approximation, as there will be even stronger
relations between background fields order by order through iteration.

The calculation of Eq.(\ref{01}) was done in the first order weak field
approximation but valid to all energies or all orders in $\alpha^{\prime}$. A
second order weak field calculation implies an even more interesting
spontaneously broken inter-mass level symmetry in string theory
\cite{Lee4,LEO}. Some implication of the corresponding stringy Ward identity
on the scattering amplitudes were discussed in \cite{JCLee,Lee4} and will be
presented in Eq.(\ref{D22}). It was then realized that \cite{KaoLee, CLYang}
the symmetry in Eq.(\ref{01}) can be reproduced from gauge transformation of
Witten string field theory (WSFT) \cite{Witten} after imposing the no ghost
conditions. It is important to note that this stringy symmetry exists only for
$D=26$ thanks to type II ZNS in the OCFQ string spectrum , which is zero norm
only when $D=26$.

Incidentally, it was well known in $2D$ string theory that the operator
products of the discrete positive norm states $\psi_{J,M}^{+}$ form a
$w_{\infty}$ algebra \cite{Winfinity,Winfinity2,Klebanov1}%

\begin{equation}
\int\frac{dz}{2\pi i}\psi_{J_{1},M_{1}}^{+}\psi_{J_{2},M_{2}}^{+}=(J_{2}%
M_{1}-J_{1}M_{2})\psi_{J_{1}+J_{2}-1,M_{1}+M_{2}}^{+}. \label{2D1}%
\end{equation}
This is in parallel with the work of Ref \cite{Ring,Ring1} where the ground
ring structure of ghost number zero operators was identified in the BRST
quantization. Interestingly, a set of discrete ZNS $G_{J,M}^{+}$ with Polyakov
momenta can be constructed (see Eq.(\ref{3.11.}) in section III.A.2) and were
also shown \cite{ChungLee1,ChungLee2} to carry the spacetime $\omega_{\infty}$
symmetry \cite{Winfinity,Winfinity2,Klebanov1} charges of $2D$ string theory
\cite{ChungLee1,ChungLee2}%
\begin{equation}
\int\frac{dz}{2\pi i}G_{J_{1},M_{1}}^{+}(z)G_{J_{2},M_{2}}^{+}(0)=(J_{2}%
M_{1}-J_{1}M_{2})G_{J_{1}+J_{2}-1,M_{1}+M_{2}}^{+}(0). \label{2D2}%
\end{equation}
The calculation above can be generalized to $2D$ superstring theory
\cite{ChungLee2}.

One can also use ZNS to calculate spacetime symmetries of string on compact
backgrounds. The existence of soliton ZNS at some moduli points was shown to
be responsible for the enhanced Kac-Moody symmetry of closed string theory. As
a simple example, for the case of $26D$ bosonic closed string compactified on
a $2$-dimensional torus $T^{2}\equiv\frac{R^{2}}{2\pi\Lambda^{2}}$, it was
found that massless ZNS (including soliton ZNS) form a representation of
enhanced Kac-Moody $SU(3)_{R}\otimes$ $SU(3)_{L}$ symmetry at the moduli point
(see section IV.A.2)%
\begin{equation}
R_{1}=R_{2}=\sqrt{2},B=\frac{1}{2},\overset{\rightarrow}{e}_{1}=\left(
\sqrt{2},0\right)  ,\overset{\rightarrow}{e}_{2}=\left(  -\sqrt{\frac{1}{2}%
},\sqrt{\frac{3}{2}}\right)
\end{equation}
where $\Lambda^{2}$ is a $2$-dimensional lattice with a basis $\left\{
R_{1}\frac{\overset{\rightarrow}{e}_{1}}{\sqrt{2}},R_{2}\frac
{\overset{\rightarrow}{e}_{2}}{\sqrt{2}}\right\}  $, and $B$ is the
antisymmetric tensor $B_{ij}=B\epsilon_{ij}$. In this calculation one has four
moduli parameters $R_{1},R_{2},B$ and $\overset{\rightarrow}{e}_{1}%
\cdot\overset{\rightarrow}{e}_{2}$with $\left\vert \overset{\rightarrow
}{e}_{i}\right\vert ^{2}=2$.\ Moreover, an infinite number of massive soliton
ZNS at any higher massive level of the spectrum were constructed in
\cite{Lee1}. Presumably, these massive soliton ZNS are responsible for
enhanced stringy symmetries of the theory.

For the case of open string compactification, unlike the closed string case
discussed above, it was found that \cite{Lee2} the soliton ZNS exist only at
massive levels. These Chan-Paton soliton ZNS correspond to the existence of
enhanced massive stringy symmetries with transformation parameters containing
both Einstein and Yang-Mills indices in the case of Heterotic string
\cite{Lee4}. In the T-dual picture, these symmetries exist only at some
discrete values of compactified radii when $N$ $D$-branes are coincident
\cite{Lee2}.

All the above results which are valid to all energies will constitute the part
I of this review paper. On the other hand, in part II of this review, we will
show that the high energy limit of the discrete ZNS $G_{J,M}^{+}$ in $2D$
string theory constructed in Eq.(\ref{2D2}) in part I approaches $\psi
_{J,M}^{+}$ in Eq.(\ref{2D1}) and thus form a high energy $w_{\infty}$
symmetry of $2D$ string. This result strongly suggests that the linear
relations obtained from decoupling of ZNS in $26D$ string theory are indeed
related to the hidden symmetry also for the $26D$ string theory.

In part II of this paper, we will review high energy, fixed angle calculations
of HSS amplitudes. The high energy, fixed angle Ward identities derived from
the decoupling of ZNS in the HSS limit, which combines the previous two key
ideas of probing stringy symmetry, were used to explicitly prove Gross's two
conjectures \cite{ChanLee,ChanLee1,ChanLee2, CHL, CHLTY1,CHLTY2}. An infinite
number of linear relations among high energy scattering amplitudes of
different string states were derived. Remarkably, these linear relations were
just good enough to fix the proportionality constants or ratios among high
energy scattering amplitudes of different string states algebraically at each
fixed mass level. The first example calculated was the ratios among HSS
amplitudes at mass level $M^{2}=4$ \cite{ChanLee,ChanLee2} (see the definition
of polarizations $e^{T}$ and $e^{L}$ after Eq.(\ref{214}) below)
\begin{equation}
\mathcal{T}_{TTT}:\mathcal{T}_{LLT}:\mathcal{T}_{(LT)}:\mathcal{T}%
_{[LT]}=8:1:-1:-1 \label{03}%
\end{equation}
which corresponds to stringy symmetries in the $\sigma$-model calculation
discussed from Eq.(\ref{BZNS}) to Eq.(\ref{02}). Eq.(\ref{03}) is presumably
valid order by order in string perturbation theory as we expect the decoupling
of ZNS is valid even for string loop amplitudes \cite{ChanLee3}.

To calculate Eq.(\ref{03}), we note that there are four ZNS at mass level
\ $M^{2}$ $=4$. For type I ZNS, there is one symmetric spin two tensor, one
vector and one scalar ZNS. In addition, there is only one vector type II ZNS.
The corresponding Ward identities for these four ZNS were calculated to be
\cite{JCLee}%

\begin{equation}
k_{\mu}\theta_{\nu\lambda}\mathcal{T}_{\chi}^{(\mu\nu\lambda)}+2\theta_{\mu
\nu}\mathcal{T}_{\chi}^{(\mu\nu)}=0,
\end{equation}%
\begin{equation}
(\frac{5}{2}k_{\mu}k_{\nu}\theta_{\lambda}^{\prime}+\eta_{\mu\nu}%
\theta_{\lambda}^{\prime})\mathcal{T}_{\chi}^{(\mu\nu\lambda)}+9k_{\mu}%
\theta_{\nu}^{\prime}\mathcal{T}_{\chi}^{(\mu\nu)}+6\theta_{\mu}^{\prime
}\mathcal{T}_{\chi}^{\mu}=0,
\end{equation}%
\begin{equation}
(\frac{1}{2}k_{\mu}k_{\nu}\theta_{\lambda}+2\eta_{\mu\nu}\theta_{\lambda
})\mathcal{T}_{\chi}^{(\mu\nu\lambda)}+9k_{\mu}\theta_{\nu}\mathcal{T}_{\chi
}^{[\mu\nu]}-6\theta_{\mu}\mathcal{T}_{\chi}^{\mu}=0, \label{D22}%
\end{equation}%
\begin{equation}
(\frac{17}{4}k_{\mu}k_{\nu}k_{\lambda}+\frac{9}{2}\eta_{\mu\nu}k_{\lambda
})\mathcal{T}_{\chi}^{(\mu\nu\lambda)}+(9\eta_{\mu\nu}+21k_{\mu}k_{\nu
})\mathcal{T}_{\chi}^{(\mu\nu)}+25k_{\mu}\mathcal{T}_{\chi}^{\mu}=0
\end{equation}
where $\theta_{\mu\nu}$ is transverse and traceless, and $\theta_{\lambda
}^{\prime}$ and $\theta_{\lambda}$ are transverse vectors. $\mathcal{T}_{\chi
}^{\prime}s$ in the above equations are the mass level $M^{2}$ $=4$, $\chi$-th
order string-loop amplitudes. In each equation, we have chosen, say,
$v_{2}(k_{2})$\ to be the vertex operators constructed from ZNS and $k_{\mu
}\equiv k_{2\mu}$. Note that Eq.(\ref{D22}) is the inter-particle Ward
identity corresponding to $D_{2}$ vector ZNS in Eq.(\ref{02}) obtained by
antisymmetrizing those terms which contain $\alpha_{-1}^{\mu}\alpha_{-2}^{\nu
}$ in the original type I and type II vector ZNS \cite{Lee}. We will use 1 and
2 for the incoming particles and 3 and 4 for the scattered particles. In the
Ward identities, 1, 3 and 4 can be any string states and we have omitted their
tensor indices for the cases of excited string states.

In the HSS limit, one enjoys many simplifications in the calculation. First,
all polarizations of the amplitudes orthogonal to the scattering plane are of
subleading order in energy, and one needs only consider polarizations on the
scattering plane. Second, to the leading order in energy, $e^{P}\simeq$
$e^{L}$ in the HSS calculation. In the end of the calculation, one ends up
with the simple linear equations for leading order amplitudes
\cite{ChanLee,ChanLee2}%

\begin{align}
\mathcal{T}_{LLT}^{5\rightarrow3}+\mathcal{T}_{(LT)}^{3}  &  =0,\label{212}\\
10\mathcal{T}_{LLT}^{5\rightarrow3}+\mathcal{T}_{TTT}^{3}+18\mathcal{T}%
_{(LT)}^{3}  &  =0,\label{213}\\
\mathcal{T}_{LLT}^{5\rightarrow3}+\mathcal{T}_{TTT}^{3}+9\mathcal{T}%
_{[LT]}^{3}  &  =0 \label{214}%
\end{align}
where $e^{P}=\frac{1}{M_{2}}(E_{2},\mathrm{k}_{2},0)=\frac{k_{2}}{M_{2}}$ the
momentum polarization, $e^{L}=\frac{1}{M_{2}}(\mathrm{k}_{2},E_{2},0)$ the
longitudinal polarization and $e^{T}=(0,0,1)$ the transverse polarization are
the three polarizations on the scattering plane. In Eq.(\ref{212}) to
Eq.(\ref{214}), we have assigned a relative energy power for each amplitude.
For each longitudinal $L$ component, the order is $E^{2}$ and for each
transverse $T$ component, the order is $E.$ This is due to the definitions of
$e_{L}$and $e_{T}$ above, where $e_{L}$ got one energy power more than
$e_{T}.$ By Eq.(\ref{213}), the naive leading order $E^{5}$ term of the energy
expansion for $\mathcal{T}_{LLT}$ is forced to be zero. As a result, the real
leading order term is $E^{3}$. Similar rule applies to $\mathcal{T}_{LLT}$ in
Eq.(\ref{212}) and Eq.(\ref{214}). The solution of these three linear
relations gives Eq.(\ref{03}). Eq.(\ref{03}) gives the first evidence of Gross
conjecture \cite{Gross,Gross1} on HSS amplitudes.

A sample calculation of scattering amplitudes for mass level $M^{2}$ $=4$
\cite{ChanLee2} justified the ratios calculated in Eq.(\ref{03}). Since the
proportionality constants in Eq.(\ref{03}) are independent of particles chosen
for vertex $v_{1,3,4}$, for simplicity, we will choose them to be tachyons.
For the string-tree level $\chi=1$, with one tensor $v_{2}$ and three tachyons
$v_{1,3,4}$, all scattering amplitudes of mass level $M_{2}^{2}$ $=4$ were
calculated to be ($s-t$ channel)%
\begin{equation}
\mathcal{T}_{TTT}=-8E^{9}\mathcal{T}(3)\sin^{3}\phi_{CM}[1+\frac{3}{E^{2}%
}+\frac{5}{4E^{4}}-\frac{5}{4E^{6}}+O(\frac{1}{E^{8}})],
\end{equation}

\begin{align}
\mathcal{T}_{LLT}  &  =-E^{9}\mathcal{T}(3)[\sin^{3}\phi_{CM}+(6\sin\phi
_{CM}\cos^{2}\phi_{CM})\frac{1}{E^{2}}\nonumber\\
&  -\sin\phi_{CM}(\frac{11}{2}\sin^{2}\phi_{CM}-6)\frac{1}{E^{4}}+O(\frac
{1}{E^{6}})],
\end{align}

\begin{align}
\mathcal{T}_{[LT]}  &  =E^{9}\mathcal{T}(3)[\sin^{3}\phi_{CM}-(2\sin\phi
_{CM}\cos^{2}\phi_{CM})\frac{1}{E^{2}}\nonumber\\
&  +\sin\phi_{CM}(\frac{3}{2}\sin^{2}\phi_{CM}-2)\frac{1}{E^{4}}+O(\frac
{1}{E^{6}})],
\end{align}

\begin{align}
\mathcal{T}_{(LT)}  &  =E^{9}\mathcal{T}(3)[\sin^{3}\phi_{CM}+\sin\phi
_{CM}(\frac{3}{2}-10\cos\phi_{CM}\nonumber\\
&  -\frac{3}{2}\cos^{2}\phi_{CM})\frac{1}{E^{2}}-\sin\phi_{CM}(\frac{1}%
{4}+10\cos\phi_{CM}+\frac{3}{4}\cos^{2}\phi_{CM})\frac{1}{E^{4}}+O(\frac
{1}{E^{6}})]
\end{align}
where $\mathcal{T}(N)\mathcal{=}\sqrt{\pi}(-1)^{N-1}2^{-n}E^{-1-2N}(\sin
\frac{\phi_{CM}}{2})^{-3}(\cos\frac{\phi_{CM}}{2})^{5-2N}\exp(-\frac{s\ln
s+t\ln t-(s+t)\ln(s+t)}{2})$ is the high energy limit of $\frac{\Gamma
(-\frac{s}{2}-1)\Gamma(-\frac{t}{2}-1)}{\Gamma(\frac{u}{2}+2)}$ with
$s+t+u=2N-8$. We thus have justified Eq.(\ref{03}) with $\mathcal{T}_{TTT}%
^{3}=-8E^{9}\mathcal{T}(3)\sin^{3}\phi_{CM}$.

The calculations based on ZNS thus relate \cite{CLYang} gauge transformation
of WSFT to high energy string symmetries of Gross. However, in the sample
calculation of \cite{GrossManes}, two of the four high energy amplitudes in
Eq.(\ref{03}) were missing, and thus the decoupling of ZNS or unitarity was
violated. This is of course due to the unawareness of the importance of ZNS in
the saddle-point calculation of \cite{GM,GM1, Gross, Gross1,GrossManes}.

The calculations for $M^{2}$ $=4$ above can be generalized to $M^{2}$ $=6$
\cite{ChanLee2}. To the leading order in energy, one ended up with $8$
equations and $9$ amplitudes. A calculation showed that \cite{ChanLee2}%

\begin{align}
\mathcal{T}_{TTTT}^{4}  &  :\mathcal{T}_{TTLL}^{4}:\mathcal{T}_{LLLL}%
^{4}:\mathcal{T}_{TTL}^{4}:\mathcal{T}_{LLL}^{4}:\widetilde{\mathcal{T}%
}_{LT,T}^{4}:\widetilde{\mathcal{T}}_{LP,P}^{4}:\mathcal{T}_{LL}%
^{4}:\widetilde{\mathcal{T}}_{LL}^{4}=\nonumber\\
16  &  :\frac{4}{3}:\frac{1}{3}:-\frac{4\sqrt{6}}{9}:-\frac{\sqrt{6}}%
{9}:-\frac{2\sqrt{6}}{3}:0:\frac{2}{3}:0. \label{055}%
\end{align}
A sample calculation of scattering amplitudes for mass level $M^{2}$ $=6$
\cite{ChanLee2} justified the ratios above calculated by solving $8$ linear
relations derived from the decoupling of high energy ZNS in the GR. The ratios
for $M^{2}$ $=8$ can be found in Eq.(\ref{M8}) in the appendix A.

The results of mass level $M^{2}$ $=4,6$ and $8$ can be generalized to
arbitrary higher mass levels. From the calculations of Eq.(\ref{212}) to
Eq.(\ref{214}), one first observes that only states of the following form
\cite{CHLTY1,CHLTY2}
\begin{equation}
\left\vert N,2m,q\right\rangle \equiv(\alpha_{-1}^{T})^{N-2m-2q}(\alpha
_{-1}^{L})^{2m}(\alpha_{-2}^{L})^{q}|0,k\rangle\label{Nmq}%
\end{equation}
are of leading order in energy in the HSS limit. The choice of only even power
$2m$ in $\alpha_{-1}^{L}$ is the result of the observation that the naive
energy order of the amplitudes will in general drop by even number of energy
power as can be seen in Eq.(\ref{212}) to Eq.(\ref{214}). Scattering
amplitudes corresponding to states with $(\alpha_{-1}^{L})^{2m+1}$\ turn out
to be of subleading order in energy. Many simplifications occur if we apply
Ward identities or decoupling of ZNS only on these high energy states in the
HSS limit. First, consider the decoupling of type I \ high energy ZNS
\begin{equation}
L_{-1}|N-1,2m-1,q\rangle\simeq M|N,2m,q\rangle+(2m-1)|N,2m-2,q+1\rangle
\end{equation}
where many terms are omitted because they are not of the form of the leading
order. This implies that
\begin{equation}
\mathcal{T}^{(N,2m,q)}=-\frac{2m-1}{M}\mathcal{T}^{(N,2m-2,q+1)}.
\end{equation}
Using this relation repeatedly, we get
\begin{equation}
\mathcal{T}^{(N,2m,q)}=\frac{(2m-1)!!}{(-M)^{m}}\mathcal{T}^{(N,0,m+q)}
\label{123}%
\end{equation}
where the double factorial is defined by $(2m-1)!!=\frac{(2m)!}{2^{m}m!}$.

Next, consider the decoupling of type II high energy ZNS
\begin{equation}
L_{-2}|N-2,0,q\rangle\simeq\frac{1}{2}|N,0,q\rangle+M|N,0,q+1\rangle.
\end{equation}
Again, irrelevant terms are omitted here. From this we deduce that
\begin{equation}
\mathcal{T}^{(N,0,q+1)}=-\frac{1}{2M}\mathcal{T}^{(N,0,q)},
\end{equation}
which leads to
\begin{equation}
\mathcal{T}^{(N,0,q)}=\frac{1}{(-2M)^{q}}\mathcal{T}^{(N,0,0)}. \label{124}%
\end{equation}

Our main result for arbitrary mass levels $M^{2}=2(N-1)$ is an immediate
deduction of the above two equations, Eq.(\ref{123}) and Eq.(\ref{124}),
\cite{CHLTY1,CHLTY2}
\begin{equation}
\frac{T^{(N,2m,q)}}{T^{(N,0,0)}}=\left(  -\frac{1}{M}\right)  ^{2m+q}\left(
\frac{1}{2}\right)  ^{m+q}(2m-1)!!. \label{04}%
\end{equation}

Exactly the same results can also be obtained by two other calculations, the
Virasoro constraint calculation and the saddle-point calculation. Here we
review the saddle-point calculation. Since the result in Eq.(\ref{04}) is
valid for all string loop order, we need only do saddle-point calculation of
the string tree level amplitudes. Without loss of generality, we choose
particles 1,3 and 4 to be tachyons, and particle 2 to be of the form of
Eq.(\ref{Nmq}). The $t-u$ channel contribution to the stringy amplitude at
tree level is
\begin{align}
\mathcal{T}^{(N,2m,q)}  &  =\int_{1}^{\infty}dxx^{(1,2)}(1-x)^{(2,3)}\left[
\frac{e^{T}\cdot k_{1}}{x}-\frac{e^{T}\cdot k_{3}}{1-x}\right]  ^{N-2m-2q}%
\nonumber\\
&  \cdot\left[  \frac{e^{P}\cdot k_{1}}{x}-\frac{e^{P}\cdot k_{3}}%
{1-x}\right]  ^{2m}\left[  -\frac{e^{P}\cdot k_{1}}{x^{2}}-\frac{e^{P}\cdot
k_{3}}{(1-x)^{2}}\right]  ^{q}%
\end{align}
where $(1,2)=k_{1}\cdot k_{2}$ etc.

In order to apply the saddle-point method, we rewrite the amplitude above into
the following form \cite{CHLTY1,CHLTY2}%
\begin{equation}
\mathcal{T}^{(N,2m,q)}(K)=\int_{1}^{\infty}dx\mbox{ }u(x)e^{-Kf(x)},
\end{equation}
where
\begin{align}
K  &  \equiv-(1,2)\rightarrow\frac{s}{2}\rightarrow2E^{2},\\
\tau &  \equiv-\frac{(2,3)}{(1,2)}\rightarrow-\frac{t}{s}\rightarrow\sin
^{2}\frac{\phi}{2},\\
f(x)  &  \equiv\ln x-\tau\ln(1-x),\\
u(x)  &  \equiv\left[  \frac{(1,2)}{M}\right]  ^{2m+q}(1-x)^{-N+2m+2q}%
(f^{\prime})^{2m}(f^{\prime\prime})^{q}(-e^{T}\cdot k_{3})^{N-2m-2q}.
\end{align}
The saddle-point for the integration of moduli, $x=x_{0}$, is defined by
\begin{equation}
f^{\prime}(x_{0})=0,
\end{equation}
and we have%
\begin{equation}
x_{0}=\frac{1}{1-\tau},\hspace{1cm}1-x_{0}=-\frac{\tau}{1-\tau},\hspace
{1cm}f^{\prime\prime}(x_{0})=(1-\tau)^{3}\tau^{-1}. \label{saddle}%
\end{equation}
It is easy to see that%
\begin{equation}
u(x_{0})=u^{\prime}(x_{0})=....=u^{(2m-1)}(x_{0})=0,
\end{equation}
and
\begin{equation}
u^{(2m)}(x_{0})=\left[  \frac{(1,2)}{M}\right]  ^{2m+q}(1-x_{0})^{-N+2m+2q}%
(2m)!(f_{0}^{\prime\prime})^{2m+q}(-e^{T}\cdot k_{3})^{N-2m-2q}.
\end{equation}

With these inputs, one can easily evaluate the Gaussian integral associated
with the four-point amplitudes
\begin{align}
&  \int_{1}^{\infty}dx\mbox{ }u(x)e^{-Kf(x)}\nonumber\\
&  =\sqrt{\frac{2\pi}{Kf_{0}^{\prime\prime}}}e^{-Kf_{0}}\left[  \frac
{u_{0}^{(2m)}}{2^{m}\ m!\ (f_{0}^{\prime\prime})^{m}\ K^{m}}+O(\frac
{1}{K^{m+1}})\right] \nonumber\\
&  =\sqrt{\frac{2\pi}{Kf_{0}^{\prime\prime}}}e^{-Kf_{0}}\left[  (-1)^{N-q}%
\frac{2^{N-2m-q}(2m)!}{m!\ {M}^{2m+q}}\ \tau^{-\frac{N}{2}}(1-\tau)^{\frac
{3N}{2}}E^{N}+O(E^{N-2})\right]  .
\end{align}
This result shows explicitly that with one tensor and three tachyons, the
energy and angle dependence for the four-point HSS amplitudes only depend on
the level $N$
\begin{align}
\lim_{E\rightarrow\infty}\frac{\mathcal{T}^{(N,2m,q)}}{\mathcal{T}^{(N,0,0)}}
&  =\frac{(-1)^{q}(2m)!}{m!(2M)^{2m+q}}\nonumber\\
&  =(-\frac{2m-1}{M})....(-\frac{3}{M})(-\frac{1}{M})(-\frac{1}{2M})^{m+q},
\end{align}
which is consistent with calculation of decoupling of high energy ZNS obtained
in Eq.(\ref{04}).

We conclude that there is only one independent component of high energy
scattering amplitude at each fixed mass level. Based on this independent
component of high energy scattering amplitude, one can then derive the general
formula of high energy scattering amplitude for four arbitrary string states,
and express them in terms of that of tachyons. This completes the general
proof \cite{ChanLee,ChanLee1,ChanLee2, CHL, CHLTY1,CHLTY2} of Gross's two
conjectures on high energy symmetry of string theory stated above.

All the above calculations can be extended to the case of hard superstring
scattering amplitudes which will be discussed in chapter XIII of this review.
However, it was found that \cite{susy} there were new HSS amplitudes for the
superstring case. The existence of these new high energy scattering amplitudes
of string states with polarizations orthogonal to the scattering plane is due
to the worldsheet fermion exchange in the correlation functions. These
worldsheet fermion exchanges do not exist in the bosonic string correlation
functions and is, presumably, related to the high energy massive spacetime
fermionic scattering amplitudes in the R-sector of the theory.

Obviously, these new high energy amplitudes create complications for a full
understanding of stringy symmetry. Nevertheless, the claim that there is only
one independent high energy scattering amplitude at each fixed mass level of
the string spectrum persists in the case of superstring theory, at least, for
the NS sector of the theory \cite{susy}.

Incidentally, it was important to discover \cite{ChanLee,ChanLee1,ChanLee2,
CHL} that the result of saddle-point calculation in Refs \cite{GM,GM1,
Gross,Gross1, GrossManes} was inconsistent with high energy stringy Ward
identities of ZNS calculation in Refs \cite{ChanLee,ChanLee1,ChanLee2, CHL}.
One simple example was the missing of two of the four amplitudes in
Eq.(\ref{03}) as has been pointed out previously. A corrected saddle-point
calculation was given in \cite{CHL}, where the missing terms of the
calculation in Refs \cite{GM,GM1, Gross,Gross1,GrossManes} were identified to
recover the stringy Ward identities.

Indeed, it was found \cite{CHL} that saddle point calculation in \cite{GM,GM1,
Gross,Gross1,GrossManes} is only valid for the tachyon amplitude. In general,
the results calculated in \cite{GM,GM1, Gross,Gross1,GrossManes} gives the
right energy exponent in the scattering amplitudes, but not the energy power
factors in front of the exponential for the cases of the \textit{excited
string states}. These energy power factors are subleading terms ignored in
\cite{GM,GM1, Gross,Gross1,GrossManes} but they are crucial if one wants to
get the linear relations among high energy scattering amplitudes conjectured
by Gross.

Interestingly, the inconsistency of the saddle point calculation discussed
above for the excited string states was also pointed out by the authors of
\cite{West2}. The source of disagreement in their so-called group theoretic
approach of stringy symmetries stems from the proper choice of local
coordinates for the worldsheet saddle points to describe the behavior of the
excited string states at high energy limit. It seems that both the ZNS
calculation and the calculation based on group theoretic approach agree with
tachyon amplitudes obtained in \cite{GM,GM1, Gross,Gross1, GrossManes} (ignore
the possible phase factors in the amplitudes to be discussed in the next few
paragraphs), but disagree with amplitudes for other excited string states.

The next interesting issues were the calculation of \textit{closed} string
scattering amplitudes and their symmetries in the HSS limit \cite{Closed}.
Historically, the open string four tachyon amplitude in the HSS limit was
first calculated in the original paper of Veneziano in 1968. On the other
hand, the $\mathcal{N}$-loop closed HSS amplitudes were calculated by the
saddle-point method in \cite{GM,GM1} in 1988. Both open and closed HSS
amplitudes exhibit the very soft exponential fall-off behaviors in contrast to
the power law behavior of the scattering amplitudes of quantum field theory.

However, an inconsistency arises if one plugs, for example, the tree level
four tachyon open and closed string HSS amplitudes\bigskip\ calculated by
these authors, into the KLT relation (1986)%
\begin{equation}
A_{\text{closed}}^{\left(  4\right)  }\left(  s,t,u\right)  =\sin\left(  \pi
k_{2}\cdot k_{3}\right)  A_{\text{open}}^{\left(  4\right)  }\left(
s,t\right)  \bar{A}_{\text{open}}^{\left(  4\right)  }\left(  t,u\right)
\label{KLT2}%
\end{equation}
which is valid for \textit{all} kinematic regimes and for \textit{all} string
states. This is due to the phase factor $\sin\left(  \pi k_{2}\cdot
k_{3}\right)  $ in the above equation which was missing in the closed string
saddle-point calculation in \cite{GM,GM1}. One clue to see the origin of this
inconsistency is to note that the saddle-point $x_{0}=\frac{1}{1-\tau}$
identified for the open string calculation in Eq.(\ref{saddle}) is in the
regime $[1,\infty)$. So only saddle point calculation for $\bar{A}%
_{\text{open}}^{\left(  4\right)  }\left(  t,u\right)  $ is reliable, but not
that of $A_{\text{open}}^{\left(  4\right)  }\left(  s,t\right)  $ and neither
that of closed string amplitude $A_{\text{closed}}^{\left(  4\right)  }\left(
s,t,u\right)  $ \cite{Closed} by the KLT relation.

Instead of using saddle-point calculation for the closed HSS amplitudes, the
above considerations led the authors of \cite{Closed} to study the
relationship between $A_{\text{open}}^{\left(  4\right)  }\left(  s,t\right)
$ and $\bar{A}_{\text{open}}^{\left(  4\right)  }\left(  t,u\right)  $ for
arbitrary string states in the HSS limit. \textit{With the help of the
infinite linear relations in Eq.(\ref{04})}, one needs only calculate
relationship between $s-t$ and $t-u$ channel HSS amplitudes for the leading
trajectory string states. They ended up with the following result in the HSS
limit (2006) \cite{Closed}%
\begin{equation}
A_{\text{open}}^{\left(  4\right)  }\left(  s,t\right)  =\frac{\sin\left(  \pi
k_{2}.k_{4}\right)  }{\sin\left(  \pi k_{1}k_{2}\right)  }\bar{A}%
_{\text{open}}^{\left(  4\right)  }\left(  t,u\right)  , \label{CLY-BCJ}%
\end{equation}
which is valid for four arbitrary string states. It is now clear that due to
the phase factor in the above equation, the saddle-point calculation of
$A_{\text{open}}^{\left(  4\right)  }\left(  s,t\right)  $ is not reliable,
neither for the closed one $A_{\text{closed}}^{\left(  4\right)  }\left(
s,t,u\right)  $ in view of the KLT relation in Eq.(\ref{KLT2}). One can now
use the reliable saddle-point calculation of $\bar{A}_{\text{open}}^{\left(
4\right)  }\left(  t,u\right)  $%
\begin{equation}
A_{\text{open}}^{(4-\text{tachyon})}\left(  t,u\right)  \simeq(stu)^{-\frac
{3}{2}}\exp\left(  -\frac{s\ln s+t\ln t+u\ln u}{2}\right)  , \label{exp}%
\end{equation}
and Eq.(\ref{CLY-BCJ}) to calculate $A_{\text{open}}^{\left(  4\right)
}\left(  s,t\right)  $ in the HSS limit. The consistent closed string
four-tachyon HSS amplitudes can then be calculated by using the KLT relation
in Eq.(\ref{KLT2}) to be \cite{Closed}%
\begin{equation}
A_{\text{closed}}^{(4-\text{tachyon})}\left(  s,t,u\right)  \simeq\frac
{\sin\left(  \pi t/2\right)  \sin\left(  \pi u/2\right)  }{\sin\left(  \pi
s/2\right)  }(stu)^{-3}\exp\left(  -\frac{s\ln s+t\ln t+u\ln u}{4}\right)
\label{open1}%
\end{equation}
The exponential factor in Eq.(\ref{exp}) was first discussed by Veneziano
\cite{Veneziano}. The result for the high energy closed string four-tachyon
amplitude in Eq.(\ref{open1}) differs from the one calculated in the
literature \cite{GM,GM1} by an oscillating factor $\frac{\sin\left(  \pi
t/2\right)  \sin\left(  \pi u/2\right)  }{\sin\left(  \pi s/2\right)  }$. One
notes here that the results of Eqs.(\ref{open1}), (\ref{exp}) and
Eq.(\ref{CLY-BCJ}) are consistent with the KLT formula, while the previous
calculation in \cite{GM,GM1} is NOT.

Indeed, one might try to use the saddle-point method to calculate the high
energy closed string scattering amplitude. The closed string four-tachyon
scattering amplitude is%
\begin{align}
A_{\text{closed}}^{(4-\text{tachyon})}\left(  s,t,u\right)   &  =\int
dxdy\exp\left(  \frac{k_{1}\cdot k_{2}}{2}\ln\left\vert z\right\vert
+\frac{k_{2}\cdot k_{3}}{2}\ln\left\vert 1-z\right\vert \right) \nonumber\\
&  \equiv\int dxdy(x^{2}+y^{2})^{-2}[(1-x)^{2}+y^{2}]^{-2}\exp\left[
-Kf(x,y)\right]
\end{align}
where $K=\frac{s}{8}$ and $f(x,y)=\ln(x^{2}+y^{2})-\tau\ln[(1-x)^{2}+y^{2}]$
with $\tau=-\frac{t}{s}$. One can then calculate the "saddle-point" of
$\ f(x,y)$ to be%
\begin{equation}
\nabla f(x,y)\mid_{x_{0}=\frac{1}{1-\tau},y_{0}=0}=0.
\end{equation}
The HSS limit of the closed string four-tachyon scattering amplitude is then
calculated to be%
\begin{equation}
A_{\text{closed}}^{(4-\text{tachyon})}\left(  s,t,u\right)  \simeq\frac{2\pi
}{K\sqrt{\det\frac{\partial^{2}f(x_{0},y_{0})}{\partial x\partial y}}}%
\exp[-Kf(x_{0},y_{0})]\simeq(stu)^{-3}\exp\left(  -\frac{s\ln s+t\ln t+u\ln
u}{4}\right)  ,
\end{equation}
which is consistent with the previous one calculated in the literature
\cite{GM,GM1}, but is different from the result in Eq.(\ref{open1}). However,
one notes that%
\begin{equation}
\frac{\partial^{2}f(x_{0},y_{0})}{\partial x^{2}}=\frac{2(1-\tau)^{3}}{\tau
}=-\frac{\partial^{2}f(x_{0},y_{0})}{\partial y^{2}},\frac{\partial^{2}%
f(x_{0},y_{0})}{\partial x\partial y}=0,
\end{equation}
which means that $(x_{0},y_{0})$ is NOT the local minimum of $f(x,y)$, and one
should not trust this saddle-point calculation. There was other evidence
pointed out by authors of \cite{Closed} to support this conclusion. Finally,
the ratios of closed HSS amplitudes turned out to be the tensor products of
two open string ratios
\begin{equation}
\frac{T^{\left(  N;2m,2m^{^{\prime}};q,q^{^{\prime}}\right)  }}{T^{\left(
N;0,0;0,0\right)  }}=\left(  -\frac{1}{M_{2}}\right)  ^{2(m+m^{^{\prime}%
})+q+q^{^{\prime}}}\left(  \frac{1}{2}\right)  ^{m+m^{^{\prime}}%
+q+q^{^{\prime}}}(2m-1)!!(2m^{\prime}-1)!!. \label{closed4}%
\end{equation}

The relationship between $s-t$ and $t-u$ channels HSS amplitudes in
Eq.(\ref{CLY-BCJ}) was later argued to be valid for \textit{all} kinematic
regime based on monodromy of integration in string amplitude calculation in
2009 \cite{BCJ2}. An explicit proof of Eq.(\ref{CLY-BCJ}) for arbitrary four
string states and all kinematic regimes was given very recently in
\cite{LLY1,LLY2}.

The motivation for the author in \cite{BCJ2} to calculate Eq.(\ref{CLY-BCJ})
was different from the discussion above which was related to the calculation
of hard closed string scattering amplitudes. The motivation in \cite{BCJ2} was
based on the field theory BCJ relation \cite{BCJ1} for Yang-Mills gluon
color-stripped scattering amplitudes $A$ which was first pointed out and
calculated in 2008 to be
\begin{equation}
sA(k_{1},k_{2},k_{3},k_{4})-uA(k_{1},k_{4},k_{2},k_{3})=0. \label{fbcj}%
\end{equation}
Note that for the supersymmetric case, there is no tachyon and the low energy
massless limit of Eq.(\ref{CLY-BCJ}) reproduces Eq.(\ref{fbcj}).

Recently the mass level dependent of Eq.(\ref{CLY-BCJ}) was calculated to be
\cite{LLY1,LLY2}%
\begin{equation}
\frac{A_{st}^{(p,r,q)}}{A_{tu}^{(p,r,q)}}=\left(  -1\right)  ^{N}%
\frac{B\left(  -M_{1}M_{2}+1,\frac{M_{1}M_{2}}{2}\right)  }{B\left(
\frac{M_{1}M_{2}}{2},\frac{M_{1}M_{2}}{2}\right)  }\simeq\frac{\sin\pi\left(
k_{2}\cdot k_{4}\right)  }{\sin\pi\left(  k_{1}\cdot k_{2}\right)  }
\label{level}%
\end{equation}
by taking the \textit{nonrelativistic} limit $|\vec{k_{2}}|<<M_{S}$ of
Eq.(\ref{CLY-BCJ}). In Eq.(\ref{level}), $B$ was the beta function, and
$k_{1}$, $k_{3}$ and $k_{4}$ were taken to be tachyons, and $k_{2}$ was the
following tensor string state%

\begin{equation}
V_{2}=(i\partial X^{T})^{p}(i\partial X^{L})^{r}(i\partial X^{P})^{q}%
e^{ik_{2}X}%
\end{equation}
where%
\begin{equation}
N=p+r+q\text{, \ }M_{2}^{2}=2(N-1)\text{, }N\geq2.
\end{equation}

The generalization of the four point function relation in Eq.(\ref{CLY-BCJ})
to higher point string amplitudes can be found in \cite{BCJ2}. It is
interesting to see that historically the four point (high energy) string BCJ
relations Eq.(\ref{CLY-BCJ}) \cite{Closed} were discovered even earlier than
the field theory BCJ relations Eq.(\ref{fbcj})! \cite{BCJ1}.

\bigskip The ratios calculated in Eq.(\ref{closed4}) persist for the case of
closed string D-particle scatterings in the HSS limit. For the simple case of
$m=0=m^{\prime}$, the ratios were first calculated to be $\left(  -\frac
{1}{2M}\right)  ^{q+q^{\prime}}$ \cite{Dscatt}. The complete ratios were\ then
calculated through a correspondence between HSS ratios and RSS ratios to be
discussed in Eq.(\ref{corresp}) below, and were found to be
\textit{factorized} \cite{LMY} (see section XIV.C)
\begin{equation}
\frac{T_{SD}^{\left(  N;2m,2m^{^{\prime}};q,q^{^{\prime}}\right)  }}%
{T_{SD}^{\left(  N;0,0;0,0\right)  }}=\left(  -\frac{1}{M_{2}}\right)
^{2(m+m^{^{\prime}})+q+q^{^{\prime}}}\left(  \frac{1}{2}\right)
^{m+m^{^{\prime}}+q+q^{^{\prime}}}(2m-1)!!(2m^{\prime}-1)!! \label{corresp2}%
\end{equation}

It is well known that the closed string-string scattering amplitudes can be
factorized into two open string-string scattering amplitudes due to the
existence of the KLT formula \cite{KLT}. On the contrary, there is no physical
picture for open string D-particle tree scattering amplitudes and thus no
factorization for closed string D-particle scatterings into two channels of
open string D-particle scatterings, and hence no KLT-like formula there.

Thus the factorized ratios in HSS regime calculated above came as a surprise.
However, these ratios are consistent with the decoupling of high energy ZNS
calculated previously in \cite{ChanLee,ChanLee1,ChanLee2,
CHL,CHLTY1,CHLTY2,CHLTY3,susy,Closed}. It will be interesting if one can
calculate the complete HSS amplitudes directly and see how the
\textit{non-factorized amplitudes} can give the result of factorized ratios.

On the other hand, in contrast to the closed string D-particle scatterings in
the HSS limit discussed above, it was shown that, instead of the exponential
fall-off behavior of the form factors with Regge-pole structure, the HSS
amplitudes of closed string scattered from D24-brane, or D-domain-wall, behave
as \textit{power-law with Regge-pole structure }\cite{Wall}. See Eq.(\ref{am})
and Eq.(\ref{pole}) in section IX.A.4. This is to be compared with the
well-known power law form factors without Regge-pole structure of the
D-instanton scatterings.

This discovery makes D-domain-wall scatterings an unique example of a hybrid
of string and field theory scatterings. Moreover, it was discovered that
\cite{Wall} the usual linear relations of HSS amplitudes at each fixed mass
level, Eq.(\ref{corresp2}), breaks down for the D-domain-wall scatterings.
This result gives a strong evidence that the existence of the infinite linear
relations, or stringy symmetries, of HSS amplitudes is responsible for the
softer, exponential fall-off HSS scatterings than the power-law field theory scatterings.

Being a consistent theory of quantum gravity, string theory is remarkable for
its soft ultraviolet structure. Presumably, this is mainly due to
\textit{three} closely related fundamental characteristics of HSS amplitudes.
The first is the softer exponential fall-off behavior of the form factors in
the HSS in contrast to the power-law field theory scatterings. The second is
the existence of infinite Regge poles in the form factor of string scattering
amplitudes. The existence of infinite linear relations discussed in part II of
the review constitutes the \textit{third} fundamental characteristics of HSS amplitudes.

It will be important to study more string scatterings, which exhibit the above
three unusual behaviors in the HSS limit. In section IX.B, we will consider
closed string scattered from O-planes. In particular we first calculate
massive closed string states at arbitrary mass levels scattered from
Orientifold planes in the HSS limit \cite{O-plane}. The scatterings of
massless states from Orientifold planes were calculated in the literature by
using the boundary states formalism \cite{Craps,Craps1,Craps2,Schnitzer}, and
on the worldsheet of real projected plane $RP_{2}$ \cite{Garousi}. Many
speculations were made about the scatterings of \textit{massive} string
states, in particular, for the case of O-domain-wall scatterings. It is one of
the purposes of section IV.B to clarify these speculations and to discuss
their relations with the three fundamental characteristics of HSS scatterings
stated above.

For the generic O$p$-planes with $p\geq0$, one expects to get the infinite
linear relations except O-domain-wall HSS. For simplicity, we consider only
the case of O-particle HSS \cite{O-plane}. For the case of O-particle
scatterings, we obtain infinite linear relations among HSS amplitudes of
different string states. We also confirm that there exist only $t$-channel
closed string Regge poles in the form factor of the O-particle scatterings
amplitudes as expected.

For the case of O-domain-wall scatterings, we find that, like the well-known
D-instanton scatterings, the amplitudes behave like field theory scatterings,
namely \textit{UV power-law without Regge pole}. In addition, we find that
there exist only finite number of $t$-channel closed string poles in the form
factor of O-domain-wall scatterings, and the masses of the poles are bounded
by the masses of the external legs \cite{O-plane}. We thus confirm that all
massive closed string states do couple to the O-domain-wall as was conjectured
previously \cite{Myers,Garousi}. This is also consistent with the boundary
state descriptions of O-planes.

For both cases of O-particle and O-domain-wall scatterings, we confirm that
there exist no $s$-channel open string Regge poles in the form factor of the
amplitudes as O-planes were known to be not dynamical. However, the usual
claim that there is a thickness of order$\sqrt{\alpha^{^{\prime}}}$ for the
O-domain-wall is misleading as the UV behavior of its scatterings is power-law
instead of exponential fall-off.

In the end of section IX.B, we summarize the Regge pole structures of closed
strings states scattered from various D-branes and O-planes in the following
table. The $s$-channel and $t$-channel scatterings for both D-branes and
O-planes are shown in the Fig. \ref{t-s}. For O-plane scatterings, the
$s$-channel open string Regge poles are not allowed since O-planes are not
dynamical. For both cases of Domain-wall scatterings, the $t$-channel closed
string Regge poles are not allowed since there is only one kinematic variable
instead of two as in the usual cases.

\begin{center}
\ \
\begin{tabular}
[c]{|c|c|c|c|}\hline
& $p=-1$ & $1\leq p\leq23$ & $p=24$\\\hline
D$p$-branes & X & C+O & O\\\hline
O$p$-planes & X & C & X\\\hline
\end{tabular}

\end{center}

In this table, "C" and "O" represent infinite Closed string Regge poles and
Open string Regge poles respectively. "X" means there are no infinite Regge poles.

In chapter X, following an old suggestion of Mende \cite{Mende}, we calculate
high energy massive scattering amplitudes of bosonic string with some
coordinates compactified on the torus \cite{Compact,Compact2}. We obtain
infinite linear relations among high energy scattering amplitudes of different
string states in the Hard scattering limit. In addition, we analyze all
possible power-law and soft exponential fall-off regimes of high energy
compactified bosonic string scatterings by comparing the scatterings with
their 26D noncompactified counterparts.

Interestingly, we discover in section X.A the existence of a power-law regime
at fixed angle and an exponential fall-off regime at small angle for high
energy compactified open string scatterings \cite{Compact2}. These new
phenomena never happen in the 26D string scatterings. The linear relations
break down as expected in all power-law regimes. The analysis can be extended
to the high energy scatterings of the compactified closed string in section
X.B, which corrects and extends the results in \cite{Compact}.

At this point, one may ask an important question for the results of
Eqs.(\ref{03}), (\ref{055}), (\ref{M8}) and (\ref{04}) above , namely, is
there any group theoretical structure of the ratios of these scattering
amplitudes? Let's consider a simple analogy from particle physics. The ratios
of the nucleon-nucleon scattering processes%
\begin{align}
(a)\text{ \ }p+p  &  \rightarrow d+\pi^{+},\nonumber\\
(b)\text{ \ }p+n  &  \rightarrow d+\pi^{0},\nonumber\\
(c)\text{ \ }n+n  &  \rightarrow d+\pi^{-} \label{05}%
\end{align}
can be calculated to be (ignore the tiny mass difference between proton and
neutron)%
\begin{equation}
T_{a}:T_{b}:T_{c}=1:\frac{1}{\sqrt{2}}:1 \label{06}%
\end{equation}
from $SU(2)$ isospin symmetry. Is there any symmetry structure which can be
used to calculate ratios in Eqs.(\ref{03}), (\ref{055}), (\ref{M8}) and
(\ref{04})? It turned out that part of the answer can be addressed by studying
another high energy regime of string scattering amplitudes, namely, the fixed
momentum transfer or Regge regime (RR)
\cite{RR1,RR2,RR3,RR4,RR5,RR6,RR7,bosonic,RRsusy,hep-th/0410131}.

In part III of this paper, we will discuss RSS amplitudes and their relations
to the fixed angle HSS amplitudes. We will find that the number of RSS
amplitudes is much more numerous than that of HSS amplitudes. For example,
there are only $4$ HSS amplitudes while there are $22$ RSS amplitudes at mass
level $M^{2}=4$ \cite{bosonic}. This is one of the reason why decoupling of
ZNS in the RR, in contrast to the GR, is not good enough to solve RSS
amplitudes in terms of one single amplitude at each mass level.

\bigskip For illustration and to identify the ratios in Eqs.(\ref{03}) from
RSS amplitudes, we will first calculate amplitudes at mass level $M^{2}=4$ in
the RR%
\begin{equation}
s\rightarrow\infty,\sqrt{-t}=\text{fixed (but }\sqrt{-t}\neq\infty).
\end{equation}
The relevant kinematics are%
\begin{subequations}
\begin{align}
e^{P}\cdot k_{1}  &  =-\frac{1}{M_{2}}\left(  \sqrt{p^{2}+M_{1}^{2}}%
\sqrt{p^{2}+M_{2}^{2}}+p^{2}\right)  \simeq-\frac{s}{2M_{2}},\\
e^{L}\cdot k_{1}  &  =-\frac{p}{M_{2}}\left(  \sqrt{p^{2}+M_{1}^{2}}%
+\sqrt{p^{2}+M_{2}^{2}}\right)  \simeq-\frac{s}{2M_{2}},\\
e^{T}\cdot k_{1}  &  =0 \label{111}%
\end{align}
and%
\end{subequations}
\begin{subequations}
\begin{align}
e^{P}\cdot k_{3}  &  =\frac{1}{M_{2}}\left(  \sqrt{q^{2}+M_{3}^{2}}\sqrt
{p^{2}+M_{2}^{2}}-pq\cos\theta\right)  \simeq-\frac{\tilde{t}}{2M_{2}}%
\equiv-\frac{t-M_{2}^{2}-M_{3}^{2}}{2M_{2}},\\
e^{L}\cdot k_{3}  &  =\frac{1}{M_{2}}\left(  p\sqrt{q^{2}+M_{3}^{2}}%
-q\sqrt{p^{2}+M_{2}^{2}}\cos\theta\right)  \simeq-\frac{\tilde{t}^{\prime}%
}{2M_{2}}\equiv-\frac{t+M_{2}^{2}-M_{3}^{2}}{2M_{2}},\\
e^{T}\cdot k_{3}  &  =-q\sin\phi\simeq-\sqrt{-{t}}.
\end{align}
Note that in contrast to the identification $e^{P}\simeq$ $e^{L}$ in the HSS
limit, $e^{P}$ \textit{does not} approach to $e^{L}$ in the RSS limit.

We will list the relevant RSS amplitudes at mass level $M^{2}=4$ which contain
polarizations $(e^{T},e^{L})$ only. It turned out that there are eight high
energy amplitudes in the RR%
\end{subequations}
\begin{align}
&  \alpha_{-1}^{T}\alpha_{-1}^{T}\alpha_{-1}^{T}|0\rangle,\alpha_{-1}%
^{L}\alpha_{-1}^{T}\alpha_{-1}^{T}|0\rangle,\alpha_{-1}^{L}\alpha_{-1}%
^{L}\alpha_{-1}^{T}|0\rangle,\alpha_{-1}^{L}\alpha_{-1}^{L}\alpha_{-1}%
^{L}|0\rangle,\nonumber\\
&  \alpha_{-1}^{T}\alpha_{-2}^{T}|0\rangle,\alpha_{-1}^{T}\alpha_{-2}%
^{L}|0\rangle,\alpha_{-1}^{L}\alpha_{-2}^{T}|0\rangle,\alpha_{-1}^{L}%
\alpha_{-2}^{L}|0\rangle.
\end{align}
Among them only four of the above amplitudes are relevant here and can be
calculated to be \cite{bosonic}%
\begin{align}
A^{TTT}  &  =\int_{0}^{1}dx\cdot x^{k_{1}\cdot k_{2}}\left(  1-x\right)
^{k_{2}\cdot k_{3}}\cdot\left(  \frac{ie^{T}\cdot k_{1}}{x}-\frac{ie^{T}\cdot
k_{3}}{1-x}\right)  ^{3}\nonumber\\
&  \simeq-i\left(  \sqrt{-t}\right)  ^{3}\frac{\Gamma\left(  -\frac{s}%
{2}-1\right)  \Gamma\left(  -\frac{\tilde{t}}{2}-1\right)  }{\Gamma\left(
\frac{u}{2}+3\right)  }\cdot\left(  -\frac{1}{8}s^{3}+\frac{1}{2}s\right)  ,
\end{align}%
\begin{align}
A^{LLT}  &  =\int_{0}^{1}dx\cdot x^{k_{1}\cdot k_{2}}\left(  1-x\right)
^{k_{2}\cdot k_{3}}\cdot\left(  \frac{ie^{T}\cdot k_{1}}{x}-\frac{ie^{T}\cdot
k_{3}}{1-x}\right)  \left(  \frac{ie^{L}\cdot k_{1}}{x}-\frac{ie^{L}\cdot
k_{3}}{1-x}\right)  ^{2}\nonumber\\
&  \simeq-i\left(  \sqrt{-t}\right)  \left(  -\frac{1}{2M_{2}}\right)
^{2}\frac{\Gamma\left(  -\frac{s}{2}-1\right)  \Gamma\left(  -\frac{\tilde{t}%
}{2}-1\right)  }{\Gamma\left(  \frac{u}{2}+3\right)  }\nonumber\\
&  \cdot\left[  \left(  \frac{1}{4}t-\frac{9}{2}\right)  s^{3}+\left(
\frac{1}{4}t^{2}+\frac{7}{2}t\right)  s^{2}+\frac{\left(  t+6\right)  ^{2}}%
{2}s\right]  ,
\end{align}%
\begin{align}
A^{TL}  &  =\int_{0}^{1}dx\cdot x^{k_{1}\cdot k_{2}}\left(  1-x\right)
^{k_{2}\cdot k_{3}}\cdot\left(  \frac{ie^{T}\cdot k_{1}}{x}-\frac{ie^{T}\cdot
k_{3}}{1-x}\right)  \left[  \frac{e^{L}\cdot k_{1}}{x^{2}}+\frac{e^{L}\cdot
k_{3}}{\left(  1-x\right)  ^{2}}\right] \nonumber\\
&  \simeq i\left(  \sqrt{-t}\right)  \left(  -\frac{1}{2M_{2}}\right)
\frac{\Gamma\left(  -\frac{s}{2}-1\right)  \Gamma\left(  -\frac{\tilde{t}}%
{2}-1\right)  }{\Gamma\left(  \frac{u}{2}+3\right)  }\nonumber\\
&  \cdot\left[  -\left(  \frac{1}{8}t+\frac{3}{4}\right)  s^{3}-\frac{1}%
{8}\left(  t^{2}-2t\right)  s^{2}-\left(  \frac{1}{4}t^{2}-t-3\right)
s\right]  ,
\end{align}
and%
\begin{align}
A^{LT}  &  =\int_{0}^{1}dx\cdot x^{k_{1}\cdot k_{2}}\left(  1-x\right)
^{k_{2}\cdot k_{3}}\cdot\left(  \frac{ie^{L}\cdot k_{1}}{x}-\frac{ie^{L}\cdot
k_{3}}{1-x}\right)  \left[  \frac{e^{T}\cdot k_{1}}{x^{2}}+\frac{e^{T}\cdot
k_{3}}{\left(  1-x\right)  ^{2}}\right] \nonumber\\
&  \simeq i\left(  \sqrt{-t}\right)  \left(  -\frac{1}{2M_{2}}\right)
\frac{\Gamma\left(  -\frac{s}{2}-1\right)  \Gamma\left(  -\frac{\tilde{t}}%
{2}-1\right)  }{\Gamma\left(  \frac{u}{2}+3\right)  }\cdot\left[  \frac{3}%
{4}s^{3}-\frac{t}{4}s^{2}-\left(  \frac{t}{2}+3\right)  s\right]  .
\end{align}
where the kinematic variables $(s,t)$ were used instead of $(E,\theta)$ used
in the GR. From the above calculation, one can easily see that all the
amplitudes are in the same leading order $\left(  \sim s^{3}\right)  $\ in the
RR. On the other hand, one notes that, for example, the terms $\sqrt{-t}%
t^{2}s^{2}$ in $A^{LLT}$ and $A^{TL}$ are in the leading order in the GR, but
are in the subleading order in the RR. On the contrary, the terms $\sqrt
{-t}s^{3}$ in $A^{LLT}$ and $A^{TL}$ are in the subleading order in the GR,
but are in the leading order in the RR. These observations suggest that the
high energy string scattering amplitudes in the GR and RR contain information
complementary to each other.

One important observation for high energy amplitudes in the RR is for those
amplitudes with the same structure as those of the GR in Eq.(\ref{Nmq}). The
amplitudes $A^{TTT}$, $A^{LLT}$ , $A^{TL}$ and $A^{LT}$ at mass level
$M^{2}=4$ are such examples. For these amplitudes, the relative ratios of the
coefficients of the highest power of $t$ in the leading order amplitudes in
the RR can be calculated to be \cite{bosonic}
\begin{align}
A^{TTT}  &  =-i\left(  \sqrt{-t}\right)  \frac{\Gamma\left(  -\frac{s}%
{2}-1\right)  \Gamma\left(  -\frac{\tilde{t}}{2}-1\right)  }{\Gamma\left(
\frac{u}{2}+3\right)  }\cdot\left(  \frac{1}{8}ts^{3}\right)  \sim\frac{1}%
{8},\\
A^{LLT}  &  =-i\left(  \sqrt{-t}\right)  \left(  -\frac{1}{2M_{2}}\right)
^{2}\frac{\Gamma\left(  -\frac{s}{2}-1\right)  \Gamma\left(  -\frac{\tilde{t}%
}{2}-1\right)  }{\Gamma\left(  \frac{u}{2}+3\right)  }\left(  \frac{1}%
{4}ts^{3}\right)  \sim\frac{1}{64},\\
A^{TL}  &  =i\left(  \sqrt{-t}\right)  \left(  -\frac{1}{2M_{2}}\right)
\frac{\Gamma\left(  -\frac{s}{2}-1\right)  \Gamma\left(  -\frac{\tilde{t}}%
{2}-1\right)  }{\Gamma\left(  \frac{u}{2}+3\right)  }\cdot\left(  -\frac{1}%
{8}ts^{3}\right)  \sim-\frac{1}{32},
\end{align}
which reproduces the ratios in the GR in Eq.(\ref{03}). Note that the
symmetrized and anti-symmetrized amplitudes are defined as%
\begin{align}
T^{\left(  TL\right)  }  &  =\frac{1}{2}\left(  T^{TL}+T^{LT}\right)  ,\\
T^{\left[  TL\right]  }  &  =\frac{1}{2}\left(  T^{TL}-T^{LT}\right)  ;
\end{align}
and similarly for the amplitudes $A^{\left(  TL\right)  }$ and $A^{\left[
TL\right]  }$ in the RR. It is interesting to see that $T^{LT}$ $\sim
(\alpha_{-1}^{L})(\alpha_{-2}^{T})|0\rangle$ in the GR is of subleading order
in energy, while $A^{LT}$ in the RR is of leading order in energy. However,
the contribution of the amplitude $A^{LT}$ to $A^{\left(  TL\right)  }$ and
$A^{\left[  TL\right]  }$ in the RR will not affect the ratios calculated above.

\bigskip From the calculation above, it was thus believed that there existed
intimate link between high energy string scattering amplitudes in the HSS
regime and those in the RSS regime. To study this link and to reproduce the
ratios in Eq.(\ref{04}) in particular, one was led to calculate RSS amplitudes
for arbitrary mass levels. To simplify the calculation, we use the simple
kinematics $e^{T}\cdot k_{1}=0$ in Eq.(\ref{04}) and the energy power counting
of the string amplitudes, and end up with the following rules
\begin{align}
&  \alpha_{-n}^{T}:\quad\text{1 term (contraction of $ik_{3}\cdot X$ with
$\varepsilon_{T}\cdot\partial^{n}X$),}\\
&  \alpha_{-n}^{L}:%
\begin{cases}
n>1,\quad\text{1 term}\\
n=1\quad\text{2 terms}\text{ (contraction of $ik_{1}\cdot X$ and $ik_{3}\cdot
X$ with $\varepsilon_{L}\cdot\partial^{n}X$).}%
\end{cases}
\end{align}
A class of the leading order high energy open string states in the RR at each
fixed mass level $N=\sum_{n,l>0}np_{n}+lr_{l}$ are%
\begin{equation}
\left\vert p_{n},r_{l}\right\rangle =\prod_{n>0}(\alpha_{-n}^{T})^{p_{n}}%
\prod_{l>0}(\alpha_{-l}^{L})^{r_{l}}|0,k\rangle.
\end{equation}
The $s-t$ channel scattering amplitudes of this state with three other
tachyonic states can be calculated to be \cite{bosonic}
\begin{align}
A^{(p_{n},q_{m})}  &  =\left(  -\frac{i}{M_{2}}\right)  ^{q_{1}}U\left(
-q_{1},\frac{t}{2}+2-q_{1},\frac{\tilde{t}^{\prime}}{2}\right)  B\left(
-1-\frac{s}{2},-1-\frac{t}{2}\right) \nonumber\\
&  \cdot\prod_{n=1}\left[  i\sqrt{-t}(n-1)!\right]  ^{p_{n}}\prod_{m=2}\left[
i\tilde{t}^{\prime}(m-1)!\left(  -\frac{1}{2M_{2}}\right)  \right]  ^{q_{m}}.
\label{Apq}%
\end{align}
In the above, $U(a,c,x)$ is the Kummer function of the second kind. It is
crucial to note that $c=\frac{t}{2}+2-q_{1},$ and is not a constant as in the
usual case, so $U$ in the above amplitude is not a solution of the Kummer
equation. On the contrary, since $a=-q_{1}$ an integer, the Kummer function in
Eq.(\ref{equality}) terminated to be a finite sum.

\bigskip It can be seen from Eq.(\ref{Apq}) that the RSS amplitudes with spin
polarizations corresponding to Eq.(\ref{Nmq}) at each fixed mass level are no
longer proportional to each other. The ratios are $t$ dependent functions and
can be calculated to be \cite{bosonic}
\begin{gather}
\frac{A^{(N,2m,q)}(s,t)}{A^{(N,0,0)}(s,t)}=(-1)^{m}\left(  -\frac{1}{2M_{2}%
}\right)  ^{2m+q}(\tilde{t}^{\prime}-2N)^{-m-q}(\tilde{t}^{\prime}%
)^{2m+q}\nonumber\\
\hphantom{\frac{A^{(N,2m,q)}(s,t)}{A^{(N,0,0)}(s,t)} =}{}\times\sum_{j=0}%
^{2m}(-2m)_{j}\left(  -1+N-\frac{\tilde{t}^{\prime}}{2}\right)  _{j}%
\frac{(-2/\tilde{t}^{\prime})^{j}}{j!}+\mathit{O}\left\{  \left(  \frac
{1}{\tilde{t}^{\prime}}\right)  ^{m+1}\right\}  ,
\end{gather}
where $(x)_{j}=x(x+1)(x+2)\cdots(x+j-1)$ is the Pochhammer symbol.

To deduce the link and ensure the following identification\/for the general
mass levels%
\begin{equation}
\lim_{\tilde{t}^{\prime}\rightarrow\infty}\frac{A^{(N,2m,q)}}{A^{(N,0,0,)}%
}=\frac{T^{(N,2m,q)}}{T^{(N,0,0)}}=\left(  -\frac{1}{M_{2}}\right)
^{2m+q}\left(  \frac{1}{2}\right)  ^{m+q}(2m-1)!! \label{corresp}%
\end{equation}
suggested by the explicit calculation for the mass level $M_{2}^{2}=4$
\cite{bosonic}, one needs the following identity
\begin{gather}
\sum_{j=0}^{2m}(-2m)_{j}\left(  -L-\frac{\tilde{t}^{\prime}}{2}\right)
_{j}\frac{(-2/\tilde{t}^{\prime})^{j}}{j!}\nonumber\\
\qquad{}=0\cdot(-\tilde{t}^{\prime})^{0}\!+0\cdot(-\tilde{t}^{\prime}%
)^{-1}\!+\dots+0\cdot(-\tilde{t}^{\prime})^{-m+1}\!+\frac{(2m)!}{m!}%
(-\tilde{t}^{\prime})^{-m}+\mathit{O}\left\{  \left(  \frac{1}{\tilde
{t}^{\prime}}\right)  ^{m+1}\right\}  \!
\end{gather}
where $L=1-N$ and is an integer. The identity was proved to be valid for any
non-negative integer $m$ and any \textit{real} number $L$ by using technique
of combinatorial number theory \cite{LYAM}. It was remarkable to first predict
\cite{bosonic} the mathematical identity above provided by string theory, and
then a rigorous mathematical proof followed \cite{LYAM}. It was also
interesting to see that the validity of the above identity includes
non-integer values of $L$ which were later shown to be realized by Regge
string scatterings in compact space \cite{HLY}. We thus have shown that the
ratios among HSS amplitudes calculated in Eqs.(\ref{03}) and (\ref{04}) can be
deduced and extracted from Kummer functions \cite{bosonic,KLY1,KLY2}
\begin{equation}
\frac{T^{(N,2m,q)}}{T^{(N,0,0)}}=\lim_{t\rightarrow\infty}\frac{A^{(N,2m,q)}%
}{A^{(N,0,0)}}=\left(  -\frac{1}{2M}\right)  ^{2m+q}2^{2m}\lim_{t\rightarrow
\infty}(-t)^{-m}U\left(  -2m\,,\,\frac{t}{2}+2-2m\,,\,\frac{t}{2}\right)  .
\label{07}%
\end{equation}
All the above calculations so far can be generalized to four classes of
superstring Regge scattering amplitudes \cite{RRsusy}. See the discussion in
chapter XII.

The next interesting issue is to study relations among RSS amplitudes of
different string states. To achieve this, one considers the more general RSS
amplitudes corresponding to three tachyons and \ one leading order high energy
open string states in the RR at each fixed mass level $N=\sum_{n,m,l>0}%
np_{n}+mq_{m}+lr_{l}$%
\begin{equation}
\left\vert p_{n},q_{m},r_{l}\right\rangle =\prod_{n>0}(\alpha_{-n}^{T}%
)^{p_{n}}\prod_{m>0}(\alpha_{-m}^{P})^{q_{m}}\prod_{l>0}(\alpha_{-l}%
^{L})^{r_{l}}|0,k\rangle. \label{RRSS}%
\end{equation}
The $s-t$ channel scattering amplitudes of this state with three other
tachyonic states can be calculated to be%
\begin{align}
A^{(p_{n};q_{m};r_{l})}  &  =\int_{0}^{1}dx\,x^{k_{1}\cdot k_{2}}%
(1-x)^{k_{2}\cdot k_{3}}\cdot\left[  \frac{e^{P}\cdot k_{1}}{x}-\frac
{e^{P}\cdot k_{3}}{1-x}\right]  ^{q_{1}}\left[  \frac{e^{L}\cdot k_{1}}%
{x}+\frac{e^{L}\cdot k_{3}}{1-x}\right]  ^{r_{1}}\nonumber\\
&  \cdot\prod_{n=1}\left[  \frac{(n-1)!e^{T}\cdot k_{3}}{(1-x)^{n}}\right]
^{p_{n}}\prod_{m=2}\left[  \frac{(m-1)!e^{P}\cdot k_{3}}{(1-x)^{m}}\right]
^{q_{m}}\prod_{l=2}\left[  \frac{(l-1)!e^{L}\cdot k_{3}}{(1-x)^{l}}\right]
^{r_{l}}.
\end{align}
Finally, the amplitudes can be written as two equivalent expressions \cite{LY}%
\begin{align}
A^{(p_{n};q_{m};r_{l})}  &  =\prod_{n>0}\left[  \left(  n-1\right)  !\sqrt
{-t}\right]  ^{p_{n}}\cdot\prod_{m>0}\left[  -\left(  m-1\right)
!\frac{\tilde{t}}{2M_{2}}\right]  ^{q_{m}}\cdot\prod_{l>1}\left[  \left(
l-1\right)  !\frac{\tilde{t}^{\prime}}{2M_{2}}\right]  ^{r_{l}}\nonumber\\
&  \quad\cdot B\left(  -\frac{s}{2}-1,-\frac{t}{2}+1\right)  \left(  \frac
{1}{M_{2}}\right)  ^{r_{1}}\nonumber\\
&  \cdot\sum_{i=0}^{q_{1}}\binom{q_{1}}{i}\left(  \frac{2}{\tilde{t}}\right)
^{i}\left(  -\frac{t}{2}-1\right)  _{i}U\left(  -r_{1},\frac{t}{2}%
+2-i-r_{1},\frac{\tilde{t}^{\prime}}{2}\right) \\
&  =\prod_{n>0}\left[  \left(  n-1\right)  !\sqrt{-t}\right]  ^{p_{n}}%
\cdot\prod_{m>1}\left[  -\left(  m-1\right)  !\frac{\tilde{t}}{2M}\right]
^{q_{m}}\cdot\prod_{l>0}\left[  \left(  l-1\right)  !\frac{\tilde{t}^{\prime}%
}{2M}\right]  ^{r_{l}}\nonumber\\
&  \cdot B\left(  -\frac{s}{2}-1,-\frac{t}{2}+1\right)  \left(  -\frac
{1}{M_{2}}\right)  ^{q_{1}}\nonumber\\
&  \cdot\sum_{j=0}^{r_{1}}\binom{r_{1}}{j}\left(  \frac{2}{\tilde{t}^{\prime}%
}\right)  ^{j}\left(  -\frac{t}{2}-1\right)  _{j}U\left(  -q_{1},\frac{t}%
{2}+2-j-q_{1},\frac{\tilde{t}}{2}\right)  . \label{kummer22}%
\end{align}
It is easy to see that, for $q_{1}=0$ or $r_{1}=0$, the RSS amplitudes can be
expressed in terms of only one single Kummer function $U\left(  -r_{1}%
,\frac{t}{2}+2-i-r_{1},\frac{\tilde{t}^{\prime}}{2}\right)  $ or $U\left(
-q_{1},\frac{t}{2}+2-j-q_{1},\frac{\tilde{t}}{2}\right)  $. In general the RSS
amplitudes can be expressed in terms of a finite sum of Kummer functions. One
can then solve these Kummer functions at each mass level and express them in
terms of RSS amplitudes. Recurrence relations of Kummer functions can then be
used to derive recurrence relations among RSS amplitudes \cite{LY}. As an
example at mass level $M^{2}=4$, the recurrence relation%
\begin{equation}
U\left(  -3,\frac{t}{2}-1,\frac{t}{2}-1\right)  +\left(  \frac{t}{2}+1\right)
U(-2,\frac{t}{2}-1,\frac{t}{2}-1)-(\frac{t}{2}-1)U\left(  -2,\frac{t}{2}%
,\frac{t}{2}-1\right)  =0
\end{equation}
leads to the following recurrence relation among Regge string scattering
amplitudes%
\begin{equation}
M\sqrt{-t}A^{PPP}-4A^{PPT}+M\sqrt{-t}A^{PPL}=0. \label{444}%
\end{equation}

In addition, the addition theorem of Kummer function \cite{Slater}%
\begin{equation}
U(a,c,x+y)=\sum_{k=0}^{\infty}\frac{1}{k!}\left(  a\right)  _{k}(-1)^{k}%
y^{k}U(a+k,c+k,x)
\end{equation}
which terminates to a finite sum for a non-positive integer $a$ can be used to
derive inter-mass level recurrence relation of RSS amplitudes. By taking, for
example, $a=-1,c=\frac{t}{2}+1,x=\frac{t}{2}-1$ and $y=1,$ the theorem gives%
\begin{equation}
U\left(  -1,\frac{t}{2}+1,\frac{t}{2}\right)  -U(-1,\frac{t}{2}+1,\frac{t}%
{2}-1)-U\left(  0,\frac{t}{2}+2,\frac{t}{2}-1\right)  =0.
\end{equation}
Note that the last arguments of Kummer functions in the above equation can be
different. It leads to an inter-mass level recurrence relation of RSS
amplitudes \cite{LY}%
\begin{equation}
M(2)(t+6)A_{2}^{TP}-2M(4)^{2}\sqrt{-t}A_{4}^{LP}+2M(4)A_{4}^{LT}=0 \label{rsr}%
\end{equation}
where \ masses $M(2)=\sqrt{2},M(4)=\sqrt{4}=2,$ and $A_{2},A_{4}$ are RSS
amplitudes for mass levels $M^{2}=2,4$ respectively. In deriving
Eq.(\ref{rsr}), it is important to use the fact that the Regge power law
behavior for each RSS amplitude in Eq.(\ref{rsr}) is universal and is mass
level independent \cite{bosonic}.

Finally, Kummer recurrence relations can also be used to explicitly prove
Regge stringy Ward identities or decoupling of ZNS in the RR, but not
vice-versa. Thus in the RR, recurrence relations are more fundamental than
linear relations derived from decoupling of Regge ZNS. However, only Ward
identities derived from the decoupling of Regge ZNS can be generalized to the
string loop amplitudes. As an example, it can be shown that, in the Regge
limit, the decoupling of the scalar type I Regge ZNS \cite{LY}%
\begin{equation}
\lbrack25(\alpha_{-1}^{P})^{3}+9\alpha_{-1}^{P}(\alpha_{-1}^{L})^{2}%
+9\alpha_{-1}^{P}(\alpha_{-1}^{T})^{2}-9\alpha_{-2}^{L}\alpha_{-1}^{L}%
-9\alpha_{-2}^{T}\alpha_{-1}^{T}-75\alpha_{-2}^{P}\alpha_{-1}^{P}%
+50\alpha_{-3}^{P}]\left\vert 0,k\right\rangle \label{ward}%
\end{equation}
can be explicitly demonstrated by using the following recurrence relations of
Kummer functions%
\begin{align}
U(a-1,c,x)-(2a-c+x)U(a,c,x)+a(1+a-c)U(a+1,c,x)  &  =0,\\
U(a,c,x)-aU(a+1,c,x)-U(a,c-1,x)  &  =0,\\
\left(  c-a-1\right)  U(a,c-1,x)-\left(  x+c-1\right)  U\left(  a,c,x\right)
+xU\left(  a,c+1,x\right)   &  =0.
\end{align}
Following the same procedure, one can construct infinite number of recurrence
relations among RSS amplitudes at arbitrary mass levels which, in general, are
independent of Regge stringy Ward identities derived from the decoupling of
Regge ZNS. However, in contrast to Ward identity derived from the decoupling
of Regge ZNS like Eq.(\ref{ward}), we have no proof at loop levels for other
ward identities derived directly from Kummer function recurrence relations.
This is the subtle difference between linear relations obtained in the GR and
the recurrence relations calculated in the RR discussed in this review.
Recurrence relations of higher spin generalization of the BPST vertex
operators \cite{RR6} can also be constructed in this way \cite{Tan}.

Since in general each RSS amplitude was expressed in terms of more than one
Kummer function, it was awkward to derive the complete recurrence relations at
arbitrary higher mass levels. More recently \cite{AppellLY}, it was shown that
each $26D$ open bosonic RSS amplitude can be expressed in terms of one
\textit{single} Appell function $F_{1}$. In fact, the $s-t$ channel RSS
amplitudes with string state in Eq.(\ref{RRSS}) and three tachyons can be
calculated as \cite{AppellLY}%
\begin{align}
A^{(p_{n};q_{m};r_{l})}  &  =\prod_{n=1}\left[  (n-1)!\sqrt{-t}\right]
^{p_{n}}\prod_{m=1}\left[  -(m-1)!\dfrac{\tilde{t}}{2M_{2}}\right]  ^{q_{m}%
}\prod_{l=1}\left[  (l-1)!\dfrac{\tilde{t}^{\prime}}{2M_{2}}\right]  ^{r_{l}%
}\nonumber\\
&  \cdot F_{1}\left(  -\tfrac{t}{2}-1,-q_{1},-r_{1},-\tfrac{s}{2};\dfrac
{s}{\tilde{t}},\dfrac{s}{\tilde{t}^{\prime}}\right)  \cdot B\left(  -\tfrac
{t}{2}-1,-\frac{s}{2}-1\right)  \label{abc}%
\end{align}
where the Appell function $F_{1}$ is one of the four extensions of the
hypergeometric function $_{2}F_{1}$ to two variables and is defined to be%
\begin{equation}
F_{1}\left(  a;b,b^{\prime};c;x,y\right)  =\sum_{m=0}^{\infty}\sum
_{n=0}^{\infty}\dfrac{\left(  a\right)  _{m+n}\left(  b\right)  _{m}\left(
b^{\prime}\right)  _{n}}{m!n!\left(  c\right)  _{m+n}}x^{m}y^{n}%
\end{equation}
where $(a)_{n}=a\cdot\left(  a+1\right)  \cdots\left(  a+n-1\right)  $ is the
rising Pochhammer symbol. Note that when $a$ or $b(b^{\prime})$ is a
non-positive integer, the Appell function truncates to a polynomial. This is
the case for the Appell function in the RSS amplitudes calculated above. It is
important to keep in mind that the expression in Eq.(\ref{abc}) is valid only
when $s$ in the arguments of $F_{1}$ goes to $\infty$.

In contrast to the calculation of a sum of Kummer functions, this result made
it easier to derive the complete infinite recurrence relations among RSS
amplitudes at arbitrary mass levels, which are conjectured to be related to
the known $SL(5,C)$ dynamical symmetry of $F_{1}$ \cite{sl5c}. For example,
the recurrence relation among RSS amplitudes \cite{AppellLY}%
\begin{equation}
\sqrt{-t}\left[  A^{(N;q_{1},r_{1})}+A^{(N;q_{1}-1,r_{1}+1)}\right]
-MA^{(N;q_{1}-1,r_{1})}=0 \label{555}%
\end{equation}
for arbitrary mass levels $M^{2}=2(N-1)$ can be derived from recurrence
relations of the Appell functions. Eq.(\ref{555}) is a generalization of
Eq.(\ref{444}) to arbitrary mass levels. More general recurrence relations can
be obtained similarly. For example, by taking the leading term of $s$ in the
Regge limit, one ends up with the recurrence relation for $b_{2}$%
\begin{align}
cx^{2}F_{1}\left(  a;b_{1},b_{2};c;x,y\right)   & \nonumber\\
+\left[  \left(  a-b_{1}-b_{2}-1\right)  xy^{2}+cx^{2}-2cxy\right]
F_{1}\left(  a;b_{1},b_{2}+1;c;x,y\right)   & \nonumber\\
-\left[  \left(  a+1\right)  x^{2}y-\left(  a-b_{2}-1\right)  xy^{2}%
-cx^{2}+cxy\right]  F_{1}\left(  a;b_{1},b_{2}+2;c;x,y\right)   & \nonumber\\
-\left(  b_{2}+2\right)  x\left(  x-y\right)  yF_{1}\left(  a;b_{1}%
,b_{2}+3;c;x,y\right)   &  =0,
\end{align}
which leads to a recurrence relation for RSS amplitudes at arbitrary mass
levels \cite{AppellLY}%
\begin{align}
\tilde{t}^{\prime2}A^{(N;q_{1},r_{1})}  & \nonumber\\
+\left[  \tilde{t}^{\prime2}+\tilde{t}\left(  t-2\tilde{t}^{\prime}%
-2q_{1}-2r_{1}+4\right)  \right]  \left(  \frac{\frac{\tilde{t}^{\prime}%
}{2M_{2}}}{\sqrt{-t}}\right)  A^{(N;q_{1},r_{1}+1)}  & \nonumber\\
+\left[  \tilde{t}^{\prime2}-\tilde{t}^{\prime}\left(  \tilde{t}+t\right)
+\tilde{t}\left(  t-2r_{1}+4\right)  \right]  \left(  \frac{\frac{\tilde
{t}^{\prime}}{2M_{2}}}{\sqrt{-t}}\right)  ^{2}A^{(N;q_{1},r_{1}+2)}  &
\nonumber\\
-2\left(  r_{1}-2\right)  \left(  \tilde{t}^{\prime}-\tilde{t}\right)  \left(
\frac{\frac{\tilde{t}^{\prime}}{2M_{2}}}{\sqrt{-t}}\right)  ^{3}%
A^{(N;q_{1},r_{1}+3)}  &  =0.
\end{align}
More higher recurrence relations which contain general number of $l\geq3$
Appell functions can be found in \cite{Wang}.

More importantly, one can show \cite{LY,AppellLY} that these recurrence
relations in the Regge limit can be systematically solved so that all RSS
amplitudes can be expressed in terms of one amplitude. All these results seem
to dual to high energy symmetries of fixed angle string scattering amplitudes
discussed in part II
\cite{ChanLee,ChanLee1,ChanLee2,CHLTY1,CHLTY2,CHLTY3,susy}.

We now proceed to show that the recurrence relations of the Appell function
$F_{1}$ \textit{in the Regge limit} can be systematically solved so that all
RSS amplitudes can be expressed in terms of one amplitude. As the first step,
we note that in \cite{LY} the RSS amplitudes was expressed in terms of finite
sum of Kummer functions. There are two equivalent expressions \cite{LY} as was
previously shown in Eq.(\ref{kummer22}). It is easy to see that, for $q_{1}=0$
or $r_{1}=0$, the RSS amplitudes can be expressed in terms of only one single
Kummer function $U\left(  -r_{1},\frac{t}{2}+2-i-r_{1},\frac{\tilde{t}%
^{\prime}}{2}\right)  $ or $U\left(  -q_{1},\frac{t}{2}+2-j-q_{1},\frac
{\tilde{t}}{2}\right)  $, which are thus related to the Appell function
$F_{1}\left(  -\frac{t}{2}-1;0,-r_{1};\frac{s}{2};-\dfrac{s}{\tilde{t}%
},-\dfrac{s}{\tilde{t}^{\prime}}\right)  $ or $F_{1}\left(  -\frac{t}%
{2}-1;-q_{1},0;\frac{s}{2};-\dfrac{s}{\tilde{t}},-\dfrac{s}{\tilde{t}^{\prime
}}\right)  $ respectively%
\begin{align}
\lim_{s\rightarrow\infty}F_{1}\left(  -\frac{t}{2}-1;0,-r_{1};\frac{s}%
{2};-\dfrac{s}{\tilde{t}},-\dfrac{s}{\tilde{t}^{\prime}}\right)   &  =\left(
\frac{2}{\tilde{t}^{\prime}}\right)  ^{r_{1}}U\left(  -r_{1},\frac{t}%
{2}+2-r_{1},\frac{\tilde{t}^{\prime}}{2}\right)  ,\\
\lim_{s\rightarrow\infty}F_{1}\left(  -\frac{t}{2}-1;-q_{1},0;\frac{s}%
{2};-\dfrac{s}{\tilde{t}},-\dfrac{s}{\tilde{t}^{\prime}}\right)   &  =\left(
\frac{2}{\tilde{t}}\right)  ^{q_{1}}U\left(  -q_{1},\frac{t}{2}+2-q_{1}%
,\frac{\tilde{t}}{2}\right)  .
\end{align}

On the other hand, it was shown in \cite{LY} that the Kummer functions ratio%
\begin{equation}
\frac{U(\alpha,\gamma,z)}{U(0,z,z)}=f(\alpha,\gamma,z),\alpha=0,-1,-2,-3,...
\end{equation}
is determined and $f(\alpha,\gamma,z)$ can be calculated by using recurrence
relations of $U(\alpha,\gamma,z)$. Note in addition that $U(0,z,z)=1$ by
explicit calculation. We thus conclude that in the Regge limit%
\begin{equation}
c=\dfrac{s}{2}\rightarrow\infty;x,y\rightarrow\infty;a,b_{1},b_{2}\text{
fixed,}%
\end{equation}
the Appell functions $F_{1}\left(  a;0,b_{2};c;x,y\right)  $ and $F_{1}\left(
a;b_{1},0;c;x,y\right)  $ are determined up to an overall factor by recurrence
relations. The next step is to derive the recurrence relation%
\begin{equation}
yF_{1}\left(  a;b_{1},b_{2};c;x,y\right)  -xF_{1}\left(  a;b_{1}%
+1,b_{2}-1;c;x,y\right)  +\left(  x-y\right)  F_{1}\left(  a;b_{1}%
+1,b_{2};c;x,y\right)  =0, \label{1531}%
\end{equation}
which can be obtained from two of the four Appell recurrence relations among
contiguous functions.

We can now show that in the Regge limit all RSS amplitudes can be expressed in
terms of one single amplitude. We will use the short notation $F_{1}\left(
a;b_{1},b_{2};c;x,y\right)  =F_{1}\left(  b_{1},b_{2}\right)  $ in the
following. For $b_{2}=-1$, by using Eq.(\ref{1531}) and the known
$F_{1}\left(  b_{1},0\right)  $ and $F_{1}\left(  0,b_{2}\right)  $, one can
easily show that $F_{1}\left(  b_{1},-1\right)  $ are determined for all
$b_{1}=-1,-2,-3...$. Similarly, $F_{1}\left(  b_{1},-2\right)  $ are
determined for all $b_{1}=-1,-2,-3...$.if one uses the result of $F_{1}\left(
b_{1},-1\right)  $ in addition to Eq.(\ref{1531}) and the known $F_{1}\left(
b_{1},0\right)  $ and $F_{1}\left(  0,b_{2}\right)  $. This process can be
continued and one ends up with the result that $F_{1}\left(  b_{1}%
,b_{2}\right)  $ are determined for all $b_{1},b_{2}=-1,-2,-3...$. This
completes the proof that the recurrence relations of the Appell function
$F_{1}$ in the Regge limit in Eq.(\ref{abc}) can be systematically solved so
that all RSS amplitudes can be expressed in terms of one amplitude.

In a very recent paper \cite{LLY2}, it was discovered that the 26D open
bosonic string scattering amplitudes (SSA) of three tachyons and one arbitrary
string state can be expressed in terms of the D-type Lauricella functions with
associated $SL(K+3;C)$ symmetry. As a result, SSA and symmetries or relations
among SSA of different string states at various limits calculated previously
can be rederived. These include the linear relations conjectured by Gross
\cite{Gross, Gross1}. and proved in \cite{ChanLee,ChanLee1,ChanLee2, CHL,
CHLTY1,CHLTY2} in the hard scattering limit, the recurrence relations in the
Regge scattering limit derived from Eq.(\ref{abc}) and the extended recurrence
relations in the nonrelativistic scattering limit \cite{LLY1} discovered
recently. Moreover, one can calculate new recurrence relations of SSA which
are valid for all energies. We expect more interesting developments on these
research directions in the near future.

In addition to the high energy string scatterings discussed in this review,
there were other related approaches in the literature discussing higher spin
dynamics of string theory. String theory includes infinitely many higher spin
massive fields with consistent mutual interactions, and can provide useful
hints on the dynamics of higher spin field theory. On the other hand, a better
understanding of higher spin dynamics could also help our comprehension of
string theory. It is widely believed that the tensionless limit of string
\cite{Sagnotti,WS,WS1,WS2,WS3,WS4,WS5} is a theory of higher spin gauge
fields. In flat spacetime a non-trivial field theory dynamics of the
tensionless limit of string theory seems to be ruled out by the theorem of
Coleman and Mandula. However, the assumptions of this theorem are violated by
the presence of a non-trivial cosmological constant, and one may expect a
consistent interacting field theory of higher spins on curved space time. One
of the most important explicit and nontrivial construction of interacting
higher spin gauge theory is Vasiliev' system in AdS space-time.

In \cite{0305052}, the spectrum of Kaluza-Klein descendants of fundamental
string excitations on $AdS_{5}\times S^{5}$ was derived and organized at the
higher spin long multiplets of the AdS supergroup $SU\left(  2,2|4\right)  $
with a rich pattern of shortenings at the higher spin enhancement point.
Furthermore, in the tensionless limit, the field equations from BRST
quantization of string theory provide a direct route toward local field
equations for higher-spin gauge fields \cite{0311257}.

Recently, in \cite{1207.4485}, one parameter families of parity violating
Vasiliev theory were formulated that preserve $N=6$ SUSY in $AdS_{4}$. The
theory was suggested to be dual to the vector model limit of the $N=6$
$U(N)_{k}\times U(M)_{-k}$ ABJ theory in the limit of large $N$ and $k$ but
finite $M$. Since the ABJ theory is also dual to type IIA string theory in
$AdS_{4}\times CP^{3}$ with flat B-field, it was speculated that the Vasiliev
theory must therefore be a limit of this string theory. Roughly speaking, the
fundamental string of string theory is simply the flux tube string of the
non-Abelian bulk Vasiliev theory. The relations between ABJ vector model,
Vasiliev theory, and type IIA string theory suggests a bulk--bulk duality
between Vasiliev theory and type IIA string field theory, which suggests a
concrete way of embedding Vasiliev theory into string theory. It is
interesting to investigate whether---and in what guise---the huge bulk gauge
symmetry of Vasiliev's description survives in the bulk string sigma model
description of the same system.

There existed other approaches of stringy symmetries which include other
studies of string collisions in the high energy, fixed momentum transfer
regime \cite{RR1,RR2,RR3,RR4,RR5,RR6,RR7}, the Hagedorn transition at high
temperature \cite{Hagedorn,Hagedorn1,hep-th/9908001}, vertex operator algebra
for compactified spacetime or on a lattice \cite{Moore,Moore1,CKT}, group
theoretical approach of string \cite{West1,West2}.

Another motivation of studying high energy string scattering is to investigate
the gravitational effect, such as black hole formation due to high energy
string collision, and to understand the nonlocal behavior of string theory.
Nevertheless, in \cite{0705.1816}, it was shown that there is no evidence that
the extendedness of strings produces any long-distance nonlocal effects in
high energy scattering, and no grounds have been found for string effects
interfering with formation of a black hole either.

\part{Stringy symmetries at all energies}

In the first part of this review, we discuss stringy symmetries which were
calculated to be valid for all energies. These include stringy symmetries
calculated by (1) $\sigma$-model approach of string theory in the weak field
approximation, (2) decoupling of ZNS and stringy Ward identities, (3) Witten's
string field theory, (4) Discrete ZNS and $w_{\infty}$ symmetry of $2D$ string
and (5) Soliton ZNS and enhanced stringy gauge symmetries. We will concentrate
on the idea of ZNS and its applications to various calculations of stringy symmetries.

In chapter I we apply ZNS to Sigma model calculation of stringy symmetries
\cite{Lee,Lee-Ov,LeePRL}. We calculate generalized stringy symmetries of
massive background fields \cite{Lee,Lee-Ov}. We discover the existence of
inter-particle, inter-spin symmetry \cite{Lee} for higher spin string
background fields. In addition, we demonstrate the decoupling of degenerate
positive-norm states by using two approaches, the $\sigma$-model calculation
\cite{LeePRL} and Witten's string field theory \cite{KaoLee}. All these
results are consistent with calculations of high energy string scattering
amplitudes which will be discussed in details in part II and part III. In
chapter II, we give a prescription to simplify the calculation of ZNS for
higher mass levels \cite{0302123}. In chapter III, we calculate
\cite{ChungLee1,ChungLee2} a set of $2D$ string ZNS with discrete Polyakov
momenta and show that its operator algebra forms the $w_{\infty}$ symmetry
algebra of $2D$ string theory. Incidentally, In chapter V of part II, the
corresponding high energy ZNS will be shown to form a high energy $w_{\infty}$
symmetry \cite{CHLTY2}. These results strongly suggest that ZNS are symmetry
charges of $26D$ string theory. In chapter IV we calculate soliton ZNS in
compact spaces for both closed \cite{Lee1} and open string \cite{Lee2}
theories and study their relations to enhanced stringy gauge symmetries.%

\setcounter{equation}{0}
\renewcommand{\theequation}{\arabic{section}.\arabic{equation}}%

\section{Zero norm states (ZNS) and Sigma model calculation of stringy
symmetries}

In the first chapter, we review the calculations of string symmetries from ZNS
without taking the high energy limit. In the OCFQ spectrum of $26D$ open
bosonic string theory, the solutions of physical state conditions include
positive-norm propagating states and two types of ZNS. The latter are
\cite{GSW}%
\begin{equation}
\text{Type I}:L_{-1}\left\vert x\right\rangle ,\text{ where }L_{1}\left\vert
x\right\rangle =L_{2}\left\vert x\right\rangle =0,\text{ }L_{0}\left\vert
x\right\rangle =0; \label{1.1}%
\end{equation}%
\begin{equation}
\text{Type II}:(L_{-2}+\frac{3}{2}L_{-1}^{2})\left\vert \widetilde{x}%
\right\rangle ,\text{ where }L_{1}\left\vert \widetilde{x}\right\rangle
=L_{2}\left\vert \widetilde{x}\right\rangle =0,\text{ }(L_{0}+1)\left\vert
\widetilde{x}\right\rangle =0. \label{1.2}%
\end{equation}
While type I states have zero-norm at any spacetime dimension, type II states
have zero-norm \emph{only} at $D=26$. It can be shown \cite{GSW} that the
string spectrum is ghost-free provided that $D=26$ and the Regge intercept
$a=1$ or $D\leq25$ and $a\leq1.$ However, there are far more ZNS for the
former case than that of the latter case. Thus the choice of $D=26$ case is
closely related to the existence of type II ZNS which is crucial in the
discussion of this paper. Eqs.(\ref{1.1}) and Eqs.(\ref{1.2}) will be
extensively used in the review. Some explicit solutions of ZNS can be found in
\cite{Lee,0302123} and will be discussed in chapter II.

In the $\sigma$-model approach of string theory, one turns on background
fields on the worldsheet energy momentum tensor $T$. Conformal invariance of
the worldsheet then requires, in addition to $D=26$, cancellation of various
$q$-number anomalies and results to equations of motion of the background
fields. A spacetime effective action can then be constructed and used to
reproduce string scattering amplitudes. This was a powerful method to study
dynamics of the string modes \cite{GSW}. On the other hand, it was suggested
that a spacetime symmetry transformation $\delta\Phi$ for a background field
$\Phi$ can be generated by a \textit{worldsheet} generator $h$ \cite{EO}
\begin{equation}
T_{\Phi}+i[h,T_{\Phi}]=T_{\Phi+\delta\Phi} \label{1.3}%
\end{equation}
where $T_{\Phi}$ is the worldsheet energy momentum tensor with background
fields $\Phi$ and $T_{\Phi+\delta\Phi}$ is the new energy momentum tensor with
new background fields $\Phi+\delta\Phi$. However, there was no systematic
prescription to calculate the worldsheet generator $h.$

It was then shown that \cite{Lee} for each \textit{spacetime} ZNS, one can
systematically construct a $\delta T_{\Phi}$ such that
\begin{equation}
T_{\Phi}+\delta T_{\Phi}=T_{\Phi+\delta\Phi} \label{1.33}%
\end{equation}
was satisfied to some order of weak field approximation in the background
fields $\beta$ function calculation. It turned out that Eqs.(\ref{1.33}) gave
the complete symmetry transformations for string modes while Eqs.(\ref{1.3})
did not. Indeed, there were many symmetry transformations which can not be
generated by a worldsheet generator $h$. One important example was the
inter-particle symmetry transformation Eqs.(\ref{01}) generated by the $D_{2}$
\textit{type II} ZNS in Eqs.(\ref{02}). In contrast to the usual $\sigma
$-model loop expansion (or $\alpha^{\prime}$ expansion) of the string
background field calculation\footnote{See section 3.4 of \cite{GSW} and
references there.}, which was nonrenormalizable for the massive background
fields, it turned out that weak field approximation was the more convenient
expansion to deal with massive background fields.

\subsection{Stringy symmetries of massive background fields}

In this section, as illustrations, we calculate stringy symmetries for $26D$
open bosonic string up to mass levels $M^{2}=4.$ All physical states including
positive-norm propagating states and two types of ZNS can be found in chapter
II. It was demonstrated in the first order weak field approximation of the
string modes that for each ZNS in the OCFQ $26D$ open bosonic string spectrum,
there corresponds an on-shell gauge transformation for the positive-norm
background field $(\alpha^{\prime}\equiv\frac{1}{2})$\cite{Lee,Lee-Ov} :%

\begin{subequations}
\begin{align}
M^{2}=0:\qquad &  \delta A_{\mu}=\partial_{\mu}\theta;\label{3a}\\
&  \partial^{2}\theta=0. \label{3b}%
\end{align}

\end{subequations}
\begin{subequations}
\begin{align}
M^{2}=2:\qquad &  \delta B_{\mu\nu}=\partial_{(\mu}\theta_{\nu)};\label{4a}\\
&  \partial^{\mu}\theta_{\mu}=0,(\partial^{2}-2)\theta_{\mu}=0. \label{4b}%
\end{align}

\end{subequations}
\begin{subequations}
\begin{align}
&  \delta B_{\mu\nu}=\frac{3}{2}\partial_{\mu}\partial_{\nu}\theta-\frac{1}%
{2}\eta_{\mu\nu}\theta;\label{5a}\\
&  (\partial^{2}-2)\theta=0. \label{5b}%
\end{align}

\end{subequations}
\begin{subequations}
\begin{align}
M^{2}=4:\qquad &  \delta C_{\mu\nu\lambda}=\partial_{(\mu}\theta_{\nu\lambda
)};\label{6a}\\
&  \partial^{\mu}\theta_{\mu\nu}=\theta_{\mu}^{\;\mu}=0,(\partial^{2}%
-4)\theta_{\mu\nu}=0. \label{6b}%
\end{align}%
\end{subequations}
\begin{subequations}
\begin{align}
&  \delta C_{(\mu\nu\lambda)}=\frac{5}{2}\partial_{(\mu}\partial_{\nu}%
\theta_{\lambda)}^{1}-\eta_{(\mu\nu}\theta_{\lambda)}^{1};\label{7a}\\
&  \partial^{\mu}\theta_{\mu}^{1}=0,(\partial^{2}-4)\theta_{\mu}^{1}=0.
\label{7b}%
\end{align}%
\end{subequations}
\begin{subequations}
\begin{align}
&  \delta C_{(\mu\nu\lambda)}=\frac{3}{5}\partial_{\mu}\partial_{\nu}%
\partial_{\lambda}\theta-\frac{1}{5}\eta_{(\mu\nu}\partial_{\lambda)}%
\theta;\label{9a}\\
&  (\partial^{2}-4)\theta=0. \label{9b}%
\end{align}
\qquad\qquad\ \ 

In the above equations, $A$, $B$, $C$ are positive-norm background fields,
$\theta s$ represent zero-norm background fields, and $\partial^{2}%
\equiv\partial^{\mu}\partial_{\mu}$. There are on-mass-shell, gauge and
traceless conditions on the transformation parameters $\theta s$, which will
correspond to BRST ghost fields in a one-to-one manner in WSFT \cite{KaoLee}.
This will be discussed in section I.D. Eqs.(\ref{3a}), (\ref{3b}) is easily
identified to be the on-shell gauge transformation of photon. Note that, for
example, Eqs.(\ref{5a}), (\ref{5b}) is the residual on-shell gauge
transformation induced by a type II ZNS at mass level $M^{2}=2$.

Similar massive stringy symmetry transformations can be constructed for
superstring. In particular, based on ZNS calculations, an infinite number of
Heterotic massive symmetry transformations \cite{Lee4} with parameters
$\theta_{\mu}^{(ab)}$, $\theta_{\lbrack\mu\nu]}^{(ab)}$ etc. containing both
Einstein and $E_{8}\otimes E_{8}$ (or $SO(32)$) Yang-Mills indices can be
constructed in the $10D$ Heterotic string theories \cite{Heter}.

\subsection{Inter-particle stringy symmetries}

It is interesting to see that an inter-particle symmetry transformation for
two high spin states at mass level $M^{2}=4$ can be generated \cite{Lee}%

\end{subequations}
\begin{equation}
\delta C_{(\mu\nu\lambda)}=(\frac{1}{2}\partial_{(\mu}\partial_{\nu}%
\theta_{\lambda)}^{2}-2\eta_{(\mu\nu}\theta_{\lambda)}^{2}),\delta C_{[\mu
\nu]}=9\partial_{\lbrack\mu}\theta_{\nu]}^{2} \label{8a}%
\end{equation}
where $\partial^{\mu}\theta_{\mu}^{2}=0,(\partial^{2}-4)\theta_{\mu}^{2}=0$
which are the on-shell conditions of the mixed type I and type II $D_{2}$
vector ZNS%

\begin{equation}
|D_{2}\rangle=[(\frac{1}{2}k_{\mu}k_{\nu}\theta_{\lambda}+2\eta_{\mu\nu}%
\theta_{\lambda}^{2})\alpha_{-1}^{\mu}\alpha_{-1}^{\nu}\alpha_{-1}^{\lambda
}+9k_{\mu}\theta_{\nu}^{2}\alpha_{-2}^{[\mu}\alpha_{-1}^{\nu]}-6\theta_{\mu
}^{2}\alpha_{-3}^{\mu}]\left\vert 0,k\right\rangle ,\text{ \ }k\cdot\theta
^{2}=0, \label{8b}%
\end{equation}
and $C_{(\mu\nu\lambda)}$ and $C_{[\mu\nu]}$ are the background fields of the
symmetric spin-three and antisymmetric spin-two states respectively at the
mass level $M^{2}=4$. It is important to note that the decoupling of the
$D_{2}$ vector ZNS, or unitarity of the theory, implies simultaneous change of
both $C_{(\mu\nu\lambda)}$ and $C_{[\mu\nu]}$ , thus they form a gauge
multiplet. This is a generic feature for background fields of higher massive
levels in the $\sigma$-model calculation of string theory. One might want to
generalize the calculation to the second order weak background fields to see
the inter-mass level symmetry. This however suffers from the so-called
non-perturbative non-renormalizability of $2d$ $\sigma$-model and one is
forced to introduce infinite number of counter-terms to preserve the
worldsheet conformal invariance \cite{Das,Das1}.

Note that $\theta_{\mu}^{2}$ in Eqs.(\ref{8a}), (\ref{8b}) are some linear
combination of the original type I and type II vector ZNS calculated by
Eqs.(\ref{1.1}), (\ref{1.2}). This inter-particle stringy symmetry is
consistent with the linear relations among high energy, fixed angle scattering
amplitudes of $C_{(\mu\nu\lambda)}$ and $C_{[\mu\nu]}$, which will be
discussed in details in part II of the review.

\subsection{Decoupling of degenerate positive-norm states}

In the even higher mass levels, $M^{2}=6$ for example, a new phenomenon begins
to show up. Indeed, there are ambiguities in defining positive-norm spin-two
and scalar states due to the existence of ZNS in the same Young
representations \cite{LeePRL}. As a result, the degenerate spin two and scalar
positive-norm states can be gauged to the higher rank fields $D_{\mu\nu
\alpha\beta}$ and mixed-symmetric $D_{\mu\nu\alpha}$ in the first order weak
field approximation. Instead of calculating the stringy gauge symmetry at
level $M^{2}=6$, we will only concentrate on the equation of motions. Take the
energy-momentum tensor on the worldsheet boundary in the first order weak
field approximation to be of the following form.%

\begin{equation}%
\begin{split}
T(\tau)=  &  -\frac{1}{2}\eta_{\mu\nu}\partial_{\tau}X^{\mu}\partial_{\tau
}X^{\nu}+D_{\mu\nu\alpha\beta}\partial_{\tau}X^{\mu}\partial_{\tau}X^{\nu
}\partial_{\tau}X^{\alpha}\partial_{\tau}X^{\beta}+D_{\mu\nu\alpha}%
\partial_{\tau}X^{\mu}\partial_{\tau}X^{\nu}\partial_{\tau}^{2}X^{\alpha}\\
&  +D_{\mu\nu}^{0}\partial_{\tau}^{2}X^{\mu}\partial_{\tau}^{2}X^{\nu}%
+D_{\mu\nu}^{1}\partial_{\tau}X^{\mu}\partial_{\tau}^{3}X^{\nu}+D_{\mu
}\partial_{\tau}^{4}X^{\mu},
\end{split}
\label{BB}%
\end{equation}
where $\tau$ is the worldsheet time, $X\equiv X(\tau)$. This is the most
general worldsheet coupling in the generalized $\sigma$-model approach
consistent with vertex operator consideration \cite{Weinberg,Sasaki}. The
conditions to cancel all $q$-number worldsheet conformal anomalous terms
correspond to cancelling all kinds of loop divergences \cite{Labas} up to the
four loop order in the $2d$ conformal field theory. It is easier to use
$T\cdot T$ operator-product calculation and the conditions read \cite{LeePRL}%

\begin{subequations}
\begin{align}
&  2\partial^{\mu}D_{\mu\nu\alpha\beta}-D_{(\nu\alpha\beta)}=0,\label{11a}\\
&  \partial^{\mu}D_{\mu\nu\alpha}-2D_{\nu\alpha}^{0}-3D_{\nu\alpha}%
^{1}=0,\label{11b}\\
&  \partial^{\mu}D_{\mu\nu}^{1}-12D_{\nu}=0,\label{11c}\\
&  3D_{\;\mu\nu\alpha}^{\mu}+\partial^{\mu}D_{\nu\alpha\mu}-3D_{(\nu\alpha
)}^{1}=0,\label{11d}\\
&  D_{\;\mu\nu}^{\mu}+4\partial^{\mu}D_{\mu\nu}^{0}-24D_{\nu}=0,\label{11e}\\
&  2D_{\mu\nu}^{\;\;\nu}+3\partial^{\nu}D_{\mu\nu}^{1}-12D_{\mu}%
=0,\label{11f}\\
&  2D_{\mu}^{0\;\mu}+3D_{\mu}^{1\;\mu}+12\partial^{\mu}D_{\mu}=0,\label{11g}\\
&  (\partial^{2}-6)\phi=0. \label{11h}%
\end{align}
Here, $\phi$ represents all background fields introduced in Eqs.(\ref{BB}). It
is now clear through Eq.(\ref{11b}) and Eq.(\ref{11d}) that both $D_{\mu\nu
}^{0}$ and $D_{(\mu\nu)}^{1}$ can be expressed in terms of $D_{\mu\nu
\alpha\beta}$ and $D_{\mu\nu\alpha}$ . $D_{[\mu\nu]}^{1}$ can be expressed in
terms of $D_{\mu\nu\alpha\beta}$ and $D_{\mu\nu\alpha}$ by Eq.(\ref{11b}%
).Eq.(\ref{11a}) and Eq.(\ref{11c}) imply that $D_{(\mu\nu\alpha)}$ and
$D_{\mu}$ can also be expressed in terms of $D_{\mu\nu\alpha\beta}$ and
mixed-symmetric $D_{\mu\nu\alpha}$ . Finally Eqs.(\ref{11e}) to (\ref{11g})
are the gauge conditions for $D_{\mu\nu\alpha\beta}$ and mixed-symmetric
$D_{\mu\nu\alpha}$ after substituting $D_{\mu\nu}^{0}$, $D_{\mu\nu}^{1}$ and
$D_{\mu}$ in terms of $D_{\mu\nu\alpha\beta}$ and mixed symmetric $D_{\mu
\nu\alpha}$ . The remaining scalar particle has automatically been gauged to
higher rank fields since Eq.(\ref{BB}) is already the most general form of
background-field coupling. This means that the degenerate spin two and scalar
positive-norm states can be gauged to the higher rank fields $D_{\mu\nu
\alpha\beta}$ and mixed-symmetric $D_{\mu\nu\alpha}$ in the first order weak
field approximation.

In fact, for instance, it can be explicitly shown \cite{Lee3} that the
scattering amplitude involving the positive-norm spin-two state can be
expressed in terms of those of spin-four and mixed-symmetric spin-three states
due to the existence of a \textit{degenerate} type I and a type II spin-two
ZNS. Although all the four-point amplitudes considered in Ref. \cite{Lee3}
contain three tachyons, the argument can be easily generalized to more general
amplitudes. This is very different from the analysis of lower massive levels
where all positive-norm states seem to have independent scattering amplitudes.

Presumably, this decoupling phenomenon comes from the ambiguity in defining
positive-norm states due to the existence of ZNS in the same Young
representations. We will justify this decoupling by WSFT in the next section.
Finally one expects this decoupling to persist even if one includes the higher
order corrections in weak field approximation, as there will be even stronger
relations between background fields order by order through iteration.

\subsection{Witten's string field theory (WSFT) calculations}

It would be much more convincing if one can rederive the stringy phenomena
discussed in the previous sections from WSFT. Not only can one compare the
first quantized string with the second quantized string, but also the old
covariant quantized string with the BRST quantized string. Although the
calculation is lengthy, the result, as we shall see, are still controllable by
utilizing the results from first quantized approach in previous sections.

There exist important consistency checks of first quantized string results
from WSFT in the literature, e.g. the rederivation of Veneziano and
Kubo-Nielson amplitudes from WSFT \cite{Giddings}. In some stringy cases,
calculations can only be done in string field theory approach. For example,
the pp-wave string amplitudes can only be calculated in the light-cone string
field theory \cite{Spradlin}. Therefore, a consistent check by both first and
second quantized approaches of any reliable string results would be of great importance.

The infinitesimal gauge transformation of WSFT is%

\end{subequations}
\begin{equation}
\delta\Phi=Q_{B}\Lambda+g_{0}(\Phi\ast\Lambda-\Lambda\ast\Phi). \label{W12}%
\end{equation}
To compare with our first quantized results in previous sections, we only need
to calculate the first term on the right hand side of Eq.(\ref{W12}). Up to
the second massive level, $\Phi$ and $\Lambda$ can be expressed as%

\begin{equation}%
\begin{split}
\Phi= \bigg\{  &  \phi(x)+iA_{\mu}(x)\alpha^{\mu}_{-1}+\alpha(x)b_{-1}%
c_{0}-B_{\mu\nu}(x)\alpha^{\mu}_{-1}\alpha^{\nu}_{-1}+iB_{\mu}(x)\alpha^{\mu
}_{-2}\\
&  +i\beta_{\mu}(x)\alpha^{\mu}_{-1}b_{-1}c_{0}+\beta^{0}(x)b_{-2}c_{0}%
+\beta^{1}(x)b_{-1}c_{-1}\\
&  -iC_{\mu\nu\lambda}(x)\alpha^{\mu}_{-1}\alpha^{\nu}_{-1}\alpha^{\lambda
}_{-1}-C_{\mu\nu}(x)\alpha^{\mu}_{-2}\alpha^{\nu}_{-1}+iC_{\mu}(x)\alpha^{\mu
}_{-3}\\
&  -\gamma_{\mu\nu}(x)\alpha^{\mu}_{-1}\alpha^{\nu}_{-1}b_{-1}c_{0}%
+i\gamma_{\mu}^{0}(x)\alpha^{\mu}_{-1}b_{-2}c_{0}+i\gamma_{\mu}^{1}%
(x)\alpha^{\mu}_{-1}b_{-1}c_{-1}+i\gamma_{\mu}^{2}(x)\alpha^{\mu}_{-2}%
b_{-1}c_{0}\\
&  +\gamma^{0}(x)b_{-3}c_{0}+\gamma^{1}(x)b_{-2}c_{-1}+\gamma^{2}%
(x)b_{-1}c_{-2} \bigg\}c_{1}\left|  k\right\rangle ,
\end{split}
\end{equation}

\begin{equation}%
\begin{split}
\Lambda=\bigg\{  &  \epsilon^{0}(x)b_{-1}-\epsilon_{\mu\nu}^{0}(x)\alpha
_{-1}^{\mu}\alpha_{-1}^{\nu}b_{-1}+i\epsilon_{\mu}^{0}(x)\alpha_{-1}^{\mu
}b_{-1}+i\epsilon_{\mu}^{1}(x)\alpha_{-2}^{\mu}b_{-1}+i\epsilon_{\mu}%
^{2}(x)\alpha_{-1}^{\mu}b_{-2}\\
&  +\epsilon^{1}(x)b_{-2}+\epsilon^{2}(x)b_{-3}+\epsilon^{3}(x)b_{-1}%
b_{-2}c_{0}\bigg\}\left\vert \Omega\right\rangle
\end{split}
\end{equation}
where $\Phi$ and $\Lambda$ are restricted to ghost number $1$ and $0$
respectively, and the BRST charge is
\begin{equation}
Q_{B}=\sum\limits_{n=-\infty}^{\infty}L_{-n}^{matt}c_{n}+\sum
\limits_{m,n=-\infty}^{\infty}\frac{m-n}{2}:c_{m}c_{n}b_{-m-n}:-c_{0}.
\label{15.}%
\end{equation}
The transformation one gets for each mass level are
\begin{subequations}%
\begin{align}
M^{2}=0,\qquad &  \delta A_{\mu}=\partial_{\mu}\epsilon^{0},\label{16a}\\
&  \delta\alpha=\frac{1}{2}\partial^{2}\epsilon^{0}; \label{16b}%
\end{align}
\end{subequations}%
%

\begin{subequations}%
\begin{align}
M^{2}=2,\qquad &  \delta B_{\mu\nu}=-\partial_{(\mu}\epsilon_{\nu)}^{0}%
-\frac{1}{2}\epsilon^{1}\eta_{\mu\nu},\label{17a}\\
&  \delta B_{\mu}=-\partial_{\mu}\epsilon^{1}+\epsilon_{\mu}^{0},\label{17b}\\
&  \delta\beta_{\mu}=\frac{1}{2}(\partial^{2}-2)\epsilon_{\mu}^{0}%
,\label{17c}\\
&  \delta\beta^{0}=\frac{1}{2}(\partial^{2}-2)\epsilon_{\mu}^{1},\label{17d}\\
&  \delta\beta^{1}=-\partial^{\mu}\epsilon_{\mu}^{0}-3\epsilon^{1};
\label{17e}%
\end{align}
\end{subequations}%
%

\begin{subequations}%
\begin{align}
M^{2}=4,\qquad &  \delta C_{\mu\nu\lambda}=-\partial_{(\mu}\epsilon
_{\nu\lambda)}^{0}-\frac{1}{2}\epsilon_{(\mu}^{2}\eta_{\nu\lambda
)},\label{18a}\\
&  \delta C_{[\mu\nu]}=-\partial_{\lbrack\nu}\epsilon_{\mu]}^{1}%
-\partial_{\lbrack\mu}\epsilon_{\nu]}^{2},\label{18b}\\
&  \delta C_{(\mu\nu)}=-\partial_{(\nu}\epsilon_{\mu)}^{1}-\partial_{(\mu
}\epsilon_{\nu)}^{2}+2\epsilon_{\mu\nu}^{0}-\epsilon^{2}\eta_{\mu\nu
},\label{18c}\\
&  \delta C_{\mu}=-\partial_{\mu}\epsilon^{2}+2\epsilon_{\mu}^{1}%
+\epsilon_{\mu}^{2},\label{18d}\\
&  \delta\gamma_{\mu\nu}=\frac{1}{2}(\partial^{2}-4)\epsilon_{\mu\nu}%
^{0}-\frac{1}{2}\epsilon^{3}\eta_{\mu\nu},\label{18e}\\
&  \delta\gamma_{\mu}^{0}=\frac{1}{2}(\partial^{2}-4)\epsilon_{\mu}%
^{2}+\partial_{\mu}\epsilon^{3},\label{18f}\\
&  \delta\gamma_{\mu}^{1}=-2\partial^{\nu}\epsilon_{\nu\mu}^{0}-2\epsilon
_{\mu}^{1}-3\epsilon_{\mu}^{2},\label{18g}\\
&  \delta\gamma_{\mu}^{2}=\frac{1}{2}(\partial^{2}-4)\epsilon_{\mu}%
^{1}-\partial_{\mu}\epsilon^{3},\label{18h}\\
&  \delta\gamma^{0}=\frac{1}{2}(\partial^{2}-4)\epsilon^{2}-\epsilon
^{3},\label{18i}\\
&  \delta\gamma^{1}=-\partial^{\mu}\epsilon_{\mu}^{2}-4\epsilon^{2}%
-2\epsilon^{3},\label{18j}\\
&  \delta\gamma^{2}=-2\partial^{\mu}\epsilon_{\mu}^{1}-5\epsilon^{2}%
+4\epsilon^{3}+\epsilon_{\mu}^{0\;\mu}. \label{18k}%
\end{align}
\end{subequations}%

It is interesting to note that Eq.(\ref{16b}) corresponds to the lifting of
on-mass-shell condition in eqs Eq.(\ref{3b}). Meanwhile Eq.(\ref{17c}) and
Eq.(\ref{17d}) correspond to on-mass-shell condition in Eq.(\ref{5b}) and
Eq.(\ref{4b}); Eq.(\ref{17e}) corresponds to the gauge condition in
Eq.(\ref{4b}). Similar correspondence applies to level $M^{2}=4$.
Eq.(\ref{18e}), Eq.(\ref{18f}), Eq.(\ref{18h}) and Eq.(\ref{18i}) correspond
to on-mass-shell conditions in Eq.(\ref{6b}), Eq.(\ref{7b}), Eq.(\ref{8b}) and
Eq.(\ref{9b}). Eq.(\ref{18g}), Eq.(\ref{18j}) and Eq.(\ref{18k}) correspond to
gauge conditions in Eq.(\ref{6b}), Eq.(\ref{7b}) and Eq.(\ref{8b}). The
traceless condition in Eq.(\ref{6b}) corresponds to the trace part of
Eq.(\ref{18e}). Also, only ZNS transformation parameters appear on the r.h.s.
of matter transformation $A$,$B$,$C$, and all ghost transformations
correspond, in a one-to one manner, to the lifting of on-shell conditions
(including on-mass-shell, gauge and traceless conditions) in the OCFQ approach.

These important observations simplify the demonstration of decoupling of
degenerate positive-norm states at higher mass levels, $M^{2}=6$ and $M^{2}=8$
more specifically in WSFT. We will present the calculation for level $M^{2}%
=6$. The calculation for $M^{2}=8$ was discussed in \cite{KaoLee}. For
$M^{2}=4$, it can be checked that only $C_{\mu\nu\lambda}$ and $C_{[\mu\nu]}$
are dynamically independent and they form a gauge multiplet, which is
consistent with result of first quantized calculation presented in the
previous sections .

We now show the decoupling phenomenon for the third massive level $M^{2}=6$,
in which $\Phi$ and $\Lambda$ can be expanded as%

\begin{equation}%
\begin{split}
\Phi_{4}=\bigg\{  &  D_{\mu\nu\alpha\beta}(x)\alpha_{-1}^{\mu}\alpha_{-1}%
^{\nu}\alpha_{-1}^{\alpha}\alpha_{-1}^{\beta}-iD_{\mu\nu\alpha}(x)\alpha
_{-1}^{\mu}\alpha_{-1}^{\nu}\alpha_{-2}^{\alpha}-D_{\mu\nu}^{0}(x)\alpha
_{-2}^{\mu}\alpha_{-2}^{\nu}-D_{\mu\nu}^{1}(x)\alpha_{-1}^{\mu}\alpha
_{-3}^{\nu}\\
&  +iD_{\mu}(x)\alpha_{-4}^{\mu}-i\xi_{\mu\nu\alpha}(x)\alpha_{-1}^{\mu}%
\alpha_{-1}^{\nu}\alpha_{-1}^{\alpha}b_{-1}c_{0}-\xi_{\mu\nu}^{0}%
(x)\alpha_{-2}^{\mu}\alpha_{-1}^{\nu}b_{-1}c_{0}-\xi_{\mu\nu}^{1}%
(x)\alpha_{-1}^{\mu}\alpha_{-1}^{\nu}b_{-2}c_{0}\\
&  -\xi_{\mu\nu}^{2}(x)\alpha_{-1}^{\mu}\alpha_{-1}^{\nu}b_{-1}c_{-1}%
+i\xi_{\mu}^{0}(x)\alpha_{-3}^{\mu}b_{-1}c_{0}+i\xi_{\mu}^{1}(x)\alpha
_{-2}^{\mu}b_{-2}c_{0}+i\xi_{\mu}^{2}(x)\alpha_{-1}^{\mu}b_{-3}c_{0}\\
&  +i\xi_{\mu}^{3}(x)\alpha_{-2}^{\mu}b_{-1}c_{-1}+i\xi_{\mu}^{4}%
(x)\alpha_{-1}^{\mu}b_{-2}c_{-1}+i\xi_{\mu}^{5}(x)\alpha_{-1}^{\mu}%
b_{-1}c_{-2}+\xi^{0}(x)b_{-4}c_{0}+\xi^{1}(x)b_{-3}c_{-1}\\
&  +\xi^{2}(x)b_{-2}c_{-2}+\xi^{3}(x)b_{-1}c_{-3}+\xi^{4}(x)b_{-2}b_{-1}%
c_{-1}c_{0}\bigg\}c_{1}\left\vert k\right\rangle ,
\end{split}
\label{19.}%
\end{equation}

\begin{equation}%
\begin{split}
\Lambda_{4}=\bigg\{  &  -i\epsilon_{\mu\nu\alpha}^{0}(x)\alpha_{-1}^{\mu
}\alpha_{-1}^{\nu}\alpha_{-1}^{\alpha}b_{-1}-\epsilon_{\mu\nu}^{1}%
(x)\alpha_{-2}^{\mu}\alpha_{-1}^{\nu}b_{-1}-\epsilon_{\mu\nu}^{2}%
(x)\alpha_{-1}^{\mu}\alpha_{-1}^{\nu}b_{-2}+i\epsilon_{\mu}^{3}(x)\alpha
_{-3}^{\mu}b_{-1}\\
&  +i\epsilon_{\mu}^{4}(x)\alpha_{-2}^{\mu}b_{-2}+i\epsilon_{\mu}^{5}%
(x)\alpha_{-1}^{\mu}b_{-3}+i\epsilon_{\mu}^{6}(x)\alpha_{-1}^{\mu}b_{-2}%
b_{-1}c_{0}+\epsilon^{4}(x)b_{-4}\\
&  +\epsilon^{5}(x)b_{-3}b_{-1}c_{0}+\epsilon^{6}(x)b_{-2}b_{-1}%
c_{-1}\bigg\}\left\vert \Omega\right\rangle ,
\end{split}
\label{20.}%
\end{equation}
The transformations for the matter part are%

\begin{subequations}%
\begin{align}
&  \delta D_{\mu\nu\alpha\beta}=-\partial_{(\beta}\epsilon_{\mu\nu\alpha)}%
^{0}-\frac{1}{2}\epsilon_{(\mu\nu}^{2}\eta_{\alpha\beta)},\label{21a}\\
&  \delta D_{\mu\nu\alpha}=-\partial_{(\mu}\epsilon_{|\alpha|\nu)}%
^{1}-\partial_{\alpha}\epsilon_{\nu\mu}^{2}+3\epsilon_{\mu\nu\alpha}^{0}%
-\frac{1}{2}\epsilon_{\alpha}^{4}\eta_{\nu\mu}-\epsilon_{(\mu}^{5}\eta
_{\nu)\alpha},\label{21b}\\
&  \delta D_{[\mu\nu]}^{1}=-\partial_{\lbrack\mu}\epsilon_{\nu]}^{3}%
-\partial_{\lbrack\nu}\epsilon_{\mu]}^{5}+2\epsilon_{\lbrack\nu\mu]}%
^{1},\label{21c}\\
&  \delta D_{(\mu\nu)}^{1}=-\partial_{(\mu}\epsilon_{\nu)}^{3}-\partial_{(\nu
}\epsilon_{\mu)}^{5}+2\epsilon_{(\nu\mu)}^{1}+2\epsilon_{\mu\nu}^{2}%
-\epsilon^{4}\eta_{\mu\nu},\label{21d}\\
&  \delta D_{\mu\nu}^{0}=-\partial_{(\mu}\epsilon_{\nu)}^{4}+\epsilon_{(\nu
\mu)}^{1}-\frac{1}{2}\epsilon^{4}\eta_{\mu\nu},\label{21e}\\
&  \delta D_{\mu}=-\partial_{\mu}\epsilon^{4}+3\epsilon_{\mu}^{3}%
+2\epsilon_{\mu}^{4}+\epsilon_{\mu}^{5}. \label{21f}%
\end{align}
\end{subequations}%

It can be checked from the above equations that only $D_{\mu\nu\alpha\beta}$
and mixed-symmetric $D_{\mu\nu\alpha}$ cannot be gauged away, which is
consistent with the result of the first quantized approach in the previous
sections. That is , the spin-two and scalar positive-norm physical propagating
modes can be gauged to $D_{\mu\nu\alpha\beta}$ and mixed symmetric $D_{\mu
\nu\alpha}$ . In fact, $D_{\mu\nu\alpha}$ , $D_{[\mu\nu]}^{1}$ , $D_{(\mu\nu
)}^{1}$, $D_{\mu\nu}^{0}$ and $D_{\mu}$ can be gauged away by $\epsilon
_{\mu\nu\alpha}^{0}$ , $\epsilon_{\lbrack\mu\nu]}^{1}$ , $\epsilon_{(\mu\nu
)}^{1}$ , $\epsilon_{\mu\nu}^{2}$ and one of the vector parameters, say
$\epsilon_{\mu}^{3}$ . The rest, $\epsilon_{\mu}^{4}$ , $\epsilon_{\mu}^{5}$
and $\epsilon^{4}$ are gauge artifacts of $D_{\mu\nu\alpha\beta}$ and
mixed-symmetric $D_{\mu\nu\alpha}$.

The transformation for the ghost part are%

\begin{subequations}%
\begin{align}
&  \delta\xi_{\mu\nu\alpha}=\frac{1}{2}(\partial^{2}-6)\epsilon_{\mu\nu\alpha
}^{0}-\frac{1}{2}\epsilon_{(\mu}^{6}\eta_{\nu\alpha)},\label{22a}\\
&  \delta\xi_{\lbrack\mu\nu]}^{0}=\frac{1}{2}(\partial^{2}-6)\epsilon
_{\lbrack\mu\nu]}^{1}-\partial_{\lbrack\mu}\epsilon_{\nu]}^{6},\label{22b}\\
&  \delta\xi_{(\mu\nu)}^{0}=\frac{1}{2}(\partial^{2}-6)\epsilon_{(\mu\nu)}%
^{1}-\partial_{(\mu}\epsilon_{\mu)}^{6}+\epsilon^{5}\eta_{\mu\nu}%
,\label{22c}\\
&  \delta\xi_{\mu\nu}^{1}=\frac{1}{2}(\partial^{2}-6)\epsilon_{\mu\nu}%
^{2}+\partial_{(\mu}\epsilon_{\mu)}^{6},\label{22d}\\
&  \delta\xi_{\mu\nu}^{2}=-3\partial^{\alpha}\epsilon_{\mu\nu\alpha}%
^{0}-2\epsilon_{(\mu\nu)}^{1}-3\epsilon_{\mu\nu}^{2}-\frac{1}{2}\epsilon
^{6}\eta_{\mu\nu},\label{22e}\\
&  \delta\xi_{\mu}^{0}=\frac{1}{2}(\partial^{2}-6)\epsilon_{\mu}^{3}%
-\partial_{\mu}\epsilon^{5}+\epsilon_{\mu}^{6},\label{22f}\\
&  \delta\xi_{\mu}^{1}=\frac{1}{2}(\partial^{2}-6)\epsilon_{\mu}^{4}%
-\epsilon_{\mu}^{6},\label{22g}\\
&  \delta\xi_{\mu}^{2}=\frac{1}{2}(\partial^{2}-6)\epsilon_{\mu}^{5}%
+\partial_{\mu}\epsilon^{5}-\epsilon_{\mu}^{6},\label{22h}\\
&  \delta\xi_{\mu}^{3}=-\partial^{\nu}\epsilon_{\mu\nu}^{1}-\partial_{\mu
}\epsilon^{6}-3\epsilon_{\mu}^{3}-3\epsilon_{\mu}^{4},\label{22i}\\
&  \delta\xi_{\mu}^{4}=2\partial^{\nu}\epsilon_{\mu\nu}^{2}+\partial_{\mu
}\epsilon^{6}-2\epsilon_{\mu}^{4}-4\epsilon_{\mu}^{5}-2\epsilon_{\mu}%
^{6},\label{22j}\\
&  \delta\xi_{\mu}^{5}=-2\partial^{\nu}\epsilon_{\mu\nu}^{1}-3\epsilon_{\mu
}^{3}-5\epsilon_{\mu}^{5}+4\epsilon_{\mu}^{6}+3\epsilon_{\mu\nu}^{0\;\nu
},\label{22k}\\
&  \delta\xi^{0}=\frac{1}{2}(\partial^{2}-6)\epsilon^{4}-2\epsilon
^{5},\label{22l}\\
&  \delta\xi^{1}=-\partial^{\mu}\epsilon_{\mu}^{5}-5\epsilon^{4}-2\epsilon
^{5}-\epsilon^{6},\label{22m}\\
&  \delta\xi^{2}=-2\partial^{\mu}\epsilon_{\mu}^{4}-6\epsilon^{4}%
-3\epsilon^{6}+\epsilon_{\mu}^{2\;\mu},\label{22n}\\
&  \delta\xi^{3}=-3\partial^{\mu}\epsilon_{\mu}^{3}-7\epsilon^{4}%
+6\epsilon^{5}+5\epsilon^{6}+2\epsilon_{\mu}^{1\;\mu},\label{22o}\\
&  \delta\xi^{4}=\frac{1}{2}(\partial^{2}-6)\epsilon^{6}+\partial^{\mu
}\epsilon_{\mu}^{6}+4\epsilon^{5}. \label{22p}%
\end{align}
\end{subequations}%

There are nine on-mass-shell conditions, which contains a symmetric spin
three, an antisymmetric spin two, two symmetric spin two, three vector and two
scalar fields, and seven gauge conditions which amounts to sixteen equations
in Eq.(\ref{22a}) to Eq.(\ref{22p}). This is consistent with counting from ZNS
listed in the table. Three traceless conditions read from ZNS corresponds to
the three equations involving $\delta\xi_{\mu\nu}^{\;\;\nu}$ , $\delta\xi
_{\mu}^{0\;\mu}$ , $\delta\xi_{\mu}^{1\;\mu}$ which are contained in
Eq.(\ref{22a}), Eq.(\ref{22c}), and Eq.(\ref{22d}).

It is important to note that the transformation for the matter parts,
Eq.(\ref{18a}) to Eq.(\ref{18d}) and Eq.(\ref{21a}) to Eq.(\ref{21f}), are the
same as the calculation \cite{Lee3} based on the chordal gauge transformation
of free covariant string field theory constructed by Banks and Peskin
\cite{Banks}. The Chordal gauge transformation can be written in the following form%

\begin{equation}
\delta\Phi\lbrack X(\sigma)]=\sum\limits_{n>0}L_{-n}\Phi_{n}[X(\sigma)]
\label{23.}%
\end{equation}
where $\Phi\lbrack X(\sigma)]$ is the string field and $\Phi_{n}[X(\sigma)]$
are gauge parameters which are functions of $X[\sigma]$ only and free of ghost
fields. This is because the pure ghost part of $Q_{B}$ in Eq.(\ref{15.}) does
not contribute to the transformation of matter background fields. It is
interesting to note that the r.h.s. of Eq.(\ref{23.}) is in the form of
off-shell spurious states \cite{GSW} in the OCFQ approach. They become ZNS on
imposing the physical and on-shell state condition.

Finally, it can be shown that the number of scalar ZNS at $n$-th massive level
$(n\geq3)$ is at least the sum of those at $(n-2)$-th and $(n-1)$-th massive
levels. So positive-norm scalar modes at $n$-th level, if they exist, will be
decoupled according to our decoupling conjecture.

The decoupling of these scalars has important implication on Sen's conjectures
on the decay of open string tachyon \cite{Sen}. Since all scalars on $D$-brane
including tachyon get non-zero vev. in the false vacuum, they will decay
together with tachyon and disappear eventually to the true closed string
vacuum. As the scalar states together with higher tensor states form a large
gauge multiplet at each mass level, and its scattering amplitudes are fixed by
the tensor fields, these tensor fields of open string ($D25$-brane) will
accompany the decay process. This means that the whole $D$-brane could
disappear to the true closed string vacuum.%

\setcounter{equation}{0}
\renewcommand{\theequation}{\arabic{section}.\arabic{equation}}%

\section{Calculation of higher massive ZNS}

Since ZNS are the most important key to generate stringy symmetries, in this
chapter we give a simplified method \cite{0302123} to generate two types of
ZNS in the old covariant first quantized (OCFQ) spectrum of open bosonic
string. ZNS up to the fourth massive level and general formulas of some
zero-norm tensor states at arbitrary mass levels will be calculated.

The vertex operator of a physical state of open bosonic string%

\begin{equation}
\left\vert \Psi\right\rangle =\sum C_{\mu_{1}...\mu_{m}}\alpha_{-n_{1}}%
^{\mu_{1}}...\alpha_{-n_{m}}^{\mu_{m}}\left\vert 0;k\right\rangle ,[\alpha
_{m}^{\mu},\alpha_{n}^{\nu}]=m\eta^{\mu\nu}\delta_{m+n} \label{2.1.}%
\end{equation}
is given by \cite{Sasaki}%

\begin{equation}
\Psi(z)=\sum C_{\mu_{1}...\mu_{m}}N_{m}:\prod(\partial_{z}^{n_{j}}x^{\mu_{j}%
})e^{ik\cdot X(z)}:, \label{2.2.}%
\end{equation}
where $N_{m}=i^{m}\prod\{(n_{j}-1)!\}^{-1}$. In the OCFQ spectrum, physical
states in Eq.(\ref{2.1.}) are subject to the following Virasoro conditions%

\begin{equation}
(L_{0}-1)\left\vert \Psi\right\rangle =0,L_{1}\left\vert \Psi\right\rangle
=L_{2}\left\vert \Psi\right\rangle =0, \label{3a,b}%
\end{equation}
where%

\begin{equation}
L_{m}=\frac{1}{2}\sum_{-\infty}^{\infty}:\alpha_{m-n}\cdot\alpha_{n}:
\end{equation}
and $\alpha_{0}\equiv k$. The solutions of Eq.(\ref{3a,b}) include
positive-norm propagating states and two types of ZNS in Eq.(\ref{1.1}) and
Eq.(\ref{1.2}) which can be derived from Kac determinant in conformal field
theory. While type I states have zero-norm at any spacetime dimension, type II
states have zero-norm \textit{only} at $D=26$. The existence of type II ZNS
signals the importance of ZNS in the structure of the theory of string. It is
straightforward to solve positive-norm state solutions of Eq.(\ref{3a,b}) for
some low-lying states, but soon becomes practically unmanageable. The authors
of Ref \cite{Manes} gave a simple prescription to solve the positive-norm
state solutions of Eq.(\ref{3a,b}). The strategy is to apply the Virasoro
conditions only to purely transverse states, so that the ZNS will be removed
at the very beginning. This prescription simplified a lot of computation
although some complexities remained for low spin states at higher levels. Our
aim in this chapter is to generate ZNS in Eq.(\ref{1.1}) and Eq.(\ref{1.2}) so
that all physical state solutions of Eq.(\ref{3a,b}) will be completed.

Let's first assume we are given positive-norm state solutions of some mass
level $n$. The number of positive-norm degree of freedom at mass level $n$ (
$M^{2}=2(n-1)$) is given by $N_{24}(n)$, where \cite{Polyakov2}%

\begin{equation}
N_{D}(n)=\frac{1}{2\pi i}%
{\displaystyle\oint}
\frac{dx}{x^{n+1}}(%
{\textstyle\prod_{k=1}^{\infty}}
\frac{1}{1-x^{k}})^{D}.
\end{equation}
On the other hand, the number of physical state degree of freedom is given by
$N_{25}(n)$ in view of the constraints in Eq.(\ref{3a,b}). The discrepancy is
of course due to physical ZNS given by solutions of Eq.(\ref{1.1}) and
Eq.(\ref{1.2}). That is, among $25$ chains of $\alpha_{m}^{\mu}$ oscillators,
one chain forms ZNS. Thus we can easily tabulate Young diagrams of ZNS at each
mass level given Young diagrams of positive-norm states at the same mass level
calculated by the simplified prescription in \cite{Manes}. For example,
positive-norm state
$\raisebox{0.06in}{\fbox{\rule[0.04cm]{0.04cm}{0cm}}}\raisebox{0.06in}{\fbox{\rule[0.04cm]{0.04cm}{0cm}}}\raisebox{0.06in}{\fbox{\rule[0.04cm]{0.04cm}{0cm}}}\raisebox{0.06in}{\fbox{\rule[0.04cm]{0.04cm}{0cm}}}$
\ at mass level $n=4$ gives ZNS
$\raisebox{0.06in}{\fbox{\rule[0.04cm]{0.04cm}{0cm}}}\raisebox{0.06in}{\fbox{\rule[0.04cm]{0.04cm}{0cm}}}\raisebox{0.06in}{\fbox{\rule[0.04cm]{0.04cm}{0cm}}}$%
+$\raisebox{0.06in}{\fbox{\rule[0.04cm]{0.04cm}{0cm}}}\raisebox{0.06in}{\fbox{\rule[0.04cm]{0.04cm}{0cm}}}$%
+$\raisebox{0.06in}{\fbox{\rule[0.04cm]{0.04cm}{0cm}}}+\bullet$ ,
positive-norm state
$\raisebox{0.06in}{\fbox{\rule[0.04cm]{0.04cm}{0cm}}}\hspace{-0.094in}%
\hspace{-0.04cm}\raisebox{-.047in}{\fbox{\rule[0.04cm]{0.04cm}{0cm}}}\hspace
{-0.006in}\hspace{-0.006in}\hspace{0.02cm}%
\raisebox{0.06in}{\fbox{\rule[0.04cm]{0.04cm}{0cm}}}$ gives ZNS
$\raisebox{0.06in}{\fbox{\rule[0.04cm]{0.04cm}{0cm}}}\hspace{-0.094in}%
\hspace{-0.04cm}%
\raisebox{-.047in}{\fbox{\rule[0.04cm]{0.04cm}{0cm}}}+\raisebox{0.06in}{\fbox{\rule[0.04cm]{0.04cm}{0cm}}}\raisebox{0.06in}{\fbox{\rule[0.04cm]{0.04cm}{0cm}}}+\raisebox{0.06in}{\fbox{\rule[0.04cm]{0.04cm}{0cm}}}$
\ and positive-norm state
$\raisebox{0.06in}{\fbox{\rule[0.04cm]{0.04cm}{0cm}}}\raisebox{0.06in}{\fbox{\rule[0.04cm]{0.04cm}{0cm}}}$
gives ZNS $\raisebox{0.06in}{\fbox{\rule[0.04cm]{0.04cm}{0cm}}}+\bullet.$ This
completes the ZNS at mass level $n=4.$ Young diagrams of ZNS up to mass level
$M^{2}=10$, together with positive-norm states calculated in \cite{Manes},
will be listed in the later part of this chapter. A consistent check of
counting of ZNS by using background ghost fields in WSFT \ was given in
\cite{KaoLee}.

To explicitly calculate ZNS is another complicated issue. Suppose we are given
some low-lying positive-norm state solutions. It is interesting to see the
similarity between Eq.(\ref{3a,b}) and Eq.(\ref{1.1}) and Eq.(\ref{1.2}) for
$\left\vert x\right\rangle $ and $\left\vert \widetilde{x}\right\rangle $. The
only difference is the \textquotedblright mass shift\textquotedblright\ of
$L_{0}$ equations. As is well-known, the $L_{1\text{ }}$and $L_{2}$ equations
give the transverse and traceless conditions on the spin polarization. It
turns out that, in many cases, the $L_{1\text{ }}$and $L_{2}$ equations will
not refer to the $L_{0}$ equation or on-mass-shell condition. In these cases,
a positive-norm state solution for $\left\vert \Psi\right\rangle $ at mass
level $n$ will give a ZNS solution $L_{-1}\left\vert x\right\rangle $ at mass
level $n+1$ simply by taking $\left\vert x\right\rangle =\left\vert
\Psi\right\rangle $ and shifting $k^{2}$ by one unit. Similarly, one can
easily get a type II ZNS $(L_{-2}+\frac{3}{2}L_{-1}^{2})\left\vert
\widetilde{x}\right\rangle $ at mass level $n+2$ simply by taking $\left\vert
\widetilde{x}\right\rangle =\left\vert \Psi\right\rangle $ and shifting
$k^{2}$ by two units. For those cases where $L_{1\text{ }}$and $L_{2}$
equations do refer to $L_{0}$ equation, our prescription needs to be modified.
We will give some examples to illustrate this method. Note that once we
generate a ZNS, it soon becomes \ a candidate of physical state $\left\vert
\Psi\right\rangle $ to generate two new ZNS at even higher levels.

1. The \ first ZNS begin at $k^{2}=0$. This state is suggested from the
positive-norm tachyon state $\left\vert 0,k\right\rangle $ with $k^{2}$ $=2.$
Taking $\left\vert x\right\rangle =\left\vert 0,k\right\rangle $ and shifting
$k^{2}$ by one unit to $k^{2}=0$, we get a type I ZNS.%

\begin{equation}
L_{-1\text{ }}\left\vert x\right\rangle =k\cdot\alpha_{-1}\left\vert
0,k\right\rangle ;\left\vert x\right\rangle =\left\vert 0,k\right\rangle
,-k^{2}=M^{2}=0. \label{2.8}%
\end{equation}

2. At the first massive level $k^{2}=-2,$ tachyon suggests a type II ZNS%

\begin{equation}
(L_{-2}+\frac{3}{2}L_{-1}^{2})\left\vert \widetilde{x}\right\rangle =[\frac
{1}{2}\alpha_{-1}\cdot\alpha_{-1}+\frac{5}{2}k\cdot\alpha_{-2}+\frac{3}%
{2}(k\cdot\alpha_{-1})^{2}]\left\vert 0,k\right\rangle ;\left\vert
\widetilde{x}\right\rangle =\left\vert 0,k\right\rangle ,-k^{2}=2. \label{2.9}%
\end{equation}
Positive-norm massless vector state suggests a type I ZNS%

\begin{equation}
L_{-1}\left\vert x\right\rangle =[\theta\cdot\alpha_{-2}+(k\cdot\alpha
_{-1})(\theta\cdot\alpha_{-1})]\left\vert 0,k\right\rangle ;\left\vert
x\right\rangle =\theta\cdot\alpha_{-1}\left\vert 0,k\right\rangle
,-k^{2}=2,\theta\cdot k=0. \label{2.10}%
\end{equation}
However, massless singlet ZNS Eq.(\ref{2.8}) does not give a type I ZNS at the
first massive level $k^{2}=-2$ since $L_{1\text{ }}$ equation on state
Eq.(\ref{2.8}) refers to $L_{0}$ equation, $k^{2}=0.$ This means that $L_{1}$
will not annihilate state Eq.(\ref{2.8}) if one shifts the mass to $k^{2}=-2.$

3. At the second massive level $k^{2}=-4,$ positive-norm massless vector state
suggests a type II ZNS%

\begin{align}
(L_{-2}+\frac{3}{2}L_{-1}^{2})\left\vert \widetilde{x}\right\rangle  &
=\{4\theta\cdot\alpha_{-3}+\frac{1}{2}(\alpha_{-1}\cdot\alpha_{-1}%
)(\theta\cdot\alpha_{-1})+\frac{5}{2}(k\cdot\alpha_{-2})(\theta\cdot
\alpha_{-1})\nonumber\\[0.01in]
&  +\frac{3}{2}(k\cdot\alpha_{-1})^{2}(\theta\cdot\alpha_{-1})+3(k\cdot
\alpha_{-1})(\theta\cdot\alpha_{-2})\}\left\vert 0,k\right\rangle ;\nonumber\\
\left\vert \widetilde{x}\right\rangle  &  =\theta\cdot\alpha_{-1}\left\vert
0,k\right\rangle ,-k^{2}=4,k\cdot\theta=0. \label{2.11}%
\end{align}
However, massless singlet ZNS Eq.(\ref{2.8}) does not give a type I ZNS at
mass level $k^{2}=-4$ for the same reason stated after Eq.(\ref{2.10}).
Positive-norm spin-two state at $k^{2}=-2$ suggests a type I ZNS%

\begin{align}
L_{-1}\left\vert x\right\rangle  &  =[2\theta_{\mu\nu}\alpha_{-1}^{\mu}%
\alpha_{-2}^{\nu}+k_{\lambda}\theta_{\mu\nu}\alpha_{-1}^{\lambda\mu\nu
}]\left\vert 0,k\right\rangle ;\left\vert x\right\rangle =\theta_{\mu\nu
}\alpha_{-1}^{\mu\nu}\left\vert 0,k\right\rangle ,-k^{2}=4,\nonumber\\
k\cdot\theta &  =\eta^{\mu\nu}\theta_{\mu\nu}=0,\theta_{\mu\nu}=\theta_{\nu
\mu}, \label{2.12}%
\end{align}
where $\alpha_{-1}^{\lambda\mu\nu}\equiv\alpha_{-1}^{\lambda}\alpha_{-1}^{\mu
}\alpha_{-1}^{\nu}.$ Similar notations will be used in the rest of this paper.
Vector ZNS with $k^{2}=-2$ in Eq.(\ref{2.10}) does not give a type I ZNS for
the same reason stated after Eq.(\ref{2.10}). In this case, however, one can
modify $\left\vert x\right\rangle $ to be%

\begin{equation}
\text{Ansatz: }\left\vert x\right\rangle =[a\theta\cdot\alpha_{-2}%
+b(k\cdot\alpha_{-1})(\theta\cdot\alpha_{-1})]\left\vert 0,k\right\rangle
;-k^{2}=4,\theta\cdot k=0, \label{2.13}%
\end{equation}
where $a,b$ are undetermined constants. $L_{0}$ equation is then trivially
satisfied and $L_{1},$ $L_{2}$ equations give $a:b=2:1$. This gives a type I ZNS%

\begin{align}
L_{-1}\left\vert x\right\rangle  &  =[\frac{1}{2}(k\cdot\alpha_{-1}%
)^{2}(\theta\cdot\alpha_{-1})+2\theta\cdot\alpha_{-3}+\frac{3}{2}(k\cdot
\alpha_{-1})(\theta\cdot\alpha_{-2})\nonumber\\
&  +\frac{1}{2}(k\cdot\alpha_{-2})(\theta\cdot\alpha_{-1})]\left\vert
0,k\right\rangle ;-k^{2}=4,\theta\cdot k=0. \label{2.14.}%
\end{align}
Similarly, we modify the singlet ZNS with $k^{2}=-2$ in Eq.(9) to be%

\begin{equation}
\text{Ansatz: }\left\vert x\right\rangle =[\frac{5}{2}ak\cdot\alpha_{-2}%
+\frac{1}{2}b\alpha_{-1}\cdot\alpha_{-1}+\frac{3}{2}c(k\cdot\alpha_{-1}%
)^{2}]\left\vert 0,k\right\rangle ;-k^{2}=4, \label{2.15}%
\end{equation}
where $a$, $b$ and $c$ are undetermined constants. $L_{1}$ and $L_{2}$
equations give%

\begin{equation}
5a+b+3k^{2}c=0,5k^{2}a+13b+\frac{3}{2}k^{2}c=0. \label{2.16}%
\end{equation}
For $k^{2}=-4$, we \ have $a:b:c=5:9:\frac{17}{6}$. This gives a type I ZNS%

\begin{align}
L_{-1}\left\vert x\right\rangle  &  =[\frac{17}{4}(k\cdot\alpha_{-1}%
)^{3}+\frac{9}{2}(k\cdot\alpha_{-1})(\alpha_{-1}\cdot\alpha_{-1}%
)+9(\alpha_{-1}\cdot\alpha_{-2})\nonumber\\
&  +21(k\cdot\alpha_{-1})(k\cdot\alpha_{-2})+25(k\cdot\alpha_{-3})]\left\vert
0,k\right\rangle ;\label{2.17}\\
-k^{2}  &  =4.\nonumber
\end{align}

This completes the four ZNS at the second massive level. It is interesting to
note that the Young tableau of ZNS at level $M^{2}=4$ are the sum of those of
all physical states at two lower levels, $M^{2}=2$ and $M^{2}=0$,
\textit{except }the singlet ZNS due to the dependence of $L_{1}$ and $L_{2}$
equations on$\ L_{0}$ condition in state Eq.(\ref{2.8}). For those cases that
$L_{1}$\ and $L_{2}$\ equations not referring to $L_{0}$\ condition, our
construction gives us a very simple way to calculate ZNS at any mass level
\ $n$\ given those of positive-norm states at lower levels constructed by the
simplified method in Ref \cite{Manes}. When the modified method was needed to
calculate a higher mass level ZNS from a lower mass level physical state like
Eq.(\ref{2.8}), an inconsistency may result and one gets no ZNS. This explains
the discrepancy of singlet ZNS at levels $M^{2}=2,4$,$8$ and a vector ZNS at
level $M^{2}=10$.

4. Similar method can be used to calculate ZNS at level $M^{2}=6.$ We will
just list some examples here. They are (from now on, unless otherwise stated,
each spin polarization is assumed to be transverse, traceless and is symmetric
with respect to each group of indices as in Ref \cite{Manes})%

\begin{equation}
L_{-1}\left\vert x\right\rangle =\theta_{\mu\nu\lambda}(k_{\beta}\alpha
_{-1}^{\mu\nu\lambda\beta}+3\alpha_{-1}^{\mu\nu}\alpha_{-2}^{\lambda
})\left\vert 0,k\right\rangle ;\left\vert x\right\rangle =\theta_{\mu
\nu\lambda}\alpha_{-1}^{\mu\nu\lambda}\left\vert 0,k\right\rangle ,
\label{2.18}%
\end{equation}

\begin{equation}
L_{-1}\left\vert x\right\rangle =[k_{\lambda}\theta_{\mu\nu}\alpha_{-1}%
^{\mu_{\lambda}}\alpha_{-2}^{\nu}+2\theta_{\mu\nu}\alpha_{-1}^{\mu}\alpha
_{-3}^{\nu}\left\vert 0,k\right\rangle ;\left\vert x\right\rangle =\theta
_{\mu\nu}\alpha_{-1}^{\mu}\alpha_{-2}^{\nu}\left\vert 0,k\right\rangle ,\text{
where }\theta_{\mu\nu}=-\theta_{\nu\mu}, \label{2.19}%
\end{equation}

\begin{align}
L_{-1}\left\vert x\right\rangle  &  =[2\theta_{\mu\nu}\alpha_{-2}^{\mu\nu
}+4\theta_{\mu\nu}\alpha_{-1}^{\mu}\alpha_{-3}^{\nu}+2(k_{\lambda}\theta
_{\mu\nu}+k_{(\lambda}\theta_{\mu\nu)})\alpha_{-1}^{\lambda\mu}\alpha
_{-2}^{\nu}+\frac{2}{3}k_{\lambda}k_{\beta}\theta_{\mu\nu}\alpha_{-1}^{\mu
\nu\lambda\beta}]\left\vert 0,k\right\rangle ;\nonumber\\
\left\vert x\right\rangle  &  =[2\theta_{\mu\nu}\alpha_{-1}^{\mu}\alpha
_{-2}^{\nu}+\frac{2}{3}k_{\lambda}\theta_{\mu\nu}\alpha_{-1}^{\mu\nu\lambda
}]\left\vert 0,k\right\rangle \label{2.20}%
\end{align}
and%

\begin{align}
(L_{-2}+\frac{3}{2}L_{-1}^{2})\left\vert \widetilde{x}\right\rangle  &
=[3\theta_{\mu\nu}\alpha_{-2}^{\mu\nu}+8\theta_{\mu\nu}\alpha_{-1}^{\mu}%
\alpha_{-3}^{\nu}+(k_{\lambda}\theta_{\mu\nu}+\frac{15}{2}k_{(\lambda}%
\theta_{\mu\nu)})\alpha_{-1}^{\lambda\mu}\alpha_{-2}^{\nu}\nonumber\\
&  +(\frac{1}{2}\eta_{\lambda\beta}\theta_{\mu\nu}+\frac{3}{2}k_{\lambda
}k_{\beta}\theta_{\mu\nu})\alpha_{-1}^{\mu\nu\lambda\beta}]\left\vert
0,k\right\rangle ;\nonumber\\
\left\vert \widetilde{x}\right\rangle  &  =\theta_{\mu\nu}\alpha_{-1}^{\mu\nu
}\left\vert 0,k\right\rangle . \label{2.21}%
\end{align}
Note that $\left\vert x\right\rangle $ in Eq.(\ref{2.20}) has been modified as
we did for Eq.(\ref{2.13}). To further illustrate our method, we calculate the
type I singlet ZNS from Eq.(\ref{2.17}) as following%

\begin{align}
\text{Ansatz}  &  :\left\vert x\right\rangle =[a(k\cdot\alpha_{-1}%
)^{3}+b(k\cdot\alpha_{-1})(\alpha_{-1}\cdot\alpha_{-1})+c(k\cdot\alpha
_{-1})(k\cdot\alpha_{-2})\nonumber\\
&  +d(\alpha_{-1}\cdot\alpha_{-2})+f(k\cdot\alpha_{-3})\left\vert
0,k\right\rangle ;\nonumber\\
-k^{2}  &  =6. \label{2.22}%
\end{align}
The $L_{1}$ and $L_{2}$ equations can be easily used to determine
$a:b:c:d:f=37:72:261:216:450.$ This gives the type I singlet ZNS%

\begin{align}
L_{-1}\left\vert x\right\rangle  &  =[a(k\cdot\alpha_{-1})^{4}+b(k\cdot
\alpha_{-1})^{2}(\alpha_{-1}\cdot\alpha_{-1})+(2b+d)(k\cdot\alpha_{-1}%
)(\alpha_{-1}\cdot\alpha_{-2})\nonumber\\
&  +(c+3a)(k\cdot\alpha_{-1})^{2}(k\cdot\alpha_{-2})+c(k\cdot\alpha_{-2}%
)^{2}+d(\alpha_{-2}\cdot\alpha_{-2})+b(k\cdot\alpha_{-2})(\alpha_{-1}%
\cdot\alpha_{-1})\nonumber\\
&  +(2c+f)(k\cdot\alpha_{-3})(k\cdot\alpha_{-1})+2d(\alpha_{-1}\cdot
\alpha_{-3})+3f(k\cdot\alpha_{-4})]\left\vert 0,k\right\rangle ,\nonumber\\
-k^{2}  &  =6. \label{2.23}%
\end{align}

5. We list relevant ZNS at level $M^{2}=8$ from the known positive-norm states
and ZNS at level $M^{2}=4,6.$ They are%

\begin{equation}
L_{-1}\left\vert x\right\rangle =(k_{\beta}\theta_{\mu\nu\lambda\gamma}%
\alpha_{-1}^{\mu\nu\lambda\gamma\beta}+4\theta_{\mu\nu\lambda\gamma}%
\alpha_{-1}^{\mu\nu\lambda}\alpha_{-2}^{\gamma}\left\vert 0,k\right\rangle
;\left\vert x\right\rangle =\theta_{\mu\nu\lambda\gamma}\alpha_{-1}^{\mu
\nu\lambda\gamma}\left\vert 0,k\right\rangle , \label{2.24}%
\end{equation}

\begin{align}
L_{-1}\left\vert x\right\rangle  &  =\theta_{\mu\nu\lambda}[\frac{3}%
{4}k_{\beta}k_{\gamma}\alpha_{-1}^{\mu\nu\lambda\gamma\beta}+3k_{\beta}%
\alpha_{-1}^{\mu\nu\beta}\alpha_{-2}^{\lambda}+3k_{\beta}\alpha_{-1}^{(\mu
\nu\lambda}\alpha_{-2}^{\beta)}+6\alpha_{-1}^{(\mu}\alpha_{-2}^{\nu\lambda
)}\nonumber\\
+6\alpha_{-1}^{(\mu\nu}\alpha_{-3}^{\lambda)}]\left\vert 0,k\right\rangle
;\text{ }\left\vert x\right\rangle  &  =\theta_{\mu\nu\lambda}(\frac{3}%
{4}k_{\beta}\alpha_{-1}^{\mu\nu\lambda\beta}+3\alpha_{-1}^{\mu\nu}\alpha
_{-2}^{\lambda})\left\vert 0,k\right\rangle , \label{2.25}%
\end{align}

\begin{align}
(L_{-2}+\frac{3}{2}L_{-1}^{2})\left\vert \widetilde{x}\right\rangle  &
=\theta_{\mu\nu\lambda}[(\frac{3}{2}k_{\beta}k_{\gamma}+\frac{1}{2}%
\eta_{\gamma\beta})\alpha_{-1}^{\mu\nu\lambda\beta\gamma}+k_{\gamma}(\frac
{1}{2}\alpha_{-1}^{\mu\nu\lambda}\alpha_{-2}^{\gamma}+8\alpha_{-1}^{(\mu
\nu\lambda}\alpha_{-2}^{\gamma)})\nonumber\\
&  +3\alpha_{-1}^{(\mu}\alpha_{-2}^{\nu\lambda)}+6\alpha_{-1}^{(\mu\nu}%
\alpha_{-3}^{\lambda)}]\left\vert 0,k\right\rangle ;\nonumber\\
\left\vert \widetilde{x}\right\rangle  &  =\theta_{\mu\nu\lambda}\alpha
_{-1}^{\mu\nu\lambda}\left\vert 0,k\right\rangle , \label{2.26}%
\end{align}

\begin{align}
L_{-1}\left\vert x\right\rangle  &  =\theta_{\mu\nu,\lambda}(k_{\gamma}%
\alpha_{-1}^{\gamma\mu\nu}\alpha_{-2}^{\lambda}+2\alpha_{-1}^{\mu}\alpha
_{-2}^{\nu\lambda}+2\alpha_{-1}^{\mu\nu}\alpha_{-3}^{\lambda})\left\vert
0,k\right\rangle ;\nonumber\\
\left\vert x\right\rangle  &  =\theta_{\mu\nu,\lambda}\alpha_{-1}^{\mu\nu
}\alpha_{-2}^{\lambda}\left\vert 0,k\right\rangle ,\text{ where }\theta
_{\mu\nu,\lambda}\text{ is mixed symmetric,} \label{2.27}%
\end{align}

\begin{align}
L_{-1}\left\vert x\right\rangle  &  =\theta_{\mu\nu}(\frac{3}{4}k_{\beta
}k_{\lambda}\alpha_{-1}^{\beta\lambda\mu}\alpha_{-2}^{\nu}+4k_{\lambda}%
\alpha_{-1}^{\lambda\mu}\alpha_{-3}^{\nu}+\frac{3}{4}k_{\lambda}\alpha
_{-1}^{\mu}\alpha_{-2}^{\nu\lambda}+2\alpha_{-2}^{\mu}\alpha_{-3}^{\nu
}+6\alpha_{-1}^{\mu}\alpha_{-4}^{\nu})\left\vert 0,k\right\rangle ;\nonumber\\
\left\vert x\right\rangle  &  =(\frac{3}{4}k_{\lambda}\alpha_{-1}^{\lambda\mu
}\alpha_{-2}^{\nu}+2\alpha_{-1}^{\mu}\alpha_{-3}^{\nu})\left\vert
0,k\right\rangle ,\text{ where }\theta_{\mu\nu}=-\theta_{\nu\mu}, \label{2.28}%
\end{align}
and%

\begin{align}
(L_{-2}+\frac{3}{2}L_{-1}^{2})\left\vert \widetilde{x}\right\rangle  &
=\theta_{\mu\nu}[(\frac{3}{2}k_{\gamma}k_{\lambda}+\frac{1}{2}\eta
_{\gamma\lambda})\alpha_{-1}^{\gamma\lambda\mu}\alpha_{-2}^{\nu}+6k_{\lambda
}\alpha_{-1}^{\lambda\mu}\alpha_{-3}^{\nu}+\frac{5}{2}k_{\lambda}\alpha
_{-1}^{\mu}\alpha_{-2}^{\nu\lambda}\nonumber\\
+2\alpha_{-2}^{\mu}\alpha_{-3}^{\nu}+\alpha_{-1}^{\mu}\alpha_{-4}^{\nu
}]\left\vert 0,k\right\rangle ,\left\vert \widetilde{x}\right\rangle  &
=\theta_{\mu\nu}\alpha_{-1}^{\mu}\alpha_{-2}^{\nu}\left\vert 0,k\right\rangle
,\text{ where }\theta_{\mu\nu}=-\theta_{\nu\mu}. \label{2.29}%
\end{align}
Note that the modified method was used in Eq.(\ref{2.25}) and Eq.(\ref{2.28}).

6. Finally, we calculate general formulas of some zero-norm tensor states at
arbitrary mass levels by making use of general formulas of some positive-norm
states listed in Ref \cite{Manes}.

a.
\begin{equation}
L_{-1}\theta_{\mu_{1}...\mu_{m}}\alpha_{-1}^{\mu_{1}...\mu_{m}}\left\vert
0,k\right\rangle =\theta_{\mu_{1}...\mu_{m}}(k_{\lambda}\alpha_{-1}%
^{\lambda\mu_{1}...\mu_{m}}+m\alpha_{-2}^{\mu_{1}}\alpha_{-1}^{\mu_{2}.\mu
_{m}})\left\vert 0,k\right\rangle , \label{2.30}%
\end{equation}
where $-k^{2}=M^{2}=2m,m=0,1,2,3...$. For example, $m=0,1$ give Eq.(\ref{2.8})
and Eq.(\ref{2.10}).

b.
\begin{align}
&  (L_{-2}+\frac{3}{2}L_{-1}^{2})\theta_{\mu_{1}...\mu_{m}}\alpha_{-1}%
^{\mu_{1}...\mu_{m}}\left\vert 0,k\right\rangle \nonumber\\
&  =\{\theta_{\mu_{1}...\mu_{m}}[(\frac{3}{2}k_{\nu}k_{\lambda}+\frac{1}%
{2}\eta_{\nu\lambda})\alpha_{-1}^{\nu\lambda\mu_{1}...\mu_{m}}+\frac{3}%
{2}m(m-1)\alpha_{-2}^{\mu_{1}\mu_{2}}\alpha_{-1}^{\mu_{3}...\mu_{m}%
}\nonumber\\
&  +(1+3m)\alpha_{-1}^{\mu_{1}...\mu_{m-1}}\alpha_{-3}^{\mu_{m}}]+[\frac{3}%
{2}(m+1)k_{(\lambda}\theta_{\mu_{1}...\mu_{m})}+\frac{3}{2}mk_{\mu_{m}}%
\theta_{\mu_{1}...\mu_{m-1\lambda})}]\nonumber\\
&  \alpha_{-1}^{\mu_{1}...\mu_{m}}\alpha_{-2}^{\lambda}\}\left\vert
0,k\right\rangle , \label{2.31}%
\end{align}
where $-k^{2}=M^{2}=2m+2,m=0,1,2...$. For example, $m=0,1$ give Eq.(\ref{2.9})
and Eq.(\ref{2.11}).

c.
\begin{align}
&  L_{-1}\theta_{\mu_{1}...\mu_{m-2},\mu_{m-1}}\alpha_{-1}^{\mu_{1}..\mu
_{m-2}}\alpha_{-2}^{\mu_{m-1}}\left\vert 0,k\right\rangle \nonumber\\
&  =\theta_{\mu_{1}...\mu_{m-2},\mu_{m-1}}[k_{\lambda}\alpha_{-1}^{\lambda
\mu_{1}...\mu_{m-2}}\alpha_{-2}^{\mu_{m-1}}+(m-2)\alpha_{-1}^{\mu_{1}\mu
_{m-3}}\alpha_{-2}^{\mu_{m-2}\mu_{m}}\nonumber\\
&  +2\alpha_{-1}^{\mu_{1}...\mu_{m-2}}\alpha_{-2}^{\mu_{m-1}}]\left\vert
0,k\right\rangle ,\hspace{0.1cm}%
\raisebox{0.06in}{\fbox{\rule[0.04cm]{0.04cm}{0cm}}}\hspace{-0.094in}%
\hspace{-0.04cm}\raisebox{-.047in}{\fbox{\rule[0.04cm]{0.04cm}{0cm}}}\hspace
{-0.006in}\hspace{-0.006in}\hspace{0.02cm}%
\raisebox{0.06in}{\fbox{......\rule[0.01cm]{0.18cm}{0cm}}}\hspace
{-0.025cm}\raisebox{0.06in}{\fbox{\rule[0.04cm]{0.04cm}{0cm}}} \label{2.32}%
\end{align}
where $-k^{2}=M^{2}=2m,m=3,4,5...$. For example, $m=3,4$ give Eq.(\ref{2.19})
and Eq.(\ref{2.27}).

d.
\begin{align}
&  (L_{-2}+\frac{3}{2}L_{-1}^{2})\theta_{\mu_{1}...\mu_{m-2},\mu_{m-1}}%
\alpha_{-1}^{\mu_{1}...\mu_{m-2}}\alpha_{-2}^{\mu_{m-1}}\left\vert
0,k\right\rangle \nonumber\\
&  =\theta_{\mu_{1}...\mu_{m-2},\mu_{m-1}}[(\frac{3}{2}k_{\lambda}k_{\nu
}+\frac{1}{2}\eta_{\lambda\nu})\alpha_{-1}^{\mu_{1}...\mu_{m-2}\lambda\nu
}\alpha_{-2}^{\mu_{m-1}}+6k_{\lambda}\alpha_{-1}^{\mu_{1}...\mu_{m-2}\lambda
}\alpha_{-3}^{\mu_{m-1}}\nonumber\\
&  +(\frac{3}{2}m-2)k_{\lambda}\alpha_{-1}^{\mu_{1}...\mu_{m-2}}\alpha
_{-2}^{\mu_{m-1\lambda}}+2(m-2)\alpha_{-1}^{\mu_{1}...\mu_{m-3}}\alpha
_{-2}^{\mu_{m-2}}\alpha_{-3}^{\mu_{m-1}}+11\alpha_{-1}^{\mu_{1}...\mu_{m-2}%
}\alpha_{-4}^{\mu_{m-1}}\nonumber\\
&  +k_{\lambda}\alpha_{-1}^{\mu_{1}...\mu_{m-3}\lambda}\alpha_{-2}^{\mu
_{m-2}\mu_{m-1}}+(m-3)\alpha_{-1}^{\mu_{1}...\mu_{m-4}}\alpha_{-2}^{\mu
_{m-3}\mu_{m-2}\mu_{m-1}}]\left\vert 0,k\right\rangle ,\hspace{0.1cm}%
\raisebox{0.06in}{\fbox{\rule[0.04cm]{0.04cm}{0cm}}}\hspace{-0.094in}%
\hspace{-0.04cm}\raisebox{-.047in}{\fbox{\rule[0.04cm]{0.04cm}{0cm}}}\hspace
{-0.006in}\hspace{-0.006in}\hspace{0.02cm}%
\raisebox{0.06in}{\fbox{......\rule[0.01cm]{0.18cm}{0cm}}}\hspace
{-0.025cm}\raisebox{0.06in}{\fbox{\rule[0.04cm]{0.04cm}{0cm}}} \label{2.33}%
\end{align}
where $-k^{2}=M^{2}=2m+2,m=3,4,5...$. For example, $m=3$ gives Eq.(\ref{2.29}).

e.
\begin{align}
&  L_{-1}\theta_{\mu_{1}...\mu_{m-4},\mu_{m-3}\mu_{m-2}}(\alpha_{-1}^{\mu
_{1}...\mu_{m-4}}\alpha_{-2}^{\mu_{m-3}\mu_{m-2}}-\frac{4}{3}\alpha_{-1}%
^{\mu_{1}...\mu_{m-3}}\alpha_{-3}^{\mu_{m-2}})\nonumber\\
&  =\theta_{\mu_{1}...\mu_{m-4},\mu_{m-3}\mu_{m-2}}[k_{\lambda}\alpha
_{-1}^{\lambda\mu_{1}...\mu_{m-4}}\alpha_{-2}^{\mu_{m-3}\mu_{m-2}}%
+(m-4)\alpha_{-1}^{\mu_{1}...\mu_{m-3}}\alpha_{-2}^{\mu_{m-4}\mu_{m-3}%
\mu_{m-2}}\nonumber\\
&  +\frac{16}{3}\alpha_{-1}^{\mu_{1}...\mu_{m-4}}\alpha_{-3}^{\mu_{m-3}}%
\alpha_{-2}^{\mu_{m-2}}+\frac{4}{3}k_{\lambda}\alpha_{-1}^{\lambda\mu
_{1}...\mu_{m-3}}\alpha_{-3}^{\mu_{m-2}}+4\alpha_{-1}^{\mu_{1}...\mu_{m-3}%
}\alpha_{-4}^{\mu_{m-4}}],\hspace{0.1cm}%
\raisebox{0.06in}{\fbox{\rule[0.04cm]{0.04cm}{0cm}}}\hspace{-0.094in}%
\hspace{-0.04cm}\raisebox{-.047in}{\fbox{\rule[0.04cm]{0.04cm}{0cm}}}\hspace
{-0.006in}\hspace{-0.006in}\hspace{0.02cm}%
\raisebox{0.06in}{\fbox{......\rule[0.01cm]{0.18cm}{0cm}}}\hspace
{-0.025cm}\raisebox{0.06in}{\fbox{\rule[0.04cm]{0.04cm}{0cm}}} \label{2.34}%
\end{align}
where $-k^{2}=M^{2}=2m,m=5,6...$.

f. The ZNS of Eq.(\ref{2.30}) can be used to generate new type I ZNS by the
modified method as following\bigskip%
\begin{align}
&  L_{-1}\theta_{\mu_{1}...\mu_{m}}(\frac{m}{m+1}k_{\lambda}\alpha
_{-1}^{\lambda\mu_{1}...\mu_{m}}+\alpha_{-2}^{\mu_{1}}\alpha_{-1}^{\mu
_{2}...\mu_{m}})\left\vert 0,k\right\rangle \nonumber\\
&  =[\frac{m}{m+1}k_{\nu}k_{\lambda}\theta_{\mu_{1}...\mu_{m}}\alpha_{-1}%
^{\nu\lambda\mu_{1}...\mu_{m}}+m(k_{(\lambda}\theta_{\mu_{1}...\mu_{m)}%
}+k_{\lambda}\theta_{\mu_{1}...\mu_{m}})\alpha_{-2}^{\mu_{1}}\alpha
_{-1}^{\lambda\mu_{2}...\mu_{m}}\nonumber\\
&  +m(m-1)\theta_{\mu_{1}...\mu_{m}}\alpha_{-2}^{\mu_{1}\mu_{2}}\alpha
_{-1}^{\mu_{3}...\mu_{m}}+2m\theta_{\mu_{1}...\mu_{m}}\alpha_{-3}^{\mu_{1}%
}\alpha_{-1}^{\mu_{2}...\mu_{m}}]\left\vert 0,k\right\rangle , \label{2.35}%
\end{align}
where $-k^{2}=M^{2}=2m+2,m=1,2,3...$. For example, $m=1,2$ and $3$ give
Eq.(\ref{2.14}), Eq.(\ref{2.20}) and Eq.(\ref{2.25}). Note that the
coefficient of the first term in Eq.(\ref{2.35}) has been modified to
$\frac{m}{m+1}.$ Similarly, new type II ZNS can also be constructed.

The Young tabulations of all physical states solutions of Eq.(\ref{3a,b}) up
to level six, including two types of ZNS solutions of Eq.(\ref{1.1}) and
Eq.(\ref{1.2}), are listed in the following table%

\begin{tabular}
[c]{|l|l|l|}\hline
massive level & positive-norm states & ZNS\\\hline
$M^{2}=-2$ & $\bullet$ & \\\hline
$M^{2}=0$ & $\raisebox{0.06in}{\fbox{\rule[0.04cm]{0.04cm}{0cm}}}$ & $\bullet$
(singlet)\\\hline
$M^{2}=2$ &
$\raisebox{0.06in}{\fbox{\rule[0.04cm]{0.04cm}{0cm}}}\raisebox{0.06in}{\fbox{\rule[0.04cm]{0.04cm}{0cm}}}$
& $\raisebox{0.06in}{\fbox{\rule[0.04cm]{0.04cm}{0cm}}},$ $\bullet$\\\hline
$M^{2}=4$ &
$\raisebox{0.06in}{\fbox{\rule[0.04cm]{0.04cm}{0cm}}}\raisebox{0.06in}{\fbox{\rule[0.04cm]{0.04cm}{0cm}}}\raisebox{0.06in}{\fbox{\rule[0.04cm]{0.04cm}{0cm}}},\raisebox{0.06in}{\fbox{\rule[0.04cm]{0.04cm}{0cm}}}\hspace
{-0.094in}\hspace{-0.04cm}%
\raisebox{-.047in}{\fbox{\rule[0.04cm]{0.04cm}{0cm}}}$ &
$\raisebox{0.06in}{\fbox{\rule[0.04cm]{0.04cm}{0cm}}}\raisebox{0.06in}{\fbox{\rule[0.04cm]{0.04cm}{0cm}}},2\times
\raisebox{0.06in}{\fbox{\rule[0.04cm]{0.04cm}{0cm}}},\bullet$\\\hline
$M^{2}=6$ &
$\raisebox{0.06in}{\fbox{\rule[0.04cm]{0.04cm}{0cm}}}\raisebox{0.06in}{\fbox{\rule[0.04cm]{0.04cm}{0cm}}}\raisebox{0.06in}{\fbox{\rule[0.04cm]{0.04cm}{0cm}}}\raisebox{0.06in}{\fbox{\rule[0.04cm]{0.04cm}{0cm}}},\raisebox{0.06in}{\fbox{\rule[0.04cm]{0.04cm}{0cm}}}\hspace
{-0.094in}\hspace{-0.04cm}%
\raisebox{-.047in}{\fbox{\rule[0.04cm]{0.04cm}{0cm}}}\hspace{-0.006in}%
\hspace{-0.006in}\hspace{0.02cm}%
\raisebox{0.06in}{\fbox{\rule[0.04cm]{0.04cm}{0cm}}},\raisebox{0.06in}{\fbox{\rule[0.04cm]{0.04cm}{0cm}}}\raisebox{0.06in}{\fbox{\rule[0.04cm]{0.04cm}{0cm}}},\bullet
$ &
$\raisebox{0.06in}{\fbox{\rule[0.04cm]{0.04cm}{0cm}}}\raisebox{0.06in}{\fbox{\rule[0.04cm]{0.04cm}{0cm}}}\raisebox{0.06in}{\fbox{\rule[0.04cm]{0.04cm}{0cm}}},\raisebox{0.06in}{\fbox{\rule[0.04cm]{0.04cm}{0cm}}}\hspace
{-0.094in}\hspace{-0.04cm}%
\raisebox{-.047in}{\fbox{\rule[0.04cm]{0.04cm}{0cm}}},2\times
\raisebox{0.06in}{\fbox{\rule[0.04cm]{0.04cm}{0cm}}}\raisebox{0.06in}{\fbox{\rule[0.04cm]{0.04cm}{0cm}}},3\times
\raisebox{0.06in}{\fbox{\rule[0.04cm]{0.04cm}{0cm}}},$ $2\times\bullet
$\\\hline
$M^{2}=8$ &
$\raisebox{0.06in}{\fbox{\rule[0.04cm]{0.04cm}{0cm}}}\raisebox{0.06in}{\fbox{\rule[0.04cm]{0.04cm}{0cm}}}\raisebox{0.06in}{\fbox{\rule[0.04cm]{0.04cm}{0cm}}}\raisebox{0.06in}{\fbox{\rule[0.04cm]{0.04cm}{0cm}}}\raisebox{0.06in}{\fbox{\rule[0.04cm]{0.04cm}{0cm}}},\raisebox{0.06in}{\fbox{\rule[0.04cm]{0.04cm}{0cm}}}\hspace
{-0.094in}\hspace{-0.04cm}%
\raisebox{-.047in}{\fbox{\rule[0.04cm]{0.04cm}{0cm}}}\hspace{-0.006in}%
\hspace{-0.006in}\hspace{0.02cm}%
\raisebox{0.06in}{\fbox{\rule[0.04cm]{0.04cm}{0cm}}}\raisebox{0.06in}{\fbox{\rule[0.04cm]{0.04cm}{0cm}}},\raisebox{0.06in}{\fbox{\rule[0.04cm]{0.04cm}{0cm}}}\hspace
{-0.094in}\hspace{-0.04cm}%
\raisebox{-.047in}{\fbox{\rule[0.04cm]{0.04cm}{0cm}}}\hspace{-0.006in}%
\hspace{-0.006in}\hspace{0.02cm}%
\raisebox{0.06in}{\fbox{\rule[0.04cm]{0.04cm}{0cm}}},\raisebox{0.06in}{\fbox{\rule[0.04cm]{0.04cm}{0cm}}}\raisebox{0.06in}{\fbox{\rule[0.04cm]{0.04cm}{0cm}}}\raisebox{0.06in}{\fbox{\rule[0.04cm]{0.04cm}{0cm}}},\raisebox{0.06in}{\fbox{\rule[0.04cm]{0.04cm}{0cm}}}\hspace
{-0.094in}\hspace{-0.04cm}%
\raisebox{-.047in}{\fbox{\rule[0.04cm]{0.04cm}{0cm}}},\raisebox{0.06in}{\fbox{\rule[0.04cm]{0.04cm}{0cm}}}$
&
$\raisebox{0.06in}{\fbox{\rule[0.04cm]{0.04cm}{0cm}}}\raisebox{0.06in}{\fbox{\rule[0.04cm]{0.04cm}{0cm}}}\raisebox{0.06in}{\fbox{\rule[0.04cm]{0.04cm}{0cm}}}\raisebox{0.06in}{\fbox{\rule[0.04cm]{0.04cm}{0cm}}},\raisebox{0.06in}{\fbox{\rule[0.04cm]{0.04cm}{0cm}}}\hspace
{-0.094in}\hspace{-0.04cm}%
\raisebox{-.047in}{\fbox{\rule[0.04cm]{0.04cm}{0cm}}}\hspace{-0.006in}%
\hspace{-0.006in}\hspace{0.02cm}%
\raisebox{0.06in}{\fbox{\rule[0.04cm]{0.04cm}{0cm}}},2\times
\raisebox{0.06in}{\fbox{\rule[0.04cm]{0.04cm}{0cm}}}\raisebox{0.06in}{\fbox{\rule[0.04cm]{0.04cm}{0cm}}}\raisebox{0.06in}{\fbox{\rule[0.04cm]{0.04cm}{0cm}}},2\times
\raisebox{0.06in}{\fbox{\rule[0.04cm]{0.04cm}{0cm}}}\hspace{-0.094in}%
\hspace{-0.04cm}\raisebox{-.047in}{\fbox{\rule[0.04cm]{0.04cm}{0cm}}},4\times
\raisebox{0.06in}{\fbox{\rule[0.04cm]{0.04cm}{0cm}}}\raisebox{0.06in}{\fbox{\rule[0.04cm]{0.04cm}{0cm}}},5\times
\raisebox{0.06in}{\fbox{\rule[0.04cm]{0.04cm}{0cm}}},3\times\bullet$\\\hline
$M^{2}=10$ & $%
\begin{array}
[c]{c}%
\raisebox{0.06in}{\fbox{\rule[0.04cm]{0.04cm}{0cm}}}\raisebox{0.06in}{\fbox{\rule[0.04cm]{0.04cm}{0cm}}}\raisebox{0.06in}{\fbox{\rule[0.04cm]{0.04cm}{0cm}}}\raisebox{0.06in}{\fbox{\rule[0.04cm]{0.04cm}{0cm}}}\raisebox{0.06in}{\fbox{\rule[0.04cm]{0.04cm}{0cm}}}\raisebox{0.06in}{\fbox{\rule[0.04cm]{0.04cm}{0cm}}},\raisebox{0.06in}{\fbox{\rule[0.04cm]{0.04cm}{0cm}}}\hspace
{-0.094in}\hspace{-0.04cm}%
\raisebox{-.047in}{\fbox{\rule[0.04cm]{0.04cm}{0cm}}}\hspace{-0.006in}%
\hspace{-0.006in}\hspace{0.02cm}%
\raisebox{0.06in}{\fbox{\rule[0.04cm]{0.04cm}{0cm}}}\raisebox{0.06in}{\fbox{\rule[0.04cm]{0.04cm}{0cm}}}\raisebox{0.06in}{\fbox{\rule[0.04cm]{0.04cm}{0cm}}},\raisebox{0.06in}{\fbox{\rule[0.04cm]{0.04cm}{0cm}}}\raisebox{0.06in}{\fbox{\rule[0.04cm]{0.04cm}{0cm}}}\raisebox{0.06in}{\fbox{\rule[0.04cm]{0.04cm}{0cm}}}\raisebox{0.06in}{\fbox{\rule[0.04cm]{0.04cm}{0cm}}},\raisebox{0.06in}{\fbox{\rule[0.04cm]{0.04cm}{0cm}}}\hspace
{-0.094in}\hspace{-0.04cm}%
\raisebox{-.047in}{\fbox{\rule[0.04cm]{0.04cm}{0cm}}}\hspace{-0.006in}%
\hspace{-0.006in}\hspace{0.02cm}%
\raisebox{0.06in}{\fbox{\rule[0.04cm]{0.04cm}{0cm}}}\raisebox{0.06in}{\fbox{\rule[0.04cm]{0.04cm}{0cm}}}\\
\raisebox{0.06in}{\fbox{\rule[0.04cm]{0.04cm}{0cm}}}\hspace{-0.094in}%
\hspace{-0.04cm}%
\raisebox{-.047in}{\fbox{\rule[0.04cm]{0.04cm}{0cm}}}\raisebox{0.06in}{\fbox{\rule[0.04cm]{0.04cm}{0cm}}}\hspace
{-0.094in}\hspace{-0.04cm}%
\raisebox{-.047in}{\fbox{\rule[0.04cm]{0.04cm}{0cm}}},\raisebox{0.06in}{\fbox{\rule[0.04cm]{0.04cm}{0cm}}}\raisebox{0.06in}{\fbox{\rule[0.04cm]{0.04cm}{0cm}}}\raisebox{0.06in}{\fbox{\rule[0.04cm]{0.04cm}{0cm}}},\raisebox{0.06in}{\fbox{\rule[0.04cm]{0.04cm}{0cm}}}\hspace
{-0.094in}\hspace{-0.04cm}%
\raisebox{-.047in}{\fbox{\rule[0.04cm]{0.04cm}{0cm}}}\hspace{-0.006in}%
\hspace{-0.006in}\hspace{0.02cm}%
\raisebox{0.06in}{\fbox{\rule[0.04cm]{0.04cm}{0cm}}},\raisebox{0.06in}{\fbox{\rule[0.04cm]{0.04cm}{0cm}}}\hspace
{-0.094in}\hspace{-0.04cm}%
\raisebox{-.047in}{\fbox{\rule[0.04cm]{0.04cm}{0cm}}}\hspace{-0.28cm}%
\raisebox{-.150in}{\fbox{\rule[0.04cm]{0.04cm}{0cm}}},\text{\ \ }%
\hspace{-0.094in}2\times
\raisebox{0.06in}{\fbox{\rule[0.04cm]{0.04cm}{0cm}}}\raisebox{0.06in}{\fbox{\rule[0.04cm]{0.04cm}{0cm}}},\raisebox{0.06in}{\fbox{\rule[0.04cm]{0.04cm}{0cm}}},\bullet
\end{array}
$ & $%
\begin{array}
[c]{c}%
\raisebox{0.06in}{\fbox{\rule[0.04cm]{0.04cm}{0cm}}}\raisebox{0.06in}{\fbox{\rule[0.04cm]{0.04cm}{0cm}}}\raisebox{0.06in}{\fbox{\rule[0.04cm]{0.04cm}{0cm}}}\raisebox{0.06in}{\fbox{\rule[0.04cm]{0.04cm}{0cm}}}\raisebox{0.06in}{\fbox{\rule[0.04cm]{0.04cm}{0cm}}},2\times
\raisebox{0.06in}{\fbox{\rule[0.04cm]{0.04cm}{0cm}}}\raisebox{0.06in}{\fbox{\rule[0.04cm]{0.04cm}{0cm}}}\raisebox{0.06in}{\fbox{\rule[0.04cm]{0.04cm}{0cm}}}\raisebox{0.06in}{\fbox{\rule[0.04cm]{0.04cm}{0cm}}},\raisebox{0.06in}{\fbox{\rule[0.04cm]{0.04cm}{0cm}}}\hspace
{-0.094in}\hspace{-0.04cm}%
\raisebox{-.047in}{\fbox{\rule[0.04cm]{0.04cm}{0cm}}}\hspace{-0.006in}%
\hspace{-0.006in}\hspace{0.02cm}%
\raisebox{0.06in}{\fbox{\rule[0.04cm]{0.04cm}{0cm}}}\raisebox{0.06in}{\fbox{\rule[0.04cm]{0.04cm}{0cm}}},3\times
\raisebox{0.06in}{\fbox{\rule[0.04cm]{0.04cm}{0cm}}}\hspace{-0.094in}%
\hspace{-0.04cm}\raisebox{-.047in}{\fbox{\rule[0.04cm]{0.04cm}{0cm}}}\hspace
{-0.006in}\hspace{-0.006in}\hspace{0.02cm}%
\raisebox{0.06in}{\fbox{\rule[0.04cm]{0.04cm}{0cm}}},\\
4\times
\raisebox{0.06in}{\fbox{\rule[0.04cm]{0.04cm}{0cm}}}\raisebox{0.06in}{\fbox{\rule[0.04cm]{0.04cm}{0cm}}}\raisebox{0.06in}{\fbox{\rule[0.04cm]{0.04cm}{0cm}}},4\times
\raisebox{0.06in}{\fbox{\rule[0.04cm]{0.04cm}{0cm}}}\hspace{-0.094in}%
\hspace{-0.04cm}\raisebox{-.047in}{\fbox{\rule[0.04cm]{0.04cm}{0cm}}},7\times
\raisebox{0.06in}{\fbox{\rule[0.04cm]{0.04cm}{0cm}}}\raisebox{0.06in}{\fbox{\rule[0.04cm]{0.04cm}{0cm}}},8\times
\raisebox{0.06in}{\fbox{\rule[0.04cm]{0.04cm}{0cm}}},6\times\bullet
\end{array}
$\\\hline
\end{tabular}
Note that the Young tabulations of ZNS at level n are subset of the sum of all
physical states at levels $n-1$ and $n-2$.%

\setcounter{equation}{0}
\renewcommand{\theequation}{\arabic{section}.\arabic{equation}}%

\section{Discrete ZNS and $w_{\infty}$ symmetry of $2D$ string theory}

For the $26D$ $(10D)$ string theory, it is difficult to do calculations for
higher mass string states and extract their symmetry structures which are
valid for all energies. This is of course due to the high dimensionality of
spacetime. One way to overcome this difficulty has been to probe high energy
regimes of the theory and simplify the calculations. This will be done in part
II and part III of this review. Another strategy was to study the toy string
model, namely, $2D$ string theory or $c=1$ $2D$ quantum gravity. The $2D$
string theory has been an important laboratory to study non-perturbative
information of string theory. In the continuum Liouville approach
\cite{2Dstring}, in addition to the massless tachyon mode, an infinite number
of massive discrete momentum physical degrees of freedom were discovered
\cite{GKN,GKN1,GKN2,GKN3,Polyakov} and the target spacetime $w_{\infty}$
symmetry and Ward identities were then identified
\cite{Winfinity,Winfinity2,Klebanov1,Ring,Ring1}.

In this chapter, we will derive the $w_{\infty}$ symmetry structure from the
ZNS point of view in the old covariant quantization scheme
\cite{ChungLee1,ChungLee2}. This is in parallel with the works of
\cite{Winfinity} and \cite{Ring,Ring1} where the ground ring structure of
ghost number zero operators were identified in the BRST quantization.
Moreover, the results we obtained will justify the idea of ZNS used in the
$26D$ (or $10D$) theories as discussed previously.

Unlike the discrete Polyakov states, we will find that there are still an
infinite number of continuum momentum ZNS in the massive levels of the $2D$
spectrum and it is very difficult to give a general formula for them just as
in the case of $26D$ theory \cite{Lee4,LEO,JCLee}. However, as far as the
dynamics of the theory is concerned, only those ZNS with Polyakov discrete
momenta are relevant. This is because all other ZNS are trivially decoupled
from the correlation functions due to kinematic reason. Hence, we will only
identify all discrete ZNS or discrete gauge states (DGS) in the spectrum. The
higher the momentum is, the more numerous the DGS are found. In particular, we
will give an explicit formula for one such set of DGS in terms of Schur
polynomials. Finally, we can show that these DGS carry the $w_{\infty}$
charges and serve as the symmetry parameters of the theory.

\subsection{$2D$ string theory}

\subsubsection{ZNS in $2D$ string theory}

We consider the two dimensional critical string action \cite{2Dstring}
\begin{equation}
S=\frac{1}{8\pi}\int d^{2}\sigma\sqrt{\hat{g}}[g^{\mu\nu}(\partial_{\mu
}X\partial_{\nu}X+\partial_{\mu}\phi\partial_{\nu}\phi)-Q\hat{R}\phi]
\end{equation}
with $\phi$ being the Liouville field. For $c=1$ theory $Q$, which represents
the background charge of the Liouville field, is set to be $2\sqrt{2}$ so that
the total anomalies cancels that from ghost contribution.

For simplicity here we consider only one of the chiral sectors, while the
other sector (denoted by $\bar{z}$) is the same. The stress energy tensor is%

\begin{equation}
T_{zz}=-\frac{1}{2}(\partial_{z}X)^{2}-\frac{1}{2}(\partial_{z}\phi)^{2}%
-\frac{1}{2}Q\partial_{z}^{2}\phi. \label{3.2.}%
\end{equation}
If we define the mode expansion of $X^{\mu}=(\phi,X)$ by
\begin{equation}
\partial_{z}X^{\mu}=-\sum\limits_{n=-\infty}^{\infty}z^{-n-1}(\alpha_{n}%
^{0},i\alpha_{n}^{1})
\end{equation}
with the Minkowski metric $\eta_{\mu\nu}=\left(
\begin{array}
[c]{cc}%
-1 & 0\\
0 & 1
\end{array}
\right)  $ , $Q^{\mu}=(2\sqrt{2},0)$ and the zero mode $\alpha_{0}^{\mu
}=f^{\mu}=(\epsilon,p)$, we find the Virasoro generators%

\begin{equation}%
\begin{split}
L_{n}  &  =\left(  \frac{n+1}{2}Q^{\mu}+f^{\mu}\right)  \alpha_{\mu,n}%
+\frac{1}{2}\sum\limits_{k\neq0}:\alpha_{\mu,-k}\alpha_{n+k}^{\mu}:\qquad
n\neq0,\\
L_{0}  &  =\frac{1}{2}\left(  Q^{\mu}+f^{\mu}\right)  f_{\mu}+\sum
\limits_{k=1}^{\infty}:\alpha_{\mu,-k}\alpha_{k}^{\mu}:.
\end{split}
\end{equation}

The vacuum $\left|  0\right\rangle $ is annihilated by all $\alpha^{\mu}_{n}$
with $n>0$. In the old covariant quantization, physical states $\left|
\psi\right\rangle $ are those satisfy the condition%

\begin{equation}%
\begin{split}
L_{n}\left\vert \psi\right\rangle  &  =0\qquad for\qquad n>0,\\
L_{0}\left\vert \psi\right\rangle  &  =\left\vert \psi\right\rangle .
\end{split}
\label{conditions}%
\end{equation}

One can easily check that the two branches of massless tachyon%
\begin{equation}
T^{\pm}(p)=e^{ipX+(\pm|p|-\sqrt{2})\phi}%
\end{equation}
are positive norm physical states. In the material gauge[6], it was also known
that there exist discrete states [3] [10] ($J=\{0,12,1...\}$ and
$M=\{{-J},-J+1,...J\}$)
\begin{equation}
\psi_{J,M}^{(\pm)}\sim(H_{-})^{J-M}\psi_{J,J}^{(\pm)}\sim(H_{+})^{J+M}%
\psi_{J,-J}^{(\pm)}, \label{JM}%
\end{equation}
which are also positive norm physical states. In Eq.(\ref{JM}) $H_{\pm}%
=\int\frac{dz}{2\pi i}T^{+}(\pm\sqrt{2})$ are the zero modes of the ladder
operators of the $SU(2)$ Kac-Moody currents at the self-dual radius in $c=1$
$2d$ conformal field theory and $\psi_{J,\pm J}^{(\pm)}=T^{(\pm)}(\pm\sqrt
{2}J)$. These exhaust all positive norm physical states. In this chapter we
are interested in the discrete ZNS or discrete gauge states (DGS), i.e., the
zero norm physical states at the same discrete momenta as those states in
Eq.(\ref{JM}). We thus no longer restrict ourselves in the material gauge, and
the Liouville field $\phi$ will play an important role in the following discussions.

In general, there are two types of ZNS in $2D$ string theory,

Type I:%

\begin{equation}
\left\vert \psi\right\rangle =L_{-1}\left\vert \chi\right\rangle \qquad
where\qquad L_{m}\left\vert \chi\right\rangle =0\qquad m\geq0, \label{I}%
\end{equation}
Type II:
\begin{equation}%
\begin{split}
\left\vert \psi\right\rangle =\left(  L_{-2}+\frac{3}{2}L_{-1}^{2}\right)
\left\vert \tilde{\chi}\right\rangle \qquad where\qquad &  L_{m}\left\vert
\tilde{\chi}\right\rangle =0\qquad m>0,\\
&  (L_{0}+1)\left\vert \tilde{\chi}\right\rangle =0.
\end{split}
\label{II}%
\end{equation}
They satisfy the physical state conditions Eq.(\ref{conditions}), and have
zero norm. It is important to note that state in Eq.(\ref{II}) is a ZNS only
when $Q=\sqrt{\frac{25-c}{3}}$, while the states in Eq.(\ref{I}) are
insensitive to this condition. In this section we will explicitly calculate
the DGS at the two lowest mass levels. At mass level one (i.e. spin
one),$f_{\mu}(f^{\mu}+Q^{\mu})=0$, only DGS of type I are found:$f_{\mu}%
\alpha_{-1}^{\mu}\left\vert f\right\rangle $, where $\left\vert f\right\rangle
=:e^{ipX+\epsilon\phi}:\left\vert 0\right\rangle $. The DGS $G_{1,0}%
^{-}=:\partial\phi e^{-2\sqrt{2}\phi}:\left\vert 0\right\rangle $ corresponds
to the momentum of $\psi_{1,0}^{-}$. There is no corresponding DGS for
$\psi_{1,0}^{+}$.

At mass level two, $f_{\mu}(f^{\mu}+Q^{\mu})=-2$ , if $e_{\mu}(f^{\mu}+Q^{\mu
})=0$ then the type I ZNS is
\begin{equation}
\left\vert \psi\right\rangle =\left[  \frac{1}{2}(f_{\mu}e_{\nu}+e_{\mu}%
f_{\nu})\alpha_{-1}^{\mu}\alpha_{-1}^{\nu}+e_{\mu}\alpha_{-2}^{\mu}\right]
\left\vert f\right\rangle ,
\end{equation}
while the type II ZNS is%

\begin{equation}
\left\vert \psi\right\rangle =\frac{1}{2}\left[  (3f_{\mu}f_{\nu}+\eta_{\mu
\nu})\alpha_{-1}^{\mu}\alpha_{-1}^{\nu}+(5f_{\mu}-Q_{\mu})\alpha_{-2}^{\mu
}\right]  \left\vert f\right\rangle .
\end{equation}
The DGS corresponding to $\psi_{\frac{3}{2},\pm\frac{1}{2}}^{-}$ are
$G_{\frac{3}{2},\pm\frac{1}{2}}^{-}$:

(type I)%

\begin{equation}
G_{\frac{3}{2},\pm\frac{1}{2}}^{-(1)}\sim\left[  \left(
\begin{array}
[c]{cc}%
\frac{5}{2} & \pm\frac{3}{2}\\
\pm\frac{3}{2} & \frac{1}{2}%
\end{array}
\right)  \alpha_{-1}^{\mu}\alpha_{-1}^{\nu}+\left(
\begin{array}
[c]{c}%
\frac{1}{\sqrt{2}}\\
\pm\frac{1}{\sqrt{2}}%
\end{array}
\right)  \alpha_{-2}^{\mu}\right]  \left\vert f_{\mu}=(-\frac{5}{2},\pm
\frac{1}{2})\right\rangle ,
\end{equation}
(type II)
\begin{equation}
G_{\frac{3}{2},\pm\frac{1}{2}}^{-(2)}\sim\frac{1}{2}\left[  \left(
\begin{array}
[c]{cc}%
\frac{73}{2} & \pm\frac{15}{2}\\
\pm\frac{15}{2} & \frac{5}{2}%
\end{array}
\right)  \alpha_{-1}^{\mu}\alpha_{-1}^{\nu}+\left(
\begin{array}
[c]{c}%
\frac{29}{\sqrt{2}}\\
\pm\frac{5}{\sqrt{2}}%
\end{array}
\right)  \alpha_{-2}^{\mu}\right]  \left\vert f_{\mu}=(-\frac{5}{2},\pm
\frac{1}{2})\right\rangle .
\end{equation}
Note that a linear combination of these two states produces a pure $\phi$ DGS%

\begin{equation}
G_{\frac{3}{2},\pm\frac{1}{2}}^{-}\sim\left[  (\partial\phi)^{2}-\frac
{1}{\sqrt{2}}\partial^{2}\phi\right]  e^{\pm\frac{i}{\sqrt{2}}X-\frac{5}%
{2}\phi}\left\vert 0\right\rangle . \label{2.14}%
\end{equation}
The DGS corresponding to discrete momenta of $\psi_{\frac{3}{2},\pm\frac{1}%
{2}}^{+}$ are degenerate, i.e., the type I and type II DGS are linearly
dependent:
\begin{equation}
G_{\frac{3}{2},\pm\frac{1}{2}}^{+}\sim\left[  \left(
\begin{array}
[c]{cc}%
-\frac{1}{2} & \pm\frac{3}{2}\\
\pm\frac{3}{2} & -\frac{5}{2}%
\end{array}
\right)  \alpha_{-1}^{\mu}\alpha_{-1}^{\nu}+\left(
\begin{array}
[c]{c}%
\frac{1}{\sqrt{2}}\\
\mp\frac{5}{\sqrt{2}}%
\end{array}
\right)  \alpha_{-2}^{\mu}\right]  \left\vert f_{\mu}=(\frac{1}{2},\pm\frac
{1}{2})\right\rangle \label{2.15.}%
\end{equation}

There is no pure $\phi$ DGS here. In general, the $\psi^{+}$ sector has fewer
DGS than the $\psi^{-}$ sector at the same discrete momenta, as a result, the
pure $\phi$ DGS only arise at the minus sector. This fact is related to the
degeneracy of the DGS in the plus sector. Historically the $\psi^{+}$ sector
discrete states arise when one considers the singular gauge transformation
constructed from the difference of the two plus gauge states
\cite{GKN,GKN1,GKN2,GKN3,Polyakov1}.

\subsubsection{Generating the Discrete ZNS}

In this section, we will give a general formula for the DGS. In general, there
are many DGS for each discrete momentum. The higher the momentum is, the more
numerous the DGS are found. We first express the discrete states in
Eq.(\ref{JM}) in terms of Schur polynomials, which are defined as follows:%

\begin{equation}
Exp\left(  \sum\limits_{k=1}^{\infty}a_{k}x^{k}\right)  =\sum\limits_{k=0}%
^{\infty}S_{k}(\{a_{k}\})x^{k}%
\end{equation}
where $S_{k}$ is the Schur polynomial, a function of $\{a_{k}\}=\{a_{i}%
:i\in\mathbb{Z}_{k}\}$. Performing the operator products in Eq.(\ref{JM}) ,
the discrete states $\psi_{J,M}^{\pm}$ can be written as%

\begin{equation}%
\begin{split}
\psi_{J,M}^{\pm}\sim &  \prod\limits_{i=1}^{J-M}\int\frac{dz_{i}}{2\pi i}%
z_{i}^{-2J}\prod\limits_{j<k}^{J-M}(z_{j}-z_{k})^{2}\\
&  Exp\left[  \sum\limits_{i=1}^{J-M}[-i\sqrt{2}X(z_{i})]+\sqrt{2}%
(iJX(0)+(-1\pm J)\phi(0))\right]  .
\end{split}
\label{3.2t}%
\end{equation}

We can write%

\begin{equation}
\prod\limits_{j<k}^{J-M}(z_{j}-z_{k})^{2}=\sum\limits_{f}\left\vert
\begin{array}
[c]{cccc}%
1 & z_{f_{1}} & \cdots & z_{f_{1}}^{J-M-1}\\
z_{f_{2}} & z_{f_{2}}^{2} & \cdots & z_{f_{2}}^{J-M-1}\\
\vdots & \vdots & \ddots & \vdots\\
z_{f_{J-M}}^{J-M-1} & z_{f_{J-M}}^{J-M} & \cdots & z_{f_{J-M}}^{2J-2M-2}%
\end{array}
\right\vert , \label{3.3t}%
\end{equation}
and Taylor expand $X(z_{i})$ around $z_{i}=0$%

\begin{equation}
e^{-i\sqrt{2}X(z_{i})}=e^{-i\sqrt{2}X(0)}\left[  \sum\limits_{k=0}^{\infty
}S_{k}\left(  \{\frac{-i\sqrt{2}}{k!}\partial^{k}X(0)\}\right)  z_{i}%
^{k}\right]  . \label{3.4t}%
\end{equation}
In Eq.(\ref{3.3t}) the sum is over all permutations $f=(f_{1},...,f_{J-M})$ of
$(1,2...,J-M)$. Putting Eq.(\ref{3.3t}) and Eq.(\ref{3.4t}) into
Eq.(\ref{3.2t}), and using the symmetry of the integrand over the index $i$,
we have%

\begin{equation}
\psi_{J,M}^{\pm}\sim\left\vert
\begin{array}
[c]{cccc}%
S_{2J-1} & S_{2J-2} & \cdots & S_{J+M}\\
S_{2J-2} & S_{2J-3} & \cdots & S_{J+M-1}\\
\vdots & \vdots & \ddots & \vdots\\
S_{J+M} & S_{J+M-1} & \cdots & S_{2M+1}%
\end{array}
\right\vert Exp\left[  \sqrt{2}(iMX(0)+(-1\pm J)\phi(0))\right]  \label{3.5.}%
\end{equation}
with $S_{k}=S_{k}\left(  \{\frac{-i\sqrt{2}}{k!}\partial^{k}X(0)\}\right)  $
and $S_{k}=0$ if $k<0$. We will denote the rank $(J-M)$ determinant in
Eq.(\ref{3.5.}) as $\Delta(J,M,-i\sqrt{2}X)$. As a by-product, comparing the
two definitions in Eq.(\ref{JM}) we can use Eq.(\ref{3.5.}) to deduce a
mathematical identity relating the determinants of rank $(J-M)$ and $(J+M)$,%

\begin{equation}
\Delta(J,M,-i\sqrt{2}X)=(-1)^{J+M+1}\Delta(J,-M,i\sqrt{2}X).
\end{equation}

We now begin to study the DGS. One first notes that the DGS in Eq.(\ref{2.14})
can be generated by $\int\frac{dz}{2\pi i}e^{-\sqrt{2}\phi(z)}\psi_{\frac
{1}{2},\pm\frac{1}{2}}^{-}(0)$. In general it is also possible to write down
explicitly one of the many ZNS for each discrete momentum in the $\psi^{-}$
sector as follows
\begin{equation}%
\begin{split}
G_{J,M}^{-}  &  \sim\left[  \int\frac{dz}{2\pi i}e^{-\sqrt{2}\phi(z)}\right]
\psi_{J-1,M}^{-}\\
&  \sim S_{2J-1}(\{\frac{-\sqrt{2}}{k!}\partial^{k}\phi\})\Delta
(J-1,M,-i\sqrt{2}X)e^{iMX+(-1-J)\phi}.
\end{split}
\label{3.7.}%
\end{equation}
Using Eq.(\ref{3.2.}), it can be verified explicitly after a length algebra
that they are primary, and are of dimension 1. For $M=J-1$ Eq.(\ref{3.7.}) are
pure $\phi$ states, but orthogonal to the pure $X$ discrete physical states at
the same momenta, and are therefore ZNS. For general M the polynomial
prefactor in Eq.(\ref{3.7.}) factorizes into pure $\phi$ and pure $X$ parts,
and are still orthogonal to the physical states at the same momenta. They are,
therefore, also ZNS. This is also suggested by the following result
\cite{Winfinity,Winfinity2,Klebanov1}%

\begin{equation}
\int\frac{dz}{2\pi i}\psi_{J_{1},M_{1}}^{-}\psi_{J_{2},M_{2}}^{-}\sim0
\end{equation}
where the r.h.s. is meant to be a DGS. We thus have explicitly obtained a DGS
for each $\psi^{-}$ discrete momentum. We stress that there are still other
DGS in this sector, for example, the states%

\begin{equation}
G_{J,M}^{\prime-}\sim\left[  \int\frac{dz}{2\pi i}e^{-\sqrt{2}\psi(z)}\right]
^{J-M}\psi_{M,M}^{-}%
\end{equation}
can be shown to be of dimension 1. Since they are pure $\phi$ states, they are
also DGS. This expression reminds us of Eq.(\ref{JM}). However, there is no
$SU(2)$ structure in the $\phi$ direction, and the usual techniques of $c=1$
2d conformal field theory cannot be applied. The pure $\phi$ DGS are only
found in the minus sector.

For the plus sector, the operator products of the discrete states defined in
Eq.(\ref{JM}) form a $w_{\infty}$ algebra
\cite{Winfinity,Winfinity2,Klebanov1}%

\begin{equation}
\int\frac{dz}{2\pi i}\psi_{J_{1},M_{1}}^{+}\psi_{J_{2},M_{2}}^{+}=(J_{2}%
M_{1}-J_{1}M_{2})\psi_{J_{1}+J_{2}-1,M_{1}+M_{2}}^{+}. \label{3.10.}%
\end{equation}
(Again, the r.h.s. is up to a DGS.) We can subtract two positive norm discrete
states to obtain a pure gauge state as following
\begin{equation}%
\begin{split}
G_{J,M}^{+}=  &  (J+M+1)^{-1}\int\frac{dz}{2\pi i}\left[  \psi_{1,-1}%
^{+}(z)\psi_{J,M+1}^{+}(0)+\psi_{J,M+1}^{+}(z)\psi_{1,-1}^{+}(0)\right] \\
\sim &  (J-M)!\Delta(J,M,-i\sqrt{2}X)Exp\left[  \sqrt{2}(iMX+(J-1)\phi)\right]
\\
&  +(-1)^{2J}\sum\limits_{j=1}^{J-M}(J-M-1)!\int\frac{dz}{2\pi i}%
\mathcal{D}(J,M,-i\sqrt{2}X(z),j)\\
\cdot &  Exp\left[  \sqrt{2}(i(M+1)X(z)+(J-1)\phi(z)-X(0))\right]
\end{split}
\label{3.11.}%
\end{equation}
where $\mathcal{D}(J,M,-i\sqrt{2}X(z),j)$ is defined as%

\begin{equation}
\mathcal{D}(J,M,-i\sqrt{2}X(z),j)=\left\vert
\begin{array}
[c]{ccccc}%
S_{2J-1} & S_{2J-2} & \cdots & \cdots & S_{J+M}\\
S_{2J-2} & S_{2J-3} & \cdots & \cdots & S_{J+M-1}\\
\vdots & \vdots & \ddots &  & \vdots\\
(-z)^{j-1-2J} & (-z)^{j-2J} &  &  & (-z)^{j-J-M-2}\\
\vdots & \vdots &  & \ddots & \vdots\\
S_{J+M} & S_{J+M-1} & \cdots & \cdots & S_{2M+1}%
\end{array}
\right\vert , \label{3.12L}%
\end{equation}
which is the same as $\Delta(J,M,-i\sqrt{2}X(z))$ except that the $j^{th}$ row
is replaced by $\{(-z)^{j-1-2J},(-z)^{j-2J}...\}$. As an example, with
Eq.(\ref{3.11.}) one easily obtains the state $G_{\frac{3}{2},\pm\frac{1}{2}%
}^{+}$ of Eq.(\ref{2.15.}).

\subsubsection{$w_{\infty}$ Charges and conclusion}

It was shown \cite{Winfinity,Winfinity2,Klebanov1} that the operators products
of the states $\psi_{J,M}^{+}$ defined in Eq.(\ref{JM}) satisfy the
$w_{\infty}$ algebra in Eq.(\ref{3.10.}). By construction Eq.(\ref{3.11.}) one
can easily see that the plus sector DGS $G_{J,M}^{+}$ carry the $w_{\infty}$
charges and can be considered as the symmetry parameters of the theory. In
fact, the operator products of the DGS $G_{J,M}^{+}$ of Eq.(\ref{3.11.}) form
the same $w_{\infty}$ algebra
\begin{equation}
\int\frac{dz}{2\pi i}G_{J_{1},M_{1}}^{+}(z)G_{J_{2},M_{2}}^{+}(0)=(J_{2}%
M_{1}-J_{1}M_{2})G_{J_{1}+J_{2}-1,M_{1}+M_{2}}^{+}(0) \label{3.13L}%
\end{equation}
where the r.h.s. is defined up to another DGS. The high energy limit of
Eq.(\ref{3.11.}) will be discussed in section V.E of part II of this review.

In summary, we have shown that the spacetime $w_{\infty}$ symmetry parameters
of $2D$ string theory come from solution of equations Eq.(\ref{I}) and
Eq.(\ref{II}). This argument is valid also in the case of $26D$ (or $10D$)
string theory although it would be very difficult to exhaust all the solutions
of the ZNS \cite{Lee4,LEO,JCLee}. This difficulty is, of course, related to
the high dimensionality of spacetime. The DGS we introduced in the old
covariant quantization in this chapter seem to be related to the ghost sectors
and the ground ring structure \cite{Ring,Ring1} in the BRST quantization of
the theory.

\subsection{2D superstring theory}

In this section, we will generalize our results in the previous section to
$N=1$ super-Liouville theory in the worldsheet supersymmetric way
\cite{ChungLee2}. We will work out the DGS of the Neveu-Schwarz sector in the
zero ghost picture. We first discuss the $N=1$ super-Liouville theory and set
up the notations. We then calculate the general formula for discrete
positive-norm states in a worldsheet superfield form. Finally a general
formula of discrete ZNS or DGS will be presented and $w_{\infty}$ charges will
then be calculated.

\subsubsection{2D super-Liouville theory}

The $N=1$ two dimensional supersymmetric Liouville action is given by
\cite{Distler}
\begin{equation}
S=\frac{1}{8\pi}\int d^{2}\mathbf{z}[g^{\alpha\beta}(\partial_{\alpha
}\mathbf{X}\partial_{\beta}\mathbf{X}+\partial_{\alpha}\mathbf{\Phi}%
\partial_{\beta}\mathbf{\Phi})-Q\hat{\mathbf{Y}}\mathbf{\Phi}]
\end{equation}
where $\mathbf{\Phi}$ is the super-Liouville field, $\hat{\mathbf{Y}}$ the
superfield curvature, $d\mathbf{z}=dzd\theta$ and with $\mathbf{X}^{\mu
}=\left(
\begin{array}
[c]{c}%
\mathbf{\Phi}\\
\mathbf{X}%
\end{array}
\right)  $,
\begin{equation}
\mathbf{X}^{\mu}(z,\theta,\bar{z},\bar{\theta})=X^{\mu}+\theta\psi^{\mu}%
+\bar{\theta}\bar{\psi}^{\mu}+\theta\bar{\theta}F^{\mu}.
\end{equation}
Bold faced variables denote superfields hereafter.

For $\hat{c}=1=\frac{2}{3}c$ theory $Q$, which represents the background
charge of the super-Liouville field, is set to be $2$ so that the total
conformal anomaly cancels that from conformal and superconformal ghost contribution.

The equations of motion show that the left and right-moving components of
$\mathbf{X}^{\mu}$ decouple, and the auxiliary fields $F^{\mu}$ vanish. As a
result, we need to consider only one of the chiral sectors, while the other
(anti-holomorphic) sector has a similar formula. The stress energy tensor is%

\begin{equation}
\mathbf{T}_{zz}=-\frac{1}{2}\mathbf{D}\mathbf{X}^{\mu}\mathbf{D}^{2}%
\mathbf{X}_{\mu}-\frac{1}{2}Q\mathbf{D}^{3}\mathbf{\Phi}=T_{F}+\theta T_{B}
\label{2.3.}%
\end{equation}

with%

\begin{equation}%
\begin{split}
T_{F}  &  =-\frac{1}{2}\partial X^{\mu}\partial X_{\mu}-\frac{1}{2}%
Q\partial^{2}X^{0}+\frac{1}{2}\psi^{\mu}\partial\psi_{\mu},\\
T_{B}  &  =-\frac{1}{2}\psi^{\mu}\partial X_{\mu}-\frac{1}{2}Q\partial\psi_{0}%
\end{split}
\end{equation}
where $\mathbf{D}=\partial_{\theta}+\theta\partial_{z}$, and now
$\mathbf{X}^{\mu}=X^{\mu}(z)+\theta\psi^{\mu}(z)$.

For the Neveu-Schwarz sector, if we define the mode expansion by
\begin{equation}
\partial_{z}X^{\mu}=-\sum\limits_{n=-\infty}^{\infty}z^{-n-1}(\alpha_{n}%
^{0},i\alpha_{n}^{1}),
\end{equation}%
\begin{equation}
\psi^{\mu}=-\sum\limits_{r\in\mathbb{Z}+\frac{1}{2}}z^{-r-\frac{1}{2}}%
(b_{r}^{0},ib_{n}^{1}),
\end{equation}
then we have
\begin{equation}
\lbrack\alpha_{m}^{\mu},\alpha_{n}^{\nu}]=n\eta^{\mu\nu}\delta_{m+n}%
,\qquad\{b_{r}^{\mu},b_{s}^{\nu}\}=\eta^{\mu\nu}\delta_{r+s}.
\end{equation}

With the Minkowski metric $\eta^{\mu\nu}= \left(
\begin{array}
[c]{cc}%
-1 & 0\\
0 & 1
\end{array}
\right)  $, $Q^{\mu}= \left(
\begin{array}
[c]{c}%
2\\
0
\end{array}
\right)  $ and the zero modes $\alpha^{\mu}_{0}=f^{\mu}= \left(
\begin{array}
[c]{c}%
\epsilon\\
p
\end{array}
\right)  $, we find the super-Virasoro generators as modes of $T_{F}$ and
$T_{B}$ ,%

\begin{equation}%
\begin{split}
L_{n}  &  =\left(  \frac{n+1}{2}Q^{\mu}+f^{\mu}\right)  \alpha_{\mu,n}%
+\frac{1}{2}\sum\limits_{k\neq0}:\alpha_{\mu,-k}\alpha_{n+k}^{\mu}:+\frac
{1}{2}\sum\limits_{r\in\mathbb{Z}+\frac{1}{2}}\left(  r+n+\frac{1}{2}\right)
:b_{-r}^{\mu}b_{n+r,\mu}:,\\
L_{0}  &  =\frac{1}{2}\left(  Q^{\mu}+f^{\mu}\right)  f_{\mu}+\sum
\limits_{k=1}^{\infty}:\alpha_{\mu,-k}\alpha_{k}^{\mu}:+\frac{1}{2}%
\sum\limits_{r\in\mathbb{Z}+\frac{1}{2}}\left(  r+\frac{1}{2}\right)
:b_{-r}^{\mu}b_{r,\mu}:,\\
G_{r}  &  =\sum\limits_{s\in\mathbb{Z}+\frac{1}{2}}\alpha_{r-s}^{\mu}b_{\mu
,s}+\left(  r+\frac{1}{2}\right)  Q^{\mu}b_{\mu,r}.
\end{split}
\end{equation}

The vacuum $\left|  0\right\rangle $ is annihilated by all $\alpha^{\mu}_{n}$
and $b^{\mu}_{r}$ with $n > 0$ and $r > 0$. In the old covariant quantization
of the theory, physical states $\left|  \psi\right\rangle $ are those
satisfying the conditions%

\begin{equation}%
\begin{split}
G_{\frac{1}{2}}\left\vert \psi\right\rangle  &  =G_{\frac{3}{2}}\left\vert
\psi\right\rangle =0\\
and\qquad L_{0}\left\vert \psi\right\rangle  &  =\frac{1}{2}\left\vert
\psi\right\rangle .
\end{split}
\label{2.9.}%
\end{equation}

\subsubsection{World-sheet superfield form of the discrete states}

With Eq.(\ref{2.3.}) one can easily check that the two branches of massless
\textquotedblleft tachyon"
\begin{equation}
T^{\pm}(p)=\int d\mathbf{z}e^{ip\mathbf{X}+(\pm|p|-1)\mathbf{\Phi}}%
\end{equation}
are positive norm physical states. It was also known that there exists
discrete momentum physical states. Writing $\int d\mathbf{z}\mathbf{\Psi
}_{J,\pm J}^{(\pm)}=T^{(\pm)}(\pm J)$, the discrete states in the
\textquotedblleft material gauge" are
\begin{equation}
\mathbf{\Psi}_{J,M}^{(\pm)}\sim(H^{-})^{J-M}\mathbf{\Psi}_{J,J}^{(\pm)}%
\sim(H^{+})^{J+M}\mathbf{\Psi}_{J,-J}^{(\pm)} \label{3.2..}%
\end{equation}
where
\begin{equation}
H^{\pm}=\sqrt{2}\int d\mathbf{z}e^{\pm i\mathbf{X}(\mathbf{z})},\qquad
H^{0}=\int d\mathbf{z}\mathbf{D}\mathbf{X}%
\end{equation}
are zero modes of the level $2$ $SU(2)_{\kappa=2}$ Kac-Moody algebra in
$\hat{c}=1$ $2d$ superconformal field theory. Here we note that the NS sector
corresponds to states with $J\in\mathbb{Z}$ while the Ramond sector
corresponds to those with $J\in\mathbb{Z}+\frac{1}{2}$.

To find the explicit expressions for the discrete states, we first define the
super-Schur polynomials,
\begin{equation}
\mathbf{S}_{k}(-i\mathbf{X})=\frac{\mathbf{D}^{k}e^{-i\mathbf{X}}}%
{[k/2]!}e^{i\mathbf{X}}%
\end{equation}
where $\left[  \frac{k}{2}\right]  $ denotes the integral part of $\frac{k}%
{2}$ , as the $N=1$ generalization to the Schur polynomial $S_{k}$ , which is
defined as
\begin{equation}
Exp\left(  \sum\limits_{k=1}^{\infty}a_{k}x^{k}\right)  =\sum\limits_{k=0}%
^{\infty}S_{k}(\{a_{k}\})x^{k}.
\end{equation}
Note that $S_{k}(\{i\partial^{m}\mathbf{X}/m!\})=\mathbf{S}_{2k}(i\mathbf{X}%
)$. Direct integration shows that
\begin{equation}%
\begin{split}
\int d\mathbf{z}_{1}\frac{1}{(z_{1}-z-\theta_{1}\theta)^{n}}f(\mathbf{X}_{1})
&  =\frac{\mathbf{D}^{2n-1}f(\mathbf{X})}{(n-1)!}\\
&  =\frac{\partial_{z}^{2n-1}(f_{z}^{\prime n}f(X)}{(n-1)!}.
\end{split}
\label{3.6.}%
\end{equation}
Using Eq.(\ref{3.6.}) we obtain
\begin{equation}%
\begin{split}
\mathbf{\Psi}_{J,J-1}^{\pm}  &  \sim\mathbf{S}_{2J-1}(-i\mathbf{X}%
)e^{i(J-1)\mathbf{X}+(\pm J-1)\mathbf{\Phi}}\\
&  =\frac{1}{(J-1)!}[-i\partial^{J-1}(e^{-iX^{1}}\psi^{1})+\theta
\partial^{J-1}e^{-iX^{1}}]e^{iJ\mathbf{X}+(\pm J-1)\mathbf{\Phi}}.
\end{split}
\end{equation}
For example, by
\begin{equation}
(-\mathbf{D}^{2r}\mathbf{\Phi}^{0},i\mathbf{D}^{2r}\mathbf{X}^{1})\rightarrow
b_{-r}^{\mu},\qquad(-\mathbf{D}^{2n}\mathbf{X}^{0},i\mathbf{D}^{2n}%
\mathbf{X}^{1})\rightarrow\alpha_{-n}^{\mu}, \label{3.8.}%
\end{equation}
we have
\begin{equation}
\mathbf{\Psi}_{1,0}^{+}=\mathbf{D}\mathbf{X}\rightarrow b_{\frac{1}{2}}%
^{1}\left\vert f^{\mu}=(0,0)\right\rangle
\end{equation}
and
\begin{equation}%
\begin{split}
\mathbf{\Psi}_{2,\pm1}^{+}  &  =[-i\mathbf{D^{3}}\mathbf{X}-\mathbf{D}%
\mathbf{X}\mathbf{D}^{2}\mathbf{X}]e^{\pm i\mathbf{X}+\mathbf{\Phi}}\\
&  \rightarrow\lbrack-b_{-\frac{3}{2}}^{1}+b_{-\frac{1}{2}}^{1}\alpha_{-1}%
^{1}]\left\vert f^{\mu}=(1,\pm1)\right\rangle .
\end{split}
\end{equation}
They can be checked to satisfy the physical state conditions in Eq.(\ref{2.9.}).

Performing the operator products in Eq.(\ref{3.2..}) , the discrete states
$\mathbf{\Psi}_{J,M}^{\pm}$ are
\begin{equation}%
\begin{split}
\mathbf{\Psi}_{J,M}^{\pm}\sim &  \prod\limits_{i=1}^{J-M}\int d\mathbf{z}%
_{i}\mathbf{z}_{i0}^{-J}\prod\limits_{j<k}^{J-M}\mathbf{z}_{jk}\\
&  Exp\left[  \sum\limits_{i=1}^{J-M}[-i\mathbf{X}(\mathbf{z}_{i}%
)]+(iJ\mathbf{X}(\mathbf{z}_{0})+(1\pm J)\mathbf{\Phi}(\mathbf{z}%
_{0}))\right]
\end{split}
\end{equation}
where $\mathbf{z}_{ab}=z_{a}-z_{b}-\theta_{a}\theta_{b}$. If we write
$\mathbf{z}_{ab}=\mathbf{z}_{a0}-\mathbf{z}_{b0}-(\theta_{a}-\theta
_{0})(\theta_{b}-\theta_{0})$, and use $\int d\mathbf{z}_{a}(\theta_{a}%
-\theta_{0})\mathbf{z}_{a0}^{-n}f(\mathbf{X}_{a})=\mathbf{D}^{2n-2}%
f(\mathbf{X}_{0})/(n-1)!$, we get, for $M=J-2$,
\begin{equation}
\mathbf{\Psi}_{J,J-2}^{\pm}\sim\lbrack2\mathbf{S}_{2J-3}\mathbf{S}%
_{2J-1}+\mathbf{S}_{2J-2}\mathbf{S}_{2J-2}]e^{i(J-2)\mathbf{X}+(\pm
J-1)\mathbf{\Phi}}. \label{3.12.}%
\end{equation}
The vertex operators correspond to the upper components of Eq.(\ref{3.12.}),
i.e.,
\begin{equation}%
\begin{split}
\int d\theta\mathbf{\Psi}_{J,J-2}^{\pm}\sim\lbrack &  (iJ\psi^{1}+(\pm
J-1)\psi^{0})(S_{J-1}^{2}+2S_{J-\frac{3}{2}}^{NS}S_{J-\frac{1}{2}}^{NS})\\
&  -2J(S_{J}S_{J-\frac{3}{2}}^{NS}-S_{J-1}S_{J-\frac{1}{2}}^{NS}%
)]e^{i(J-2)X^{1}+(\pm J-1)X^{0}}%
\end{split}
\end{equation}
where $S_{J}=S_{J}(\{-i\partial^{m}\mathbf{X}/m!\})$ and $S_{k+\frac{1}{2}%
}^{NS}=\sum\limits_{m=0}^{k}\frac{-iS_{m}\partial^{k-m}\psi^{1}}{(k-m)!}$.
Using Eq.(\ref{3.8.}) and Eq.(\ref{3.12.}) it is found that
\begin{equation}
\mathbf{\Psi}_{2,0}^{+}\rightarrow\lbrack2b_{-\frac{1}{2}}^{1}b_{-\frac{3}{2}%
}^{1}+\alpha_{-1}^{1}\alpha_{-1}^{1}]\left\vert f^{\mu}=(1,0)\right\rangle .
\end{equation}
It can be checked that it satisfies the physical state conditions
Eq.(\ref{2.9.})

For $M = J - 3$, a straighforward calculation gives%

\begin{equation}%
\begin{split}
\mathbf{\Psi}_{J,J-3}^{\pm}\sim\lbrack &  3!\mathbf{S}_{2J-1}\mathbf{S}%
_{2J-3}\mathbf{S}_{2J-5}+3!\mathbf{S}_{2J-2}\mathbf{S}_{2J-3}\mathbf{S}%
_{2J-4}\\
&  -\frac{3!}{1!2!}\mathbf{S}_{2J-1}\mathbf{S}_{2J-4}^{2}-\frac{3!}%
{2!1!}\mathbf{S}_{2J-2}^{2}\mathbf{S}_{2J-5}]e^{i(J-3)\mathbf{X}+(\pm
J-1)\mathbf{\Phi}}.
\end{split}
\end{equation}

It is now easy to write down an expression for general $M$ ,
\begin{equation}
\mathbf{\Psi}_{J,M}^{\pm}\sim\left\vert
\begin{array}
[c]{cccc}%
\mathbf{S}_{2J-1} & \mathbf{S}_{2J-2} & \cdots & \mathbf{S}_{J+M}\\
\mathbf{S}_{2J-2} & \mathbf{S}_{2J-3} & \cdots & \mathbf{S}_{J+M-1}\\
\vdots & \vdots & \ddots & \vdots\\
\mathbf{S}_{J+M} & \mathbf{S}_{J+M-1} & \cdots & \mathbf{S}_{2M+1}%
\end{array}
\right\vert ^{\prime}Exp[(iM\mathbf{X}(\mathbf{z}_{0})+(-1\pm J)\mathbf{\Phi
}(\mathbf{z}_{0}))] \label{3.16.}%
\end{equation}
with $\mathbf{S}_{k}=\mathbf{S}_{k}(-i\mathbf{X}(\mathbf{z}_{0}))$ and
$\mathbf{S}_{k}=0$ if $k<0$. We will denote the rank $(J-M)$ \textquotedblleft
primed"-determinant in Eq.(\ref{3.16.}) as $\mathbf{\Delta}^{\prime
}(J,M,-i\mathbf{X})$, which (by definition) has all the signed terms in the
normal determinant, except with a multiplicity of the multinomial coefficient
$\frac{(J-M)!}{n_{a}!n_{b}!\ldots}$ for the term $\mathbf{S}_{a}^{n_{a}%
}\mathbf{S}_{b}^{n_{b}}\ldots$ (where $\sum\limits_{a}n_{a}=J-M$).

\subsubsection{Discrete ZNS and $w_{\infty}$ charges}

It was known \cite{Bouwknegt,Itoh,Itoh1} that the discrete states in
Eq.(\ref{3.2..}) satisfy the $w_{\infty}$ algebra
\begin{equation}
\int d\mathbf{z}\mathbf{\Psi}_{J_{1},M_{1}}^{+}(\mathbf{z})\mathbf{\Psi
}_{J_{2},M_{2}}^{+}(\mathbf{0})=(J_{2}M_{1}-J_{1}M_{2})\mathbf{\Psi}%
_{J_{1}+J_{2}-1,M_{1}+M_{2}}^{+}(\mathbf{0}),
\end{equation}%
\begin{equation}
\int d\mathbf{z}\mathbf{\Psi}_{J_{1},M_{1}}^{-}(\mathbf{z})\mathbf{\Psi
}_{J_{2},M_{2}}^{-}(\mathbf{0})\sim0 \label{4.2.}%
\end{equation}
where the RHS is defined up to a DGS.

In general, there are two types of ZNS in the old covariant quantization of
the theory,

Type I:%

\begin{equation}
\left\vert \psi\right\rangle =G_{-\frac{1}{2}}\left\vert \chi\right\rangle
\qquad where\qquad G_{\frac{1}{2}}\left\vert \chi\right\rangle =G_{\frac{3}%
{2}}\left\vert \chi\right\rangle =L_{0}\left\vert \tilde{\chi}\right\rangle
=0. \label{4.3.}%
\end{equation}

Type II:
\begin{equation}%
\begin{split}
\left\vert \psi\right\rangle =\left(  G_{-\frac{3}{2}}+2L_{-1}G_{-\frac{1}{2}%
}\right)  \left\vert \tilde{\chi}\right\rangle \qquad where\qquad &
G_{\frac{1}{2}}\left\vert \tilde{\chi}\right\rangle =G_{\frac{3}{2}}\left\vert
\tilde{\chi}\right\rangle =0\\
&  (L_{0}+1)\left\vert \tilde{\chi}\right\rangle =0.
\end{split}
\label{4.4.}%
\end{equation}
They satisfy the physical state conditions Eq.(\ref{2.9.}), and have zero
norm. There is an infinite number of continuum momentum ZNS solutions for
Eq.(\ref{4.3.}) and Eq.(\ref{4.4.}) . However, as far as the dynamics is
concerned, we are only interested in those with discrete momentum.

At mass level one, $f_{\mu}(f^{\mu}+Q^{\mu})=0$, only ZNS of type I are found:
$f_{\mu}\alpha_{-1}^{\mu}\left\vert f\right\rangle $, where $\left\vert
f\right\rangle =:e^{ip\mathbf{X}+\epsilon\mathbf{\Phi}}:\left\vert
0\right\rangle $. The DGS $G_{1,0}^{-}=:\mathbf{D}\mathbf{\Phi}e^{-2\Phi
}:\left\vert 0\right\rangle $ corresponds to the momentum of $\mathbf{\Psi
}_{1,0}^{-}$. There is no corresponding DGS for $\mathbf{\Psi}_{1,0}%
^{+}=:\mathbf{D}\mathbf{X}:$.

At the next mass level, $f_{\mu}(f^{\mu}+Q^{\mu})=-2$, $N_{\mu\nu}=-N_{\nu\mu
}$ and $M_{\mu}=2N_{\mu\nu}(f^{\nu}+Q^{\nu})$, the type I ZNS is found to be
\begin{equation}
\left\vert \psi\right\rangle =[(M_{\mu}f_{\nu}\alpha_{-1}^{\mu}b_{-\frac{1}%
{2}}^{\nu}+M_{\mu}b_{-\frac{3}{2}}^{\nu}+2N_{\mu\nu}\alpha_{-1}^{\mu}%
b_{-\frac{1}{2}}^{\nu}]\left\vert f\right\rangle ,
\end{equation}
while the type II state is%

\begin{equation}
\left\vert \psi\right\rangle =[(2f_{\mu}f_{\nu}+\eta_{\mu\nu})\alpha_{-1}%
^{\mu}b_{-\frac{1}{2}}^{\nu}+(3f_{\mu}-Q_{\mu})b_{-\frac{3}{2}}^{\mu
}]\left\vert f\right\rangle .
\end{equation}
As in the bosonic Liouville theory \cite{ChungLee1}, the ZNS corresponding to
the discrete momenta of $\mathbf{\Psi}_{2,\pm1}^{+}$ are degenerate, i.e., the
type I and type II gauge states are linearly dependent:%

\begin{equation}
G_{2,\pm1}^{+}\sim\left[  \left(
\begin{array}
[c]{cc}%
1 & \mp2\\
\mp2 & 3
\end{array}
\right)  \alpha_{-1}^{\mu}b_{-\frac{1}{2}}^{\nu}+\left(
\begin{array}
[c]{c}%
-1\\
\pm3
\end{array}
\right)  b_{-\frac{3}{2}}^{\mu}\right]  \left\vert f^{\mu}=(1,\pm
1)\right\rangle . \label{4.7.}%
\end{equation}
For the minus sector, type I DGS is
\begin{equation}
G_{2,\pm1}^{-,I}\sim\left[  \left(
\begin{array}
[c]{cc}%
3 & \pm2\\
\pm2 & 1
\end{array}
\right)  \alpha_{-1}^{\mu}b_{-\frac{1}{2}}^{\nu}+\left(
\begin{array}
[c]{c}%
1\\
\pm1
\end{array}
\right)  b_{-\frac{3}{2}}^{\mu}\right]  \left\vert f^{\mu}=(-3,\pm
1)\right\rangle , \label{4.8.}%
\end{equation}
and type II DGS is
\begin{equation}
G_{2,\pm1}^{-,II}\sim\left[  \left(
\begin{array}
[c]{cc}%
17 & \pm6\\
\pm6 & 3
\end{array}
\right)  \alpha_{-1}^{\mu}b_{-\frac{1}{2}}^{\nu}+\left(
\begin{array}
[c]{c}%
11\\
\pm3
\end{array}
\right)  b_{-\frac{3}{2}}^{\mu}\right]  \left\vert f^{\mu}=(-3,\pm
1)\right\rangle . \label{4.9.}%
\end{equation}
Note that $3G_{2,\pm1}^{-,I}-G_{2,\pm1}^{-,II}$ is a \textquotedblleft pure
$\mathbf{\Phi}$" DGS, similar to the DGS in the bosonic Liouville theory.

We now apply the scheme used in \cite{ChungLee1} to derive a general formula
for the DGS. From Eq.(\ref{4.2.}), the DGS in the minus sector can be written
down explicitly as follows%

\begin{equation}%
\begin{split}
\mathbf{G}_{J,M}^{-}  &  \sim\left[  \int d\mathbf{z}e^{-\Phi(\mathbf{z}%
)}\right]  \mathbf{\Psi}_{J-1,M}^{-}(\mathbf{0})\\
&  \sim\mathbf{S}_{2J-1}(-\mathbf{\Phi})\mathbf{\Delta}^{\prime\lbrack
iM\mathbf{X}+(-1-J)\mathbf{\Phi}]}.
\end{split}
\end{equation}

We thus have explicitly obtained a DGS for each $\mathbf{\Psi}^{-}$ discrete
momentum. However, there are still other DGS in this sector, for example, the
states
\begin{equation}
\mathbf{G}_{J,M}^{\prime-}\sim\left[  \int d\mathbf{z}e^{-\Phi(\mathbf{z}%
)}\right]  ^{J-M}\mathbf{\Psi}_{M,M}^{-}(\mathbf{0})
\end{equation}
can be shown to satisfy the physical state conditions. Since they are
\textquotedblleft pure $\mathbf{\Phi}$" states, they are also DGS. For
example, $\mathbf{G}_{1,0}^{-}=\mathbf{D}\mathbf{\Phi}e^{-\mathbf{\Phi}}$ and
$\mathbf{G}_{2,\pm1}^{-}=[-\mathbf{D}^{3}\mathbf{\Phi}+\mathbf{D}\mathbf{\Phi
}\mathbf{D}^{2}\mathbf{\Phi}]e^{\pm i\mathbf{X}-3\mathbf{\Phi}}$, which is a
linear combination of Eq.(\ref{4.8.}) and Eq.(\ref{4.9.}).

For the plus sector, we can subtract two (distinct) positive norm discrete
states at the same momentum to obtain a pure DGS
\begin{equation}
\mathbf{G}_{J,M}^{+}=(J+M+1)^{-1}\int d\mathbf{z}\left[  \mathbf{\Psi}%
_{1,-1}^{+}(\mathbf{z})\mathbf{\Psi}_{J,M+1}^{+}(\mathbf{0})-\mathbf{\Psi
}_{J,M+1}^{+}(\mathbf{z})\mathbf{\Psi}_{1,-1}^{+}(\mathbf{0})\right]  .
\label{4.12.}%
\end{equation}
As an example, with Eq.(\ref{4.12.}) one finds
\begin{equation}%
\begin{split}
\mathbf{G}_{2,\pm1}^{+}=[  &  \pm3i\mathbf{D}^{3}\mathbf{X}+\mathbf{D}%
^{3}\mathbf{\Phi}+3i\mathbf{D}^{2}\mathbf{X}\mathbf{D}\mathbf{X}\\
&  \pm2i\mathbf{D}^{2}\mathbf{X}\mathbf{D}\mathbf{\Phi}\pm2i\mathbf{D}%
\mathbf{X}\mathbf{D}^{2}\mathbf{\Phi}+\mathbf{D}\mathbf{\Phi}\mathbf{D}%
^{2}\mathbf{\Phi}]e^{\pm i\mathbf{X}+\mathbf{\Phi}}%
\end{split}
\end{equation}
which is exactly the state we found in Eq.(\ref{4.7.}). We thus have
explicitly obtained a DGS for each $\mathbf{\Psi}^{+}$ momentum.

By construction in Eq.(\ref{4.12.}) one can see that $\mathbf{G}_{J,M}^{+}$
carry the $w_{\infty}$ charges and serve as the symmetry parameters of the
theory. In fact, their operator products form the same $w_{\infty}$ algebra
\begin{equation}
\int d\mathbf{z}\mathbf{G}_{J_{1},M_{1}}^{+}(\mathbf{z})\mathbf{G}%
_{J_{2},M_{2}}^{+}(\mathbf{0})=(J_{2}M_{1}-J_{1}M_{2})\mathbf{G}_{J_{1}%
+J_{2}-1,M_{1}+M_{2}}^{+}(\mathbf{0})
\end{equation}
where the RHS is defined up to another DGS.

We have demonstrated that the spacetime $w_{\infty}$ symmetry parameters in
the $2D$ superstring theory come from solution of equations Eq.(\ref{4.3.})
and Eq.(\ref{4.4.}). This phenomenon should survive in the more realistic high
dimensional string theory \cite{Lee4,LEO,KaoLee} , although it would be
difficult to find the general solution of ZNS (due to the high dimensionality
of spacetime). The DGS in the old covariant quantization of the theory is
related to the ground ring structure in the BRST approach.%

\setcounter{equation}{0}
\renewcommand{\theequation}{\arabic{section}.\arabic{equation}}%

\section{Soliton ZNS in compact space and enhanced gauge symmetry}

In this chapter we calculate ZNS in the spectrum of compactified closed and
open string theories \cite{Lee1,Lee2}. For simplicity we will only do the
calculations on torus compactifications. The programs can certainly be
generalized to more complicated background geometries. We will see that there
exist soliton ZNS which generate various enhanced stringy symmetries of the theories.

\subsection{Compactified closed string}

In this section, we study soliton gauge states in the spectrum of bosonic
string compactified on torus. The enhanced Kac-Moody gauge symmetry, and thus
T-duality, is shown to be related to the existence of these soliton ZNS in
some moduli points.

\subsubsection{Soliton ZNS on $R^{25}\otimes T^{1}$}

In the simplest torus compactification, one coordinate of the string was
compactified on a circle of radius $R$%

\begin{equation}
X^{25}\left(  \sigma+2\pi,\pi\right)  =X^{25}\left(  \sigma,\pi\right)  +2\pi
R_{n} \label{2.1s}%
\end{equation}
The single valued condition of the wave function then restricts the allowed
momenta to be $p^{25}=m/R$ with $m,n\in Z$. The mode expansion of the
compactified coordinate for right (left) mover is%
\begin{align}
X_{R}^{25}  &  =\frac{1}{2}x^{25}+\left(  p^{25}-\frac{1}{2}nR\right)  \left(
\tau-\sigma\right)  +i\sum_{r\neq0}\frac{1}{r}\alpha_{r}^{25}e^{-ir\left(
\tau-\sigma\right)  },\label{2.2s}\\
X_{L}^{25}  &  =\frac{1}{2}x^{25}+\left(  p^{25}+\frac{1}{2}nR\right)  \left(
\tau+\sigma\right)  +i\sum_{r\neq0}\frac{1}{r}\overset{\thicksim}{\alpha}%
_{r}^{25}e^{-ir\left(  \tau+\sigma\right)  }. \label{2.3s}%
\end{align}
We have normalized the string tension to be $\frac{1}{4\pi T}=1$ or
$\alpha^{\prime}=2$ .\mathstrut The Virasoro operators can be written as%
\begin{align}
L_{0}  &  =\frac{1}{2}\left(  p^{25}-\frac{1}{2}nR\right)  +\frac{1}{2}%
p^{\mu^{2}}+\sum_{n=1}^{\infty}\alpha_{-n}\cdot\alpha_{n},\label{2.4s}\\
\overset{\thicksim}{L_{0}}  &  =\frac{1}{2}\left(  p^{25}+\frac{1}%
{2}nR\right)  +\frac{1}{2}p^{\mu^{2}}+\sum_{n=1}^{\infty}\overset{\thicksim
}{\alpha}_{-n}\cdot\overset{\thicksim}{\alpha}_{n}, \label{2.5s}%
\end{align}
and%

\begin{align}
L_{m}  &  =\frac{1}{2}\alpha_{0}^{2}+\sum_{-\infty}^{\infty}\alpha_{m-n}%
\cdot\alpha_{n},\label{2.6s}\\
\overset{\thicksim}{L}_{m}  &  =\frac{1}{2}\overset{\thicksim}{\alpha}_{0}%
^{2}+\sum_{-\infty}^{\infty}\overset{\thicksim}{\alpha}_{m-n}\cdot
\overset{\thicksim}{\alpha}_{n}\text{ }\left(  m\neq0\right)  \label{2.7s}%
\end{align}
where%

\begin{align}
\alpha_{0}^{25}  &  =p^{25}-\frac{1}{2}nR\equiv p_{R}^{25},\label{2.8s}\\
\overset{\thicksim}{\alpha}_{0}^{25}  &  =p^{25}+\frac{1}{2}nR\equiv
p_{L}^{25}, \label{2.9s}%
\end{align}
and the $25d$ momentum is $\alpha_{0}^{\mu}=\overset{\thicksim}{\alpha}%
_{0}^{\mu}=p^{\mu}\equiv k^{\mu}$. In the old covariant quantization of the
theory, in addition to the physical propagating states, there are four types
of ZNS in the spectrum%

\begin{equation}
I.a\text{ \ }\left\vert \psi\right\rangle =L_{-1}\left\vert \chi\right\rangle
\text{ where }L_{m}\left\vert \chi\right\rangle =0,\left(  \overset{\thicksim
}{L}_{m}-\delta_{m}\right)  \left\vert \chi\right\rangle =0,\text{ }\left(
m=0,1,2,\ldots\right)  , \label{2.10s}%
\end{equation}

\begin{equation}
II.a\text{ \ }\left\vert \psi\right\rangle =\left(  L_{-2}+\frac{3}{2}%
L_{-1}^{2}\right)  \left\vert \chi\right\rangle \text{ where }\left(
L_{m}+\delta_{m}\right)  \left\vert \chi\right\rangle =0,\left(
\overset{\thicksim}{L}_{m}-\delta_{m}\right)  \left\vert \chi\right\rangle
=0,\text{ }\left(  m=0,1,2,\ldots\right)  , \label{2.11s}%
\end{equation}
and by interchanging all left and right mover operators, one gets $I.b$ and
$II.b$ states. Type II states are ZNS only at critical space-time dimension.
We will only calculate type $a$ states. Similar results can be easily obtained
for type $b$ states. For type $I.a$ state, the $m=0$ constraint of
Eq.(\ref{2.10s}) gives%

\begin{equation}
M^{2}=\frac{m^{2}}{R^{2}}+\frac{1}{4}n^{2}R^{2}+N+\overset{\thicksim}{N}-1,
\label{2.12s}%
\end{equation}

\begin{equation}
N-\overset{\thicksim}{N}=mn-1 \label{2.13s}%
\end{equation}
where $N\equiv\sum\limits_{n=1}^{\infty}\alpha_{-n}\cdot\alpha_{n}$ and
$\overset{\thicksim}{N\text{ }}\equiv\sum\limits_{n=1}^{\infty}%
\overset{\thicksim}{\alpha}_{-n}\cdot\overset{\thicksim}{\alpha}_{n}$. For
massless $M^{2}=0$ states, $N+\overset{\thicksim}{N}=0$ or $1$. The solutions
of Eq.(\ref{2.12s}) and Eq.(\ref{2.13s}) are%

\begin{equation}
N=0,\overset{\thicksim}{N}=1,m=n=0\text{ (\textit{any R})} \label{2.14s}%
\end{equation}
or%

\begin{equation}
N=\overset{\thicksim}{N}=0,m=n=\pm1,R=\sqrt{2}. \label{2.15s}%
\end{equation}
Eq.(\ref{2.15s}) gives us our first soliton ZNS. It is easy to write down the
explicit form of $\left\vert \chi\right\rangle $ and $\left\vert
\psi\right\rangle $, and impose the $m\neq0$ constraints of Eq.(\ref{2.10s}).
There are also a vector and a scalar ZNS in Eq.(\ref{2.14s}). Similar results
can be obtained for the type $I.b$ state. In this case, $m=-n=\pm1$. There is
no type II solution in the massless case. We note that there are massless
soliton ZNS only when $R=\sqrt{2}$ which is known as self-dual point in the
moduli space. The vertex operators of all ZNS are calculated to be%

\begin{equation}
k_{\mu}\theta_{\nu}\partial X_{R}^{\mu}\overset{\_}{\partial}X_{L}^{\nu
}e^{ikx};\text{ }L\leftrightarrow R, \label{2.16s}%
\end{equation}

\begin{align}
&  k_{\mu}\partial X_{R}^{\mu}\overset{\_}{\partial}X_{L}^{25}e^{ikx}%
,\label{2.17s}\\
&  k_{\mu}\overset{\_}{\partial}X_{L}^{\mu}\partial X_{R}^{25}e^{ikx}%
,\label{2.18s}\\
&  k_{\mu}\partial X_{R}^{\mu}e^{\pm i\sqrt{2}X_{L}^{25}}e^{ikx}%
,\label{2.19s}\\
&  k_{\mu}\partial X_{L}^{\mu}e^{\pm i\sqrt{2}X_{R}^{25}}e^{ikx}.
\label{2.20s}%
\end{align}
It is easy to see that the three ZNS of Eq.(\ref{2.18s}) and (Eq.(\ref{2.20s})
form a representation of $SU(2)_{R}$ Kac-Moody algebra. Similarly, Eqs.
Eq.(\ref{2.17s}) and Eq.(\ref{2.19s}) form a representation of $SU(2)_{L}$
Kac-Moody algebra. The vector ZNS in Eq.(\ref{2.16s}) are responsible for the
gauge symmetry of graviton and antisymmetric tensor field. We see that the
self-dual point $R=\sqrt{2}$ is very special even from the gauge sector point
of view.

\subsubsection{Soliton ZNS on $R^{26-D}\otimes T^{D}$}

In this section we compactify $D$ coordinates on a $D$-dimensional torus
$T^{D}\equiv\frac{R^{D}}{2\pi\Lambda^{D}}$%

\begin{equation}
\overset{\rightarrow}{X}\left(  \sigma+2\pi,\pi\right)  =\overset{\rightarrow
}{X}\left(  \sigma,\pi\right)  +2\pi\overset{\rightarrow}{L} \label{3.1s}%
\end{equation}
with

\mathstrut%
\begin{equation}
\overset{\rightarrow}{L}=\sum_{i=1}^{D}n_{i}\left(  R_{i}\frac
{\overset{\rightarrow}{e}_{i}}{\sqrt{2}}\right)  \in\left(  \Lambda
^{D}\right)  \label{3.2s}%
\end{equation}
where $\Lambda^{D}$ is a $D$-dimensional lattice with a basis $\left\{
R_{1}\frac{\overset{\rightarrow}{e}_{1}}{\sqrt{2}},R_{2}\frac
{\overset{\rightarrow}{e}_{2}}{\sqrt{2}},\ldots,R_{D}\frac
{\overset{\rightarrow}{e}_{D}}{\sqrt{2}}\right\}  $. We have chosen
$\left\vert \overset{\rightarrow}{e}_{i}\right\vert ^{2}=2$. The allowed
momenta $\overset{\rightarrow}{p}$ take values on the dual lattice of
$\Lambda^{D}$%

\begin{equation}
\overset{\rightarrow}{p}=\sum_{i=1}^{D}m_{i}\left(  \frac{1}{R_{i}}\sqrt
{2}\overset{\rightarrow}{e}_{i}^{\star}\right)  \in\left(  \Lambda^{D}\right)
^{\star}. \label{3.3s}%
\end{equation}
The basis of $\left(  \Lambda^{D}\right)  ^{\star}$ is $\left\{  \frac{1}%
{R1}\sqrt{2}\overset{\rightarrow}{e}_{1}^{\star},\frac{1}{R_{2}}\sqrt
{2}\overset{\rightarrow}{e}_{2}^{\star},\ldots,\frac{1}{R_{D}}\sqrt
{2}\overset{\rightarrow}{e}_{D}^{\star}\right\}  $ and we have
$\overset{\rightarrow}{e}_{i}\cdot\overset{\rightarrow}{e}_{i}^{\star}%
=\delta_{ij}$. The mode expansion of the compactified coordinates is%

\begin{align}
\overset{\rightarrow}{X}_{R}  &  =\frac{1}{2}\overset{\rightarrow}{x}+\left(
\overset{\rightarrow}{p}-\frac{1}{2}\overset{\rightarrow}{L}\right)  \left(
\tau-\sigma\right)  +i\sum_{r\neq0}\frac{1}{r}\alpha_{r}^{25}e^{-ir\left(
\tau-\sigma\right)  },\label{3.4s}\\
\overset{\rightarrow}{X}_{L}  &  =\frac{1}{2}\overset{\rightarrow}{x}+\left(
\overset{\rightarrow}{p}+\frac{1}{2}\overset{\rightarrow}{L}\right)  \left(
\tau+\sigma\right)  +i\sum_{r\neq0}\frac{1}{r}\alpha_{r}^{25}e^{-ir\left(
\tau+\sigma\right)  }. \label{3.5s}%
\end{align}
The right and left momenta are defined to be $\overset{\rightarrow}{p}%
_{R}=\left(  \overset{\rightarrow}{p}-\frac{1}{2}\overset{\rightarrow
}{L}\right)  $ and $\overset{\rightarrow}{p}_{L}=\left(  \overset{\rightarrow
}{p}+\frac{1}{2}\overset{\rightarrow}{L}\right)  $. It can be shown that the
$2D$-vector $\left(  \overset{\rightarrow}{p}_{R},\overset{\rightarrow}{p}%
_{L}\right)  $ build an even self-dual Lorentzian lattice $\Gamma_{D,D}$,
which guarantees the string one loop modular invariance of the theory
\cite{Narain,Narain1}. The moduli space of the theory is \cite{Giveon}%

\begin{equation}
\mu=\frac{SO\left(  D,D\right)  }{SO\left(  D\right)  \times SO\left(
D\right)  }/O\left(  D,D,Z\right)  \label{3.6s}%
\end{equation}
where $O(D,D,Z)$ is the discrete T-duality group and $dim$ $\mu=D^{2}$. To
complete the parametrization of the moduli space, one needs to introduce an
antisymmetric tensor field $B_{ij}$ in the bosonic string action. This will
modify the right (left) momenta to be%

\begin{equation}
\overset{\rightarrow}{p}_{R}=\left(  \overset{\rightarrow}{p}_{B}-\frac{1}%
{2}\overset{\rightarrow}{L}\right)  , \label{3.7s}%
\end{equation}

\begin{equation}
\overset{\rightarrow}{p}_{L}=\left(  \overset{\rightarrow}{p}_{B}+\frac{1}%
{2}\overset{\rightarrow}{L}\right)  \label{3.8s}%
\end{equation}
where%

\begin{equation}
\overset{\rightarrow}{p}_{B}=\sum_{i,j}\left(  m_{i}\frac{1}{R_{i}}\sqrt
{2}\overset{\rightarrow}{e}_{i}^{\star}-n_{j}\frac{1}{\sqrt{2}R_{i}}%
B_{ij}\overset{\rightarrow}{e}_{i}^{\star}\right)  . \label{3.9s}%
\end{equation}
We are now ready to discuss the gauge state. As a first step, we restrict
ourselves to moduli space with $B_{ij}=0$ . For the type $I.a$ state, the
$m=0$ constraint of Eq.(\ref{2.10s}) for massless states gives%

\begin{equation}
N+\overset{\thicksim}{N}+\overset{\rightarrow}{p}^{2}+\frac{1}{4}%
\overset{\rightarrow}{L}^{2}=1, \label{3.10s}%
\end{equation}

\begin{equation}
N-\overset{\thicksim}{N}=\sum_{i}m_{i}n_{i}-1. \label{3.11s}%
\end{equation}
It is easy to see $N+\overset{\thicksim}{N}=0$ or $1$. For
$N+\overset{\thicksim}{N}=1$, $m_{i}=n_{i}=0$, we have trivial ZNS solutions.
Soliton ZNS exists for the case $N+\overset{\thicksim}{N}=0$ and the following
moduli points%

\begin{equation}
R_{i}=\sqrt{2},e_{i}^{I}=\sqrt{2}\delta_{i}^{I}\text{ }\left(  i=1,2,\ldots
,d\right)  \label{3.12s}%
\end{equation}
with $m_{i}=n_{i}=\pm1$, and $m_{j}=n_{j}=0$ for $d<j\leq D$. In each case,
the ZNS and soliton ZNS form a representation of $SU(2)^{d}$ algebra. Similar
results can be easily obtained for the type $I.b$ soliton ZNS. As in section
II, there is no massless type II soliton ZNS. We now discuss $B_{ij}\neq0$
case. For illustration, we choose $D=2$. In this case $B_{ij}=B\epsilon_{ij}$,
and one has four moduli parameters $R_{1},R_{2},B,$ and $\overset{\rightarrow
}{e}_{1}\cdot\overset{\rightarrow}{e}_{2}$. For type $I.a$ state, the $m=0$
constraint of Eq.(\ref{2.9s}) gives%

\begin{equation}
N+\overset{\thicksim}{N}+\overset{\rightarrow}{p}_{B}^{2}+\frac{1}%
{4}\overset{\rightarrow}{L}^{2}=1, \label{3.13s}%
\end{equation}

\begin{equation}
N-\overset{\thicksim}{N}=m_{1}n_{1}+m_{2}n_{2}-1. \label{3.14s}%
\end{equation}
soliton ZNS exists only for $N+\overset{\thicksim}{N}=0$. For the moduli point%

\begin{equation}
R_{1}=R_{2}=\sqrt{2},B=\frac{1}{2},\overset{\rightarrow}{e}_{1}=\left(
\sqrt{2},0\right)  ,\overset{\rightarrow}{e}_{2}=\left(  -\sqrt{\frac{1}{2}%
},\sqrt{\frac{3}{2}}\right)  , \label{3.15s}%
\end{equation}
one gets six soliton ZNS with momenta $\overset{\rightarrow}{p}_{R}$ being the
six root vectors of $SU(3)_{R}$. Together with two other trivial ZNS
corresponding to $N=0,\overset{\thicksim}{N}=1$ , they form the
Frenkel-Kac-Segal \cite{Goddard} representation of $SU(3)_{k=1}$ Kac-Moody
algebra. Note that $\overset{\rightarrow}{e}_{1},\overset{\rightarrow}{e}_{2}$
are the two simple roots of $SU(3)$ and $\overset{\rightarrow}{e}_{1}^{\star
}=\left(  \sqrt{\frac{1}{2}},\sqrt{\frac{1}{6}}\right)  ,\overset{\rightarrow
}{e}_{2}^{\star}=\left(  0,\sqrt{\frac{2}{3}}\right)  $. The six sets of
winding number are $\left(  m_{1},n_{1},m_{2},n_{2}\right)  =\left(
1,1,0,0\right)  ,\left(  -1,-1,0,0\right)  ,\left(  0,0,1,1\right)  ,\left(
0,0,-1,-1\right)  ,\left(  1,1,1,0\right)  ,\left(  -1,-1,-1,0\right)  $.
Similar results can be obtained for type $I.b$ soliton ZNS. The ZNS (including
soliton ZNS) thus form a representation of enhanced $SU(3)_{R}\otimes$
$SU(3)_{L}$ at the moduli point of Eq.(\ref{3.15s}). In general, we expect
that all enhanced Kac-Moody gauge symmetry at any moduli point should have a
realization on soliton ZNS.

\subsubsection{ Massive soliton ZNS}

In this section we derive the massive soliton ZNS at the first massive level
$M^{2}=2$. We will find that soliton ZNS exists at infinite number of moduli
points. One can also show that they exist at an infinite number of massive
level. The existence of these massive soliton ZNS implies that there is an
infinite enhanced gauge symmetry structure of compactified string theory. For
type $I.a$ state, the $m=0$ constraint of Eq.(\ref{2.10s}) gives%

\begin{equation}
\frac{m^{2}}{R^{2}}+\frac{1}{4}n^{2}R^{2}+N+\overset{\thicksim}{N}=3,
\label{4.1s}%
\end{equation}

\begin{equation}
N-\overset{\thicksim}{N}=mn-1, \label{4.2s}%
\end{equation}
which implies $N+\overset{\thicksim}{N}=0,1,2,3$. Eq.(\ref{4.1s}) and
Eq.(\ref{4.2s}) can be easily solved as following:

1. $N+\overset{\thicksim}{N}=3:$%

\begin{equation}
m=n=0,N=1,\overset{\thicksim}{N}=2,\text{\textit{any R.}} \label{4.3s}%
\end{equation}

2. $N+\overset{\thicksim}{N}=2:$%

\begin{align}
mn  &  =1,N=\overset{\thicksim}{N}=1,R=\sqrt{2},\nonumber\\
mn  &  =-1,N=0,\overset{\thicksim}{N}=2,R=\sqrt{2}. \label{4.4s}%
\end{align}

3. $N+\overset{\thicksim}{N}=1:$%

\begin{align}
mn  &  =2,N=1,\overset{\thicksim}{N}=0,R=2,1.\text{ }\left(  T-duality\right)
,\nonumber\\
mn  &  =0,N=0,\overset{\thicksim}{N}=1,R=\frac{\left\vert m\right\vert }%
{\sqrt{2}},\frac{2\sqrt{2}}{\left\vert m\right\vert }.\text{ }\left(
T-duality\right)  . \label{4.5s}%
\end{align}

4. $N+\overset{\thicksim}{N}=0:$%

\begin{equation}
mn=1,N=\overset{\thicksim}{N}=1,R=2\pm\sqrt{2}.\text{ }\left(
T-duality\right)  \label{4.6s}%
\end{equation}
where we have included a T-duality transformation $R\rightarrow\frac{2}{R}$
for some moduli points. Note that Eq.(\ref{4.5s}) tells us that massive
soliton ZNS exists at an infinite number of moduli point. For type $II.a$
state, the $m=0$ constraint of Eq.(\ref{2.11s}) gives%

\begin{equation}
\frac{m^{2}}{R^{2}}+\frac{1}{4}n^{2}R^{2}+N+\overset{\thicksim}{N}=2,
\label{4.7s}%
\end{equation}

\begin{equation}
N-\overset{\thicksim}{N}=mn-2, \label{4.8s}%
\end{equation}
which implies $N+\overset{\thicksim}{N}=0,1,2$. Eq.(\ref{4.7s}) and
Eq.(\ref{4.8s}) can be solved as following:

1. $N+\overset{\thicksim}{N}=2:$%

\begin{equation}
m=n=0,N=0,\overset{\thicksim}{N}=2,\text{\textit{any R.}} \label{4.9s}%
\end{equation}

2. $N+\overset{\thicksim}{N}=1:$%

\begin{equation}
mn=1,N=0,\overset{\thicksim}{N}=1,R=\sqrt{2}. \label{4.10s}%
\end{equation}

3. $N+\overset{\thicksim}{N}=0:$%

\begin{equation}
mn=2,N=\overset{\thicksim}{N}=0,R=2,1.\text{ }\left(  T-duality\right)  .
\label{4.11s}%
\end{equation}
The vertex operators of all soliton ZNS can be easily calculated and written
down. Similar results can be obtained for type $b$ ZNS. One can also calculate
propagating soliton states by using the same technique. We summarize the
moduli points which exist soliton state and soliton ZNS as following:

\mathstrut a. Soliton ZNS\textit{ :}%

\begin{equation}
R=\sqrt{2},2\pm\sqrt{2},\frac{\left\vert m\right\vert }{\sqrt{2}},\frac
{2\sqrt{2}}{\left\vert m\right\vert },2,1. \label{4.12s}%
\end{equation}

\mathstrut b. \textit{Soliton states :}%

\begin{equation}
R=\sqrt{2},2\pm\sqrt{2},\frac{\left\vert m\right\vert }{\sqrt{2}},\frac
{2\sqrt{2}}{\left\vert m\right\vert },\frac{\left\vert m\right\vert }{2}%
,\frac{4}{\left\vert m\right\vert }. \label{4.13s}%
\end{equation}
In Eq.(\ref{4.12s}) and Eq.(\ref{4.13s}), $m\in Z_{+}$. There is one
interesting remark we would like to point out by the end of this section. One
notes that in the second case of Eq.(\ref{4.5s}), instead of specifying
$M^{2}=2$, in general we have%

\begin{equation}
\frac{m^{2}}{R^{2}}+\frac{1}{4}n^{2}R^{2}=M^{2} \label{4.14s}%
\end{equation}
with $mn=0$. For say $R=\sqrt{2}$, one gets $M^{2}=\frac{m^{2}}{2}\left(
n=0\right)  $. This means that we have an infinite number of massive soliton
ZNS at any higher massive level of the spectrum. One can even explicitly write
down the vertex operators of these soliton ZNS. We conjecture that the
$w_{\infty}$ symmetry of $2D$ string theory \cite{ChungLee1,2Dstring} can be
realized in these soliton ZNS. Other moduli points also consist of higher
massive soliton ZNS in the spectrum.

Many known spacetime symmetries of string theory can be shown to be related to
the existence of ZNS in the spectrum. The Heterotic ZNS for the $10D$
Heterotic string \cite{Lee4} and the discrete ZNS \cite{ChungLee1,2Dstring}
for the toy $2D$ string are such examples. We have introduced soliton ZNS for
compactified closed string in this chapter, and have related them to the
enhanced Kaluza-Klein Kac-Moody symmetries in the theory. In many cases,
especially for the massive states, it is easier to study stringy symmetries in
the ZNS sector than in the propagating spectrum directly.

Since the discrete T-duality symmetry group for bosonic closed string is the
Weyl subgroup of the enhanced gauge group, it can also be considered as due to
the existence of soliton ZNS. It is not clear whether other discrete duality
symmetry group can be understood in this way. Finally, it would be interesting
to consider more complicated compactification, e.g. orbifold and Calabi-Yau
compactifications and study the relation between soliton ZNS and duality symmetries.

\subsection{Compactified open string}

In this section, we study the mechanism of enhanced gauge symmetry of bosonic
open string compactified on torus by analyzing the ZNS (nonzero winding of
Wilson line) in the spectrum \cite{Lee2}. Unlike the closed string case
discussed in the previous section, we will find that the soliton ZNS exist
only at massive levels.

These soliton ZNS correspond to the existence of enhanced massive stringy
symmetries with transformation parameters containing both Einstein and
Yang-Mills indices in the case of Heterotic string \cite{Lee4}. In the T-dual
picture, these symmetries exist only at some discrete values of compactified
radii when $N$ $D$-branes are coincident.

\subsubsection{Chan-Paton ZNS}

We first discuss ZNS of uncompactified open string with Chan-Paton factor and
its implication on on-shell symmetry and Ward identity. For simplicity, we
consider the oriented $U\left(  N\right)  $ case. The vertex operators of
massless gauge state is%

\begin{equation}
\theta^{a}\lambda_{ij}^{a}k\cdot\partial xe^{ikx}%
\end{equation}
where $\lambda\in U\left(  N\right)  ,i\in N,j\in\overset{\_}{N}$ and $a\in$
adjoint representation of $U\left(  N\right)  $. The on-shell conformal
deformation and $U\left(  N\right)  $ gauge symmetry to lowest order in the
weak background field approximation are $\left(  \Box\theta^{a}=0,\Box
\equiv\partial_{\mu}\partial^{\mu}\right)  $%

\begin{equation}
\delta T=\lambda_{ij}^{a}\partial_{\mu}\theta^{a}\partial x^{\mu},
\end{equation}
and%

\begin{equation}
\delta A_{\mu}^{a}=\partial_{\mu}\theta^{a}%
\end{equation}
with T the energy momentum tensor and $A_{\mu}^{a}$ the massless gauge field.

One can verify the corresponding Ward identity by calculating e.g., $1$-vector
and $3$-tachyons four point correlators. The amplitude is calculated to be%

\begin{align}
T_{\mu}^{abcd}  &  =\int\prod_{i=1}^{4}dx_{i}\left\langle e^{ik_{1}x_{1}%
}\partial x_{\mu}e^{ik_{2}x_{2}}e^{ik_{3}x_{3}}e^{ik_{4}x_{4}}\right\rangle
T_{r}\left(  \lambda^{a}\lambda^{b}\lambda^{c}\lambda^{d}\right) \\
&  =\frac{\Gamma\left(  -\frac{s}{2}-1\right)  \Gamma\left(  -\frac{t}%
{2}-1\right)  }{\Gamma\left(  \frac{u}{2}+1\right)  }\left[  k_{3\mu}\left(
\frac{s}{2}+1\right)  -k_{1\mu}\left(  \frac{t}{2}+1\right)  \right]  \times
T_{r}\left(  \lambda^{a}\lambda^{b}\lambda^{c}\lambda^{d}\right) \nonumber
\end{align}
In Eq.(2.4), $s,t$ and $u$ are the usual Mandelstam variables. One can then
verify the Ward identity%

\begin{equation}
\theta^{b}k_{2}^{\mu}T_{\mu}^{abcd}=0.
\end{equation}

We now discuss the massive ZNS. The vertex operator of type I massive vector
ZNS is%

\begin{equation}
\theta_{\mu}^{a}\lambda_{ij}^{a}\left[  k\cdot\partial x\partial x^{\mu
}+\partial^{2}x^{\mu}\right]  e^{ikx}.
\end{equation}
We note that the ZNS polarization contains both Einstein and Yang-Mills
indices. This is very similar to the $10d$ closed Heterotic string case
\cite{Lee4,LEO}. The only difference is that in the Heterotic string, one
could have more than one Yang-Mills index. The on-shell conformal deformation
and the mixed Einstein-Yang-Mills-type symmetry to lowest order weak field
approximation are $\left(  \left(  \Box-2\right)  \theta_{\mu}^{a}%
=\partial\cdot\theta^{a}=0\right)  $%

\begin{equation}
\delta T=\lambda_{ij}^{a}\partial_{\mu}\theta_{\nu}^{a}\partial x^{\mu
}\partial x^{\nu}+\lambda_{ij}^{a}\theta_{\mu}^{a}\partial^{2}x^{\mu}%
\end{equation}

and%

\begin{equation}
\delta M_{\mu\nu}^{a}=\partial_{\mu}\theta_{\nu}^{a}+\partial_{\nu}\theta
_{\mu}^{a}.
\end{equation}

One can also derive the corresponding massive Ward identity by calculating the
decay rate of one massive state to three tachyons. The most general amplitude
is calculated to be%

\begin{equation}
A^{abcd}=\varepsilon^{a}\varepsilon^{c}\varepsilon^{d}\left(  \varepsilon
_{\mu\nu}^{b}T^{\mu\nu}+\varepsilon_{\mu}^{b}T^{\mu}\right)  Tr\left(
\lambda^{a}\lambda^{b}\lambda^{c}\lambda^{d}\right)  \label{9c}%
\end{equation}

where%

\begin{equation}
T^{\mu\nu}=\frac{\Gamma\left(  -\frac{s}{2}-1\right)  \Gamma\left(  -\frac
{t}{2}-1\right)  }{\Gamma\left(  \frac{u}{2}+2\right)  }\left\{  \frac{s}%
{2}\left(  \frac{s}{2}+1\right)  k_{3}^{\mu}k_{3}^{\nu}+\frac{t}{2}\left(
\frac{t}{2}+1\right)  k_{1}^{\mu}k_{1}^{\nu}-2\left(  \frac{s}{2}+1\right)
\left(  \frac{t}{2}+1\right)  k_{1}^{\mu}k_{3}^{\nu}\right\}
\end{equation}

and%

\begin{equation}
T^{\mu}=\frac{\Gamma\left(  -\frac{s}{2}-1\right)  \Gamma\left(  -\frac{t}%
{2}-1\right)  }{\Gamma\left(  \frac{u}{2}+2\right)  }\left\{  -k_{3}^{\mu
}\frac{s}{2}\left(  \frac{s}{2}+1\right)  -k_{1}^{\mu}\frac{t}{2}\left(
\frac{t}{2}+1\right)  \right\}  .
\end{equation}
In Eq.(\ref{9c}) $\varepsilon^{a}$ etc. are polarizations corresponding to
tachyons and $\left(  \varepsilon_{\mu\nu}^{b},\varepsilon_{\mu}^{b}\right)  $
is polarization of the massive state. The above amplitude satisfies the
following ward identity%

\begin{equation}
k_{\mu}\theta_{\nu}^{a}T^{\mu\nu}+\theta_{\mu}^{a}T^{\mu}=0
\end{equation}

Similar consideration can be applied to the following type II massive scalar
gauge state%

\begin{equation}
\left[  \frac{1}{2}\alpha_{-1}\cdot\alpha_{-1}+\frac{5}{2}k\cdot\alpha
_{-2}+\frac{3}{2}\left(  k\cdot\alpha_{-1}\right)  ^{2}\right]  \left|
k,l=0,i,j\right\rangle
\end{equation}

which corresponds to a \emph{massive} $U\left(  N\right)  $ symmetry.

\subsubsection{Chan-Paton soliton ZNS on $R^{25}\otimes T^{1}$}

We now discuss soliton ZNS on torus compactification of bosonic open string.
As is well known, the massless $U\left(  N\right)  $ gauge symmetry will be
broken in general after compactification unless N D-branes, in the T-dual
picture, are coincident. We will see that when $D$-branes are coincident, one
has enhancement of (unwinding) ZNS and the massless $U\left(  N\right)  $
symmetry will be recovered. These ZNS can be considered as charges or symmetry
parameters of $U\left(  N\right)  $ group.

In the discussion of open string compactification, one needs to turn on the
Wilson line or nonzero background gauge field in the compact direction. This
will effect the momentum in the compact direction, and the Virasoro operators become%

\begin{align}
L_{0}  &  =\frac{1}{2}\left(  \frac{2\pi l-\theta_{j}+\theta_{i}}{2\pi
R}\right)  ^{2}+\frac{1}{2}\left(  k^{\mu}\right)  ^{2}+\sum_{n=1}^{\infty
}\left(  \alpha_{-n}^{\mu}\alpha_{n}^{\mu}+\alpha_{-n}^{25}\alpha_{n}%
^{25}\right)  ,\label{3.1c}\\
L_{m}  &  =\frac{1}{2}\sum_{-\infty}^{\infty}\overset{\rightharpoonup}{\alpha
}_{m-n}\cdot\overset{\rightharpoonup}{\alpha}_{n}. \label{3.2c}%
\end{align}
Note that in Eq.(\ref{3.2c}), $\alpha_{0}^{25}\equiv p^{25}$ which also
appears in the first term in Eq.(\ref{3.1c}). k is the 25d momentum.
$\theta_{i},R$ are the gauge and space-time moduli respectively and $l$ is the
winding number in the compact direction. The spectrums of type I and type II
ZNS become%

\begin{equation}
M^{2}=\left(  \frac{2\pi l-\theta_{j}+\theta_{i}}{2\pi R}\right)  ^{2}+2I
\label{3.3c}%
\end{equation}
and%

\begin{equation}
M^{2}=\left(  \frac{2\pi l-\theta_{j}+\theta_{i}}{2\pi R}\right)
^{2}+2\left(  I+1\right)  \label{3.4c}%
\end{equation}
where $I=\sum\limits_{n=1}^{\infty}\left(  \alpha_{-n}^{\mu}\alpha_{n}^{\mu
}+\alpha_{-n}^{25}\alpha_{n}^{25}\right)  $.

For the massless case $I=l=0$, one gets $N^{2}$ massless solution from
equation Eq.(\ref{3.3c})%

\begin{equation}
k_{\mu}\alpha_{-1}^{\mu}\left\vert k,l=0,i,j\right\rangle \label{3.5c}%
\end{equation}
if all $\theta_{i}$ are equal, or in the T-dual picture when $N$ $D$-branes
are coincident. These $N^{2}$ massless ZNS correspond to the charges of
massless $U\left(  N\right)  $ gauge symmetry. There is no type II massless
solution in Eq.(\ref{3.4c}).

We are now ready to discuss the interesting massive case. For $M^{2}=2$ and
general moduli $\left(  R,\theta_{i}\right)  $,

1. $I=1,l=0$, one gets two ZNS solutions from Eq.(\ref{3.3c}):%

\begin{equation}
\left[  \left(  \varepsilon\cdot\alpha_{-1}\right)  \left(  k\cdot\alpha
_{-1}\right)  +\varepsilon\cdot\alpha_{-2}\right]  \left\vert
k,l=0,i,i\right\rangle ,\text{ }\varepsilon\cdot k=0 \label{3.6c}%
\end{equation}
and%

\begin{equation}
\left(  k\cdot\alpha_{-1}\alpha_{-1}^{25}+\alpha_{-2}^{25}\right)  \left\vert
k,l=0,i,i\right\rangle . \label{3.7c}%
\end{equation}

If all $\theta_{i}$ are equal, the $\left(  i,i\right)  $ is enhanced to
$\left(  i,j\right)  $. Eq.(\ref{3.7c}) implies a massive $U\left(  N\right)
$ symmetry with transformation parameter $\theta^{a}$. Eq.(\ref{3.6c}) implies
a massive Einstein-Yang-Mills-type symmetry with transformation parameter
$\theta_{\mu}^{a}$

2. $I=0,\frac{2\pi l-\theta_{j}+\theta_{i}}{2\pi R}=\pm\sqrt{2}$, one gets
solution from Eq.(\ref{3.3c})%

\begin{equation}
\left(  k\cdot\alpha_{-1}\pm\sqrt{2}\alpha_{-1}^{25}\right)  \left\vert
k,l,i,j\right\rangle . \label{3.8c}%
\end{equation}
Now since $\left\vert \theta_{i}-\theta_{j}\right\vert <2\pi$, for any given
$R$, there is at most one solution of $\left(  \left\vert l\right\vert
,\left\vert \theta_{i}-\theta_{j}\right\vert \right)  $. One is tempted to
consider the case%

\begin{equation}
\left(  k\cdot\alpha_{-1}\pm\sqrt{2}\alpha_{-1}^{25}\right)  \left\vert
k,l=\pm\sqrt{2}R,i,i\right\rangle . \label{3.9c}%
\end{equation}
That means in the moduli $\left(  R=\sqrt{2}n,\theta_{i}\right)  $ with $n\in
Z^{+}$, one has \emph{soliton} ZNS which imply a \emph{massive} $U\left(
1\right)  ^{N}$ symmetry. If all $\theta_{i}$ are equal, the $\left(
i,i\right)  $ is enhanced to $\left(  i,j\right)  $. Eq.(\ref{3.9c}) implies a
massive $U\left(  N\right)  $ symmetry at the \emph{discrete} values of moduli
point $R=\sqrt{2}n$. For example, in the T-dual picture, for $R=\sqrt{2}%
,l=\pm2$, and if all $D$-branes are coincident, we have an enhanced massive
$U\left(  N\right)  $ symmetry. This phenomenon is very different from the
massless case, where one gets enhanced $U\left(  N\right)  $ symmetry at
\emph{any} radius $R$ when $N$ $D$-branes are coincident.

We would like to point out that similar Einstein-Yang-Mills-type symmetry was
discovered before in the closed Heterotic string theory. There, however, one
could have more than one Yang-Mills indices on the transformation parameters.
For the type II states with $M^{2}=2$ in Eq.(\ref{3.4c}), $I=l=0$. One gets
one more $U\left(  N\right)  $ ZNS%

\begin{equation}
\left[  \frac{1}{2}\alpha_{-1}\cdot\alpha_{-1}+\frac{1}{2}\alpha_{-1}%
^{25}\alpha_{-1}^{25}+\frac{5}{2}k\cdot\alpha_{-2}+\frac{3}{2}\left(
k\cdot\alpha_{-1}\right)  ^{2}\right]  \left\vert k,l=0,i,j\right\rangle
\label{3.10c}%
\end{equation}
if all $\theta_{i}$ are equal.

For the general mass level, choosing $I=0$ and $i,j$ in Eq.(\ref{3.2c}), we
have $l/R=\pm M$. For say $R=\sqrt{2}$ and $l=\pm\sqrt{2}M$, which implies%

\begin{equation}
M^{2}=2n^{2},\text{ }n=0,1,2,\ldots\label{3.11c}%
\end{equation}
So we have Chan-Paton soliton ZNS at any higher massive level of the spectrum.
Similar result was found in Eq.(\ref{4.14s}) for the closed string case.

\part{Stringy symmetries of hard string scattering amplitudes}

As we mentioned in the introduction at the beginning of this review, there are
two main key ideas to probe symmetry of string theory. These are high energy
limit of string scatterings and the decoupling of ZNS in the OCFQ string
spectrum. In the part I of this review, we used only the idea of ZNS to
calculate stringy symmetries in various approaches. Although the results we
obtained are valid to all energies, only very limited stringy symmetries of
low mass level states can be calculated except for $2D$ strings. It will be,
for example, very complicated to do calculations for states with low spin at
higher mass levels. In the part II of this review, we will combine both
crucial two ideas to simplify the calculation.

The high energy Ward identities derived from the decoupling of 26D open
bosonic string ZNS, which combines the two key ideas of probing stringy
symmetry, will be used to explicitly prove \cite{ChanLee1,ChanLee2, CHL,
CHLTY1,CHLTY2,CHLTY3} Gross's two conjectures. An infinite number of linear
relations among high energy, fixed angle string scattering amplitudes of
different string states can be derived. Moreover, these linear relations can
be used to fix the proportionality constants or ratios among high energy,
fixed angle scattering amplitudes of different string states algebraically at
each fixed mass level.

The part II of this review is organized as following. Chapter V is one of the
main part of this review. We will use three different methods to explicitly
prove Gross conjectures \cite{ChanLee1,ChanLee2, CHL, CHLTY1,CHLTY2}. These
are the decoupling of high energy ZNS, the high energy Virasoro constraints
and a saddle-point calculation. In addition, we show that the high energy
limit of the discrete ZNS in $2D$ string theory constructed in part I form a
high energy $w_{\infty}$ symmetry. This result strongly suggests that the
linear relations obtained from decoupling of ZNS in $2D$ string theory are
indeed related to the hidden symmetry also for the $26D$ string theory.

In chapter VI, in addition to analyze ZNS in the helicity basis in the OCFQ
string spectrum, we will work out ZNS in the light-cone DDF construction of
string spectrum and ZNS in the BRST WSFT\cite{CLYang}. In chapter VII, we
discuss hard closed string scatterings \cite{Closed}. The KLT relation
\cite{KLT} will be extensively used. We also discuss string BCJ relation
\cite{BCJ1,BCJ2,BCJ3,BCJ4,BCJ5}, which is of much interest in the recent
development of calculation of field theory scattering amplitudes. In chapter
VIII, we calculate four classes of hard superstring scattering amplitudes and
derive the ratios among them \cite{susy}. In chapter IX, we discuss hard
string scattering from D-branes/ O-planes, and closed string decays to open
string \cite{Dscatt,O-plane,Decay}. Finally in chapter X, we discuss both hard
open and closed string scatterings in the compact spaces
\cite{Compact,Compact2}.%

\setcounter{equation}{0}
\renewcommand{\theequation}{\arabic{section}.\arabic{equation}}%

\section{Infinite linear relations among high energy, fixed angle string
scattering amplitudes}

In this chapter, we will use three different methods to explicitly prove Gross
conjectures for $26D$ bosonic open string theory. We will also show that the
high energy limit of discrete ZNS of $2D$ string constructed in part I form a
high energy $w_{\infty}$ symmetry. We begin with an example
\cite{ChanLee,ChanLee1,ChanLee2} to do the calculation.

\subsection{The first example}

For our purpose here, there are four ZNS at mass level \ $M^{2}$ $=4$. The
complete list of ZNS were calculated in chapter II, and the corresponding Ward
identities were calculated to be \cite{JCLee}%

\begin{equation}
k_{\mu}\theta_{\nu\lambda}\mathcal{T}_{\chi}^{(\mu\nu\lambda)}+2\theta_{\mu
\nu}\mathcal{T}_{\chi}^{(\mu\nu)}=0, \label{2.4.}%
\end{equation}%
\begin{equation}
(\frac{5}{2}k_{\mu}k_{\nu}\theta_{\lambda}^{\prime}+\eta_{\mu\nu}%
\theta_{\lambda}^{\prime})\mathcal{T}_{\chi}^{(\mu\nu\lambda)}+9k_{\mu}%
\theta_{\nu}^{\prime}\mathcal{T}_{\chi}^{(\mu\nu)}+6\theta_{\mu}^{\prime
}\mathcal{T}_{\chi}^{\mu}=0, \label{2.5.}%
\end{equation}%
\begin{equation}
(\frac{1}{2}k_{\mu}k_{\nu}\theta_{\lambda}+2\eta_{\mu\nu}\theta_{\lambda
})\mathcal{T}_{\chi}^{(\mu\nu\lambda)}+9k_{\mu}\theta_{\nu}\mathcal{T}_{\chi
}^{[\mu\nu]}-6\theta_{\mu}\mathcal{T}_{\chi}^{\mu}=0, \label{2.6.}%
\end{equation}%
\begin{equation}
(\frac{17}{4}k_{\mu}k_{\nu}k_{\lambda}+\frac{9}{2}\eta_{\mu\nu}k_{\lambda
})\mathcal{T}_{\chi}^{(\mu\nu\lambda)}+(9\eta_{\mu\nu}+21k_{\mu}k_{\nu
})\mathcal{T}_{\chi}^{(\mu\nu)}+25k_{\mu}\mathcal{T}_{\chi}^{\mu}=0,
\label{2.7.}%
\end{equation}
where $\theta_{\mu\nu}$ is transverse and traceless, and $\theta_{\lambda
}^{\prime}$ and $\theta_{\lambda}$ are transverse vectors. They are
polarizations of ZNS. In each equation, we have chosen, say, $v_{2}(k_{2}%
)$\ to be the vertex operators constructed from ZNS and $k_{\mu}\equiv
k_{2\mu}$. Note that Eq.(\ref{2.6.}) is the inter-particle Ward identity
corresponding to $D_{2}$ vector ZNS in Eq.(\ref{8b}) obtained by
antisymmetrizing those terms which contain $\alpha_{-1}^{\mu}\alpha_{-2}^{\nu
}$ in the original type I and type II vector ZNS \cite{Lee}. We will use 1 and
2 for the incoming particles and 3 and 4 for the scattered particles. In
Eq.(\ref{2.4.}) to Eq.(\ref{2.7.}), 1,3 and 4 can be any string states
(including ZNS) and we have omitted their tensor indices for the cases of
excited string states. For example, one can choose $v_{1}(k_{1})$\ to be the
vertex operator constructed from another ZNS which generates an inter-particle
Ward identity of the third massive level. The resulting Ward-identity of
Eq.(\ref{2.6.}) then relates scattering amplitudes of particles at different
mass level. $\mathcal{T}_{\chi}^{\prime}s$ in Eqs (\ref{2.4.}-\ref{2.7.}) are
the mass level $M^{2}$ $=4$, $\chi$-th order string-loop amplitudes.

At this point, \{$\mathcal{T}_{\chi}^{(\mu\nu\lambda)},\mathcal{T}_{\chi
}^{(\mu\nu)},\mathcal{T}_{\chi}^{\mu}$\} is identified to be the
\emph{amplitude triplet} \cite{Lee} of the spin-three state. $\mathcal{T}%
_{\chi}^{[\mu\nu]}$ is obviously identified to be the scattering amplitude of
the antisymmetric spin-two state with the same momenta as $\mathcal{T}_{\chi
}^{(\mu\nu\lambda)}$. Eq.(\ref{2.6.}) thus relates the scattering amplitudes
of two different string states at mass level $M^{2}$ $=4$. Note that
Eq.(\ref{2.4.}) to Eq.(\ref{2.7.}) are valid order by order and are
\emph{automatically} of the identical form in string perturbation theory. This
is consistent with Gross's argument through the calculation of high energy
scattering amplitudes. However, it is important to note that Eq.(\ref{2.4.})
to Eq.(\ref{2.7.}) are, in contrast to the high energy $\alpha^{\prime
}\rightarrow\infty$ result of Gross, valid to \emph{all} energy $\alpha
^{\prime}$ and their coefficients do depend on the center of mass scattering
angle $\phi_{CM}$ , which is defined to be the angle between
$\overrightarrow{k}_{1}$ and $\overrightarrow{k}_{3}$, through the dependence
of momentum $k$ .

We will calculate high energy limit of Eq.(\ref{2.4.}) to Eq.(\ref{2.7.})
without referring to the saddle point calculation in
\cite{GM,GM1,Gross,Gross1,GrossManes}. Let's define the normalized
polarization vectors%

\begin{equation}
e_{P}=\frac{1}{m_{2}}(E_{2},\mathrm{k}_{2},0)=\frac{k_{2}}{m_{2}},
\label{2.8.}%
\end{equation}

\begin{equation}
e_{L}=\frac{1}{m_{2}}(\mathrm{k}_{2},E_{2},0), \label{2.9..}%
\end{equation}%
\begin{equation}
e_{T}=(0,0,1) \label{2.10.}%
\end{equation}
in the CM frame contained in the plane of scattering. They satisfy the
completeness relation%

\begin{equation}
\eta^{\mu\nu}=\underset{\alpha,\beta}{%
{\textstyle\sum}
}e_{\alpha}^{\mu}e_{\beta}^{\nu}\eta^{\alpha\beta} \label{2.11.}%
\end{equation}
where $\mu,\nu=0,1,2$ and $\alpha,\beta=P,L,T.$ $Diag$ $\eta^{\mu\nu
}=(-1,1,1).$ One can now transform all $\mu,\nu$ coordinates in Eq.(\ref{2.4.}%
) to Eq.(\ref{2.7.}) to coordinates $\alpha,\beta$. For Eq.(\ref{2.4.}), we
have $\theta^{\mu\nu}=e_{L}^{\mu}e_{L}^{\nu}-e_{T}^{\mu}e_{T}^{\nu}$ or
$\theta^{\mu\nu}=e_{L}^{\mu}e_{T}^{\nu}+e_{T}^{\mu}e_{L}^{\nu}$ . In the high
energy $E\rightarrow$ $\infty,$ fixed angle $\phi_{CM}$ limit, one identifies
$e_{P}=e_{L}$ and Eq.(\ref{2.4.}) gives ( we drop loop order $\chi$ here to
simplify the notation)%

\begin{equation}
\mathcal{T}_{LLL}^{6\rightarrow4}-\mathcal{T}_{LTT}^{4}+\mathcal{T}_{(LL)}%
^{4}-\mathcal{T}_{(TT)}^{2}=0, \label{2.12.}%
\end{equation}%
\begin{equation}
\mathcal{T}_{LLT}^{5\rightarrow3}+\mathcal{T}_{(LT)}^{3}=0. \label{2.13.}%
\end{equation}

In Eq.(\ref{2.12.}) and Eq.(\ref{2.13.}), we have assigned a relative energy
power for each amplitude. For each longitudinal $L$ component, the order is
$E^{2}$ and for each transverse $T$ component, the order is $E.$ This is due
to the definitions of $e_{L}$and $e_{T}$ in Eq.(\ref{2.9..}) and
Eq.(\ref{2.10.}), where $e_{L}$ got one energy power more than $e_{T}.$ By
Eq.(\ref{2.12.}), the $E^{6}$ term of the energy expansion for $\mathcal{T}%
_{LLL}$ is forced to be zero. As a result, the possible leading order term is
$E^{4}$. Similar rule applies to $\mathcal{T}_{LLT}$ in Eq.(\ref{2.13.}). For
Eq.(\ref{2.5.}), we have $\theta^{\prime\mu}=e_{L}^{\mu}$ or $\theta
^{\prime\mu}=e_{T}^{\mu}$ and one gets, in the high energy limit,%

\begin{equation}
10\mathcal{T}_{LLL}^{6\rightarrow4}+\mathcal{T}_{LTT}^{4}+18\mathcal{T}%
_{(LL)}^{4}+6\mathcal{T}_{L}^{2}=0, \label{2.14..}%
\end{equation}%
\begin{equation}
10\mathcal{T}_{LLT}^{5\rightarrow3}+\mathcal{T}_{TTT}^{3}+18\mathcal{T}%
_{(LT)}^{3}+6\mathcal{T}_{T}^{1}=0. \label{2.15..}%
\end{equation}
For the $D_{2}$ Ward identity, Eq.(\ref{2.6.}), we have $\theta^{\mu}%
=e_{L}^{\mu}$ or $\theta^{\mu}=e_{T}^{\mu}$ and one gets, in the high energy limit,%

\begin{equation}
\mathcal{T}_{LLL}^{6\rightarrow4}+\mathcal{T}_{LTT}^{4}+9\mathcal{T}%
_{[LL]}^{4\rightarrow2}-3\mathcal{T}_{L}^{2}=0, \label{2.16..}%
\end{equation}%

\begin{equation}
\mathcal{T}_{LLT}^{5\rightarrow3}+\mathcal{T}_{TTT}^{3}+9\mathcal{T}%
_{[LT]}^{3}-3\mathcal{T}_{T}^{1}=0. \label{2.17..}%
\end{equation}

It is important to note that $\mathcal{T}_{[LL]}$ in Eq.(\ref{2.16..})
originate from the high energy limit of $\mathcal{T}_{[PL]}$, and the
antisymmetric property of the tensor forces the leading $E^{4}$ term to be
zero. Finally the singlet zero norm state Ward identity, Eq.(\ref{2.7.}),
imply, in the high energy limit,%

\begin{equation}
34\mathcal{T}_{LLL}^{6\rightarrow4}+9\mathcal{T}_{LTT}^{4}+84\mathcal{T}%
_{(LL)}^{4}+9\mathcal{T}_{(TT)}^{2}+50\mathcal{T}_{L}^{2}=0. \label{2.18..}%
\end{equation}
One notes that all components of high energy amplitudes of symmetric spin
three and antisymmetric spin two states appear at least once in
Eq.(\ref{2.12.}) to Eq.(\ref{2.18..}). It is now easy to see that the naive
leading order amplitudes corresponding to $E^{4}$ \ appear in Eq.(\ref{2.12.}%
), (\ref{2.14..}), Eq.(\ref{2.16..}) and Eq.(\ref{2.18..}). However, a simple
calculation shows that $\mathcal{T}_{LLL}^{4}=\mathcal{T}_{LTT}^{4}%
=\mathcal{T}_{(LL)}^{4}=0.$ So the real leading order amplitudes correspond to
$E^{3}$, which appear in Eq.(\ref{2.13.}), Eq.(\ref{2.15..}) and
Eq.(\ref{2.17..}). A simple calculation shows that \cite{ChanLee,ChanLee2}%

\begin{equation}
\mathcal{T}_{TTT}^{3}:\mathcal{T}_{LLT}^{3}:\mathcal{T}_{(LT)}^{3}%
:\mathcal{T}_{[LT]}^{3}=8:1:-1:-1. \label{2.19..}%
\end{equation}

Note that these proportionality constants are, as conjectured by Gross
\cite{Gross,Gross1}, independent of the scattering angle $\phi_{CM}$ and the
loop order $\chi$ of string perturbation theory. They are also independent of
particles chosen for vertex $v_{1,3,4}$. The ratios in Eq.(\ref{2.19..})
should be \textit{measurable} if the energy scale of string theory is not
Planckian. \textit{Most importantly, we now understand that the ratios
originate from ZNSs in the OCFQ spectrum of the string!}

The subleading order amplitudes corresponding to $E^{2}$ appear in
Eq.(\ref{2.12.}), (\ref{2.14..}), Eq.(\ref{2.16..}) and Eq.(\ref{2.18..}). One
has $6$ unknown amplitudes and $4$ equations. Presumably, they are not
proportional to each other or the proportional coefficients do depend on the
scattering angle $\phi_{CM}$. We will justify this point later in our sample
calculation. Our calculation here is purely algebraic without any integration
and is independent of saddle point calculation in
\cite{GM,GM1,Gross,Gross1,GrossManes}.

It is important to note that our result in Eq.(\ref{2.19..}) is gauge
invariant as it should be since we derive it from Ward identities
Eq.(\ref{2.4.}) to Eq.(\ref{2.7.}). On the other hand, the result obtained in
\cite{GrossManes} with $\mathcal{T}_{TTT}^{3}\propto\mathcal{T}_{[LT]}^{3},$
and $\mathcal{T}_{LLT}^{3}=0$ in the leading order energy at this mass level
is, on the contrary, \textit{not} gauge invariant. In fact, with
$\mathcal{T}_{LLT}^{3}=0$, an \textit{inconsistency} arises
\cite{ChanLee,ChanLee1,ChanLee2}, for example, between Eq.(\ref{2.13.}) and
Eq.(\ref{2.15..}).

We give one example here \cite{ChanLee,ChanLee1,ChanLee2} to illustrate the
meaning of the massive gauge invariant amplitude. To be more specific, we will
use two different gauge choices to calculate the high energy scattering
amplitude of symmetric spin three state. The first gauge choice is%

\begin{equation}
(\epsilon_{\mu\nu\lambda}\alpha_{-1}^{\mu\nu\lambda}+\epsilon_{(\mu\nu)}%
\alpha_{-1}^{\mu}\alpha_{-2}^{\nu})\left\vert 0,k\right\rangle ;\epsilon
_{(\mu\nu)}=-\frac{3}{2}k^{\lambda}\epsilon_{\mu\nu\lambda},k^{\mu}k^{\nu
}\epsilon_{\mu\nu\lambda}=0,\eta^{\mu\nu}\epsilon_{\mu\nu\lambda}=0.
\label{2.20..}%
\end{equation}
In the high energy limit, using the helicity decomposition and writing
$\epsilon_{\mu\nu\lambda}=\Sigma_{\mu,\nu,\lambda}e_{\mu}^{\alpha}e_{\nu
}^{\beta}e_{\lambda}^{\delta}u_{\alpha\beta\delta};\alpha,\beta,\delta=P,L,T,$
we get%

\begin{align}
(\epsilon_{\mu\nu\lambda}\alpha_{-1}^{\mu\nu\lambda}+\epsilon_{(\mu\nu)}%
\alpha_{-1}^{\mu}\alpha_{-2}^{\nu})\left\vert 0,k\right\rangle  &
=[u_{PLT}(6\alpha_{-1}^{PLT}+6\alpha_{-1}^{(L}\alpha_{-2}^{T)})\nonumber\\
&  +u_{TTP}(3\alpha_{-1}^{TTP}-3\alpha_{-1}^{LLP}+3\alpha_{-1}^{(T}\alpha
_{-2}^{T)}-3\alpha_{-1}^{(L}\alpha_{-2}^{L)})\nonumber\\
&  +u_{TTL}(3\alpha_{-1}^{TTL}-\alpha_{-1}^{LLL})+u_{TTT}(\alpha_{-1}%
^{TTT}-3\alpha_{-1}^{LLT})]\left\vert 0,k\right\rangle . \label{2.21..}%
\end{align}

The second gauge choice is%

\begin{equation}
\widetilde{\varepsilon}_{\mu\nu\lambda}\alpha_{-1}^{\mu\nu\lambda}\left\vert
0,k\right\rangle ;k^{\mu}\widetilde{\varepsilon}_{\mu\nu\lambda}=0,\eta
^{\mu\nu}\widetilde{\varepsilon}_{\mu\nu\lambda}=0. \label{2.22..}%
\end{equation}
In the high energy limit, similar calculation gives%

\begin{equation}
\widetilde{\varepsilon}_{\mu\nu\lambda}\alpha_{-1}^{\mu\nu\lambda}\left\vert
0,k\right\rangle =[\widetilde{u}_{TTL}(3\alpha_{-1}^{TTL}-\alpha_{-1}%
^{LLL})+\widetilde{u}_{TTT}(\alpha_{-1}^{TTT}-3\alpha_{-1}^{LLT})]\left\vert
0,k\right\rangle . \label{2.23..}%
\end{equation}

It is now easy to see that the first and second terms of Eq.(\ref{2.21..})
will not contribute to the high energy scattering amplitude of the symmetric
spin three state due to the spin two Ward identities Eq.(\ref{2.13.}) and
Eq.(\ref{2.12.}) if we identify $e_{P}=e_{L}.$ Thus the two different gauge
choices Eq.(\ref{2.20..}) and Eq.(\ref{2.22..}) give the same high energy
scattering amplitude. It can be shown that this massive gauge symmetry is
valid to all energy and is the result of the decoupling of massive spin two
ZNS at mass level $M^{2}$ $=4.$ Note that the $\alpha_{-1}^{LLT}$ term of
Eq.(\ref{2.23..}), which corresponds to the amplitude $\mathcal{T}_{LLT}^{3},$
was missing in the calculation of Ref \cite{GrossManes}. The issue\ was
discussed in details in \cite{CHL}. To further justify our result, we give a
sample calculation in the next section.

\subsubsection{A sample calculation of mass level $M^{2}$ $=4$}

In this section, we give a detailed calculation of a set of sample scattering
amplitudes to explicitly justify our results presented in the last section.
Since the proportionality constants in Eq.(\ref{2.19..}) are independent of
particles chosen for vertex $v_{1,3,4}$, for simplicity, we will choose them
to be tachyons. For the string-tree level $\chi=1$, with one tensor $v_{2}$
and three tachyons $v_{1,3,4}$, all scattering amplitudes of mass level
$M_{2}^{2}$ $=4$ were calculated in \cite{JCLee}. They are ( $s-t$ channel only)%

\begin{align}
\mathcal{T}^{\mu\nu\lambda}  &  =%
{\textstyle\int}
{\textstyle\prod_{i=1}^{4}}
dx_{i}<e^{ik_{1}X}\partial X^{\mu}\partial X^{\nu}\partial X^{\lambda
}e^{ik_{2}X}e^{ik_{3}X}e^{ik_{4}X}>\nonumber\\
&  =\frac{\Gamma(-\frac{s}{2}-1)\Gamma(-\frac{t}{2}-1)}{\Gamma(\frac{u}{2}%
+2)}[{-{t/2}({t^{2}}/4-1)k_{1}^{\mu}k_{1}^{\nu}k_{1}^{\lambda}%
+3(s/2+1)t/2(t/2+1)k_{1}^{(\mu}k_{1}^{\nu}k_{3}^{\lambda)}}\nonumber\\
&  {-3s/2(s/2+1)(t/2+1)k_{1}^{(\mu}k_{3}^{\nu}k_{3}^{\lambda)}+s/2({s^{2}%
}/4-1)k_{3}^{\mu}k_{3}^{\nu}k_{3}^{\lambda}]}, \label{3.1..}%
\end{align}

\begin{align}
\mathcal{T}^{(\mu\nu)}  &  =%
{\textstyle\int}
{\textstyle\prod_{i=1}^{4}}
dx_{i}<e^{ik_{1}X}\partial^{2}X^{(\mu}\partial X^{\nu)}e^{ik_{2}X}e^{ik_{3}%
X}e^{ik_{4}X}>\nonumber\\
&  =\frac{\Gamma(-\frac{s}{2}-1)\Gamma(-\frac{t}{2}-1)}{\Gamma(\frac{u}{2}%
+2)}[{t/2({t^{2}}/4-1)k_{1}^{\mu}k_{1}^{\nu}-(s/2+1)t/2(t/2+1)k_{1}^{(\mu
}k_{3}^{\nu)}}\nonumber\\
&  {+\newline s/2(s/2+1)(t/2+1)k_{3}^{(\mu}k_{1}^{\nu)}-s/2({s^{2}}%
/4-1)k_{3}^{\mu}k_{3}^{\nu}]}, \label{3..2..}%
\end{align}

\begin{align}
\mathcal{T}^{\mu}  &  =\frac{1}{2}%
{\textstyle\int}
{\textstyle\prod_{i=1}^{4}}
dx_{i}<e^{ik_{1}X}\partial^{3}X^{\mu}e^{ik_{2}X}e^{ik_{3}X}e^{ik_{4}%
X}>\nonumber\\
&  =\frac{\Gamma(-\frac{s}{2}-1)\Gamma(-\frac{t}{2}-1)}{\Gamma(\frac{u}{2}%
+2)}[s/2({s^{2}}/4-1)k_{3}^{\mu}-t/2({t^{2}}/4-1)k_{1}^{\mu}], \label{3.3.}%
\end{align}

\begin{align}
\mathcal{T}^{[\mu\nu]}  &  =%
{\textstyle\int}
{\textstyle\prod_{i=1}^{4}}
dx_{i}<e^{ik_{1}X}\partial^{2}X^{[\mu}\partial X^{\nu]}e^{ik_{2}X}e^{ik_{3}%
X}e^{ik_{4}X}>\nonumber\\
&  =\frac{\Gamma(-\frac{s}{2}-1)\Gamma(-\frac{t}{2}-1)}{\Gamma(\frac{u}{2}%
+2)}[(\frac{s+t}{2})(s/2+1)(t/2+1)k_{3}^{[\mu}k_{1}^{\nu]}] \label{3.4.}%
\end{align}
where $%
s=-\left(  k_{1}+k_{2}\right)  ^{2}%
$, $%
t=-\left(  k_{2}+k_{3}\right)  ^{2}%
$and $%
u=-\left(  k_{1}+k_{3}\right)  ^{2}%
$ are the Mandelstam variables. In deriving Eq.(\ref{3.1..}) to Eq.(\ref{3.4.}%
), we have made the $SL(2,R)$ gauge fixing by choosing $x_{1}=0,0\leqq
x_{2}\leqq1,x_{3}=1,x_{4}=\infty.$

To calculate the high energy expansions $(s,t\rightarrow\infty,\frac{s}{t}=$
fixed $)$ of these scattering amplitudes, one needs the following energy
expansion formulas \cite{ChanLee2}
\begin{equation}
e_{P}.k_{1}=(\frac{-2E^{2}}{M_{2}})[1-(\frac{M_{2}^{2}-2}{4})\frac{1}{E^{2}}],
\end{equation}

\begin{equation}
e_{L}.k_{1}=(\frac{-2E^{2}}{M_{2}})[1-(\frac{M_{2}^{2}-2}{4})\frac{1}{E^{2}%
}+(\frac{M_{2}^{2}}{4})\frac{1}{E^{4}}+(\frac{M_{2}^{4}-2M_{2}^{2}}{16}%
)\frac{1}{E^{6}}+O(\frac{1}{E^{8}})],
\end{equation}

\begin{equation}
e_{T}.k_{1}=0,
\end{equation}

\begin{equation}
e_{P}.k_{3}=(\frac{E^{2}}{M_{2}})\left\{  2\xi^{2}+[\frac{M_{2}^{2}}{2}%
\eta^{2}+(3\xi^{2}-1)]\frac{1}{E^{2}}+(2\xi^{2}-1)(\frac{M_{2}^{2}+2}{4}%
)^{2}\frac{1}{E^{6}}+O(\frac{1}{E^{8}})\right\}  ,
\end{equation}

\begin{equation}
e_{L}.k_{3}=(\frac{E^{2}}{M_{2}})\left\{
\begin{array}
[c]{c}%
2\xi^{2}+[-\frac{M_{2}^{2}}{2}\eta^{2}+(3\xi^{2}-1)]\frac{1}{E^{2}}%
+(\frac{M_{2}^{2}}{2}\xi^{2})\frac{1}{E^{4}}\\
+(\frac{M_{2}^{4}-4M_{2}^{2}\xi^{2}+8\xi^{2}-4}{16})\frac{1}{E^{6}}+O(\frac
{1}{E^{8}})
\end{array}
\right\}  ,
\end{equation}

\begin{equation}
e_{T}.k_{3}=(-2\xi\eta)E-(\frac{2\xi\eta}{E})+(\frac{\xi\eta}{E^{3}}%
)-(\frac{\xi\eta}{E^{5}})+O(\frac{1}{E^{7}})
\end{equation}
where $\xi=\sin\frac{\phi_{CM}}{2}$ and $\eta=\cos\frac{\phi_{CM}}{2}.$ The
high energy expansions of Mandelstam variables are given by%

\begin{equation}
s=(E_{1}+E_{2})^{2}=4E^{2},
\end{equation}

\begin{equation}
t=(-4\xi^{2})E^{2}+(M_{2}^{2}-6)\xi^{2}+\frac{1}{8}(M_{2}^{2}+2)^{2}%
(1-2\xi^{2})\frac{1}{E^{4}}+O(\frac{1}{E^{6}}).
\end{equation}

We can now explicitly calculate all amplitudes in Eq.(\ref{2.19..}). After
some algebra, we get%

\begin{equation}
\mathcal{T}_{TTT}=-8E^{9}\mathcal{T}(3)\sin^{3}\phi_{CM}[1+\frac{3}{E^{2}%
}+\frac{5}{4E^{4}}-\frac{5}{4E^{6}}+O(\frac{1}{E^{8}})],
\end{equation}

\begin{align}
\mathcal{T}_{LLT}  &  =-E^{9}\mathcal{T}(3)[\sin^{3}\phi_{CM}+(6\sin\phi
_{CM}\cos^{2}\phi_{CM})\frac{1}{E^{2}}\nonumber\\
&  -\sin\phi_{CM}(\frac{11}{2}\sin^{2}\phi_{CM}-6)\frac{1}{E^{4}}+O(\frac
{1}{E^{6}})],
\end{align}

\begin{align}
\mathcal{T}_{[LT]}  &  =E^{9}\mathcal{T}(3)[\sin^{3}\phi_{CM}-(2\sin\phi
_{CM}\cos^{2}\phi_{CM})\frac{1}{E^{2}}\nonumber\\
&  +\sin\phi_{CM}(\frac{3}{2}\sin^{2}\phi_{CM}-2)\frac{1}{E^{4}}+O(\frac
{1}{E^{6}})],
\end{align}

\begin{align}
\mathcal{T}_{(LT)}  &  =E^{9}\mathcal{T}(3)[\sin^{3}\phi_{CM}+\sin\phi
_{CM}(\frac{3}{2}-10\cos\phi_{CM}\nonumber\\
&  -\frac{3}{2}\cos^{2}\phi_{CM})\frac{1}{E^{2}}-\sin\phi_{CM}(\frac{1}%
{4}+10\cos\phi_{CM}+\frac{3}{4}\cos^{2}\phi_{CM})\frac{1}{E^{4}}+O(\frac
{1}{E^{6}})]
\end{align}
where%
\begin{equation}
\mathcal{T}(N)\mathcal{=}\sqrt{\pi}(-1)^{N-1}2^{-n}E^{-1-2N}(\sin\frac
{\phi_{CM}}{2})^{-3}(\cos\frac{\phi_{CM}}{2})^{5-2N}\exp(-\frac{s\ln s+t\ln
t-(s+t)\ln(s+t)}{2}) \label{ttnn}%
\end{equation}
is the high energy limit of $\frac{\Gamma(-\frac{s}{2}-1)\Gamma(-\frac{t}%
{2}-1)}{\Gamma(\frac{u}{2}+2)}$ with $s+t+u=2N-8$, and we have calculated it
up to the next leading order in $E$. We thus have justified Eq.(\ref{2.19..})
with $\mathcal{T}_{TTT}^{3}=-8E^{9}\mathcal{T}(3)\sin^{3}\phi_{CM}$ and
$\mathcal{T}_{LLT}^{5}=0.$ We have also checked that $\mathcal{T}_{LLL}%
^{6}=\mathcal{T}_{LLL}^{4}=\mathcal{T}_{LTT}^{4}=\mathcal{T}_{(LL)}^{4}=0$ as
claimed in the previous section.

Note that, unlike the leading $\ E^{9}$ order, the angular dependences of
$E^{7}$ order are different for each amplitudes. The subleading order
amplitudes corresponding to $\mathcal{T}^{2}$ ($E^{8}$ order) appear in
Eq.(\ref{2.12.}), (\ref{2.14..}), Eq.(\ref{2.16..}) and Eq.(\ref{2.18..}). One
has $6$ unknown amplitudes. An explicit sample calculation gives%

\begin{equation}
\mathcal{T}_{LLL}^{2}=-4E^{8}\sin\phi_{CM}\cos\phi_{CM}\mathcal{T}(3),
\end{equation}

\begin{equation}
\mathcal{T}_{LTT}^{2}=-8E^{8}\sin^{2}\phi_{CM}\cos\phi_{CM}\mathcal{T}(3),
\end{equation}
which show that their angular dependences are indeed different or the
proportional coefficients do depend on the scattering angle $\phi_{CM}$.

\subsubsection{Results of mass level $M^{2}$ $=6$}

The calculations for $M^{2}$ $=4$ in the previous section can be generalized
to $M^{2}$ $=6$ \cite{ChanLee2}. The calculation was however much more
tedious, and to the leading order in energy one ended up with $8$ equations
and $9$ amplitudes. A calculation showed that \cite{ChanLee2}%

\begin{align}
\mathcal{T}_{TTTT}^{4}  &  :\mathcal{T}_{TTLL}^{4}:\mathcal{T}_{LLLL}%
^{4}:\mathcal{T}_{TTL}^{4}:\mathcal{T}_{LLL}^{4}:\widetilde{\mathcal{T}%
}_{LT,T}^{4}:\widetilde{\mathcal{T}}_{LP,P}^{4}:\mathcal{T}_{LL}%
^{4}:\widetilde{\mathcal{T}}_{LL}^{4}=\nonumber\\
16  &  :\frac{4}{3}:\frac{1}{3}:-\frac{4\sqrt{6}}{9}:-\frac{\sqrt{6}}%
{9}:-\frac{2\sqrt{6}}{3}:0:\frac{2}{3}:0. \label{1111}%
\end{align}
Note that these proportionality constants are again, as conjectured by Gross,
independent of the scattering angle $\phi_{CM}$ and the loop order $\chi$ of
string perturbation theory. A sample calculation of scattering amplitudes for
mass level $M^{2}$ $=6$ \cite{ChanLee2} justified the ratios above calculated
by solving $8$ linear relations derived from the decoupling of high energy ZNS
in the GR. There are two $0$ amplitudes in Eq.(\ref{1111}), which mean they
are subleading order amplitudes in energy.

It was remarkable to see that the linear relations obtained by high energy
limit of stringy Ward identities or decoupling of ZNS were just good enough to
solve all the high energy amplitudes in terms of one amplitude! It was even
more remarkable to see that the ratios obtained by solving these linear
relations matched exactly with the sample calculations for the high energy
string amplitudes. However, the calculation soon becomes too complicated to
manage when one goes to even higher mass levels. In the next section, we will
adopt another strategy to generalize the calculations to \textit{arbitrary}
mass levels.

\subsection{Decoupling of high energy ZNS at arbitrary mass levels}

\begin{figure}[ptb]
\label{scattering} \setlength{\unitlength}{3pt}
\par
\begin{center}
\begin{picture}(100,70)(-50,-30)
{\large
\put(45,0){\vector(-1,0){42}} \put(-45,0){\vector(1,0){42}}
\put(2,2){\vector(1,1){30}} \put(-2,-2){\vector(-1,-1){30}}
\put(25,2){$k_1$} \put(-27,2){$k_2$} \put(11,20){$-k_3$}
\put(-24,-15){$-k_4$}
\put(40,0){\vector(0,-1){10}} \put(-40,0){\vector(0,1){10}}
\put(26,26){\vector(-1,1){7}} \put(-26,-26){\vector(1,-1){7}}
\put(36,-16){$e^{T}(1)$} \put(-44,15){$e^{T}(2)$}
\put(15,36){$e^{T}(3)$} \put(-18,-35){$e^{T}(4)$}
\qbezier(10,0)(10,4)(6,6) \put(12,4){$\phi$}
}
\end{picture}
\end{center}
\caption{Kinematic variables in the center of mass frame}%
\label{Kinematic}%
\end{figure}
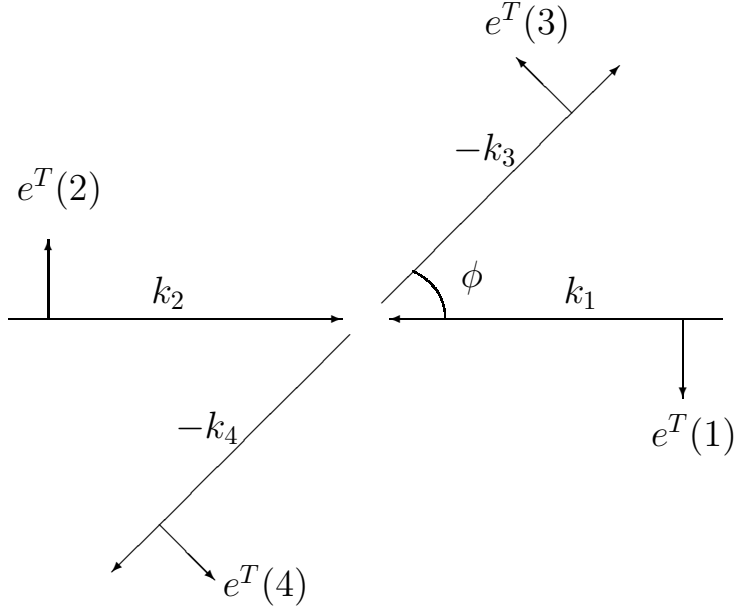

In the following three sections, we will use three methods to generalize our
calculations in the previous section to arbitrary mass levels. In this section
we will first use method of decoupling of high energy ZNS. We will focus on
4-point functions in this section, although our discussion can be generalized
to higher point correlation functions. Due to Poincare symmetry, a 4-point
function is a function of merely two parameters. Viewing a 4-point function as
the scattering amplitude of a two-body scattering process, one can choose the
two parameters to be $E$ (one half of the center of mass energy for the
incoming particles i.e., particles 1 and 2 in Fig. \ref{Kinematic}, and $\phi$
(the scattering angle between particles 1 and 3). For convenience we will take
the center of mass frame and put the momenta of particles 1 and 2 along the
$X^{1}$-direction, with the momenta of particles 3 and 4 on the $X^{1}-X^{2}$
plane. The momenta of the particles are
\begin{align}
k_{1}  &  =(\sqrt{p^{2}+M_{1}^{2}},-p,0),\\
k_{2}  &  =(\sqrt{p^{2}+M_{2}^{2}},p,0),\\
k_{3}  &  =(-\sqrt{q^{2}+M_{3}^{2}},-q\cos\phi,-q\sin\phi),\\
k_{4}  &  =(-\sqrt{q^{2}+M_{4}^{2}},q\cos\phi,q\sin\phi).
\end{align}
They satisfy $k_{i}^{2}=-m_{i}^{2}$. In the high energy limit, the Mandelstam
variables are
\begin{align}
s  &  \equiv-(k_{1}+k_{2})^{2}=4E^{2}+\mathcal{O}(1/E^{2}),\\
t  &  \equiv-(k_{2}+k_{3})^{2}=-4\left(  E^{2}-\frac{\sum_{i=1}^{4}M_{i}^{2}%
}{4}\right)  \sin^{2}\frac{\phi}{2}+\mathcal{O}(1/E^{2}),\\
u  &  \equiv-(k_{1}+k_{3})^{2}=-4\left(  E^{2}-\frac{\sum_{i=1}^{4}M_{i}^{2}%
}{4}\right)  \cos^{2}\frac{\phi}{2}+\mathcal{O}(1/E^{2}),
\end{align}
where $E$ is related to $p$ and $q$ as
\begin{equation}
E^{2}=p^{2}+\frac{M_{1}^{2}+M_{2}^{2}}{2}=q^{2}+\frac{M_{3}^{2}+M_{4}^{2}}{2}.
\end{equation}
The polarization bases for the 4 particles are
\begin{align}
e^{L}(1)=\frac{1}{M_{1}}(p,-\sqrt{p^{2}+M_{1}^{2}},0),  &  e^{T}%
(1)=(0,0,-1),\\
e^{L}(2)=\frac{1}{M_{2}}(p,\sqrt{p^{2}+M_{2}^{2}},0),  &  e^{T}(2)=(0,0,1),\\
e^{L}(3)=\frac{1}{M_{3}}(-q,-\sqrt{q^{2}+M_{3}^{2}}\cos\phi,-\sqrt{q^{2}%
+M_{3}^{2}}\sin\phi),  &  e^{T}(3)=(0,-\sin\phi,\cos\phi),\\
e^{L}(4)=\frac{1}{M_{4}}(-q,\sqrt{q^{2}+M_{4}}\cos\phi,\sqrt{q^{2}+M_{4}^{2}%
}\sin\phi),  &  e^{T}(4)=(0,\sin\phi,-\cos\phi).
\end{align}
The high energy limit under consideration is
\begin{equation}
\alpha^{\prime}E^{2}\rightarrow\infty,\quad\phi=\text{fixed}.
\end{equation}
Based on the saddle-point approximation of Gross and Mende \cite{GM,GM1},
Gross and Manes \cite{GrossManes} computed the high energy limit of 4-point
functions in the bosonic open string theory. To explain their result, let us
first define our notations and conventions. For a particle of momentum $k$, we
define an orthonormal basis of polarizations $\{e^{P},e^{L},e^{T_{i}}\}$. The
momentum polarization $e^{P}$ is proportional to $k$, the longitudinal
polarization $e^{L}$ is the space-like unit vector whose spatial component is
proportional to that of $k$, and $e^{T_{i}}$ are the space-like unit-vectors
transverse to the spatial momentum. As an example, for $k$ pointing along the
$X^{1}$-direction,
\begin{equation}
k=(k^{0},k^{1},k^{2},\cdots,k^{25})=(E,p,0,\cdots,0),\hspace{1cm}p>0,
\end{equation}
the basis of polarization is
\begin{equation}
e^{P}=\frac{1}{M}(\sqrt{p^{2}+M^{2}},p,0,0,\cdots,0),\hspace{0.3cm}e^{L}%
=\frac{1}{M}(p,\sqrt{p^{2}+M^{2}},0,0,\cdots,0),\hspace{0.3cm}e^{T_{i}%
}=(0,0,\cdots,1,\cdots),
\end{equation}
where $M$ is the mass of the particle. In general, $e^{T_{i}}$ (for
$i=3,\cdots,25$) is just the unit vector in the $X^{i}$-direction, and the
definitions of $e^{P},e^{L}$ and $e^{T_{2}}$ will depend on the motion of the
particle. For $e^{T_{2}}$, which is parallel to the scattering plane, we
denote it by $e^{T}$ (see Fig. \ref{Kinematic}). The orientations of
$e^{T_{2}}$ for each particle are fixed by the right-hand rule, $\vec{k}\times
e^{T_{2}}=e^{T_{3}}$, where $\vec{k}$ is the spatial momentum of 4-vector $k$.
We will use the notation $\partial^{n}X^{A}\equiv e^{A}\cdot\partial^{n}X$ for
$A=P,L,T,T_{i}$.

Each vertex is a polynomial of $\{\partial^{n}X^{A}\}$ times the exponential
factor $\exp(ik\cdot X)$. Among all possible choices of polarizations for the
4 vertices in a 4-point function, we now argue that only the polarizations $L$
and $T$ need to be considered. The polarization $P$ can be gauged away using
ZNS \cite{CLYang}. To see why we can ignore all $T_{i}$'s except $T$, we note
that a prefactor $\partial^{n}X^{A}$ can be contracted with the exponent
$ik\cdot X$ of another vertex. The contribution of this contraction to a
scattering amplitude is proportional to $k^{A}\sim E$. If $k^{A}\neq0$ (i.e.,
if $A=L$ or $T$), this is much more important in the high energy limit than a
contraction with another prefactor $\partial^{m}X^{B}$, which gives $\eta
^{AB}\sim E^{0}$. Therefore, if all polarizations in all vertices are chosen
to be either $L$ or $T$, the resulting 4-point function will dominate over
other choices of polarizations.

The central idea behind the algebraic approach used in
\cite{ChanLee,ChanLee1,ChanLee2} and \cite{CHL} was the decoupling of ZNS
(i.e., the requirement of stringy gauge invariance). A crucial step in the
derivation is to replace the polarization $e^{P}$ by $e^{L}$ in the ZNS. It is
assumed that, while ZNS decouple at all energies, the replacement leads to
states that are decoupled at high energies.

\subsubsection{Main results}

For brevity, we will refer to all 4-point functions different from each other
by a single vertex at the same mass level as a ``\emph{family}''. When we
compare members of a family, we only need to specify the vertex which is changed.

A 4-point function will be said to be \emph{at the leading order} if it is not
subleading to any of its \emph{siblings}. We will ignore those that are not at
the leading order. Our aim is to find the numerical ratios of all 4-point
functions in the same family at the leading order. Apparently, there are more
4-point functions at the leading order at higher mass levels. Our goal may
seem insurmountable at first sight.

Saving the derivation for later, we give our main results here. A 4-point
function is at the leading order if and only if the vertex $V$ under
comparison is a linear combination of vertices of the form
\begin{equation}
V^{(N,m,q)}(k)=\left(  \partial X^{T}\right)  ^{N-m-2q}\left(  \partial
X^{L}\right)  ^{2m}\left(  \partial^{2}X^{L}\right)  ^{q}e^{ik\cdot X},
\end{equation}
where
\begin{equation}
\quad N\geq2m+2q,\quad m,q\geq0.
\end{equation}
The corresponding states are of the form
\begin{equation}
\left(  \alpha_{-1}^{T}\right)  ^{N-2m-2q}\left(  \alpha_{-1}^{L}\right)
^{2m}\left(  \alpha_{-2}^{L}\right)  ^{q}|0,k\rangle. \label{nono}%
\end{equation}
The mass squared is $2(N-1)$. All other states involving $\alpha_{-2}%
^{T},\alpha_{-3}^{A},\cdots$ are subleading.

Using the notation\footnote{More rigorously, $V_{2}$ needs to be a physical
state in order for the correlation function to be well-defined. We should keep
in mind that our results should be applied to suitable linear combinations of
Eq.(\ref{nono}), possibly together with subleading states, to satisfy Virasoro
constraints.}
\begin{equation}
\mathcal{T}^{(N,m,q)}=\langle V_{1}V^{(N,m,q)}(k)V_{3}V_{4}\rangle,
\end{equation}
all linear relations among different choices of $V^{(N,m,q)}$ (obtained from
the decoupling of high energy ZNS) can be solved by the simple expression
\begin{align}
\lim_{E\rightarrow\infty}\frac{\mathcal{T}^{(N,2m,q)}}{\mathcal{T}^{(N,0,0)}}
&  =\left(  -\frac{1}{M}\right)  ^{2m+q}\left(  \frac{1}{2}\right)
^{m+q}(2m-1)!!,\label{mainA}\\
\lim_{E\rightarrow\infty}\frac{\mathcal{T}^{(N,2m+1,q)}}{\mathcal{T}%
^{(N,0,0)}}  &  =0,
\end{align}
where $M=\sqrt{2(N-1)}$. This formula tells us how to trade $\partial X^{L}$
and $\partial^{2}X^{L}$ for $\partial X^{T}$, so that all 4-point functions
can be related to the one involving only $\partial X^{T}$ in $V_{2}$. The
formula above applies equally well to all vertices.

Since we know the value of a representative 4-point function
\cite{ChanLee,ChanLee2, CHL}
\begin{equation}
\mathcal{T}_{N_{1}N_{2}N_{3}N_{4}}^{T^{1}\cdot\cdot T^{2}\cdot\cdot T^{3}%
\cdot\cdot T^{4}\cdot\cdot}=(-1)^{N_{2}+N_{4}}[2E^{3}\sin\phi_{CM}]^{\Sigma
N_{i}}\mathcal{T}(\Sigma N_{i}), \label{TN}%
\end{equation}
where $\mathcal{T}(N)$ is the high-energy limit of $\frac{\Gamma(-\frac{s}%
{2}-1)\Gamma(-\frac{t}{2}-1)}{\Gamma(\frac{u}{2}+2)}$ with $s+t+u=2\Sigma
N_{i}-8$, and we have calculated it up to the next leading order in $E$ in
Eq.(\ref{ttnn}), we can immediately write down the explicit expression of a
4-point function if all vertices are nontrivial at the leading order. In
Eq.(\ref{TN}), $N_{i}$ is the number of $T^{i}$ of the $i-th$ vertex operators
and $T^{i}$ is the transverse direction of the $i-th$ particle.

\subsubsection{Decoupling of high energy ZNS}

Before we go on, we recall some terminology used in the old covariant
quantization. A state $|\psi\rangle$ in the Hilbert space is \emph{physical}
if it satisfies the Virasoro constraints
\begin{equation}
\left(  L_{n}-\delta_{n}^{0}\right)  |\psi\rangle=0,\quad n\geq0.
\end{equation}
Since $L_{n}^{\dagger}=L_{-n}$, states of the form
\begin{equation}
L_{-n}|\chi\rangle
\end{equation}
are orthogonal to all physical states, and they are called spurious states.
ZNS are spurious states that are also physical. They correspond to gauge
symmetries. In the old covariant first quantization spectrum of open bosonic
string theory, the solutions of physical state conditions include
positive-norm propagating states and two types of ZNS. In this section we
derive the linear relations among all amplitudes in the same family by taking
the high energy limit of ZNS (HZNS). The first step in the derivation is to
identify the class of states that are relevant, i.e., those at the leading
order. As we explained before, we only need to consider the polarizations
$e^{T}$ and $e^{L}$.

To get a rough idea about how each vertex operator scales with $E$ in the high
energy limit, we associate a naive dimension to each prefactor $\partial
^{m}X^{A}$ according to the following rule
\begin{equation}
\partial^{m}X^{T}\rightarrow1,\quad\partial^{m}X^{L}\rightarrow2.
\end{equation}
The reason is the following. Each factor of $\partial^{m}X^{\mu}$ has the
possibility of contracting with the exponent $ik_{i}\cdot X$ of another vertex
operator so that it scales like $E$ in the high energy limit. Furthermore,
components of the polarization vectors $e^{T}$ and $e^{L}$ scale with $E$ like
$E^{0}$ and $E^{1}$, respectively.

When we compare vertex operators at the same mass level, the sum of all the
integers $m$ in $\partial^{m}X^{A}$ is fixed. Roughly speaking, it is
advantageous to have many $\partial X^{A}$ than having fewer number of
$\partial^{m}X^{A}$ with $m>1$. For example, at the first massive level, the
vertex operator $\partial X^{T}\partial X^{T}e^{ik\cdot X}$ has a larger naive
dimension than $\partial^{2}X^{T}e^{ik\cdot X}$.

The counting of the naive dimension does not take into consideration the
possibility that the coefficient of the leading order term happens to vanish
by cancellation. The true dimension of a vertex operator can be lower than its
naive dimension, although the reverse never happens.

Through experiences accumulated from explicit computations
\cite{ChanLee,ChanLee1, ChanLee2, CHL}, we find that the highest spin vertex
\begin{equation}
(\partial X^{T})^{N}e^{ik\cdot X}\leftrightarrow\left(  \alpha_{-1}%
^{T}\right)  ^{N}|0,k\rangle
\end{equation}
is always at the leading order in its family. Since the naive dimension of
this state equals its true dimension, any state with a lower naive dimension
than this vertex operator can be ignored. This implies that we can immediately
throw away a lot of vertex operators at each mass level, but there are still
many left. The problem is that, although there are disadvantages to have
$\partial^{m}X^{T}$ with $m\geq2$ or $\partial^{m}X^{L}$ with $m\geq3$
compared with having $\left(  \partial X^{T}\right)  ^{m}$, it may be possible
that having extra factors of $\partial X^{L}$, which has a higher naive
dimension than $\partial X^{T}$, can compensate the disadvantage of these
factors. However, explicit computations at the first few massive levels showed
that this never happens.

We will now argue why this is generically true, and show in this section that
the only states that will survive the high energy limit at level $N$ are of
the form
\begin{equation}
|N,2m,q\rangle\equiv\left(  \alpha_{-1}^{T}\right)  ^{N-2m-2q}\left(
\alpha_{-1}^{L}\right)  ^{2m}\left(  \alpha_{-2}^{L}\right)  ^{q}|0;k\rangle.
\label{relevant}%
\end{equation}

Our argument is essentially based on the decoupling of ZNS in the high energy
limit. Thanks to the Virasoro algebra, we only need two Virasoro operators
\begin{align}
L_{-1}  &  =\frac{1}{2}\sum_{n\in%
\mathbb{Z}
}\alpha_{-1+n}\cdot\alpha_{-n}=M\alpha_{-1}^{P}+\alpha_{-2}\cdot\alpha
_{-1}+\cdots,\\
L_{-2}  &  =\frac{1}{2}\sum_{n\in%
\mathbb{Z}
}\alpha_{-2+n}\cdot\alpha_{-n}=\frac{1}{2}\alpha_{-1}\cdot\alpha_{-1}%
+M\alpha_{-2}^{P}+\alpha_{-3}\cdot\alpha_{-1}+\cdots
\end{align}
to generate all high energy ZNS or HZNS. Here $M$ is the mass operator, i.e.,
$M^{2}=-k^{2}$ when acting on the state $|0,k\rangle$.

\paragraph{Irrelevance of other states}

To prove that only states of the form Eq.(\ref{relevant}) are at the leading
order, we shall prove that $(i)$ any state which has an odd number of
$\alpha_{-1}^{L}$ is irrelevant (i.e., subleading in the high energy limit),
and $(ii)$ any state involving a creation operator whose naive dimension is
less than its mode index $n$, i.e., states belonging to
\begin{equation}
\{\alpha_{-n}^{L},\;\;n>2;\quad\alpha_{-m}^{T},\;\;m>1\} \label{irrset}%
\end{equation}
is also irrelevant. We proceed by mathematical induction.

First we prove that any state which has a single factor of $\alpha_{-1}^{L}$
is irrelevant, and that any state with two $\alpha_{-1}^{L}$'s is irrelevant
if it contains an operator of naive dimension less than its index.

Consider the HZNS $L_{-1}\chi$ where $\chi$ is any state without any
$\alpha_{-1}^{L}$, and it is at level $(N-1)$. Note that, except $\alpha
_{-1}^{L}$, the naive dimension of an operator is always less than or equal to
its index (we exclude $\alpha_{-1}^{P}$ as mentioned above). This means that
the naive dimension of $\chi$ is less than or equal to $(N-1)$. Since we know
that at level $N$, the state Eq.(\ref{relevant}) has true dimension $N$, when
computing $L_{-1}\chi$ in the high energy limit, we can ignore everything with
naive dimension less than $N$. This means that we need $L_{-1}$ to increase
the naive dimension of $\chi$ by no less than 1. In the high energy limit of
$L_{-1}$
\begin{equation}
L_{-1}\rightarrow M\alpha_{-1}^{L}+\alpha_{-2}^{L}a_{-1}^{L}+\alpha_{-2}%
^{T}\alpha_{-1}^{T}+\cdots, \label{HL-1}%
\end{equation}
only the first term will increase the naive dimension of $\chi$ by 1. All the
rest do not change the naive dimension. This means that, to the leading
order,
\begin{equation}
L_{-1}\chi\sim M\alpha_{-1}^{L}\chi.
\end{equation}
This is a state with a single factor of $\alpha_{-1}^{L}$ and it is a HZNS, so
it should be decoupled in the high energy limit.

Now consider an arbitrary state $\chi$ at level $(N-1)$ which has a single
factor of $\alpha_{-1}^{L}$. If $\chi$ involves any operator whose naive
dimension is less than its index, the naive dimension of $\chi$ is at most
$(N-1)$. In the high energy limit
\begin{equation}
L_{-1}\chi\rightarrow M\alpha_{-1}^{L}\chi+\alpha_{-2}^{L}\alpha_{-1}^{L}%
\chi+\cdots, \label{L-1chi}%
\end{equation}
except the first two terms, all other terms are irrelevant because they
contain a single factor of $\alpha_{-1}^{L}$. As the second term has a naive
dimension $(n-1)$ and can be ignored, we conclude that $\alpha_{-1}^{L}\chi$
is irrelevant.

The next step in mathematical induction is to show that if (a) states with
$(2m-1)$ factors of $\alpha_{-1}^{L}$ are irrelevant, and (b) states with $2m$
factors of $\alpha_{-1}^{L}$ are still irrelevant if it also contains any of
the operators in Eq.(\ref{irrset}), then we can prove that both statements are
also valid for $m\rightarrow m+1$.

Suppose $\chi$ is an arbitrary state at level $(N-1)$ which has $2m$ factors
of $\alpha_{-1}^{L}$'s. The high energy limit of $L_{-1}\chi$ is given by
Eq.(\ref{L-1chi}). The second term has $(2m-1)$ factors of $\alpha_{-1}^{L}$
and is irrelevant. The rest of the terms, except the first, are irrelevant
because they contains at least one operator from the set Eq.(\ref{irrset}).
Hence the first term is a HZNS and is irrelevant. We have proved our first
claim for $(m+1)$, i.e., a state with $(2m+1)$ factors of $\alpha_{-1}^{L}$
decouple at high energies.

Similarly, consider the case when $\chi$ is at level $(N-1)$ and has $(2m-1)$
factors of $\alpha_{-1}^{L}$. Furthermore we assume that it involves operators
from the set Eq.(\ref{irrset}). Then the first term in Eq.(\ref{L-1chi}) is
what we want to prove to be irrelevant. The second term is irrelevant because
we have just proved that a state with $(2m+1)$ factors of $\alpha_{-1}^{L}$ is
irrelevant. The rest of the terms are irrelevant because they have $(2m-1)$
$\alpha_{-1}^{L}$'s. Thus we conclude that both claims are correct for $m+1$
as well. The mathematical induction is complete.

\paragraph{Examples of high energy ZNS at low-lying mass levels}

In this section, we explicitly calculate high energy ZNS (HZNS) of some
low-lying mass level. We will also show that the decoupling of these HZNS can
be used to derive the desired linear relations. In the old covariant first
quantization spectrum of open bosonic string theory, the solutions of physical
state conditions include positive-norm propagating states and two types of
ZNS. Based on a simplified calculation of higher mass level positive-norm
states in \cite{Manes} , some general solutions of ZNS of Eqs.(\ref{1.1}) and
(\ref{1.2}) at arbitrary mass level were calculated in \cite{0302123}.
Eqs.(\ref{1.1}) and (\ref{1.2}) can be derived from Kac determinant in
conformal field theory. While type I states have zero-norm at any spacetime,
type II states have zero-norm \textit{only} at $D=26$.

The solutions of Eqs.(\ref{1.1}) and (\ref{1.2}) up to the mass level
$M^{2}=4$ were calculated in part I of this review. For illustration, let's
repeat them and list as follows :

1. $M^{2}=-k^{2}=0:$%
\begin{equation}
L_{-1\text{ }}\left\vert x\right\rangle =k\cdot\alpha_{-1}\left\vert
0,k\right\rangle ;\left\vert x\right\rangle =\left\vert 0,k\right\rangle
;\left\vert x\right\rangle =\left\vert 0,k\right\rangle . \label{D.3}%
\end{equation}

2. $M^{2}=-k^{2}=2:$%
\begin{equation}
(L_{-2}+\frac{3}{2}L_{-1}^{2})\left\vert \widetilde{x}\right\rangle =[\frac
{1}{2}\alpha_{-1}\cdot\alpha_{-1}+\frac{5}{2}k\cdot\alpha_{-2}+\frac{3}%
{2}(k\cdot\alpha_{-1})^{2}]\left\vert 0,k\right\rangle ;\left\vert
\widetilde{x}\right\rangle =\left\vert 0,k\right\rangle , \label{D.4}%
\end{equation}%
\begin{equation}
L_{-1}\left\vert x\right\rangle =[\theta\cdot\alpha_{-2}+(k\cdot\alpha
_{-1})(\theta\cdot\alpha_{-1})]\left\vert 0,k\right\rangle ;\left\vert
x\right\rangle =\theta\cdot\alpha_{-1}\left\vert 0,k\right\rangle ,\theta\cdot
k=0. \label{D.5}%
\end{equation}

3. $M^{2}=-k^{2}=4:$%
\begin{align}
(L_{-2}+\frac{3}{2}L_{-1}^{2})\left\vert \widetilde{x}\right\rangle  &
=\{4\theta\cdot\alpha_{-3}+\frac{1}{2}(\alpha_{-1}\cdot\alpha_{-1}%
)(\theta\cdot\alpha_{-1})+\frac{5}{2}(k\cdot\alpha_{-2})(\theta\cdot
\alpha_{-1})\nonumber\\[0.01in]
&  +\frac{3}{2}(k\cdot\alpha_{-1})^{2}(\theta\cdot\alpha_{-1})+3(k\cdot
\alpha_{-1})(\theta\cdot\alpha_{-2})\}\left\vert 0,k\right\rangle ;\nonumber\\
\left\vert \widetilde{x}\right\rangle  &  =\theta\cdot\alpha_{-1}\left\vert
0,k\right\rangle ,k\cdot\theta=0, \label{D.6}%
\end{align}%
\begin{align}
L_{-1}\left\vert x\right\rangle  &  =[2\theta_{\mu\nu}\alpha_{-1}^{\mu}%
\alpha_{-2}^{\nu}+k_{\lambda}\theta_{\mu\nu}\alpha_{-1}^{\lambda}\alpha
_{-1}^{\mu}\alpha_{-1}^{\nu}]\left\vert 0,k\right\rangle ;\nonumber\\
\left\vert x\right\rangle  &  =\theta_{\mu\nu}\alpha_{-1}^{\mu\nu}\left\vert
0,k\right\rangle ,k\cdot\theta=\eta^{\mu\nu}\theta_{\mu\nu}=0,\theta_{\mu\nu
}=\theta_{\nu\mu}, \label{D.7}%
\end{align}%
\begin{align}
L_{-1}\left\vert x\right\rangle  &  =[\frac{1}{2}(k\cdot\alpha_{-1}%
)^{2}(\theta\cdot\alpha_{-1})+2\theta\cdot\alpha_{-3}+\frac{3}{2}(k\cdot
\alpha_{-1})(\theta\cdot\alpha_{-2})\nonumber\\
&  +\frac{1}{2}(k\cdot\alpha_{-2})(\theta\cdot\alpha_{-1})]\left\vert
0,k\right\rangle ;\text{ \ }\nonumber\\
\text{\ }\left\vert x\right\rangle  &  =[2\theta\cdot\alpha_{-2}+(k\cdot
\alpha_{-1})(\theta\cdot\alpha_{-1})]\left\vert 0,k\right\rangle ,\theta\cdot
k=0, \label{D.8}%
\end{align}%
\begin{align}
L_{-1}\left\vert x\right\rangle  &  =[\frac{17}{4}(k\cdot\alpha_{-1}%
)^{3}+\frac{9}{2}(k\cdot\alpha_{-1})(\alpha_{-1}\cdot\alpha_{-1}%
)+9(\alpha_{-1}\cdot\alpha_{-2})\nonumber\\
&  +21(k\cdot\alpha_{-1})(k\cdot\alpha_{-2})+25(k\cdot\alpha_{-3})]\left\vert
0,k\right\rangle ;\nonumber\\
\left\vert x\right\rangle  &  =[\frac{25}{2}k\cdot\alpha_{-2}+\frac{9}%
{2}\alpha_{-1}\cdot\alpha_{-1}+\frac{17}{4}(k\cdot\alpha_{-1})^{2}]\left\vert
0,k\right\rangle . \label{D.9}%
\end{align}

Note that there are two degenerate vector ZNS, Eq.(\ref{D.6}) for type II and
Eq.(\ref{D.8}) for type I, at mass level $M^{2}=4$. For mass level $M^{2}=2,$
the high energy limit of Eqs. (\ref{D.5}) and (\ref{D.4}) are calculated to
be
\begin{align}
L_{-1}(\theta\cdot\alpha_{-1})\left\vert 0\right\rangle  &  \rightarrow
\sqrt{2}\alpha_{-1}^{L}\alpha_{-1}^{L}+\alpha_{-2}^{L}\left\vert
0\right\rangle ;\label{D.10}\\
(L_{-2}+\frac{3}{2}L_{-1}^{2})\left\vert 0\right\rangle  &  \rightarrow
(\sqrt{2}\alpha_{-2}^{L}+\frac{1}{2}\alpha_{-1}^{T}\alpha_{-1}^{T})\left\vert
0\right\rangle \label{D.11}\\
&  +\frac{3}{2}(2\alpha_{-1}^{L}\alpha_{-1}^{L}+\sqrt{2}\alpha_{-2}%
^{L})\left\vert 0\right\rangle . \label{D.12}%
\end{align}
Note that Eq.(\ref{D.12}) is the high energy limit of the second term of type
II ZNS. It is easy to see that the decoupling of (\ref{D.10}) implies the
decoupling of (\ref{D.12}). So one can neglect the effect of (\ref{D.12}) even
though it is of leading order in energy. It turns out that this phenomena
persists to any higher mass level as well\textit{.} By solving Eqs.(\ref{D.10}%
) and (\ref{D.11}), we get the desired linear relation, $\mathcal{T}%
_{TT}:\mathcal{T}_{L}:\mathcal{T}_{LL}=4:-\sqrt{2}:1$. \ Similarly, the high
energy limit of Eqs.(\ref{D.6})-(\ref{D.9}) are calculated to be%
\begin{align}
(L_{-2}+\frac{3}{2}L_{-1}^{2})\left\vert 0\right\rangle  &  \rightarrow
(4\alpha_{-1}^{(T}\alpha_{-2}^{L)}+\frac{1}{2}\alpha_{-1}^{T}\alpha_{-1}%
^{T}\alpha_{-1}^{T})\left\vert 0\right\rangle \label{D.13}\\
&  +\frac{3}{2}(4\alpha_{-1}^{(L}\alpha_{-1}^{L}\alpha_{-1}^{T)}+4\alpha
_{-1}^{(T}\alpha_{-2}^{L)})\left\vert 0\right\rangle ;\label{D.14}\\
L_{-1}(\theta_{\mu\nu}\alpha_{-1}^{\mu\nu})\left\vert 0\right\rangle  &
\rightarrow\lbrack2\alpha_{-1}^{(T}\alpha_{-2}^{L)}+2\alpha_{-1}^{(L}%
\alpha_{-1}^{L}\alpha_{-1}^{T)}]\left\vert 0\right\rangle ;\label{D.15}\\
L_{-1}[2\theta\cdot\alpha_{-2}+(k\cdot\alpha_{-1})(\theta\cdot\alpha
_{-1})]\left\vert 0\right\rangle  &  \rightarrow(4\alpha_{-1}^{(L}\alpha
_{-1}^{L}\alpha_{-1}^{T)}+4\alpha_{-1}^{(T}\alpha_{-2}^{L)})\left\vert
0\right\rangle ;\label{D.16}\\
L_{-1}[\frac{25}{2}k\cdot\alpha_{-2}+\frac{9}{2}\alpha_{-1}\cdot\alpha
_{-1}+\frac{17}{4}(k\cdot\alpha_{-1})^{2}]\left\vert 0\right\rangle  &
\rightarrow0. \label{D.17}%
\end{align}

It is easy to see that the decoupling of Eq.(\ref{D.15}) or (\ref{D.16})
implies the decoupling of Eq.(\ref{D.14}). By solving the equations, one gets
$\mathcal{T}_{TTT}:\mathcal{T}_{LLT}:\mathcal{T}_{(LT)}:\mathcal{T}%
_{[LT]}=8:1:-1:-1$ which agrees with Eq.(\ref{2.19..}).

\paragraph{Linear relations for general mass level}

According to the previous section, only states of the form (\ref{relevant})
are relevant in the high energy limit. The mass of the state is $\sqrt
{2(N-1)}$. The 4-point function associated with $|N,m,q\rangle$ will be
denoted $\mathcal{T}^{(N,m,q)}$. The aim of this section is to find the ratio
between a generic $\mathcal{T}^{(N,m,q)}$ and the reference 4-point function,
which is taken to be $\mathcal{T}^{(N,0,0)}$.

Consider the type I HZNS calculated from Eq.(\ref{1.1})
\begin{equation}
L_{-1}|N-1,2m-1,q\rangle\simeq M|N,2m,q\rangle+(2m-1)|N,2m-2,q+1\rangle,
\end{equation}
where many terms are omitted because they are not of the form (\ref{relevant}%
). This implies that
\begin{equation}
\mathcal{T}^{(N,2m,q)}=-\frac{2m-1}{M}\mathcal{T}^{(N,2m-2,q+1)}.
\end{equation}
Using this relation repeatedly, we get
\begin{equation}
\mathcal{T}^{(N,2m,q)}=\frac{(2m-1)!!}{(-M)^{m}}\mathcal{T}^{(N,0,m+q)},
\label{L1.}%
\end{equation}
where the double factorial is defined by $(2m-1)!!=\frac{(2m)!}{2^{m}m!}$.

Next, consider another class of HZNS calculated from type II ZNS in
Eq.(\ref{1.2})
\begin{equation}
L_{-2}|N-2,0,q\rangle\simeq\frac{1}{2}|N,0,q\rangle+M|N,0,q+1\rangle.
\end{equation}
Again, irrelevant terms are omitted here. From this we deduce that
\begin{equation}
\mathcal{T}^{(N,0,q+1)}=-\frac{1}{2M}\mathcal{T}^{(N,0,q)},
\end{equation}
which leads to
\begin{equation}
\mathcal{T}^{(N,0,q)}=\frac{1}{(-2M)^{q}}\mathcal{T}^{(N,0,0)}. \label{L2.}%
\end{equation}

Our main result Eq.(\ref{mainA}) is an immediate result of combining
Eq.(\ref{L1.}) and Eq.(\ref{L2.}).

\subsection{High energy Virasoro constraints}

In this section we will establish a \textquotedblleft dual
description\textquotedblright\ of our approach explained above. The notion
dual to the decoupling of high energy ZNS is Virasoro constraints.

Let us briefly explain how to proceed. First write down a state at a given
mass level as linear combination of states of the form Eq.(\ref{relevant})
with undetermined coefficients, which are interpreted as the Fourier
components of spacetime fields. Requiring that the Virasoro generators $L_{1}$
and $L_{2}$ annihilate the state implies several linear relations on the
coefficients. The linear relations can then be solved to obtain ratios among
all fields.

To compare the results of the two dual descriptions, we note that the
correlation functions can be interpreted as source terms for the particle
corresponding to a chosen vertex. Thus the ratios among sources should be the
same as the ratios among the fields, since all fields of the same mass have
the same propagator. However, some care is needed for the normalization of the
field variables. One should use BPZ conjugates to determine the norm of a
state and normalize the fields accordingly.

\subsubsection{Examples}

To illustrate how Virasoro constraints can be used to derive linear relations
among scattering amplitudes at high energies, we give some explicit examples
in this section. We will calculate the proportionality constants among high
energy scattering amplitudes of different string states up to mass levels
$M^{2}=8$. The results are of course consistent with those of previous work
\cite{ChanLee,ChanLee1,ChanLee2} using high energy ZNS.

At the mass level $M^{2}=4$, the most general form of physical states at mass
level $M^{2}=4$ are given by
\begin{equation}
\lbrack\epsilon_{\mu\nu\lambda}\alpha_{-1}^{\mu}\alpha_{-1}^{\nu}\alpha
_{-1}^{\lambda}+\epsilon_{(\mu\nu)}\alpha_{-1}^{\mu}\alpha_{-2}^{\nu}%
+\epsilon_{\lbrack\mu\nu]}\alpha_{-1}^{\mu}\alpha_{-2}^{\nu}+\epsilon_{\mu
}\alpha_{-3}^{\mu}]|0,k\rangle. \label{49}%
\end{equation}
The Virasoro constraints are
\begin{align}
\epsilon_{(\mu\nu)}+\frac{3}{2}k^{\lambda}\epsilon_{\mu\nu\lambda}  &
=0,\label{50}\\
-k^{\nu}\epsilon_{\lbrack\mu\nu]}+3\epsilon_{\mu}-\frac{3}{2}k^{\nu}%
k^{\lambda}\epsilon_{\mu\nu\lambda}  &  =0,\label{51}\\
2k^{\nu}\epsilon_{\lbrack\mu\nu]}+3\epsilon_{\mu}-3(k^{\nu}k^{\lambda}%
-\eta^{\nu\lambda})\epsilon_{\mu\nu\lambda}  &  =0. \label{52}%
\end{align}
By replacing $P$ by $L$, and ignoring irrelevant states, one easily gets
\begin{equation}
\epsilon_{TTT}:\epsilon_{(LLT)}:\epsilon_{(LT)}:\epsilon_{\lbrack
LT]}=8:1:-3:-3. \label{53}%
\end{equation}
After including the normalization factor of the field variables \footnote{The
normalization factors are determined by the inner product of a state with its
BPZ conjugate.} and the appropriate symmetry factors, one ends up with%
\begin{align}
&  \mathcal{T}_{TTT}:\mathcal{T}_{(LLT)}:\mathcal{T}_{(LT)}:\mathcal{T}%
_{[LT]}\nonumber\\
&  =6\epsilon_{TTT}:6\epsilon_{(LLT)}:-2\epsilon_{(LT)}:-2\epsilon_{\lbrack
LT]}=8:1:-1:-1, \label{54}%
\end{align}
which agrees with Eq.(\ref{2.19..}). It also agree with the results of
Eq.(\ref{mainA}) after Young tableaux decomposition. Here the definitions of
$\mathcal{T}_{TTT},\mathcal{T}_{(LLT)},\mathcal{T}_{(LT)},\mathcal{T}_{[LT]}$
and similar amplitudes hereafter can be found in
\cite{ChanLee,ChanLee1,ChanLee2} and the result obtained is consistent with
the previous ZNS calculation in \cite{ChanLee,ChanLee1} or Eq.(\ref{mainA}).

The ratios for $M^{2}=6$ and $M^{2}=8$ can be obtained similarly in Appendix
\ref{ratios in GR}. At $M^{2}=6$,%
\begin{align}
&  \mathcal{T}_{(TTTT)}:\mathcal{T}_{(TTLL)}:\mathcal{T}_{(LLLL)}%
:\mathcal{T}_{TT,L}:\mathcal{T}_{(TTL)}:\mathcal{T}_{(LLL)}:\mathcal{T}%
_{(LL)}\nonumber\\
&  =4!\mathcal{\epsilon}_{(TTTT)}:4!\mathcal{\epsilon}_{(TTLL)}%
:4!\mathcal{\epsilon}_{(LLLL)}:-4\mathcal{\epsilon}_{TT,L}:-4\mathcal{\epsilon
}_{(TTL)}:-4\mathcal{\epsilon}_{(LLL)}:8\mathcal{\epsilon}_{(LL)}%
^{(2)}\nonumber\\
&  =16:\frac{4}{3}:\frac{1}{3}:-\frac{2\sqrt{6}}{3}:-\frac{4\sqrt{6}}%
{9}:-\frac{\sqrt{6}}{9}:\frac{2}{3}, \label{61}%
\end{align}
which is consistent with the previous ZNS calculation in \cite{ChanLee2} or
Eq.(\ref{1111}). It also agree with the results of Eq.(\ref{mainA}) after
Young tableaux decomposition. At $M^{2}=8$,%
\begin{align}
&  \mathcal{T}_{(TTTTT)}:\mathcal{T}_{(TTTL)}:\mathcal{T}_{(TTTLL)}%
:\mathcal{T}_{(TLLL)}:\mathcal{T}_{(TLLLL)}:\mathcal{T}_{(TLL)}:\mathcal{T}%
_{T,LL}:\mathcal{T}_{TLL,L}:\mathcal{T}_{TTT,L}\nonumber\\
&  =5!\mathcal{\epsilon}_{(TTTTT)}:3!\times2\mathcal{\epsilon}_{(TTTL)}%
:5!\mathcal{\epsilon}_{(TTTLL)}:3!\times2\mathcal{\epsilon}_{(TLLL)}%
:5!\mathcal{\epsilon}_{(TLLLL)}\nonumber\\
&  :8\mathcal{\epsilon}_{(TLL)}:8\mathcal{\epsilon}_{T,LL}:3!\times
2\mathcal{\epsilon}_{TLL,L}:3!\times2\mathcal{\epsilon}_{TTT,L}\nonumber\\
&  =32:\sqrt{2}:2:\frac{3\sqrt{2}}{16}:\frac{3}{8}:\frac{1}{3}:\frac{2}%
{3}:\frac{\sqrt{2}}{16}:3\sqrt{2}, \label{mass8}%
\end{align}
which can be checked to be remarkably consistent with the results of
Eq.(\ref{mainA}) after Young tableaux decomposition.

\subsubsection{General mass levels}

In this section we calculate the ratios of string scattering amplitudes in the
high energy limit for general mass levels by imposing Virasoro constraints.
The final result will, of course, be exactly the same as what we obtained by
requiring the decoupling of high energy ZNS. In the presentation here we use
the notation of Young's tableaux.

We consider the general mass level $M^{2}=2\left(  N-1\right)  $. The most
general state can be written as%
\begin{align}
\left\vert N\right\rangle  &  =\sum_{\left\{  m_{j}\right\}  }\left[  \left(
\frac{1}{1^{m_{1}}m_{1}!}%
\begin{tabular}
[c]{|c|c|c|}\hline
$\mu_{1}^{1}$ & $\cdots$ & $\mu_{m_{1}}^{1}$\\\hline
\end{tabular}
\alpha_{-1}^{\mu_{1}^{1}}\cdots\alpha_{-1}^{\mu_{m_{1}}^{1}}\right)
\otimes\left(  \frac{1}{2^{m_{2}}m_{2}!}%
\begin{tabular}
[c]{|c|c|c|}\hline
$\mu_{1}^{2}$ & $\cdots$ & $\mu_{m_{k}}^{2}$\\\hline
\end{tabular}
\alpha_{-2}^{\mu_{1}^{2}}\cdots\alpha_{-2}^{\mu_{m_{2}}^{2}}\right)
\otimes\cdots\right]  \left\vert 0,k\right\rangle \nonumber\\
&  =\sum_{\left\{  m_{j}\right\}  }\left[  \overset{N}{\underset{j=1}{\otimes
}}\left(  \frac{1}{j^{m_{j}}m_{j}!}%
\begin{tabular}
[c]{|c|c|c|}\hline
$\mu_{1}^{j}$ & $\cdots$ & $\mu_{m_{j}}^{j}$\\\hline
\end{tabular}
\alpha_{-j}^{\mu_{1}^{j}\cdots\mu_{m_{j}}^{j}}\right)  \right]  \left\vert
0,k\right\rangle \label{general state}%
\end{align}
where $1/\left(  j^{m_{j}}m_{j}!\right)  $ are the normalization factors and
we defined the abbreviation
\begin{equation}
\alpha_{-j}^{\mu_{1}^{j}\cdots\mu_{m_{j}}^{j}}\equiv\alpha_{-j}^{\mu_{1}^{j}%
}\cdots\alpha_{-j}^{\mu_{m_{j}}^{j}},
\end{equation}
with $m_{j}$ is the number of the operator $\alpha_{-j}$. The summation runs
over all possible combinations of $m_{j}$'s with the constraints
\begin{equation}
\sum_{j=1}^{N}jm_{j}=N\quad\text{ and }\quad0\leq m_{j}\leq N,
\end{equation}
so that the total mass is $N$. Since the upper indices $\left\{  \mu_{1}%
^{j}\cdots\mu_{m_{j}}^{j}\right\}  $ in $\alpha_{-j}^{\mu_{1}^{j}}\cdots
\alpha_{-j}^{\mu_{m_{j}}^{j}}$ are symmetric, we used the Young tableaux
notation to denote the coefficients in Eq.(\ref{general state}). The direct
product $\otimes$ acts on the Young tableaux in the standard way, for example
\begin{equation}%
\begin{tabular}
[c]{|c|c|}\hline
$1$ & $2$\\\hline
\end{tabular}
\otimes%
\begin{tabular}
[c]{|c|}\hline
$3$\\\hline
\end{tabular}
=%
\begin{tabular}
[c]{|c|c|c|}\hline
$1$ & $2$ & $3$\\\hline
\end{tabular}
\oplus%
\begin{tabular}
[c]{|c|c}\hline
$1$ & \multicolumn{1}{|c|}{$2$}\\\hline
$3$ & \\\cline{1-1}%
\end{tabular}
\oplus%
\begin{tabular}
[c]{|c|c}\hline
$2$ & \multicolumn{1}{|c|}{$1$}\\\hline
$3$ & \\\cline{1-1}%
\end{tabular}
.
\end{equation}
To be clear, for example $n=4$, the state can be written as
\begin{align}
\left\vert 4\right\rangle  &  =\left\{  \frac{1}{4!}%
\begin{tabular}
[c]{|c|c|c|c|}\hline
$\mu_{1}^{1}$ & $\mu_{2}^{1}$ & $\mu_{3}^{1}$ & $\mu_{4}^{1}$\\\hline
\end{tabular}
\alpha_{-1}^{\mu_{1}^{1}}\alpha_{-1}^{\mu_{2}^{1}}\alpha_{-1}^{\mu_{3}^{1}%
}\alpha_{-1}^{\mu_{4}^{1}}+\frac{1}{2\cdot2!}%
\begin{tabular}
[c]{|c|c|}\hline
$\mu_{1}^{1}$ & $\mu_{2}^{1}$\\\hline
\end{tabular}
\otimes%
\begin{tabular}
[c]{|c|}\hline
$\mu_{1}^{2}$\\\hline
\end{tabular}
\alpha_{-1}^{\mu_{1}^{1}}\alpha_{-1}^{\mu_{2}^{1}}\alpha_{-2}^{\mu_{1}^{2}%
}\right. \nonumber\\
&  \left.  +\frac{1}{3}%
\begin{tabular}
[c]{|c|}\hline
$\mu_{1}^{1}$\\\hline
\end{tabular}
\otimes%
\begin{tabular}
[c]{|c|}\hline
$\mu_{1}^{3}$\\\hline
\end{tabular}
\alpha_{-1}^{\mu_{1}^{1}}\alpha_{-3}^{\mu_{1}^{3}}+\frac{1}{2^{2}\cdot2!}%
\begin{tabular}
[c]{|c|c|}\hline
$\mu_{1}^{2}$ & $\mu_{2}^{2}$\\\hline
\end{tabular}
\alpha_{-2}^{\mu_{1}^{2}}\alpha_{-2}^{\mu_{2}^{2}}+\frac{1}{4}%
\begin{tabular}
[c]{|c|}\hline
$\mu_{1}^{4}$\\\hline
\end{tabular}
\alpha_{-4}^{\mu_{1}^{4}}\right\}  \left\vert 0,k\right\rangle .
\end{align}
Next, we will apply the Virasoro constraints to the state
Eq.(\ref{general state} ). The only Virasoro constraints which need to be
considered are%
\begin{equation}
L_{1}\left\vert N\right\rangle =L_{2}\left\vert N\right\rangle =0,
\end{equation}
with $L_{m}$ the standard Virasoro operator%
\begin{equation}
L_{m}=\frac{1}{2}\sum_{n=-\infty}^{\infty}\alpha_{m+n}\cdot\alpha_{-n}.
\end{equation}
After taking care the symmetries of the Young tableaux, the Virasoro
constraints become
\begin{subequations}%
%

\begin{align}
L_{1}\left\vert N\right\rangle  &  =\sum_{\left\{  m_{j}\right\}  }\left[
k^{\mu_{1}^{1}}\overset{N}{\underset{j=1}{\otimes}}%
\begin{tabular}
[c]{|c|c|c|}\hline
$\mu_{1}^{j}$ & $\cdots$ & $\mu_{m_{j}}^{j}$\\\hline
\end{tabular}
\right. \nonumber\\
&  +\sum_{i=2}^{m_{1}}%
\begin{tabular}
[c]{|c|c|c|c|c|}\hline
$\mu_{2}^{1}$ & $\cdots$ & $\hat{\mu}_{i}^{1}$ & $\cdots$ & $\mu_{m_{1}}^{1}%
$\\\hline
\end{tabular}
\otimes%
\begin{tabular}
[c]{|c|c|c|c|}\hline
$\mu_{i}^{1}$ & $\mu_{1}^{2}$ & $\cdots$ & $\mu_{m_{2}}^{2}$\\\hline
\end{tabular}
\overset{N}{\underset{j\neq1,2}{\otimes}}%
\begin{tabular}
[c]{|c|c|c|}\hline
$\mu_{1}^{j}$ & $\cdots$ & $\mu_{m_{j}}^{j}$\\\hline
\end{tabular}
\nonumber\\
&  +\sum_{l=3}^{N}\left(  l-1\right)
\begin{tabular}
[c]{|c|c|c|}\hline
$\mu_{2}^{1}$ & $\cdots$ & $\mu_{m_{1}}^{1}$\\\hline
\end{tabular}
\otimes\sum_{i=1}^{m_{l-1}}%
\begin{tabular}
[c]{|c|c|c|c|c|}\hline
$\mu_{1}^{l-1}$ & $\cdots$ & $\hat{\mu}_{i}^{l-1}$ & $\cdots$ & $\mu_{m_{l-1}%
}^{l-1}$\\\hline
\end{tabular}
\nonumber\\
&  \left.  \otimes%
\begin{tabular}
[c]{|c|c|c|c|}\hline
$\mu_{i}^{l-1}$ & $\mu_{1}^{l}$ & $\cdots$ & $\mu_{m_{l}}^{l}$\\\hline
\end{tabular}
\overset{N}{\underset{j\neq1,l,l-1}{\otimes}}%
\begin{tabular}
[c]{|c|c|c|}\hline
$\mu_{1}^{j}$ & $\cdots$ & $\mu_{m_{j}}^{j}$\\\hline
\end{tabular}
\right] \nonumber\\
&  \frac{1}{\left(  m_{1}-1\right)  !}\alpha_{-1}^{\mu_{2}^{1}\cdots\mu
_{m_{1}}^{1}}\prod_{j\neq1}^{N}\left(  \frac{1}{j^{m_{j}}m_{j}!}\alpha
_{-j}^{\mu_{1}^{j}\cdots\mu_{m_{j}}^{j}}\right)  |0,k\rangle=0, \label{L1}%
\end{align}
and
\begin{align}
L_{2}\left\vert N\right\rangle  &  =\sum_{\left\{  m_{j}\right\}  }\left[
\frac{1}{2}\eta^{\mu_{1}^{1}\mu_{2}^{1}}\overset{N}{\underset{j=1}{\otimes}}%
\begin{tabular}
[c]{|c|c|c|}\hline
$\mu_{1}^{j}$ & $\cdots$ & $\mu_{m_{j}}^{j}$\\\hline
\end{tabular}
\right. \nonumber\\
&  +%
\begin{tabular}
[c]{|c|c|c|}\hline
$\mu_{3}^{1}$ & $\cdots$ & $\mu_{m_{1}}^{1}$\\\hline
\end{tabular}
\otimes%
\begin{tabular}
[c]{|c|c|c|}\hline
$\mu_{1}^{2}$ & $\cdots$ & $\mu_{m_{2}+1}^{2}$\\\hline
\end{tabular}
k^{\mu_{m_{2}+1}^{2}}\overset{N}{\underset{j\neq1,2}{\otimes}}%
\begin{tabular}
[c]{|c|c|c|}\hline
$\mu_{1}^{j}$ & $\cdots$ & $\mu_{m_{j}}^{j}$\\\hline
\end{tabular}
\nonumber\\
&  +\sum_{i=3}^{m_{1}}%
\begin{tabular}
[c]{|c|c|c|c|c|}\hline
$\mu_{3}^{1}$ & $\cdots$ & $\hat{\mu}_{i}^{1}$ & $\cdots$ & $\mu_{m_{1}}^{1}%
$\\\hline
\end{tabular}
\otimes%
\begin{tabular}
[c]{|c|c|c|c|}\hline
$\mu_{i}^{1}$ & $\mu_{1}^{3}$ & $\cdots$ & $\mu_{m_{3}}^{3}$\\\hline
\end{tabular}
\overset{N}{\underset{j\neq1,3}{\otimes}}%
\begin{tabular}
[c]{|c|c|c|}\hline
$\mu_{1}^{j}$ & $\cdots$ & $\mu_{m_{j}}^{j}$\\\hline
\end{tabular}
\nonumber\\
&  +\sum_{l=4}^{N}\left(  l-2\right)
\begin{tabular}
[c]{|c|c|c|}\hline
$\mu_{3}^{1}$ & $\cdots$ & $\mu_{m_{1}}^{1}$\\\hline
\end{tabular}
\otimes\sum_{i=1}^{m_{l-2}}%
\begin{tabular}
[c]{|c|c|c|c|c|}\hline
$\mu_{1}^{l-2}$ & $\cdots$ & $\hat{\mu}_{i}^{l-2}$ & $\cdots$ & $\mu_{m_{l}%
}^{l-2}$\\\hline
\end{tabular}
\nonumber\\
&  \left.  \otimes%
\begin{tabular}
[c]{|c|c|c|c|}\hline
$\mu_{i}^{l-2}$ & $\mu_{1}^{l}$ & $\cdots$ & $\mu_{m_{l}}^{l}$\\\hline
\end{tabular}
\overset{N}{\underset{j\neq1,l,l-2}{\otimes}}%
\begin{tabular}
[c]{|c|c|c|}\hline
$\mu_{1}^{j}$ & $\cdots$ & $\mu_{m_{j}}^{j}$\\\hline
\end{tabular}
\right] \nonumber\\
&  \frac{1}{\left(  m_{1}-2\right)  !}\alpha_{-1}^{\mu_{3}^{1}\cdots\mu
_{m_{1}}^{1}}\prod_{j\neq1}^{N}\left(  \frac{1}{j^{m_{j}}m_{j}!}\alpha
_{-j}^{\mu_{1}^{j}\cdots\mu_{m_{j}}^{j}}\right)  |0,k\rangle=0. \label{L2}%
\end{align}%
\end{subequations}%
A hat on an index means that the index is skipped there (and it should appear
somewhere else). In the above derivation we have used the identity for the
Young tableaux
\begin{align}%
\begin{tabular}
[c]{|c|c|c|}\hline
$1$ & $\cdots$ & $p$\\\hline
\end{tabular}
&  =\frac{1}{p}\left[  1+\sigma_{\left(  21\right)  }+\sigma_{\left(
321\right)  }+\cdots\sigma_{\left(  p\cdots1\right)  }\right]
\begin{tabular}
[c]{|c|c|c|}\hline
$2$ & $\cdots$ & $p$\\\hline
\end{tabular}
\otimes%
\begin{tabular}
[c]{|c|}\hline
$1$\\\hline
\end{tabular}
\nonumber\\
&  =\frac{1}{p}\sum_{i=1}^{p}\sigma_{\left(  i1\right)  }%
\begin{tabular}
[c]{|c|c|c|}\hline
$2$ & $\cdots$ & $p$\\\hline
\end{tabular}
\otimes%
\begin{tabular}
[c]{|c|}\hline
$1$\\\hline
\end{tabular}
,
\end{align}
where $\sigma_{\left(  i\cdots j\right)  }$ are permutation operators.

\paragraph{High energy limit of Virasoro constraints}

\label{High Energy}

States which satisfy the Virasoro constraints are physical states. What we are
going to show in the following is that, in the high energy limit, the Virasoro
constraints turn out to be strong enough to give the linear relationship among
the physical states. To take the high energy limit for the Virasoro
constraints, we replace the indices $\left(  \mu_{i},\nu_{i}\right)  $ by $L$
or $T$ with%
\begin{equation}
k^{\mu_{i}}\rightarrow Me^{L}\text{, }\eta^{\mu_{1}\mu_{2}}\rightarrow
e^{T}e^{T},
\end{equation}
where $M$ is the mass operator.

The Virasoro constraints (\ref{L1}) and (\ref{L2}) at high energies become
(see Appendix \ref{Virasoro} for detail)%
\begin{subequations}%
\begin{align}
\underset{n-2q-2-2m}{\underbrace{%
\begin{tabular}
[c]{|l|l|l|}\hline
$T$ & $\cdots\cdots$ & $T$\\\hline
\end{tabular}
}}\underset{2m+2}{\underbrace{%
\begin{tabular}
[c]{|l|l|l|}\hline
$L$ & $\cdots$ & $L$\\\hline
\end{tabular}
}}\otimes\underset{q}{\underbrace{%
\begin{tabular}
[c]{|l|l|l|}\hline
$L$ & $\cdots$ & $L$\\\hline
\end{tabular}
}}  &  =-\frac{2m+1}{M}\underset{n-2q-2-2m}{\underbrace{%
\begin{tabular}
[c]{|l|l|l|}\hline
$T$ & $\cdots\cdots$ & $T$\\\hline
\end{tabular}
}}\underset{2m}{\underbrace{%
\begin{tabular}
[c]{|l|l|l|}\hline
$L$ & $\cdots$ & $L$\\\hline
\end{tabular}
}}\otimes\underset{q+1}{\underbrace{%
\begin{tabular}
[c]{|l|l|l|}\hline
$L$ & $\cdots$ & $L$\\\hline
\end{tabular}
}},\label{L1.1}\\
\underset{n-2q-2-2m}{\underbrace{%
\begin{tabular}
[c]{|l|l|l|}\hline
$T$ & $\cdots\cdots$ & $T$\\\hline
\end{tabular}
}}\underset{2m}{\underbrace{%
\begin{tabular}
[c]{|l|l|l|}\hline
$L$ & $\cdots$ & $L$\\\hline
\end{tabular}
}}\otimes\underset{q+1}{\underbrace{%
\begin{tabular}
[c]{|l|l|l|}\hline
$L$ & $\cdots$ & $L$\\\hline
\end{tabular}
}}  &  =-\frac{1}{2M}\underset{n-2q-2m}{\underbrace{%
\begin{tabular}
[c]{|l|l|l|}\hline
$T$ & $\cdots$ & $T$\\\hline
\end{tabular}
}}\underset{2m}{\underbrace{%
\begin{tabular}
[c]{|l|l|l|}\hline
$L$ & $\cdots$ & $L$\\\hline
\end{tabular}
}}\otimes\underset{q}{\underbrace{%
\begin{tabular}
[c]{|l|l|l|}\hline
$L$ & $\cdots$ & $L$\\\hline
\end{tabular}
}}, \label{L2.1}%
\end{align}
where we have renamed $m_{2}\rightarrow q$ and $m_{1}\rightarrow N-2q$.%
\end{subequations}
By mathematical recursion, Eq.(\ref{L1.1}) and Eq.(\ref{L2.1}) lead to
\begin{subequations}%
\begin{align}
\underset{N-2q-2m}{\underbrace{%
\begin{tabular}
[c]{|l|l|l|}\hline
$T$ & $\cdots$ & $T$\\\hline
\end{tabular}
}}\underset{2m}{\underbrace{%
\begin{tabular}
[c]{|l|l|l|}\hline
$L$ & $\cdots$ & $L$\\\hline
\end{tabular}
}}\otimes\underset{q}{\underbrace{%
\begin{tabular}
[c]{|l|l|l|}\hline
$L$ & $\cdots$ & $L$\\\hline
\end{tabular}
}}  &  =\frac{\left(  2m-1\right)  !!\left(  -M\right)  ^{q}}{\left(
2m+2q-1\right)  !!}\underset{N-2q-2m}{\underbrace{%
\begin{tabular}
[c]{|l|l|l|}\hline
$T$ & $\cdots$ & $T$\\\hline
\end{tabular}
}}\underset{2m+2q}{\underbrace{%
\begin{tabular}
[c]{|l|l|l|}\hline
$L$ & $\cdots$ & $L$\\\hline
\end{tabular}
}},\label{L1.2}\\
\underset{N-2q-2m}{\underbrace{%
\begin{tabular}
[c]{|l|l|l|}\hline
$T$ & $\cdots$ & $T$\\\hline
\end{tabular}
}}\underset{2m}{\underbrace{%
\begin{tabular}
[c]{|l|l|l|}\hline
$L$ & $\cdots$ & $L$\\\hline
\end{tabular}
}}\otimes\underset{q}{\underbrace{%
\begin{tabular}
[c]{|l|l|l|}\hline
$L$ & $\cdots$ & $L$\\\hline
\end{tabular}
}}  &  =\left(  -\frac{1}{2M}\right)  ^{q}\underset{N-2m}{\underbrace{%
\begin{tabular}
[c]{|l|l|l|}\hline
$T$ & $\cdots$ & $T$\\\hline
\end{tabular}
}}\underset{2m}{\underbrace{%
\begin{tabular}
[c]{|l|l|l|}\hline
$L$ & $\cdots$ & $L$\\\hline
\end{tabular}
}}. \label{L2.2}%
\end{align}%
\end{subequations}%
Combining equations (\ref{L1.2}) and (\ref{L2.2}), we get%
\begin{equation}
\underset{N-2q-2m}{\underbrace{%
\begin{tabular}
[c]{|l|l|l|}\hline
$T$ & $\cdots$ & $T$\\\hline
\end{tabular}
}}\underset{2m}{\underbrace{%
\begin{tabular}
[c]{|l|l|l|}\hline
$L$ & $\cdots$ & $L$\\\hline
\end{tabular}
}}\otimes\underset{q}{\underbrace{%
\begin{tabular}
[c]{|l|l|l|}\hline
$L$ & $\cdots$ & $L$\\\hline
\end{tabular}
}}=\left(  -\frac{1}{2M}\right)  ^{q}\frac{\left(  2k-1\right)  !!}%
{4^{m}\left(  N-1\right)  ^{m}}\underset{N}{\underbrace{%
\begin{tabular}
[c]{|l|l|l|}\hline
$T$ & $\cdots$ & $T$\\\hline
\end{tabular}
}},
\end{equation}
which is equivalent to Eq.(\ref{mainA}).

To get the ratio for the specific physical states, we make the Young tableaux
decomposition%
\begin{equation}
\underset{N-2q-2m}{\underbrace{%
\begin{tabular}
[c]{|l|l|l|}\hline
$T$ & $\cdots$ & $T$\\\hline
\end{tabular}
}}\underset{2m}{\underbrace{%
\begin{tabular}
[c]{|l|l|l|}\hline
$L$ & $\cdots$ & $L$\\\hline
\end{tabular}
}}\otimes\underset{q}{\underbrace{%
\begin{tabular}
[c]{|l|l|l|}\hline
$L$ & $\cdots$ & $L$\\\hline
\end{tabular}
}}=\sum_{l=0}^{q}%
\begin{tabular}
[c]{|c|rcccccc}\hline
$T$ & $\cdots$ &  &  & \multicolumn{1}{|c}{$T$} & \multicolumn{1}{|c}{$L$} &
\multicolumn{1}{|c}{$\cdots$} & \multicolumn{1}{|c|}{$L$}\\\hline
$L$ & \multicolumn{1}{|c}{$\cdots$} & \multicolumn{1}{|c}{$L$} &
\multicolumn{1}{|c}{} &  &  &  & \\\cline{1-3}%
\end{tabular}
\cdot\left(  l!C_{q}^{l}C_{N-2q-2m}^{l}\right)  ,
\end{equation}
where $C_{q}^{l}=\frac{q!}{l!(q-l)!}$ and we have $(N-2q-2m)$ $T$'s and
$(2m+q-l)$ $L$'s in the first column, $(l)$ $L$'s in the second column in the
second line of the above equation. Therefore, we obtain
\begin{equation}%
\begin{tabular}
[c]{|c|rcccccc}\hline
$T$ & $\cdots$ &  &  & \multicolumn{1}{|c}{$T$} & \multicolumn{1}{|c}{$L$} &
\multicolumn{1}{|c}{$\cdots$} & \multicolumn{1}{|c|}{$L$}\\\hline
$L$ & \multicolumn{1}{|c}{$\cdots$} & \multicolumn{1}{|c}{$L$} &
\multicolumn{1}{|c}{} &  &  &  & \\\cline{1-3}%
\end{tabular}
=\frac{1}{\sum_{l=0}^{q}l!C_{q}^{l}C_{N-2q-2m}^{l}}\left(  -\frac{1}%
{2M}\right)  ^{q}\frac{\left(  2m-1\right)  !!}{4^{m}\left(  N-1\right)  ^{m}%
}\underset{N}{\underbrace{%
\begin{tabular}
[c]{|l|l|l|}\hline
$T$ & $\cdots$ & $T$\\\hline
\end{tabular}
}},\nonumber
\end{equation}
which is consistent with the ratios Eqs.(\ref{2.19..}), (\ref{1111}), and
(\ref{mass8}) for $M^{2}=4,6,8$ respectively.

\subsection{Saddle-point calculation}

In previous sections, we have identified the leading high energy amplitudes
and derived the ratios among high energy amplitudes for members of a family at
given mass levels, based on decoupling principle. While deductive arguments
help to clarify the underlying assumptions and solidify the validity of
decoupling principle, it is instructive to compare it with a different
approach, such as the saddle-point approximation \cite{CHL}. Therefore, we
shall perform direct calculations to check the results obtained above and make
comparisons between these two approaches.

In this section, we give a direct verification of the ratios among leading
high energy amplitudes based on the saddle-point method. The four-point
amplitudes to be calculated consist of one massive tensor and three tachyons.
Since we have shown that in the high energy limit the only relevant states are
those corresponding to
\begin{equation}
(\alpha_{-1}^{T})^{N-2m-2q}(\alpha_{-1}^{L})^{2m}(\alpha_{-2}^{L}%
)^{q}|0,k\rangle,\hspace{1cm}-k^{2}=2(N-1),
\end{equation}
we only need to calculate the following four-point amplitude
\begin{equation}
\mathcal{T}^{(N,2m,q)}\equiv\int\prod_{i=1}^{4}dx_{i}\langle V_{1}%
V_{2}^{(N,2m,q)}V_{3}V_{4}\rangle,
\end{equation}
where
\begin{align}
V_{2}^{(N,2m,q)}  &  \equiv(\partial X^{T})^{N-2m-2q}(\partial X^{P}%
)^{2m}(\partial^{2}X^{P})^{q}e^{ik_{2}X_{2}},\\
V_{i}  &  \equiv e^{ik_{i}X_{i}},\hspace{2cm}i=1,3,4.
\end{align}
Notice that here for leading high energy amplitudes we replace the
polarization $L$ by $P$.

Using either path-integral or operator formalism, after $SL(2,R)$ gauge
fixing, we obtain the $s-t$ channel contribution to the stringy amplitude at
tree level
\begin{align}
\mathcal{T}^{(N,2m,q)}  &  \Rightarrow\int_{0}^{1}dxx^{(1,2)}(1-x)^{(2,3)}%
\left[  \frac{e^{T}\cdot k_{1}}{x}-\frac{e^{T}\cdot k_{3}}{1-x}\right]
^{N-2m-2q}\nonumber\\
&  \cdot\left[  \frac{e^{P}\cdot k_{1}}{x}-\frac{e^{P}\cdot k_{3}}%
{1-x}\right]  ^{2m}\left[  -\frac{e^{P}\cdot k_{1}}{x^{2}}-\frac{e^{P}\cdot
k_{3}}{(1-x)^{2}}\right]  ^{q}. \label{singletensor}%
\end{align}

In order to apply the saddle-point method, we need to rewrite the amplitude
above into the \textquotedblleft canonical form\textquotedblright. That is,
\begin{equation}
\mathcal{T}^{(N,2m,q)}(K)=\int_{0}^{1}dx\mbox{ }u(x)e^{-Kf(x)},
\label{integral}%
\end{equation}
where
\begin{align}
K  &  \equiv-(1,2)\rightarrow\frac{s}{2}\rightarrow2E^{2},\\
\tau &  \equiv-\frac{(2,3)}{(1,2)}\rightarrow-\frac{t}{s}\rightarrow\sin
^{2}\frac{\phi}{2},\\
f(x)  &  \equiv\ln x-\tau\ln(1-x),\\
u(x)  &  \equiv\left[  \frac{(1,2)}{M}\right]  ^{2m+q}(1-x)^{-N+2m+2q}%
(f^{\prime})^{2m}(f^{\prime\prime})^{q}(-e^{T}\cdot k_{3})^{N-2m-2q}.
\end{align}
The saddle-point for the integration of moduli, $x=x_{0}$, is defined by
\begin{equation}
f^{\prime}(x_{0})=0,
\end{equation}
and we have
\begin{equation}
x_{0}=\frac{1}{1-\tau},\hspace{1cm}1-x_{0}=-\frac{\tau}{1-\tau},\hspace
{1cm}f^{\prime\prime}(x_{0})=(1-\tau)^{3}\tau^{-1}.
\end{equation}
{}From the definition of $u(x)$, it is easy to see that
\begin{equation}
u(x_{0})=u^{\prime}(x_{0})=....=u^{(2m-1)}(x_{0})=0,
\end{equation}
and
\begin{equation}
u^{(2m)}(x_{0})=\left[  \frac{(1,2)}{M}\right]  ^{2m+q}(1-x_{0})^{-N+2m+2q}%
(2m)!(f_{0}^{\prime\prime})^{2m+q}(-e^{T}\cdot k_{3})^{N-2m-2q}.
\end{equation}

With these inputs, one can easily evaluate the Gaussian integral associated
with the four-point amplitudes, Eq.(\ref{integral}),
\begin{align}
&  \int_{0}^{1}dx\mbox{ }u(x)e^{-Kf(x)}\nonumber\\
&  =\sqrt{\frac{2\pi}{Kf_{0}^{\prime\prime}}}e^{-Kf_{0}}\left[  \frac
{u_{0}^{(2m)}}{2^{m}\ m!\ (f_{0}^{\prime\prime})^{m}\ K^{m}}+O(\frac
{1}{K^{m+1}})\right] \nonumber\\
&  =\sqrt{\frac{2\pi}{Kf_{0}^{\prime\prime}}}e^{-Kf_{0}}\left[  (-1)^{N-q}%
\frac{2^{N-2m-q}(2m)!}{m!\ {M}^{2m+q}}\ \tau^{-\frac{N}{2}}(1-\tau)^{\frac
{3N}{2}}E^{N}+O(E^{N-2})\right]  . \label{leading}%
\end{align}
This result shows explicitly that with one tensor and three tachyons, the
energy and angle dependence for the high energy four-point amplitudes only
depend on the level $N$, and we can solve for the ratios among high energy
amplitudes within the same family,
\begin{equation}
\lim_{E\rightarrow\infty}\frac{\mathcal{T}^{(N,2m,q)}}{\mathcal{T}^{(N,0,0)}%
}=\frac{(-1)^{q}(2m)!}{m!(2M)^{2m+q}}=\left(  -\frac{2m-1}{M}\right)
....\left(  -\frac{3}{M}\right)  \left(  -\frac{1}{M}\right)  \left(
-\frac{1}{2M}\right)  ^{m+q},
\end{equation}
which is consistent with Eq.(\ref{mainA}).

We conclude this section with three remarks. Firstly, from the saddle-point
approach, it is easy to see why the product of $\alpha_{-1}^{P}$ oscillators
induce energy suppression. Their contribution to the stringy amplitude is
proportional to powers of $f^{\prime}(x_{0})$, which is zero in the leading
order calculation. Secondly, one can also understand why only even numbers of
$\alpha_{-1}^{P}$ oscillators will survive for high energy amplitudes based on
the structure of Gaussian integral in Eq.(\ref{integral}). While for a vertex
operator containing $(2m+1)$ $\alpha_{-1}^{P}$'s, we have $u(x_{0})=u^{\prime
}(x_{0})=....=u^{(2m)}(x_{0})=0$, and the leading contribution comes from
$u^{(2m+1)}(x_{0})(x-x_{0})^{2m+1}$, this gives zero since the odd-power
moments of Gaussian integral vanish. Finally, for the alert readers, since we
only discuss the $s-t$ channel contribution to the scattering amplitudes, the
integration range for the $x$ variable seems to devoid of a direct application
of saddle-point method. We will discuss this issue in chapter VII.

\subsection{$2D$ string at high energies}

Although we have shown that there exist infinitely many linear relations among
4-point functions which uniquely fix their ratios in the high-energy limit, it
is not totally clear that there is a hidden symmetry responsible for it.
However, we would like to claim that these linear relations are indeed the
manifestation of the long-sought hidden symmetry of string theory, and that we
are on the right track of understanding the symmetry. To persuade the readers,
we test our claim on a toy model of string theory --the $2D$ string theory.

While the hidden symmetry of the $26D$ bosonic string theory is still at
large, the $w_{\infty}$ symmetry of the $2D$ string theory is much better
understood. It is known to be associated with the discrete Polyakov states
discussed in chapter III of part I of this review. Let us now check whether
the $w_{\infty}$ symmetry is generated by the high energy limit of ZNS. In
\cite{ChungLee1} , explicit expression for a class of discrete ZNS with
Polyakov momenta was given in Eq.(\ref{3.11.}). For illustration, let's repeat
it here
\begin{equation}%
\begin{split}
G_{J,M}^{+}=  &  (J+M+1)^{-1}\int\frac{dz}{2\pi i}\left[  \psi_{1,-1}%
^{+}(z)\psi_{J,M+1}^{+}(0)+\psi_{J,M+1}^{+}(z)\psi_{1,-1}^{+}(0)\right] \\
\sim &  (J-M)!\Delta(J,M,-i\sqrt{2}X)Exp\left[  \sqrt{2}(iMX+(J-1)\phi)\right]
\\
&  +(-1)^{2J}\sum\limits_{j=1}^{J-M}(J-M-1)!\int\frac{dz}{2\pi i}%
\mathcal{D}(J,M,-i\sqrt{2}X(z),j)\\
&  \cdot Exp\left[  \sqrt{2}(i(M+1)X(z)+(J-1)\phi(z)-X(0))\right]  .
\end{split}
\label{G+}%
\end{equation}
Here $\Delta(J,M,-i\sqrt{2}X)$ is defined by
\begin{equation}
\Delta(J,M,-i\sqrt{2}X)=\left\vert
\begin{array}
[c]{cccc}%
S_{2J-1} & S_{2J-2} & \cdots & S_{J+M}\\
S_{2J-2} & S_{2J-3} & \cdots & S_{J+M-1}\\
\cdots & \cdots & \cdots & \cdots\\
S_{J+M} & S_{J+M-1} & \cdots & S_{2M+1}%
\end{array}
\right\vert , \label{Delta}%
\end{equation}
where
\begin{equation}
S_{k}=S_{k}\left(  \left\{  \frac{-i\sqrt{2}}{k!}\partial^{k}X(0)\right\}
\right)  ,\quad\mbox{and}\quad S_{k}=0\quad\mbox{if}\quad k<0,
\end{equation}
and $S_{k}(\{a_{i}\})$'s denote the Schur polynomial defined by
\begin{equation}
\exp\left(  \sum_{k=1}^{\infty}a_{k}x^{k}\right)  =\sum_{k=0}^{\infty}%
S_{k}(\{a_{i}\})x^{k}.
\end{equation}
$\mathcal{D}(J,M,-i\sqrt{2}X(z),j)$ is defined by a similar expression as
Eq.(\ref{Delta}), but with the $j$-th row replaced by $\{(-z)^{j}%
-1-2J,(-z)^{j}-2J,\cdots,(-z)^{j-J-M-2}\}$. It was shown \cite{ChungLee1} that
ZNS in Eq.(\ref{G+}) generate a $w_{\infty}$ algebra.

In the high energy limit, the factors $\partial^{k}X^{A}$ are generically
proportional to a linear combination of the momenta of other vertices, so it
scales with energy $E$. Thus $\mathcal{D}(J,M,-i\sqrt{2}X,j)$ is subleading to
$\Delta(J,M,-i\sqrt{2}X)$. Ignoring the second term in Eq.(\ref{G+}) for this
reason, we see that these ZNS indeed approach to the discrete states
$\psi_{JM}^{+}$ in Eq.(\ref{3.5.})! Thus, the $w_{\infty}$ algebra generated
by Eq.(\ref{G+}) is identified to $w_{\infty}$ symmetry in Eq.(\ref{3.10.}).
This result strongly suggests that the linear relations among correlation
functions obtained from ZNS are indeed related to the hidden symmetry also for
the $26D$ strings.

In chapter XV of part III of this review, we will address a similar issue in
the RR of $26D$ string theory, where high energy spacetime symmetry\ is shown
to be related to $SL(5,C)$ of the Appell function $F_{1}$. Although we still
do not know what is the exact symmetry group of $26D$ strings, or how it acts
on states, these works shed new light on the road to finding the answers.%

\setcounter{equation}{0}
\renewcommand{\theequation}{\arabic{section}.\arabic{equation}}%

\section{ZNS in DDF construction and WSFT}

In this chapter, in addition to the OCFQ scheme, we will identify and
calculate \cite{CLYang} the counterparts of ZNS in two other quantization
schemes of 26D open bosonic string theory, namely, the light-cone DDF
\cite{Giudice,DDF1,DDF2} ZNS and the off-shell BRST ZNS (with ghost) in WSFT.
In particular, special attentions are paid to the inter-particle ZNS in all
quantization schemes. For the case of off-shell BRST ZNS, we impose the no
ghost conditions and exactly recover two types of on-shell ZNS in the OCFQ
string spectrum for the first few low-lying mass levels. We then show that
off-shell gauge transformations of WSFT are identical to the on-shell stringy
gauge symmetries generated by two types of ZNS in the OCFQ string theory.

Our calculations in this chapter serve as the first step to study stringy
symmetries in light-cone DDF and BRST string theories, and to bridge the links
between different quantization schemes for both on-shell and off-shell string
theories. In section A, we first review the calculations of ZNS in OCFQ
spectrum. The most general spectrum analysis in the helicity basis, including
ZNS, is then given to discuss the inter-particle $D_{2}$ ZNS
\cite{LeePRL,Lee,Lee4,LEO} at mass level $M^{2}=4$. We will see that one can
use polarization of either one of the two positive-norm states to represent
the polarization of the inter-particle ZNS.

In section B, we calculate both type I and type II ZNS in the light-cone DDF
string up to mass level $M^{2}=4$. In section C, we first calculate off-shell
ZNS with ghosts from linearized gauge transformation of WSFT. After imposing
the no ghost conditions on these ZNS, we can exactly reproduce two types of
ZNS in OCFQ spectrum for the first few low-lying mass levels. We then show
that off-shell gauge transformations of WSFT are identical to the on-shell
stringy gauge symmetries generated by two types of ZNS in the generalized
massive $\sigma$-model approach \cite{LeePRL,Lee} of string theory.

Based on the ZNS calculations \cite{ChanLee,ChanLee1,ChanLee2,CHL}, we thus
have related gauge symmetry of WSFT \cite{Witten} to the high energy stringy
symmetry conjectured by Gross \cite{GM,GM1,Gross,Gross1,GrossManes}.

\subsection{ZNS in the OCFQ spectrum}

\subsubsection{ZNS with constraints\bigskip}

In the OCFQ spectrum of open bosonic string theory, the solutions of physical
states conditions include positive-norm propagating states and two types of
ZNSs. The solutions of ZNS up to the mass level $M^{2}=4$ were calculated in
chapter II. We re-list them in the following :

1. $M^{2}=-k^{2}=0:$%

\begin{equation}
L_{-1\text{ }}\left\vert x\right\rangle =k\cdot\alpha_{-1}\left\vert
0,k\right\rangle ;\left\vert x\right\rangle =\left\vert 0,k\right\rangle
;\left\vert x\right\rangle =\left\vert 0,k\right\rangle . \label{2.3q}%
\end{equation}

2. $M^{2}=-k^{2}=2:$%
\begin{subequations}%
%

\begin{equation}
\left(  L_{-2}+\frac{3}{2}L_{-1}^{2}\right)  \left\vert \widetilde{x}%
\right\rangle =\left[  \frac{1}{2}\alpha_{-1}\cdot\alpha_{-1}+\frac{5}%
{2}k\cdot\alpha_{-2}+\frac{3}{2}(k\cdot\alpha_{-1})^{2}\right]  \left\vert
0,k\right\rangle ;\left\vert \widetilde{x}\right\rangle =\left\vert
0,k\right\rangle , \label{2.4q}%
\end{equation}

\begin{equation}
L_{-1}\left\vert x\right\rangle =[\theta\cdot\alpha_{-2}+(k\cdot\alpha
_{-1})(\theta\cdot\alpha_{-1})]\left\vert 0,k\right\rangle ;\left\vert
x\right\rangle =\theta\cdot\alpha_{-1}\left\vert 0,k\right\rangle ,\theta\cdot
k=0. \label{2.5q}%
\end{equation}%
\end{subequations}%

3. $M^{2}=-k^{2}=4:$%
\begin{subequations}%
%

\begin{align}
\left(  L_{-2}+\frac{3}{2}L_{-1}^{2}\right)  \left\vert \widetilde{x}%
\right\rangle  &  =\left[  4\theta\cdot\alpha_{-3}+\frac{1}{2}(\alpha
_{-1}\cdot\alpha_{-1})(\theta\cdot\alpha_{-1})+\frac{5}{2}(k\cdot\alpha
_{-2})(\theta\cdot\alpha_{-1})\right. \nonumber\\[0.01in]
&  \left.  +\frac{3}{2}(k\cdot\alpha_{-1})^{2}(\theta\cdot\alpha
_{-1})+3(k\cdot\alpha_{-1})(\theta\cdot\alpha_{-2})\right]  \left\vert
0,k\right\rangle ;\nonumber\\
\left\vert \widetilde{x}\right\rangle  &  =\theta\cdot\alpha_{-1}\left\vert
0,k\right\rangle ,k\cdot\theta=0, \label{2.6q}%
\end{align}

\begin{align}
L_{-1}\left\vert x\right\rangle  &  =[2\theta_{\mu\nu}\alpha_{-1}^{\mu}%
\alpha_{-2}^{\nu}+k_{\lambda}\theta_{\mu\nu}\alpha_{-1}^{\lambda}\alpha
_{-1}^{\mu}\alpha_{-1}^{\nu}]\left\vert 0,k\right\rangle ;\nonumber\\
\left\vert x\right\rangle  &  =\theta_{\mu\nu}\alpha_{-1}^{\mu\nu}\left\vert
0,k\right\rangle ,k\cdot\theta=\eta^{\mu\nu}\theta_{\mu\nu}=0,\theta_{\mu\nu
}=\theta_{\nu\mu}, \label{2.7q}%
\end{align}

\begin{align}
L_{-1}\left\vert x\right\rangle  &  =\left[  \frac{1}{2}(k\cdot\alpha
_{-1})^{2}(\theta\cdot\alpha_{-1})+2\theta\cdot\alpha_{-3}+\frac{3}{2}%
(k\cdot\alpha_{-1})(\theta\cdot\alpha_{-2})\right. \nonumber\\
&  \left.  +\frac{1}{2}(k\cdot\alpha_{-2})(\theta\cdot\alpha_{-1})\right]
\left\vert 0,k\right\rangle ;\text{ \ }\nonumber\\
\text{\ }\left\vert x\right\rangle  &  =[2\theta\cdot\alpha_{-2}+(k\cdot
\alpha_{-1})(\theta\cdot\alpha_{-1})]\left\vert 0,k\right\rangle ,\theta\cdot
k=0, \label{2.8q}%
\end{align}

\begin{align}
L_{-1}\left\vert x\right\rangle  &  =\left[  \frac{17}{4}(k\cdot\alpha
_{-1})^{3}+\frac{9}{2}(k\cdot\alpha_{-1})(\alpha_{-1}\cdot\alpha
_{-1})+9(\alpha_{-1}\cdot\alpha_{-2})\right. \nonumber\\
&  +21(k\cdot\alpha_{-1})(k\cdot\alpha_{-2})+25(k\cdot\alpha_{-3})]\left\vert
0,k\right\rangle ;\nonumber\\
\left\vert x\right\rangle  &  =\left[  \frac{25}{2}k\cdot\alpha_{-2}+\frac
{9}{2}\alpha_{-1}\cdot\alpha_{-1}+\frac{17}{4}(k\cdot\alpha_{-1})^{2}\right]
\left\vert 0,k\right\rangle . \label{2.9q}%
\end{align}%
\end{subequations}%
Note that there are two degenerate vector ZNSs, Eq.(\ref{2.6q}) for type II
and Eq.(\ref{2.8q}) for type I, at mass level $M^{2}=4$. We define $D_{2}$
vector ZNS by antisymmetrizing those terms which contain $\alpha_{-1}^{\mu
}\alpha_{-2}^{\nu}$ in Eq.(\ref{2.6q}) and Eq.(\ref{2.8q}) as following%

\begin{equation}
|D_{2}\rangle=\left[  \left(  \frac{1}{2}k_{\mu}k_{\nu}\theta_{\lambda}%
+2\eta_{\mu\nu}\theta_{\lambda}\right)  \alpha_{-1}^{\mu}\alpha_{-1}^{\nu
}\alpha_{-1}^{\lambda}+9k_{\mu}\theta_{\nu}\alpha_{-2}^{[\mu}\alpha_{-1}%
^{\nu]}-6\theta_{\mu}\alpha_{-3}^{\mu}\right]  \left\vert 0,k\right\rangle
,\text{ \ }k\cdot\theta=0. \label{2.10q}%
\end{equation}
Similarly $D_{1}$ vector ZNS is defined by symmetrizing those terms which
contain $\alpha_{-1}^{\mu}\alpha_{-2}^{\nu}$ in Eq.(\ref{2.6q}) and
Eq.(\ref{2.8q})
\begin{equation}
|D_{1}\rangle=\left[  \left(  \frac{5}{2}k_{\mu}k_{\nu}\theta_{\lambda}%
+\eta_{\mu\nu}\theta_{\lambda}\right)  \alpha_{-1}^{\mu}\alpha_{-1}^{\nu
}\alpha_{-1}^{\lambda}+9k_{\mu}\theta_{\nu}\alpha_{-2}^{(\mu}\alpha_{-1}%
^{\nu)}+6\theta_{\mu}\alpha_{-3}^{\mu}\right]  \left\vert 0,k\right\rangle
\text{, \ }k\cdot\theta=0. \label{2.11q}%
\end{equation}
In general, \textit{an inter-particle ZNS} can be defined to be $D_{2}+\alpha
D_{1}$, where $\alpha$ is an arbitrary constant.

\subsubsection{ZNS in the helicity basis}

In this section, we are going to do the most general spectrum analysis which
naturally includes ZNS. We will then \textit{solve} the Virasoro constraints
in the helicity basis and recover the ZNS listed above. In particular, this
analysis will make it clear how $D_{2\text{ }}$ZNS in Eq.(\ref{2.10q}) can
induce the inter-particular symmetry transformation for two propagating states
at the mass level $M^{2}=4$.

We begin our discussion for the mass level $M^{2}=2$. At this mass level, the
general expression for the physical states can be written as
\begin{equation}
\lbrack\epsilon_{\mu\nu}\alpha_{-1}^{\mu}\alpha_{-1}^{\nu}+\epsilon_{\mu
}\alpha_{-2}^{\mu}]|0,k\rangle. \label{2.13q}%
\end{equation}
In the OCFQ of string theory, physical states satisfy the mass shell
condition
\begin{equation}
(L_{0}-1)|phys\rangle=0\Rightarrow k^{2}=-2; \label{2.14q}%
\end{equation}
and the Virasoro constraints $L_{1}|phys\rangle=L_{2}|phys\rangle=0$ which
give
\begin{align}
\epsilon_{\mu}  &  =-\epsilon_{\mu\nu}k^{\nu},\label{2.15q}\\
\eta^{\mu\nu}\epsilon_{\mu\nu}  &  =2\epsilon_{\mu\nu}k^{\mu}k^{\nu}.
\label{2.16q}%
\end{align}
In order to solve for the constraints Eq.(\ref{2.15q}) and Eq.(\ref{2.16q}) in
a covariant way, it is convenient to make the following change of basis,%
\begin{subequations}%
\begin{align}
e_{P}  &  \equiv\frac{1}{m}(E,0,....,\text{k})\label{2.17q}\\
e_{L}  &  \equiv\frac{1}{m}(\text{k},0,....,E)\label{2.18q}\\
e_{T_{i}}  &  \equiv(0,0,....,1(\text{i-th spatial direction}),....,0),\hspace
{0.5cm}i=1,2,....,24. \label{2.19q}%
\end{align}%
\end{subequations}%
The 2nd rank tensor $\epsilon_{\mu\nu}$ can be written in the helicity basis
Eq.(\ref{2.17q}) to Eq.(\ref{2.19q}) as
\begin{equation}
\epsilon_{\mu\nu}=\sum_{A,B}u_{AB}e_{\mu}^{A}e_{\nu}^{B},\text{ \ \ \ \ }%
A,B=P,L,T_{i}. \label{2.20q}%
\end{equation}
In this new representation, the second Virasoro constraint Eq.(\ref{2.16q})
reduces to a simple algebraic relation, and one can solve it
\begin{equation}
u_{PP}=\frac{1}{5}\left(  u_{LL}+\sum_{i=1}^{24}u_{T_{i}T_{i}}\right)  .
\label{2.21q}%
\end{equation}
In order to perform an irreducible decomposition of the spin-two state into
the trace and traceless parts, we define the following variables
\begin{align}
x  &  \equiv\frac{1}{25}\left(  u_{LL}+\sum_{i=1}^{24}u_{T_{i}T_{i}}\right)
,\label{2.22q}\\
y  &  \equiv\frac{1}{25}\left(  u_{LL}-\frac{1}{24}\sum_{i=1}^{24}%
u_{T_{i}T_{i}}\right)  . \label{2.23q}%
\end{align}
We can then write down the complete decompositions of the spin-two
polarization tensor as
\begin{align}
\epsilon_{\mu\nu}  &  =x\left(  5e_{\mu}^{P}e_{\nu}^{P}+e_{\mu}^{L}e_{\nu}%
^{L}+\sum_{i=1}^{24}e_{\mu}^{T_{i}}e_{\nu}^{T_{i}}\right)  +y\sum_{i=1}%
^{24}\left(  e_{\mu}^{L}e_{\nu}^{L}-e_{\mu}^{T_{i}}e_{\nu}^{T_{i}}\right)
+\sum_{i,j}(u_{T_{i}T_{j}}-\frac{\delta_{ij}}{24}\sum_{l=1}^{24}u_{T_{l}T_{l}%
})e_{\mu}^{T_{i}}e_{\nu}^{T_{j}}\nonumber\\
&  +u_{PL}(e_{\mu}^{P}e_{\nu}^{L}+e_{\mu}^{L}e_{\nu}^{P})+\sum_{i=1}%
^{24}u_{PT_{i}}(e_{\mu}^{P}e_{\nu}^{T_{i}}+e_{\mu}^{T_{i}}e_{\nu}^{P}%
)+\sum_{i=1}^{24}u_{LT_{i}}(e_{\mu}^{L}e_{\nu}^{T_{i}}+e_{\mu}^{T_{i}}e_{\nu
}^{L}). \label{2.24q}%
\end{align}
The first Virasoro constraint Eq.(\ref{2.15q}) implies that $\epsilon_{\mu}$
vector is not an independent variable, and is related to the spin-two
polarization tensor $\epsilon_{\mu\nu}$ as follows
\begin{equation}
\epsilon_{\mu}=5\sqrt{2}xe_{\mu}^{P}+\sqrt{2}u_{PL}e_{\mu}^{L}+\sqrt{2}%
\sum_{i=1}^{24}u_{PT_{i}}e_{\mu}^{T_{i}}. \label{2.25q}%
\end{equation}
Finally, combining the results of Eq.(\ref{2.23q}),Eq.(\ref{2.24q}) and
Eq.(\ref{2.25q}), we get the complete solution for physical states at mass
level $M^{2}=2$
\begin{align}
&  [\epsilon_{\mu\nu}\alpha_{-1}^{\mu}\alpha_{-1}^{\nu}+\epsilon_{\mu}%
\alpha_{-2}^{\mu}]|0,k\rangle\nonumber\\
&  =x\left(  5\alpha_{-1}^{P}\alpha_{-1}^{P}+\alpha_{-1}^{L}\alpha_{-1}%
^{L}+\sum_{i=1}^{24}\alpha_{-1}^{T_{i}}\alpha_{-1}^{T_{i}}+5\sqrt{2}%
\alpha_{-2}^{P}\right)  |0,k\rangle\label{2.26q}\\
&  +y\sum_{i=1}^{24}\left(  \alpha_{-1}^{L}\alpha_{-1}^{L}-\alpha_{-1}^{T_{i}%
}\alpha_{-1}^{T_{i}}\right)  |0,k\rangle\label{2.27q}\\
&  +\sum_{i,j}\left(  u_{T_{i}T_{j}}-\frac{\delta_{ij}}{24}\sum_{l=1}%
^{24}u_{T_{l}T_{l}}\right)  \alpha_{-1}^{T_{i}}\alpha_{-1}^{T_{j}}%
|0,k\rangle\label{2.28q}\\
&  +u_{PL}\left(  2\alpha_{-1}^{P}\alpha_{-1}^{L}+\sqrt{2}\alpha_{-2}%
^{L}\right)  |0,k\rangle\label{2.29q}\\
&  +\sum_{i=1}^{24}u_{PT_{i}}\left(  2\alpha_{-1}^{P}\alpha_{-1}^{T_{i}}%
+\sqrt{2}\alpha_{-2}^{T_{i}}\right)  |0,k\rangle\label{2.30q}\\
&  +2\sum_{i=1}^{24}u_{LT_{i}}\alpha_{-1}^{L}\alpha_{-1}^{T_{i}}|0,k\rangle,
\label{2.31q}%
\end{align}
where the oscillator creation operators $\alpha_{-1}^{P},\alpha_{-1}%
^{L},\alpha_{-1}^{T_{i}},$ etc., are defined as
\begin{equation}
\alpha_{-n}^{A}\equiv e_{\mu}^{A}\cdot\alpha_{-n}^{\mu},\hspace{1cm}n\in
N,\hspace{1cm}A=P,L,T_{i}. \label{2.32q}%
\end{equation}

In comparison with the standard expressions for ZNS in section A, we find that
Eq.(\ref{2.26q}), Eq.(\ref{2.29q}) and Eq.(\ref{2.30q}) are identical to the
type II singlet and type I vector ZNS for the mass level $M^{2}$ $=2$%
\begin{subequations}%
\begin{align}
\text{Eq.(\ref{2.26q})}  &  =2x\left[  \left(  \frac{1}{2}\eta_{\mu\nu}%
+\frac{3}{2}k_{\mu}k_{\nu}\right)  \alpha_{-1}^{\mu}\alpha_{-1}^{\nu}+\frac
{5}{2}k_{\mu}\alpha_{-2}^{\mu}\right]  |0,k\rangle,\\
\text{Eq.(\ref{2.29q})}  &  =\sqrt{2}u_{PL}[e_{\mu}^{L}k_{\nu}\alpha_{-1}%
^{\mu}\alpha_{-1}^{\nu}+e_{\mu}^{L}\alpha_{-2}^{\mu}]|0,k\rangle,\\
\text{Eq.(\ref{2.30q})}  &  =\sum_{i=1}^{24}\sqrt{2}u_{PT_{i}}\left[  e_{\mu
}^{T_{i}}k_{\nu}\alpha_{-1}^{\mu}\alpha_{-1}^{\nu}+e_{\mu}^{T_{i}}\alpha
_{-2}^{\mu}\right]  |0,k\rangle.
\end{align}%
\end{subequations}%
In addition, one can clearly see from our covariant decomposition how ZNS
generate gauge transformations on positive-norm states. While a nonzero value
for $x$ induces a gauge transformation along the type II singlet ZNS
direction, the coefficients $u_{PL},u_{PT_{i}}$ parametrize the type I vector
gauge transformations with polarization vectors $\theta=e^{L}$ and
$\theta=e^{T_{i}}$, respectively. Finally, by a simple counting of degrees of
freedom, one can identify Eq.(\ref{2.27q}), Eq.(\ref{2.28q}) and
Eq.(\ref{2.31q}) as the singlet ($1$), (traceless) tensor ($299$), and vector
($24$) positive-norm states, respectively. These positive-norm states are in a
one-to-one correspondence with the degrees of freedom in the light-cone
quantization scheme.

We now turn to the analysis of $M^{2}=4$ spectrum. Due to the complexity of
our calculations, we shall present the calculations in three steps. We shall
first write down all of physical states (including both positive-norm and ZNS)
in the simplest gauge choices in the helicity basis. We then calculate the
spin-$3$ state decomposition in the most general gauge choice. Finally, the
complete analysis will be given to see how $D_{2\text{ }}$ZNS in
Eq.(\ref{2.10q}) can induce the inter-particle symmetry transformation for two
propagating states at the mass level $M^{2}=4$.

\paragraph{Physical states in the simplest gauge choices}

To begin with, let us first analyses the positive-norm states. There are two
particles at the mass level $M^{2}=4$, a totally symmetric spin-three particle
and an antisymmetric spin-two particle. The canonical representation of the
spin-three state is usually chosen as
\begin{equation}
\epsilon_{\mu\nu\lambda}\ \alpha_{-1}^{\mu}\alpha_{-1}^{\nu}\alpha
_{-1}^{\lambda}|0,k\rangle,\hspace{1cm}k^{2}=-4, \label{2.34q}%
\end{equation}
where the totally symmetric polarization tensor $\epsilon_{\mu\nu\lambda}$ can
be expanded in the helicity basis as
\begin{equation}
\epsilon_{\mu\nu\lambda}=\sum_{A,B,C}\ \tilde{u}_{ABC}e_{\mu}^{A}e_{\nu}%
^{B}e_{\lambda}^{C},\hspace{1cm}A,B,C=P,L,T_{i}. \label{2.35q}%
\end{equation}
The Virasoro conditions on the polarization tensor can be solved as follows%

\begin{align}
k^{\lambda}\epsilon_{\mu\nu\lambda}=0\Rightarrow &  \tilde{u}_{PAB}%
=0,\hspace{0.5cm}\forall A,B=P,L,T_{i},\label{2.36q}\\
\eta^{\nu\lambda}\epsilon_{\mu\nu\lambda}=0\Rightarrow &  \tilde{u}_{LLL}%
+\sum_{i}\tilde{u}_{T_{i}T_{i}L}=0,\nonumber\\
&  \tilde{u}_{LLT_{i}}+\sum_{j}\tilde{u}_{T_{j}T_{j}T_{i}}=0. \label{2.37q}%
\end{align}
If we choose to keep the minimal number of $L$ components in the expansion
coefficients $\tilde{u}_{ABC}$ for the spin-three particle, we get the
following canonical decomposition
\begin{align}
|A(\epsilon)\rangle &  \equiv(\epsilon_{\mu\nu\lambda}\alpha_{-1}^{\mu}%
\alpha_{-1}^{\nu}\alpha_{-1}^{\lambda})|0,k\rangle=|A(\tilde{u})\rangle
\nonumber\\
&  =\sum_{i}\tilde{u}_{T_{i}T_{i}T_{i}}(\alpha_{-1}^{T_{i}}\alpha_{-1}^{T_{i}%
}\alpha_{-1}^{T_{i}}-3\alpha_{-1}^{L}\alpha_{-1}^{L}\alpha_{-1}^{T_{i}%
})|0,k\rangle\nonumber\\
&  +\sum_{i\neq j}\ 3\ \tilde{u}_{T_{j}T_{j}T_{i}}(\alpha_{-1}^{T_{j}}%
\alpha_{-1}^{T_{j}}\alpha_{-1}^{T_{i}}-\alpha_{-1}^{L}\alpha_{-1}^{L}%
\alpha_{-1}^{T_{i}})|0,k\rangle\nonumber\\
&  +\sum_{(i\neq j\neq k)}6\ \tilde{u}_{T_{i}T_{j}T_{k}}(\alpha_{-1}^{T_{i}%
}\alpha_{-1}^{T_{j}}\alpha_{-1}^{T_{k}})|0,k\rangle\nonumber\\
&  +\sum_{i}\tilde{u}_{LT_{i}T_{i}}(3\ \alpha_{-1}^{L}\alpha_{-1}^{T_{i}%
}\alpha_{-1}^{T_{i}}-\alpha_{-1}^{L}\alpha_{-1}^{L}\alpha_{-1}^{L}%
)|0,k\rangle\nonumber\\
&  +\sum_{(i\neq j)}6\ \tilde{u}_{LT_{i}T_{j}}(\alpha_{-1}^{L}\alpha
_{-1}^{T_{i}}\alpha_{-1}^{T_{j}})|0,k\rangle. \label{2.38q}%
\end{align}
It is easy to check that the $2900$ independent degrees of freedom of the
spin-three particle decompose into $24+552+2024+24+276$ in the above representation.

Similarly, for the antisymmetric spin-two particle, we have the following
canonical representation
\begin{equation}
\epsilon_{\lbrack\mu,\nu]}\alpha_{-1}^{\mu}\alpha_{-2}^{\nu}|0,k\rangle.
\label{2.39q}%
\end{equation}
Rewriting the polarization tensor $\epsilon_{\lbrack\mu,\nu]}$ in the helicity
basis
\begin{equation}
\epsilon_{\lbrack\mu,\nu]}=\sum_{A,B}v_{[A,B]}e_{\mu}^{A}e_{\nu}^{B},
\label{2.40q}%
\end{equation}
and solving the Virasoro constraints
\begin{equation}
k^{\nu}\epsilon_{\lbrack\mu,\nu]}=2v_{[P,L]}e_{\mu}^{L}+2\sum_{i=1}%
^{24}v_{[P,T_{i}]}e_{\mu}^{T_{i}}=0, \label{2.41q}%
\end{equation}
we obtain the following decomposition for the spin-two state
\begin{align}
|B(\epsilon)\rangle &  \equiv\epsilon_{\lbrack\mu,\nu]}\alpha_{-1}^{\mu}%
\alpha_{-2}^{\nu}|0,k\rangle=|B(v)\rangle\nonumber\\
&  =\sum_{i}v_{[L,T_{i}]}(\alpha_{-1}^{L}\alpha_{-2}^{T_{i}}-\alpha
_{-1}^{T_{i}}\alpha_{-2}^{L})|0,k\rangle+\sum_{(i\neq j)}v_{[T_{i},T_{j}%
]}(\alpha_{-1}^{T_{i}}\alpha_{-2}^{T_{j}}-\alpha_{-1}^{T_{j}}\alpha
_{-2}^{T_{i}})|0,k\rangle. \label{2.42q}%
\end{align}
Finally, one can check that the $300$ independent degrees of freedom of the
spin-two particle decompose into $24+276$ in the above expression.

For the ZNS at $M^{2}=4$, we have the following decompositions

\begin{enumerate}
\item Spin-two tensor
\begin{align}
|C(\theta)\rangle &  \equiv(k_{\lambda}\theta_{\mu\nu}\alpha_{-1}^{\mu}%
\alpha_{-1}^{\nu}\alpha_{-1}^{\lambda}+2\ \theta_{\mu\nu}\alpha_{-1}^{\mu
}\alpha_{-2}^{\nu})|0,k\rangle\nonumber\\
&  =\sum_{i}2\ \theta_{T_{i}T_{i}}(\alpha_{-1}^{T_{i}}\alpha_{-1}^{T_{i}%
}\alpha_{-1}^{P}-\alpha_{-1}^{L}\alpha_{-1}^{L}\alpha_{-1}^{P}+\alpha
_{-1}^{T_{i}}\alpha_{-2}^{T_{i}}-\alpha_{-1}^{L}\alpha_{-2}^{L})|0,k\rangle
\nonumber\\
&  +\sum_{(i\neq j)}2\ \theta_{T_{i}T_{j}}(2\ \alpha_{-1}^{T_{i}}\alpha
_{-1}^{T_{j}}\alpha_{-1}^{P}+\alpha_{-1}^{T_{i}}\alpha_{-2}^{T_{j}}%
+\alpha_{-1}^{T_{j}}\alpha_{-2}^{T_{i}})|0,k\rangle\nonumber\\
&  +\sum_{i}2\ \theta_{LT_{i}}(2\ \alpha_{-1}^{L}\alpha_{-1}^{T_{i}}%
\alpha_{-1}^{P}+\alpha_{-1}^{L}\alpha_{-2}^{T_{i}}+\alpha_{-1}^{T_{i}}%
\alpha_{-2}^{L})|0,k\rangle, \label{2.43q}%
\end{align}
where we have solved the Virasoro constraints on the polarization tensor
$\theta_{\mu\nu}$
\begin{subequations}%
\begin{align}
\theta_{\mu\nu}  &  =\sum_{A,B}\theta_{AB}e_{\mu}^{A}e_{\nu}^{B}%
,\label{2.44q}\\
\eta^{\mu\nu}\theta_{\mu\nu}  &  =-\theta_{PP}+\theta_{LL}+\sum_{i}%
\theta_{T_{i}T_{i}}=0,\label{2.45q}\\
k^{\nu}\theta_{\mu\nu}  &  =-2\theta_{PP}e_{\mu}^{P}-2\theta_{PL}e_{\nu}%
^{L}-2\sum_{i}\theta_{PT_{i}}e_{\mu}^{T_{i}}=0. \label{2.46q}%
\end{align}%
\end{subequations}%
The $324$ degrees of freedom of on-shell $\theta_{\mu\nu}$decompose into
$24+276+24$ in Eq.(\ref{2.43q}).

\item Spin-one vector (with polarization vector $\theta\cdot k=0$,
$\theta_{\mu}=$ $\sum_{A}\theta_{A}e_{\mu}^{A},\ A$ $=L,T_{i}$)
\begin{align}
|D_{1}(\theta)\rangle &  \equiv\left[  \left(  \frac{5}{2}k_{\mu}k_{\nu}%
\theta_{\lambda}+\eta_{\mu\nu}\theta_{\lambda}\right)  \alpha_{-1}^{\mu}%
\alpha_{-1}^{\nu}\alpha_{-1}^{\lambda}+9k_{(\mu}\theta_{\nu)}\alpha_{-1}^{\mu
}\alpha_{-2}^{\nu}+6\theta_{\mu}\alpha_{-3}^{\mu}\right]  |0,k\rangle
\nonumber\\
&  =\sum_{A}\theta_{A}\left[  9\alpha_{-1}^{P}\alpha_{-1}^{P}\alpha_{-1}%
^{A}+\alpha_{-1}^{L}\alpha_{-1}^{L}\alpha_{-1}^{A}+\sum_{i}\alpha_{-1}^{T_{i}%
}\alpha_{-1}^{T_{i}}\alpha_{-1}^{A}\right. \nonumber\\
&  \left.  +9(\alpha_{-1}^{P}\alpha_{-2}^{A}+\alpha_{-1}^{A}\alpha_{-2}%
^{P})+6\alpha_{-3}^{A}\right]  |0,k\rangle. \label{2.48q}%
\end{align}

\item Spin-one vector (with polarization vector $\theta\cdot k=0$,
$\theta_{\mu}=$ $\sum_{A}\theta_{A}e_{\mu}^{A},\ A$ $=L,T_{i}$)%

\begin{align}
|D_{2}(\theta)\rangle &  \equiv\left[  \left(  \frac{1}{2}k_{\mu}k_{\nu}%
\theta_{\lambda}+2\eta_{\mu\nu}\theta_{\lambda}\right)  \alpha_{-1}^{\mu
}\alpha_{-1}^{\nu}\alpha_{-1}^{\lambda}-9k_{[\mu}\theta_{\nu]}\alpha_{-1}%
^{\mu}\alpha_{-2}^{\nu}-6\theta_{\mu}\alpha_{-3}^{\mu}\right]  |0,k\rangle
\label{2.49q}\\
&  =\sum_{A}\theta_{A}\left[  2\alpha_{-1}^{L}\alpha_{-1}^{L}\alpha_{-1}%
^{A}+2\sum_{j}\alpha_{-1}^{T_{j}}\alpha_{-1}^{T_{j}}\alpha_{-1}^{A}%
-9(\alpha_{-1}^{P}\alpha_{-2}^{A}-\alpha_{-1}^{A}\alpha_{-2}^{P})-6\alpha
_{-3}^{A}\right]  |0,k\rangle. \label{2.50q}%
\end{align}

\item spin-zero singlet
\begin{align}
|E\rangle &  \equiv\left[  \left(  \frac{17}{4}k_{\mu}k_{\nu}k_{\lambda}%
+\frac{9}{2}\eta_{\mu\nu}k_{\lambda}\right)  \alpha_{-1}^{\mu}\alpha_{-1}%
^{\nu}\alpha_{-1}^{\lambda}+(21k_{\mu}k_{\nu}+9\eta_{\mu\nu})\alpha_{-1}^{\mu
}\alpha_{-2}^{\nu}+25k_{\mu}\alpha_{-3}^{\mu}\right]  |0,k\rangle\nonumber\\
&  =\left[  25\left(  \alpha_{-1}^{P}\alpha_{-1}^{P}\alpha_{-1}^{P}%
+3\alpha_{-1}^{P}\alpha_{-2}^{P}+2\alpha_{-3}^{P}\right)  +9\alpha_{-1}%
^{L}\alpha_{-1}^{L}\alpha_{-1}^{P}+9\alpha_{-1}^{L}\alpha_{-2}^{L}+9\sum
_{i}\left(  \alpha_{-1}^{T_{i}}\alpha_{-1}^{T_{i}}\alpha_{-1}^{P}+\alpha
_{-1}^{T_{i}}\alpha_{-2}^{T_{i}}\right)  \right]  |0,k\rangle. \label{2.52q}%
\end{align}

\end{enumerate}

\paragraph{ \bigskip Spin-three state in the most general gauge choice}

In this section, we study the most general gauge choice associated with the
totally symmetric spin-three state
\begin{equation}
\lbrack\varepsilon_{\mu\nu\lambda}\alpha_{-1}^{\mu}\alpha_{-1}^{\nu}%
\alpha_{-1}^{\lambda}+\varepsilon_{(\mu\nu)}\alpha_{-1}^{\mu}\alpha_{-2}^{\nu
}+\varepsilon_{\mu}\alpha_{-3}^{\mu}]|0,k\rangle, \label{2.53q}%
\end{equation}
where Virasoro constraints imply%
\begin{subequations}
\begin{align}
\varepsilon_{(\mu\nu)}  &  =-\frac{3}{2}k^{\lambda}\varepsilon_{\mu\nu\lambda
},\label{2.54q}\\
\varepsilon_{\mu}  &  =\frac{1}{2}k^{\nu}k^{\lambda}\varepsilon_{\mu\nu
\lambda},\label{2.55q}\\
2\eta^{\mu\nu}\varepsilon_{\mu\nu\lambda}  &  =k^{\mu}k^{\nu}\varepsilon
_{\mu\nu\lambda}. \label{2.56q}%
\end{align}%
\end{subequations}%
Eq.(\ref{2.54q}) and Eq.(\ref{2.55q}) imply that both $\varepsilon_{(\mu\nu)}$
and $\varepsilon_{\mu}$ are not independent variables, and Eq.(\ref{2.56q})
stands for the constraint on the polarization $\varepsilon_{\mu\nu\lambda}$.
In the helicity basis, we define
\begin{equation}
\varepsilon_{\mu\nu\lambda}=\sum_{A,B,C}u_{ABC}\ e_{\mu}^{A}e_{\nu}%
^{B}e_{\lambda}^{C},\hspace{1cm}A,B,C=P,L,T_{i}. \label{2.57q}%
\end{equation}
Eq.(\ref{2.56q}) then gives
\begin{equation}
\sum_{A,B}\eta^{AB}u_{ABC}=2u_{PPC},\hspace{1cm}A,B,C=P,L,T_{i}, \label{2.58q}%
\end{equation}
which implies%
\begin{equation}
3u_{PPC}-u_{LLC}-\sum_{j}u_{T_{j}T_{j}C}=0,\hspace{1cm}C=P,L,T_{i}.
\label{2.59q}%
\end{equation}
Eliminating $u_{LLP},u_{LLL}$ and $u_{LLT_{i}}$ from above equations, we have
the solution for $\varepsilon_{\mu\nu\lambda}$, $\varepsilon_{(\mu\nu)}$ and
$\varepsilon_{\mu}$
\begin{align}
\varepsilon_{\mu\nu\lambda}  &  =u_{PPP}\ [e_{\mu}^{P}e_{\nu}^{P}e_{\lambda
}^{P}+3(e_{\mu}^{L}e_{\nu}^{L}e_{\lambda}^{P}+\text{per.})]+u_{PPL}\ [(e_{\mu
}^{P}e_{\nu}^{P}e_{\lambda}^{L}+\text{per.})+3e_{\mu}^{L}e_{\nu}^{L}%
e_{\lambda}^{L}]\nonumber\\
&  +\sum_{i}u_{PPT_{i}}\ [(e_{\mu}^{P}e_{\nu}^{P}e_{\lambda}^{T_{i}%
}+\text{per.})+3(e_{\mu}^{L}e_{\nu}^{L}e_{\lambda}^{T_{i}}+\text{per.}%
)]+\sum_{i}u_{PT_{i}T_{i}}\ [(e_{\mu}^{P}e_{\nu}^{T_{i}}e_{\lambda}^{T_{i}%
}+\text{per.})-(e_{\mu}^{L}e_{\nu}^{L}e_{\lambda}^{P}+\text{per.})]\nonumber\\
&  +\sum_{(i\neq j)}u_{PT_{i}T_{j}}\ [e_{\mu}^{P}e_{\nu}^{T_{i}}e_{\lambda
}^{T_{j}}+\text{per.}]+\sum_{i}u_{PLT_{i}}\ [e_{\mu}^{P}e_{\nu}^{L}e_{\lambda
}^{T_{i}}+\text{per.}]+\sum_{i}u_{LT_{i}T_{i}}\ [(e_{\mu}^{L}e_{\nu}^{T_{i}%
}e_{\lambda}^{T_{i}}+\text{per.})-e_{\mu}^{L}e_{\nu}^{L}e_{\lambda}%
^{L}]\nonumber\\
&  +\sum_{(i\neq j)}u_{LT_{i}T_{j}}\ [e_{\mu}^{L}e_{\nu}^{T_{i}}e_{\lambda
}^{T_{j}}+\text{per.}]+\sum_{i}u_{T_{i}T_{i}T_{i}}\ [e_{\mu}^{T_{i}}e_{\nu
}^{T_{i}}e_{\lambda}^{T_{i}}-(e_{\mu}^{L}e_{\nu}^{L}e_{\lambda}^{T_{i}%
}+\text{per.})]\nonumber\\
&  +\sum_{i\neq j}u_{T_{j}T_{j}T_{i}}\ [(e_{\mu}^{T_{j}}e_{\nu}^{T_{j}%
}e_{\lambda}^{T_{i}}+\text{per.})-(e_{\mu}^{L}e_{\nu}^{L}e_{\lambda}^{T_{i}%
}+\text{per.})]\nonumber\\
&  +\sum_{(i\neq j\neq k)}u_{T_{i}T_{j}T_{k}}\ [e_{\mu}^{T_{i}}e_{\nu}^{T_{j}%
}e_{\lambda}^{T_{k}}+\text{per.}], \label{2.60q}%
\end{align}%
\begin{align}
\frac{1}{3}\varepsilon_{(\mu\nu)}  &  =u_{PPP}(e_{\mu}^{P}e_{\nu}^{P}+3e_{\mu
}^{L}e_{\nu}^{L})+u_{PPL}(e_{\mu}^{P}e_{\nu}^{L}+e_{\mu}^{L}e_{\nu}^{P}%
)+\sum_{i}u_{PPT_{i}}(e_{\mu}^{P}e_{\nu}^{T_{i}}+e_{\mu}^{T_{i}}e_{\nu}%
^{P})\nonumber\\
&  +\sum_{i}u_{PLT_{i}}(e_{\mu}^{L}e_{\nu}^{T_{i}}+e_{\mu}^{T_{i}}e_{\nu}%
^{L})+\sum_{i}u_{PT_{i}T_{i}}(e_{\mu}^{T_{i}}e_{\nu}^{T_{i}}-e_{\mu}^{L}%
e_{\nu}^{L})+\sum_{(i\neq j)}u_{PT_{i}T_{j}}(e_{\mu}^{T_{i}}e_{\nu}^{T_{j}%
}+e_{\nu}^{T_{j}}e_{\mu}^{T_{j}}), \label{2.61q}%
\end{align}

\begin{equation}
\frac{1}{2}\varepsilon_{\mu}=[u_{PPP}\ e_{\mu}^{P}+u_{PPL}\ e_{\mu}^{L}%
+\sum_{i}u_{PPT_{i}}\ e_{\mu}^{T_{i}}]. \label{2.62q}%
\end{equation}
Putting all these polarizations back to the general form of physical states
Eq.(\ref{2.53q}), we get
\begin{align}
&  [\varepsilon_{\mu\nu\lambda}\alpha_{-1}^{\mu}\alpha_{-1}^{\nu}\alpha
_{-1}^{\lambda}+\varepsilon_{(\mu\nu)}\alpha_{-1}^{\mu}\alpha_{-2}^{\nu
}+\varepsilon_{\mu}\alpha_{-3}^{\mu}]|0,k\rangle=|A(\tilde{u})\rangle
+|C(\theta)\rangle\nonumber\\
&  +\left[  \frac{1}{9}(u_{LLL}+\sum_{i}u_{T_{i}T_{i}L})\right]  |D_{1}%
(e^{L})\rangle\nonumber\\
&  +\sum_{i}\left[  \frac{1}{9}(u_{LLT_{i}}+\sum_{j}u_{T_{j}T_{j}T_{i}%
})\right]  |D_{1}(e^{T_{i}})\rangle\nonumber\\
&  +\frac{1}{75}\left[  u_{LLP}+\sum_{i}u_{PT_{i}T_{i}}\right]  |E\rangle.
\label{2.63q}%
\end{align}
For the first two terms on the right hand side of Eq.(\ref{2.63q}), we need to
make the following replacements. For the positive-norm state $|A(\tilde
{u})\rangle$ in Eq.(\ref{2.38q})
\begin{align}
&  \tilde{u}_{T_{i}T_{i}T_{i}}\rightarrow u_{T_{i}T_{i}T_{i}}-\frac{1}%
{3}u_{PPT_{i}},\hspace{1cm}\tilde{u}_{T_{j}T_{j}T_{i}}\rightarrow
u_{T_{j}T_{j}T_{i}}-\frac{1}{9}u_{PPT_{i}},\nonumber\\
&  \tilde{u}_{T_{i}T_{j}T_{k}}\rightarrow u_{T_{i}T_{j}T_{k}}\hspace
{0.7cm}\tilde{u}_{LT_{i}T_{i}}\rightarrow u_{LT_{i}T_{i}}-\frac{1}{9}%
u_{PPL},\hspace{0.7cm}\tilde{u}_{LT_{i}T_{j}}\rightarrow u_{LT_{i}T_{j}}.
\label{2.64q}%
\end{align}
For the spin-two ZNS $|C(\theta)\rangle$ in Eq.(\ref{2.43q}), the replacement
is given by
\begin{equation}
2\theta_{LT_{i}}\rightarrow3u_{PLT_{i}},\ \ \ 2\theta_{T_{i}T_{j}}%
\rightarrow3u_{PT_{i}T_{j}},\ \text{for }i\neq j,\ \ \ 2\theta_{T_{i}T_{i}%
}\rightarrow3(u_{PT_{i}T_{i}}-\frac{3}{25}u_{PPP}). \label{2.65q}%
\end{equation}

It is important to note that for the spin-three gauge multiplet, only
spin-two, singlet and $D_{1}$ vector ZNS appear in the decomposition
Eq.(\ref{2.63q}). In the next section, we will see how one can include the
missing $D_{2}$ ZNS in the analysis.

\paragraph{Complete spectrum analysis and the $D_{2}$ ZNS}

After all these preparations, we are ready for a complete analysis of the most
general decomposition of physical states at $M^{2}=4$. The most general form
of physical states at this mass level are given by
\begin{equation}
\lbrack\epsilon_{\mu\nu\lambda}\alpha_{-1}^{\mu}\alpha_{-1}^{\nu}\alpha
_{-1}^{\lambda}+\epsilon_{(\mu\nu)}\alpha_{-1}^{\mu}\alpha_{-2}^{\nu}%
+\epsilon_{\lbrack\mu\nu]}\alpha_{-1}^{\mu}\alpha_{-2}^{\nu}+\epsilon_{\mu
}\alpha_{-3}^{\mu}]|0,k\rangle. \label{2.66q}%
\end{equation}
The Virasoro constraints are%
\begin{subequations}
\begin{align}
\epsilon_{(\mu\nu)}  &  =-\frac{3}{2}k^{\lambda}\epsilon_{\mu\nu\lambda
},\label{2.67q}\\
-k^{\nu}\epsilon_{\lbrack\mu\nu]}+3\epsilon_{\mu}  &  =\frac{3}{2}k^{\nu
}k^{\lambda}\epsilon_{\mu\nu\lambda},\label{2.68q}\\
2k^{\nu}\epsilon_{\lbrack\mu\nu]}+3\epsilon_{\mu}  &  =3(k^{\nu}k^{\lambda
}-\eta^{\nu\lambda})\epsilon_{\mu\nu\lambda}. \label{2.69q}%
\end{align}%
\end{subequations}%
The solutions to Eq.(\ref{2.68q}) and Eq.(\ref{2.69q}) are given by
\begin{align}
k^{\nu}\epsilon_{\lbrack\mu\nu]}  &  =\left(  \frac{1}{2}k^{\nu}k^{\lambda
}-\eta^{\nu\lambda}\right)  \epsilon_{\mu\nu\lambda},\label{2.70q}\\
3\epsilon_{\mu}  &  =(2k^{\nu}k^{\lambda}-\eta^{\nu\lambda})\epsilon_{\mu
\nu\lambda}. \label{2.71q}%
\end{align}
In contrast to the previous discussion Eq.(\ref{2.54q}) and Eq.(\ref{2.55q})
where both $\epsilon_{(\mu\nu)}$ and $\epsilon_{\mu}$ are completely fixed by
the leading spin-three polarization tensor $\epsilon_{\mu\nu\lambda}$, we now
have a new contribution from $k^{\nu}\epsilon_{\lbrack\mu\nu]}$. It will
become clear that this extra term includes the inter-particle ZNS $D_{2}$,
Eq.(\ref{2.49q}) or Eq.(\ref{2.50q}). Furthermore, it should be clear that the
antisymmetric spin-two positive-norm physical states are defined by requiring
$\epsilon_{\mu\nu\lambda}=\epsilon_{(\mu\nu)}=0$ and $\epsilon_{\mu}=k^{\nu
}\epsilon_{\lbrack\mu\nu]}=0$. In the following, for the sake of clarity, we
shall focus on the effects of the new contribution induced by the
$\epsilon_{\lbrack\mu\nu]}$ only.

The two independent polarization tensors of the most general representation
for physical states Eq.(\ref{2.66q}) are given in the helicity basis by
\begin{align}
\epsilon_{\mu\nu\lambda}  &  =\sum_{ABC}U_{ABC}\ e_{\mu}^{A}e_{\nu}%
^{B}e_{\lambda}^{C},\hspace{1cm}A,B,C=P,L,T_{i};\label{2.72q}\\
\epsilon_{\lbrack\mu\nu]}  &  =\sum_{A,B}V_{[AB]}\ e_{\mu}^{A}e_{\nu}^{B}.
\label{2.73q}%
\end{align}
The Virasoro constraint Eq.(\ref{2.70q}) demands that%
\begin{subequations}
\begin{align}
3U_{PPP}-U_{LLP}-\sum_{i}U_{PT_{i}T_{i}}  &  =0,\label{2.74q}\\
3U_{PPL}-U_{LLL}-\sum_{i}U_{LT_{i}T_{i}}  &  =2V_{[PL]},\label{2.75q}\\
3U_{PPT_{i}}-U_{LLT_{i}}-\sum_{j}U_{T_{j}T_{j}T_{i}}  &  =2V_{[PT_{i}]}.
\label{2.76q}%
\end{align}%
\end{subequations}%
In contrast to Eq.(\ref{2.59q}), the solution to the above equations become
\begin{align}
U_{PPL}=U_{PPL}^{(1)}+U_{PPL}^{(2)},  &  \text{ \ where \ }U_{PPL}^{(1)}%
=\frac{1}{3}(U_{LLL}+\sum_{i}U_{T_{i}T_{i}L}),\text{ }U_{PPL}^{(2)}=\frac
{2}{3}V_{[PL]}\text{ ;}\label{2.77q}\\
U_{PPT_{i}}=U_{PPT_{i}}^{(1)}+U_{PPT_{i}}^{(2)},  &  \text{ \ where
\ \ }U_{PPT_{i}}^{(1)}=\frac{1}{3}(U_{LLT_{i}}+\sum_{j}U_{T_{j}T_{j}T_{i}%
}),\text{ }U_{PPT_{i}}^{(2)}=\frac{2}{3}V_{[PT_{i}]}. \label{2.78q}%
\end{align}
It is clear from the expressions above that only $U_{PPL}^{(2)}$ and
$U_{PPT_{i}}^{(2)}$ give new contributions to our previous analysis in the
last section, so we can simply write down all these new terms%
\begin{subequations}
\begin{align}
\delta\epsilon_{\mu\nu\lambda}  &  =\frac{2}{3}[V_{[PL]}(e_{\mu}^{P}e_{\nu
}^{P}e_{\lambda}^{L}+\text{per.})+\sum_{i}V_{[PT_{i}]}(e_{\mu}^{P}e_{\nu}%
^{P}e_{\lambda}^{T_{i}}+\text{per.})],\label{2.79q}\\
\delta\epsilon_{\lbrack\mu\nu]}  &  =V_{[PL]}(e_{\mu}^{P}e_{\nu}%
^{L}-\text{per.})+\sum_{i}V_{[PT_{i}]}(e_{\mu}^{P}e_{\nu}^{T_{i}}%
-\text{per.})\nonumber\\
&  +\sum_{i}V_{[T_{i}L]}(e_{\mu}^{T_{i}}e_{\nu}^{L}-\text{per.})+\sum_{i\neq
j}V_{[T_{j}T_{i}]}(e_{\mu}^{T_{j}}e_{\nu}^{T_{i}}-\text{per.}),\label{2.80q}\\
\delta\epsilon_{(\mu\nu)}  &  =2[V_{[PL]}(e_{\mu}^{P}e_{\nu}^{L}%
+\text{per.})+\sum_{i}V_{[PT_{i}]}(e_{\mu}^{P}e_{\nu}^{T_{i}}+\text{per.}%
)],\label{2.81q}\\
\delta\epsilon_{\mu}  &  =2[V_{[PL]}e_{\mu}^{L}+\sum_{i}V_{[PT_{i}]}e_{\mu
}^{T_{i}}]. \label{2.82q}%
\end{align}%
\end{subequations}%
Finally, the complete decomposition of physical states Eq.(\ref{2.66q}) in the
helicity basis becomes%
\begin{subequations}
\begin{align}
&  [\epsilon_{\mu\nu\lambda}\alpha_{-1}^{\mu}\alpha_{-1}^{\nu}\alpha
_{-1}^{\lambda}+\epsilon_{(\mu\nu)}\alpha_{-1}^{\mu}\alpha_{-2}^{\nu}%
+\epsilon_{\lbrack\mu\nu]}\alpha_{-1}^{\mu}\alpha_{-2}^{\nu}+\epsilon_{\mu
}\alpha_{-3}^{\mu}]|0,k\rangle\nonumber\\
&  =|A(U_{CBA})\rangle+|B(V_{[T_{i}A]})\rangle+|C(U_{PBA})\rangle
\label{2.83q}\\
&  +\sum_{A=L,T_{i}}[\frac{1}{9}(U_{LLA}+\sum_{i}U_{T_{i}T_{i}A})]|D_{1}%
(e^{A})\rangle\label{2.84q}\\
&  -\frac{1}{9}\sum_{A=L,T_{i}}V_{[PA]}|D_{2}^{\prime}(e^{A})\rangle
\label{2.85q}\\
&  +\frac{1}{75}[U_{LLP}+\sum_{i}U_{PT_{i}T_{i}}]|E\rangle. \label{2.86q}%
\end{align}%
\end{subequations}%
In Eq.(\ref{2.83q}), $|A(U_{CBA})\rangle$ is given by Eq.(\ref{2.38q}) with
$\tilde{u}_{CBA}$ given by Eq.(\ref{2.64q}) and we have replaced $u$ by $U$ on
the r.h.s. of Eq.(\ref{2.64q}). The antisymmetric spin-two positive-norm state
$|B(V_{[T_{i}A]})\rangle$ is given by Eq.(\ref{2.42q}) and we have replaced
$v$ by $V$ \ in Eq.(\ref{2.42q}). Finally, $|C(U_{PBA})\rangle$ is given by
Eq.(\ref{2.43q}) with $\theta$ given by Eq.(\ref{2.65q}) and we have replaced
$u$ by $U$ on the r.h.s. of Eq.(\ref{2.65q}). In Eq.(\ref{2.85q}),
$|D_{2}^{\prime}(e^{A})\rangle\equiv|D_{2}(e^{A})\rangle-2|D_{1}(e^{A}%
)\rangle$ is the inter-particle ZNS introduced in the end of section A with
$\alpha=-2$. Note that the value of $\alpha$ is a choice of convention fixed
by the parametrization of the polarizations. It can always be adjusted to be
zero. In view of Eq.(\ref{2.77q}) and Eq.(\ref{2.78q}), we see that one can
use either $V_{[PA]}$ or $U_{PPA}^{(2)}$ ( $A=L,T_{i})$ to represent the
polarization of the $|D_{2}^{\prime}(e^{A})\rangle$ inter-particle ZNS.

We conclude that once we turn on the antisymmetric spin-two positive-norm
state in the general representation of physical states Eq.(\ref{2.66q}), it is
naturally accompanied by the $D_{2}^{\prime}$ inter-particle ZNS. The
polarization of the $D_{2}^{\prime}$ inter-particle ZNS can be represented by
either $V_{[PA]}$ or $U_{PPA}^{(2)}$ ( $A=L,T_{i})$ in Eq.(\ref{2.72q}) and
Eq.(\ref{2.73q}). Thus this inter-particle ZNS will generate an inter-particle
symmetry transformation in the $\sigma$-model calculation considered in
chapter I. Note that, in contrast to the high-energy symmetry of Gross, this
symmetry is valid to all orders in $\alpha^{\prime}$.

\subsection{Light-cone ZNS in DDF construction}

In the usual light-cone quantization of bosonic string theory, one solves the
Virasoro constraints to get rid of two string coordinates $X^{\pm}$. Only $24$
string coordinates $\alpha_{n}^{i},$ $i=1,...,,24,$ remain, and there are no
ZNS in the spectrum. However, there existed another similar quantization
scheme, the DDF quantization, which did \textit{include the ZNS} in the
spectrum. In the light-cone DDF quantization of open bosonic string
\cite{Giudice,DDF1,DDF2}, one constructs transverse physical states with
discrete momenta%

\begin{equation}
p^{\mu}=p_{0}^{\mu}-Nk_{0}^{\mu}=(1,0.....,-1+N), \label{3.1f}%
\end{equation}
where $X^{\pm}$ $\equiv\frac{1}{\sqrt{2}}(X^{0}$ $\pm X^{25})$ and
$\ p^{+}=1,$ $p^{-}=-1+N.$ In Eq.(\ref{3.1f}), $M^{2}=-p^{2}=2(N-1)$ and
$p_{0}^{\mu}\equiv(1,0...,-1)$, $k_{0}^{\mu}\equiv(0,0...,-1)$, respectively.
All other states can be reached by Lorentz transformations. The DDF operators
are given by \cite{Giudice,DDF1,DDF2}%
\begin{equation}
A_{n}^{i}=\frac{1}{2\pi}\int_{0}^{2\pi}\dot{X}^{i}(\tau)e^{inX^{+}(\tau)}%
d\tau,\text{ }i=1,...,,24, \label{3.2f}%
\end{equation}
where the massless vertex operator $V^{i}(nk_{0},\tau)=\dot{X}^{i}%
(\tau)e^{inX^{+}(\tau)}$ is a primary field with conformal dimension one, and
is periodic in the worldsheet time $\tau$ if one chooses $k^{\mu}=$
$nk_{0}^{\mu}$ with $n\in Z.$ It is then easy to show that%
\begin{align}
\lbrack L_{m},A_{n}^{i}]  &  =0,\label{3.3f}\\
\lbrack A_{m}^{i},A_{n}^{j}]  &  =m\delta_{ij}\delta_{m+n}. \label{3.4f}%
\end{align}

In addition to sharing the same algebra, Eq.(\ref{3.4f}), with string
coordinates $\alpha_{n}^{i},$ the DDF operators $A_{n}^{i}$ possess a nicer
property Eq.(\ref{3.3f}), which enables us to easily write down a general
formula for the positive-norm physical states as following%
\begin{equation}
(A_{-1}^{j})^{i_{1}}(A_{-2}^{k})^{i_{2}}....(A_{-m}^{l})^{i_{m}}\mid
0,p_{0}>,\text{ }i_{r}\in Z, \label{3.5f}%
\end{equation}
where $\mid0,p_{0}>$ is the tachyon ground state and $N=%
{\textstyle\sum_{r=1}^{m}}
ri_{r}$ is the level of the state. Historically, DDF operators were used to
prove no-ghost (negative-norm states) theorem for $D=26$ string theory. Here
we are going to use them to analyses ZNS. It turns out that ZNS can be
generated by%
\begin{equation}
\tilde{A}_{n}^{-}=A_{n}^{-}-%
{\textstyle\sum_{m=1}^{\infty}}
{\textstyle\sum_{i=1}^{D-2}}
:A_{m}^{i}A_{n-m}^{i}:, \label{3.6f}%
\end{equation}
where $A_{n}^{-}$ is given by%
\begin{equation}
A_{n}^{-}=\frac{1}{2\pi}\int_{0}^{2\pi}\left[  :\dot{X}^{-}e^{inX^{+}}%
:-\frac{1}{2}in\frac{d}{d\tau}(\log\dot{X}^{+})e^{inX^{+}}\right]  d\tau.
\label{3.7f}%
\end{equation}

It can be shown that $\tilde{A}_{n}^{-}$ commute with $L_{m}$ and satisfy the
following algebra%
\begin{align}
\lbrack\tilde{A}_{m}^{-},A_{n}^{i}]  &  =0,\label{3.8f}\\
\lbrack\tilde{A}_{m}^{-},\tilde{A}_{n}^{-}]  &  =(m-n)\tilde{A}_{m+n}%
^{-}+\frac{26-D}{12}m^{3}\delta_{m+n}. \label{3.9f}%
\end{align}
Eq.(\ref{3.4f}), Eq.(\ref{3.8f}) and Eq.(\ref{3.9f}) constitute the spectrum
generating algebra for the open bosonic string including ZNS. The ground state
$\left\vert 0,p_{0}\right\rangle \equiv\left\vert 0\right\rangle $ satisfies
the following conditions%
\begin{align}
A_{n}^{i}\left\vert 0\right\rangle  &  =\ \tilde{A}_{n}^{-}\left\vert
0\right\rangle =0,\text{ \ }n>0,\label{3.10f}\\
\tilde{A}_{0}^{-}\left\vert 0\right\rangle  &  =-\frac{26-D}{24},\text{
\ }A_{0}^{i}\left\vert 0\right\rangle =0. \label{3.11f}%
\end{align}

We are now ready to construct ZNS in the DDF formalism.

1. $M^{2}=0:$ One has only one scalar $\tilde{A}_{-1}^{-}\left\vert
0\right\rangle $, which has zero-norm for any $D$.

2. $M^{2}=2:$ One has a light-cone vector $A_{-1}^{i}\tilde{A}_{-1}%
^{-}\left\vert 0\right\rangle $, which has zero-norm for any $D$, and two
scalars, whose norms are calculated to be%
\begin{equation}
\parallel(a\tilde{A}_{-1}^{-}\tilde{A}_{-1}^{-}+b\tilde{A}_{-2}^{-})\left\vert
0\right\rangle \parallel=\frac{26-D}{2}b^{2}. \label{3.12f}%
\end{equation}
For $b=0,$ one has a "pure type I" ZNS, $\tilde{A}_{-1}^{-}\tilde{A}_{-1}%
^{-}\left\vert 0\right\rangle $, which has zero-norm for any $D$. By combining
with the light-cone vector $A_{-1}^{i}\tilde{A}_{-1}^{-}\left\vert
0\right\rangle ,$ one obtains a vector ZNS with $25$ degrees of freedom, which
corresponds to Eq.(\ref{2.5q}) in the OCFQ approach. For $b\neq0$, one obtains
a type II scalar ZNS for $D=26,$ which corresponds to Eq.(\ref{2.4q}) in the
OCFQ approach.

3. $M^{2}=4:$

I. A spin-two tensor $A_{-1}^{i}A_{-1}^{j}\tilde{A}_{-1}^{-}\left\vert
0\right\rangle $, which has zero-norm for any $D$.

II. Three light-cone vectors, whose norms are calculated to be%
\begin{equation}
\parallel(aA_{-1}^{i}\tilde{A}_{-1}^{-}\tilde{A}_{-1}^{-}+bA_{-2}^{i}\tilde
{A}_{-1}^{-}+cA_{-1}^{i}\tilde{A}_{-2}^{-})\left\vert 0\right\rangle
\parallel=\frac{26-D}{2}c^{2}. \label{3.13f}%
\end{equation}
\qquad\qquad

III. Three scalars, whose norms are calculated to be%
\begin{equation}
\parallel(d\tilde{A}_{-1}^{-}\tilde{A}_{-1}^{-}\tilde{A}_{-1}^{-}+e\tilde
{A}_{-1}^{-}\tilde{A}_{-2}^{-}+f\tilde{A}_{-3}^{-})\left\vert 0\right\rangle
\parallel=2(26-D)(e+f)^{2}. \label{3.14f}%
\end{equation}
For $c=0$ in Eq.(\ref{3.13f}), one has two "pure type I" light-cone vector
ZNS. For $e+f=0$ in Eq.\ref{3.14f}), one has two "pure type I" scalar ZNS. One
of the two type I light-cone vectors, when combining with the spin-two state
in I, gives the type I spin-two tensor which corresponds to Eq.(\ref{2.7q}) in
the OCFQ approach. The other type I light-cone vector, when combining with one
of the two type I scalar, gives the type I vector ZNS which corresponds to
Eq.(\ref{2.8q}) in the OCFQ approach. The other type I scalar corresponds to
Eq.(\ref{2.9q}). Finally, for $c\neq0$ and $e+f\neq0$, one obtains the type II
vector ZNS for $D=26$, which corresponds to Eq.(\ref{2.6q}) in the OCFQ
approach. It is easy to see that a special linear combination of $b$ and $e$
will give the inter-particle vector ZNS which corresponds to the
inter-particle $D_{2}$ ZNS in Eq.(\ref{2.10q}). This completes the analysis of
ZNS for $M^{2}=4.$

Note that the exact mapping of ZNS in the light-cone DDF formalism and the
OCFQ approach depends on the exact relation between operators $(\tilde{A}%
_{n}^{-},A_{n}^{i},L_{n})$ and $\alpha_{n}^{\mu}$, which has not been worked
out in the literature.

\subsection{BRST ZNS in WSFT}

In this section, we calculate BRST ZNS in the formulation of WSFT. In
addition, we apply the results to demonstrate that off-shell gauge
transformations of WSFT are indeed identical to the on-shell stringy gauge
symmetries generated by two types of ZNS in the generalized massive $\sigma
$-model approach \cite{LeePRL,Lee} of string theory. In section I.D
\cite{KaoLee}, the background ghost transformations in the gauge
transformations of WSFT \cite{Witten} were shown to correspond, in a
one-to-one manner, to the lifting of on-shell conditions of ZNS in the OCFQ
approach. Here we go one step further to demonstrate the correspondence of
stringy symmetries induced by ZNS in OCFQ and BRST approaches.

Cubic string field theory is defined on a disk with the action
\begin{equation}
S=-\frac{1}{g_{0}}\left(  \frac{1}{2}\int\Phi\ast Q_{\text{B}}\Phi+\frac{1}%
{3}\int\Phi\ast\Phi\ast\Phi\right)  , \label{4.1f}%
\end{equation}
where $Q_{\text{B}}$ is the BRST charge%
\begin{equation}
Q_{\text{B}}=\sum_{n=-\infty}^{\infty}L_{-n}^{\text{m}}c_{n}+\sum
_{m,n=-\infty}^{\infty}\dfrac{m-n}{2}:c_{m}c_{n}b_{-m-n}:-c, \label{4.2f}%
\end{equation}
and $\Phi$ is the string field with ghost number $1$ and $b,c$ are conformal
ghosts. Since the ghost number of vacuum on a disk is $-3$, the total ghost
number of this action is 0 as expected. The string field can be expanded as
\begin{equation}
\Phi=\sum_{k,m,n}A_{\mu\cdots,k\cdots m\cdots n\cdots}\left(  x\right)
\alpha_{k}^{\mu}\cdots b_{m}\cdots c_{n}\cdots\left\vert \Omega\right\rangle ,
\end{equation}
where the string ground states $\left\vert \Omega\right\rangle $\ are
\begin{equation}
\left\vert \Omega\right\rangle =c_{1}\left\vert 0\right\rangle . \label{4.3f}%
\end{equation}
The gauge transformation for string field can be written as
\begin{equation}
\delta\Phi=Q_{\text{B}}\Lambda+g\left(  \Phi\ast\Lambda-\Lambda\ast
\Phi\right)  . \label{4.4f}%
\end{equation}
where $\Lambda$ is the a string field with ghost number $0$.

For the purpose of discussion in this chapter, we are going to consider the
linearized gauge transformation
\begin{equation}
\delta\Phi=Q_{\text{B}}\Lambda, \label{4.5f}%
\end{equation}
where $Q_{\text{B}}\Lambda$ is just the off-shell ZNS. In the following, we
will explicitly show that the solutions of Eq.(\ref{4.5f}) are in one-to-one
correspondence to the ZNS obtained in OCFQ approach in section VI.A level by
level for the first several mass levels.

There is no ZNS in the lowest string mass level with $M^{2}=-2$, so our
analysis will start with the mass level of $M^{2}=0$.

\bigskip

\noindent\underline{$M^{2}=0$:}

The string field can be expanded as%
\begin{align}
\Phi &  =\left\{  iA_{\mu}\left(  x\right)  \alpha_{-1}^{\mu}+\alpha\left(
x\right)  b_{-1}c_{0}\right\}  \left\vert \Omega\right\rangle ,\label{4.6f}\\
\Lambda &  =\left\{  \epsilon^{0}\left(  x\right)  b_{-1}\right\}  \left\vert
\Omega\right\rangle . \label{4.7f}%
\end{align}
The gauge transformation is then%
\begin{equation}
Q_{\text{B}}\Lambda=\left\{  -\dfrac{1}{2}\alpha_{0}^{2}\epsilon^{0}%
b_{-1}c_{0}+\epsilon^{0}\alpha_{0}\cdot\alpha_{-1}\right\}  \left\vert
\Omega\right\rangle . \label{4.8f}%
\end{equation}
The nilpotency of BRST charge $Q_{\text{B}}$ gives%
\begin{equation}
Q_{\text{B}}^{2}\Lambda=0, \label{4.9f}%
\end{equation}
which can be easily checked to be valid for any $D$. Thus Eq.(\ref{4.8f}) can
be interpreted as a type I ZNS. To compare it with the ZNS obtained in OCFQ
approach in section II, we need to reduce our Hilbert space by removing the
ghosts states. In particular, the coefficients of terms with ghost operators
must vanish. For the state in Eq.(\ref{4.8f}), it is%
\begin{equation}
\alpha_{0}^{2}\epsilon^{0}=0, \label{4.10f}%
\end{equation}
which give the on-shell condition $k^{2}=0$ and the following ZNS%
\begin{equation}
Q_{\text{B}}\Lambda=\epsilon^{0}\alpha_{0}\cdot\alpha_{-1}\left\vert
\Omega\right\rangle . \label{4.11f}%
\end{equation}
This is the same as the scalar ZNS obtained in OCFQ approach.

\bigskip

\noindent\underline{$M^{2}=2$:}

The string fields expansion are%
\begin{align}
\Phi &  =\left\{  -B_{\mu\nu}\left(  x\right)  \alpha_{-1}^{\mu}\alpha
_{-1}^{\nu}+iB_{\mu}\left(  x\right)  \alpha_{-2}^{\mu}\right. \nonumber\\
&  \text{ \ \ \ \ \ }\left.  +i\beta_{\mu}\left(  x\right)  \alpha_{-1}^{\mu
}b_{-1}c_{0}+\beta^{0}\left(  x\right)  b_{-2}c_{0}+\beta^{1}\left(  x\right)
b_{-1}c_{-1}\right\}  \left\vert \Omega\right\rangle ,\label{4.12f}\\
\Lambda &  =\left\{  i\epsilon_{\mu}^{0}\left(  x\right)  \alpha_{-1}^{\mu
}b_{-1}+\epsilon^{1}\left(  x\right)  b_{-2}\right\}  \left\vert
\Omega\right\rangle . \label{4.13f}%
\end{align}
The off-shell ZNS are calculated to be%
\begin{align}
Q_{\text{B}}\Lambda &  =\left\{  \left(  i\alpha_{0\mu}\epsilon_{\nu}%
^{0}+\frac{1}{2}\epsilon^{1}\eta_{\mu\nu}\right)  \alpha_{-1}^{\mu}\alpha
_{-1}^{\nu}+\left(  i\epsilon^{0}+\epsilon^{1}\alpha_{0}\right)  \cdot
\alpha_{-2}\right. \nonumber\\
&  \text{ \ \ \ \ \ }-i\dfrac{1}{2}\left(  \alpha_{0}^{2}+2\right)  \left(
\epsilon^{0}\cdot\alpha_{-1}\right)  b_{-1}c_{0}-\dfrac{1}{2}\left(
\alpha_{0}^{2}+2\right)  \epsilon^{1}b_{-2}c_{0}\nonumber\\
&  \text{ \ \ \ \ \ }\left.  -\left(  i\epsilon^{0}\cdot\alpha_{0}%
+3\epsilon^{1}\right)  b_{-1}c_{-1}\right\}  \left\vert \Omega\right\rangle .
\label{4.14f}%
\end{align}
Nilpotency condition requires%
\begin{equation}
Q_{\text{B}}^{2}\Lambda=\dfrac{D-26}{2}\epsilon^{1}c_{-2}\left\vert
\Omega\right\rangle =0. \label{4.15f}%
\end{equation}
There are two solutions of Eq.(\ref{4.15f}), which correspond to the type I
and type II ZNS, respectively.

\begin{enumerate}
\item Type I: in this case $D$ is not restricted to the critical string
dimension in Eq.(\ref{4.15f}), i.e. $D\neq26$. Thus
\begin{equation}
\epsilon^{1}=0. \label{4.16f}%
\end{equation}
The no-ghost conditions of Eq.(\ref{4.14f}) lead to the on-shell constraints%
\begin{align}
\alpha_{0}^{2}+2  &  =0,\label{4.17f}\\
\epsilon^{0}\cdot\alpha_{0}  &  =0. \label{4.18f}%
\end{align}
The off-shell ZNS in Eq.(\ref{4.14f}) then reduces to an on-shell vector ZNS%
\begin{equation}
Q_{\text{B}}\Lambda=i\left\{  \left(  \epsilon^{0}\cdot\alpha_{-1}\right)
\left(  \alpha_{0}\cdot\alpha_{-1}\right)  +\epsilon^{0}\cdot\alpha
_{-2}\right\}  \left\vert \Omega\right\rangle \label{4.19f}%
\end{equation}

\item Type II: in this case $D$ is restricted to the critical string
dimension, i.e. $D=26$. Then $\epsilon^{1}$ can be arbitrary constant. The
no-ghost conditions then lead to the on-shell constraints%
\begin{align}
\alpha_{0}^{2}+2  &  =0,\label{4.20f}\\
i\epsilon^{0}\cdot\alpha_{0}+3\epsilon^{1}  &  =0. \label{4.21f}%
\end{align}
The second condition can be solved by a special solution%
\begin{equation}
\epsilon_{\mu}^{0}=-\dfrac{3i}{2}\epsilon^{1}\alpha_{0\mu}, \label{4.22f}%
\end{equation}
which leads to an on-shell scalar ZNS%
\begin{equation}
Q_{\text{B}}\Lambda=\epsilon^{1}\left\{  \dfrac{3}{2}\left(  \alpha_{0}%
\cdot\alpha_{-1}\right)  ^{2}+\frac{1}{2}\left(  \alpha_{-1}\cdot\alpha
_{-1}\right)  +\dfrac{5}{2}\left(  \alpha_{0}\cdot\alpha_{-2}\right)
\right\}  \left\vert \Omega\right\rangle \label{4.23f}%
\end{equation}

\end{enumerate}

Again, up to a constant factor, the ZNS Eq.(\ref{4.19f}) and Eq.(\ref{4.23f})
are the same as Eq.(\ref{2.5q}) and Eq.(\ref{2.4q}) calculated in the OCFQ approach.

\noindent\underline{$M^{2}=4$:}

The string fields are expanded as%
\begin{align}
\Phi &  =\left\{  -iC_{\mu\nu\lambda}\left(  x\right)  \alpha_{-1}^{\mu}%
\alpha_{-1}^{\nu}\alpha_{-1}^{\lambda}-C_{\mu\nu}\left(  x\right)  \alpha
_{-2}^{\mu}\alpha_{-1}^{\nu}+iC_{\mu}\left(  x\right)  \alpha_{-3}^{\mu
}\right. \nonumber\\
&  \text{ \ \ \ \ \ }-\gamma_{\mu\nu}\left(  x\right)  \alpha_{-1}^{\mu}%
\alpha_{-1}^{\nu}b_{-1}c_{0}+i\gamma_{\mu}^{0}\left(  x\right)  \alpha
_{-1}^{\mu}b_{-2}c_{0}+i\gamma_{\mu}^{1}\left(  x\right)  \alpha_{-1}^{\mu
}b_{-1}c_{-1}\nonumber\\
&  \text{ \ \ \ \ \ }\left.  +i\gamma_{\mu}^{2}\left(  x\right)  \alpha
_{-2}^{\mu}b_{-1}c_{0}+\gamma^{0}\left(  x\right)  b_{-3}c_{0}+\gamma
^{1}\left(  x\right)  b_{-2}c_{-1}+\gamma^{2}\left(  x\right)  b_{-1}%
c_{-2}\right\}  \left\vert \Omega\right\rangle ,\label{4.24f}\\
\Lambda &  =\left\{  -\epsilon_{\mu\nu}\left(  x\right)  \alpha_{-1}^{\mu
}\alpha_{-1}^{\nu}b_{-1}+i\epsilon_{\mu}^{1}\left(  x\right)  \alpha_{-2}%
^{\mu}b_{-1}\right.  \left\vert \Omega\right\rangle \nonumber\\
&  \text{ \ \ \ \ \ }\left.  +i\epsilon_{\mu}^{2}\left(  x\right)  \alpha
_{-1}^{\mu}b_{-2}+\epsilon^{2}\left(  x\right)  b_{-3}+\epsilon^{3}\left(
x\right)  b_{-1}b_{-2}c_{0}\right\}  \left\vert \Omega\right\rangle .
\label{4.25f}%
\end{align}
The off-shell ZNS are%
\begin{align}
Q_{\text{B}}\Lambda &  =\left\{  \left(  -\alpha_{0\left(  \mu\right.
}\epsilon_{\left.  \nu\lambda\right)  }+\dfrac{i}{2}\epsilon_{\left(
\mu\right.  }^{2}\eta_{\left.  \nu\lambda\right)  }\right)  \alpha_{-1}^{\mu
}\alpha_{-1}^{\nu}\alpha_{-1}^{\lambda}+\left(  i\alpha_{0\mu}\epsilon_{\nu
}^{2}+i\alpha_{0\nu}\epsilon_{\mu}^{1}-2\epsilon_{\mu\nu}+\epsilon^{2}%
\eta_{\mu\nu}\right)  \alpha_{-2}^{\mu}\alpha_{-1}^{\nu}\right. \nonumber\\
&  \text{ \ \ \ \ }+\left(  \alpha_{0\mu}\epsilon^{2}+2i\epsilon_{\mu}%
^{1}+i\epsilon_{\mu}^{2}\right)  \alpha_{-3}^{\mu}+\left[  \dfrac{1}{2}\left(
\alpha_{0}^{2}+4\right)  \epsilon_{\mu\nu}+\dfrac{1}{2}\epsilon^{3}\eta
_{\mu\nu}\right]  \alpha_{-1}^{\mu}\alpha_{-1}^{\nu}b_{-1}c_{0}\nonumber\\
&  \text{ \ \ \ \ }+\left[  -\dfrac{i}{2}\left(  \alpha_{0}^{2}+4\right)
\epsilon_{\mu}^{2}-\alpha_{0\mu}\epsilon^{3}\right]  \alpha_{-1}^{\mu}%
b_{-2}c_{0}+\left(  2\alpha_{0}^{\nu}\epsilon_{\nu\mu}-2i\epsilon_{\mu}%
^{1}-3i\epsilon_{\mu}^{2}\right)  \alpha_{-1}^{\mu}b_{-1}c_{-1}\nonumber\\
&  \text{ \ \ \ \ }+\left[  -\dfrac{i}{2}\left(  \alpha_{0}^{2}+4\right)
\epsilon_{\mu}^{1}+\alpha_{0\mu}\epsilon^{3}\right]  \alpha_{-2}^{\mu}%
b_{-1}c_{0}+\left[  -\dfrac{1}{2}\left(  \alpha_{0}^{2}+4\right)  \epsilon
^{2}-\epsilon^{3}\right]  b_{-3}c_{0}\nonumber\\
&  \text{ \ \ \ \ }\left.  +\left(  -i\alpha_{0}^{\mu}\epsilon_{\mu}%
^{2}-4\epsilon^{2}-2\epsilon^{3}\right)  b_{-2}c_{-1}+\left(  -2i\alpha
_{0}^{\mu}\epsilon_{\mu}^{1}-5\epsilon^{2}+4\epsilon^{3}+\epsilon_{\mu}^{\mu
}\right)  b_{-1}c_{-2}\right\}  \left\vert \Omega\right\rangle . \label{4.26f}%
\end{align}
Nilpotency condition requires%
\begin{equation}
Q_{\text{B}}^{2}\Lambda=\left(  D-26\right)  \left[  \dfrac{i}{2}\epsilon
_{\mu}^{2}\alpha_{-1}^{\mu}c_{-2}+2\epsilon^{2}c_{-3}-\dfrac{1}{2}\epsilon
^{3}b_{-1}c_{-2}c_{0}\right]  =0. \label{4.27f}%
\end{equation}
Similarly, we classify the solutions of Eq.(\ref{4.27f}) by type I and type II
in the following:

\begin{enumerate}
\item Type I: $D\neq26$. This leads to%
\begin{equation}
\epsilon^{2}=\epsilon^{3}=\epsilon_{\mu}^{2}=0, \label{4.28f}%
\end{equation}
The no-ghost conditions lead to the on-shell constraints%
\begin{subequations}
\begin{align}
\alpha_{0}^{2}+4  &  =0,\label{4.29f}\\
\alpha_{0}^{\nu}\epsilon_{\nu\mu}-i\epsilon_{\mu}^{1}  &  =0,\label{4.30f}\\
-2i\left(  \alpha_{0}\cdot\epsilon^{1}\right)  +\epsilon_{\mu}^{\mu}  &  =0.
\label{4.31f}%
\end{align}
There are three independent solutions to the above equations, which correspond
to the three type I on-shell ZNS:
\end{subequations}
\begin{itemize}
\item Tensor ZNS%
\begin{equation}
\epsilon_{\mu}^{1}=0\text{, \ \ }\alpha_{0}^{\nu}\epsilon_{\mu\nu}=0\text{,
\ \ }\epsilon_{\mu}^{\mu}=0, \label{4.32f}%
\end{equation}%
\begin{equation}
Q_{\text{B}}\Lambda=-\left\{  \alpha_{0\mu}\epsilon_{\nu\lambda}\alpha
_{-1}^{\mu}\alpha_{-1}^{\nu}\alpha_{-1}^{\lambda}+2\epsilon_{\mu\nu}%
\alpha_{-2}^{\mu}\alpha_{-1}^{\nu}\right\}  \left\vert \Omega\right\rangle .
\label{4.33f}%
\end{equation}

\item Vector ZNS%
\begin{equation}
\alpha_{0}\cdot\epsilon^{1}=0\text{, \ \ }\epsilon_{\mu\nu}=-\dfrac{i}%
{4}\left(  \alpha_{0\nu}\epsilon_{\mu}^{1}+\alpha_{0\mu}\epsilon_{\nu}%
^{1}\right)  . \label{4.34f}%
\end{equation}%
\begin{align}
Q_{\text{B}}\Lambda &  =\left\{  i\dfrac{1}{2}\left(  \alpha_{0}\cdot
\alpha_{-1}\right)  ^{2}\left(  \epsilon^{1}\cdot\alpha_{-1}\right)
+2i\left(  \epsilon^{1}\cdot\alpha_{-3}\right)  \right. \nonumber\\
&  \text{ \ \ \ \ \ }\left.  +\dfrac{3}{2}\left(  \alpha_{0}\cdot\alpha
_{-1}\right)  \left(  \epsilon^{1}\cdot\alpha_{-2}\right)  +\dfrac{1}%
{2}\left(  \alpha_{0}\cdot\alpha_{-2}\right)  \left(  \epsilon^{1}\cdot
\alpha_{-1}\right)  \right\}  \left\vert \Omega\right\rangle . \label{4.35f}%
\end{align}

\item Scalar ZNS%
\begin{equation}
\epsilon_{\mu}^{1}=\dfrac{i\left(  D-1\right)  }{9}\theta\alpha_{0\mu}\text{,
\ \ }\epsilon_{\mu\nu}=\theta\eta_{\mu\nu}+\dfrac{\left(  8+D\right)  }%
{36}\theta\alpha_{0\mu}\alpha_{0\nu}. \label{4.36f}%
\end{equation}%
\begin{align}
Q_{\text{B}}\Lambda &  =-\dfrac{2}{9}\theta\left\{  \dfrac{\left(  8+D\right)
}{8}\left(  \alpha_{0}\cdot\alpha_{-1}\right)  ^{3}+\dfrac{9}{2}\left(
\alpha_{0}\cdot\alpha_{-1}\right)  \left(  \alpha_{-1}\cdot\alpha_{-1}\right)
+9\left(  \alpha_{-1}\cdot\alpha_{-2}\right)  \right. \nonumber\\
&  \text{ \ \ \ }\left.  +\dfrac{3\left(  D+2\right)  }{4}\left(  \alpha
_{0}\cdot\alpha_{-1}\right)  \left(  \alpha_{0}\cdot\alpha_{-2}\right)
+\left(  D-1\right)  \left(  \alpha_{0}\cdot\alpha_{-3}\right)  \right\}
\left\vert \Omega\right\rangle . \label{4.37f}%
\end{align}
If we set $D=26$, then%
\begin{align}
Q_{\text{B}}\Lambda &  =-\dfrac{2}{9}\theta\left\{  \dfrac{17}{4}\left(
\alpha_{0}\cdot\alpha_{-1}\right)  ^{3}+\dfrac{9}{2}\left(  \alpha_{0}%
\cdot\alpha_{-1}\right)  \left(  \alpha_{-1}\cdot\alpha_{-1}\right)  +9\left(
\alpha_{-1}\cdot\alpha_{-2}\right)  \right. \nonumber\\
&  \text{ \ \ \ }\left.  +21\left(  \alpha_{0}\cdot\alpha_{-1}\right)  \left(
\alpha_{0}\cdot\alpha_{-2}\right)  +25\left(  \alpha_{0}\cdot\alpha
_{-3}\right)  \right\}  \left\vert \Omega\right\rangle , \label{4.38f}%
\end{align}
where $\theta$ is an arbitrary constant.
\end{itemize}

\item Type II: $D=26$ in Eq.(\ref{4.27f}), and $\epsilon^{2},\epsilon^{3}$ and
$\epsilon_{\mu}^{2}$ are arbitrary constants. The no-ghost conditions lead to
the on-shell constraints%
\begin{subequations}
\begin{align}
\alpha_{0}^{2}+4  &  =0,\label{4.39f}\\
\epsilon^{3}  &  =0,\label{4.40f}\\
2\alpha_{0}^{\nu}\epsilon_{\nu\mu}-2i\epsilon_{\mu}^{1}-3i\epsilon_{\mu}^{2}
&  =0,\label{4.41f}\\
i\alpha_{0}^{\mu}\epsilon_{\mu}^{2}+4\epsilon^{2}  &  =0,\label{4.42f}\\
-2i\alpha_{0}^{\mu}\epsilon_{\mu}^{1}-5\epsilon^{2}+\epsilon_{\mu}^{\mu}  &
=0. \label{4.43f}%
\end{align}
A special solution of above equations is%
\end{subequations}
\begin{subequations}
\begin{align}
\epsilon^{2}  &  =-\dfrac{i}{4}\left(  \alpha_{0}\cdot\epsilon^{2}\right)
=0,\label{4.44f}\\
\epsilon_{\mu\nu}  &  =-C\left(  \alpha_{0\mu}\epsilon_{\nu}^{2}+\alpha_{0\nu
}\epsilon_{\mu}^{2}\right)  ,\label{4.45f}\\
\epsilon_{\mu}^{1}  &  =\dfrac{8iC-3}{2}\epsilon_{\mu}^{2}, \label{4.46f}%
\end{align}
which gives an on-shell vector ZNS%
\end{subequations}
\begin{align}
Q_{\text{B}}\Lambda &  =i\left\{  \left(  8iC-2\right)  \left(  \epsilon
^{2}\cdot\alpha_{-3}\right)  +\dfrac{1}{2}\left(  \alpha_{-1}\cdot\alpha
_{-1}\right)  \left(  \epsilon^{2}\cdot\alpha_{-1}\right)  \right. \nonumber\\
&  \text{ \ \ \ \ \ \ }+\left(  2iC+1\right)  \left(  \alpha_{0}\cdot
\alpha_{-2}\right)  \left(  \epsilon^{2}\cdot\alpha_{-1}\right)  +2iC\left(
\alpha_{0}\cdot\alpha_{-1}\right)  ^{2}\left(  \epsilon^{2}\cdot\alpha
_{-1}\right) \nonumber\\
&  \text{ \ \ \ \ \ \ }\left.  +\dfrac{12iC-3}{2}\left(  \alpha_{0}\cdot
\alpha_{-1}\right)  \left(  \epsilon^{2}\cdot\alpha_{-2}\right)  \right\}
\left\vert \Omega\right\rangle . \label{4.47f}%
\end{align}
For a special value of $C=-3i/4$, Eq.(\ref{4.47f}) becomes%
\begin{align}
Q_{\text{B}}\Lambda &  =i\left\{  4\left(  \epsilon^{2}\cdot\alpha
_{-3}\right)  +\dfrac{1}{2}\left(  \alpha_{-1}\cdot\alpha_{-1}\right)  \left(
\epsilon^{2}\cdot\alpha_{-1}\right)  +\dfrac{5}{2}\left(  \alpha_{0}%
\cdot\alpha_{-2}\right)  \left(  \epsilon^{2}\cdot\alpha_{-1}\right)  \right.
\nonumber\\
&  \text{ \ \ \ \ \ \ }\left.  +\dfrac{3}{2}\left(  \alpha_{0}\cdot\alpha
_{-1}\right)  ^{2}\left(  \epsilon^{2}\cdot\alpha_{-1}\right)  +3\left(
\alpha_{0}\cdot\alpha_{-1}\right)  \left(  \epsilon^{2}\cdot\alpha
_{-2}\right)  \right\}  \left\vert \Omega\right\rangle . \label{4.48f}%
\end{align}

Up to a constant factor, ZNS in Eq.(\ref{4.33f}), Eq.(\ref{4.35f}),
Eq.(\ref{4.38f}) and Eq.(\ref{4.48f}) are exactly the same as Eq.(\ref{2.7q}),
Eq.(\ref{2.8q}), Eq.(\ref{2.9q}) and Eq.(\ref{2.6q}) calculated in the OCFQ
approach. In addition, it can be checked that for $C=-5i/8$ and $-i/16$ in
Eq.(\ref{4.47f}), one gets $D_{1}$ and $D_{2}$ ZNS of OCFQ approach in
Eq.(\ref{2.11q}) and Eq.(\ref{2.10q}) respectively.

\qquad In section I.D \cite{KaoLee}, the background ghost transformations in
the gauge transformations of WSFT \cite{Witten} were shown to correspond, in a
one-to-one manner, to the lifting of on-shell conditions of ZNS in the OCFQ
approach. For the rest of this section, we are going to go one step further
and apply the results calculated above to demonstrate that off-shell gauge
transformations of WSFT are indeed identical to the on-shell stringy gauge
symmetries generated by two types of ZNS in the generalized massive $\sigma
$-model approach \cite{LeePRL,Lee} of string theory. For the mass level
$M^{2}=2$, by using Eq.(\ref{4.12f}) and Eq.(\ref{4.13f}), the linearized
gauge transformation of WSFT in Eq.(\ref{4.5f}) gives
\begin{subequations}
\begin{align}
\delta B_{\mu\nu}  &  =-\partial_{(\mu}\epsilon_{\nu)}^{0}-\frac{1}{2}%
\epsilon^{1}\eta_{\mu\nu},\label{4.49f}\\
\delta B_{\mu}  &  =-\partial_{\mu}\epsilon^{1}+\frac{1}{2}\epsilon_{\mu}%
^{0},\label{4.50f}\\
\delta\beta_{\mu}  &  =\frac{1}{2}(\partial^{2}-2)\epsilon_{\mu}%
^{0},\label{4.51f}\\
\delta\beta^{0}  &  =\frac{1}{2}(\partial^{2}-2)\epsilon^{1},\label{4.52f}\\
\delta\beta^{1}  &  =-\partial^{\mu}\epsilon_{\mu}^{0}-3\epsilon^{1}
\label{4.53f}%
\end{align}
For the type I gauge transformation induced by ZNS in Eq.(\ref{4.19f}), one
can use Eq.(\ref{4.16f}) to Eq.(\ref{4.18f}) to eliminate the background ghost
transformations Eq.(\ref{4.51f}) to Eq.(\ref{4.53f}). Finally, conditions of
worldsheet conformal invariance in the presence of weak background fields
\cite{LeePRL,Lee} can be used to express $B_{\mu}$ in terms of $B_{\mu\nu}$,
and one ends up with the following on-shell gauge transformation by
Eq.(\ref{4.49f})%
\end{subequations}
\begin{equation}
\delta B_{\mu\nu}=\partial_{(\mu}\epsilon_{\nu)}^{0};\text{ \ \ }\partial
^{\mu}\epsilon_{\mu}^{0}=0,\text{ \ }(\partial^{2}-2)\epsilon_{\mu}^{0}=0.
\label{4.54f}%
\end{equation}
Similarly, one can apply the same procedure to type II ZNS in Eq.(\ref{4.23f}%
), and derive the following type II gauge transformation%
\begin{equation}
\delta B_{\mu\nu}=\frac{3}{2}\partial_{\mu}\partial_{\nu}\epsilon^{1}-\frac
{1}{2}\eta_{\mu\nu}\epsilon^{1},\text{ \ }(\partial^{2}-2)\epsilon^{1}=0.
\label{4.55f}%
\end{equation}
Eq.(\ref{4.54f}) and Eq.(\ref{4.55f}) are consistent with the massive $\sigma
$-model calculation in the OCFQ string theory in.
\end{enumerate}

\qquad For the mass level $M^{2}=4$, by using Eq.(\ref{4.24f}) and
Eq.(\ref{4.25f}), the linearized gauge transformation of WSFT in
Eq.(\ref{4.5f}) gives
\begin{subequations}
\begin{align}
\delta C_{\mu\nu\lambda}  &  =-\partial_{(\mu}\epsilon_{\nu\lambda)}^{0}%
-\frac{1}{2}\epsilon_{(\mu}^{2}\eta_{\mu\nu)},\label{4.56f}\\
\delta C_{[\mu\nu]}  &  =-\partial_{\lbrack\nu}\epsilon_{\mu]}^{1}%
-\partial_{\lbrack\mu}\epsilon_{\nu]}^{2},\label{4.57f}\\
\delta C_{(\mu\nu)}  &  =-\partial_{(\nu}\epsilon_{\mu)}^{1}-\partial_{(\mu
}\epsilon_{\nu)}^{2}+2\epsilon_{\mu\nu}^{0}-\epsilon^{2}\eta_{\mu\nu
},\label{4.58f}\\
\delta C_{\mu}  &  =-\partial_{\mu}\epsilon^{2}+2\epsilon_{\mu}^{1}%
+\epsilon_{\mu}^{2},\label{4.59f}\\
\delta\gamma_{\mu\nu}  &  =\frac{1}{2}(\partial^{2}-4)\epsilon_{\mu\nu}%
^{0}-\frac{1}{2}\epsilon^{3}\eta_{\mu\nu},\label{4.60f}\\
\delta\gamma_{\mu}^{0}  &  =\frac{1}{2}(\partial^{2}-4)\epsilon_{\mu}%
^{2}+\partial_{\mu}\epsilon^{3},\label{4.61f}\\
\delta\gamma_{\mu}^{1}  &  =-2\partial^{\nu}\epsilon_{\nu\mu}^{0}%
-2\epsilon_{\mu}^{1}-3\epsilon_{\mu}^{2},\label{4.62f}\\
\delta\gamma_{\mu}^{2}  &  =\frac{1}{2}(\partial^{2}-4)\epsilon_{\mu}%
^{1}-\partial_{\mu}\epsilon^{3},\label{4.63f}\\
\delta\gamma^{0}  &  =\frac{1}{2}(\partial^{2}-4)\epsilon^{2}-\epsilon
^{3},\label{4.64f}\\
\delta\gamma^{1}  &  =-\partial^{\mu}\epsilon_{\mu}^{2}-4\epsilon
^{2}-2\epsilon^{3},\label{4.65f}\\
\delta\gamma^{2}  &  =-2\partial^{\mu}\epsilon_{\mu}^{1}-5\epsilon
^{2}+4\epsilon^{3}+\epsilon_{\mu}^{0\mu}. \label{4.66f}%
\end{align}
For the gauge transformation induced by $D_{2}$ ZNS in Eq.(\ref{4.48f}), for
example, one can use Eq.(\ref{4.39f}) to Eq.(\ref{4.46f}) with $C=-i/16$ to
eliminate Eq.(\ref{4.60f}) to Eq.(\ref{4.66f}). One can then use the fact that
background fields $C_{(\mu\nu)}$ and $C_{\mu}$ are gauge artifacts of
$C_{\mu\nu\lambda}$ in the $\sigma$-model calculation, and deduce from
Eq.(\ref{4.56f}) to Eq.(\ref{4.59f}) the inter-particle symmetry
transformation
\end{subequations}
\begin{equation}
\delta C_{\mu\nu\lambda}=\frac{1}{2}\partial_{(\mu}\partial_{\nu}%
\epsilon_{\lambda)}^{(D_{2})}-2\eta_{(\mu\nu}\epsilon_{\lambda)}^{(D_{2}%
)},\text{ \ \ }\delta C_{[\mu\nu]}=9\partial_{\lbrack\mu}\epsilon_{\nu
]}^{(D_{2})}, \label{4.67f}%
\end{equation}
where $\partial^{\lambda}\epsilon_{\lambda}^{(D_{2})}=0,(\partial
^{2}-4)\epsilon_{\lambda}^{(D_{2})}=0$. The other three gauge transformations
corresponding to three other ZNS, the spin-two, $D_{1}$, and scalar can be
similarly constructed from Eq.(\ref{4.56f}) to Eq.(\ref{4.66f}). One gets%
\begin{equation}
\delta C_{\mu\nu\lambda}=\partial_{(\mu}\epsilon_{\nu\lambda)};\text{
\ }\partial^{\mu}\epsilon_{\mu\nu}=0,\text{ }(\partial^{2}-4)\epsilon_{\mu\nu
}=0, \label{4.68f}%
\end{equation}%
\begin{equation}
\delta C_{\mu\nu\lambda}=\frac{5}{2}\partial_{(\mu}\partial_{\nu}%
\epsilon_{\lambda)}^{(D_{1})}-\eta_{(\mu\nu}\epsilon_{\lambda)}^{(D_{1}%
)};\text{ \ }\partial^{\lambda}\epsilon_{\lambda}^{(D_{1})}=0,\text{
}(\partial^{2}-4)\epsilon_{\lambda}^{(D_{1})}=0, \label{4.69f}%
\end{equation}%
\begin{equation}
\delta C_{\mu\nu\lambda}=\frac{17}{4}\partial_{\mu}\partial_{\nu}%
\partial_{\lambda}\theta-\frac{9}{2}\eta_{(\mu\nu}\theta_{\lambda)};\text{
}(\partial^{2}-4)\theta=0. \label{4.70f}%
\end{equation}
Eq.(\ref{4.67f}) to Eq.(\ref{4.70f}) are exactly the same as those calculated
by the generalized massive $\sigma$-model approach of string theory
\cite{LeePRL,Lee}.

We thus have shown in this section that off-shell gauge transformations of
WSFT are identical to the on-shell stringy gauge symmetries generated by two
types of ZNS in the OCFQ string theory. The high energy limit of these stringy
gauge symmetries generated by ZNS was recently used to fix the proportionality
constants among high energy scattering amplitudes of different string states
conjectured by Gross \cite{Gross,Gross1}. Based on the ZNS calculations in
\cite{ChanLee,ChanLee1,ChanLee2,CHL} and the calculations in this section, we
thus have related gauge symmetry of WSFT \cite{Witten} to the high energy
stringy symmetry conjectured by Gross \cite{GM,GM1,Gross,Gross1,GrossManes}.

In conclusion of this chapter, we have calculated ZNS in the OCFQ string, the
light-cone DDF string and the off-shell BRST string theories. In the OCFQ
string, we have solved the Virasoro constraints for all physical states (
including ZNS) in the helicity basis. Much attention was paid to discuss the
inter-particle ZNS at the mass level $M^{2}=4$. We found that\ one can use
polarization of either one of the two positive-norm states to represent the
polarization of the inter-particle ZNS. This justified why one can derive the
inter-particle symmetry transformation for the two massive modes in the weak
field massive $\sigma$-model calculation \cite{LeePRL,Lee}.

In the light-cone DDF string, one can easily write down the general formula
for all ZNS in the spectrum. We have identified type I and Type II ZNS up to
the mass level $M^{2}=4.$ An analysis for the general mass levels should be
easy to generalize.

Finally, we have calculated off-shell ZNS in the WSFT. After imposing the no
ghost conditions, we can recover two types of on-shell ZNS in the OCFQ string.
We then show that off-shell gauge transformations of WSFT are identical to the
on-shell stringy gauge symmetries generated by two types of ZNS in the
generalized massive $\sigma$-model approach of string theory. Based on these
ZNS calculations, we thus have related gauge symmetry of WSFT \cite{Witten} to
the high energy stringy symmetry of Gross \cite{Gross,Gross1}.%

\setcounter{equation}{0}
\renewcommand{\theequation}{\arabic{section}.\arabic{equation}}%

\section{Hard closed string scatterings, KLT and string BCJ relations}

In this chapter, we generalize the calculations in chapter V to high energy
closed string scattering amplitudes \cite{Closed}. We will find that the
methods of decoupling of high energy ZNS and the high energy Virasoro
constraints, which were adopted in chapter V to calculate the ratios among
high energy open string scattering amplitudes of different string states,
persist for the case of closed string. The result is simply the tensor product
of two pieces of open string ratios of high energy scattering amplitudes.

However, we clarify the previous saddle-point calculation for high energy open
string scattering amplitudes and claim that only $(t,u)$ channel of the
amplitudes is suitable for saddle-point calculation. We then discuss three
evidences to show that saddle-point calculation for high energy closed string
scattering amplitudes is not reliable. By using the relation of tree-level
closed and open string scattering amplitudes of Kawai, Lewellen and Tye (KLT)
\cite{KLT,1106.0033}, we calculate the tree-level high energy closed string
scattering amplitudes for \textit{arbitrary} mass levels. For the case of high
energy closed string four-tachyon amplitude, our result differs from the
previous one of Gross and Mende \cite{GM,GM1}, which is NOT consistent with
KLT formula, by an oscillating factor. See also \cite{1108.2381,1205.6369}.
One interesting application of this result is the string BCJ relations
\cite{BCJ1,BCJ2,BCJ3,BCJ4,BCJ5} which will be discussed in section D.

\subsection{Decoupling of high energy ZNS}

In this section, we calculate the ratios among high energy closed string
scattering amplitudes of different string states by the decoupling of high
energy closed string ZNS. Since the calculation is similar to that of open
string in chapter V, we will, for simplicity, work on the first massive level
$M^{2}=8(N-1)=8$ only. At this mass level, the corresponding open string Ward
identities are ($M^{2}=2$ for open string, $\alpha_{\text{closed}}^{\prime
}=4\alpha_{\text{open}}^{\prime}=2$) \cite{ChanLee3}
\begin{subequations}
\begin{align}
k_{\mu}\theta_{\nu}\mathcal{T}^{\mu\nu}+\theta_{\mu}\mathcal{T}^{\mu}  &
=0,\label{1a}\\
\left(  \frac{3}{2}k_{\mu}k_{\nu}+\frac{1}{2}\eta_{\mu\nu}\right)
\mathcal{T}^{\mu\nu}+\frac{5}{2}k_{\mu}\mathcal{T}^{\mu}  &  =0, \label{2a}%
\end{align}
where $\theta_{\nu}$ is a transverse vector. In Eq.(\ref{1a}) and
Eq.(\ref{2a}), we have chosen, say, the second vertex $V_{2}(k_{2})$ to be the
vertex operators constructed from ZNS and $k_{\mu}\equiv k_{2\mu}$. The other
three vertices can be any string states. Note that Eq.(\ref{1a}) is the type I
Ward identity while Eq.(\ref{2a}) is the type II Ward identity which is valid
only at $D=26$. The high energy limits of Eq.(\ref{1a}) and Eq.(\ref{2a}) were
calculated to be
\end{subequations}
\begin{subequations}
\begin{align}
M\mathcal{T}_{TP}^{3\rightarrow1}+\mathcal{T}_{T}^{1}  &  =0,\label{3aa}\\
M\mathcal{T}_{LL}^{4\rightarrow2}+\mathcal{T}_{L}^{2}  &  =0,\label{4aa}\\
3M^{2}\mathcal{T}_{LL}^{4\rightarrow2}+\mathcal{T}_{TT}^{2}+5M\mathcal{T}%
_{L}^{2}  &  =0. \label{5aa}%
\end{align}
Note that since $\mathcal{T}_{TP}^{1}$ is of subleading order in energy, in
general $\mathcal{T}_{TP}^{1}\neq\mathcal{T}_{TL}^{1}$. A simple calculation
of Eq.(\ref{3aa}) to Eq.(\ref{5aa}) shows that \cite{ChanLee3}
\end{subequations}
\begin{equation}
\mathcal{T}_{TP}^{1}:\mathcal{T}_{T}^{1}=1:-\sqrt{2}=1:-M. \label{9aa}%
\end{equation}%
\begin{equation}
\mathcal{T}_{TT}^{2}:\mathcal{T}_{LL}^{2}:\mathcal{T}_{L}^{2}\text{ }=\text{
}4:1:-\sqrt{2}=2M^{2}:1:-M. \label{10aa}%
\end{equation}
It is interesting to see that, in addition to the leading order amplitudes in
Eq.(\ref{10aa}), the subleading order amplitudes in Eq.(\ref{9aa}) are also
proportional to each other. This does not seem to happen at higher mass level.

We are now back to the closed string calculation. The OCFQ closed string
spectrum at this mass level are
$(\raisebox{0.06in}{\fbox{\rule[0.04cm]{0.04cm}{0cm}}}\raisebox{0.06in}{\fbox{\rule[0.04cm]{0.04cm}{0cm}}}+\raisebox{0.06in}{\fbox{\rule[0.04cm]{0.04cm}{0cm}}}+\bullet
)\otimes
(\raisebox{0.06in}{\fbox{\rule[0.04cm]{0.04cm}{0cm}}}\raisebox{0.06in}{\fbox{\rule[0.04cm]{0.04cm}{0cm}}}+\raisebox{0.06in}{\fbox{\rule[0.04cm]{0.04cm}{0cm}}}+\bullet
)^{^{\prime}}.$ In addition to the spin-four positive-norm state
$\raisebox{0.06in}{\fbox{\rule[0.04cm]{0.04cm}{0cm}}}\raisebox{0.06in}{\fbox{\rule[0.04cm]{0.04cm}{0cm}}}\otimes
\raisebox{0.06in}{\fbox{\rule[0.04cm]{0.04cm}{0cm}}}\raisebox{0.06in}{\fbox{\rule[0.04cm]{0.04cm}{0cm}}}^{\prime
}$, one has 8 ZNS, each of which gives a Ward identity. In the high energy
limit, we have $\theta^{\mu\nu}=e_{L}^{\mu}e_{L}^{\nu}-e_{T}^{\mu}e_{T}^{\nu}$
or $\theta^{\mu\nu}=e_{L}^{\mu}e_{T}^{\nu}+e_{T}^{\mu}e_{L}^{\nu}$,
$\theta^{\mu}=e_{L}^{\mu}$ or $e_{T}^{\mu}$ and one replace $\eta_{\mu\nu}$ by
$e_{T}^{\mu}e_{T}^{\nu}$. In the following, we list only high energy Ward
identities which relate amplitudes with even-energy power in the high energy
expansion :

\noindent\bigskip1.
$\raisebox{0.06in}{\fbox{\rule[0.04cm]{0.04cm}{0cm}}}\raisebox{0.06in}{\fbox{\rule[0.04cm]{0.04cm}{0cm}}}\otimes
\raisebox{0.06in}{\fbox{\rule[0.04cm]{0.04cm}{0cm}}}^{^{\prime}}:$%
\begin{align}
M(\mathcal{T}_{LL,LL}-\mathcal{T}_{TT,LL})+\mathcal{T}_{LL,L}-\mathcal{T}%
_{TT,L}  &  =0,\label{11aa}\\
M\mathcal{T}_{LT,PT}+\mathcal{T}_{LT,T}  &  =0. \label{12aa}%
\end{align}
2.
$\raisebox{0.06in}{\fbox{\rule[0.04cm]{0.04cm}{0cm}}}\raisebox{0.06in}{\fbox{\rule[0.04cm]{0.04cm}{0cm}}}\otimes
\bullet^{^{\prime}}:$%
\begin{equation}
3M^{2}(\mathcal{T}_{LL,LL}-\mathcal{T}_{TT,LL})+(\mathcal{T}_{LL,TT}%
-\mathcal{T}_{TT,TT})+5M(\mathcal{T}_{LL,L}-\mathcal{T}_{TT,L})=0.
\label{13aa}%
\end{equation}
3. $\raisebox{0.06in}{\fbox{\rule[0.04cm]{0.04cm}{0cm}}}\otimes
\raisebox{0.06in}{\fbox{\rule[0.04cm]{0.04cm}{0cm}}}\raisebox{0.06in}{\fbox{\rule[0.04cm]{0.04cm}{0cm}}}^{^{\prime
}}:$%
\begin{align}
M(\mathcal{T}_{LL,LL}-\mathcal{T}_{LL,TT})+\mathcal{T}_{L,LL}-\mathcal{T}%
_{L,TT}  &  =0,\label{14aa}\\
M\mathcal{T}_{PT,LT}+\mathcal{T}_{T,LT}  &  =0. \label{15aa}%
\end{align}
4. $\raisebox{0.06in}{\fbox{\rule[0.04cm]{0.04cm}{0cm}}}\otimes
\raisebox{0.06in}{\fbox{\rule[0.04cm]{0.04cm}{0cm}}}^{^{\prime}}:$%
\begin{align}
M^{2}\mathcal{T}_{LL,LL}+M\mathcal{T}_{LL,L}+M\mathcal{T}_{L,LL}%
+\mathcal{T}_{L,L}  &  =0,\label{16aa}\\
M^{2}\mathcal{T}_{PT,PT}+M\mathcal{T}_{PT,T}+M\mathcal{T}_{T,PT}%
+\mathcal{T}_{T,T}  &  =0. \label{17aa}%
\end{align}
5. $\raisebox{0.06in}{\fbox{\rule[0.04cm]{0.04cm}{0cm}}}\otimes\bullet
^{^{\prime}}:$%
\begin{equation}
3M^{3}\mathcal{T}_{LL,LL}+M\mathcal{T}_{LL,TT}+5M^{2}\mathcal{T}_{LL,L}%
+3M^{2}\mathcal{T}_{L,LL}+\mathcal{T}_{L,TT}+5M^{2}\mathcal{T}_{L,L}=0.
\label{18aa}%
\end{equation}
6. $\bullet\otimes
\raisebox{0.06in}{\fbox{\rule[0.04cm]{0.04cm}{0cm}}}\raisebox{0.06in}{\fbox{\rule[0.04cm]{0.04cm}{0cm}}}^{\prime
}:$%
\begin{equation}
3M^{2}(\mathcal{T}_{LL,LL}-\mathcal{T}_{LL,TT})+(\mathcal{T}_{TT,LL}%
-\mathcal{T}_{TT,TT})+5M(\mathcal{T}_{L,LL}-\mathcal{T}_{L,TT})=0.
\label{19aa}%
\end{equation}
7. $\bullet\otimes
\raisebox{0.06in}{\fbox{\rule[0.04cm]{0.04cm}{0cm}}}^{^{\prime}}:$%
\begin{equation}
3M^{3}\mathcal{T}_{LL,LL}+M\mathcal{T}_{TT,LL}+5M^{2}\mathcal{T}_{L,LL}%
+3M^{2}\mathcal{T}_{LL,L}+\mathcal{T}_{TT,L}+5M^{2}\mathcal{T}_{L,L}=0.
\label{20aa}%
\end{equation}
8. $\bullet\otimes\bullet^{^{\prime}}:$%
\begin{align}
&  9M^{4}\mathcal{T}_{LL,LL}+3M^{2}\mathcal{T}_{LL,TT}+3M^{2}\mathcal{T}%
_{TT,LL}+15M^{3}\mathcal{T}_{LL,L}\nonumber\\
&  +15M^{3}\mathcal{T}_{L,LL}+5M\mathcal{T}_{TT,L}+5M\mathcal{T}%
_{L,TT}+25M^{2}\mathcal{T}_{L,L}+\mathcal{T}_{TT,TT}=0. \label{21aa}%
\end{align}
Those Ward identities which relate amplitudes with odd-energy power in the
high energy expansion are omitted as they are subleading order in energy. The
mass $M$ in Eq.(\ref{11aa}) to Eq.(\ref{21aa}) should now be interpreted as
the closed string mass $M^{2}=8$. Eq.(\ref{12aa}),Eq.(\ref{15aa}) and
Eq.(\ref{17aa}) are subleading order amplitudes, and one can then solve the
other $8$ equations to give the ratios%
\begin{align}
&  \mathcal{T}_{TT,TT}:\mathcal{T}_{TT,LL}:\mathcal{T}_{LL,TT}:\mathcal{T}%
_{LL,LL}:\mathcal{T}_{TT,L}:\mathcal{T}_{L,TT}:\mathcal{T}_{LL,L}%
:\mathcal{T}_{L,LL}:\mathcal{T}_{L,L}\nonumber\\
=  &  1:\frac{1}{2M^{2}}:\frac{1}{2M^{2}}:\frac{1}{4M^{4}}:-\frac{1}%
{2M}:-\frac{1}{2M}:-\frac{1}{4M^{3}}:-\frac{1}{4M^{3}}:\frac{1}{4M^{2}}.
\label{22aa}%
\end{align}
Eq.(\ref{22aa}) is exactly the tensor product of two pieces of open string
ratios calculated in Eq.(\ref{10aa}).

\subsection{Virasoro constraints}

We consider the mass level $M^{2}=8$. The most general state is%
\begin{align}
\left\vert 2\right\rangle  &  =\left\{  \frac{1}{2!}%
\begin{tabular}
[c]{|c|c|}\hline
$\mu_{1}^{1}$ & $\mu_{2}^{1}$\\\hline
\end{tabular}
\alpha_{-1}^{\mu_{1}^{1}}\alpha_{-1}^{\mu_{2}^{1}}+\frac{1}{2}%
\begin{tabular}
[c]{|c|}\hline
$\mu_{1}^{2}$\\\hline
\end{tabular}
\alpha_{-2}^{\mu_{1}^{2}}\right\}  \otimes\left\{  \frac{1}{2!}%
\begin{tabular}
[c]{|c|c|}\hline
$\tilde{\mu}_{1}^{1}$ & $\tilde{\mu}_{2}^{1}$\\\hline
\end{tabular}
\tilde{\alpha}_{-1}^{\tilde{\mu}_{1}^{1}}\tilde{\alpha}_{-1}^{\tilde{\mu}%
_{2}^{1}}+\frac{1}{2}%
\begin{tabular}
[c]{|c|}\hline
$\tilde{\mu}_{1}^{2}$\\\hline
\end{tabular}
\tilde{\alpha}_{-2}^{\tilde{\mu}_{1}^{2}}\right\}  \left\vert 0,k\right\rangle
\nonumber\\
&  =\frac{1}{4}\left\{
\begin{tabular}
[c]{|c|c|}\hline
$\mu_{1}^{1}$ & $\mu_{2}^{1}$\\\hline
\end{tabular}
\alpha_{-1}^{\mu_{1}^{1}}\alpha_{-1}^{\mu_{2}^{1}}+%
\begin{tabular}
[c]{|c|}\hline
$\mu_{1}^{2}$\\\hline
\end{tabular}
\alpha_{-2}^{\mu_{1}^{2}}\right\}  \otimes\left\{
\begin{tabular}
[c]{|c|c|}\hline
$\tilde{\mu}_{1}^{1}$ & $\tilde{\mu}_{2}^{1}$\\\hline
\end{tabular}
\tilde{\alpha}_{-1}^{\tilde{\mu}_{1}^{1}}\tilde{\alpha}_{-1}^{\tilde{\mu}%
_{2}^{1}}+%
\begin{tabular}
[c]{|c|}\hline
$\tilde{\mu}_{1}^{2}$\\\hline
\end{tabular}
\tilde{\alpha}_{-2}^{\tilde{\mu}_{1}^{2}}\right\}  \left\vert 0,k\right\rangle
.
\end{align}
The Virasoro constraints are%
\begin{subequations}
\begin{align}
L_{1}\left\vert 2\right\rangle  &  \sim\left\{  k^{\mu_{1}^{1}}%
\begin{tabular}
[c]{|c|c|}\hline
$\mu_{1}^{1}$ & $\mu_{2}^{1}$\\\hline
\end{tabular}
\alpha_{-1}^{\mu_{2}^{1}}+%
\begin{tabular}
[c]{|c|}\hline
$\mu_{1}^{2}$\\\hline
\end{tabular}
\alpha_{-1}^{\mu_{1}^{2}}\right\}  \otimes\left\{
\begin{tabular}
[c]{|c|c|}\hline
$\tilde{\mu}_{1}^{1}$ & $\tilde{\mu}_{2}^{1}$\\\hline
\end{tabular}
\tilde{\alpha}_{-1}^{\tilde{\mu}_{1}^{1}}\tilde{\alpha}_{-1}^{\tilde{\mu}%
_{2}^{1}}+%
\begin{tabular}
[c]{|c|}\hline
$\tilde{\mu}_{1}^{2}$\\\hline
\end{tabular}
\tilde{\alpha}_{-2}^{\tilde{\mu}_{1}^{2}}\right\}  =0,\\
\tilde{L}_{1}\left\vert 2\right\rangle  &  \sim\left\{
\begin{tabular}
[c]{|c|c|}\hline
$\mu_{1}^{1}$ & $\mu_{2}^{1}$\\\hline
\end{tabular}
\alpha_{-1}^{\mu_{1}^{1}}\alpha_{-1}^{\mu_{2}^{1}}+%
\begin{tabular}
[c]{|c|}\hline
$\mu_{1}^{2}$\\\hline
\end{tabular}
\alpha_{-2}^{\mu_{1}^{2}}\right\}  \otimes\left\{  k^{\mu_{1}^{1}}%
\begin{tabular}
[c]{|c|c|}\hline
$\tilde{\mu}_{1}^{1}$ & $\tilde{\mu}_{2}^{1}$\\\hline
\end{tabular}
\tilde{\alpha}_{-1}^{\tilde{\mu}_{2}^{1}}+%
\begin{tabular}
[c]{|c|}\hline
$\tilde{\mu}_{1}^{2}$\\\hline
\end{tabular}
\tilde{\alpha}_{-1}^{\tilde{\mu}_{1}^{2}}\right\}  =0,\\
L_{2}\left\vert 2\right\rangle  &  \sim\left\{
\begin{tabular}
[c]{|c|c|}\hline
$\mu_{1}^{1}$ & $\mu_{2}^{1}$\\\hline
\end{tabular}
\eta^{\mu_{1}^{1}\mu_{2}^{1}}+2k^{\mu_{1}^{2}}%
\begin{tabular}
[c]{|c|}\hline
$\mu_{1}^{2}$\\\hline
\end{tabular}
\right\}  \otimes\left\{
\begin{tabular}
[c]{|c|c|}\hline
$\tilde{\mu}_{1}^{1}$ & $\tilde{\mu}_{2}^{1}$\\\hline
\end{tabular}
\tilde{\alpha}_{-1}^{\tilde{\mu}_{1}^{1}}\tilde{\alpha}_{-1}^{\tilde{\mu}%
_{2}^{1}}+%
\begin{tabular}
[c]{|c|}\hline
$\tilde{\mu}_{1}^{2}$\\\hline
\end{tabular}
\tilde{\alpha}_{-2}^{\tilde{\mu}_{1}^{2}}\right\}  =0,\\
\tilde{L}_{2}\left\vert 2\right\rangle  &  \sim\left\{
\begin{tabular}
[c]{|c|c|}\hline
$\mu_{1}^{1}$ & $\mu_{2}^{1}$\\\hline
\end{tabular}
\alpha_{-1}^{\mu_{1}^{1}}\alpha_{-1}^{\mu_{2}^{1}}+%
\begin{tabular}
[c]{|c|}\hline
$\mu_{1}^{2}$\\\hline
\end{tabular}
\alpha_{-2}^{\mu_{1}^{2}}\right\}  \otimes\left\{
\begin{tabular}
[c]{|c|c|}\hline
$\tilde{\mu}_{1}^{1}$ & $\tilde{\mu}_{2}^{1}$\\\hline
\end{tabular}
\eta^{\tilde{\mu}_{12}^{1\tilde{\mu}1}}+2k^{\tilde{\mu}_{1}^{2}}%
\begin{tabular}
[c]{|c|}\hline
$\tilde{\mu}_{1}^{2}$\\\hline
\end{tabular}
\right\}  =0.
\end{align}
Taking the high energy limit in the above equations by letting $\left(
\mu_{i},\nu_{i}\right)  \rightarrow\left(  L,T\right)  $, and%
\end{subequations}
\begin{equation}
k^{\mu_{i}}\rightarrow Me^{L}\text{, }\eta^{\mu_{1}\mu_{2}}\rightarrow
e^{T}e^{T},
\end{equation}
we obtain
\begin{subequations}
\begin{align}
\left\{  M%
\begin{tabular}
[c]{|c|c|}\hline
$L$ & $\mu$\\\hline
\end{tabular}
+%
\begin{tabular}
[c]{|c|}\hline
$\mu$\\\hline
\end{tabular}
\right\}  \alpha_{-1}^{\mu}\otimes\left\{
\begin{tabular}
[c]{|c|c|}\hline
$\tilde{\mu}_{1}^{1}$ & $\tilde{\mu}_{2}^{1}$\\\hline
\end{tabular}
\tilde{\alpha}_{-1}^{\tilde{\mu}_{1}^{1}}\tilde{\alpha}_{-1}^{\tilde{\mu}%
_{2}^{1}}+%
\begin{tabular}
[c]{|c|}\hline
$\tilde{\mu}_{1}^{2}$\\\hline
\end{tabular}
\tilde{\alpha}_{-2}^{\tilde{\mu}_{1}^{2}}\right\}   &  =0,\\
\left\{
\begin{tabular}
[c]{|c|c|}\hline
$\mu_{1}^{1}$ & $\mu_{2}^{1}$\\\hline
\end{tabular}
\alpha_{-1}^{\mu_{1}^{1}}\alpha_{-1}^{\mu_{2}^{1}}+%
\begin{tabular}
[c]{|c|}\hline
$\mu_{1}^{2}$\\\hline
\end{tabular}
\alpha_{-2}^{\mu_{1}^{2}}\right\}  \otimes\left\{  M%
\begin{tabular}
[c]{|c|c|}\hline
$L$ & $\tilde{\mu}$\\\hline
\end{tabular}
+%
\begin{tabular}
[c]{|c|}\hline
$\tilde{\mu}$\\\hline
\end{tabular}
\right\}  \tilde{\alpha}_{-1}^{\tilde{\mu}}  &  =0,\\
\left\{
\begin{tabular}
[c]{|c|c|}\hline
$T$ & $T$\\\hline
\end{tabular}
+2M%
\begin{tabular}
[c]{|c|}\hline
$L$\\\hline
\end{tabular}
\right\}  \otimes\left\{
\begin{tabular}
[c]{|c|c|}\hline
$\tilde{\mu}_{1}^{1}$ & $\tilde{\mu}_{2}^{1}$\\\hline
\end{tabular}
\tilde{\alpha}_{-1}^{\tilde{\mu}_{1}^{1}}\tilde{\alpha}_{-1}^{\tilde{\mu}%
_{2}^{1}}+%
\begin{tabular}
[c]{|c|}\hline
$\tilde{\mu}_{1}^{2}$\\\hline
\end{tabular}
\tilde{\alpha}_{-2}^{\tilde{\mu}_{1}^{2}}\right\}   &  =0,\\
\left\{
\begin{tabular}
[c]{|c|c|}\hline
$\mu_{1}^{1}$ & $\mu_{2}^{1}$\\\hline
\end{tabular}
\alpha_{-1}^{\mu_{1}^{1}}\alpha_{-1}^{\mu_{2}^{1}}+%
\begin{tabular}
[c]{|c|}\hline
$\mu_{1}^{2}$\\\hline
\end{tabular}
\alpha_{-2}^{\mu_{1}^{2}}\right\}  \otimes\left\{
\begin{tabular}
[c]{|c|c|}\hline
$T$ & $T$\\\hline
\end{tabular}
+2M%
\begin{tabular}
[c]{|c|}\hline
$L$\\\hline
\end{tabular}
\right\}   &  =0,
\end{align}
which lead to the following equations%
\end{subequations}
\begin{subequations}
\begin{align}
\left\{  M%
\begin{tabular}
[c]{|c|c|}\hline
$L$ & $\mu$\\\hline
\end{tabular}
+%
\begin{tabular}
[c]{|c|}\hline
$\mu$\\\hline
\end{tabular}
\right\}  \otimes%
\begin{tabular}
[c]{|c|c|}\hline
$\tilde{\mu}_{1}^{1}$ & $\tilde{\mu}_{2}^{1}$\\\hline
\end{tabular}
=0,\\
\left\{  M%
\begin{tabular}
[c]{|c|c|}\hline
$L$ & $\mu$\\\hline
\end{tabular}
+%
\begin{tabular}
[c]{|c|}\hline
$\mu$\\\hline
\end{tabular}
\right\}  \otimes%
\begin{tabular}
[c]{|c|}\hline
$\tilde{\mu}_{1}^{2}$\\\hline
\end{tabular}
=0,\\%
\begin{tabular}
[c]{|c|c|}\hline
$\mu_{1}^{1}$ & $\mu_{2}^{1}$\\\hline
\end{tabular}
\otimes\left\{  M%
\begin{tabular}
[c]{|c|c|}\hline
$L$ & $\tilde{\mu}$\\\hline
\end{tabular}
+%
\begin{tabular}
[c]{|c|}\hline
$\tilde{\mu}$\\\hline
\end{tabular}
\right\}   &  =0,\\%
\begin{tabular}
[c]{|c|}\hline
$\mu_{1}^{2}$\\\hline
\end{tabular}
\otimes\left\{  M%
\begin{tabular}
[c]{|c|c|}\hline
$L$ & $\tilde{\mu}$\\\hline
\end{tabular}
+%
\begin{tabular}
[c]{|c|}\hline
$\tilde{\mu}$\\\hline
\end{tabular}
\right\}  =0,\\
\left\{
\begin{tabular}
[c]{|c|c|}\hline
$T$ & $T$\\\hline
\end{tabular}
+2M%
\begin{tabular}
[c]{|c|}\hline
$L$\\\hline
\end{tabular}
\right\}  \otimes%
\begin{tabular}
[c]{|c|c|}\hline
$\tilde{\mu}_{1}^{1}$ & $\tilde{\mu}_{2}^{1}$\\\hline
\end{tabular}
&  =0,\\
\left\{
\begin{tabular}
[c]{|c|c|}\hline
$T$ & $T$\\\hline
\end{tabular}
+2M%
\begin{tabular}
[c]{|c|}\hline
$L$\\\hline
\end{tabular}
\right\}  \otimes%
\begin{tabular}
[c]{|c|}\hline
$\tilde{\mu}_{1}^{2}$\\\hline
\end{tabular}
&  =0,\\%
\begin{tabular}
[c]{|c|c|}\hline
$\mu_{1}^{1}$ & $\mu_{2}^{1}$\\\hline
\end{tabular}
\otimes\left\{
\begin{tabular}
[c]{|c|c|}\hline
$T$ & $T$\\\hline
\end{tabular}
+2M%
\begin{tabular}
[c]{|c|}\hline
$L$\\\hline
\end{tabular}
\right\}   &  =0,\\%
\begin{tabular}
[c]{|c|}\hline
$\mu_{1}^{2}$\\\hline
\end{tabular}
\otimes\left\{
\begin{tabular}
[c]{|c|c|}\hline
$T$ & $T$\\\hline
\end{tabular}
+2M%
\begin{tabular}
[c]{|c|}\hline
$L$\\\hline
\end{tabular}
\right\}   &  =0.
\end{align}
The remaining indices $\mu,\tilde{\mu}$ in the above equations can be set to
be $T$ or $L$, and we obtain%
\end{subequations}
\begin{subequations}
\begin{align}
M%
\begin{tabular}
[c]{|c|c|}\hline
$L$ & $L$\\\hline
\end{tabular}
\otimes%
\begin{tabular}
[c]{|c|c|}\hline
$L$ & $L$\\\hline
\end{tabular}
+%
\begin{tabular}
[c]{|c|}\hline
$L$\\\hline
\end{tabular}
\otimes%
\begin{tabular}
[c]{|c|c|}\hline
$L$ & $L$\\\hline
\end{tabular}
&  =0,\label{1}\\
M%
\begin{tabular}
[c]{|c|c|}\hline
$L$ & $L$\\\hline
\end{tabular}
\otimes%
\begin{tabular}
[c]{|c|c|}\hline
$T$ & $T$\\\hline
\end{tabular}
+%
\begin{tabular}
[c]{|c|}\hline
$L$\\\hline
\end{tabular}
\otimes%
\begin{tabular}
[c]{|c|c|}\hline
$T$ & $T$\\\hline
\end{tabular}
&  =0,\label{2}\\
M%
\begin{tabular}
[c]{|c|c|}\hline
$T$ & $L$\\\hline
\end{tabular}
\otimes%
\begin{tabular}
[c]{|c|c|}\hline
$T$ & $L$\\\hline
\end{tabular}
+%
\begin{tabular}
[c]{|c|}\hline
$T$\\\hline
\end{tabular}
\otimes%
\begin{tabular}
[c]{|c|c|}\hline
$T$ & $L$\\\hline
\end{tabular}
&  =0, \label{3}%
\end{align}%
\end{subequations}
\begin{subequations}
\begin{align}
M%
\begin{tabular}
[c]{|c|c|}\hline
$L$ & $L$\\\hline
\end{tabular}
\otimes%
\begin{tabular}
[c]{|c|}\hline
$L$\\\hline
\end{tabular}
+%
\begin{tabular}
[c]{|c|}\hline
$L$\\\hline
\end{tabular}
\otimes%
\begin{tabular}
[c]{|c|}\hline
$L$\\\hline
\end{tabular}
&  =0,\label{4}\\
M%
\begin{tabular}
[c]{|c|c|}\hline
$T$ & $L$\\\hline
\end{tabular}
\otimes%
\begin{tabular}
[c]{|c|}\hline
$T$\\\hline
\end{tabular}
+%
\begin{tabular}
[c]{|c|}\hline
$T$\\\hline
\end{tabular}
\otimes%
\begin{tabular}
[c]{|c|}\hline
$T$\\\hline
\end{tabular}
&  =0, \label{5}%
\end{align}%
\end{subequations}
\begin{subequations}
\begin{align}
M%
\begin{tabular}
[c]{|c|c|}\hline
$L$ & $L$\\\hline
\end{tabular}
\otimes%
\begin{tabular}
[c]{|c|c|}\hline
$L$ & $L$\\\hline
\end{tabular}
+%
\begin{tabular}
[c]{|c|c|}\hline
$L$ & $L$\\\hline
\end{tabular}
\otimes%
\begin{tabular}
[c]{|c|}\hline
$L$\\\hline
\end{tabular}
&  =0,\label{6}\\
M%
\begin{tabular}
[c]{|c|c|}\hline
$T$ & $T$\\\hline
\end{tabular}
\otimes%
\begin{tabular}
[c]{|c|c|}\hline
$L$ & $L$\\\hline
\end{tabular}
+%
\begin{tabular}
[c]{|c|c|}\hline
$T$ & $T$\\\hline
\end{tabular}
\otimes%
\begin{tabular}
[c]{|c|}\hline
$L$\\\hline
\end{tabular}
&  =0,\label{7}\\
M%
\begin{tabular}
[c]{|c|c|}\hline
$T$ & $L$\\\hline
\end{tabular}
\otimes%
\begin{tabular}
[c]{|c|c|}\hline
$T$ & $L$\\\hline
\end{tabular}
+%
\begin{tabular}
[c]{|c|c|}\hline
$T$ & $L$\\\hline
\end{tabular}
\otimes%
\begin{tabular}
[c]{|c|}\hline
$T$\\\hline
\end{tabular}
&  =0, \label{8}%
\end{align}%
\end{subequations}
\begin{subequations}
\begin{align}
M%
\begin{tabular}
[c]{|c|}\hline
$L$\\\hline
\end{tabular}
\otimes%
\begin{tabular}
[c]{|c|c|}\hline
$L$ & $L$\\\hline
\end{tabular}
+%
\begin{tabular}
[c]{|c|}\hline
$L$\\\hline
\end{tabular}
\otimes%
\begin{tabular}
[c]{|c|}\hline
$L$\\\hline
\end{tabular}
&  =0,\label{9}\\
M%
\begin{tabular}
[c]{|c|}\hline
$T$\\\hline
\end{tabular}
\otimes%
\begin{tabular}
[c]{|c|c|}\hline
$T$ & $L$\\\hline
\end{tabular}
+%
\begin{tabular}
[c]{|c|}\hline
$T$\\\hline
\end{tabular}
\otimes%
\begin{tabular}
[c]{|c|}\hline
$T$\\\hline
\end{tabular}
&  =0, \label{10}%
\end{align}%
\end{subequations}
\begin{subequations}
\begin{align}%
\begin{tabular}
[c]{|c|c|}\hline
$T$ & $T$\\\hline
\end{tabular}
\otimes%
\begin{tabular}
[c]{|c|c|}\hline
$L$ & $L$\\\hline
\end{tabular}
+2M%
\begin{tabular}
[c]{|c|}\hline
$L$\\\hline
\end{tabular}
\otimes%
\begin{tabular}
[c]{|c|c|}\hline
$L$ & $L$\\\hline
\end{tabular}
&  =0,\label{11}\\%
\begin{tabular}
[c]{|c|c|}\hline
$T$ & $T$\\\hline
\end{tabular}
\otimes%
\begin{tabular}
[c]{|c|c|}\hline
$T$ & $T$\\\hline
\end{tabular}
+2M%
\begin{tabular}
[c]{|c|}\hline
$L$\\\hline
\end{tabular}
\otimes%
\begin{tabular}
[c]{|c|c|}\hline
$T$ & $T$\\\hline
\end{tabular}
&  =0, \label{12}%
\end{align}%
\end{subequations}
\begin{equation}%
\begin{tabular}
[c]{|c|c|}\hline
$T$ & $T$\\\hline
\end{tabular}
\otimes%
\begin{tabular}
[c]{|c|}\hline
$L$\\\hline
\end{tabular}
+2M%
\begin{tabular}
[c]{|c|}\hline
$L$\\\hline
\end{tabular}
\otimes%
\begin{tabular}
[c]{|c|}\hline
$L$\\\hline
\end{tabular}
=0, \label{13}%
\end{equation}%
\begin{subequations}
\begin{align}%
\begin{tabular}
[c]{|c|c|}\hline
$L$ & $L$\\\hline
\end{tabular}
\otimes%
\begin{tabular}
[c]{|c|c|}\hline
$T$ & $T$\\\hline
\end{tabular}
+2M%
\begin{tabular}
[c]{|c|c|}\hline
$L$ & $L$\\\hline
\end{tabular}
\otimes%
\begin{tabular}
[c]{|c|}\hline
$L$\\\hline
\end{tabular}
&  =0,\label{14}\\%
\begin{tabular}
[c]{|c|c|}\hline
$T$ & $T$\\\hline
\end{tabular}
\otimes%
\begin{tabular}
[c]{|c|c|}\hline
$T$ & $T$\\\hline
\end{tabular}
+2M%
\begin{tabular}
[c]{|c|c|}\hline
$T$ & $T$\\\hline
\end{tabular}
\otimes%
\begin{tabular}
[c]{|c|}\hline
$L$\\\hline
\end{tabular}
&  =0, \label{15}%
\end{align}%
\end{subequations}
\begin{equation}%
\begin{tabular}
[c]{|c|}\hline
$L$\\\hline
\end{tabular}
\otimes%
\begin{tabular}
[c]{|c|c|}\hline
$T$ & $T$\\\hline
\end{tabular}
+2M%
\begin{tabular}
[c]{|c|}\hline
$L$\\\hline
\end{tabular}
\otimes%
\begin{tabular}
[c]{|c|}\hline
$L$\\\hline
\end{tabular}
=0. \label{16}%
\end{equation}
Since the transverse component of the highest spin state $\alpha_{-1}%
^{T}\cdots\alpha_{-1}^{T}\otimes\tilde{\alpha}_{-1}^{T}\cdots\tilde{\alpha
}_{-1}^{T}$ at each fixed mass level gives the leading order scattering
amplitude, there should have even number of $T$ at each fixed mass level. Thus
Eqs.(\ref{3}), (\ref{5}), (\ref{8}) and (\ref{10}) are subleading order in
energy and are therefore irrelevant. Set $%
\begin{tabular}
[c]{|c|c|}\hline
$T$ & $T$\\\hline
\end{tabular}
\otimes%
\begin{tabular}
[c]{|c|c|}\hline
$T$ & $T$\\\hline
\end{tabular}
=1$, we can solve the ratios from the remaining equations. The final result is

\begin{center}
\noindent%
\begin{tabular}
[c]{|c|c|}\hline
$\epsilon_{TT,TT}$ & $1$\\\hline
$\epsilon_{TT,LL}=\epsilon_{LL,TT}$ & $1/\left(  2M^{2}\right)  $\\\hline
$\epsilon_{LL,LL}$ & $1/\left(  4M^{4}\right)  $\\\hline
$\epsilon_{TT,L}=\epsilon_{L,TT}$ & $-1/\left(  2M\right)  $\\\hline
$\epsilon_{LL,L}=\epsilon_{L,LL}$ & $-1/\left(  4M^{3}\right)  $\\\hline
$\epsilon_{L,L}$ & $1/\left(  4M^{2}\right)  $\\\hline
\end{tabular}

\end{center}

\noindent which is exactly the tensor product of two pieces of open string
ratios. This result is consistent with Eq.(\ref{22aa}) calculated from the
decoupling of high energy ZNS in the previous section.

\subsection{Saddle-point calculation}

In this section, we calculate the tree-level high energy closed string
scattering amplitudes for arbitrary mass levels. We first review the
calculation of high energy open string scattering amplitude. The $(s,t)$
channel scattering amplitude with $V_{2}=\alpha_{-1}^{\mu_{1}}\alpha_{-1}%
^{\mu_{2}}..\alpha_{-1}^{\mu_{n}}\mid0,k>$, the highest spin state at mass
level $M^{2}$ $=2(N-1),$ and three tachyons $V_{1,3,4}$ is \cite{CHL}%
\begin{equation}
\mathcal{T}_{N;st}^{\mu_{1}\mu_{2}\cdot\cdot\mu_{n}}%
=\overset{N}{\underset{l=0}{\sum}}(-)^{l}\binom{N}{l}B\left(  -\frac{s}%
{2}-1+l,-\frac{t}{2}-1+N-l\right)  k_{1}^{(\mu_{1}}..k_{1}^{\mu_{n-l}}%
k_{3}^{\mu_{n-l+1}}..k_{3}^{\mu_{N})}, \label{B}%
\end{equation}
where $B(u,v)=\int_{0}^{1}dxx^{u-1}(1-x)^{v-1}$ is the Euler beta function. It
is now easy to calculate the general high energy scattering amplitude at
the$M^{2}$ $=2(N-1)$ level
\begin{equation}
\mathcal{T}_{n}^{TTT\cdot\cdot}(s,t)\simeq\lbrack-2E^{3}\sin\phi_{c.m.}%
]^{N}\mathcal{T}_{N}(s,t) \label{T}%
\end{equation}
where $\mathcal{T}_{N}(s,t)$ is the high energy limit of $\frac{\Gamma
(-\frac{s}{2}-1)\Gamma(-\frac{t}{2}-1)}{\Gamma(\frac{u}{2}+2)}$ with
$s+t+u=2N-8$, and was previously \cite{ChanLee,ChanLee1,CHL} miscalculated to
be%
\begin{align}
\mathcal{\tilde{T}}_{N;st}  &  \simeq\sqrt{\pi}(-1)^{N-1}2^{-N}E^{-1-2N}%
\left(  \sin\frac{\phi_{c.m.}}{2}\right)  ^{-3}\left(  \cos\frac{\phi_{c.m.}%
}{2}\right)  ^{5-2N}\nonumber\\
&  \times\exp\left[  -\frac{s\ln s+t\ln t-(s+t)\ln(s+t)}{2}\right]  \label{st}%
\end{align}
One can now generalize this result to multi-tensors. The $(s,t)$ channel of
open string high energy scattering amplitude at mass level $(N_{1},N_{2}%
,N_{3},N_{4})$ was calculated to be \cite{ChanLee,ChanLee1,CHL}%

\begin{equation}
\mathcal{T}_{N_{1}N_{2}N_{3}N_{4};st}^{T^{1}\cdot\cdot T^{2}\cdot\cdot
T^{3}\cdot\cdot T^{4}\cdot\cdot}=[-2E^{3}\sin\phi_{c.m.}]^{\Sigma N_{i}%
}\mathcal{T}_{\Sigma N_{i}}(s,t). \label{AM}%
\end{equation}
In the above calculations, the scattering angle $\phi_{c.m.}$ in the center of
mass frame is defined to be the angle between $\overrightarrow{k}_{1}$ and
$\overrightarrow{k}_{3}$. $s=-(k_{1}+k_{2})^{2}$, $t=-(k_{2}+k_{3})^{2}$ and
$u=-(k_{1}+k_{3})^{2}$ are the Mandelstam variables. $M_{i}^{2}=2(N_{i}-1)$
with $N_{i}$ the mass level of the $i$th vertex. $T^{i}$ in Eq.(\ref{AM}) is
the transverse polarization of the $i$th vertex defined in Eq.(8). All other
4-point functions at mass level $(N_{1},N_{2},N_{3},N_{4})$ were shown to be
proportional to Eq.(\ref{AM}).

The corresponding $(t,u)$ channel scattering amplitudes of Eqs.(\ref{T}) and
(\ref{AM}) can be obtained by replacing $(s,t)$ in Eq.(\ref{st}) by $(t,u)$
\begin{align}
\mathcal{T}_{N}(t,u)  &  \simeq\sqrt{\pi}(-1)^{N-1}2^{-N}E^{-1-2N}\left(
\sin\frac{\phi_{c.m.}}{2}\right)  ^{-3}\left(  \cos\frac{\phi_{c.m.}}%
{2}\right)  ^{5-2N}\nonumber\\
&  \times\exp\left[  -\frac{t\ln t+u\ln u-(t+u)\ln(t+u)}{2}\right]  .
\label{tu}%
\end{align}

We now claim that only $(t,u)$ channel of the amplitude, Eq.(\ref{tu}), is
suitable for saddle-point calculation. The previous saddle-point calculation
for the $(s,t)$ channel amplitude, Eq.(\ref{st}), in the high energy expansion
is misleading. The corrected high energy calculation of the $(s,t)$ channel
amplitude will be given in Eq.(\ref{corr}). The reason is as following. When
calculating Eq.(\ref{T}) from Eq.(\ref{B}), one calculates the high energy
limit of%
\begin{equation}
\frac{\Gamma(-\frac{s}{2}-1)\Gamma(-\frac{t}{2}-1)}{\Gamma(\frac{u}{2}%
+2)},s+t+u=2N-8,
\end{equation}
in Eq.(\ref{B}) by expanding the $\Gamma$ function with the Stirling formula%
\begin{equation}
\Gamma\left(  x\right)  \sim\sqrt{2\pi}x^{x-1/2}e^{-x}.
\end{equation}
However, the above expansion is not suitable for negative real $x$ as there
are poles for $\Gamma\left(  x\right)  $ at $x=-N$, negative integers.
Unfortunately, our high energy limit
\begin{subequations}
\label{high energy limit}%
\begin{align}
s  &  \sim4E^{2}\gg0,\\
t  &  \sim-4E^{2}\sin^{2}\left(  \frac{\phi_{c.m.}}{2}\right)  \ll0,\\
u  &  \sim-4E^{2}\cos^{2}\left(  \frac{\phi_{c.m.}}{2}\right)  \ll0,
\label{43.}%
\end{align}
contains this dangerous situation in the $(s,t)$ channel calculation of
Eq.(\ref{st}). On the other hand, the corresponding high energy expansion of
$(t,u)$ channel scattering amplitude in Eq.(\ref{tu}) is well defined. Another
evidence for this point is the following. When one uses the saddle point
method to calculate the high energy open string scattering amplitudes in the
$(s,t)$ channel, the saddle-point we identified was \cite{CHL,CHLTY1,CHLTY2}%
\end{subequations}
\begin{equation}
x_{0}=\frac{s}{s+t}=\frac{1}{1-\sin^{2}\left(  \phi/2\right)  }>1, \label{44.}%
\end{equation}
which is out of the integration range $\left(  0,1\right)  $. Therefore, we
can not trust the saddle point calculation for the $(s,t)$ channel scattering
amplitude. On the other hand, the corresponding saddle-point calculation for
the $(t,u)$ channel scattering amplitude is safe since the saddle-point
$x_{0}$ is within the integration range $\left(  1,\infty\right)  $. This
subtle situation becomes crucial and relevant when one tries to calculate the
high energy closed string scatterings amplitude and compare them with the open
string ones.

We now discuss the high energy closed string scattering amplitudes. There
exists a celebrated formula by Kawai, Lewellen and Tye (KLT), which expresses
the relation between tree amplitudes of closed and open string $(\alpha
_{\text{closed}}^{\prime}=4\alpha_{\text{open}}^{\prime}=2)$
\begin{equation}
A_{\text{closed}}^{\left(  4\right)  }\left(  s,t,u\right)  =\sin\left(  \pi
k_{2}\cdot k_{3}\right)  A_{\text{open}}^{\left(  4\right)  }\left(
s,t\right)  \bar{A}_{\text{open}}^{\left(  4\right)  }\left(  t,u\right)
\label{KLT}%
\end{equation}
To calculate the high energy closed string scattering amplitudes, one
encounters the difficulty of calculation of high energy open string amplitude
in the $(s,t)$ channel discussed above. To avoid this difficulty, we can use
the well known formula
\begin{equation}
\Gamma\left(  x\right)  =\frac{\pi}{\sin\left(  \pi x\right)  \Gamma\left(
1-x\right)  }%
\end{equation}
to calculate the large negative $x$ expansion of the $\Gamma$ function. We
first discuss the high energy four-tachyon scattering amplitude which already
existed in the literature. We can express the open string $(s,t)$ channel
amplitude in terms of the $(t,u)$ channel amplitude,
\begin{align}
A_{\text{open}}^{\left(  4\text{-tachyon}\right)  }\left(  s,t\right)   &
=\frac{\Gamma\left(  -\frac{s}{2}-1\right)  \Gamma\left(  -\frac{t}%
{2}-1\right)  }{\Gamma\left(  \frac{u}{2}+2\right)  }\nonumber\\
&  =\frac{\sin\left(  \pi u/2\right)  }{\sin\left(  \pi s/2\right)  }%
\frac{\Gamma\left(  -\frac{t}{2}-1\right)  \Gamma\left(  -\frac{u}%
{2}-1\right)  }{\Gamma\left(  \frac{s}{2}+2\right)  }\nonumber\\
&  \equiv\frac{\sin\left(  \pi u/2\right)  }{\sin\left(  \pi s/2\right)
}A_{\text{open}}^{\left(  4\text{-tachyon}\right)  }\left(  t,u\right)  ,
\label{Tach}%
\end{align}
which we know how to calculate the high energy limit. \ Note that for the
four-tachyon case,$\ \bar{A}_{\text{open}}^{\left(  4\right)  }\left(
t,u\right)  =A_{\text{open}}^{\left(  4\right)  }\left(  t,u\right)  $ in
Eq.(\ref{KLT}). The KLT formula, Eq.(\ref{KLT}), can then be used to express
the closed string four-tachyon scattering amplitude in terms of that of open
string in the $(t,u)$ channel
\begin{equation}
A_{\text{closed}}^{\left(  4\text{-tachyon}\right)  }\left(  s,t,u\right)
=\frac{\sin\left(  \pi t/2\right)  \sin\left(  \pi u/2\right)  }{\sin\left(
\pi s/2\right)  }A_{\text{open}}^{\left(  4\text{-tachyon}\right)  }\left(
t,u\right)  A_{\text{open}}^{\left(  4\text{-tachyon}\right)  }\left(
t,u\right)  .
\end{equation}
The high energy limit of open string four-tachyon amplitude in the $(t,u)$
channel can be easily calculated to be%
\begin{equation}
A_{\text{open}}^{(4-\text{tachyon})}\left(  t,u\right)  \simeq(stu)^{-\frac
{3}{2}}\exp\left(  -\frac{s\ln s+t\ln t+u\ln u}{2}\right)  , \label{open}%
\end{equation}
which gives the corresponding amplitude in the $(s,t)$ channel%
\begin{equation}
A_{\text{open}}^{(4-\text{tachyon})}\left(  s,t\right)  \simeq\frac
{\sin\left(  \pi u/2\right)  }{\sin\left(  \pi s/2\right)  }(stu)^{-\frac
{3}{2}}\exp\left(  -\frac{s\ln s+t\ln t+u\ln u}{2}\right)  \label{sttachyon}%
\end{equation}
The high energy limit of closed string four-tachyon scattering amplitude can
then be calculated, through the KLT formula, to be%
\begin{equation}
A_{\text{closed}}^{(4-\text{tachyon})}\left(  s,t,u\right)  \simeq\frac
{\sin\left(  \pi t/2\right)  \sin\left(  \pi u/2\right)  }{\sin\left(  \pi
s/2\right)  }(stu)^{-3}\exp\left(  -\frac{s\ln s+t\ln t+u\ln u}{4}\right)
\label{new}%
\end{equation}
The exponential factor in Eq.(\ref{open}) was first discussed by Veneziano
\cite{Veneziano}. Our result for the high energy closed string four-tachyon
amplitude in Eq.(\ref{new}) differs from the one calculated in the literature
\cite{GM,GM1} by an oscillating factor $\frac{\sin\left(  \pi t/2\right)
\sin\left(  \pi u/2\right)  }{\sin\left(  \pi s/2\right)  }$. We stress here
that our results for Eqs.(\ref{open}), (\ref{sttachyon}) and (\ref{new}) are
consistent with the KLT formula, while the previous calculation in
\cite{GM,GM1} is NOT.

One might try to use the saddle-point method to calculate the high energy
closed string scattering amplitude. The closed string four-tachyon scattering
amplitude is%
\begin{align}
A_{\text{closed}}^{(4-\text{tachyon})}\left(  s,t,u\right)   &  =\int
dxdy\exp\left(  \frac{k_{1}\cdot k_{2}}{2}\ln\left\vert z\right\vert
+\frac{k_{2}\cdot k_{3}}{2}\ln\left\vert 1-z\right\vert \right) \nonumber\\
&  =\int dxdy(x^{2}+y^{2})^{-2}[(1-x)^{2}+y^{2}]^{-2}\nonumber\\
&  \cdot\exp\left\{  -\frac{s}{8}\ln(x^{2}+y^{2})-\frac{t}{8}\ln
[(1-x)^{2}+y^{2}]\right\} \nonumber\\
&  \equiv\int dxdy(x^{2}+y^{2})^{-2}[(1-x)^{2}+y^{2}]^{-2}\exp\left[
-Kf(x,y)\right]
\end{align}
where $K=\frac{s}{8}$ and $f(x,y)=\ln(x^{2}+y^{2})-\tau\ln[(1-x)^{2}+y^{2}]$
with $\tau=-\frac{t}{s}$. One can then calculate the "saddle-point" of
$\ f(x,y)$ to be%
\begin{equation}
\nabla f(x,y)\mid_{x_{0}=\frac{1}{1-\tau},y_{0}=0}=0.
\end{equation}
The high energy limit of the closed string four-tachyon scattering amplitude
is then calculated to be%
\begin{equation}
A_{\text{closed}}^{(4-\text{tachyon})}\left(  s,t,u\right)  \simeq\frac{2\pi
}{K\sqrt{\det\frac{\partial^{2}f(x_{0},y_{0})}{\partial x\partial y}}}%
\exp[-Kf(x_{0},y_{0})]\simeq(stu)^{-3}\exp\left(  -\frac{s\ln s+t\ln t+u\ln
u}{4}\right)  ,
\end{equation}
which is consistent with the previous one calculated in the literature
\cite{GM,GM1}, but is different from our result in Eq.(\ref{new}). However,
one notes that%
\begin{equation}
\frac{\partial^{2}f(x_{0},y_{0})}{\partial x^{2}}=\frac{2(1-\tau)^{3}}{\tau
}=-\frac{\partial^{2}f(x_{0},y_{0})}{\partial y^{2}},\frac{\partial^{2}%
f(x_{0},y_{0})}{\partial x\partial y}=0, \label{55.}%
\end{equation}
which means that $(x_{0},y_{0})$ is NOT the local minimum of $f(x,y)$, and one
should not trust this saddle-point calculation. This is the third evidence to
see that there is no clear definition of saddle-point in the calculation of
the high energy open string scattering amplitude in the $(s,t)$ channel, and
thus the invalid saddle-point calculation of high energy closed string
scattering amplitude.

Finally we calculate the high energy closed string scattering amplitudes for
arbitrary mass levels. The $(t,u)$ channel open string scattering amplitude
with $V_{2}=\alpha_{-1}^{\mu_{1}}\alpha_{-1}^{\mu_{2}}..\alpha_{-1}^{\mu_{n}%
}\mid0,k>$, the highest spin state at mass level$M^{2}$ $=2(N-1)$, and three
tachyons $V_{1,3,4}$ can be calculated to be%
\begin{equation}
\mathcal{T}_{N;tu}^{\mu_{1}\mu_{2}\cdot\cdot\mu_{n}}%
=\overset{N}{\underset{l=0}{\sum}}\binom{N}{l}B\left(  -\frac{t}%
{2}+N-l-1,-\frac{u}{2}-1\right)  k_{1}^{(\mu_{1}}..k_{1}^{\mu_{N-l}}k_{3}%
^{\mu_{N-l+1}}k_{3}^{\mu_{N})}. \label{ntu.}%
\end{equation}
In calculating Eq.(\ref{ntu.}), we have used the Mobius transformation
$y=\frac{x-1}{x}$ to change the integration region from $\left(
1,\infty\right)  $ to $\left(  0,1\right)  $. One notes that Eq.(\ref{ntu.})
is NOT the same as Eq.(\ref{B}) with $(s,t)$ replaced by $(t,u)$, as one would
have expected from the four-tachyon case discussed in the paragraph after
Eq.(\ref{KLT}) \ In the high energy limit, one easily sees that%
\begin{equation}
\mathcal{T}_{N}(s,t)\simeq(-)^{N}\frac{\sin\left(  \pi u/2\right)  }%
{\sin\left(  \pi s/2\right)  }\mathcal{T}_{N}(t,u), \label{corr}%
\end{equation}
which is the generalization of Eq.(\ref{Tach}) to arbitrary mass levels.
Eq.(\ref{corr}) is part of the string BCJ relations which will be discussed in
the next section. Eq.(\ref{corr}) is the correction of Eqs.(\ref{T}) and
(\ref{st}) as claimed in the paragraph after Eq.(\ref{tu}). The $(s,t)$
channel of high energy open string scattering amplitudes at mass level
$(n_{1},n_{2},n_{3},n_{4})$ can then be written as, apart from an overall
constant,%
\begin{align}
A_{\text{open}}^{\left(  4\right)  }\left(  s,t\right)   &  \simeq(-)^{\Sigma
N_{i}}\frac{\sin\left(  \pi u/2\right)  }{\sin\left(  \pi s/2\right)
}[-2E^{3}\sin\phi_{c.m.}]^{\Sigma N_{i}}\mathcal{T}_{\Sigma N_{i}%
}(t,u)\nonumber\\
&  \simeq(-)^{\Sigma N_{i}}\frac{\sin\left(  \pi u/2\right)  }{\sin\left(  \pi
s/2\right)  }(stu)^{\frac{\Sigma N_{i}-3}{2}}\exp\left(  -\frac{s\ln s+t\ln
t+u\ln u}{2}\right)  . \label{openst}%
\end{align}

Finally the total high energy open string scattering amplitude is the sum of
$\left(  s,t\right)  $, $\left(  t,u\right)  $ and $\left(  u,s\right)  $
channel amplitudes, and can be calculated to be%
\begin{equation}
A_{\text{open}}^{\left(  4\right)  }\simeq(-)^{\Sigma N_{i}}\frac{\sin\left(
\pi s/2\right)  +\sin\left(  \pi t/2\right)  +\sin\left(  \pi u/2\right)
}{\sin\left(  \pi s/2\right)  }(stu)^{\frac{\Sigma N_{i}-3}{2}}\exp\left(
-\frac{s\ln s+t\ln t+u\ln u}{2}\right)  . \label{openall}%
\end{equation}
By using Eqs.(\ref{KLT}) and (\ref{corr}), the high energy closed string
scattering amplitude at mass level $(N_{1},N_{2},N_{3},N_{4})$ is calculated
to be, apart from an overall constant,%
\begin{align}
A_{\text{closed}}^{\left(  4\right)  }\left(  s,t,u\right)   &  \simeq
(-)^{\Sigma N_{i}}\frac{\sin\left(  \pi t/2\right)  \sin\left(  \pi
u/2\right)  }{\sin\left(  \pi s/2\right)  }[-2E^{3}\sin\phi_{c.m.}]^{2\Sigma
N_{i}}\mathcal{T}_{\Sigma N_{i}}(t,u)^{2}\nonumber\\
&  \simeq(-)^{\Sigma N_{i}}\frac{\sin\left(  \pi t/2\right)  \sin\left(  \pi
u/2\right)  }{\sin\left(  \pi s/2\right)  }(stu)^{\Sigma N_{i}-3}\exp\left(
-\frac{s\ln s+t\ln t+u\ln u}{4}\right)  , \label{SA}%
\end{align}
where $\mathcal{T}_{\Sigma N_{i}}(t,u)$ is given by Eq.(\ref{tu}). For the
case of four-tachyon scattering amplitude at mass level $(0,0,0,0)$,
Eq.(\ref{SA}) reduces to Eq.(\ref{new}). All other high energy closed string
scattering amplitudes at mass level $(N_{1},N_{2},N_{3},N_{4})$ are
proportional to Eq.(\ref{SA}). The proportionality constants are the tensor
product of two pieces of open string ratios.

\subsection{String BCJ relations}

In 2008, the four point BCJ relations \cite{BCJ1,BCJ2,BCJ3,BCJ4,BCJ5} for
Yang-Mills gluon color-stripped scattering amplitudes $A$ were pointed out and
calculated to be%
\begin{align}
tA(k_{1},k_{4},k_{2},k_{3})-sA(k_{1},k_{3},k_{4},k_{2})  &  =0,\nonumber\\
sA(k_{1},k_{2},k_{3},k_{4})-uA(k_{1},k_{4},k_{2},k_{3})  &  =0,\nonumber\\
\text{ }uA(k_{1},k_{3},k_{4},k_{2})-tA(k_{1},k_{2},k_{3},k_{4})  &  =0,
\label{BCJ}%
\end{align}
which relates field theory scattering amplitudes in the $s$, $t$ and $u$
channels. In the following, we will discuss the relation for $s$ and $u$
channel amplitudes only. Other relations can be similarly discussed.

For string theory, in contrast to the field theory BCJ relations, one has to
deal with scattering amplitudes of infinite number of string states. For the
tachyon state, the string BCJ relation was first calculated in 2006 to be
Eq.(\ref{Tach}) \cite{Closed} which can be rewritten as%
\begin{equation}
A_{\text{open}}^{\left(  4\text{-tachyon}\right)  }\left(  s,t\right)
\equiv\frac{\sin\left(  \pi k_{2}.k_{4}\right)  }{\sin\left(  \pi k_{1}%
k_{2}\right)  }A_{\text{open}}^{\left(  4\text{-tachyon}\right)  }\left(
t,u\right)  . \label{teq}%
\end{equation}
This relation for tachyon is valid for all energies. For $all$ other higher
spin string states at arbitrary mass levels, the high energy limit of string
BCJ relation was worked out to be Eq.(\ref{corr}) \cite{Closed} and can be
rewritten as%
\begin{equation}
\mathcal{T}_{N}(s,t)\simeq\frac{\sin\left(  \pi k_{2}.k_{4}\right)  }%
{\sin\left(  \pi k_{1}.k_{2}\right)  }\mathcal{T}_{N}(t,u). \label{BCJ2}%
\end{equation}
Note that unlike the case of tachyon in Eq.(\ref{teq}), this relation was
proved only for high energy limit. The result of Eq.(\ref{BCJ2}) was based on
two calculations. The first calculation was done for amplitudes in
Eq.(\ref{ntu.}) and Eq.(\ref{B}). Although the calculations in Eq.(\ref{ntu.})
and Eq.(\ref{B}) were done only for three tachyons and one leading Regge
trajectory higher spin state in the second vertex, it can be easily extended
to three arbitrary string states and one leading Regge trajectory higher spin
state in the high energy limit, and Eq.(\ref{BCJ2}) is still valid. The second
calculation was based on the fact that high energy, fixed angle amplitudes for
states differ from leading Regge trajectory higher spin state in the second
vertex are all proportional to each other at each fixed mass level as were
shown in Eq.(\ref{mainA}).

The two relations in Eq.(\ref{teq}) and Eq.(\ref{BCJ2}) can be written as the
four point \textit{string BCJ relation} which are valid to all energies as%
\begin{equation}
A_{\text{open}}^{(4)}\left(  s,t\right)  =\frac{\sin\left(  \pi k_{2}%
.k_{4}\right)  }{\sin\left(  \pi k_{1}.k_{2}\right)  }\bar{A}_{\text{open}%
}^{\left(  4\right)  }\left(  t,u\right)  \label{BCJ3}%
\end{equation}
if one can generalize the proof of Eq.(\ref{corr}) to all energies. This was
done in a paper based on monodromy of integration for string amplitudes
published in 2009 \cite{BCJ2}. The motivation for the author in \cite{BCJ2} to
calculate Eq.(\ref{BCJ3}) was different from the calculation done in
Eq.(\ref{BCJ2}). It was based on the field theory BCJ relation \cite{BCJ1}. An
explicit proof of Eq.(\ref{BCJ3}) for arbitrary four string states and all
kinematic regimes was given very recently in \cite{LLY1,LLY2}.

Note that for the supersymmetric case, there is no tachyon and the low energy
massless limit of Eq.(\ref{BCJ3}) reproduces the second equation of
Eq.(\ref{BCJ}). Recently the mass level dependent of Eq.(\ref{BCJ3}) was
calculated to be \cite{LLY1,LLY2}%
\begin{equation}
\frac{A_{st}^{(p,r,q)}}{A_{tu}^{(p,r,q)}}=\left(  -1\right)  ^{N}%
\frac{B\left(  -M_{1}M_{2}+1,\frac{M_{1}M_{2}}{2}\right)  }{B\left(
\frac{M_{1}M_{2}}{2},\frac{M_{1}M_{2}}{2}\right)  }\simeq\frac{\sin\pi\left(
k_{2}\cdot k_{4}\right)  }{\sin\pi\left(  k_{1}\cdot k_{2}\right)  }\nonumber
\end{equation}
by taking the \textit{nonrelativistic} limit $|\vec{k_{2}}|<<M_{S}$ of
Eq.(\ref{BCJ3}). In Eq.(\ref{level}), $B$ was the beta function, and $k_{1}$,
$k_{3}$ and $k_{4}$ were taken to be tachyons, and $k_{2}$ was the following
tensor string state%

\begin{equation}
V_{2}=(i\partial X^{T})^{p}(i\partial X^{L})^{r}(i\partial X^{P})^{q}%
e^{ik_{2}X}%
\end{equation}
where%
\begin{equation}
N=p+r+q\text{, \ }M_{2}^{2}=2(N-1).
\end{equation}

The generalization of the four point function relation in Eq.(\ref{BCJ3}) to
higher point string amplitudes can be found in \cite{BCJ2}. It is interesting
to see that historically the four point (high energy) string BCJ relations
Eq.(\ref{Tach}) and Eq.(\ref{corr}) \cite{Closed} were discovered even earlier
than the field theory BCJ relations Eq.(\ref{BCJ})! \cite{BCJ1}.

In conclusion of this chapter, we have used the methods of decoupling of high
energy ZNS and the high energy Virasoro constraints to calculate the ratios
among high energy closed string scattering amplitudes of different string
states. The result is exactly the tensor product of two pieces of open string
ratios calculated before. However, we clarify the previous saddle-point
calculation for high energy open string scattering amplitudes and show that
only $(t,u)$ channel of the amplitudes is suitable for saddle-point
calculation. We also discuss three evidences, Eq.(\ref{43.}), Eq.(\ref{44.})
and Eq.(\ref{55.}), to show that saddle-point calculation for high energy
closed string scattering amplitudes is not reliable. Instead of using
saddle-point calculation adopted before, we then propose to use the formula of
Kawai, Lewellen and Tye (KLT) to calculate the high energy closed string
scattering amplitudes for \textit{arbitrary} mass levels.

For the case of high energy closed string four-tachyon amplitude, our result
differs from the previous one of Gross and Mende, which is NOT consistent with
KLT formula, by an oscillating factor. The oscillating prefactors in
Eqs.(\ref{openall}) and (\ref{SA}) imply the existence of infinitely many
zeros and poles in the string scattering amplitudes even in the high energy
limit. Physically, the presence of poles simply reflects the fact that there
are infinite number of resonances in the string spectrum \cite{GSW}, and the
presence of zeros reflects the coherence of string scattering. In addition,
the oscillating prefactors are crucial to discuss the string BCJ relations.%

\setcounter{equation}{0}
\renewcommand{\theequation}{\arabic{section}.\arabic{equation}}%

\section{Hard superstring scatterings}

In this chapter, we consider high energy scattering amplitudes for the NS
sector of $10D$ open superstring theory \cite{susy}. Based on the calculations
of $26D$ bosonic open string \cite{CHLTY1,CHLTY2,CHLTY3}, all the three
independent calculations of bosonic string, namely the decoupling of high
energy ZNS (HZNS), the Virasoro constraints and the saddle-point calculation
can be generalized to scattering amplitudes of string states with
polarizations on the scattering plane of superstring. All three methods give
the consistent results \cite{susy}.

In addition, we discover new leading order high energy scattering amplitudes,
which are still proportional to the previous ones, with polarizations
\textit{orthogonal} to the scattering plane \cite{susy}. These scattering
amplitudes are of subleading order in energy for the case of $26D$ open
bosonic string theory. The existence of these new high energy scattering
amplitudes is due to the worldsheet fermion exchange in the correlation
functions and is, presumably, related to the high energy massive fermionic
scattering amplitudes in the R-sector of the theory. We thus conjecture that
the validity of Gross's two conjectures on high energy stringy symmetry
persists for superstring theory.

\subsection{Decoupling of high energy ZNS}

We will first consider high energy scattering amplitudes of string states with
polarizations on the scattering plane. Those with polarizations orthogonal to
the scattering plane will be discussed in section VIII.D. It can be argued
that there are four types of high energy scattering amplitudes for states in
the NS sector with even GSO parity \cite{susy}%
\begin{equation}
\left\vert n,2m,q\right\rangle \otimes\left\vert b_{-\frac{1}{2}}%
^{T}\right\rangle \equiv(\alpha_{-1}^{T})^{n-2m-2q}(\alpha_{-1}^{L}%
)^{2m}(\alpha_{-2}^{L})^{q}(b_{-\frac{1}{2}}^{T})\left\vert 0,k\right\rangle ,
\label{T/2}%
\end{equation}%
\begin{equation}
\left\vert n,2m+1,q\right\rangle \otimes\left\vert b_{-\frac{1}{2}}%
^{L}\right\rangle \equiv(\alpha_{-1}^{T})^{n-2m-2q-1}(\alpha_{-1}^{L}%
)^{2m+1}(\alpha_{-2}^{L})^{q}(b_{-\frac{1}{2}}^{L})\left\vert 0,k\right\rangle
, \label{L/2}%
\end{equation}%
\begin{equation}
\left\vert n,2m,q\right\rangle \otimes\left\vert b_{-\frac{3}{2}}%
^{L}\right\rangle \equiv(\alpha_{-1}^{T})^{n-2m-2q}(\alpha_{-1}^{L}%
)^{2m}(\alpha_{-2}^{L})^{q}(b_{-\frac{3}{2}}^{L})\left\vert 0,k\right\rangle ,
\label{L/3}%
\end{equation}%
\begin{equation}
\left\vert n,2m,q\right\rangle \otimes\left\vert b_{-\frac{1}{2}}^{T}%
b_{-\frac{1}{2}}^{L}b_{-\frac{3}{2}}^{L}\right\rangle \equiv(\alpha_{-1}%
^{T})^{n-2m-2q}(\alpha_{-1}^{L})^{2m}(\alpha_{-2}^{L})^{q}(b_{-\frac{1}{2}%
}^{T})(b_{-\frac{1}{2}}^{L})(b_{-\frac{3}{2}}^{L})\left\vert 0,k\right\rangle
\label{TLL/3}%
\end{equation}
Note that the number of $\alpha_{-1}^{L}$ operator in Eq.(\ref{L/2}) is odd.
In the OCFQ spectrum of open superstring, the solutions of physical states
conditions include positive-norm propagating states and two types of ZNS. In
the NS sector, the latter are \cite{GSW}%
\begin{equation}
\text{Type I}:G_{-\frac{1}{2}}\left\vert x\right\rangle ,\text{ where
}G_{\frac{1}{2}}\left\vert x\right\rangle =G_{\frac{3}{2}}\left\vert
x\right\rangle =0,\text{ }L_{0}\left\vert x\right\rangle =0; \label{S1}%
\end{equation}%
\begin{equation}
\text{Type II}:(G_{-\frac{3}{2}}+2G_{-\frac{1}{2}}L_{-1})\left\vert
\widetilde{x}\right\rangle ,\text{ where }G_{\frac{1}{2}}\left\vert
\widetilde{x}\right\rangle =G_{\frac{3}{2}}\left\vert \widetilde{x}%
\right\rangle =0,\text{ }(L_{0}+1)\left\vert \widetilde{x}\right\rangle =0.
\label{S2}%
\end{equation}
While Type I states have zero-norm at any space-time dimension, Type II states
have zero-norm \emph{only} at $D=10$. We will show that, for each fixed mass
level, all high energy scattering amplitudes corresponding to states in
Eqs.(\ref{T/2})-(\ref{TLL/3}) are proportional to each other, and the
proportionality constants can be determined from the decoupling of two types
of ZNS, Eqs.(\ref{S1}) and (\ref{S2}) in the high energy limit. For
simplicity, based on the result of Eq.(\ref{mainA}), one needs only calculate
the proportionality constants among the scattering amplitudes of the following
four lower mass level states%
\begin{equation}
\left\vert 2,0,0\right\rangle \otimes\left\vert b_{-\frac{1}{2}}%
^{T}\right\rangle \equiv(\alpha_{-1}^{T})^{2}(b_{-\frac{1}{2}}^{T})\left\vert
0,k\right\rangle , \label{fT/2}%
\end{equation}%
\begin{equation}
\left\vert 2,1,0\right\rangle \otimes\left\vert b_{-\frac{1}{2}}%
^{L}\right\rangle \equiv(\alpha_{-1}^{T})(\alpha_{-1}^{L})(b_{-\frac{1}{2}%
}^{L})\left\vert 0,k\right\rangle , \label{fL/2}%
\end{equation}%
\begin{equation}
\left\vert 1,0,0\right\rangle \otimes\left\vert b_{-\frac{3}{2}}%
^{L}\right\rangle \equiv(\alpha_{-1}^{T})(b_{-\frac{3}{2}}^{L})\left\vert
0,k\right\rangle , \label{fL/3}%
\end{equation}%
\begin{equation}
\left\vert 0,0,0\right\rangle \otimes\left\vert b_{-\frac{1}{2}}^{T}%
b_{-\frac{1}{2}}^{L}b_{-\frac{3}{2}}^{L}\right\rangle \equiv(b_{-\frac{1}{2}%
}^{T})(b_{-\frac{1}{2}}^{L})(b_{-\frac{3}{2}}^{L})\left\vert 0,k\right\rangle
\label{fTLL/3}%
\end{equation}
Other proportionality constants for higher mass level can be obtained through
Eqs.(\ref{mainA}) and (\ref{fT/2})-(\ref{fTLL/3}). To calculate the ratio
among the high energy scattering amplitudes corresponding to states in
Eqs.(\ref{fL/2}) and (\ref{fL/3}), we use the decoupling of the Type I HZNS at
mass level $M^{2}=2$
\begin{equation}
G_{-\frac{1}{2}}(\alpha_{-1}^{L})\left\vert 0,k\right\rangle =[M(\alpha
_{-1}^{L})(b_{-\frac{1}{2}}^{L})+(b_{-\frac{3}{2}}^{L})]\left\vert
0,k\right\rangle . \label{22}%
\end{equation}
Eq.(\ref{22}) gives the ratio for states at mass level $M^{2}=4$
\begin{equation}
(\alpha_{-1}^{T})(b_{-\frac{3}{2}}^{L})\left\vert 0,k\right\rangle
:(\alpha_{-1}^{T})(\alpha_{-1}^{L})(b_{-\frac{1}{2}}^{L})\left\vert
0,k\right\rangle =M:-1. \label{23..}%
\end{equation}
We have used an abbreviated notation for the scattering amplitudes on the
l.h.s. of Eq.(\ref{23..}). The HZNS in Eq.(\ref{22}) is the high energy limit
of the vector ZNS at mass level $M^{2}=2$%
\begin{equation}
G_{-\frac{1}{2}}\left\vert x\right\rangle =[k_{(\mu}\theta_{\nu)}\alpha
_{-1}^{\mu}b_{-\frac{1}{2}}^{\nu}+\theta\cdot b_{-\frac{3}{2}}]\left\vert
0,k\right\rangle , \label{24}%
\end{equation}
where%
\begin{equation}
\left\vert x\right\rangle =[\theta\cdot\alpha_{-1}+\frac{1}{2}k\cdot
b_{-\frac{1}{2}}\theta\cdot b_{-\frac{1}{2}}]\left\vert 0,k\right\rangle
,k\cdot\theta=0 \label{25}%
\end{equation}
In fact, in the high energy limit, $\theta=$ $e^{L},$ so $\left\vert
x\right\rangle \rightarrow(\alpha_{-1}^{L})\left\vert 0,k\right\rangle $ and
Eq.(\ref{24}) reduces to Eq.(\ref{22}). To calculate the ratio among the high
energy scattering amplitudes corresponding to states in Eqs.(\ref{fT/2}) and
(\ref{fL/3}), we use the decoupling of the Type II HZNS at mass level
$M^{2}=4$
\begin{equation}
G_{-\frac{3}{2}}(\alpha_{-1}^{T})\left\vert 0,k\right\rangle =[M(\alpha
_{-1}^{T})(b_{-\frac{3}{2}}^{L})+(\alpha_{-1}^{T})^{2}(b_{-\frac{1}{2}}%
^{T})]\left\vert 0,k\right\rangle . \label{26}%
\end{equation}
Eq.(\ref{26}) gives the ratio%
\begin{equation}
(\alpha_{-1}^{T})(b_{-\frac{3}{2}}^{L})\left\vert 0,k\right\rangle
:(\alpha_{-1}^{T})^{2}(b_{-\frac{1}{2}}^{T})\left\vert 0,k\right\rangle =1:-M.
\label{27.}%
\end{equation}
Finally, To calculate the ratio among the high energy scattering amplitudes
corresponding to states in Eqs.(\ref{fT/2}) and (\ref{fTLL/3}), we use the
decoupling of the Type II HZNS at mass level $M^{2}=4$
\begin{equation}
G_{-\frac{3}{2}}(b_{-\frac{1}{2}}^{T})(b_{-\frac{1}{2}}^{L})\left\vert
0,k\right\rangle \equiv\lbrack M(b_{-\frac{1}{2}}^{T})(b_{-\frac{1}{2}}%
^{L})(b_{-\frac{3}{2}}^{L})+(\alpha_{-2}^{L})(b_{-\frac{1}{2}}^{T})]\left\vert
0,k\right\rangle . \label{28.}%
\end{equation}
Eq.(\ref{28.}) gives the ratio%
\begin{equation}
(b_{-\frac{1}{2}}^{T})(b_{-\frac{1}{2}}^{L})(b_{-\frac{3}{2}}^{L})\left\vert
0,k\right\rangle :(\alpha_{-2}^{L})(b_{-\frac{1}{2}}^{T})\left\vert
0,k\right\rangle =1:-M. \label{29}%
\end{equation}
On the other hand, Eq.(\ref{L2.}) gives%
\begin{equation}
(\alpha_{-2}^{L})(b_{-\frac{1}{2}}^{T})\left\vert 0,k\right\rangle
:(\alpha_{-1}^{T})^{2}(b_{-\frac{1}{2}}^{T})\left\vert 0,k\right\rangle
=1:-2M. \label{30}%
\end{equation}
We conclude that%
\begin{equation}
(b_{-\frac{1}{2}}^{T})(b_{-\frac{1}{2}}^{L})(b_{-\frac{3}{2}}^{L})\left\vert
0,k\right\rangle :(\alpha_{-1}^{T})^{2}(b_{-\frac{1}{2}}^{T})\left\vert
0,k\right\rangle =1:2M^{2}. \label{31}%
\end{equation}
Eqs.(\ref{23..}),(\ref{27.}) and (\ref{31}) give the proportionality constants
among high energy scattering amplitudes corresponding to states in
Eqs.(\ref{fT/2})-(\ref{fTLL/3}). Finally, by using Eq.(\ref{mainA}), one can
then easily calculate the proportionality constants among high energy
scattering amplitudes corresponding to states in Eqs.(\ref{T/2})-(\ref{TLL/3}%
). The results will be presented in Eqs.(\ref{SS1})-(\ref{SS4}) of chapter XII

\subsection{Virasoro constraints}

In this section, we will use the method of Virasoro constrains to derive the
ratios between the physical states in the NS sector. In the superstring
theory, the physical state $\left\vert \phi\right\rangle $ in the NS sector
should satisfy the following conditions:
\begin{align}
\left(  L_{0}-\dfrac{1}{2}\right)  \left\vert \phi\right\rangle  &  =0,\\
L_{m}\left\vert \phi\right\rangle  &  =0\text{, }m=1,2,3,\cdots,\label{Lm}\\
G_{r}\left\vert \phi\right\rangle  &  =0\text{, }r=\dfrac{1}{2},\dfrac{3}%
{2},\dfrac{5}{2},\cdots, \label{Gr}%
\end{align}
where the $L_{m}$ and $G_{r}$ are super Virasoro operators in the NS sector,%
\begin{align}
L_{m}  &  =\frac{1}{2}\sum_{n}:\alpha_{m-n}\cdot\alpha_{n}:+\frac{1}{4}%
\sum_{r}\left(  2r-m\right)  :\psi_{m-r}\cdot\psi_{r}:,\\
G_{r}  &  =\sum_{n}\alpha_{n}\cdot\psi_{r-n}.
\end{align}
These super Virasoro operators satisfy the following superconformal algebra,%
\begin{align}
\left[  L_{m},L_{n}\right]   &  =\left(  m-n\right)  L_{m+n}+\dfrac{1}%
{8}D\left(  m^{3}-m\right)  \delta_{m+n},\nonumber\\
\left[  L_{m},G_{r}\right]   &  =\left(  \dfrac{1}{2}m-r\right)
G_{m+r},\nonumber\\
\left\{  G_{r},G_{s}\right\}   &  =2L_{r+s}+\dfrac{1}{2}D\left(  r^{2}%
-\dfrac{1}{4}\right)  \delta_{r+s}.
\end{align}
Using the above superconformal algebra, the Virasoro conditions (\ref{Lm}) and
(\ref{Gr}) reduce to the following simple form,%
\begin{align}
G_{1/2}\left\vert \phi\right\rangle  &  =0,\label{G1/2}\\
G_{3/2}\left\vert \phi\right\rangle  &  =0. \label{G3/2}%
\end{align}
In the following, we will use the reduced Virasoro conditions (\ref{G1/2}) and
(\ref{G3/2}) to determine the ratios between the physical states in the NS
sector in the high energy limit.

To warm up, let us consider the mass level at $M^{2}=2$ first. The most
general state in the NS sector at this mass level can be written as%
\begin{equation}
\left\vert 2\right\rangle =\left\{
\begin{tabular}
[c]{|c|}\hline
$\mu$\\\hline
\end{tabular}
\psi_{-\frac{3}{2}}^{\mu}+%
\begin{tabular}
[c]{|c|}\hline
$\mu$\\\hline
\end{tabular}
\otimes%
\begin{tabular}
[c]{|c|}\hline
$\nu$\\\hline
\end{tabular}
\alpha_{-1}^{\mu}\psi_{-\frac{1}{2}}^{\nu}+%
\begin{tabular}
[c]{|c|}\hline
$\mu$\\\hline
$\nu$\\\hline
$\sigma$\\\hline
\end{tabular}
\psi_{-\frac{1}{2}}^{\mu}\psi_{-\frac{1}{2}}^{\nu}\psi_{-\frac{1}{2}}^{\sigma
}\right\}  \left\vert 0\right\rangle _{NS}, \label{|2>}%
\end{equation}
where we use the Young tableaux to represent the coefficients of different
tensors. The properties of symmetry and anti-symmetry can be easily and
clearly described in this representation.

We then apply the reduced Virasoro conditions (\ref{G1/2}) and (\ref{G3/2}) to
the state (\ref{|2>}). It is easy to obtain%
\begin{subequations}%
\begin{align}
G_{1/2}\left\vert 2\right\rangle  &  =\alpha_{-1}^{\mu}\left\{
\begin{tabular}
[c]{|c|}\hline
$\mu$\\\hline
\end{tabular}
+k^{\nu}%
\begin{tabular}
[c]{|c|}\hline
$\mu$\\\hline
\end{tabular}
\otimes%
\begin{tabular}
[c]{|c|}\hline
$\nu$\\\hline
\end{tabular}
\right\}  +\psi_{-\frac{1}{2}}^{\mu}\psi_{-\frac{1}{2}}^{\nu}\left\{
\begin{tabular}
[c]{|c|}\hline
$\mu$\\\hline
\end{tabular}
\otimes%
\begin{tabular}
[c]{|c|}\hline
$\nu$\\\hline
\end{tabular}
-%
\begin{tabular}
[c]{|c|}\hline
$\nu$\\\hline
\end{tabular}
\otimes%
\begin{tabular}
[c]{|c|}\hline
$\mu$\\\hline
\end{tabular}
+3k^{\sigma}%
\begin{tabular}
[c]{|c|}\hline
$\mu$\\\hline
$\nu$\\\hline
$\sigma$\\\hline
\end{tabular}
\right\}  ,\\
G_{3/2}\left\vert 2\right\rangle  &  =%
\begin{tabular}
[c]{|c|}\hline
$\mu$\\\hline
\end{tabular}
k^{\mu}+%
\begin{tabular}
[c]{|c|}\hline
$\mu$\\\hline
\end{tabular}
\otimes%
\begin{tabular}
[c]{|c|}\hline
$\nu$\\\hline
\end{tabular}
\eta^{\mu\nu},
\end{align}%
\end{subequations}%
which leads to the following equations%
\begin{subequations}%
,%
\begin{align}%
\begin{tabular}
[c]{|c|}\hline
$\mu$\\\hline
\end{tabular}
+k^{\nu}%
\begin{tabular}
[c]{|c|}\hline
$\mu$\\\hline
\end{tabular}
\otimes%
\begin{tabular}
[c]{|c|}\hline
$\nu$\\\hline
\end{tabular}
&  =0,\\%
\begin{tabular}
[c]{|c|}\hline
$\mu$\\\hline
\end{tabular}
\otimes%
\begin{tabular}
[c]{|c|}\hline
$\nu$\\\hline
\end{tabular}
-%
\begin{tabular}
[c]{|c|}\hline
$\nu$\\\hline
\end{tabular}
\otimes%
\begin{tabular}
[c]{|c|}\hline
$\mu$\\\hline
\end{tabular}
+3k^{\sigma}%
\begin{tabular}
[c]{|c|}\hline
$\mu$\\\hline
$\nu$\\\hline
$\sigma$\\\hline
\end{tabular}
&  =0,\\%
\begin{tabular}
[c]{|c|}\hline
$\mu$\\\hline
\end{tabular}
k^{\mu}+%
\begin{tabular}
[c]{|c|}\hline
$\mu$\\\hline
\end{tabular}
\otimes%
\begin{tabular}
[c]{|c|}\hline
$\nu$\\\hline
\end{tabular}
\eta^{\mu\nu}  &  =0.
\end{align}%
\end{subequations}%
To solve the above equation, we first take the high energy limit by letting
$\mu\rightarrow\left(  L,T\right)  $ and%
\begin{equation}
k^{\mu}\rightarrow M\left(  e^{L}\right)  ^{\mu}\text{, }\eta^{\mu\nu
}\rightarrow\left(  e^{T}\right)  ^{\mu}\left(  e^{T}\right)  ^{\nu}.
\end{equation}
The above equations reduce to%
\begin{align}%
\begin{tabular}
[c]{|c|}\hline
$\mu$\\\hline
\end{tabular}
+M%
\begin{tabular}
[c]{|c|}\hline
$\mu$\\\hline
\end{tabular}
\otimes%
\begin{tabular}
[c]{|c|}\hline
$L$\\\hline
\end{tabular}
&  =0,\\%
\begin{tabular}
[c]{|c|}\hline
$\mu$\\\hline
\end{tabular}
\otimes%
\begin{tabular}
[c]{|c|}\hline
$\nu$\\\hline
\end{tabular}
-%
\begin{tabular}
[c]{|c|}\hline
$\nu$\\\hline
\end{tabular}
\otimes%
\begin{tabular}
[c]{|c|}\hline
$\mu$\\\hline
\end{tabular}
&  =0,\\
M%
\begin{tabular}
[c]{|c|}\hline
$L$\\\hline
\end{tabular}
+%
\begin{tabular}
[c]{|c|}\hline
$T$\\\hline
\end{tabular}
\otimes%
\begin{tabular}
[c]{|c|}\hline
$T$\\\hline
\end{tabular}
&  =0.
\end{align}
At this mass level, the terms with odd number of $T$'s will be sub-leading in
the high energy limit and be ignored, the resulting equations contain only
terms will even number of $T$' as following,
\begin{align}%
\begin{tabular}
[c]{|c|}\hline
$L$\\\hline
\end{tabular}
+M%
\begin{tabular}
[c]{|c|}\hline
$L$\\\hline
\end{tabular}
\otimes%
\begin{tabular}
[c]{|c|}\hline
$L$\\\hline
\end{tabular}
&  =0,\\
M%
\begin{tabular}
[c]{|c|}\hline
$L$\\\hline
\end{tabular}
+%
\begin{tabular}
[c]{|c|}\hline
$T$\\\hline
\end{tabular}
\otimes%
\begin{tabular}
[c]{|c|}\hline
$T$\\\hline
\end{tabular}
&  =0.
\end{align}
The ratio of the coefficients then can be obtained as%

\begin{equation}%
\begin{tabular}
[c]{|l|c|}\hline
$\varepsilon_{TT}$ & $M^{2}\left(  =2\right)  $\\\hline
$\varepsilon_{LL}$ & $1$\\\hline
$\varepsilon_{L}$ & $-M\left(  =-\sqrt{2}\right)  $\\\hline
\end{tabular}
. \label{ratio.}%
\end{equation}
Now we will consider the general mass level at $M^{2}=2(N-1)$. At this mass
level, the most general state can be written as%
\begin{equation}
\left\vert N\right\rangle =\sum_{\left\{  m_{j},m_{r}\right\}  }\left[
\overset{N}{\underset{j=1}{\otimes}}\frac{1}{j^{m_{j}}m_{j}!}%
\begin{tabular}
[c]{|c|c|c|}\hline
$\mu_{1}^{j}$ & $\cdots$ & $\mu_{m_{j}}^{j}$\\\hline
\end{tabular}
\alpha_{-j}^{\mu_{1}^{j}\cdots\mu_{m_{j}}^{j}}%
\overset{N-1/2}{\underset{r=1/2}{\otimes}}\frac{1}{m_{r}!}%
\begin{tabular}
[c]{|c|c|c|}\hline
$\nu_{1}^{r}$ & $\cdots$ & $\nu_{m_{r}}^{r}$\\\hline
\end{tabular}
^{T}\psi_{-r}^{\nu_{1}^{r}\cdots\nu_{m_{r}}^{r}}\right]  \left\vert
0,k\right\rangle , \label{|n>}%
\end{equation}
where%
\begin{equation}%
\begin{tabular}
[c]{|c|c|c|}\hline
$\nu_{1}^{r}$ & $\cdots$ & $\nu_{m_{r}}^{r}$\\\hline
\end{tabular}
^{T}=%
\begin{tabular}
[c]{|c|}\hline
$\nu_{1}^{r}$\\\hline
$\vdots$\\\hline
$\nu_{m_{r}}^{r}$\\\hline
\end{tabular}
,
\end{equation}
and we have defined the abbreviation%
\begin{equation}
\alpha_{-j}^{\mu_{1}^{j}\cdots\mu_{m_{j}}^{j}}\equiv\alpha_{-j}^{\mu_{1}^{j}%
}\cdots\alpha_{-j}^{\mu_{m_{j}}^{j}}\text{ and }\psi_{-r}^{\nu_{1}^{r}%
\cdots\nu_{m_{r}}^{r}}\equiv\psi_{-r}^{\nu_{1}^{r}}\cdots\psi_{-r}^{\nu
_{m_{r}}^{r}},
\end{equation}
with $m_{j}\left(  m_{r}\right)  $ is the number of the operator $\alpha
_{-j}^{\mu}\left(  \psi_{-r}^{\nu}\right)  $ for $j\in Z$ and $r\in Z+1/2$.
The summation runs over all possible $m_{j}\left(  m_{r}\right)  $ with the
constrain%
\begin{equation}
\sum_{j=1}^{N}jm_{j}+\sum_{r=1/2}^{N-1/2}rm_{r}=N-\frac{1}{2}\text{ with
}m_{j},m_{r}\geq0,
\end{equation}
so that the total mass square is $2\left(  N-1\right)  $.

Solving the constraints (\ref{G1/2}) and (\ref{G3/2}) in the high energy
limit, the ratios between the physical states in the NS sector are obtained as
(see Appendix \ref{Virasoro}\ for detail)%
\begin{align}
&  \underset{N-2m_{2}-2k}{\underbrace{%
\begin{tabular}
[c]{|l|l|l|}\hline
$T$ & $\cdots$ & $T$\\\hline
\end{tabular}
}}\underset{2k}{\underbrace{%
\begin{tabular}
[c]{|l|l|l|}\hline
$L$ & $\cdots$ & $L$\\\hline
\end{tabular}
}}\otimes\underset{m_{2}}{\underbrace{%
\begin{tabular}
[c]{|l|l|l|}\hline
$L$ & $\cdots$ & $L$\\\hline
\end{tabular}
}}\otimes%
\begin{tabular}
[c]{|c|}\hline
$0$\\\hline
\end{tabular}
\otimes%
\begin{tabular}
[c]{|c|}\hline
$L$\\\hline
\end{tabular}
\nonumber\\
&  =\left(  -\frac{1}{2M}\right)  ^{m_{2}}\left(  -\frac{1}{2M}\right)
^{k}\frac{\left(  2k-1\right)  !!}{\left(  -M\right)  ^{k}}%
\underset{N}{\underbrace{%
\begin{tabular}
[c]{|l|l|l|}\hline
$T$ & $\cdots$ & $T$\\\hline
\end{tabular}
}}\otimes%
\begin{tabular}
[c]{|c|}\hline
$0$\\\hline
\end{tabular}
\otimes%
\begin{tabular}
[c]{|c|}\hline
$0$\\\hline
\end{tabular}
\otimes%
\begin{tabular}
[c]{|c|}\hline
$L$\\\hline
\end{tabular}
,
\end{align}%
\begin{align}
&  \underset{N-2m_{2}-2k}{\underbrace{%
\begin{tabular}
[c]{|l|l|l|}\hline
$T$ & $\cdots$ & $T$\\\hline
\end{tabular}
}}\underset{2k+1}{\underbrace{%
\begin{tabular}
[c]{|l|l|l|}\hline
$L$ & $\cdots$ & $L$\\\hline
\end{tabular}
}}\otimes\underset{m_{2}}{\underbrace{%
\begin{tabular}
[c]{|l|l|l|}\hline
$L$ & $\cdots$ & $L$\\\hline
\end{tabular}
}}\otimes%
\begin{tabular}
[c]{|c|}\hline
$L$\\\hline
\end{tabular}
\otimes%
\begin{tabular}
[c]{|c|}\hline
$0$\\\hline
\end{tabular}
\nonumber\\
&  =\left(  -\frac{1}{2M}\right)  ^{m_{2}}\left(  -\frac{1}{2M}\right)
^{k}\frac{\left(  2k+1\right)  !!}{\left(  -M\right)  ^{k+1}}%
\underset{N}{\underbrace{%
\begin{tabular}
[c]{|l|l|l|}\hline
$T$ & $\cdots$ & $T$\\\hline
\end{tabular}
}}\otimes%
\begin{tabular}
[c]{|c|}\hline
$0$\\\hline
\end{tabular}
\otimes%
\begin{tabular}
[c]{|c|}\hline
$0$\\\hline
\end{tabular}
\otimes%
\begin{tabular}
[c]{|c|}\hline
$L$\\\hline
\end{tabular}
,
\end{align}%
\begin{align}
&  \underset{N-2m_{2}-2k+1}{\underbrace{%
\begin{tabular}
[c]{|l|l|l|}\hline
$T$ & $\cdots$ & $T$\\\hline
\end{tabular}
}}\underset{2k}{\underbrace{%
\begin{tabular}
[c]{|l|l|l|}\hline
$L$ & $\cdots$ & $L$\\\hline
\end{tabular}
}}\otimes\underset{m_{2}}{\underbrace{%
\begin{tabular}
[c]{|l|l|l|}\hline
$L$ & $\cdots$ & $L$\\\hline
\end{tabular}
}}\otimes%
\begin{tabular}
[c]{|c|}\hline
$T$\\\hline
\end{tabular}
\otimes%
\begin{tabular}
[c]{|c|}\hline
$0$\\\hline
\end{tabular}
\nonumber\\
&  =\left(  -\frac{1}{2M}\right)  ^{m_{2}}\left(  -\frac{1}{2M}\right)
^{k}\frac{\left(  2k-1\right)  !!}{\left(  -M\right)  ^{k-1}}%
\underset{N}{\underbrace{%
\begin{tabular}
[c]{|l|l|l|}\hline
$T$ & $\cdots$ & $T$\\\hline
\end{tabular}
}}\otimes%
\begin{tabular}
[c]{|c|}\hline
$0$\\\hline
\end{tabular}
\otimes%
\begin{tabular}
[c]{|c|}\hline
$0$\\\hline
\end{tabular}
\otimes%
\begin{tabular}
[c]{|c|}\hline
$L$\\\hline
\end{tabular}
,
\end{align}%
\begin{align}
&  \underset{N-2m_{2}-2k+1}{\underbrace{%
\begin{tabular}
[c]{|l|l|l|}\hline
$T$ & $\cdots$ & $T$\\\hline
\end{tabular}
}}\underset{2k}{\underbrace{%
\begin{tabular}
[c]{|l|l|l|}\hline
$L$ & $\cdots$ & $L$\\\hline
\end{tabular}
}}\otimes\underset{m_{2}-1}{\underbrace{%
\begin{tabular}
[c]{|l|l|l|}\hline
$L$ & $\cdots$ & $L$\\\hline
\end{tabular}
}}\otimes%
\begin{tabular}
[c]{|c|c|}\hline
$T$ & $L$\\\hline
\end{tabular}
^{T}\otimes%
\begin{tabular}
[c]{|c|}\hline
$L$\\\hline
\end{tabular}
\nonumber\\
&  =\left(  -\frac{1}{2M}\right)  ^{m_{2}}\left(  -\frac{1}{2M}\right)
^{k}\frac{\left(  2k-1\right)  !!}{\left(  -M\right)  ^{k}}%
\underset{N}{\underbrace{%
\begin{tabular}
[c]{|l|l|l|}\hline
$T$ & $\cdots$ & $T$\\\hline
\end{tabular}
}}\otimes%
\begin{tabular}
[c]{|c|}\hline
$0$\\\hline
\end{tabular}
\otimes%
\begin{tabular}
[c]{|c|}\hline
$0$\\\hline
\end{tabular}
\otimes%
\begin{tabular}
[c]{|c|}\hline
$L$\\\hline
\end{tabular}
,
\end{align}
which are exactly consistent with the results obtained by using the decoupling
of high energy ZNS in the previous section and the saddle-point calculation
which will be discussed in the following section.

\subsection{Saddle-point approximation}

In this section, we shall calculate the high energy limits of various
scattering amplitudes based on saddle-point approximation. Since the
decoupling of ZNS holds true for arbitrary physical processes, in order to
check the ratios among scattering amplitudes at the same mass level, it is
helpful to choose low-lying states to simplify calculations. For instance, in
the case of 4-point amplitudes, we fix the first vertex to be a $M^{2}=0$
photon with polarization vector $\epsilon^{\mu}$ (in the $-1$ ghost picture,
and $\phi$ is the bosonized ghost operator),
\begin{equation}
V_{1}\equiv\epsilon^{\mu}\psi_{\mu}e^{-\phi}e^{ik_{1}X_{1}},\hspace
{1cm}\epsilon\cdot k_{1}=k_{1}^{2}=0; \label{63}%
\end{equation}
and the third and fourth vertices to be $M^{2}=-1$ tachyon (in the $0$ ghost
picture),
\begin{equation}
V_{3,4}\equiv k_{3,4}^{\mu}\psi_{\mu}e^{ik_{3,4}X_{3,4}},\hspace{1cm}%
k_{3,4}^{2}=-1. \label{64}%
\end{equation}
We shall vary the second vertex at the same level and compare the scattering
amplitudes to obtain the proportional constants.

\subsubsection{$M^{2}=2$}

The second vertex operators at mass level $M^{2}=2$, are given by (in the $-1$
ghost picture),
\begin{align}
(\alpha_{-1}^{T})(b_{-\frac{1}{2}}^{T})\left\vert 0,k\right\rangle  &
\Rightarrow\psi^{T}\partial X^{T}e^{-\phi}e^{ikX},\label{65}\\
(\alpha_{-1}^{L})(b_{-\frac{1}{2}}^{L})\left\vert 0,k\right\rangle  &
\Rightarrow\psi^{L}\partial X^{L}e^{-\phi}e^{ikX},\label{66}\\
(b_{-\frac{3}{2}}^{L})\left\vert 0,k\right\rangle  &  \Rightarrow\partial
\psi^{L}e^{-\phi}e^{ikX}. \label{67}%
\end{align}
Here we have used the polarization basis to specify the particle spins,
e.g.,$\psi^{T}\equiv e_{\mu}^{T}\cdot\psi^{\mu}$.

To illustrate the procedure, we take the first state, Eq.(\ref{65}), as an
example to calculate the scattering amplitude among one massive tensor
$(M^{2}=2)$ with one photon $(V_{1})$ and two tachyons $(V_{3},V_{4})$. As in
the case of open bosonic string theory, we list the contributions of $s-t$
channel only. The 4-point function is given by
\begin{equation}
\int_{0}^{1}dx_{2}\langle(\psi_{1}^{T_{1}}e^{-\phi_{1}}e^{ik_{1}X_{1}}%
)(\psi_{2}^{T_{2}}\partial X_{2}^{T_{2}}e^{-\phi_{2}}e^{ik_{2}X_{2}%
})(k_{3\lambda}\psi_{3}^{\lambda}e^{ik_{3}X_{3}})(k_{4\sigma}\psi_{4}^{\sigma
}e^{ik_{4}X_{4}})\rangle, \label{68}%
\end{equation}
where we have suppressed the $SL(2,R)$ gauge-fixed world-sheet coordinates
$x_{1}=0,x_{3}=1,x_{4}=\infty$. Notice that in both the first and second
vertices, it is possible to allow fermion operators $\psi^{\mu}$ to have
polarization in transverse direction $T_{i}$ out of the scattering plane. As
we shall see in next section that this leads to a new feature of
supersymmetric stringy amplitudes in the high energy limit. At this moment, we
only choose the polarization vector to be in the $P,L,T$ directions for a
comparison with results obtained by the previous two methods.

A direct application of Wick contraction among fermions $\psi$, ghosts $\phi$,
and bosons $X$ leads to the following result
\begin{equation}
\int_{0}^{1}dx\left[  \frac{(3,4)(e^{T_{1}}\cdot e^{T_{2}})}{x}-(e^{T_{1}%
}\cdot k_{3})(e^{T_{2}}\cdot k_{4})+\frac{(e^{T_{2}}\cdot k_{3})(e^{T_{1}%
}\cdot k_{4})}{1-x}\right]  \frac{1}{x}\left[  \frac{e^{T_{2}}\cdot k_{3}%
}{1-x}\right]  x^{(1,2)}(1-x)^{(2,3)}, \label{69}%
\end{equation}
where we have used the short-hand notation, $(3,4)\equiv k_{3}\cdot k_{4}$.
Based on the kinematic variables and the master formula for saddle-point
approximation,
\begin{align}
&  \int dx\hspace{0.3cm}u(x)\exp^{-Kf(x)}\nonumber\\
&  =u_{0}e^{-Kf_{0}}\sqrt{\frac{2\pi}{Kf_{0}^{\prime\prime}}}\left\{
1+\left[  \frac{u_{0}^{\prime\prime}}{2u_{0}f_{0}^{\prime\prime}}-\frac
{u_{0}^{\prime}f^{(3)}}{2u_{0}(f_{0}^{\prime\prime})^{2}}-\frac{f_{0}^{(4)}%
}{8(f_{0}^{\prime\prime})^{2}}+\frac{5[f^{(3)}]^{2}}{24(f_{0}^{\prime\prime
})^{3}}\right]  \frac{1}{K}+O(\frac{1}{{K}^{2}})\right\}  , \label{70}%
\end{align}
where $u_{0},f_{0},u_{0}^{\prime},f_{0}^{\prime\prime},$ etc, stand for the
values of functions and their derivatives evaluated at the saddle point
$f^{\prime}(x_{0})=0$. In order to apply this master formula to calculate
stringy amplitudes, we need the following substitutions $\left(
\alpha^{\prime}=1/2\right)  $
\begin{align}
K  &  \equiv2E^{2},\label{71}\\
f(x)  &  \equiv\ln(x)-\tau\ln(1-x),\label{72}\\
\tau &  \equiv-\frac{(2,3)}{(1,2)}\rightarrow\sin^{2}\frac{\theta}{2},
\label{73}%
\end{align}
where $\theta$ is the scattering angle in center of momentum frame and the
saddle point for the integration of moduli is $x_{0}=\frac{1}{1-\tau}$. In the
first scattering amplitude corresponding to Eq.(\ref{65}), we can identify the
$u(x)$ function as
\begin{equation}
u_{I}(x)\equiv\left[  \frac{(3,4)(e^{T_{1}}\cdot e^{T_{2}})}{x}-(e^{T_{1}%
}\cdot k_{3})(e^{T_{2}}\cdot k_{4})+\frac{(e^{T_{2}}\cdot k_{3})(e^{T_{1}%
}\cdot k_{4})}{1-x}\right]  \frac{1}{x}\left[  \frac{e^{T_{2}}\cdot k_{3}%
}{1-x}\right]  . \label{74}%
\end{equation}
Equipped with this, we obtain the high energy limit of the first amplitude,
\begin{align}
&  2E^{2}(1-\tau)(e^{T}\cdot k_{3})x_{0}^{(1,2)-1}(1-x_{0})^{(2,3)-1}%
\sqrt{\frac{\pi\tau}{E^{2}(1-\tau)^{3}}}\nonumber\\
&  =4\sqrt{\pi}E^{2}(1-\tau)^{2}x_{0}^{(1,2)}(1-x_{0})^{(2,3)}. \label{75}%
\end{align}
Next, we replace the second vertex operator in Eq.(\ref{68}) by Eq.(\ref{66}),
and the 4-point function is given by
\begin{equation}
\int_{0}^{1}dx\frac{1}{M^{2}}\left[  (e^{T}\cdot k_{3})(2,4)-\frac{(e^{T}\cdot
k_{4})(2,3)}{1-x}\right]  \frac{1}{x}\left[  \frac{(1,2)}{x}-\frac{(2,3)}%
{1-x}\right]  x^{(1,2)}(1-x)^{(2,3)}. \label{76}%
\end{equation}
Here we can identify the $u(x)$ function for saddle-point master formula,
Eq.(\ref{70})
\begin{equation}
u_{II}(x)\equiv\frac{(e^{T}\cdot k_{3})(1,2)}{M^{2}x}\left[  (2,4)+\frac
{(2,3)}{1-x}\right]  f^{\prime}(x). \label{77}%
\end{equation}
One can check that $u_{II}(x_{0})=u_{II}^{\prime}(x_{0})=0$, and
\begin{equation}
u_{II}^{\prime\prime}(x_{0})=\frac{2(1,2)(2,3)(e^{T}\cdot k_{3})}%
{M^{2}x(1-x)^{2}}f^{\prime\prime}(x_{0}). \label{78}%
\end{equation}
Thus, the amplitude associated with the massive state, Eq.(\ref{66}), is given
by
\begin{align}
&  -\frac{2}{M^{2}}E^{2}\tau(e^{T}\cdot k_{3})x_{0}^{(1,2)-1}(1-x_{0}%
)^{(2,3)-2}\sqrt{\frac{\pi\tau}{E^{2}(1-\tau)^{3}}}\nonumber\\
&  =\frac{4}{M^{2}}\sqrt{\pi}E^{2}(1-\tau)^{2}x_{0}^{(1,2)}(1-x_{0})^{(2,3)}.
\label{79}%
\end{align}
In the third case, after replacing the second vertex operator in Eq.(\ref{68})
by Eq.(\ref{67}), we get the Wick contraction
\begin{equation}
\int_{0}^{1}dx\frac{1}{M}\left[  -\frac{(e^{T}\cdot k_{4})(2,3)}{(1-x)^{2}%
}\right]  \frac{1}{x}x^{(1,2)}(1-x)^{(2,3)}. \label{80}%
\end{equation}
The high energy limit of this amplitude, after applying the master formula of
saddle-point approximation, is
\begin{align}
&  \frac{2}{M}E^{2}\tau(e^{T}\cdot k_{3})x^{(1,2)-1}(1-x)^{(2,3)-2}\sqrt
{\frac{\pi\tau}{E^{2}(1-\tau)^{3}}}\nonumber\\
&  =-\frac{4}{M}\sqrt{\pi}E^{2}(1-\tau)^{2}x_{0}^{(1,2)}(1-x_{0})^{(2,3)}.
\label{81}%
\end{align}

In conclusion, from these results, Eqs.(\ref{75}),(\ref{79}),(\ref{81}), we
find the ratios between the 4-point amplitudes associated with $(\alpha
_{-1}^{T})(b_{-\frac{1}{2}}^{T})\left\vert 0,k\right\rangle $, $(\alpha
_{-1}^{L})(b_{-\frac{1}{2}}^{L})\left\vert 0,k\right\rangle $, and
$(b_{-\frac{3}{2}}^{L})\left\vert 0,k\right\rangle $ to be $1:\frac{1}{M^{2}%
}:-\frac{1}{M}$, in perfect agreement with Eqs.(\ref{23..}),(\ref{27.}) and
Eq. (\ref{ratio.}).

\subsubsection{$M^{2}=4$}

Our previous examples only involve one fermion operator $b_{-\frac{1}{2}}^{T}%
$, $b_{-\frac{1}{2}}^{L}$, $b_{-\frac{3}{2}}^{L}$. Since in the 4-point
functions with the fixed states $V_{1}\rightarrow$ photon, $V_{3,4}%
\rightarrow$ tachyons, the maximum fermion number of the second vertex is
three, it is of interest to see the pattern of stringy amplitudes associated
with the next massive vertices at $M^{2}=4$.

At this mass level, the relevant states and the vertex operators are (in the
$-1$ ghost picture)
\begin{align}
(b_{-\frac{1}{2}}^{T})(b_{-\frac{1}{2}}^{L})(b_{-\frac{3}{2}}^{L})\left\vert
0,k\right\rangle  &  \Rightarrow\psi^{T}\psi^{L}\partial\psi^{L}e^{-\phi
}e^{ikX},\label{82}\\
(\alpha_{-1}^{T})(\alpha_{-1}^{T})(b_{-\frac{1}{2}}^{T})\left\vert
0,k\right\rangle  &  \Rightarrow\psi^{T}\partial X^{T}\partial X^{T}e^{-\phi
}e^{ikX}. \label{83}%
\end{align}
To calculate 4-point functions, we can fix the first vertex $(V_{1})$ to be a
photon state in the $-1$ ghost picture, Eq.(\ref{63}), and the third and the
fourth vertices to be tachyon state in the $0$ ghost picture, Eq.(\ref{64}).

Since the applications of saddle-point approximation is essentially identical
to previous cases, we simply list the results of our calculations
\begin{align}
&  (b_{-\frac{1}{2}}^{T})(b_{-\frac{1}{2}}^{L})(b_{-\frac{3}{2}}%
^{L})\left\vert 0,k\right\rangle \nonumber\\
&  \Rightarrow\int_{0}^{1}dx_{2}\langle(\psi_{1}^{T_{1}}e^{-\phi_{1}}%
e^{ik_{1}X_{1}})(\psi_{2}^{T_{2}}\psi_{2}^{L_{2}}\partial\psi_{2}^{L_{2}%
}e^{-\phi_{2}}e^{ik_{2}X_{2}})(k_{3\lambda}\psi_{3}^{\lambda}e^{ik_{3}X_{3}%
})(k_{4\sigma}\psi_{4}^{\sigma}e^{ik_{4}X_{4}})\rangle\nonumber\\
&  =\frac{4\sqrt{\pi}}{M^{2}}E^{3}\tau^{-\frac{1}{2}}(1-\tau)^{\frac{7}{2}%
}x_{0}^{(1,2)}(1-x_{0})^{(2,3)}, \label{84}%
\end{align}%
\begin{align}
&  (\alpha_{-1}^{T})(\alpha_{-1}^{T})(b_{-\frac{1}{2}}^{T})\left\vert
0,k\right\rangle \nonumber\\
&  \Rightarrow\int_{0}^{1}dx_{2}\langle(\psi_{1}^{T_{1}}e^{-\phi_{1}}%
e^{ik_{1}X_{1}})(\psi_{2}^{T_{2}}X_{2}^{L_{2}}X_{2}^{L_{2}}e^{-\phi_{2}%
}e^{ik_{2}X_{2}})(k_{3\lambda}\psi_{3}^{\lambda}e^{ik_{3}X_{3}})(k_{4\sigma
}\psi_{4}^{\sigma}e^{ik_{4}X_{4}})\rangle\nonumber\\
&  =8\sqrt{\pi}E^{3}\tau^{-\frac{1}{2}}(1-\tau)^{\frac{7}{2}}x_{0}%
^{(1,2)}(1-x_{0})^{(2,3)}. \label{85}%
\end{align}
Combining these results, we conclude that the ratio between the $M^{2}=4$
vertices is given by
\begin{equation}
(b_{-\frac{1}{2}}^{T})(b_{-\frac{1}{2}}^{L})(b_{-\frac{3}{2}}^{L})\left\vert
0,k\right\rangle :(\alpha_{-1}^{T})(\alpha_{-1}^{T})(b_{-\frac{1}{2}}%
^{L})\left\vert 0,k\right\rangle =\frac{1}{M^{2}}:2=1:8, \label{86}%
\end{equation}
which agrees with Eq.(\ref{31}).

\subsubsection{GSO odd vertices at $M^{2}=5$}

In addition to the stringy amplitudes associated with GSO even vertices we
have calculated in the previous sections, we can also apply the same method to
those associated with the GSO odd vertices. While it is a common practice to
project out the GSO odd states in order to maintain spacetime supersymmetry,
it turns out that we do find linear relation among these amplitudes. This
seems to suggest a hidden structure of superstring theory in the high energy limit.

To see this, we examine the vertices of odd GSO parity, at the mass level
$M^{2}=5$. Based on the power-counting rule as in the bosonic string case, we
can identify the relevant vertices and the associated vertex operators as
follows
\begin{align}
(\alpha_{-1}^{T})(b_{-\frac{1}{2}}^{T})(b_{-\frac{3}{2}}^{L})\left\vert
0,k\right\rangle  &  \Rightarrow\psi^{T}\partial\psi^{L}\partial X^{T}%
e^{-\phi}e^{ikX},\label{87}\\
(\alpha_{-1}^{L})(b_{-\frac{1}{2}}^{L})(b_{-\frac{3}{2}}^{L})\left\vert
0,k\right\rangle  &  \Rightarrow\psi^{L}\partial\psi^{L}\partial X^{L}%
e^{-\phi}e^{ikX}. \label{88}%
\end{align}
To calculate 4-point functions, we can fix the first vertex $(V_{1})$ to be a
tachyon state in the $-1$ ghost picture,
\begin{equation}
V_{1}=e^{-\phi_{1}}e^{ik_{1}\cdot X_{1}}, \label{89}%
\end{equation}
and the third and the fourth vertices to be tachyon state in the $0$ ghost
picture, as Eq.(\ref{64}).

Since the applications of saddle-point approximation is essentially identical
to previous cases, we simply list the results of our calculations
\begin{align}
&  (\alpha_{-1}^{T})(b_{-\frac{1}{2}}^{T})(b_{-\frac{3}{2}}^{L})\left\vert
0,k\right\rangle \nonumber\\
&  \Rightarrow\int_{0}^{1}dx_{2}\langle(e^{-\phi_{1}}e^{ik_{1}X_{1}})(\psi
_{2}^{T_{2}}\partial\psi_{2}^{L_{2}}\partial X_{2}^{T_{2}}e^{-\phi_{2}%
}e^{ik_{2}X_{2}})(k_{3\lambda}\psi_{3}^{\lambda}e^{ik_{3}X_{3}})(k_{4\sigma
}\psi_{4}^{\sigma}e^{ik_{4}X_{4}})\rangle,\nonumber\\
&  =-\frac{8\sqrt{\pi}}{M}E^{3}\tau^{-\frac{1}{2}}(1-\tau)^{\frac{7}{2}}%
x_{0}^{(1,2)}(1-x_{0})^{(2,3)}, \label{90}%
\end{align}%
\begin{align}
&  (\alpha_{-1}^{L})(b_{-\frac{1}{2}}^{L})(b_{-\frac{3}{2}}^{L})\left\vert
0,k\right\rangle \nonumber\\
&  \Rightarrow\int_{0}^{1}dx_{2}\langle(e^{-\phi_{1}}e^{ik_{1}X_{1}})(\psi
_{2}^{L_{2}}\partial\psi_{2}^{L_{2}}\partial X_{2}^{L_{2}}e^{-\phi_{2}%
}e^{ik_{2}X_{2}})(k_{3\lambda}\psi_{3}^{\lambda}e^{ik_{3}X_{3}})(k_{4\sigma
}\psi_{4}^{\sigma}e^{ik_{4}X_{4}})\rangle,\nonumber\\
&  =-\frac{4\sqrt{\pi}}{M^{3}}E^{3}\tau^{-\frac{1}{2}}(1-\tau)^{\frac{7}{2}%
}x_{0}^{(1,2)}(1-x_{0})^{(2,3)}. \label{91}%
\end{align}
It is worth noting that in the second calculations, we need to include both
$u^{\prime\prime}(x_{0})$ and $u^{(3)}(x_{0})$ terms of the first order
corrections in saddle-point approximation, Eq.(\ref{70}), to get the correct answer.

Combining these results, we conclude that the ratio between the $M^{2}=3$
vertices is given by
\begin{equation}
(\alpha_{-1}^{T})(b_{-\frac{1}{2}}^{T})(b_{-\frac{3}{2}}^{L})\left\vert
0,k\right\rangle :(\alpha_{-1}^{L})(b_{-\frac{1}{2}}^{L})(b_{-\frac{3}{2}}%
^{L})\left\vert 0,k\right\rangle =2M^{2}:1=10:1. \label{92}%
\end{equation}
Notice that here we also find an interesting connection between GSO even
$M^{2}=4$ amplitudes and those of GSO odd parity at $M^{2}=5$. The high energy
limits of the four amplitudes, Eqs.(\ref{84}),(\ref{85}),(\ref{90}%
),(\ref{91}), are proportional to each other. and their ratios are $\sqrt
{5}:8\sqrt{5}:(-8):-\frac{4}{5}$.

\subsection{Polarizations orthogonal to the scattering plane}

In this section we consider high energy scattering amplitudes of string states
with polarizations $e_{T^{i}},i=3,4...,25,$ orthogonal to the scattering
plane. We will present some examples with saddle-point calculations and
compare them with those calculated in the previous section. We will find that
they are all proportional to the previous ones considered before. These
scattering amplitudes are of subleading order in energy for the case of $26D$
open bosonic string theory. The existence of these new high energy scattering
amplitudes is due to the worldsheet fermion exchange in the correlation
functions as we will see in the following examples. Our first example is to
consider Eq.(\ref{68}) and replace $\psi_{1}^{T_{1}}$ and $\psi_{2}^{T_{2}}$
by $\psi_{1}^{T_{1}^{i}}$ and $\psi_{2}^{T_{2}^{i}}$ respectively%

\begin{equation}
\int_{0}^{1}dx_{2}\langle(\psi_{1}^{T_{1}^{i}}e^{-\phi_{1}}e^{ik_{1}X_{1}%
})(\psi_{2}^{T_{2}^{i}}\partial X_{2}^{T_{2}}e^{-\phi_{2}}e^{ik_{2}X_{2}%
})(k_{3\lambda}\psi_{3}^{\lambda}e^{ik_{3}X_{3}})(k_{4\sigma}\psi_{4}^{\sigma
}e^{ik_{4}X_{4}})\rangle. \label{93}%
\end{equation}
The calculation of Eq.(\ref{93}) is similar to that of Eq.(\ref{68}) except
that, for this new case, one ends up with only the first term in
Eq.(\ref{69}), and the second and the third terms vanish. Remarkably, the
final answer is%

\begin{align}
&  -2E^{2}(1-\tau)(e^{T}\cdot k_{3})x_{0}^{(1,2)-1}(1-x_{0})^{(2,3)-1}%
\sqrt{\frac{\pi\tau}{E^{2}(1-\tau)^{3}}}\nonumber\\
&  =-4\sqrt{\pi}E^{2}(1-\tau)^{2}x_{0}^{(1,2)}(1-x_{0})^{(2,3)}, \label{94}%
\end{align}
which is proportional to Eq.(\ref{75}). Our second example is again to replace
$\psi_{1}^{T_{1}}$ and $\psi_{2}^{T_{2}}$ in Eq.(\ref{84}) by $\psi_{1}%
^{T_{1}^{i}}$ and $\psi_{2}^{T_{2}^{i}}$ respectively. One gets exactly the
same answer as Eq.(\ref{84}). The two examples above seem to suggest that high
energy scattering of string states with polarizations $e_{T^{i}}$ are the same
as that of polarization $e_{T}$ up to a sign. Let's consider the third example
to justify this point. It is straightforward to show the following%

\begin{align}
&  \int_{0}^{1}dx_{2}\langle(\psi_{1}^{L}\psi_{1}^{T_{1}}\psi_{1}^{T_{1}^{i}%
}e^{-\phi_{1}}e^{ik_{1}X_{1}})(\psi_{2}^{L}\psi_{2}^{T_{2}}\psi_{2}^{T_{2}%
^{i}}\partial X_{2}^{T_{2}}e^{-\phi_{2}}e^{ik_{2}X_{2}})(k_{3\lambda}\psi
_{3}^{\lambda}e^{ik_{3}X_{3}})(k_{4\sigma}\psi_{4}^{\sigma}e^{ik_{4}X_{4}%
})\rangle\nonumber\\
&  =N[4E^{4}(1-\tau)-4E^{4}(1-\tau)^{2}-4E^{4}\tau(1-\tau)]=0 \label{95}%
\end{align}
On the other hand, if we assume the symmetry for all transverse polarization
vectors $T,T^{i}$ in the scattering amplitudes, one can easily derive the same
conclusion without detailed calculations. Since replacing $T^{i}$ polarization
vectors of both vertices in Eq.(\ref{95}) by $T$ will naturally leads to a
null result due to anti-commuting property of fermions.

It is clear from the above calculations that the existence of these new high
energy scattering amplitudes of string states with polarizations $e_{T^{i}}$
orthogonal to the scattering plane is due to the worldsheet fermion exchange
in the correlation functions. These fermion exchanges do not exist in the
bosonic string correlation functions and is, presumably, related to the high
energy massive spacetime fermionic scattering amplitudes in the R-sector of
the theory.

Physically, the high energy scattering amplitudes of spacetime fermion contain
the symmetry of rotations among different polarizations in the spin space and
our results here seem to justify this observation. If this conjecture turns
out to be true, then the list of vertices we considered in Eq.(\ref{fT/2}) to
Eq.(\ref{fTLL/3}) for high energy stringy amplitudes should be extended and
includes the cases with $b_{-\frac{1}{2}}^{T}$ replaced by $b_{-\frac{1}{2}%
}^{T^{i}}$. Obviously, these new high energy amplitudes create complications
for a full understanding of stringy symmetry. Nevertheless, the claim that
there is only one independent high energy scattering amplitude at each fixed
mass level of the string spectrum persists in the case of superstring theory,
at least, for the NS sector of the theory.

In conclusion of this chapter we have explicitly calculated all high energy
scattering amplitudes of string states with polarizations on the scattering
plane of open superstring theory. In particular, the proportionality constants
among high energy scattering amplitudes of different string states at each
fixed but arbitrary mass level are determined by using three different
methods. These constants are shown to originate from ZNS in the spectrum as in
the case of open bosonic string theory.

In addition, we discover new high energy scattering amplitudes, which are
still proportional to the previous ones, with polarizations
\textit{orthogonal} to the scattering plane. We conjecture the existence of a
symmetry among high energy scattering amplitudes with polarizations $e_{T^{i}%
}$ and $e_{T}$. These scattering amplitudes are subleading order in energy for
the case of open bosonic string theory. The existence of these new high energy
scattering amplitudes is due to the worldsheet fermion exchange in the
correlation functions and is argued to be related to the high energy massive
spacetime fermionic scattering amplitudes in the R-sector of the theory.
Finally, our study also suggests that the nature of GSO projection in
superstring theory might be simplified in the high energy limit. Hopefully,
this is in connection with the conjecture that supersymmetry is realized in
broken phase without GSO projection in the open string theory \cite{GSO,GSO1}.

It would be of crucial importance to calculate high energy massive fermion
scattering amplitudes in the R-sector to complete the proof of Gross's two
conjectures on high energy symmetry of superstring theory. The construction of
general \textit{massive} spacetime fermion vertex, involving picture changing,
will be the first step toward understanding of the high energy behavior of
superstring theory.%

\setcounter{equation}{0}
\renewcommand{\theequation}{\arabic{section}.\arabic{equation}}%

\section{Hard string scatterings from D-branes/O-planes}

In this chapter, we study scatterings of bosonic closed strings from D-branes
\cite{Dscatt} in section A, and O-planes \cite{O-plane} in section B. In
particular, we will discuss hard strings scattered from D-particle
\cite{Dscatt} and D-domain-wall \cite{Wall}. We will also study hard strings
scattered from O-particle \cite{O-plane} and O-domain-wall \cite{O-plane}. In
addition, in section C, we calculate the absorption amplitudes \cite{Decay} of
a closed string state at arbitrary mass level leading to two open string
states on the D-brane at high energies.

\subsection{Scatterings from D-branes}

In this section we study the general structure of an arbitrary incoming closed
string state scatters from D-brane and ends up with an arbitrary spin outgoing
closed string states at arbitrary mass levels \cite{Dscatt}. The scattering of
massless string states from D-brane has been well studied in the literature
and can be found in
\cite{Klebanov,Myers,Klebanov3,barbon1996d,bachas1999high,hirano1997scattering}
Here we extend the calculations of massless closed string states to massive
closed string states at arbitrary mass levels. Since the mass of D-brane
scales as the inverse of the string coupling constant $1/g$, we will assume
that it is infinitely heavy to leading order in $g$ and does not recoil. We
will first show that, for the $(0\rightarrow1)$ and $(1\rightarrow\infty)$
channels, all the scattering amplitudes can be expressed in terms of the beta
functions, thanks to the \textit{momentum conservation on the D-brane.}

Alternatively, the Kummer relation of the hypergeometric function $_{2}F_{1}$
can be used to reduce the scattering amplitudes to the usual beta function
\cite{Dscatt}. After summing up the $(0\rightarrow1)$ and $(1\rightarrow
\infty)$ channels, we discover that all the scattering amplitudes can be
expressed in terms of the generalized hypergeometric function $_{3}F_{2}$ with
special arguments, which terminates to a finite sum and, as a result, the
whole scattering amplitudes consistently reduce to the usual beta function.

For the simple case of D-particle, we explicitly calculate \cite{Dscatt} high
energy limit of a set of scattering amplitudes for arbitrary mass levels, and
derive infinite linear relations among them for each fixed mass level. Since
the calculation of decoupling of high energy ZNS remains the same as the case
of scatterings without D-brane, the ratios of these high energy scattering
amplitudes are found to be consistent with the decoupling of high energy ZNS
in Chapter V. The cases of RR strings scattered from D-particle will be
discussed in chapter XIV where the complete ratios among GR scattering
amplitudes will be calculated.

We will first begin with the simple case of tachyon to tachyon scattering and
then generalize to scatterings of states at arbitrary mass levels. The
standard propagators of the left and right moving fields are%
\begin{align}
\left\langle X^{\mu}\left(  z\right)  X^{\nu}\left(  w\right)  \right\rangle
&  =-\eta^{\mu\nu}\log\left(  z-w\right)  ,\label{D1}\\
\left\langle \tilde{X}^{\mu}\left(  \bar{z}\right)  \tilde{X}^{\nu}\left(
\bar{w}\right)  \right\rangle  &  =-\eta^{\mu\nu}\log\left(  \bar{z}-\bar
{w}\right)  . \label{D2}%
\end{align}
In addition, there are also nontrivial correlator between the right and left
moving fields as well%
\begin{equation}
\left\langle X^{\mu}\left(  z\right)  \tilde{X}^{\nu}\left(  \bar{w}\right)
\right\rangle =-D^{\mu\nu}\log\left(  z-\bar{w}\right)  \label{DD}%
\end{equation}
as a result of the boundary condition at the real axis. Propagator
Eq.(\ref{DD}) has the standard form Eq.(\ref{D1}) for the fields satisfying
Neumann boundary condition, while matrix $D$ reverses the sign for the fields
satisfying Dirichlet boundary condition. We will follow the standard notation
and make the following replacement%
\begin{equation}
\tilde{X}^{\mu}\left(  \bar{z}\right)  \rightarrow D_{\text{ \ }\nu}^{\mu
}X^{\nu}\left(  \bar{z}\right)
\end{equation}
which allows us to use the standard correlators Eq.(\ref{D1}) throughout our
calculations. As we will see, the existence of the Propagator Eq.(\ref{DD})
has far-reaching effect on the string scatterings from D-brane.

\subsubsection{Tachyon to tachyon}

In this section, we consider the tachyon to tachyon scattering amplitude%

\begin{align}
A_{tach}  &  =\int d^{2}z_{1}d^{2}z_{2}\left\langle V_{1}\left(  z_{1},\bar
{z}_{1}\right)  V_{2}\left(  z_{2},\bar{z}_{2}\right)  \right\rangle
\nonumber\\
&  =\int d^{2}z_{1}d^{2}z_{2}\left\langle V\left(  k_{1},z_{1}\right)
\tilde{V}\left(  k_{1},\bar{z}_{1}\right)  V\left(  k_{2},z_{2}\right)
\tilde{V}\left(  k_{2},\bar{z}_{2}\right)  \right\rangle \nonumber\\
&  =\int d^{2}z_{1}d^{2}z_{2}\left\langle e^{ik_{1}X\left(  z_{1}\right)
}e^{ik_{1}\tilde{X}\left(  \bar{z}_{1}\right)  }e^{ik_{2}X\left(
z_{2}\right)  }e^{ik_{2}\tilde{X}\left(  \bar{z}_{2}\right)  }\right\rangle
\nonumber\\
&  =\int d^{2}z_{1}d^{2}z_{2}\left(  z_{1}-\bar{z}_{1}\right)  ^{k_{1}\cdot
D\cdot k_{1}}\left(  z_{2}-\bar{z}_{2}\right)  ^{k_{2}\cdot D\cdot k_{2}%
}\left\vert z_{1}-z_{2}\right\vert ^{2k_{1}\cdot k_{2}}\left\vert z_{1}%
-\bar{z}_{2}\right\vert ^{2k_{1}\cdot D\cdot k_{2}}.
\end{align}
To fix the $SL\left(  2,R\right)  $ invariance, we set $z_{1}=iy$ and
$z_{2}=i$. Introducing the $SL\left(  2,R\right)  $ Jacobian%
\begin{equation}
d^{2}z_{1}d^{2}z_{2}=4\left(  1-y^{2}\right)  dy,
\end{equation}
we have, for the $(0\rightarrow1)$ channel,
\begin{align}
A_{tach}^{(0\rightarrow1)}  &  =4\left(  2i\right)  ^{k_{1}\cdot D\cdot
k_{1}+k_{2}\cdot D\cdot k_{2}}\int_{0}^{1}dy\text{ }y^{k_{2}\cdot D\cdot
k_{2}}\left(  1-y\right)  ^{2k_{1}\cdot k_{2}+1}\left(  1+y\right)
^{2k_{1}\cdot D\cdot k_{2}+1}\nonumber\\
&  =4\left(  2i\right)  ^{2a_{0}}\int_{0}^{1}dy\text{ }y^{a_{0}}\left(
1-y\right)  ^{b_{0}}\left(  1+y\right)  ^{c_{0}}\nonumber\\
&  =4\left(  2i\right)  ^{2a_{0}}\frac{\Gamma\left(  a_{0}+1\right)
\Gamma\left(  b_{0}+1\right)  }{\Gamma\left(  a_{0}+b_{0}+2\right)  }\text{
}_{2}F_{1}(-c_{0},a_{0}+1,a_{0}+b_{0}+2,-1)\label{ta3}\\
&  =4\left(  2i\right)  ^{2a_{0}}\frac{\Gamma\left(  a_{0}+1\right)
\Gamma\left(  b_{0}+1\right)  }{\Gamma\left(  a_{0}+b_{0}+2\right)
}2^{-2a_{0}-1-N}\text{ }_{2}F_{1}(N-a_{0},b_{0}+1,a_{0}+b_{0}+2,-1)\nonumber\\
&  =4\left(  2i\right)  ^{2a_{0}}2^{-2a_{0}-1-N}\int_{0}^{1}dt\text{ }%
t^{b_{0}}\left(  1-t\right)  ^{a_{0}}\left(  1+t\right)  ^{a_{0}+N}.
\label{ta5}%
\end{align}
In the above calculations, we have defined
\begin{align}
a_{0}  &  =k_{1}\cdot D\cdot k_{1}=k_{2}\cdot D\cdot k_{2},\\
b_{0}  &  =2k_{1}\cdot k_{2}+1,\\
c_{0}  &  =2k_{1}\cdot D\cdot k_{2}+1,
\end{align}
so that%
\begin{equation}
2a_{0}+b_{0}+c_{0}+2=4N_{1}\equiv-N,
\end{equation}
and $-k_{1}^{2}=M^{2}\equiv\frac{M_{closed}^{2}}{2\alpha_{closed}^{\prime}%
}=2(N_{1}-1)$, $N_{1}=0$ for tachyon. We have also used the integral
representation of the hypergeometric function
\begin{equation}
_{2}F_{1}\left(  \alpha,\beta,\gamma,z\right)  =\frac{\Gamma\left(
\gamma\right)  }{\Gamma\left(  \beta\right)  \Gamma\left(  \gamma
-\beta\right)  }\int_{0}^{1}dt\text{ }t^{\beta-1}\left(  1-t\right)
^{\gamma-\beta-1}\left(  1-zt\right)  ^{-\alpha}, \label{gama}%
\end{equation}
and the following identity%
\begin{equation}
_{2}F_{1}(\alpha,\beta,\gamma;x)=(1-x)^{\gamma-\alpha-\beta}\text{ }_{2}%
F_{1}(\gamma-\alpha,\gamma-\beta,\gamma;x), \label{half}%
\end{equation}
which we discuss in the section \ref{review of F}. In addition, the momentum
conservation on the D-brane%
\begin{equation}
D\cdot k_{1}+k_{1}+D\cdot k_{2}+k_{2}=0 \label{con}%
\end{equation}
is crucial to get the final result Eq.(\ref{ta5}).\ Finally, by using change
of variable $\tilde{t}=t^{2}$, Eq.(\ref{ta5}) can be further reduced to the
beta function%
\begin{equation}
A_{tach}^{(0\rightarrow1)}\simeq\frac{\Gamma\left(  a_{0}+1\right)
\Gamma\left(  \frac{b_{0}+1}{2}\right)  }{\Gamma\left(  a_{0}+\frac{b_{0}}%
{2}+\frac{3}{2}\right)  }=B\left(  a_{0}+1,\frac{b_{0}+1}{2}\right)
\label{final}%
\end{equation}
where we have omitted an irrelevant factor.

For the $(1\rightarrow\infty)$ channel, we use the change of variable
$y=\frac{1+t}{1-t}$ and end up with the same result%
\begin{align}
A_{tach}^{(1\rightarrow\infty)}  &  =4\left(  2i\right)  ^{k_{1}\cdot D\cdot
k_{1}+k_{2}\cdot D\cdot k_{2}}\int_{1}^{\infty}dy\text{ }y^{k_{2}\cdot D\cdot
k_{2}}\left(  y-1\right)  ^{2k_{1}\cdot k_{2}+1}\left(  1+y\right)
^{2k_{1}\cdot D\cdot k_{2}+1}\nonumber\\
&  =4\left(  2i\right)  ^{2a_{0}}2^{-2a_{0}-1-N}\int_{0}^{1}dt\text{ }%
t^{b_{0}}\left(  1-t\right)  ^{a_{0}+N}\left(  1+t\right)  ^{a_{0}}\nonumber\\
&  \simeq\frac{\Gamma\left(  a_{0}+1\right)  \Gamma\left(  \frac{b_{0}+1}%
{2}\right)  }{\Gamma\left(  a_{0}+\frac{b_{0}}{2}+\frac{3}{2}\right)
}=B\left(  a_{0}+1,\frac{b_{0}+1}{2}\right)
\end{align}
since $N=0$ for the case of tachyon.

Alternatively, one can use the Kummer formula of hypergeometric function%
\begin{equation}
_{2}F_{1}(\alpha,\beta,1+\alpha-\beta,-1)=\frac{\Gamma(1+\alpha-\beta
)\Gamma(1+\frac{\alpha}{2})}{\Gamma(1+\alpha)\Gamma(1+\frac{\alpha}{2}-\beta)}%
\end{equation}
and%
\begin{equation}
\Gamma\left(  \frac{1+\alpha}{2}\right)  =\frac{2^{-\alpha}\sqrt{\pi}%
\Gamma\left(  1+\alpha\right)  }{\Gamma\left(  1+\frac{\alpha}{2}\right)  },
\end{equation}
to reduce Eq.(\ref{ta3}) to the final result Eq.(\ref{final}). In this
calculation, we have used the \textit{Kummer condition}%
\begin{equation}
\gamma=1+\alpha-\beta,
\end{equation}
which is equivalent to the momentum conservation on the D-brane Eq.(\ref{con}).

\subsubsection{Tensor to tensor}

In this section, we generalize the previous calculation to general tensor to
tensor scatterings. In this case, we define%
\begin{align}
a  &  =k_{1}\cdot D\cdot k_{1}+n_{a}\equiv a_{0}+n_{a},\\
b  &  =2k_{1}\cdot k_{2}+1+n_{b}\equiv b_{0}+n_{b},\\
c  &  =2k_{1}\cdot D\cdot k_{2}+1+n_{c}\equiv c_{0}+n_{c},
\end{align}
where $n_{a}$, $n_{b}$ and $n_{c}$ are integer and%
\[
N^{\prime}=-\left(  2n_{a}+n_{b}+n_{c}\right)  ,
\]
so that%
\begin{equation}
2a+b+c+2+N^{\prime}=4N_{1}\Longrightarrow2a+b+c+2=4N_{1}-N^{\prime}\equiv-N
\label{int}%
\end{equation}
where $k_{1}^{2}=2(N_{1}-1)$ and $N_{1}$ is now the mass level of $k_{1}$.
After a similar calculation as the previous section, it is easy to see that a
typical term in the expression of the general tensor to tensor scattering
amplitudes can be reduced to the following integral%

\begin{align}
I_{\left(  0\rightarrow1\right)  }  &  =\int_{0}^{1}dt\text{ }t^{a}\left(
1-t\right)  ^{b}\left(  1+t\right)  ^{c},\nonumber\\
&  =\frac{\Gamma\left(  a+1\right)  \Gamma\left(  b+1\right)  }{\Gamma\left(
a+b+2\right)  }\text{ }_{2}F_{1}\left(  -c,a+1,a+b+2,-1\right) \nonumber\\
&  =2^{b+c+1}\frac{\Gamma\left(  a+1\right)  \Gamma\left(  b+1\right)
}{\Gamma\left(  a+b+2\right)  }\text{ }_{2}F_{1}\left(
-a-N,b+1,a+b+2,-1\right) \nonumber\\
&  =2^{-2a-1-N}\int_{0}^{1}dt\text{ }t^{b}\left(  1-t\right)  ^{a}\left(
1+t\right)  ^{a+N}.
\end{align}
Similarly, for the $(1\rightarrow\infty)$ channel, one gets
\begin{align}
I_{\left(  1\rightarrow\infty\right)  }  &  =\int_{1}^{\infty}dy\text{ }%
y^{a}\left(  y-1\right)  ^{b}\left(  1+y\right)  ^{c}\nonumber\\
&  =2^{-2a-1-N}\int_{0}^{1}dt\text{ }t^{b}\left(  1-t\right)  ^{a+N}\left(
1+t\right)  ^{a}.
\end{align}
The sum of the two channels gives%
\begin{align}
I  &  =I_{\left(  0\rightarrow1\right)  }+I_{\left(  1\rightarrow
\infty\right)  }\nonumber\\
&  =2^{-2a-1-N}\int_{0}^{1}dt\text{ }t^{b}\left(  1-t\right)  ^{a}\left(
1+t\right)  ^{a}\left[  \left(  1+t\right)  ^{N}+\left(  1-t\right)
^{N}\right] \nonumber\\
&  =2^{-2a-1-N}\sum_{m=0}^{N}\left[  1+\left(  -1\right)  ^{m}\right]
\binom{N}{m}\int_{0}^{1}dt\text{ }t^{b+m}\left(  1-t\right)  ^{a}\left(
1+t\right)  ^{a}\nonumber\\
&  =2^{-2a-2-N}\sum_{m=0}^{N}\left[  1+\left(  -1\right)  ^{m}\right]
\binom{N}{m}\cdot\frac{\Gamma\left(  a+1\right)  \Gamma\left(  \frac{b+1}%
{2}+\frac{m}{2}\right)  }{\Gamma\left(  a+\frac{b+3}{2}+\frac{m}{2}\right)
}\nonumber\\
&  =2^{-2a-1-N}\frac{\Gamma\left(  a+1\right)  \Gamma\left(  \frac{b+1}%
{2}\right)  }{\Gamma\left(  a+\frac{b+3}{2}\right)  }\sum_{n=0}^{\left[
\frac{N}{2}\right]  }\binom{N}{2n}\dfrac{\left(  \frac{b+1}{2}\right)  _{n}%
}{\left(  a+\frac{b+3}{2}\right)  _{n}}\nonumber\\
&  =2^{-2a-1-N}\cdot B\left(  a+1,\frac{b+1}{2}\right)  \cdot\text{ }_{3}%
F_{2}\left(  \frac{b+1}{2},-\left[  \frac{N}{2}\right]  ,\dfrac{1}{2}-\left[
\frac{N}{2}\right]  ;a+\frac{b+3}{2},\dfrac{1}{2};1\right)  \label{master}%
\end{align}
where the generalized hypergeometric function $_{3}F_{2}$ is defined to be%
\begin{equation}
_{3}F_{2}(\alpha_{1},\alpha_{2},\alpha_{3};\gamma_{1},\gamma_{2};x)=\sum
_{n=0}^{\infty}\frac{(\alpha_{1})_{n}(\alpha_{2})_{n}(\alpha_{3})_{n}}%
{(\gamma_{1})_{n}(\gamma_{2})_{n}}\frac{x^{n}}{n!}.
\end{equation}
Note that the energy dependence of the prefactor $4\left(  2i\right)
^{k_{1}\cdot D\cdot k_{1}+k_{2}\cdot D\cdot k_{2}}$ in the scattering
amplitude cancels, apart from an irrelevant factor, the energy dependence of
$2^{-2a-1-N}$ by using Eq.(\ref{int}). For $N=0$, one recovers the result of
tachyon scattering amplitude Eq.(\ref{ta5}). For the special arguments of
$_{3}F_{2}$ in Eq.(\ref{master}), the hypergeometric function
\textit{terminates to a finite sum} and, as a result, the whole scattering
amplitudes consistently reduce to the usual beta function. The explicit forms
of $_{3}F_{2}$ for some integer $N$ are given in the section \ref{review of F}.

\subsubsection{Hard strings scattered from D-particle}

In this section, we will calculate the high energy limit of string scattered
from D-brane \cite{Dscatt}. In particular, we will calculate the ratios among
scattering amplitudes of different string states at high energies. For our
purpose here, for simplicity, we will only consider the string states with the
form of Eq.(\ref{relevant}) with $m=0$ (the ratios for the $m\neq0$ case will
be discussed in chapter XIV). The reason is as following. It was shown that
\cite{ChanLee,ChanLee1,ChanLee2,CHL} the leading order amplitudes containing
this component will drop from energy order $E^{4m}$ to $E^{2m}$, and one needs
to calculate the complicated naive subleading contraction terms between
$\partial X$ and $\partial X$ for the \textit{multi}-tensor scattering in
order to get the real leading order scattering amplitudes. For our
\textit{closed} string scattering calculation here, even for the case of one
tachyon and one tensor scattering, one encounters the similar complicated
nonzero contraction terms in Eq.(\ref{DD}) due to the D-brane. So we will omit
high energy scattering amplitudes of string states containing this
$(\alpha_{-1}^{L})^{2m}$ component. On the other hand, we will also need the
result that the high energy closed string ratios are the tensor product of two
pieces of open string ratios \cite{Closed}.

To simplify the kinematics, we consider the case of D0 brane or D-particle
scatterings \cite{Dscatt}. The momentum of the incident particle $k_{2}$ is
along the $-X$ direction and particle $k_{1}$ is scattered at an angle $\phi$.
We will consider the general case of an incoming tensor state $\left(
\alpha_{-1}^{T}\right)  ^{n-2q}\left(  \alpha_{-2}^{L}\right)  ^{q}%
\otimes\left(  \tilde{\alpha}_{-1}^{T}\right)  ^{n-2q^{\prime}}\left(
\tilde{\alpha}_{-2}^{L}\right)  ^{q^{\prime}}\left\vert 0\right\rangle $ and
an outgoing tachyon state. Our result can be easily generalized to the more
general two tensor cases. The kinematic setup is%
\begin{align}
e^{P}  &  =\frac{1}{M}\left(  -E,-\mathrm{k}_{2},0\right)  =\frac{k_{2}}{M},\\
e^{L}  &  =\frac{1}{M}\left(  -\mathrm{k}_{2},-E,0\right)  ,\\
e^{T}  &  =\left(  0,0,1\right)  ,\\
k_{1}  &  =\left(  E,\mathrm{k}_{1}\cos\phi,-\mathrm{k}_{1}\sin\phi\right)
,\\
k_{2}  &  =\left(  -E,-\mathrm{k}_{2},0\right)  .
\end{align}
For the scattering of D-particle $D_{ij}=-\delta_{ij}$, and it is easy to
calculate%
\begin{align}
e^{T}\cdot k_{2}  &  =e^{L}\cdot k_{2}=0,\\
e^{T}\cdot k_{1}  &  =-\mathrm{k}_{1}\sin\phi\sim-E\sin\phi,\\
e^{T}\cdot D\cdot k_{1}  &  =\mathrm{k}_{1}\sin\phi\sim E\sin\phi,\\
e^{T}\cdot D\cdot k_{2}  &  =0,\\
e^{L}\cdot k_{1}  &  =\frac{1}{M}\left[  \mathrm{k}_{2}E-\mathrm{k}_{1}%
E\cos\phi\right]  \sim\frac{E^{2}}{M}\left(  1-\cos\phi\right)  ,\\
e^{L}\cdot D\cdot k_{1}  &  =\frac{1}{M}\left[  \mathrm{k}_{2}E+\mathrm{k}%
_{1}E\cos\phi\right]  \sim\frac{E^{2}}{M}\left(  1+\cos\phi\right)  ,\\
e^{L}\cdot D\cdot k_{2}  &  =\frac{1}{M}\left[  -\mathrm{k}_{2}E-\mathrm{k}%
_{2}E\right]  \sim-\frac{2E^{2}}{M},
\end{align}
and%
\begin{align}
a_{0}  &  =k_{1}\cdot D\cdot k_{1}=-E^{2}-\mathrm{k}_{1}^{2}\sim-2E^{2},\\
b_{0}  &  =2k_{1}\cdot k_{2}+1=2\left(  E^{2}-\mathrm{k}_{1}\mathrm{k}_{2}%
\cos\phi\right)  +1\sim2E^{2}\left(  1-\cos\phi\right)  ,\\
c_{0}  &  =2k_{1}\cdot D\cdot k_{2}+1=2\left(  E^{2}+\mathrm{k}_{1}%
\mathrm{k}_{2}\cos\phi\right)  +1\sim2E^{2}\left(  1+\cos\phi\right)  .
\end{align}
The high energy scattering amplitude is then calculated to be%

\begin{align}
A_{D-Par}  &  =\varepsilon_{T^{n-2q}L^{q},T^{n-2q^{\prime}}L^{q^{\prime}}}\int
d^{2}z_{1}d^{2}z_{2}\left\langle V_{1}\left(  z_{1},\bar{z}_{1}\right)
V_{2}^{T^{n-2q}L^{q},T^{n-2q^{\prime}}L^{q^{\prime}}}\left(  z_{2},\bar{z}%
_{2}\right)  \right\rangle \nonumber\\
&  =\varepsilon_{T^{n-2q}L^{q},T^{n-2q^{\prime}}L^{q^{\prime}}}\int d^{2}%
z_{1}d^{2}z_{2}\cdot\left\langle e^{ik_{1}X}\left(  z_{1}\right)
e^{ik_{1}\tilde{X}}\left(  \bar{z}_{1}\right)  \right. \nonumber\\
&  \left.  \left(  \partial X^{T}\right)  ^{n-2q}\left(  i\partial^{2}%
X^{L}\right)  ^{q}e^{ik_{2}X}\left(  z_{2}\right)  \left(  \bar{\partial
}\tilde{X}^{T}\right)  ^{n-2q^{\prime}}\left(  i\bar{\partial}^{2}\tilde
{X}^{L}\right)  ^{q^{\prime}}e^{ik_{2}\tilde{X}}\left(  \bar{z}_{2}\right)
\right\rangle \nonumber\\
&  =\left(  -1\right)  ^{q+q^{\prime}}\int d^{2}z_{1}d^{2}z_{2}\left(
z_{1}-\bar{z}_{1}\right)  ^{k_{1}\cdot D\cdot k_{1}}\left(  z_{2}-\bar{z}%
_{2}\right)  ^{k_{2}\cdot D\cdot k_{2}}\left\vert z_{1}-z_{2}\right\vert
^{2k_{1}\cdot k_{2}}\left\vert z_{1}-\bar{z}_{2}\right\vert ^{2k_{1}\cdot
D\cdot k_{2}}\nonumber\\
&  \cdot\left[  \frac{ie^{T}\cdot k_{1}}{z_{1}-z_{2}}+\frac{ie^{T}\cdot D\cdot
k_{1}}{\bar{z}_{1}-z_{2}}+\frac{ie^{T}\cdot D\cdot k_{2}}{\bar{z}_{2}-z_{2}%
}\right]  ^{n-2q}\nonumber\\
&  \cdot\left[  \frac{ie^{T}\cdot D\cdot k_{1}}{z_{1}-\bar{z}_{2}}%
+\frac{ie^{T}\cdot k_{1}}{\bar{z}_{1}-\bar{z}_{2}}+\frac{ie^{T}\cdot D\cdot
k_{2}}{z_{2}-\bar{z}_{2}}\right]  ^{n-2q^{\prime}}\nonumber\\
&  \cdot\left[  \frac{e^{L}\cdot k_{1}}{\left(  z_{1}-z_{2}\right)  ^{2}%
}+\frac{e^{L}\cdot D\cdot k_{1}}{\left(  \bar{z}_{1}-z_{2}\right)  ^{2}}%
+\frac{e^{L}\cdot D\cdot k_{2}}{\left(  \bar{z}_{2}-z_{2}\right)  ^{2}%
}\right]  ^{q}\nonumber\\
&  \cdot\left[  \frac{e^{L}\cdot D\cdot k_{1}}{\left(  z_{1}-\bar{z}%
_{2}\right)  ^{2}}+\frac{e^{L}\cdot k_{1}}{\left(  \bar{z}_{1}-\bar{z}%
_{2}\right)  ^{2}}+\frac{e^{L}\cdot D\cdot k_{2}}{\left(  z_{2}-\bar{z}%
_{2}\right)  ^{2}}\right]  ^{q^{\prime}}.
\end{align}
Set $z_{1}=iy$ and $z_{2}=i$ to fix the $SL(2,R)$ invariance, we have%
\begin{align}
A_{D-Par}^{\left(  0\rightarrow1\right)  }  &  =4\left(  2i\right)
^{k_{1}\cdot D\cdot k_{1}+k_{2}\cdot D\cdot k_{2}}\int_{0}^{1}dy\text{
}y^{k_{1}\cdot D\cdot k_{1}}\left(  1-y\right)  ^{2k_{1}\cdot k_{2}+1}\left(
1+y\right)  ^{2k_{1}\cdot D\cdot k_{2}+1}\nonumber\\
&  \cdot\left[  -\frac{e^{T}\cdot k_{1}}{1-y}-\frac{e^{T}\cdot D\cdot k_{1}%
}{1+y}-\frac{e^{T}\cdot D\cdot k_{2}}{2}\right]  ^{n-2q}\nonumber\\
&  \cdot\left[  \frac{e^{T}\cdot D\cdot k_{1}}{1+y}+\frac{e^{T}\cdot k_{1}%
}{1-y}+\frac{e^{T}\cdot D\cdot k_{2}}{2}\right]  ^{n-2q^{\prime}}\nonumber\\
&  \cdot\left[  \frac{e^{L}\cdot k_{1}}{\left(  1-y\right)  ^{2}}+\frac
{e^{L}\cdot D\cdot k_{1}}{\left(  1+y\right)  ^{2}}+\frac{e^{L}\cdot D\cdot
k_{2}}{4}\right]  ^{q}\nonumber\\
&  \cdot\left[  \frac{e^{L}\cdot D\cdot k_{1}}{\left(  1+y\right)  ^{2}}%
+\frac{e^{L}\cdot k_{1}}{\left(  1-y\right)  ^{2}}+\frac{e^{L}\cdot D\cdot
k_{2}}{4}\right]  ^{q^{\prime}}\nonumber\\
&  =\left(  -1\right)  ^{n}4\left(  2i\right)  ^{k_{1}\cdot D\cdot k_{1}%
+k_{2}\cdot D\cdot k_{2}}\left(  E\sin\phi\right)  ^{2n-2\left(  q+q^{\prime
}\right)  }\left(  \frac{E^{2}}{M}\right)  ^{q+q^{\prime}}\nonumber\\
&  \cdot\int_{0}^{1}dy\text{ }y^{k_{1}\cdot D\cdot k_{1}}\left(  1-y\right)
^{2k_{1}\cdot k_{2}+1}\left(  1+y\right)  ^{2k_{1}\cdot D\cdot k_{2}%
+1}\nonumber\\
&  \cdot\left[  \frac{1}{1-y}-\frac{1}{1+y}\right]  ^{2n-2\left(  q+q^{\prime
}\right)  }\cdot\left[  \frac{1-\cos\phi}{\left(  1-y\right)  ^{2}}%
+\frac{1+\cos\phi}{\left(  1+y\right)  ^{2}}-\dfrac{1}{2}\right]
^{q+q^{\prime}}\nonumber\\
&  =\left(  -1\right)  ^{n}4\left(  2i\right)  ^{k_{1}\cdot D\cdot k_{1}%
+k_{2}\cdot D\cdot k_{2}}\left(  2E\sin\phi\right)  ^{2n}\left(  -\dfrac
{1}{8M\sin^{2}\phi}\right)  ^{q+q^{\prime}}\nonumber\\
&  \cdot\sum_{i=0}^{q+q^{\prime}}\sum_{j=0}^{i}\binom{q+q^{\prime}}{i}%
\binom{i}{j}\left(  -2\right)  ^{i}\left(  1-\cos\phi\right)  ^{j}\left(
1+\cos\phi\right)  ^{i-j}\nonumber\\
&  \int_{0}^{1}dy\text{ }y^{k_{1}\cdot D\cdot k_{1}}\left(  1-y\right)
^{2k_{1}\cdot k_{2}+1}\left(  1+y\right)  ^{2k_{1}\cdot D\cdot k_{2}%
+1}\nonumber\\
&  \cdot\left[  \frac{y}{\left(  1-y\right)  \left(  1+y\right)  }\right]
^{2n-2\left(  q+q^{\prime}\right)  }\left[  \frac{1}{1-y}\right]  ^{2j}\left[
\frac{1}{1+y}\right]  ^{2\left(  i-j\right)  }.
\end{align}
Now in the high energy limit, the master formula Eq.(\ref{master}) reduces to%
\begin{align}
I  &  =I_{\left(  0\rightarrow1\right)  }+I_{\left(  1\rightarrow
\infty\right)  }\nonumber\\
&  \simeq2^{-2a-2-N}B\left(  a+1,\frac{b+1}{2}\right)  \left[  \left(
1+\sqrt{\left\vert \dfrac{b}{2a+b}\right\vert }\right)  ^{N}+\left(
1-\sqrt{\left\vert \dfrac{b}{2a+b}\right\vert }\right)  ^{N}\right]
\nonumber\\
&  \equiv2^{-2a-2-N}B\left(  a+1,\frac{b+1}{2}\right)  F_{N},
\end{align}
where%
\begin{align}
n_{a}  &  =2n-2\left(  q+q^{\prime}\right)  ,\\
n_{b}  &  =-2n+2\left(  q+q^{\prime}\right)  -2j,\\
n_{c}  &  =-2n+2\left(  q+q^{\prime}\right)  -2\left(  i-j\right)  ,\\
N  &  =-\left(  2n_{a}+n_{b}+n_{c}\right)  =2i,\\
2a_{0}  &  =k_{1}\cdot D\cdot k_{1}+k_{2}\cdot D\cdot k_{2},\\
F_{N}  &  =\left(  1+\sqrt{\dfrac{1-\cos\phi}{1+\cos\phi}}\right)
^{N}+\left(  1-\sqrt{\dfrac{1-\cos\phi}{1+\cos\phi}}\right)  ^{N}.
\end{align}
The total high energy scattering amplitude can then be calculated to be%
\begin{align}
&  A_{D-Par}=A_{D-Par}^{\left(  0\rightarrow1\right)  }+A_{D-Par}^{\left(
1\rightarrow\infty\right)  }\nonumber\\
&  \simeq\left(  -1\right)  ^{n}4\left(  2i\right)  ^{2a_{0}}\left(
2E\sin\phi\right)  ^{2n}\left(  -\dfrac{1}{8M\sin^{2}\phi}\right)
^{q+q^{\prime}}\nonumber\\
&  \cdot\sum_{i=0}^{q+q^{\prime}}\sum_{j=0}^{i}\binom{q+q^{\prime}}{i}%
\binom{i}{j}\left(  -2\right)  ^{i}\left(  1-\cos\phi\right)  ^{j}\left(
1+\cos\phi\right)  ^{i-j}\cdot2^{-2a-2-N}B\left(  a+1,\frac{b+1}{2}\right)
F_{N}.
\end{align}
The high energy limit of the beta function is%
\begin{equation}
B\left(  a+1,\frac{b+1}{2}\right)  \simeq B\left(  a_{0}+1,\frac{b_{0}+1}%
{2}\right)  \dfrac{a_{0}^{n_{a}}\left(  \dfrac{b_{0}}{2}\right)  ^{n_{b}/2}%
}{\left(  a_{0}+\dfrac{b_{0}}{2}\right)  ^{n_{a}+n_{b}/2}}.
\end{equation}
Finally we get the high energy scattering amplitudes at mass level
$M^{2}=2(n-1)$%
\begin{align}
A_{D-Par}  &  =A_{D-Par}^{\left(  0\rightarrow1\right)  }+A_{D-Par}^{\left(
1\rightarrow\infty\right)  }\nonumber\\
&  =\left(  -1\right)  ^{a_{0}}E^{2n}\left(  -\dfrac{1}{2M}\right)
^{q+q^{\prime}}B\left(  a_{0}+1,\frac{b_{0}+1}{2}\right) \nonumber\\
&  \cdot\sum_{i=0}^{q+q^{\prime}}\sum_{j=0}^{i}\binom{q+q^{\prime}}{i}%
\binom{i}{j}\left(  -2\right)  ^{-i}\left(  1+\cos\phi\right)  ^{i}\left(
-1\right)  ^{j}F_{N}\nonumber\\
&  =2\left(  -1\right)  ^{a_{0}}E^{2n}\left(  -\dfrac{1}{2M}\right)
^{q+q^{\prime}}B\left(  a_{0}+1,\frac{b_{0}+1}{2}\right) \nonumber\\
&  ==2\left(  -1\right)  ^{a_{0}}E^{2n}\left(  -\dfrac{1}{2M}\right)
^{q+q^{\prime}}\frac{\Gamma(a_{0}+1)\Gamma(\frac{b_{0}+1}{2})}{\Gamma
(a_{0}+\frac{b_{0}}{2}+\frac{3}{2})} \label{D-particle}%
\end{align}
where the high energy limit of $B\left(  a_{0}+1,\frac{b_{0}+1}{2}\right)  $
is independent of $q+q^{\prime}$. We thus have explicitly shown that there is
only one independent high energy scattering amplitude at each fixed mass
level. It is a remarkable result that the ratios $\left(  -\dfrac{1}%
{2M}\right)  ^{q+q^{\prime}}$for different high energy scattering amplitudes
at each fixed mass level is consistent with Eq.(\ref{mainA}) for the
scattering without D-brane as expected. In general, for an incoming tensor
state and an outgoing tensor state scatterings, the ratios are $\Sigma
_{i=1}^{2}\left(  -\dfrac{1}{2M_{i}}\right)  ^{(q_{i}+q_{i}^{\prime})}.$

Finally, one notes that the exponential fall-off behavior in energy $E$ is
hidden in the high energy beta function. Since the arguments of $\Gamma
(a_{0}+1)$ and $\Gamma(a_{0}+\frac{b_{0}}{2}+\frac{3}{2})$ in
Eq.(\ref{D-particle}) are negative in the high-energy limit, one needs to use
the well known formula%
\begin{equation}
\Gamma\left(  x\right)  =\frac{\pi}{\sin\left(  \pi x\right)  \Gamma\left(
1-x\right)  }%
\end{equation}
to calculate the large negative $x$ expansion of these $\Gamma$ functions, and
obtain the Regge-pole structure \cite{Closed} of the amplitude. This is to be
compared with the power-law behavior with Regge-pole structure for the
D-domain-wall scattering to be discussed in the next section.

\subsubsection{Hard strings scattered from D-domain-wall}

We have shown, in the last section, that the linear relations for
string/string scatterings persist for the string/D$p$-brane scatterings with
$p\geqslant0.$ In particular, the linear relations for the D-particle
scatterings \cite{Dscatt} were explicitly demonstrated. All the high energy
string/D$p$-brane scattering amplitudes with $p\geqslant0$ behave as
exponential fall-off as was claimed in
\cite{Klebanov,Myers,Klebanov3,barbon1996d,bachas1999high,hirano1997scattering}
In this section, in contrast to the common wisdom, we show that \cite{Wall},
instead of the exponential fall-off behavior of the form factors with
Regge-pole structure, the high energy scattering amplitudes of string
scattered from D$24$-brane, or Domain-wall, behave as \textit{power-law} with
Regge-pole structure. This is to be compared with the well-known power-law
form factors without Regge-pole structure of the D-instanton scatterings to be
discussed in Eq.(\ref{st.}) below.

This discovery makes Domain-wall scatterings an unique example of a hybrid of
string and field theory scatterings. Our calculation will be done for bosonic
string scatterings of arbitrary massive string states from D$24$-brane.
Moreover, we discover that the usual linear relations \cite{Dscatt} of high
energy string scattering amplitudes at each fixed mass level,
Eq.(\ref{D-particle}), breaks down for the Domain-wall scatterings
\cite{Wall}. This result gives a strong evidence that the existence of the
infinite linear relations, or stringy symmetries, of high energy string
scattering amplitudes is responsible for the softer, exponential fall-off high
energy string scatterings than the power-law field theory scatterings.

We consider an incoming tachyon closed string state with momentum $k_{1}$ and
an angle of incidence $\phi$ and an outgoing massive closed string state
$\left(  \alpha_{-1}^{T}\right)  ^{n-2q}\left(  \alpha_{-2}^{L}\right)
^{q}\otimes\left(  \tilde{\alpha}_{-1}^{T}\right)  ^{n-2q^{\prime}}\left(
\tilde{\alpha}_{-2}^{L}\right)  ^{q^{\prime}}\left\vert 0\right\rangle $ with
momentum $k_{2}$ and an angle of reflection $\theta$. The kinematic setup is%
\begin{align}
e^{P}  &  =\frac{1}{M}\left(  -E,\mathrm{k}_{2}\cos\theta,-\mathrm{k}_{2}%
\sin\theta\right)  =\frac{k_{2}}{M},\\
e^{L}  &  =\frac{1}{M}\left(  -\mathrm{k}_{2},E\cos\theta,-E\sin\theta\right)
,\\
e^{T}  &  =\left(  0,\sin\theta,\cos\theta\right)  ,\\
k_{1}  &  =\left(  E,-\mathrm{k}_{1}\cos\phi,-\mathrm{k}_{1}\sin\phi\right)
,\\
k_{2}  &  =\left(  -E,\mathrm{k}_{2}\cos\theta,-\mathrm{k}_{2}\sin
\theta\right)  .
\end{align}
In the high energy limit, the angle of incidence $\phi$ is identified to the
angle of reflection $\theta$, and $e^{P}$ approaches $e^{L}$, $\mathrm{k}%
_{1},\mathrm{k}_{2}\simeq E$. For the case of Domain-wall scattering $Diag$
$D_{\mu\nu}=(-1,1,-1)$, and we have%
\begin{align}
a_{0}  &  \equiv k_{1}\cdot D\cdot k_{1}\sim-2E^{2}\sin^{2}\phi-2M_{1}^{2}%
\cos^{2}\phi+M_{1}^{2},\\
b_{0}  &  \equiv2k_{1}\cdot k_{2}+1\nonumber\\
&  \sim4E^{2}\sin^{2}\phi+4M_{1}^{2}\cos^{2}\phi-\left(  M_{1}^{2}%
+M^{2}\right)  +1,
\end{align}
The scattering amplitude can be calculated to be%

\begin{align}
A_{D-Wall}  &  =\left(  -1\right)  ^{q+q^{\prime}}\int d^{2}z_{1}d^{2}%
z_{2}\left(  z_{1}-\bar{z}_{1}\right)  ^{k_{1}\cdot D\cdot k_{1}}\left(
z_{2}-\bar{z}_{2}\right)  ^{k_{2}\cdot D\cdot k_{2}}\left\vert z_{1}%
-z_{2}\right\vert ^{2k_{1}\cdot k_{2}}\left\vert z_{1}-\bar{z}_{2}\right\vert
^{2k_{1}\cdot D\cdot k_{2}}\nonumber\\
&  \cdot\left[  \frac{ie^{T}\cdot k_{1}}{z_{1}-z_{2}}+\frac{ie^{T}\cdot D\cdot
k_{1}}{\bar{z}_{1}-z_{2}}+\frac{ie^{T}\cdot D\cdot k_{2}}{\bar{z}_{2}-z_{2}%
}\right]  ^{n-2q}\nonumber\\
&  \cdot\left[  \frac{ie^{T}\cdot D\cdot k_{1}}{z_{1}-\bar{z}_{2}}%
+\frac{ie^{T}\cdot k_{1}}{\bar{z}_{1}-\bar{z}_{2}}+\frac{ie^{T}\cdot D\cdot
k_{2}}{z_{2}-\bar{z}_{2}}\right]  ^{n-2q^{\prime}}\nonumber\\
&  \cdot\left[  \frac{e^{L}\cdot k_{1}}{\left(  z_{1}-z_{2}\right)  ^{2}%
}+\frac{e^{L}\cdot D\cdot k_{1}}{\left(  \bar{z}_{1}-z_{2}\right)  ^{2}}%
+\frac{e^{L}\cdot D\cdot k_{2}}{\left(  \bar{z}_{2}-z_{2}\right)  ^{2}%
}\right]  ^{q}\left[  \frac{e^{L}\cdot D\cdot k_{1}}{\left(  z_{1}-\bar{z}%
_{2}\right)  ^{2}}+\frac{e^{L}\cdot k_{1}}{\left(  \bar{z}_{1}-\bar{z}%
_{2}\right)  ^{2}}+\frac{e^{L}\cdot D\cdot k_{2}}{\left(  z_{2}-\bar{z}%
_{2}\right)  ^{2}}\right]  ^{q^{\prime}}.
\end{align}
Set $z_{1}=iy$ and $z_{2}=i$ to fix the $SL(2,R)$ gauge, and include the
Jacobian $d^{2}z_{1}d^{2}z_{2}\rightarrow4\left(  1-y^{2}\right)  dy$, we
have, for the $\left(  0\rightarrow1\right)  $ channel,%
\begin{align}
&  A_{D-Wall}^{(0\rightarrow1)}\nonumber\\
&  \simeq4\left(  2i\right)  ^{k_{1}\cdot D\cdot k_{1}+k_{2}\cdot D\cdot
k_{2}}\left(  \frac{E\sin2\phi}{2}\right)  ^{2n}\left(  \frac{1}{2M\cos
^{2}\phi}\right)  ^{q+q^{\prime}}\nonumber\\
&  \cdot\sum_{i=0}^{q+q^{\prime}}\binom{q+q^{\prime}}{i}2^{i}\int_{0}%
^{1}dy\text{ }y^{k_{2}\cdot D\cdot k_{2}}\left(  1-y\right)  ^{2k_{1}\cdot
k_{2}+1}\nonumber\\
&  \cdot\left(  1+y\right)  ^{2k_{1}\cdot D\cdot k_{2}+1}\left[  \frac
{1+y}{1-y}\right]  ^{2n-\left(  q+q^{\prime}\right)  }\left(  \frac{1}%
{1-y}\right)  ^{i}\nonumber\\
&  \simeq\left(  \frac{E\sin2\phi}{2}\right)  ^{2n}\left(  \frac{1}{2M\cos
^{2}\phi}\right)  ^{q+q^{\prime}}\nonumber\\
&  \cdot\sum_{i=0}^{q+q^{\prime}}\binom{q+q^{\prime}}{i}\cdot B\left(
a_{0}+1,\frac{b+1}{2}\right)  F_{i} \label{27}%
\end{align}
where%
\begin{align}
b  &  =b_{0}+n_{b}=b_{0}-2n+\left(  q+q^{\prime}\right)  -i,\label{28}\\
F_{i}  &  \equiv\left(  1+\sqrt{\left\vert \dfrac{b}{2a_{0}+b}\right\vert
}\right)  ^{i}+\left(  1-\sqrt{\left\vert \dfrac{b}{2a_{0}+b}\right\vert
}\right)  ^{i}\label{fi2}\\
\simeq &  \left[  \left(  1+2C_{i}E\sin\phi\right)  ^{i}+\left(  1-2C_{i}%
E\sin\phi\right)  ^{i}\right]  \label{fi}%
\end{align}
with%
\begin{equation}
C_{i}\equiv\sqrt{\left\vert \frac{1}{M_{1}^{2}-M^{2}+1-2n+\left(  q+q^{\prime
}\right)  -i}\right\vert }.
\end{equation}
$F_{i}$ in Eq.(\ref{fi2}) is the high energy limit of the generalized
hypergeometric function $_{3}F_{2}\left(  \frac{b+1}{2},-\left[  \frac{i}%
{2}\right]  ,\dfrac{1}{2}-\left[  \frac{i}{2}\right]  ;a_{0}+\frac{b+3}%
{2},\dfrac{1}{2};1\right)  $ \cite{Dscatt}.

At this stage, it is crucial to note that%
\begin{equation}
b\simeq b_{0}\simeq-2a_{0} \label{match}%
\end{equation}
in the high energy limit for the Domain-wall scatterings. As a result, $F_{i}$
reduces to the form of Eq.(\ref{fi}), and depends on the energy $E.$ Thus in
contrast to the generic D$p$-brane scatterings with $p\geq0$, which contain
two independent kinematic variables, there is only \textit{one} kinematic
variable for the special case of Domain-wall scatterings. It thus becomes
meaningless to study high energy, fixed angle scattering process for the
Domain-wall scatterings. As we will see in the following calculation, this
peculiar property will reduce the high energy beta function in Eq.(\ref{27})
from exponential to power-law behavior and, simultaneously, breaks down the
linear relations as we had in Eq.(\ref{D-particle}) for the D-particle scatterings.

Finally, the scattering amplitude for the $(0\rightarrow1)$ channel can be
calculated to be (similar result can be obtained for the $(1\rightarrow
\infty)$ channel)%
\begin{align}
&  A_{D-Wall}^{(0\rightarrow1)}\nonumber\\
&  \simeq\left(  \frac{E\sin2\phi}{2}\right)  ^{2n}\left(  \frac{1}{2M\cos
^{2}\phi}\right)  ^{q+q^{\prime}}B\left(  a_{0}+1,\frac{b_{0}+1}{2}\right)
\nonumber\\
&  \cdot\sum_{i=0}^{q+q^{\prime}}\binom{q+q^{\prime}}{i}\cdot\dfrac{\left(
a_{0}\right)  _{0}\left(  \dfrac{b_{0}}{2}\right)  _{n_{b}/2}}{\left(
a_{0}+\dfrac{b_{0}}{2}\right)  _{n_{b}/2}}\left(  1+2C_{i}E\sin\phi\right)
^{i}\label{frac}\\
&  \simeq\left(  \frac{\cos\phi}{\sqrt{2}}\right)  ^{2n}\left(  \frac
{E\sin\phi}{M\sqrt{\left\vert M_{1}^{2}-2M^{2}-1\right\vert }\cos^{2}\phi
}\right)  ^{q+q^{\prime}}\nonumber\\
&  \cdot\frac{\Gamma(a_{0}+1)\Gamma(\frac{b_{0}+1}{2})}{\Gamma(a_{0}%
+\frac{b_{0}}{2}+\frac{3}{2})}\dfrac{1}{\left(  \frac{M_{1}^{2}-M^{2}+1}%
{2}\right)  _{-n}} \label{am}%
\end{align}
where $(\alpha)_{n}\equiv\frac{\Gamma(\alpha+n)}{\Gamma(\alpha)}$ for integer
$n$. On the other hand, since the argument of $\Gamma(a_{0}+1)$ in
Eq.(\ref{am}) is negative in the high energy limit, we have, by using
Eq.(\ref{gama}) and Eq.(\ref{match}),%
\begin{align}
\frac{\Gamma(a_{0}+1)\Gamma(\frac{b_{0}+1}{2})}{\Gamma(a_{0}+\frac{b_{0}}%
{2}+\frac{3}{2})}  &  \simeq\frac{\pi}{\sin\left(  \pi a_{0}\right)
\Gamma\left(  -a_{0}\right)  }\frac{\Gamma(\frac{b_{0}+1}{2})}{\Gamma
(\frac{M_{1}^{2}-M^{2}+1}{2})}\nonumber\\
&  \sim\frac{1}{\sin\left(  \pi a_{0}\right)  }\frac{1}{(E\sin\phi)^{2(n-1)}}.
\label{pole}%
\end{align}
Note that the $\sin\left(  \pi a_{0}\right)  $ factor in the denominator of
Eq.(\ref{pole}) gives the Regge-pole structure, and the energy dependence
$E^{-2(n-1)}$ gives the power-law behavior in the high energy limit. As a
result, the scattering amplitude for the Domain-wall in Eq.(\ref{am}) behaves
like \textit{power-law with the Regge-pole structure}.

The crucial differences between the Domain-wall scatterings in Eq.(\ref{am})
and the D-particle scatterings (or any other D$p$-brane scatterings except
Domain-wall and D-instanton scatterings) in Eq.(\ref{D-particle}) is the
kinematic relation Eq.(\ref{match}). For the case of D-particle scatterings
\cite{Dscatt}, the corresponding factors for both $F_{i}$ in Eq.(\ref{fi2})
and the fraction in Eq.(\ref{frac}) are independent of energy in the high
energy limit, and, as a result, the amplitudes contain no $q+q^{\prime}$
dependent energy power factor. So one gets the high energy linear relations
for the D-particle scattering amplitudes. On the contrary, for the case of
Domain-wall scatterings, both $F_{i}$ in Eq.(\ref{fi}) and the fraction in
Eq.(\ref{frac}) depend on energy due to the condition Eq.(\ref{match}). The
summation in Eq.(\ref{frac}) is then dominated by the term $i=q+q^{\prime}$,
and the whole scattering amplitude Eq.(\ref{am}) contains a $q+q^{\prime}$
dependent energy power factor. As a result, the usual linear relations for the
high energy scattering amplitudes break down for the Domain-wall scatterings.

It is crucial to note that the mechanism, Eq.(\ref{match}), to drive the
exponential fall-off form factor of the D-particle scatterings to the
power-law one of the Domain-wall scatterings is exactly the same as the
mechanism to break down the expected linear relations for the domain-wall
scatterings in the high energy limit. In conclusion, this result gives a
strong evidence that the existence of the infinite linear relations, or
stringy symmetries, of high energy string scattering amplitudes is responsible
for the softer, exponential fall-off high energy string scatterings than the
power-law field theory scatterings.

Another interesting case of D-brane scatterings is the massless form factor of
scatterings of D-instanton
\cite{Klebanov,Myers,Klebanov3,barbon1996d,bachas1999high,hirano1997scattering}%
\begin{equation}
\frac{\Gamma(s)\Gamma(t)}{\Gamma(s+t+1)}\rightarrow\frac{1}{st},\text{ as
}s\rightarrow0\text{,} \label{st.}%
\end{equation}
which contains no Regge-pole structure. In Eq.(\ref{st.}), $s,t$ are the
Mandelstam variables. Eq.(\ref{st.}) can be easily generalized to the
scatterings of arbitrary massive string states in the high energy limit. To
compare the D-instanton scatterings with the Domain-wall scatterings in
Eq.(\ref{pole}), one notes that in both cases there is only \textit{one}
kinematic variable and, as a result, behave as power-law at high energies
\cite{Decay}.

On the other hand, since $t$ is large negative in the high energy limit
\cite{Closed}, the application of Eq.(\ref{gama}) to Eq.(\ref{st.}) produces
no $\sin\left(  \pi a_{0}\right)  $ factor in contrast to the Domain-wall
scatterings. So there is \textit{no Regge-pole structure} for the D-instanton
scatterings. We conclude that the very condition of Eq.(\ref{match}) makes
Domain-wall scatterings an unique example of a hybrid of string and field
theory scatterings.

\subsubsection{A brief review of $_{2}F_{1}$ and $_{3}F_{2}\label{review of F}%
$}

In this section, we review the definitions and some formulas of hypergeometric
function $_{2}F_{1}$ and generalized hypergeometric function $_{3}F_{2}$ which
we used in the text. hypergeometric functions form an important class of
special functions. Many elementary special functions are special cases of
$_{2}F_{1}$. The hypergeometric function $_{2}F_{1}$ is defined to be
($\alpha,\beta,\gamma$ constant)%

\begin{align}
_{2}F_{1}(\alpha,\beta,\gamma;x)  &  =1+\frac{\alpha\beta}{\gamma}\frac{x}%
{1!}+\frac{\alpha(\alpha+1)\beta(\beta+1)}{\gamma(\gamma+1)}\frac{x^{2}}%
{2!}+\cdot\cdot\cdot=\sum_{n=0}^{\infty}\frac{(\alpha)_{n}(\beta)_{n}}%
{(\gamma)_{n}}\frac{x^{n}}{n!}\nonumber\\
&  =\frac{\Gamma(\gamma)}{\Gamma(\alpha)\Gamma(\beta)}\sum_{n=0}^{\infty}%
\frac{\Gamma(\alpha+n)\Gamma(\beta+n)}{\Gamma(\gamma+n)}\frac{x^{n}}{n!}%
\end{align}
where%
\begin{equation}
(\alpha)_{0}=1,(\alpha)_{n}=\alpha(\alpha+1)(\alpha+2)\cdot\cdot\cdot
(\alpha+n-1)=\frac{(\alpha+n-1)!}{(\alpha-1)!}.
\end{equation}
The hypergeometric function $_{2}F_{1}$ is a solution, at the singular point
$x=0$ with indicial root $r=0$, of the Gauss's hypergeometric differential
equation%
\begin{equation}
x(1-x)u^{\prime\prime}+\left[  \gamma-(\alpha+\beta+1)\right]  u^{\prime
}-\alpha\beta u=0, \label{DE}%
\end{equation}
which contains three regular singularities $x=0,1,\infty$. The second solution
of Eq. (\ref{DE}) with indicial root $r=1-\gamma$ can be expressed in terms of
$_{2}F_{1}$ as following ($\gamma\neq$ integer)%
\begin{equation}
u_{2}(x)=x^{1-\gamma}2F_{1}(\alpha-\gamma+1,\beta-\gamma+1,2-\gamma,x).
\end{equation}
Other solutions of Eq. (\ref{DE}), which corresponds to singularities
$x=1,\infty$, can also be expressed in terms of the hypergeometric function
$_{2}F_{1}$. The following identity
\begin{equation}
_{2}F_{1}(\alpha,\beta,\gamma;x)=(1-x)^{\gamma-\alpha-\beta}\text{ }_{2}%
F_{1}(\gamma-\alpha,\gamma-\beta,\gamma;x),
\end{equation}
which we used in the text can then be derived.

$_{2}F_{1\text{ }}$has an integral representation%
\begin{equation}
_{2}F_{1}(\alpha,\beta,\gamma;x)=\frac{\Gamma(\gamma)}{\Gamma(\beta
)\Gamma(\gamma-\beta)}\int_{0}^{1}dy\text{ }y^{\beta-1}\left(  1-y\right)
^{\gamma-\beta-1}\left(  1-yx\right)  ^{-\alpha}, \label{Key}%
\end{equation}
which can be used to do analytic continuation. Eq.(\ref{Key}) with $x=-1$ was
repeatedly used in the text in our calculations of string scattering
amplitudes with D-brane.

There exists interesting relations among hypergeometric function $_{2}F_{1}$
with different arguments%
\begin{equation}
x^{-p}(1-x)^{-q}2F_{1}(\alpha,\beta,\gamma;x)=t^{-p^{\prime}}(1-t)^{-q^{\prime
}}\text{ }_{2}F_{1}(\alpha^{\prime},\beta^{\prime},\gamma^{\prime};t),
\end{equation}
where $x=\varphi(t)$ is an algebraic function with degree up to six. As an
example, the quadratic transformation formula%
\begin{equation}
_{2}F_{1}(\alpha,\beta,1+\alpha-\beta;x)=(1-x)^{-\alpha}\text{ }_{2}%
F_{1}\left(  \frac{\alpha}{2},\frac{1+\alpha-2\beta}{2},1+\alpha-\beta
;\frac{-4x}{(1-x)^{2}}\right)  ,
\end{equation}
can be used to derive the Kummer's relation%
\begin{equation}
_{2}F_{1}(\alpha,\beta,1+\alpha-\beta,-1)=\frac{\Gamma(1+\alpha-\beta
)\Gamma(1+\frac{\alpha}{2})}{\Gamma(1+\alpha)\Gamma(1+\frac{\alpha}{2}-\beta
)},
\end{equation}
which is crucial to reduce the scattering amplitudes of string from D-brane to
the usual beta function.

In summing up the $(0\rightarrow1)$ and $(1\rightarrow\infty)$ channel
scattering amplitudes, we have used the master formula%

\begin{align}
I  &  =I_{\left(  0\rightarrow1\right)  }+I_{\left(  1\rightarrow
\infty\right)  }\nonumber\\
&  =2^{-2a-1-N}\int_{0}^{1}dt\text{ }t^{b}\left(  1-t\right)  ^{a}\left(
1+t\right)  ^{a}\left[  \left(  1+t\right)  ^{N}+\left(  1-t\right)
^{N}\right] \nonumber\\
&  =2^{-2a-1-N}\sum_{n=0}^{\left[  \frac{N}{2}\right]  }\binom{N}{2n}\cdot
B\left(  a+1,\frac{b+1}{2}+n\right) \nonumber\\
&  =2^{-2a-1-N}\cdot B\left(  a+1,\frac{b+1}{2}\right)  \cdot\text{ }_{3}%
F_{2}\left(  \frac{b+1}{2},-\left[  \frac{N}{2}\right]  ,\dfrac{1}{2}-\left[
\frac{N}{2}\right]  ;a+\frac{b+3}{2},\dfrac{1}{2};1\right)  \label{3F2}%
\end{align}
In Eq.(\ref{3F2}), $B$ is the beta function and $_{3}F_{2}$ is the generalized
hypergeometric function, which is defined to be%
\begin{equation}
_{3}F_{2}(\alpha_{1},\alpha_{2},\alpha_{3};\gamma_{1},\gamma_{2};x)=\sum
_{n=0}^{\infty}\frac{(\alpha_{1})_{n}(\alpha_{2})_{n}(\alpha_{3})_{n}}%
{(\gamma_{1})_{n}(\gamma_{2})_{n}}\frac{x^{n}}{n!}.
\end{equation}
For those arguments of $_{3}F_{2}$ in Eq. (\ref{3F2}), the series of the
generalized hypergeometric function $_{3}F_{2}$ terminates to a finite sum.
For example,%
\begin{align}
N  &  =0:\text{ }_{3}F_{2}=1,\nonumber\\
N  &  =1:\text{ }_{3}F_{2}=1,\nonumber\\
N  &  =2:\text{ }_{3}F_{2}=\dfrac{a+b+2}{a+\frac{b+3}{2}},\nonumber\\
N  &  =3:\text{ }_{3}F_{2}=\dfrac{a+2b+3}{a+\frac{b+3}{2}},\nonumber\\
N  &  =4:\text{ }_{3}F_{2}=\dfrac{a^{2}+4ab+2b^{2}+7a+12b+12}{\left(
a+\frac{b+3}{2}\right)  \left(  a+\frac{b+5}{2}\right)  },\nonumber\\
N  &  =5:\text{ }_{3}F_{2}=\dfrac{a^{2}+6ab+9b^{2}+4a+22b+20}{\left(
a+\frac{b+3}{2}\right)  \left(  a+\frac{b+5}{2}\right)  }.
\end{align}

\subsection{Scatterings from O-planes}

Being a consistent theory of quantum gravity, string theory is remarkable for
its soft ultraviolet structure. This is mainly due to two closely related
fundamental characteristics of high-energy string scattering amplitudes. The
first is the softer exponential fall-off behavior of the form factors of
high-energy string scatterings in contrast to the power-law field theory
scatterings. The second is the existence of infinite Regge poles in the form
factor of string scattering amplitudes. The existence of infinite linear
relations discussed in part II of the review constitutes the \textit{third}
fundamental characteristics of high energy string scatterings.

In the previous section, we showed that these linear relations persist
\cite{Dscatt} for string scattered from generic D$p$-brane \cite{Klebanov}
except D-instanton and D-domain-wall. For the scattering of D-instanton, the
form factor exhibits the well-known power-law behavior without Regge pole
structure, and thus resembles a field theory amplitude. For the special case
of D-domain-wall scattering \cite{Myers}, it was discovered \cite{Wall} that
its form factor behaves as\textit{ power-law} with infinite \textit{open}
Regge pole structure at high energies. This discovery makes D-domain-wall
scatterings an unique example of a hybrid of string and field theory scatterings.

Moreover, it was shown \cite{Wall} that the linear relations break down for
the D-domain-wall scattering due to this unusual power-law behavior. This
result seems to imply the coexistence of linear relations and soft UV
structure of string scatterings. In order to further uncover the mysterious
relations among these three fundamental characteristics of string scatterings,
namely, the soft UV structure, the existence of infinite Regge poles and the
newly discovered linear relations stated above, it will be important to study
more string scatterings, which exhibit the unusual behaviors in the high
energy limit.

In this section, we calculate massive closed string states at arbitrary mass
levels scattered from Orientifold planes in the high energy, fixed angle limit
\cite{O-plane}. The scatterings of massless states from Orientifold planes
were calculated previously by using the boundary states formalism
\cite{Craps,Craps1,Craps2,Schnitzer}, and on the worldsheet of real projected
plane $RP_{2}$ \cite{Garousi}. Many speculations were made about the
scatterings of \textit{massive} string states, in particular, for the case of
O-domain-wall scatterings. It is one of the purposes of this section to
clarify these speculations and to discuss their relations with the three
fundamental characteristics of high energy string scatterings.

For the generic O$p$-planes with $p\geq0$, one expects to get the infinite
linear relations except O-domain-wall scatterings. For simplicity, we consider
only the case of O-particle scatterings \cite{O-plane}. For the case of
O-particle scatterings, we will obtain infinite linear relations among high
energy scattering amplitudes of different string states. We also confirm that
there exist only $t$-channel closed string Regge poles in the form factor of
the O-particle scatterings amplitudes as expected.

For the case of O-domain-wall scatterings, we find that, like the well-known
D-instanton scatterings, the amplitudes behave like field theory scatterings,
namely \textit{UV power-law without Regge pole}. In addition, we will show
that there exist only finite number of $t$-channel closed string poles in the
form factor of O-domain-wall scatterings, and the masses of the poles are
bounded by the masses of the external legs \cite{O-plane}. We thus confirm
that all massive closed string states do couple to the O-domain-wall as was
conjectured previously \cite{Myers,Garousi}. This is also consistent with the
boundary state descriptions of O-planes.

For both cases of O-particle and O-domain-wall scatterings, we confirm that
there exist no $s$-channel open string Regge poles in the form factor of the
amplitudes as O-planes were known to be not dynamical. However, the usual
claim that there is a thickness of order$\sqrt{\alpha^{^{\prime}}}$ for the
O-domain-wall is misleading as the UV behavior of its scatterings is power-law
instead of exponential fall-off.

\subsubsection{Hard strings scattered from O-particle}

We use the real projected plane $RP_{2}$ as the worldsheet diagram for the
scatterings of Orientifold planes \cite{O-plane}. The standard propagators of
the left and right moving fields are%
\begin{align}
\left\langle X^{\mu}\left(  z\right)  X^{\nu}\left(  w\right)  \right\rangle
&  =-\eta^{\mu\nu}\log\left(  z-w\right)  ,\\
\left\langle \tilde{X}^{\mu}\left(  \bar{z}\right)  \tilde{X}^{\nu}\left(
\bar{w}\right)  \right\rangle  &  =-\eta^{\mu\nu}\log\left(  \bar{z}-\bar
{w}\right)  .
\end{align}
In addition, there are also nontrivial correlator between the right and left
moving fields as well%
\begin{equation}
\left\langle X^{\mu}\left(  z\right)  \tilde{X}^{\nu}\left(  \bar{w}\right)
\right\rangle =-D^{\mu\nu}\ln\left(  1+z\bar{w}\right)  .
\end{equation}

As in the usual convention
\cite{Klebanov,Myers,Klebanov3,barbon1996d,bachas1999high,hirano1997scattering}%
, the matrix $D$ reverses the sign for fields satisfying Dirichlet boundary
condition. The wave functions of a tensor at general mass level can be written
as%
\begin{equation}
T_{\mu_{1}\cdots\mu_{n}}=\dfrac{1}{2}\left[  \varepsilon_{\mu_{1}\cdots\mu
_{n}}e^{ik\cdot x}+\left(  D\cdot\varepsilon\right)  _{\mu_{1}}\cdots\left(
D\cdot\varepsilon\right)  _{\mu_{n}}e^{iD\cdot k\cdot x}\right]
\end{equation}
where%
\begin{equation}
\varepsilon_{\mu_{1}\cdots\mu_{n}}\equiv\varepsilon_{\mu_{1}}\cdots
\varepsilon_{\mu_{n}}.
\end{equation}

The vertex operators corresponding to the above wave functions are%
\begin{equation}
V\left(  \varepsilon,k,z,\bar{z}\right)  =\dfrac{1}{2}\left[  \varepsilon
_{\mu_{1}\cdots\mu_{n}}V^{\mu_{1}\cdots\mu_{n}}\left(  k,z,\bar{z}\right)
+\left(  D\cdot\varepsilon\right)  _{\mu_{1}}\cdots\left(  D\cdot
\varepsilon\right)  _{\mu_{n}}V^{\mu_{1}\cdots\mu_{n}}\left(  D\cdot
k,z,\bar{z}\right)  \right]  .
\end{equation}

For simplicity, we are going to calculate one tachyon and one massive closed
string state scattered from the O-particle in the high energy limit. One
expects to get similar results for the generic O$p$-plane scatterings with
$p\geq0$ except O-domain-wall scatterings, which will be discussed in the next
section. For this case $D_{\mu\nu}=-\delta_{\mu\nu}$, and the kinematic setup
are%
\begin{align}
e^{P}  &  =\frac{1}{M}\left(  -E,-\mathrm{k}_{2},0\right)  =\frac{k_{2}}{M},\\
e^{L}  &  =\frac{1}{M}\left(  -\mathrm{k}_{2},-E,0\right)  ,\\
e^{T}  &  =\left(  0,0,1\right)  ,\\
k_{1}  &  =\left(  E,\mathrm{k}_{1}\cos\phi,-\mathrm{k}_{1}\sin\phi\right)
,\\
k_{2}  &  =\left(  -E,-\mathrm{k}_{2},0\right)
\end{align}
where $e^{P}$, $e^{L}$ and $e^{T}$ are polarization vectors of the tensor
state $k_{2}$ on the high energy scattering plane. One can easily calculate
the following kinematic relations in the high energy limit%
\begin{align}
e^{T}\cdot k_{2}  &  =e^{L}\cdot k_{2}=0,\\
e^{T}\cdot k_{1}  &  =-\mathrm{k}_{1}\sin\phi\sim-E\sin\phi,\\
e^{T}\cdot D\cdot k_{1}  &  =\mathrm{k}_{1}\sin\phi\sim E\sin\phi,\\
e^{T}\cdot D\cdot k_{2}  &  =0,\\
e^{L}\cdot k_{1}  &  =\frac{1}{M}\left[  \mathrm{k}_{2}E-\mathrm{k}_{1}%
E\cos\phi\right]  \sim\frac{E^{2}}{M}\left(  1-\cos\phi\right)  ,\\
e^{L}\cdot D\cdot k_{1}  &  =\frac{1}{M}\left[  \mathrm{k}_{2}E+\mathrm{k}%
_{1}E\cos\phi\right]  \sim\frac{E^{2}}{M}\left(  1+\cos\phi\right)  ,\\
e^{L}\cdot D\cdot k_{2}  &  =\frac{1}{M}\left[  -\mathrm{k}_{2}E-\mathrm{k}%
_{2}E\right]  \sim-\frac{2E^{2}}{M}.
\end{align}

We define%
\begin{align}
a_{0}  &  \equiv k_{1}\cdot D\cdot k_{1}=-E^{2}-\mathrm{k}_{1}^{2}\sim
-2E^{2},\\
a_{0}^{\prime}  &  \equiv k_{2}\cdot D\cdot k_{2}=-E^{2}-\mathrm{k}_{2}%
^{2}\sim-2E^{2},\\
b_{0}  &  \equiv k_{1}\cdot k_{2}=\left(  E^{2}-\mathrm{k}_{1}\mathrm{k}%
_{2}\cos\phi\right)  \sim E^{2}\left(  1-\cos\phi\right)  ,\\
c_{0}  &  \equiv k_{1}\cdot D\cdot k_{2}=\left(  E^{2}+\mathrm{k}%
_{1}\mathrm{k}_{2}\cos\phi\right)  \sim E^{2}\left(  1+\cos\phi\right)  ,
\end{align}
and the Mandelstam variables can be calculated to be%
\begin{align}
t  &  \equiv-\left(  k_{1}+k_{2}\right)  ^{2}=M_{1}^{2}+M_{2}^{2}-2k_{1}\cdot
k_{2}=M_{2}^{2}-2\left(  1+b_{0}\right)  ,\\
s  &  \equiv\dfrac{1}{2}k_{1}\cdot D\cdot k_{1}=\dfrac{1}{2}a_{0},\\
u  &  =-2k_{1}\cdot D\cdot k_{2}=-2c_{0}.
\end{align}

In the high energy limit, we will consider an incoming tachyon state $k_{1}$
and an outgoing tensor state $k_{2}$ of the following form%
\begin{equation}
\left(  \alpha_{-1}^{T}\right)  ^{n-2q}\left(  \alpha_{-2}^{L}\right)
^{q}\otimes\left(  \tilde{\alpha}_{-1}^{T}\right)  ^{n-2q^{\prime}}\left(
\tilde{\alpha}_{-2}^{L}\right)  ^{q^{\prime}}\left\vert 0\right\rangle .
\label{dodo}%
\end{equation}
For simplicity, we have omitted above a possible high energy vertex
$(\alpha_{-1}^{L})^{r}\otimes(\tilde{\alpha}_{-1}^{L})^{r^{\prime}}$
\cite{Dscatt,Compact}. For this case, with momentum conservation on the
O-planes, we have%
\begin{equation}
a_{0}+b_{0}+c_{0}=M_{1}^{2}=-2. \label{conserve}%
\end{equation}

The high energy scattering amplitude can then be written as
\begin{align*}
A^{RP_{2}}  &  =\int d^{2}z_{1}d^{2}z_{2}\dfrac{1}{2}\left[  V\left(
k_{1},z_{1}\right)  \tilde{V}\left(  k_{1},\bar{z}_{1}\right)  +V\left(
D\cdot k_{1},z_{1}\right)  \tilde{V}\left(  D\cdot k_{1},\bar{z}_{1}\right)
\right] \\
&  \cdot\dfrac{1}{2}\varepsilon_{T^{n-2q}L^{q},T^{n-2q^{\prime}}L^{q^{\prime}%
}}V^{T^{n-2q}L^{q}}\left(  k_{2},z_{2}\right)  \tilde{V}^{T^{n-2q^{\prime}%
}L^{q^{\prime}}}\left(  k_{2},\bar{z}_{2}\right) \\
&  +\left(  D\cdot\varepsilon_{T}\right)  ^{n-2q}\left(  D\cdot\varepsilon
_{L}\right)  ^{q}\left(  D\cdot\tilde{\varepsilon}_{T}\right)  ^{n-2q^{\prime
}}\left(  D\cdot\tilde{\varepsilon}_{L}\right)  ^{q^{\prime}}V^{T^{n-2q}L^{q}%
}\left(  D\cdot k_{2},z_{2}\right) \\
&  \cdot\tilde{V}^{T^{n-2q^{\prime}}L^{q^{\prime}}}\left(  D\cdot k_{2}%
,\bar{z}_{2}\right) \\
&  =A_{1}+A_{2}+A_{3}+A_{4}%
\end{align*}
where%
\begin{align}
A_{1}  &  =\dfrac{1}{4}\varepsilon_{T^{n-2q}L^{q},T^{n-2q^{\prime}%
}L^{q^{\prime}}}\int d^{2}z_{1}d^{2}z_{2}\nonumber\\
&  \cdot\left\langle V\left(  k_{1},z_{1}\right)  \tilde{V}\left(  k_{1}%
,\bar{z}_{1}\right)  V^{T^{n-2q}L^{q}}\left(  k_{2},z_{2}\right)  \tilde
{V}^{T^{n-2q^{\prime}}L^{q^{\prime}}}\left(  k_{2},\bar{z}_{2}\right)
\right\rangle ,\\
A_{2}  &  =\dfrac{1}{4}\varepsilon_{T^{n-2q}L^{q},T^{n-2q^{\prime}%
}L^{q^{\prime}}}\int d^{2}z_{1}d^{2}z_{2}\nonumber\\
&  \cdot\left\langle V\left(  D\cdot k_{1},z_{1}\right)  \tilde{V}\left(
D\cdot k_{1},\bar{z}_{1}\right)  V^{T^{n-2q}L^{q}}\left(  k_{2},z_{2}\right)
\tilde{V}^{T^{n-2q^{\prime}}L^{q^{\prime}}}\left(  k_{2},\bar{z}_{2}\right)
\right\rangle ,\\
A_{3}  &  =\dfrac{1}{4}\left(  D\cdot\varepsilon_{T}\right)  ^{n-2q}\left(
D\cdot\varepsilon_{L}\right)  ^{q}\left(  D\cdot\tilde{\varepsilon}%
_{T}\right)  ^{n-2q^{\prime}}\left(  D\cdot\tilde{\varepsilon}_{L}\right)
^{q^{\prime}}\nonumber\\
&  \cdot\int d^{2}z_{1}d^{2}z_{2}\left\langle V\left(  k_{1},z_{1}\right)
\tilde{V}\left(  k_{1},\bar{z}_{1}\right)  V^{T^{n-2q}L^{q}}\left(  D\cdot
k_{2},z_{2}\right)  \tilde{V}^{T^{n-2q^{\prime}}L^{q^{\prime}}}\left(  D\cdot
k_{2},\bar{z}_{2}\right)  \right\rangle ,\\
A_{4}  &  =\dfrac{1}{4}\left(  D\cdot\varepsilon_{T}\right)  ^{n-2q}\left(
D\cdot\varepsilon_{L}\right)  ^{q}\left(  D\cdot\tilde{\varepsilon}%
_{T}\right)  ^{n-2q^{\prime}}\left(  D\cdot\tilde{\varepsilon}_{L}\right)
^{q^{\prime}}\nonumber\\
&  \cdot\int d^{2}z_{1}d^{2}z_{2}\left\langle V\left(  D\cdot k_{1}%
,z_{1}\right)  \tilde{V}\left(  D\cdot k_{1},\bar{z}_{1}\right)
V^{T^{n-2q}L^{q}}\left(  D\cdot k_{2},z_{2}\right)  \tilde{V}^{T^{n-2q^{\prime
}}L^{q^{\prime}}}\left(  D\cdot k_{2},\bar{z}_{2}\right)  \right\rangle .
\end{align}
One can easily see that%
\begin{equation}
A_{1}=A_{4},A_{2}=A_{3}.
\end{equation}
We will choose to calculate $A_{1}$ and $A_{2}$. For the case of $A_{1}$, we
have
\begin{align}
4A_{1}  &  =\varepsilon_{T^{n-2q}L^{q},T^{n-2q^{\prime}}L^{q^{\prime}}}\int
d^{2}z_{1}d^{2}z_{2}\cdot\nonumber\\
&  \left\langle e^{ik_{1}X}\left(  z_{1}\right)  e^{ik_{1}\tilde{X}}\left(
\bar{z}_{1}\right)  \left(  \partial X^{T}\right)  ^{n-2q}\left(
i\partial^{2}X^{L}\right)  ^{q}e^{ik_{2}X}\left(  z_{2}\right)  \left(
\bar{\partial}\tilde{X}^{T}\right)  ^{n-2q^{\prime}}\left(  i\bar{\partial
}^{2}\tilde{X}^{L}\right)  ^{q^{\prime}}e^{ik_{2}\tilde{X}}\left(  \bar{z}%
_{2}\right)  \right\rangle \nonumber\\
&  =\left(  -1\right)  ^{q+q^{\prime}}\int d^{2}z_{1}d^{2}z_{2}\left(
1+z_{1}\bar{z}_{1}\right)  ^{a_{0}}\left(  1+z_{2}\bar{z}_{2}\right)
^{a_{0}^{\prime}}\left\vert z_{1}-z_{2}\right\vert ^{2b_{0}}\left\vert
1+z_{1}\bar{z}_{2}\right\vert ^{2c_{0}}\nonumber\\
&  \cdot\left[  \frac{ie^{T}\cdot k_{1}}{z_{1}-z_{2}}-\frac{ie^{T}\cdot D\cdot
k_{1}}{1+\bar{z}_{1}z_{2}}\bar{z}_{1}-\frac{ie^{T}\cdot D\cdot k_{2}}%
{1+\bar{z}_{2}z_{2}}\bar{z}_{2}\right]  ^{n-2q}\nonumber\\
&  \cdot\left[  -\frac{ie^{T}\cdot D\cdot k_{1}}{1+z_{1}\bar{z}_{2}}%
z_{1}+\frac{ie^{T}\cdot k_{1}}{\bar{z}_{1}-\bar{z}_{2}}-\frac{ie^{T}\cdot
D\cdot k_{2}}{1+z_{2}\bar{z}_{2}}z_{2}\right]  ^{n-2q^{\prime}}\nonumber\\
&  \cdot\left[  \frac{e^{L}\cdot k_{1}}{\left(  z_{1}-z_{2}\right)  ^{2}%
}+\frac{e^{L}\cdot D\cdot k_{1}}{\left(  1+\bar{z}_{1}z_{2}\right)  ^{2}}%
\bar{z}_{1}^{2}+\frac{e^{L}\cdot D\cdot k_{2}}{\left(  1+\bar{z}_{2}%
z_{2}\right)  ^{2}}\bar{z}_{2}^{2}\right]  ^{q}\nonumber\\
&  \cdot\left[  \frac{e^{L}\cdot D\cdot k_{1}}{\left(  1+z_{1}\bar{z}%
_{2}\right)  ^{2}}z_{1}^{2}+\frac{e^{L}\cdot k_{1}}{\left(  \bar{z}_{1}%
-\bar{z}_{2}\right)  ^{2}}+\frac{e^{L}\cdot D\cdot k_{2}}{\left(  1+z_{2}%
\bar{z}_{2}\right)  ^{2}}z_{2}^{2}\right]  ^{q^{\prime}}.
\end{align}
To fix the modulus group on $RP_{2}$, choosing $z_{1}=r$ and $z_{2}=0$ and we
have%
\begin{align}
4A_{1}  &  =\left(  -1\right)  ^{n}\int_{0}^{1}dr^{2}\left(  1+r^{2}\right)
^{a_{0}}r^{2b_{0}}\nonumber\\
&  \cdot\left[  \frac{e^{T}\cdot k_{1}}{r}-\frac{e^{T}\cdot D\cdot k_{1}}%
{1}r\right]  ^{n-2q}\cdot\left[  -\frac{e^{T}\cdot D\cdot k_{1}}{1}%
r+\frac{e^{T}\cdot k_{1}}{r}\right]  ^{n-2q^{\prime}}\nonumber\\
&  \cdot\left[  \frac{e^{L}\cdot k_{1}}{r^{2}}+\frac{e^{L}\cdot D\cdot k_{1}%
}{1}r^{2}\right]  ^{q}\cdot\left[  \frac{e^{L}\cdot D\cdot k_{1}}{1}%
r^{2}+\frac{e^{L}\cdot k_{1}}{r^{2}}\right]  ^{q^{\prime}}\nonumber\\
&  =\left(  -1\right)  ^{n}\left(  E\sin\phi\right)  ^{2n}\left(  \frac
{2\cos^{2}\dfrac{\phi}{2}}{M\sin^{2}\phi}\right)  ^{q+q^{\prime}}\sum
_{i=0}^{q+q^{\prime}}\binom{q+q^{\prime}}{i}\left(  \dfrac{\sin^{2}\dfrac
{\phi}{2}}{\cos^{2}\dfrac{\phi}{2}}\right)  ^{i}\nonumber\\
&  \cdot\int_{0}^{1}dr^{2}\left(  1+r^{2}\right)  ^{a_{0}+2n-2\left(
q+q^{\prime}\right)  }\cdot\left(  r^{2}\right)  ^{b_{0}-n+2\left(
q+q^{\prime}\right)  -2i}.
\end{align}

Similarly, for the case of $A_{2}$, we have%
\begin{align}
4A_{2}  &  =\left(  -1\right)  ^{n}\int_{0}^{1}dr^{2}\left(  1+r^{2}\right)
^{a_{0}}r^{2c_{0}}\nonumber\\
&  \cdot\left[  \frac{e^{T}\cdot D\cdot k_{1}}{r}-\frac{e^{T}\cdot k_{1}}%
{1}r\right]  ^{n-2q}\cdot\left[  -\frac{e^{T}\cdot k_{1}}{1}r+\frac{e^{T}\cdot
D\cdot k_{1}}{r}\right]  ^{n-2q^{\prime}}\nonumber\\
&  \cdot\left[  \frac{e^{L}\cdot D\cdot k_{1}}{r^{2}}+\frac{e^{L}\cdot k_{1}%
}{1}r^{2}\right]  ^{q}\cdot\left[  \frac{e^{L}\cdot k_{1}}{1}r^{2}+\frac
{e^{L}\cdot D\cdot k_{1}}{r^{2}}\right]  ^{q^{\prime}}\nonumber\\
&  =\left(  -1\right)  ^{n}\left(  E\sin\phi\right)  ^{2n}\left(  \frac
{2\cos^{2}\dfrac{\phi}{2}}{M\sin^{2}\phi}\right)  ^{q+q^{\prime}}\sum
_{i=0}^{q+q^{\prime}}\binom{q+q^{\prime}}{i}\left(  \dfrac{\sin^{2}\dfrac
{\phi}{2}}{\cos^{2}\dfrac{\phi}{2}}\right)  ^{i}\nonumber\\
&  \cdot\int_{0}^{1}dr^{2}\text{ }\left(  1+r^{2}\right)  ^{a_{0}+2n-2\left(
q+q^{\prime}\right)  }\left(  r^{2}\right)  ^{c_{0}-n+2i}.
\end{align}
The scattering amplitude on $RP_{2}$ can therefore be calculated to be%
\begin{align}
A^{RP_{2}}  &  =A_{1}+A_{2}+A_{3}+A_{4}\nonumber\\
&  =\dfrac{1}{2}\left(  -1\right)  ^{n}\left(  E\sin\phi\right)  ^{2n}\left(
\frac{2\cos^{2}\dfrac{\phi}{2}}{M\sin^{2}\phi}\right)  ^{q+q^{\prime}}%
\sum_{i=0}^{q+q^{\prime}}\binom{q+q^{\prime}}{i}\left(  \dfrac{\sin^{2}%
\dfrac{\phi}{2}}{\cos^{2}\dfrac{\phi}{2}}\right)  ^{i}\nonumber\\
&  \cdot\int_{0}^{1}dr^{2}\left(  1+r^{2}\right)  ^{a_{0}+2n-2\left(
q+q^{\prime}\right)  }\cdot\left[  \left(  r^{2}\right)  ^{b_{0}-n+2\left(
q+q^{\prime}\right)  -2i}+\left(  r^{2}\right)  ^{c_{0}-n+2i}\right]  .
\label{RP2}%
\end{align}
The integral in Eq.(\ref{RP2}) can be calculated as following%
\begin{align}
&  \int_{0}^{1}dr^{2}\left(  1+r^{2}\right)  ^{a_{0}+2n-2\left(  q+q^{\prime
}\right)  }\cdot\left[  \left(  r^{2}\right)  ^{b_{0}-n+2\left(  q+q^{\prime
}\right)  -2i}+\left(  r^{2}\right)  ^{c_{0}-n+2i}\right] \nonumber\\
&  =[\dfrac{2^{1+a_{0}+2n-2\left(  q+q^{\prime}\right)  }}{1+b_{0}-n+2\left(
q+q^{\prime}\right)  -2i}]\nonumber\\
&  \cdot F\left(  2+a_{0}+b_{0}+n-2i,1,2+b_{0}-n+2\left(  q+q^{\prime}\right)
-2i,-1\right) \nonumber\\
&  +[\dfrac{2^{1+a_{0}+2n-2\left(  q+q^{\prime}\right)  }}{1+c_{0}%
-n+2i}]F\left(  2+a_{0}+c_{0}+n-2\left(  q+q^{\prime}\right)  +2i,1,2+c_{0}%
-n+2i,-1\right)
\end{align}
where we have used the following identities of the hypergeometric function
$F\left(  \alpha,\beta,\gamma,x\right)  $
\begin{align}
F(\alpha,\beta,\gamma;x)  &  =\frac{\Gamma(\gamma)}{\Gamma(\beta)\Gamma
(\gamma-\beta)}\int_{0}^{1}dy\text{ }y^{\beta-1}\left(  1-y\right)
^{\gamma-\beta-1}\left(  1-yx\right)  ^{-\alpha},\\
F\left(  \alpha,\beta,\gamma,x\right)   &  =2^{\gamma-\alpha-\beta}F\left(
\gamma-\alpha,\gamma-\beta,\gamma,x\right)  .
\end{align}

To further reduce the scattering amplitude into beta function, we use the
momentum conservation in Eq.(\ref{conserve}) and the identity%
\begin{align}
&  \left(  1+\alpha\right)  F\left(  -\alpha,1,2+\beta,-1\right)  +\left(
1+\beta\right)  F\left(  -\beta,1,2+\alpha,-1\right) \nonumber\\
&  =2^{1+\alpha+\beta}\dfrac{\Gamma\left(  \alpha+2\right)  \Gamma\left(
\beta+2\right)  }{\Gamma\left(  \alpha+\beta+2\right)  }%
\end{align}
to get
\begin{align}
\lbrack &  \dfrac{2^{1+a_{0}+2n-2\left(  q+q^{\prime}\right)  }}%
{1+b_{0}-n+2\left(  q+q^{\prime}\right)  -2i}]F\left(  -c_{0}+n-2i,1,2+b_{0}%
-n+2\left(  q+q^{\prime}\right)  -2i,-1\right) \nonumber\\
&  +[\dfrac{2^{1+a_{0}+2n-2\left(  q+q^{\prime}\right)  }}{1+c_{0}%
-n+2i}]F\left(  -b_{0}+n-2\left(  q+q^{\prime}\right)  +2i,1,2+c_{0}%
-n+2i,-1\right) \nonumber\\
&  =\dfrac{\Gamma\left(  1+c_{0}-n+2i\right)  \Gamma\left(  1+b_{0}-n+2\left(
q+q^{\prime}\right)  -2i\right)  }{\Gamma\left(  2+b_{0}+c_{0}-2n+2\left(
q+q^{\prime}\right)  \right)  }\nonumber\\
&  \sim B\left(  1+b_{0},1+c_{0}\right)  \dfrac{\left(  1+c_{0}\right)
^{-n+2i}\left(  1+b_{0}\right)  ^{-n+2\left(  q+q^{\prime}\right)  -2i}%
}{\left(  2+b_{0}+c_{0}\right)  ^{-2n+2\left(  q+q^{\prime}\right)  }%
}\nonumber\\
&  \sim B\left(  1+b_{0},1+c_{0}\right)  \left(  \cos^{2}\dfrac{\phi}%
{2}\right)  ^{-n+2i}\left(  \sin^{2}\dfrac{\phi}{2}\right)  ^{-n+2\left(
q+q^{\prime}\right)  -2i}.
\end{align}

We finally end up with%
\begin{align}
A^{RP_{2}}  &  =A_{1}+A_{2}+A_{3}+A_{4}\nonumber\\
&  =\dfrac{1}{2}\left(  -1\right)  ^{n}\left(  E\sin\phi\right)  ^{2n}\left(
\frac{2\cos^{2}\dfrac{\phi}{2}}{M\sin^{2}\phi}\right)  ^{q+q^{\prime}}%
\sum_{i=0}^{q+q^{\prime}}\binom{q+q^{\prime}}{i}\left(  \dfrac{\sin^{2}%
\dfrac{\phi}{2}}{\cos^{2}\dfrac{\phi}{2}}\right)  ^{i}\nonumber\\
&  \cdot B\left(  1+b_{0},1+c_{0}\right)  \left(  \cos^{2}\dfrac{\phi}%
{2}\right)  ^{-n+2i}\left(  \sin^{2}\dfrac{\phi}{2}\right)  ^{-n+2\left(
q+q^{\prime}\right)  -2i}\nonumber\\
&  =\dfrac{1}{2}\left(  -1\right)  ^{n}\left(  2E\right)  ^{2n}\left(
\frac{\sin^{2}\dfrac{\phi}{2}}{2M}\right)  ^{q+q^{\prime}}B\left(
1+b_{0},1+c_{0}\right)  \sum_{i=0}^{q+q^{\prime}}\binom{q+q^{\prime}}%
{i}\left(  \dfrac{\cos^{2}\dfrac{\phi}{2}}{\sin^{2}\dfrac{\phi}{2}}\right)
^{i}\nonumber\\
&  =\dfrac{1}{2}\left(  -1\right)  ^{n}\left(  2E\right)  ^{2n}\left(
\frac{1}{2M}\right)  ^{q+q^{\prime}}B\left(  1+b_{0},1+c_{0}\right)
\nonumber\\
&  \sim\dfrac{1}{2}\left(  -1\right)  ^{n}\left(  2E\right)  ^{2n}\left(
\frac{1}{2M}\right)  ^{q+q^{\prime}}B\left(  -\dfrac{t}{2},-\dfrac{u}%
{2}\right)  . \label{O-particle}%
\end{align}

From Eq.(\ref{O-particle}) we see that the UV behavior of O-particle
scatterings is exponential fall-off and one gets infinite linear relations
among string scattering amplitudes of different string states at each fixed
mass level. Note that both $t$ and $u$ correspond to the closed string channel
poles, while $s$ corresponds to the open string channel poles. It can be seen
from Eq.(\ref{O-particle}) that an infinite closed string Regge poles exist in
the form factor of O-particle scatterings. Furthermore, there are no
$s$-channel open string Regge poles as expected since O-planes are not
dynamical. This is in contrast to the D-particle scatterings discussed in the
last section where both infinite $s$-channel open string Regge poles and
$t$-channel closed string Regge poles exist in the form factor. We will see
that the fundamental characteristics of O-domain-wall scatterings are very
different from those of O-particle scatterings as we will now discuss in the
next section.

\subsubsection{Hard strings scattered from O-domain-wall}

For this case the kinematic setup is%
\begin{align}
e^{P}  &  =\frac{1}{M}\left(  -E,\mathrm{k}_{2}\cos\theta,-\mathrm{k}_{2}%
\sin\theta\right)  =\frac{k_{2}}{M},\\
e^{L}  &  =\frac{1}{M}\left(  -\mathrm{k}_{2},E\cos\theta,-E\sin\theta\right)
,\\
e^{T}  &  =\left(  0,\sin\theta,\cos\theta\right)  ,\\
k_{1}  &  =\left(  E,-\mathrm{k}_{1}\cos\phi,-\mathrm{k}_{1}\sin\phi\right)
,\\
k_{2}  &  =\left(  -E,\mathrm{k}_{2}\cos\theta,-\mathrm{k}_{2}\sin
\theta\right)  .
\end{align}
In the high energy limit, the angle of incidence $\phi$ is identical to the
angle of reflection $\theta$ and $Diag$ $D_{\mu\nu}=(-1,1,-1)$. The following
kinematic relations can be easily calculated%
\begin{align}
e^{T}\cdot k_{2}  &  =e^{L}\cdot k_{2}=0,\\
e^{T}\cdot k_{1}  &  =-2\mathrm{k}_{1}\sin\phi\cos\phi\sim-E\sin2\phi,\\
e^{T}\cdot D\cdot k_{1}  &  =0,\\
e^{T}\cdot D\cdot k_{2}  &  =2\mathrm{k}_{2}\sin\phi\cos\phi\sim E\sin2\phi,\\
e^{L}\cdot k_{1}  &  =\frac{1}{M}\left[  \mathrm{k}_{2}E-\mathrm{k}%
_{1}E\left(  \cos^{2}\phi-\sin^{2}\phi\right)  \right]  \sim\frac{2E^{2}}%
{M}\sin^{2}\phi,\\
e^{L}\cdot D\cdot k_{1}  &  =0,\\
e^{L}\cdot D\cdot k_{2}  &  =\frac{1}{M}\left[  -\mathrm{k}_{2}E+\mathrm{k}%
_{2}E\left(  \cos^{2}\phi-\sin^{2}\phi\right)  \right]  \sim-\frac{2E^{2}}%
{M}\sin^{2}\phi.
\end{align}

We define%
\begin{align}
a_{0}  &  \equiv k_{1}\cdot D\cdot k_{1}\sim-2E^{2}\sin^{2}\phi-2M_{1}^{2}%
\cos^{2}\phi+M_{1}^{2},\\
a_{0}^{\prime}  &  \equiv k_{2}\cdot D\cdot k_{2}=-E^{2}-\mathrm{k}_{2}%
^{2}\sim-2E^{2},\\
b_{0}  &  \equiv k_{1}\cdot k_{2}\sim2E^{2}\sin^{2}\phi+2M_{1}^{2}\cos^{2}%
\phi-\dfrac{1}{2}\left(  M_{1}^{2}+M^{2}\right)  ,\\
c_{0}  &  \equiv k_{1}\cdot D\cdot k_{2}=E^{2}-\mathrm{k}_{1}\mathrm{k}%
_{2}\sim\dfrac{1}{2}\left(  M_{1}^{2}+M^{2}\right)  , \label{c0}%
\end{align}
and the Mandelstam variables can be calculated to be%
\begin{align}
t  &  \equiv-\left(  k_{1}+k_{2}\right)  ^{2}=M_{1}^{2}+M_{2}^{2}-2k_{1}\cdot
k_{2}=M_{2}^{2}-2\left(  1+b_{0}\right)  ,\\
s  &  \equiv\dfrac{1}{2}k_{1}\cdot D\cdot k_{1}=\dfrac{1}{2}a_{0},\\
u  &  =-2k_{1}\cdot D\cdot k_{2}=-2c_{0}.
\end{align}

The first term of high energy scatterings from O-domain-wall is%
\begin{align}
4A_{1}  &  =\left(  -1\right)  ^{n}\int_{0}^{1}dr^{2}\left(  1+r^{2}\right)
^{a_{0}}r^{2b_{0}}\nonumber\\
&  \cdot\left[  \frac{e^{T}\cdot k_{1}}{r}-\frac{e^{T}\cdot D\cdot k_{1}}%
{1}r\right]  ^{n-2q}\cdot\left[  -\frac{e^{T}\cdot D\cdot k_{1}}{1}%
r+\frac{e^{T}\cdot k_{1}}{r}\right]  ^{n-2q^{\prime}}\nonumber\\
&  \cdot\left[  \frac{e^{L}\cdot k_{1}}{r^{2}}+\frac{e^{L}\cdot D\cdot k_{1}%
}{1}r^{2}\right]  ^{q}\cdot\left[  \frac{e^{L}\cdot D\cdot k_{1}}{1}%
r^{2}+\frac{e^{L}\cdot k_{1}}{r^{2}}\right]  ^{q^{\prime}}\nonumber\\
&  \sim\left(  -1\right)  ^{n}\left(  E\sin2\phi\right)  ^{2n}\left(  \frac
{1}{2M\cos^{2}\phi}\right)  ^{q+q^{\prime}}\int_{0}^{1}dr^{2}\left(
1+r^{2}\right)  ^{a_{0}}\left(  r^{2}\right)  ^{b_{0}-n}.
\end{align}
The second term can be similarly calculated to be%
\begin{align}
4A_{2}  &  =\left(  -1\right)  ^{n}\int_{0}^{1}dr^{2}\left(  1+r^{2}\right)
^{a_{0}}r^{2c_{0}}\nonumber\\
&  \cdot\left[  \frac{e^{T}\cdot D\cdot k_{1}}{r}-\frac{e^{T}\cdot k_{1}}%
{1}r\right]  ^{n-2q}\cdot\left[  -\frac{e^{T}\cdot k_{1}}{1}r+\frac{e^{T}\cdot
D\cdot k_{1}}{r}\right]  ^{n-2q^{\prime}}\nonumber\\
&  \cdot\left[  \frac{e^{L}\cdot D\cdot k_{1}}{r^{2}}+\frac{e^{L}\cdot k_{1}%
}{1}r^{2}\right]  ^{q}\cdot\left[  \frac{e^{L}\cdot k_{1}}{1}r^{2}+\frac
{e^{L}\cdot D\cdot k_{1}}{r^{2}}\right]  ^{q^{\prime}}\nonumber\\
&  \sim\left(  -1\right)  ^{n}\left(  E\sin2\phi\right)  ^{2n-2\left(
q+q^{\prime}\right)  }\left(  \frac{2E^{2}}{M}\sin^{2}\phi\right)
^{q+q^{\prime}}\int_{0}^{1}dr^{2}\text{ }\left(  1+r^{2}\right)  ^{a_{0}%
}\left(  r^{2}\right)  ^{c_{0}+n}.
\end{align}

The scattering amplitudes of O-domain-wall on $RP_{2}$ can therefore be
calculated to be%
\begin{align}
A^{RP_{2}}  &  =A_{1}+A_{2}+A_{3}+A_{4}\nonumber\\
&  =\dfrac{1}{2}\left(  -1\right)  ^{n}\left(  E\sin2\phi\right)  ^{2n}\left(
\frac{1}{2M\cos^{2}\phi}\right)  ^{q+q^{\prime}}\nonumber\\
&  \cdot\int_{0}^{1}dr^{2}\left(  1+r^{2}\right)  ^{a_{0}}\left[  \left(
r^{2}\right)  ^{b_{0}-n}+\left(  r^{2}\right)  ^{c_{0}+n}\right]  .
\end{align}
By using the similar technique for the case of O-particle scatterings, the
integral above can be calculated to be%
\begin{align}
&  \int dr^{2}\left(  1+r^{2}\right)  ^{a_{0}}\left[  \left(  r^{2}\right)
^{b_{0}-n}+\left(  r^{2}\right)  ^{c_{0}+n}\right] \nonumber\\
&  =\dfrac{F\left(  -a_{0},1+b_{0}-n,2+b_{0}-n,-1\right)  }{1+b_{0}-n}%
+\dfrac{F\left(  -a_{0},1+c_{0}+n,2+c_{0}+n,-1\right)  }{1+c_{0}+n}\nonumber\\
&  =\dfrac{2^{2+a_{0}+b_{0}+c_{0}}}{\left(  1+b_{0}-n\right)  \left(
1+c_{0}+n\right)  }\dfrac{\Gamma\left(  2+c_{0}+n\right)  \Gamma\left(
2+b_{0}-n\right)  }{\Gamma\left(  2+b_{0}+c_{0}\right)  }\nonumber\\
&  =\dfrac{\Gamma\left(  1+c_{0}+n\right)  \Gamma\left(  1+b_{0}-n\right)
}{\Gamma\left(  2+b_{0}+c_{0}\right)  }.
\end{align}

One thus ends up with%
\begin{align}
A^{RP_{2}}  &  =A_{1}+A_{2}+A_{3}+A_{4}\nonumber\\
&  =\dfrac{1}{2}\left(  -1\right)  ^{n}\left(  E\sin2\phi\right)  ^{2n}\left(
\frac{1}{2M\cos^{2}\phi}\right)  ^{q+q^{\prime}}\dfrac{\Gamma\left(
c_{0}+n+1\right)  \Gamma\left(  b_{0}-n+1\right)  }{\Gamma\left(  b_{0}%
+c_{0}+2\right)  }. \label{pole.}%
\end{align}

Some crucial points of this result are in order. First, since $c_{0}$ is a
constant in the high energy limit, the UV behavior of the O-domain-wall
scatterings is power-law instead of the usual exponential fall-off in other
O-plane scatterings.

Second, there exist only \textit{finite} number of closed string poles in the
form factor. Note that although we only look at the high energy kinematic
regime of the scattering amplitudes, it is easy to see that there exists no
infinite closed string Regge poles in the scattering amplitudes for the whole
kinematic regime. This is because there is only one kinematic variable for the
O-domain-wall scatterings. In fact, the structure of poles in Eq$.$%
(\ref{pole.}) can be calculated to be
\begin{align}
&  \dfrac{\Gamma\left(  1+c_{0}+n\right)  \Gamma\left(  1+b_{0}-n\right)
}{\Gamma\left(  2+b_{0}+c_{0}\right)  }\nonumber\\
&  =\dfrac{\Gamma\left(  1+M^{2}\right)  \Gamma\left(  1+b_{0}-n\right)
}{\Gamma\left(  b_{0}+n\right)  }\nonumber\\
&  =\Gamma\left(  1+M^{2}\right)  \dfrac{\left(  b_{0}-n\right)  !}{\left(
b_{0}+n-1\right)  !}\nonumber\\
&  =\Gamma\left(  1+M^{2}\right)
{\displaystyle\prod\limits_{k=1-n}^{n-1}}
\dfrac{1}{b_{0}-k}%
\end{align}
where we have used $c_{0}\equiv\dfrac{1}{2}\left(  M_{1}^{2}+M^{2}\right)  $
in the high energy limit. It is easy to see that the larger the mass $M$ of
the external leg is, the more numerous the closed string poles are. We thus
confirm that all massive string states do couple to the O-domain-wall as was
conjectured previously \cite{Myers, Garousi}. This is also consistent with the
boundary state descriptions of O-planes.

However, the claim that there is a thickness of order$\sqrt{\alpha^{^{\prime}%
}}$ for the O-domain-wall is misleading as the UV behavior of its scatterings
is power-law instead of exponential fall-off. This concludes that, in contrast
to the usual behavior of high energy, fixed angle string scattering
amplitudes, namely soft UV, linear relations and the existence of infinite
Regge poles, O-domain-wall scatterings, like the well-known D-instanton
scatterings, behave like field theory scatterings.%

\begin{figure}[t]%
\centering
\includegraphics[
height=2.6247in,
width=5.047in
]%
{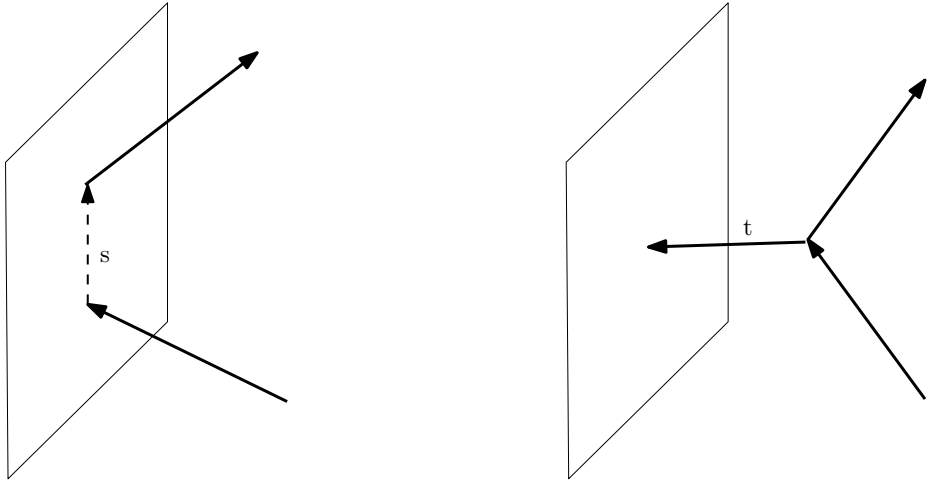}%
\caption{There are two possible channals for closed strings scattered from
D-branes/O-planes. The diagram on the left hand side corresponds to the
s-channel scatterings, and the diagram on the right hand side is the t-channel
scatterings.}%
\label{t-s}%
\end{figure}

We summarize the Regge pole structures of closed strings states scattered from
various D-branes and O-planes in the table. The $s$-channel and $t$-channel
scatterings for both D-branes and O-planes are shown in the Fig. \ref{t-s}.
For O-plane scatterings, the $s$-channel open string Regge poles are not
allowed since O-planes are not dynamical. For both cases of Domain-wall
scatterings, the $t$-channel closed string Regge poles are not allowed since
there is only one kinematic variable instead of two as in the usual cases.

\begin{center}
\ \
\begin{tabular}
[c]{|c|c|c|c|}\hline
& $p=-1$ & $1\leq p\leq23$ & $p=24$\\\hline
D$p$-branes & X & C+O & O\\\hline
O$p$-planes & X & C & X\\\hline
\end{tabular}

\end{center}

In this table, "C" and "O" represent infinite Closed string Regge poles and
Open string Regge poles respectively. "X" means there are no infinite Regge poles.

\subsection{Hard closed strings decay to open strings}

In this section, we calculate the absorption amplitudes \cite{Decay} of a
closed string state at arbitrary mass level leading to two open string states
on the D-brane at high energies. The corresponding simple case of absorption
amplitude for massless closed string state was calculated in \cite{Decay1}
(The discussion on massless string states scattered from D-brane can be found
in
\cite{Klebanov,Myers,Klebanov3,barbon1996d,bachas1999high,hirano1997scattering}%
). The inverse of this process can be used to describe Hawking radiation in
the D-brane picture.

As in the case of Domain-wall scattering discussed above, this process
contains \textit{one} kinematic variable (energy E) and thus occupies an
intermediate position between the conventional three-point and four-point
amplitudes. However, in contrast to the power-law behavior of high energy
Domain-wall scattering which contains only one kinematic variable (energy E),
its form factor behaves as exponential fall-off at high energies.

It is thus of interest to investigate whether the usual linear relations of
high energy amplitudes persist for this case or not. As will be shown in this
section, after identifying the geometric parameter of the kinematic, one can
derive the linear relations (of the kinematic variable) and ratios among the
high energy amplitudes corresponding to absorption of different closed string
states for each fixed mass level by D-brane. This result is consistent with
the coexistence \cite{Wall} of the linear relations and exponential fall-off
behavior of high energy string/D-brane amplitudes.

To study the high energy process of D$p$ brane $\left(  2\leq p\leq24\right)
$ absorbs (emits) a massive closed string state leading to two open strings on
the D$p$ brane, we set up the kinematic for the massive closed string state to
be%
\begin{align}
e^{P}  &  =\frac{1}{M}\left(  E,\mathrm{k}_{c}\cos\phi,-\mathrm{k}_{c}\sin
\phi,0\right)  =\frac{k_{c}}{M},\\
e^{L}  &  =\frac{1}{M}\left(  \mathrm{k}_{c},E\cos\phi,-E\sin\phi,0\right)
,\\
e^{T}  &  =\left(  0,\sin\phi,\cos\phi,0\right)  ,\\
k_{c}  &  =\left(  E,\mathrm{k}_{c}\cos\phi,-\mathrm{k}_{c}\sin\phi,0\right)
.
\end{align}
For simplicity, we chose the open string excitation to be two tachyons with
momenta (see Fig.\ref{Kdecay})

\begin{figure}[ptb]
\label{scattering1} \setlength{\unitlength}{3pt}
\par
\begin{center}
\begin{picture}(100,60)(-50,-15)
{
\put(-70,-20){\line(1,0){100}} \put(-30,20){\line(1,0){100}}
\put(-70,-20){\line(1,1){40}} \put(30,-20){\line(1,1){40}}
\put(-60,-18){D-brane}
\put(-40,0){\vector(1,0){80}} \put(41,0){$x$}
\put(0,0){\vector(0,1){40}} \put(0,42){$y$}
\put(18,18){\vector(-1,-1){35}} \put(-19,-19){$z$}
\put(-24,32){\vector(3,-4){24}} \put(-18,15){$k_c$}
\put(33,11){\vector(-3,-1){33}} \put(18,9){$k_2$}
\put(24,-12){\vector(-2,1){24}} \put(10,-10){$k_1$}
\put(-18,24){\vector(4,3){10}} \put(-13,31){$e^T$}
\qbezier(-5,0)(-5,3)(-3,4) \qbezier(4,-2)(7,1)(6,2)
\put(-8,2){$\phi$} \put(11,1){$\theta$} \put(9,-3){$\theta$}
}
\end{picture}
\end{center}
\caption{Kinematic setting up for a closed string decaying to two open strings
on a D-brane.}%
\label{Kdecay}%
\end{figure}
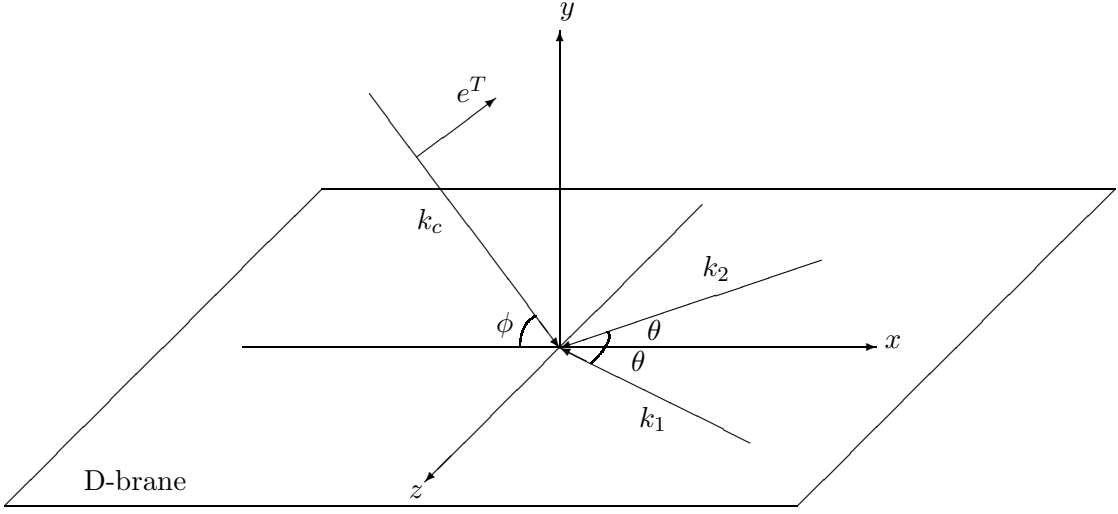%
\begin{align}
k_{1}  &  =\left(  -\frac{E}{2},-\frac{\mathrm{k}_{op}}{2}\cos\theta
,0,-\frac{\mathrm{k}_{op}}{2}\sin\theta\right)  ,\\
k_{2}  &  =\left(  -\frac{E}{2},-\frac{\mathrm{k}_{op}}{2}\cos\theta
,0,+\frac{\mathrm{k}_{op}}{2}\sin\theta\right)  .
\end{align}
Our final results, however, will remain the same for arbitrary two open string
excitation at high energies. Conservation of momentum on the D-brane implies%
\begin{equation}
\underset{\left(  k_{c}\right)  _{//}}{\underbrace{\frac{1}{2}\left(
k_{c}+D\cdot k_{c}\right)  }}+k_{1}+k_{2}=0\Rightarrow\mathrm{k}_{c}\cos
\phi=\mathrm{k}_{op}\cos\theta, \label{momen}%
\end{equation}
where $D_{\mu\nu}=$diag$\left\{  -1,1,-1,1\right\}  $. It is crucial to note
that, in the high energy limit, $\mathrm{k}_{c}=\mathrm{k}_{op}$ and the
scattering angle $\theta$ is identical to the incident angle $\phi$. One can
calculate%
\begin{align}
e^{T}\cdot k_{1}  &  =e^{T}\cdot k_{2}=e^{T}\cdot D\cdot k_{1}=e^{T}\cdot
D\cdot k_{2}\nonumber\\
&  =-\frac{\mathrm{k}_{op}\cos\theta\sin\phi}{2}=-\frac{\mathrm{k}_{c}\sin
\phi\cos\phi}{2},\label{relate1}\\
e^{L}\cdot k_{1}  &  =e^{L}\cdot k_{2}=e^{L}\cdot D\cdot k_{1}=e^{L}\cdot
D\cdot k_{2}\nonumber\\
&  =\frac{1}{M}\left[  \frac{\mathrm{k}_{c}E}{2}-\frac{\mathrm{k}_{op}E}%
{2}\cos\theta\cos\phi\right]  =\frac{\mathrm{k}_{c}E}{2M}\sin^{2}%
\phi,\label{relate2}\\
e^{T}\cdot D\cdot k_{c}  &  =2\mathrm{k}_{c}\sin\phi\cos\phi,\label{relate3}\\
e^{L}\cdot D\cdot k_{c}  &  =-\frac{2\mathrm{k}_{c}E}{M}\sin^{2}\phi,
\label{relate4}%
\end{align}
which will be useful for later calculations. We define the kinematic
invariants%
\begin{align}
t  &  \equiv-\left(  k_{1}+k_{2}\right)  ^{2}=M_{1}^{2}+M_{2}^{2}-2k_{1}\cdot
k_{2}=-2\left(  2+k_{1}\cdot k_{2}\right) \nonumber\\
&  =2k_{1}\cdot k_{c}=2k_{2}\cdot k_{c},\\
s  &  \equiv4k_{1}\cdot k_{2}=2M_{1}^{2}+2M_{2}^{2}+2\left(  k_{1}%
+k_{2}\right)  ^{2}=-2\left(  4+t\right)  , \label{sss}%
\end{align}
and calculate the following identities%
\begin{align}
k_{1}\cdot k_{c}+k_{2}\cdot D\cdot k_{c}  &  =k_{2}\cdot k_{c}+k_{1}\cdot
D\cdot k_{c}=t,\\
k_{c}\cdot D\cdot k_{c}  &  =M^{2}-2t.
\end{align}
Note that there is only one kinematic variable as $s$ and $t$ are related in
Eq.(\ref{sss}) \cite{Decay1}. On the other hand, since the scattering angle
$\theta$ is fixed by the incident angle $\phi$, $\phi$ and $\theta$ are not
the dynamical variables in the usual sense.

Following Eq.(\ref{dodo}), we consider an incoming high energy massive closed
state to be \cite{Dscatt,Wall} $\left(  \alpha_{-1}^{T}\right)  ^{n-m-2q}%
\left(  \alpha_{-1}^{L}\right)  ^{m}\left(  \alpha_{-2}^{L}\right)
^{q}\otimes\left(  \tilde{\alpha}_{-1}^{T}\right)  ^{n-m^{\prime}-2q^{\prime}%
}\left(  \tilde{\alpha}_{-1}^{L}\right)  ^{m^{\prime}}\left(  \tilde{\alpha
}_{-2}^{L}\right)  ^{q^{\prime}}\left\vert 0\right\rangle $ with
\underline{$m=m^{\prime}=0$}. The amplitude of the absorption process can be
calculated to be%

\begin{align}
A  &  =\int dx_{1}dx_{2}d^{2}z\cdot\left(  x_{1}-x_{2}\right)  ^{k_{1}\cdot
k_{2}}\left(  z-\bar{z}\right)  ^{k_{c}\cdot D\cdot k_{c}}\left(
x_{1}-z\right)  ^{k_{1}\cdot k_{c}}\nonumber\\
\cdot &  \left(  x_{1}-\bar{z}\right)  ^{k_{1}\cdot D\cdot k_{c}}\left(
x_{2}-z\right)  ^{k_{2}\cdot k_{c}}\left(  x_{2}-\bar{z}\right)  ^{k_{2}\cdot
D\cdot k_{c}}\nonumber\\
&  \cdot\exp\left\{  \left\langle \left[  ik_{1}X\left(  x_{1}\right)
+ik_{2}X\left(  x_{2}\right)  +ik_{c}\tilde{X}\left(  \bar{z}\right)  \right]
\left[  \left(  n-2q\right)  \varepsilon_{T}^{\left(  1\right)  }\partial
X^{T}+iq\varepsilon_{L}^{\left(  1\right)  }\partial^{2}X^{L}\right]  \left(
z\right)  \right\rangle \right. \nonumber\\
&  +\left\langle \left[  ik_{1}X\left(  x_{1}\right)  +ik_{2}X\left(
x_{2}\right)  +ik_{c}X\left(  z\right)  \right]  \left[  \left(  n-2q^{\prime
}\right)  \varepsilon_{T}^{\left(  2\right)  }\bar{\partial}\tilde{X}%
^{T}+iq^{\prime}\varepsilon_{L}^{\left(  2\right)  }\bar{\partial}^{2}%
\tilde{X}^{L}\right]  \left(  \bar{z}\right)  \right\rangle _{\text{linear
terms}}\nonumber\\
&  =\left(  -1\right)  ^{q+q^{\prime}}\int dx_{1}dx_{2}d^{2}z\cdot\left(
x_{1}-x_{2}\right)  ^{k_{1}\cdot k_{2}}\left(  z-\bar{z}\right)  ^{k_{c}\cdot
D\cdot k_{c}}\left(  x_{1}-z\right)  ^{k_{1}\cdot k_{c}}\nonumber\\
\cdot &  \left(  x_{1}-\bar{z}\right)  ^{k_{1}\cdot D\cdot k_{c}}\left(
x_{2}-z\right)  ^{k_{2}\cdot k_{c}}\left(  x_{2}-\bar{z}\right)  ^{k_{2}\cdot
D\cdot k_{c}}\nonumber\\
&  \cdot\left[  \frac{ie^{T}\cdot k_{1}}{x_{1}-z}+\frac{ie^{T}\cdot k_{2}%
}{x_{2}-z}+\frac{ie^{T}\cdot D\cdot k_{c}}{\bar{z}-z}\right]  ^{n-2q}%
\cdot\left[  \frac{ie^{T}\cdot D\cdot k_{1}}{x_{1}-\bar{z}}+\frac{ie^{T}\cdot
D\cdot k_{2}}{x_{2}-\bar{z}}+\frac{ie^{T}\cdot D\cdot k_{c}}{z-\bar{z}%
}\right]  ^{n-2q^{\prime}}\nonumber\\
&  \cdot\left[  \frac{e^{L}\cdot k_{1}}{\left(  x_{1}-z\right)  ^{2}}%
+\frac{e^{L}\cdot k_{2}}{\left(  x_{2}-z\right)  ^{2}}+\frac{e^{L}\cdot D\cdot
k_{c}}{\left(  \bar{z}-z\right)  ^{2}}\right]  ^{q}\cdot\left[  \frac
{e^{L}\cdot D\cdot k_{1}}{\left(  x_{1}-\bar{z}\right)  ^{2}}+\frac{e^{L}\cdot
D\cdot k_{2}}{\left(  x_{2}-\bar{z}\right)  ^{2}}+\frac{e^{L}\cdot D\cdot
k_{c}}{\left(  z-\bar{z}\right)  ^{2}}\right]  ^{q^{\prime}}.
\end{align}
Set $\left\{  x_{1},x_{2},z\right\}  =\left\{  -x,x,i\right\}  $ to fix the
$SL(2,R)$ gauge and use Eq.(\ref{relate1}-\ref{relate4}), we have%
\begin{align}
A  &  =\left(  -1\right)  ^{n+M^{2}/2+t/2}2^{M^{2}-2-5t/2}\cdot\int_{-\infty
}^{+\infty}dx\cdot x^{-t/2-2}\left(  1-ix\right)  ^{t+1}\left(  1+ix\right)
^{t+1}\nonumber\\
&  \cdot\left[  \frac{-\frac{\mathrm{k}_{c}\sin\phi\cos\phi}{2}}{1-ix}%
+\frac{-\frac{\mathrm{k}_{c}\sin\phi\cos\phi}{2}}{1+ix}+\frac{2\mathrm{k}%
_{c}\sin\phi\cos\phi}{2}\right]  ^{n-2q}\nonumber\\
&  \cdot\left[  \frac{-\frac{\mathrm{k}_{c}\sin\phi\cos\phi}{2}}{1+ix}%
+\frac{-\frac{\mathrm{k}_{c}\sin\phi\cos\phi}{2}}{1-ix}+\frac{2\mathrm{k}%
_{c}\sin\phi\cos\phi}{2}\right]  ^{n-2q^{\prime}}\nonumber\\
&  \cdot\left[  \frac{\frac{\mathrm{k}_{c}E}{2M}\sin^{2}\phi}{\left(
1-ix\right)  ^{2}}+\frac{\frac{\mathrm{k}_{c}E}{2M}\sin^{2}\phi}{\left(
1+ix\right)  ^{2}}+\frac{-\frac{2\mathrm{k}_{c}E}{M}\sin^{2}\phi}{4}\right]
^{q}\cdot\left[  \frac{\frac{\mathrm{k}_{c}E}{2M}\sin^{2}\phi}{\left(
1+ix\right)  ^{2}}+\frac{\frac{\mathrm{k}_{c}E}{2M}\sin^{2}\phi}{\left(
1-ix\right)  ^{2}}+\frac{-\frac{2\mathrm{k}_{c}E}{M}\sin^{2}\phi}{4}\right]
^{q^{\prime}}\nonumber\\
&  =\left(  -1\right)  ^{n+M^{2}/2+t/2}2^{M^{2}-2-5t/2}\cdot\left(
\mathrm{k}_{c}\sin\phi\cos\phi\right)  ^{2n-2\left(  q+q^{\prime}\right)
}\left(  -\frac{\mathrm{k}_{c}E\sin^{2}\phi}{2M}\right)  ^{q+q^{\prime}%
}\nonumber\\
&  \cdot\int_{-\infty}^{+\infty}dx\cdot x^{-t/2-2}\left(  1+x^{2}\right)
^{t+1}\left[  \frac{x^{2}}{1+x^{2}}\right]  ^{2n-2\left(  q+q^{\prime}\right)
}\left[  1-\frac{2\left(  1-x^{2}\right)  }{\left(  1+x^{2}\right)  ^{2}%
}\right]  ^{q+q^{\prime}}.
\end{align}
By using the binomial expansion, we get%
\begin{align}
A  &  =\left(  -1\right)  ^{n+M^{2}/2+t/2}2^{M^{2}-2-5t/2}\cdot\left(
E\sin\phi\cos\phi\right)  ^{2n}\left(  -\frac{1}{2M\cos^{2}\phi}\right)
^{q+q^{\prime}}\nonumber\\
&  \cdot\sum_{i=0}^{q+q^{\prime}}\sum_{j=0}^{i}\binom{q+q^{\prime}}{i}%
\binom{i}{j}\left(  -2\right)  ^{i}\left(  -1\right)  ^{j}\nonumber\\
&  \int_{0}^{\infty}d\left(  x^{2}\right)  \cdot\left(  x^{2}\right)
^{-t/4-3/2+2n-2\left(  q+q^{\prime}\right)  +j}\left(  1+x^{2}\right)
^{t+1-2n+2\left(  q+q^{\prime}\right)  -2i}.
\end{align}
Finally, to reduce the integral to the standard beta function, we do the
linear fractional transformation $x^{2}=\frac{1-y}{y}$ to get
\begin{align}
&  A=\left(  -1\right)  ^{n+M^{2}/2+t/2}2^{M^{2}-2-5t/2}\cdot\left(  E\sin
\phi\cos\phi\right)  ^{2n}\left(  -\frac{1}{2M\cos^{2}\phi}\right)
^{q+q^{\prime}}\nonumber\\
&  \cdot\sum_{i=0}^{q+q^{\prime}}\sum_{j=0}^{i}\binom{q+q^{\prime}}{i}%
\binom{i}{j}\left(  -2\right)  ^{i}\left(  -1\right)  ^{j}\int_{0}^{1}dy\cdot
y^{-3t/4-3/2+2i-j}\cdot\left(  1-y\right)  ^{-t/4-3/2+2n-2\left(  q+q^{\prime
}\right)  +j}\nonumber\\
&  =\left(  -1\right)  ^{n+M^{2}/2+t/2}2^{M^{2}-2-5t/2}\cdot\left(  E\sin
\phi\cos\phi\right)  ^{2n}\left(  -\frac{1}{2M\cos^{2}\phi}\right)
^{q+q^{\prime}}\nonumber\\
&  \cdot\frac{\Gamma\left(  -\frac{3t}{4}-\frac{1}{2}\right)  \Gamma\left(
-\frac{t}{4}-\frac{1}{2}\right)  }{\Gamma\left(  -t-1\right)  }\sum
_{i=0}^{q+q^{\prime}}\sum_{j=0}^{i}\binom{q+q^{\prime}}{i}\binom{i}{j}\left(
-2\right)  ^{i}\left(  -1\right)  ^{j}\left(  \frac{3}{4}\right)
^{2i-j}\left(  \frac{1}{4}\right)  ^{2n-2\left(  q+q^{\prime}\right)
+j}\nonumber\\
&  =\left(  -1\right)  ^{n+M^{2}/2+t/2}2^{M^{2}-2-5t/2}\cdot\left(
\frac{E\sin\phi\cos\phi}{4}\right)  ^{2n}\nonumber\\
&  \cdot\left(  -\frac{2}{M\cos^{2}\phi}\right)  ^{q+q^{\prime}}\frac
{\Gamma\left(  -\frac{3t}{4}-\frac{1}{2}\right)  \Gamma\left(  -\frac{t}%
{4}-\frac{1}{2}\right)  }{\Gamma\left(  -t-1\right)  }. \label{main}%
\end{align}

In addition to an exponential fall-off factor, the energy $E$ dependence of
Eq.(\ref{main}) contains a pre-power factor in the high energy limit. To
obtain the linear relations for the amplitudes at each fixed mass level, we
rewrite Eq.(\ref{main}) in the following form
\begin{equation}
\frac{\mathcal{T}^{(n,0,q;n,0,q^{\prime})}}{\mathcal{T}^{(n,0,0;n,0,0)}%
}=\left(  -\frac{2}{M\cos^{2}\phi}\right)  ^{q+q^{\prime}}. \label{final.}%
\end{equation}
One first notes that Eq.(\ref{final.}) does not contradict with
Eq.(\ref{mainA}), which predict the ratios $\left(  -\frac{1}{2M}\right)
^{q+q^{\prime}}$. This is because for the absorption process we are
considering, there is only one kinematic variable and the usual Ward identity
calculations do not apply. To compare Eq.(\ref{final.}) with the "ratios" of
the Domain-wall scattering \cite{Wall}%
\begin{equation}
\frac{\mathcal{T}^{(n,0,q;n,0,q^{\prime})}}{\mathcal{T}^{(n,0,0;n,0,0)}}%
\mid_{Domain}=\left(  \frac{E\sin\phi}{M\sqrt{\left\vert M_{1}^{2}%
-2M^{2}-1\right\vert }\cos^{2}\phi}\right)  ^{q+q^{\prime}}, \label{Dw}%
\end{equation}
one sees that, in addition to the incident angle $\phi$, there is an energy
dependent power factor within the bracket of $q+q^{\prime}$ in Eq.(\ref{Dw})
Thus there is no linear relations for the Domain-wall scatterings. On the
contrary, Eq.(\ref{final.}) gives the linear relations (of the kinematic
variable $E$) and ratios among the high energy amplitudes corresponding to
absorption of different closed string states for each fixed mass level $n$ by D-brane.

Note that since the scattering angle $\theta$ is fixed by the incident angle
$\phi$, $\phi$ is not a dynamical variable in the usual sense. Another way to
see this is through the relation of $s$ and $t$ in Eq.(\ref{sss}). We will
call such an angle a \textit{geometrical parameter} in contrast to the usual
dynamical variable. This kind of geometrical parameter shows up in closed
string state scattered from generic D$p$-brane (except D-instanton and
D-particle) \cite{Dscatt,Wall}. This is because one has only two dynamical
variables for the scatterings, but needs more than two variables to set up the
kinematic due to the relative geometry between the D-brane and the scattering
plane at high energies.

We emphasize that our result in Eq.(\ref{final.}) is consistent with the
coexistence \cite{Wall} of the linear relations and exponential fall-off
behavior of high energy string/D-brane amplitudes. That is, linear relations
of the amplitudes are responsible for the softer, exponential fall-off high
energy string/D-brane scatterings than the power-law field theory scatterings.%

\setcounter{equation}{0}
\renewcommand{\theequation}{\arabic{section}.\arabic{equation}}%

\section{Hard scatterings in compact spaces}

In this chapter, following an old suggestion of Mende \cite{Mende}, we
calculate high energy massive scattering amplitudes of bosonic string with
some coordinates compactified on the torus \cite{Compact,Compact2}. We obtain
infinite linear relations among high energy scattering amplitudes of different
string states in the Gross kinematic regime (GR). This result is reminiscent
of the existence of an infinite number of massive ZNS in the compactified
closed and open string spectrums constructed in chapter IV \cite{Lee1,Lee2}.

In addition, we analyze all possible power-law and soft exponential fall-off
regimes of high energy compactified bosonic string scatterings by comparing
the scatterings with their 26D noncompactified counterparts. In particular, we
discover in section X.A the existence of a power-law regime at fixed angle and
an exponential fall-off regime at small angle for high energy compactified
open string scatterings \cite{Compact2}. These new phenomena never happen in
the 26D string scatterings. The linear relations break down as expected in all
power-law regimes. The analysis can be extended to the high energy scatterings
of the compactified closed string in section X.B, which corrects and extends
the results in \cite{Compact}.

\subsection{Open string compactified on torus}

\subsubsection{High energy Scatterings}

We consider \cite{Compact2} hard scatterings of 26D open bosonic string with
one coordinate compactified on $S^{1}$ with radius $R$. As we will see later,
it is straightforward to generalize our calculation to more compactified
coordinates. The mode expansion of the compactified coordinate is%
\begin{equation}
X^{25}\left(  \sigma,\tau\right)  =x^{25}+K^{25}\tau+i\sum_{k\neq0}%
\frac{\alpha_{k}^{25}}{k}e^{-ik\tau}\cos n\sigma
\end{equation}
where $K^{25}$ is the canonical momentum in the $X^{25}$ direction%
\begin{equation}
K^{25}=\frac{2\pi l-\theta_{j}+\theta_{i}}{2\pi R}.
\end{equation}
Note that $l$ is the quantized momentum and we have included a nontrivial
Wilson line with $U(n)$ Chan-Paton factors, $i,j=1,2...n.$, which will be
important in the later discussion. The mass spectrum of the theory is
\begin{equation}
M^{2}=\left(  K^{25}\right)  ^{2}+2\left(  N-1\right)  \equiv\left(
\frac{2\pi l-\theta_{j}+\theta_{i}}{2\pi R}\right)  ^{2}+M^{2}%
\end{equation}
where we have defined level mass as$M^{2}=2\left(  N-1\right)  $ and
$N=\sum_{k\neq0}\alpha_{-k}^{25}\alpha_{k}^{25}+\alpha_{-k}^{\mu}\alpha
_{k}^{\mu},\mu=0,1,2...24.$ We are going to consider 4-point correlation
function in this chapter. In the center of momentum frame, the kinematic can
be set up to be \cite{Compact}%

\begin{align}
k_{1}  &  =\left(  +\sqrt{p^{2}+M_{1}^{2}},-p,0,-K_{1}^{25}\right)  ,\\
k_{2}  &  =\left(  +\sqrt{p^{2}+M_{2}^{2}},+p,0,+K_{2}^{25}\right)  ,\\
k_{3}  &  =\left(  -\sqrt{q^{2}+M_{3}^{2}},-q\cos\phi,-q\sin\phi,-K_{3}%
^{25}\right)  ,\\
k_{4}  &  =\left(  -\sqrt{q^{2}+M_{4}^{2}},+q\cos\phi,+q\sin\phi,+K_{4}%
^{25}\right)
\end{align}
where $p$ is the incoming momentum, $q$ is the outgoing momentum and $\phi$ is
the center of momentum scattering angle. In the high energy limit, one
includes only momenta on the scattering plane, and we have included the fourth
component for the compactified direction as the internal momentum. The
conservation of the fourth component of the momenta implies%
\begin{equation}
K_{1}^{25}-K_{2}^{25}+K_{3}^{25}-K_{4}^{25}=0.
\end{equation}
Note that%
\begin{equation}
k_{i}^{2}=K_{i}^{2}-M_{i}^{2}=-M_{i}^{2}.
\end{equation}

The center of mass energy $E$ is defined as (for large $p,q$)%
\begin{equation}
E=\dfrac{1}{2}\left(  \sqrt{p^{2}+M_{1}^{2}}+\sqrt{p^{2}+M_{2}^{2}}\right)
=\dfrac{1}{2}\left(  \sqrt{q^{2}+M_{3}^{2}}+\sqrt{q^{2}+M_{4}^{2}}\right)  .
\end{equation}
We have%
\begin{align}
-k_{1}\cdot k_{2}  &  =\sqrt{p^{2}+M_{1}^{2}}\cdot\sqrt{p^{2}+M_{2}^{2}}%
+p^{2}+K_{1}^{25}K_{2}^{25}\nonumber\\
&  =\dfrac{1}{2}\left(  s+k_{1}^{2}+k_{2}^{2}\right)  =\dfrac{1}{2}s-\frac
{1}{2}\left(  M_{1}^{2}+M_{2}^{2}\right)  ,\label{k1k2}\\
-k_{2}\cdot k_{3}  &  =-\sqrt{p^{2}+M_{2}^{2}}\cdot\sqrt{q^{2}+M_{3}^{2}%
}+pq\cos\phi+K_{2}^{25}K_{3}^{25}\nonumber\\
&  =\dfrac{1}{2}\left(  t+k_{2}^{2}+k_{3}^{2}\right)  =\dfrac{1}{2}t-\frac
{1}{2}\left(  M_{2}^{2}+M_{3}^{2}\right)  ,\label{t}\\
-k_{1}\cdot k_{3}  &  =-\sqrt{p^{2}+M_{1}^{2}}\cdot\sqrt{q^{2}+M_{3}^{2}%
}-pq\cos\phi-K_{1}^{25}K_{3}^{25}\nonumber\\
&  =\dfrac{1}{2}\left(  u+k_{1}^{2}+k_{3}^{2}\right)  =\dfrac{1}{2}u-\frac
{1}{2}\left(  M_{1}^{2}+M_{3}^{2}\right)  \label{u}%
\end{align}
where $s,t$ and $u$ are the Mandelstam variables with%
\begin{equation}
s+t+u=\sum_{i}M_{i}^{2}\sim2\left(  N-4\right)  .
\end{equation}
Note that the Mandelstam variables defined above are not the usual
$25$-dimensional Mandelstam variables in the scattering process since we have
included the internal momentum $K_{i}^{25}$ in the definition of $k_{i}$. We
are now ready to calculate the high energy scattering amplitudes.

In the high energy limit, we define the polarizations on the scattering plane
to be%

\begin{align}
e^{P}  &  =\frac{1}{M_{2}}\left(  \sqrt{p^{2}+M_{2}^{2}},p,0,0\right)  ,\\
e^{L}  &  =\frac{1}{M_{2}}\left(  p,\sqrt{p^{2}+M_{2}^{2}},0,0\right)  ,\\
e^{T}  &  =\left(  0,0,1,0\right)
\end{align}
where the fourth component refers to the compactified direction. It is easy to
calculate the following relations%
\begin{align}
e^{P}\cdot k_{1}  &  =-\frac{1}{M_{2}}\left(  \sqrt{p^{2}+M_{1}^{2}}%
\sqrt{p^{2}+M_{2}^{2}}+p^{2}\right)  ,\label{MR}\\
e^{P}\cdot k_{3}  &  =\frac{1}{M_{2}}\left(  \sqrt{q^{2}+M_{3}^{2}}\sqrt
{p^{2}+M_{2}^{2}}-pq\cos\phi\right)  ,
\end{align}%
\begin{align}
e^{L}\cdot k_{1}  &  =-\frac{p}{M_{2}}\left(  \sqrt{p^{2}+M_{1}^{2}}%
+\sqrt{p^{2}+M_{2}^{2}}\right)  ,\\
e^{L}\cdot k_{3}  &  =\frac{1}{M_{2}}\left(  p\sqrt{q^{2}+M_{3}^{2}}%
-q\sqrt{p^{2}+M_{2}^{2}}\cos\phi\right)  , \label{GR}%
\end{align}%
\begin{equation}
e^{T}\cdot k_{1}=0\text{, \ \ }e^{T}\cdot k_{3}=-q\sin\phi.
\end{equation}

In this chapter, we will consider the case of a tensor state \cite{Compact}%

\begin{equation}
\left(  \alpha_{-1}^{T}\right)  ^{N-2r}\left(  \alpha_{-2}^{L}\right)
^{r}\left\vert k_{2},l_{2},i,j\right\rangle
\end{equation}
at a general mass level $M_{2}^{2}=2\left(  N-1\right)  $ scattered with three
"tachyon" states (with $M_{1}^{2}=M_{3}^{2}=M_{4}^{2}=-2$). In general, we
could have considered the more general high energy state%

\begin{equation}
\left(  \alpha_{-1}^{T}\right)  ^{N-2r-2m-\sum_{n}ns_{n}}\left(  \alpha
_{-1}^{L}\right)  ^{2m}\left(  \alpha_{-2}^{L}\right)  ^{r}\prod
\limits_{n}\left(  \alpha_{-n}^{25}\right)  ^{s_{n}}\left\vert k_{2}%
,l_{2},i,j\right\rangle .
\end{equation}
However, for our purpose here and for simplicity, we will not consider the
general vertex in this chapter. The $s-t$ channel of the high energy
scattering amplitude can be calculated to be (We will ignore the trace factor
due to Chan-Paton in the scattering amplitude calculation . This does not
affect our final results in this chapter)
\begin{align}
A  &  =\int d^{4}x\left\langle e^{ik_{1}X}\left(  x_{1}\right)  \left(
\partial X^{T}\right)  ^{N-2r}\left(  i\partial^{2}X^{L}\right)  ^{r}%
e^{ik_{2}X}\left(  x_{2}\right)  e^{ik_{3}X}\left(  x_{3}\right)  e^{ik_{4}%
X}\left(  x_{4}\right)  \right\rangle \nonumber\\
&  =\int d^{4}x\cdot\prod\limits_{i<j}\left(  x_{i}-x_{j}\right)  ^{k_{i}\cdot
k_{j}}\nonumber\\
&  \cdot\left[  \frac{ie^{T}\cdot k_{1}}{x_{1}-x_{2}}+\frac{ie^{T}\cdot k_{3}%
}{x_{3}-x_{2}}+\frac{ie^{T}\cdot k_{4}}{x_{4}-x_{2}}\right]  ^{N-2r}%
\cdot\left[  \frac{e^{L}\cdot k_{1}}{\left(  x_{1}-x_{2}\right)  ^{2}}%
+\frac{e^{L}\cdot k_{3}}{\left(  x_{3}-x_{2}\right)  ^{2}}+\frac{e^{L}\cdot
k_{4}}{\left(  x_{4}-x_{2}\right)  ^{2}}\right]  ^{r}.
\end{align}
After fixing the $SL(2,R)$ gauge and using the kinematic relations derived
previously, we have%
\begin{align}
A  &  =i^{N}\left(  -1\right)  ^{k_{1}\cdot k_{2}+k_{1}\cdot k_{3}+k_{2}\cdot
k_{3}}\left(  q\sin\phi\right)  ^{N-2r}\left(  \frac{1}{M_{2}}\right)
^{r}\cdot\int_{0}^{1}dx\cdot x^{k_{1}\cdot k_{2}}\left(  1-x\right)
^{k_{2}\cdot k_{3}}\cdot\left[  \frac{1}{1-x}\right]  ^{N-2r}\nonumber\\
&  \cdot\left[  \frac{\left(  \sqrt{p^{2}+M_{1}^{2}}+\sqrt{p^{2}+M_{2}^{2}%
}\right)  }{x^{2}}-\frac{\left(  p\sqrt{q^{2}+M_{3}^{2}}-q\sqrt{p^{2}%
+M_{2}^{2}}\cos\phi\right)  }{\left(  1-x\right)  ^{2}}\right]  ^{r}%
\nonumber\\
&  =\left(  -1\right)  ^{k_{1}\cdot k_{2}+k_{1}\cdot k_{3}+k_{2}\cdot k_{3}%
}\left(  iq\sin\phi\right)  ^{N}\left(  -\frac{\left(  p\sqrt{q^{2}+M_{3}^{2}%
}-q\sqrt{p^{2}+M_{2}^{2}}\cos\phi\right)  }{M_{2}q^{2}\sin^{2}\phi}\right)
^{r}\nonumber\\
&  \cdot\sum_{i=0}^{r}\binom{r}{i}\left[  -\frac{\left(  \sqrt{p^{2}+M_{1}%
^{2}}+\sqrt{p^{2}+M_{2}^{2}}\right)  }{\left(  p\sqrt{q^{2}+M_{3}^{2}}%
-q\sqrt{p^{2}+M_{2}^{2}}\cos\phi\right)  }\right]  ^{i}\cdot\int_{0}%
^{1}dx\cdot x^{k_{1}\cdot k_{2}-2i}\left(  1-x\right)  ^{k_{2}\cdot
k_{3}-N+2i}\nonumber\\
&  =\left(  -iq\sin\phi\right)  ^{N}\left(  -\frac{\left(  p\sqrt{q^{2}%
+M_{3}^{2}}-q\sqrt{p^{2}+M_{2}^{2}}\cos\phi\right)  }{M_{2}q^{2}\sin^{2}\phi
}\right)  ^{r}\nonumber\\
&  \cdot\sum_{i=0}^{r}\binom{r}{i}\left[  -\frac{\left(  \sqrt{p^{2}+M_{1}%
^{2}}\sqrt{p^{2}+M_{2}^{2}}+p^{2}\right)  }{\left(  p\sqrt{q^{2}+M_{3}^{2}%
}-q\sqrt{p^{2}+M_{2}^{2}}\cos\phi\right)  }\right]  ^{i}\cdot B\left(
-\frac{1}{2}s+N-2i-1,-\frac{1}{2}t+2i-1\right)
\end{align}
where $B(u,v)$ is the Euler beta function. We can do the high energy
approximation of the gamma function $\Gamma\left(  x\right)  $ then do the
summation, and end up with
\begin{align}
A  &  =\left(  -iq\sin\phi\right)  ^{N}\left(  -\frac{\left(  p\sqrt
{q^{2}+M_{3}^{2}}-q\sqrt{p^{2}+M_{2}^{2}}\cos\phi\right)  }{M_{2}q^{2}\sin
^{2}\phi}\right)  ^{r}\nonumber\\
&  \cdot\sum_{i=0}^{r}\binom{r}{i}\left[  -\frac{\left(  \sqrt{p^{2}+M_{1}%
^{2}}+\sqrt{p^{2}+M_{2}^{2}}\right)  }{\left(  p\sqrt{q^{2}+M_{3}^{2}}%
-q\sqrt{p^{2}+M_{2}^{2}}\cos\phi\right)  }\right]  ^{i}\cdot\frac
{\Gamma\left(  -1-\frac{1}{2}s+N-2i\right)  \Gamma\left(  -1-\frac{1}%
{2}t+2i\right)  }{\Gamma\left(  2+\frac{1}{2}u\right)  }\nonumber\\
&  \simeq\left(  -iq\sin\phi\right)  ^{N}\left(  -\frac{\left(  p\sqrt
{q^{2}+M_{3}^{2}}-q\sqrt{p^{2}+M_{2}^{2}}\cos\phi\right)  }{M_{2}q^{2}\sin
^{2}\phi}\right)  ^{r}\nonumber\\
&  \cdot\sum_{i=0}^{r}\binom{r}{i}\left[  -\frac{\left(  \sqrt{p^{2}+M_{1}%
^{2}}+\sqrt{p^{2}+M_{2}^{2}}\right)  }{\left(  p\sqrt{q^{2}+M_{3}^{2}}%
-q\sqrt{p^{2}+M_{2}^{2}}\cos\phi\right)  }\right]  ^{i}\nonumber\\
&  \cdot B\left(  -1-\dfrac{1}{2}s,-1-\frac{1}{2}t\right)  \left(
-1-\dfrac{1}{2}s\right)  ^{N-2i}\left(  -1-\frac{1}{2}t\right)  ^{2i}\left(
2+\dfrac{1}{2}u\right)  ^{-N}\nonumber\\
&  =\left(  -iq\frac{\sin\frac{\phi}{2}}{\cos\frac{\phi}{2}}\right)
^{N}\left(  -\frac{1}{M_{2}}\right)  ^{r}\cdot B\left(  -1-\frac{1}%
{2}s,-1-\frac{1}{2}t\right) \nonumber\\
&  \cdot\left[  \frac{\left(  p\sqrt{q^{2}+M_{3}^{2}}-q\sqrt{p^{2}+M_{2}^{2}%
}\cos\phi\right)  }{q^{2}\sin^{2}\phi}-\frac{\left(  \sqrt{p^{2}+M_{1}^{2}%
}+\sqrt{p^{2}+M_{2}^{2}}\right)  }{q^{2}\sin^{2}\phi}\left(  \frac{t}%
{s}\right)  ^{2}\right]  ^{r}. \label{power}%
\end{align}

\subsubsection{Classification of Compactified String Scatterings}

It is well known that there are two kinematic regimes for the high energy
string scatterings in 26D open bosonic string theory. The UV behavior of the
finite and fixed angle scatterings in the GR is soft exponential fall-off.
Moreover, there exist infinite linear relations among scatterings of different
string states in this regime \cite{ChanLee,ChanLee1,ChanLee2,
CHL,CHLTY1,CHLTY2,CHLTY3,susy,Closed,Dscatt,Decay}. On the other hand, the UV
behavior of the small angle scatterings in the Regge regime is hard power-law.
The linear relations break down in the Regge regime. As we will see soon, the
UV structure of the compactified open string scatterings is more richer.

In this section, we systematically analyze all possible power-law regimes of
high energy compactified open string scatterings by comparing the scatterings
with their noncompactified counterparts. In particular, we show that all hard
power-law regimes of high energy compactified open string scatterings can be
traced back to the Regge regime of the 26D high energy string scatterings. The
linear relations break down as expected in all power-law regimes. The analysis
can be extended to the high energy scatterings of the compactified closed
string in section X.B, which corrects and extends the results in
\cite{Compact}.

\paragraph{Gross Regime - Linear Relations}

In the Gross regime, $p^{2}\simeq q^{2}\gg K_{i}^{2}$ and $p^{2}\simeq
q^{2}\gg N$, Eq.(\ref{power}) reduces to%
\begin{equation}
A\simeq\left(  -iE\frac{\sin\frac{\phi}{2}}{\cos\frac{\phi}{2}}\right)
^{N}\left(  -\frac{1}{2M_{2}}\right)  ^{r}\cdot B\left(  -1-\frac{1}%
{2}s,-1-\frac{1}{2}t\right)  . \label{beta}%
\end{equation}
For each fixed mass level $N$, we have the linear relation for the scattering
amplitudes%
\begin{equation}
\frac{\mathcal{T}^{\left(  n,r\right)  }}{\mathcal{T}^{\left(  n,0\right)  }%
}=\left(  -\frac{1}{2M_{2}}\right)  ^{r} \label{linear}%
\end{equation}
with coefficients consistent with our previous results
\cite{ChanLee,ChanLee1,ChanLee2,
CHL,CHLTY1,CHLTY2,CHLTY3,susy,Closed,Dscatt,Decay}. Note that in
Eq.(\ref{beta}) there is an exponential fall-off factor in the high energy
expansion of the beta function. The infinite linear relation in
Eq.(\ref{linear}) "soften" the high energy behavior of string scatterings in
the GR.

\paragraph{Classification of compactified open string}

Since our definitions of the Mandelstam variables $s,t$ and $u$ in
Eq.(\ref{k1k2}) to Eq.(\ref{u}) include the compactified coordinates, we can
analyze the UV structure of the compactified string scatterings by comparing
the scatterings with their simpler 26D counterparts. We introduce the space
part of the momentum vectors%

\begin{align}
\mathbf{k}_{1}  &  =\left(  -p,0,-K_{1}^{25}\right)  ,\\
\mathbf{k}_{2}  &  =\left(  +p,0,+K_{2}^{25}\right)  ,\\
\mathbf{k}_{3}  &  =\left(  -q\cos\phi,-q\sin\phi,-K_{3}^{25}\right)  ,\\
\mathbf{k}_{4}  &  =\left(  +q\cos\phi,+q\sin\phi,+K_{4}^{25}\right)  ,
\end{align}
and define the "26D scattering angle" $\widetilde{\phi}$ as following%
\begin{equation}
\mathbf{k}_{1}\cdot\mathbf{k}_{3}=\left\vert \mathbf{k}_{1}\right\vert
\left\vert \mathbf{k}_{3}\right\vert \cos\widetilde{\phi}. \label{26D}%
\end{equation}
It is then easy to see that the UV behavior of the compactified string
scatterings is power-law if and only if $\widetilde{\phi}$ is small. This
criterion can be used to classify all possible power-law and exponential
fall-off kinematic regimes of high energy compactified open string scatterings.

\paragraph{Compactified 25D scatterings}

We first consider the high energy scatterings with only one coordinate compactified.

$\boldsymbol{I.}$ For the case of $\phi=$ finite, the only choice to achieve
UV power-law behavior is to require (we choose $K_{1}^{25}\simeq K_{2}%
^{25}\simeq K_{3}^{25}\simeq K_{4}^{25}$ and $p\simeq q$ in the following
discussion)%
\begin{equation}
\left(  K_{i}^{25}\right)  ^{2}\gg p^{2}\simeq q^{2}\gg N. \label{newpower}%
\end{equation}
By the criterion of Eq.(\ref{26D}), this is a power-law regime. To explicitly
show that this choice of kinematic regime does lead to UV power-law behavior,
we will show that it implies
\begin{equation}
s=\text{ constant} \label{const}%
\end{equation}
in the open string scattering amplitudes, which in turn gives the desire
power-law behavior of high energy compactified open string scattering in
Eq.(\ref{power}). On the other hand, it can be shown that the linear relations
break down as expected in this regime. For the choice of kinematic regime in
Eq.(\ref{newpower}) , Eq.(\ref{k1k2}) and Eq.(\ref{const}) imply
\begin{equation}
\lim_{p\rightarrow\infty}\frac{\sqrt{p^{2}+M_{1}^{2}}\cdot\sqrt{p^{2}%
+M_{2}^{2}}+p^{2}}{K_{1}^{25}K_{2}^{25}}=\lim_{p\rightarrow\infty}\frac
{\sqrt{p^{2}+M_{1}^{2}}\cdot\sqrt{p^{2}+M_{2}^{2}}+p^{2}}{\left(  \frac{2\pi
l_{1}-\theta_{j,1}+\theta_{i,1}}{2\pi R}\right)  \left(  \frac{2\pi
l_{2}-\theta_{j,2}+\theta_{i,2}}{2\pi R}\right)  }=-1. \label{condition}%
\end{equation}
For finite momenta $l_{1}$and $l_{2}$, Eq.(\ref{condition}) can be achieved by
scattering of string states with "super-highly" winding nontrivial Wilson
lines%
\begin{equation}
(\theta_{i,1}-\theta_{j,1})\rightarrow\infty\text{, \ }(\theta_{i,2}%
-\theta_{j,2})\rightarrow-\infty. \label{wilson}%
\end{equation}
A careful analysis for this choice gives%
\begin{equation}
(\lambda_{1}+\lambda_{2})^{2}=0 \label{lamda}%
\end{equation}
where signs of $\lambda_{1}=\frac{p}{K_{1}^{25}}$ and $\lambda_{2}=-\frac
{p}{K_{2}^{25}}$ are chosen to be the same. It can be seen now that the
kinematic regime in Eq.(\ref{newpower}) does solve Eq.(\ref{lamda}).

We now consider the second possible regime for the case of $\phi=$ finite,
namely%
\begin{equation}
\left(  K_{i}^{25}\right)  ^{2}\simeq p^{2}\simeq q^{2}\gg N. \label{oldmende}%
\end{equation}
By the criterion of Eq.(\ref{26D}), this is an exponential fall-off regime. To
explicitly show that this choice of kinematic regime does lead to UV
exponential fall-off behavior, we see that, for this regime, it is impossible
to achieve Eq.(\ref{const}) since Eq.(\ref{lamda}) has no nontrivial solution.
Note that there are no linear relations in this regime. Although
$\widetilde{\phi}=$ finite in this regime, it is different from the GR in the
26D scatterings since $K_{i}^{25}$ is as big as the scattering energy $p.$ In
conclusion, we have discovered a $\phi=$ finite regime with UV power-law
behavior for the high energy compactified open string scatterings. This new
phenomenon never happens in the 26D string scatterings. The linear relations
break down as expected in this regime.

$\bigskip\boldsymbol{II.}$ For the case of small angle $\phi\simeq0$
scattering, we consider the first power-law regime%
\begin{equation}
qK_{1}^{25}=-pK_{3}^{25}\text{ and }\left(  K_{i}^{25}\right)  ^{2}\simeq
p^{2}\simeq q^{2}\gg N. \label{uconst}%
\end{equation}
By the criterion of Eq.(\ref{26D}), this is a power-law regime. To explicitly
show that this choice of kinematic regime does lead to UV power-law behavior,
we will show that it implies
\begin{equation}
u=\text{ constant} \label{uconstant}%
\end{equation}
in the open string scattering amplitudes, which in turn gives the desire
power-law behavior of high energy compactified open string scattering in
Eq.(\ref{power}). For this choice of kinematic regime, Eq.(\ref{u}) and
Eq.(\ref{uconst}) imply
\begin{equation}
\lim_{p\rightarrow\infty}\frac{\sqrt{p^{2}+M_{1}^{2}}\cdot\sqrt{q^{2}%
+M_{3}^{2}}+pq}{K_{1}^{25}K_{3}^{25}}=-1. \label{ratio}%
\end{equation}
By choosing different sign for $K_{1}^{25}$ and $K_{3}^{25}$, Eq.(\ref{ratio})
can be solved for any real number $\lambda\equiv\frac{p}{K_{1}^{25}}=-\frac
{q}{K_{3}^{25}}.$

The second choice for the power-law regime is the same as Eq.(\ref{newpower})
in the $\phi=$ finite regime. The proof to show that it is indeed a power-law
regime is similar to the proof in section $I.$

The last choice for the power-law regime is%
\begin{equation}
\left(  K_{i}^{25}\right)  ^{2}\ll p^{2}\simeq q^{2}\gg N.
\end{equation}
It is easy to show that this is indeed a power-law regime.

The last kinematic regime for the case of small angle $\phi\simeq0$ scattering
is%
\begin{equation}
qK_{1}^{25}\neq-pK_{3}^{25}\text{ and }\left(  K_{i}^{25}\right)  ^{2}\simeq
p^{2}\simeq q^{2}\gg N.
\end{equation}
By the criterion of Eq.(\ref{26D}), this is an exponential fall-off regime. We
give one example here. Let's choose $\lambda\equiv\frac{p}{K_{1}^{25}}%
\neq-\frac{q}{K_{3}^{25}}=2\lambda.$ By choosing different sign for
$K_{1}^{25}$ and $K_{3}^{25}$, Eq.(\ref{ratio}) reduces to%
\begin{equation}
\lambda^{2}=0,
\end{equation}
which has no nontrivial solution for $\lambda$, and one can not achieve the
power-law condition Eq.(\ref{uconstant}). So this is an exponential fall-off
regime. In conclusion, we have discovered a $\phi\simeq0$ regime with UV
exponential fall-off behavior for the high energy compactified open string
scatterings. This new phenomenon never happens in the 26D string scatterings.
This completes the classification of all kinematic regimes for compactified
25D scatterings.

\subparagraph{Compactified 24D (or less) scatterings}

For this case, we need to introduce another parameter to classify the UV
behavior of high energy scatterings, namely the angle $\delta$ between
$\vec{K}_{1}$ and $\vec{K}_{2}$, $\vec{K}_{1}\cdot\vec{K}_{2}=\left\vert
K_{1}\right\vert \left\vert K_{2}\right\vert \cos\delta.$ Similar results can
be easily derived through the same method used in the compactified 25D
scatterings. The classification is independent of the details of the moduli
space of the compact spaces. We summarize the results in the following table:

\begin{center}%
\begin{tabular}
[c]{|c|c|c|c|c|}\hline
$\phi$ & $\tilde{\phi}$ & UV Behavior & Examples of the Kinematic Regimes &
Linear Relations\\\hline
&  &  & \multicolumn{1}{|l|}{$\vec{K}_{i}^{2}\ll p^{2}\simeq q^{2}\gg N$} &
Yes\\\cline{4-5}%
finite & finite & Exponential fall-off & \multicolumn{1}{|l|}{$\vec{K}_{i}%
^{2}\simeq p^{2}\simeq q^{2}\gg N$} & \\\cline{4-4}
&  &  & \multicolumn{1}{|l|}{$\vec{K}_{i}^{2}\gg p^{2}\simeq q^{2}\gg N$ and
$\cos\delta\neq0$} & No\\\cline{2-4}
& $\tilde{\phi}\simeq0$ & Power-law & \multicolumn{1}{|l|}{$\vec{K}_{i}^{2}\gg
p^{2}\simeq q^{2}\gg N$ and $\cos\delta=0$} & \\\hline
&  &  & \multicolumn{1}{|l|}{$\vec{K}_{i}^{2}\ll p^{2}\simeq q^{2}\gg N$} &
\\\cline{4-4}
& $\tilde{\phi}\simeq0$ & Power-law & \multicolumn{1}{|l|}{$\vec{K}_{i}%
^{2}\simeq p^{2}\simeq q^{2}\gg N$ and $q\vec{K}_{1}=-p\vec{K}_{3}$} &
\\\cline{4-4}%
$\phi\simeq0$ &  &  & \multicolumn{1}{|l|}{$\vec{K}_{i}^{2}\gg p^{2}\simeq
q^{2}\gg N$ and $\cos\delta=0$} & No.\\\cline{2-4}
& finite & Exponential fall-off & \multicolumn{1}{|l|}{$\vec{K}_{i}^{2}\simeq
p^{2}\simeq q^{2}\gg N$ and $q\vec{K}_{1}\neq-p\vec{K}_{3}$} & \\\cline{4-4}
&  &  & \multicolumn{1}{|l|}{$\vec{K}_{i}^{2}\gg p^{2}\simeq q^{2}\gg N$ and
$\cos\delta\neq0$} & \\\hline
\end{tabular}

\end{center}

\subsection{Closed string compactified on torus}

In this section, we consider hard scatterings of 26D closed bosonic string
\cite{Compact} with one coordinate compactified on $S^{1}$ with radius $R$. As
we will see later, it is straightforward to generalize our calculation to more
compactified coordinates.

\subsubsection{Winding string and kinematic setup}

The closed string boundary condition for the compactified coordinate is (we
use the notation in \cite{GSW})%

\begin{equation}
X^{25}(\sigma+2\pi,\tau)=X^{25}(\sigma,\tau)+2\pi Rn,
\end{equation}
where $n$ is the winding number. The momentum in the $X^{25}$ direction is
then quantized to be%
\begin{equation}
K=\frac{m}{R},
\end{equation}
where $m$ is an integer. The mode expansion of the compactified coordinate is%
\begin{equation}
X^{25}\left(  \sigma,\tau\right)  =X_{R}^{25}\left(  \sigma-\tau\right)
+X_{L}^{25}\left(  \sigma+\tau\right)  ,
\end{equation}
where%
\begin{align}
X_{R}^{25}\left(  \sigma-\tau\right)   &  =\frac{1}{2}x+K_{R}\left(
\sigma-\tau\right)  +i\sum_{k=0}\frac{1}{k}\alpha_{k}^{25}\text{
}e^{-2ik\left(  \sigma-\tau\right)  },\\
X_{L}^{25}\left(  \sigma+\tau\right)   &  =\frac{1}{2}x+K_{L}\left(
\sigma+\tau\right)  +i\sum_{k=0}\frac{1}{k}\tilde{\alpha}_{k}^{25}\text{
}e^{-2ik\left(  \sigma+\tau\right)  }.
\end{align}
The left and right momenta are defined to be%
\begin{equation}
K_{L,R}=K\pm L=\frac{m}{R}\pm\dfrac{1}{2}nR\Rightarrow K=\dfrac{1}{2}\left(
K_{L}+K_{R}\right)  ,
\end{equation}
and the mass spectrum can be calculated to be%
\begin{equation}
\left\{
\begin{array}
[c]{c}%
M^{2}=\left(  \dfrac{m^{2}}{R^{2}}+\dfrac{1}{4}n^{2}R^{2}\right)  +N_{R}%
+N_{L}-2\equiv K_{L}^{2}+M_{L}^{2}\equiv K_{R}^{2}+M_{R}^{2}\\
N_{R}-N_{L}=mn
\end{array}
\right.  , \label{mass}%
\end{equation}
where $N_{R}$ and $N_{L}$ are the number operators for the right and left
movers, which include the counting of the compactified coordinate. We have
also introduced the left and the right level masses as%
\begin{equation}
M_{L,R}^{2}\equiv2\left(  N_{L,R}-1\right)  . \label{level mass}%
\end{equation}
Note that for the compactified closed string $N_{R}$ and $N_{L}$ are
correlated through the winding modes.

In the center of momentum frame, the kinematic can be set up to be%

\begin{align}
k_{1L,R}  &  =\left(  +\sqrt{p^{2}+M_{1}^{2}},-p,0,-K_{1L,R}\right)  ,\\
k_{2L,R}  &  =\left(  +\sqrt{p^{2}+M_{2}^{2}},+p,0,+K_{2L,R}\right)  ,\\
k_{3L,R}  &  =\left(  -\sqrt{q^{2}+M_{3}^{2}},-q\cos\phi,-q\sin\phi
,-K_{3L,R}\right)  ,\\
k_{4L,R}  &  =\left(  -\sqrt{q^{2}+M_{4}^{2}},+q\cos\phi,+q\sin\phi
,+K_{4L,R}\right)
\end{align}
where $p\equiv\left\vert \mathrm{\vec{p}}\right\vert $ and $q\equiv\left\vert
\mathrm{\vec{q}}\right\vert $ and%
\begin{align}
k_{i}  &  \equiv\dfrac{1}{2}\left(  k_{iR}+k_{iL}\right)  ,\\
k_{i}^{2}  &  =K_{i}^{2}-M_{i}^{2},\\
k_{iL,R}^{2}  &  =K_{iL,R}^{2}-M_{i}^{2}\equiv-M_{iL,R}^{2}.
\end{align}
With this setup, the center of mass energy $E$ is%
\begin{equation}
E=\dfrac{1}{2}\left(  \sqrt{p^{2}+M_{1}^{2}}+\sqrt{p^{2}+M_{2}^{2}}\right)
=\dfrac{1}{2}\left(  \sqrt{q^{2}+M_{3}^{2}}+\sqrt{q^{2}+M_{4}^{2}}\right)  .
\label{COM}%
\end{equation}
The conservation of momentum on the compactified direction gives%
\begin{equation}
m_{1}-m_{2}+m_{3}-m_{4}=0, \label{kk}%
\end{equation}
and T-duality symmetry implies conservation of winding number%
\begin{equation}
n_{1}-n_{2}+n_{3}-n_{4}=0. \label{wind}%
\end{equation}
One can easily calculate the following kinematic relations%
\begin{align}
-k_{1L,R}\cdot k_{2L,R}  &  =\sqrt{p^{2}+M_{1}^{2}}\cdot\sqrt{p^{2}+M_{2}^{2}%
}+p^{2}+\vec{K}_{1L,R}\cdot\vec{K}_{2L,R}\label{k1*k2}\\
&  =\dfrac{1}{2}\left(  s_{L,R}+k_{1L,R}^{2}+k_{2L,R}^{2}\right)  =\dfrac
{1}{2}s_{L,R}-\frac{1}{2}\left(  M_{1L,R}^{2}+M_{2L,R}^{2}\right)  ,
\end{align}%
\begin{align}
-k_{2L,R}\cdot k_{3L,R}  &  =-\sqrt{p^{2}+M_{2}^{2}}\cdot\sqrt{q^{2}+M_{3}%
^{2}}+pq\cos\phi+\vec{K}_{2L,R}\cdot\vec{K}_{3L,R}\\
&  =\dfrac{1}{2}\left(  t_{L,R}+k_{2L,R}^{2}+k_{3L,R}^{2}\right)  =\dfrac
{1}{2}t_{L,R}-\frac{1}{2}\left(  M_{2L,R}^{2}+M_{3L,R}^{2}\right)  ,
\end{align}%
\begin{align}
-k_{1L,R}\cdot k_{3L,R}  &  =-\sqrt{p^{2}+M_{1}^{2}}\cdot\sqrt{q^{2}+M_{3}%
^{2}}-pq\cos\phi-\vec{K}_{1L,R}\cdot\vec{K}_{3L,R}\\
&  =\dfrac{1}{2}\left(  u_{L,R}+k_{1L,R}^{2}+k_{3L,R}^{2}\right)  =\dfrac
{1}{2}u_{L,R}-\frac{1}{2}\left(  M_{1L,R}^{2}+M_{3L,R}^{2}\right)
\end{align}
where the left and the right Mandelstam variables are defined to be%
\begin{align}
s_{L,R}  &  \equiv-(k_{1L,R}+k_{2L,R})^{2},\\
t_{L,R}  &  \equiv-(k_{2L,R}+k_{3L,R})^{2},\\
u_{L,R}  &  \equiv-(k_{1L,R}+k_{3L,R})^{2},
\end{align}
with%
\begin{equation}
s_{L,R}+t_{L,R}+u_{L,R}=\sum_{i}M_{iL,R}^{2}. \label{sum}%
\end{equation}

\subsubsection{Four-tachyon scatterings with $N_{R}=N_{L}=0$}

We are now ready to calculate the string scattering amplitudes. Let's first
calculate the case with $N_{R}+N_{L}=0$ (or $N_{R}=N_{L}=0$),%
\begin{align}
&  A_{\text{closed}}^{(N_{R}+N_{L}=0)}\left(  s,t,u\right) \nonumber\\
&  =\int d^{2}z\exp\left\{  k_{1L}\cdot k_{2L}\ln z+k_{1R}\cdot k_{2R}\ln
\bar{z}+k_{2L}\cdot k_{3L}\ln\left(  1-z\right)  +k_{2R}\cdot k_{3R}\ln\left(
1-\bar{z}\right)  \right\} \nonumber\\
&  =\int d^{2}z\exp\left\{  2k_{1R}\cdot k_{2R}\ln\left\vert z\right\vert
+2k_{2R}\cdot k_{3R}\ln\left\vert 1-z\right\vert \right. \nonumber\\
&  \text{ \ \ \ \ \ \ \ \ \ \ \ \ \ \ \ \ }\left.  +\left(  k_{1L}\cdot
k_{2L}-k_{1R}\cdot k_{2R}\right)  \ln z+\left(  k_{2L}\cdot k_{3L}-k_{2R}\cdot
k_{3R}\right)  \ln\left(  1-z\right)  \right\}
\end{align}
where we have used $\alpha^{\prime}=2$ for closed string propagators%
\begin{align}
\left\langle X\left(  z\right)  X\left(  z^{\prime}\right)  \right\rangle  &
=-\dfrac{\alpha^{\prime}}{2}\ln\left(  z-z^{\prime}\right)  ,\\
\left\langle \tilde{X}\left(  \bar{z}\right)  \tilde{X}\left(  \bar{z}%
^{\prime}\right)  \right\rangle  &  =-\dfrac{\alpha^{\prime}}{2}\ln\left(
\bar{z}-\bar{z}^{\prime}\right)  .
\end{align}
Note that for this simple case, Eq.(\ref{mass}) implies either $m=0$ or $n=0$.
However, we will keep track of the general values of $(m,n)$ here for the
reference of future calculations. By using the formula \cite{Heter}%
\begin{align}
I  &  =\int\frac{d^{2}z}{\pi}\left\vert z\right\vert ^{\alpha}\left\vert
1-z\right\vert ^{\beta}z^{n}\left(  1-z\right)  ^{m}\nonumber\\
&  =\frac{\Gamma\left(  -1-\frac{1}{2}\alpha-\frac{1}{2}\beta\right)
\Gamma\left(  1+n+\frac{1}{2}\alpha\right)  \Gamma\left(  1+m+\frac{1}{2}%
\beta\right)  }{\Gamma\left(  -\frac{1}{2}\alpha\right)  \Gamma\left(
-\frac{1}{2}\beta\right)  \Gamma\left(  2+n+m+\frac{1}{2}\alpha+\frac{1}%
{2}\beta\right)  }, \label{Heter}%
\end{align}
we obtain%
\begin{align}
&  A_{\text{closed}}^{(N_{R}+N_{L}=0)}\left(  s,t,u\right) \nonumber\\
&  =\pi\frac{\Gamma\left(  -1-k_{1R}\cdot k_{2R}-k_{2R}\cdot k_{3R}\right)
\Gamma\left(  1+k_{1L}\cdot k_{2L}\right)  \Gamma\left(  1+k_{2L}\cdot
k_{3L}\right)  }{\Gamma\left(  -k_{1R}\cdot k_{2R}\right)  \Gamma\left(
-k_{2R}\cdot k_{3R}\right)  \Gamma\left(  2+k_{1L}\cdot k_{2L}+k_{2L}\cdot
k_{3L}\right)  }\nonumber\\
&  =\frac{\sin\left(  -\pi k_{1R}\cdot k_{2R}\right)  \sin\left(  -\pi
k_{2R}\cdot k_{3R}\right)  }{\sin\left(  -\pi-\pi k_{1R}\cdot k_{2R}-\pi
k_{2R}\cdot k_{3R}\right)  }\nonumber\\
&  \text{ \ \ \ }\cdot\frac{\Gamma\left(  1+k_{1R}\cdot k_{2R}\right)
\Gamma\left(  1+k_{2R}\cdot k_{3R}\right)  }{\Gamma\left(  2+k_{1R}\cdot
k_{2R}+k_{2R}\cdot k_{3R}\right)  }\frac{\Gamma\left(  1+k_{1L}\cdot
k_{2L}\right)  \Gamma\left(  1+k_{2L}\cdot k_{3L}\right)  }{\Gamma\left(
2+k_{1L}\cdot k_{2L}+k_{2L}\cdot k_{3L}\right)  }\nonumber\\
&  \simeq\frac{\sin\left(  \pi s_{L}/2\right)  \sin\left(  \pi t_{R}/2\right)
}{\sin\left(  \pi u_{L}/2\right)  }\frac{\Gamma\left(  -1-\frac{t_{R}}%
{2}\right)  \Gamma\left(  -1-\frac{u_{R}}{2}\right)  }{\Gamma\left(
2+\frac{s_{R}}{2}\right)  }\frac{\Gamma\left(  -1-\frac{t_{L}}{2}\right)
\Gamma\left(  -1-\frac{u_{L}}{2}\right)  }{\Gamma\left(  2+\frac{s_{L}}%
{2}\right)  }, \label{4-tachyon}%
\end{align}
where we have used $M_{iL,R}^{2}=-2$ for $i=1,2,3,4$. In the above
calculation, we have used the following well known formula for gamma function%
\begin{equation}
\Gamma\left(  x\right)  =\frac{\pi}{\sin\left(  \pi x\right)  \Gamma\left(
1-x\right)  }. \label{gamma}%
\end{equation}

\subsubsection{High energy massive scatterings for general $N_{R}+N_{L}$}

We now proceed to calculate the high energy scattering amplitudes for general
higher mass levels with fixed $N_{R}+N_{L}$. With one compactified coordinate,
the mass spectrum of the second vertex of the amplitude is%
\begin{equation}
M_{2}^{2}=\left(  \dfrac{m_{2}^{2}}{R^{2}}+\dfrac{1}{4}n_{2}^{2}R^{2}\right)
+N_{R}+N_{L}-2.
\end{equation}%
\begin{figure}[ptb]%
\centering
\includegraphics[
height=2.258in,
width=3.589in
]%
{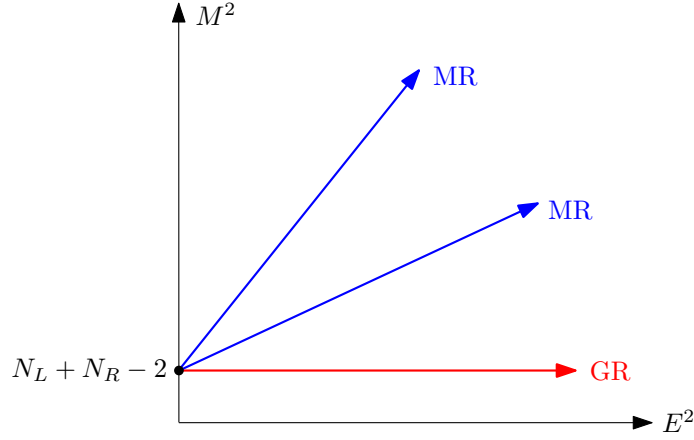}%
\caption{Different regimes of "high energy limit". The high energy regime
defined by $E^{2}\simeq M_{2}^{2}$ $\gg$ $N_{R}+N_{L}$ will be called Mende
regime (MR). The high energy regime defined by $E^{2}\gg M_{2}^{2}\simeq
N_{R}+N_{L}$ will be called Gross region (GR).}%
\label{GR-MR}%
\end{figure}
We now have more mass parameters to define the "high energy limit". So let's
first clear and redefine the concept of "high energy limit" in our following
calculations. We are going to use three quantities $E^{2},M_{2}^{2}$ and
$N_{R}+N_{L}$ to define different regimes of "high energy limit". See FIG.
\ref{GR-MR}. The high energy regime defined by $E^{2}\simeq M_{2}^{2}$ $\gg$
$N_{R}+N_{L}$ will be called Mende regime (MR). The high energy regime defined
by $E^{2}\gg M_{2}^{2}$, $E^{2}\gg$ $N_{R}+N_{L}$ will be called Gross region
(GR). In the high energy limit, the polarizations on the scattering plane for
the second vertex operator are defined to be%

\begin{align}
e^{p}  &  =\frac{1}{M_{2}}\left(  \sqrt{p^{2}+M_{2}^{2}},p,0,0\right)  ,\\
e^{L}  &  =\frac{1}{M_{2}}\left(  p,\sqrt{p^{2}+M_{2}^{2}},0,0\right)  ,\\
e^{T}  &  =\left(  0,0,1,0\right)
\end{align}
where the fourth component refers to the compactified direction. One can
calculate the following formulas in the high energy limit%
\begin{align}
e^{p}\cdot k_{1L}  &  =e^{p}\cdot k_{1R}=-\frac{1}{M_{2}}\left(  \sqrt
{p^{2}+M_{1}^{2}}\sqrt{p^{2}+M_{2}^{2}}+p^{2}\right) \nonumber\\
&  =-\frac{p^{2}}{M_{2}}\left(  2+\frac{M_{1}^{2}}{2p^{2}}+\frac{M_{2}^{2}%
}{2p^{2}}\right)  +\mathcal{O}(p^{-2}),\label{K1}\\
e^{p}\cdot k_{3L}  &  =e^{p}\cdot k_{3R}=\frac{1}{M_{2}}\left(  \sqrt
{q^{2}+M_{3}^{2}}\sqrt{p^{2}+M_{2}^{2}}-pq\cos\phi\right) \nonumber\\
&  =\frac{pq}{M_{2}}\left[  1-\cos\phi+\frac{M_{2}^{2}}{2p^{2}}+\frac
{M_{3}^{2}}{2q^{2}}\right]  +\mathcal{O}(p^{-2}),
\end{align}%
\begin{align}
e^{L}\cdot k_{1L}  &  =e^{L}\cdot k_{1R}=-\frac{p}{M_{2}}\left(  \sqrt
{p^{2}+M_{1}^{2}}+\sqrt{p^{2}+M_{2}^{2}}\right) \nonumber\\
&  =-\frac{p^{2}}{M_{2}}\left(  2+\frac{M_{1}^{2}}{2p^{2}}+\frac{M_{2}^{2}%
}{2p^{2}}\right)  +\mathcal{O}(p^{-2}),\\
e^{L}\cdot k_{3L}  &  =e^{L}\cdot k_{3R}=\frac{1}{M_{2}}\left(  p\sqrt
{q^{2}+M_{3}^{2}}-q\sqrt{p^{2}+M_{2}^{2}}\cos\phi\right) \nonumber\\
&  =\frac{pq}{M_{2}}\left[  1+\frac{M_{3}^{2}}{2q^{2}}-\left(  1+\frac
{M_{2}^{2}}{2p^{2}}\right)  \cos\phi\right]  +\mathcal{O}(p^{-2}), \label{K8}%
\end{align}%
\begin{align}
e^{T}\cdot k_{1L}  &  =e^{T}\cdot k_{1R}=0,\\
e^{T}\cdot k_{3L}  &  =e^{T}\cdot k_{3R}=-q\sin\phi, \label{K2}%
\end{align}
which will be useful in the calculations of high energy string scattering amplitudes.

For the noncompactified open string, it was shown that
\cite{CHLTY1,CHLTY2,CHLTY3}, at each fixed mass level $M_{op}^{2}=2(N-1)$, a
four-point function is at the leading order in high energy (GR) only for
states of the following form
\begin{equation}
\left\vert N,2l,q\right\rangle \equiv(\alpha_{-1}^{T})^{N-2l-2q}(\alpha
_{-1}^{L})^{2l}(\alpha_{-2}^{L})^{q}\left\vert 0,k\right\rangle
\end{equation}
where $N\geqslant2l+2q,l,q\geqslant0.$To avoid the complicated subleading
order calculation due to the $\alpha_{-1}^{L}$ operator, we will choose the
simple case $l=0$. We made a similar choice when dealing with the high energy
string/D-brane scatterings \cite{Dscatt, Wall, Decay}. There is still one
complication in the case of compactified string due to the possible choices of
$\alpha_{-n}^{25}$ and $\tilde{\alpha}_{-m}^{25}$ in the vertex operator.
However, it can be easily shown that for each fixed mass level with given
quantized and winding momenta $(\frac{m}{R},\frac{1}{2}nR)$, and thus fixed
$N_{R}+N_{L}$ level, vertex operators containing $\alpha_{-n}^{25}$ or
$\tilde{\alpha}_{-m}^{25}$ are subleading order in energy in the high energy
expansion compared to other choices $\alpha_{-1}^{T}(\tilde{\alpha}_{-1}^{T})$
and $\alpha_{-2}^{L}$ $(\tilde{\alpha}_{-2}^{L})$ on the noncompact
directions. In conclusion, in the calculation of compactified closed string in
the GR, we are going to consider tensor state of the form%
\begin{equation}
\left\vert N_{L,R},q_{L,R}\right\rangle \equiv\left(  \alpha_{-1}^{T}\right)
^{N_{L}-2q_{L}}\left(  \alpha_{-2}^{L}\right)  ^{q_{L}}\otimes\left(
\tilde{\alpha}_{-1}^{T}\right)  ^{N_{R}-2q_{R}}\left(  \tilde{\alpha}_{-2}%
^{L}\right)  ^{q_{R}}\left\vert 0\right\rangle , \label{tensor}%
\end{equation}
at general $N_{R}+N_{L}$ level scattered from three other tachyon states with
$N_{R}+N_{L}=0$.

Note that, in the GR, one can identify $e^{p}$ with $e^{L}$ as usual
\cite{ChanLee,ChanLee1,ChanLee2}. However, in the MR, one can not identify
$e^{p}$ with $e^{L}$. This can be seen from Eq.(\ref{K1}) to Eq.(\ref{K8}). In
the MR, instead of using the tensor vertex in Eq.(\ref{tensor}), we will use%
\begin{equation}
\left\vert N_{L,R},q_{L,R}\right\rangle \equiv\left(  \alpha_{-1}^{T}\right)
^{N_{L}-2q_{L}}\left(  \alpha_{-2}^{P}\right)  ^{q_{L}}\otimes\left(
\tilde{\alpha}_{-1}^{T}\right)  ^{N_{R}-2q_{R}}\left(  \tilde{\alpha}_{-2}%
^{P}\right)  ^{q_{R}}\left\vert 0\right\rangle , \label{new.}%
\end{equation}
as the second vertex operator in the calculation of high energy scattering
amplitudes. Note also that, in the MR, states in Eq.(\ref{new.}) may not be
the only states which contribute to the high energy scattering amplitudes as
in the GR. However, we will just choose these states to calculate the
scattering amplitudes in order to compare with the corresponding high energy
scattering amplitudes in the GR.

The high energy scattering amplitudes in the MR can be calculated to be
\begin{align}
A  &  =\varepsilon_{T^{N_{L}-2q_{L}}P^{q_{L}},T^{N_{R}-2q_{R}}P^{q_{R}}}\int
d^{2}z_{1}d^{2}z_{2}d^{2}z_{3}d^{2}z_{4}\nonumber\\
&  \cdot\left\langle V_{1}\left(  z_{1},\bar{z}_{1}\right)  V_{2}%
^{T^{N_{L}-2q_{L}}P^{q_{L}},T^{N_{R}-2q_{R}}P^{q_{R}}}\left(  z_{2},\bar
{z}_{2}\right)  V_{3}\left(  z_{3},\bar{z}_{3}\right)  V_{4}\left(  z_{4}%
,\bar{z}_{4}\right)  \right\rangle \nonumber\\
&  =\varepsilon_{T^{N_{L}-2q_{L}}P^{q_{L}},T^{N_{R}--2q_{R}}P^{q_{R}}}\int
d^{2}z_{1}d^{2}z_{2}d^{2}z_{3}d^{2}z_{4}\left\langle e^{ik_{1L}X}\left(
z_{1}\right)  e^{ik_{1R}\tilde{X}}\left(  \bar{z}_{1}\right)  \right.
\nonumber\\
&  \left.  \cdot\left(  \partial X^{T}\right)  ^{N_{L}-2q_{L}}\left(
i\partial^{2}X^{P}\right)  ^{q_{L}}e^{ik_{2L}X}\left(  z_{2}\right)  \left(
\bar{\partial}\tilde{X}^{T}\right)  ^{N_{R}-2q_{R}}\left(  i\bar{\partial}%
^{2}\tilde{X}^{P}\right)  ^{q_{R}}e^{ik_{2R}\tilde{X}}\left(  \bar{z}%
_{2}\right)  \right. \nonumber\\
&  \left.  e^{ik_{3L}X}\left(  z_{3}\right)  e^{ik_{3R}\tilde{X}}\left(
\bar{z}_{3}\right)  e^{ik_{4L}X}\left(  z_{4}\right)  e^{ik_{4R}\tilde{X}%
}\left(  \bar{z}_{4}\right)  \right\rangle \nonumber\\
&  =\int d^{2}z_{1}d^{2}z_{2}d^{2}z_{3}d^{2}z_{4}\cdot\left[  \prod
\limits_{i<j}\left(  z_{i}-z_{j}\right)  ^{k_{iL}\cdot k_{jL}}\left(  \bar
{z}_{i}-\bar{z}_{j}\right)  ^{k_{iR}\cdot k_{jR}}\right] \nonumber\\
&  \cdot\left[  \frac{ie^{T}\cdot k_{1L}}{z_{1}-z_{2}}+\frac{ie^{T}\cdot
k_{3L}}{z_{3}-z_{2}}+\frac{ie^{T}\cdot k_{4L}}{z_{4}-z_{2}}\right]
^{N_{L}-2q_{L}}\cdot\left[  \frac{e^{p}\cdot k_{1L}}{\left(  z_{1}%
-z_{2}\right)  ^{2}}+\frac{e^{p}\cdot k_{3L}}{\left(  z_{3}-z_{2}\right)
^{2}}+\frac{e^{p}\cdot k_{4L}}{\left(  z_{4}-z_{2}\right)  ^{2}}\right]
^{q_{L}}\nonumber\\
&  \cdot\left[  \frac{ie^{T}\cdot k_{1R}}{\bar{z}_{1}-\bar{z}_{2}}%
+\frac{ie^{T}\cdot k_{3R}}{\bar{z}_{3}-\bar{z}_{2}}+\frac{ie^{T}\cdot k_{4R}%
}{\bar{z}_{4}-\bar{z}_{2}}\right]  ^{N_{R}-2q_{R}}\cdot\left[  \frac
{e^{p}\cdot k_{1R}}{\left(  \bar{z}_{1}-\bar{z}_{2}\right)  ^{2}}+\frac
{e^{p}\cdot k_{3R}}{\left(  \bar{z}_{3}-\bar{z}_{2}\right)  ^{2}}+\frac
{e^{p}\cdot k_{4R}}{\left(  \bar{z}_{4}-\bar{z}_{2}\right)  ^{2}}\right]
^{q_{R}}.
\end{align}
\ After the standard $SL(2,C)$ gauge fixing, we get%
\begin{align}
A  &  \simeq\left(  -1\right)  ^{k_{1L}\cdot k_{2L}+k_{1R}\cdot k_{2R}%
+k_{1L}\cdot k_{3L}+k_{1R}\cdot k_{3R}+k_{2L}\cdot k_{3L}+k_{2R}\cdot k_{3R}%
}\nonumber\\
&  \cdot\int d^{2}z\cdot z^{k_{1L}\cdot k_{2L}}\cdot\bar{z}^{k_{1R}\cdot
k_{2R}}\cdot\left(  1-z\right)  ^{k_{2L}\cdot k_{3L}}\left(  1-\bar{z}\right)
^{k_{2R}\cdot k_{3R}}\nonumber\\
&  \cdot\left[  \frac{ie^{T}\cdot k_{1L}}{z}-\frac{ie^{T}\cdot k_{3L}}%
{1-z}\right]  ^{N_{L}-2q_{L}}\cdot\left[  \frac{ie^{T}\cdot k_{1R}}{\bar{z}%
}-\frac{ie^{T}\cdot k_{3R}}{1-\bar{z}}\right]  ^{N_{R}-2q_{R}}\nonumber\\
&  \cdot\left[  \frac{e^{p}\cdot k_{1L}}{z^{2}}+\frac{e^{p}\cdot k_{3L}%
}{\left(  1-z\right)  ^{2}}\right]  ^{q_{L}}\cdot\left[  \frac{e^{p}\cdot
k_{1R}}{\bar{z}^{2}}+\frac{e^{p}\cdot k_{3R}}{\left(  1-\bar{z}\right)  ^{2}%
}\right]  ^{q_{R}}.
\end{align}
By using Eqs.(\ref{K1}) to (\ref{K2}), the amplitude can be written as%

\begin{align*}
A  &  \sim\left(  -1\right)  ^{n+q+q^{\prime}+k_{1L}\cdot k_{2L}+k_{1R}\cdot
k_{2R}+k_{1L}\cdot k_{3L}+k_{1R}\cdot k_{3R}+k_{2L}\cdot k_{3L}+k_{2R}\cdot
k_{3R}}\left(  q\sin\phi\right)  ^{N_{L}+N_{R}-2q_{L}-2q_{R}}\\
&  \cdot\int d^{2}z\cdot z^{k_{1L}\cdot k_{2L}}\cdot\bar{z}^{k_{1R}\cdot
k_{2R}}\cdot\left(  1-z\right)  ^{k_{2L}\cdot k_{3L}}\left(  1-\bar{z}\right)
^{k_{2R}\cdot k_{3R}}\cdot\left[  \frac{1}{1-z}\right]  ^{N_{L}-2q_{L}}\left[
\frac{1}{1-\bar{z}}\right]  ^{N_{R}-2q_{R}}\\
&  \cdot\left[  \frac{-\frac{1}{M_{2}}\left(  \sqrt{p^{2}+M_{1}^{2}}%
\sqrt{p^{2}+M_{2}^{2}}+p^{2}\right)  }{z^{2}}+\frac{\frac{1}{M_{2}}\left(
\sqrt{q^{2}+M_{3}^{2}}\sqrt{p^{2}+M_{2}^{2}}-pq\cos\phi\right)  }{\left(
1-z\right)  ^{2}}\right]  ^{q_{L}}\\
&  \cdot\left[  \frac{-\frac{1}{M_{2}}\left(  \sqrt{p^{2}+M_{1}^{2}}%
\sqrt{p^{2}+M_{2}^{2}}+p^{2}\right)  }{\bar{z}^{2}}+\frac{\frac{1}{M_{2}%
}\left(  \sqrt{q^{2}+M_{3}^{2}}\sqrt{p^{2}+M_{2}^{2}}-pq\cos\phi\right)
}{\left(  1-\bar{z}\right)  ^{2}}\right]  ^{q_{R}}%
\end{align*}

\begin{align}
&  =\left(  -1\right)  ^{k_{1L}\cdot k_{2L}+k_{1R}\cdot k_{2R}+k_{1L}\cdot
k_{3L}+k_{1R}\cdot k_{3R}+k_{2L}\cdot k_{3L}+k_{2R}\cdot k_{3R}}\left(
q\sin\phi\right)  ^{N_{L}+N_{R}}\left(  \frac{1}{M_{2}q^{2}\sin^{2}\phi
}\right)  ^{q_{L}+q_{R}}\nonumber\\
&  \cdot\int d^{2}z\cdot z^{k_{1L}\cdot k_{2L}}\cdot\bar{z}^{k_{1R}\cdot
k_{2R}}\cdot\left(  1-z\right)  ^{k_{2L}\cdot k_{3L}}\left(  1-\bar{z}\right)
^{k_{2R}\cdot k_{3R}}\cdot\left[  \frac{1}{1-z}\right]  ^{N_{L}-2q_{L}}\left[
\frac{1}{1-\bar{z}}\right]  ^{N_{R}-2q_{R}}\nonumber\\
&  \cdot\sum_{i=0}^{q}\sum_{j=0}^{q^{\prime}}\binom{q}{i}\binom{q^{\prime}}%
{j}\left(  \frac{\sqrt{p^{2}+M_{1}^{2}}\sqrt{p^{2}+M_{2}^{2}}+p^{2}}{z^{2}%
}\right)  ^{i}\left(  \frac{\sqrt{p^{2}+M_{1}^{2}}\sqrt{p^{2}+M_{2}^{2}}%
+p^{2}}{\bar{z}^{2}}\right)  ^{j}\nonumber\\
&  =\left(  -1\right)  ^{k_{1L}\cdot k_{2L}+k_{1R}\cdot k_{2R}+k_{1L}\cdot
k_{3L}+k_{1R}\cdot k_{3R}+k_{2L}\cdot k_{3L}+k_{2R}\cdot k_{3R}}\left(
q\sin\phi\right)  ^{N_{L}+N_{R}}\nonumber\\
&  \cdot\left(  -\frac{\sqrt{q^{2}+M_{3}^{2}}\sqrt{p^{2}+M_{2}^{2}}-pq\cos
\phi}{M_{2}q^{2}\sin^{2}\phi}\right)  ^{q_{L}+q_{R}}\nonumber\\
&  \cdot\sum_{i=0}^{q_{L}}\sum_{j=0}^{q_{R}}\binom{q_{L}}{i}\binom{q_{R}}%
{j}\left(  \frac{\sqrt{p^{2}+M_{1}^{2}}\sqrt{p^{2}+M_{2}^{2}}+p^{2}}%
{-\sqrt{q^{2}+M_{3}^{2}}\sqrt{p^{2}+M_{2}^{2}}+pq\cos\phi}\right)
^{i+j}\nonumber\\
&  \cdot\frac{\sin\left[  -\pi\left(  k_{1R}\cdot k_{2R}-2j\right)  \right]
\sin\left[  -\pi\left(  k_{2R}\cdot k_{3R}-N_{R}+2j\right)  \right]  }%
{\sin\left[  -\pi\left(  1+k_{1R}\cdot k_{2R}+k_{2R}\cdot k_{3R}-N_{R}\right)
\right]  }\nonumber\\
&  \cdot\frac{\Gamma\left(  1+k_{1R}\cdot k_{2R}-2j\right)  \Gamma\left(
1+k_{2R}\cdot k_{3R}-N_{R}+2j\right)  }{\Gamma\left(  2+k_{1R}\cdot
k_{2R}+k_{2R}\cdot k_{3R}-N_{R}\right)  }\nonumber\\
&  \cdot\frac{\Gamma\left(  1+k_{1L}\cdot k_{2L}-2i\right)  \Gamma\left(
1+k_{2L}\cdot k_{3L}+2i-N_{L}\right)  }{\Gamma\left(  2+k_{1L}\cdot
k_{2L}+k_{2L}\cdot k_{3L}-N_{L}\right)  }%
\end{align}
where, as in the calculation of section B.2 for the GR, we have used
Eq.(\ref{Heter}) to do the integration. It is easy to do the following
approximations for the gamma functions%
\begin{align}
&  A\simeq\left(  -1\right)  ^{k_{1L}\cdot k_{2L}+k_{1R}\cdot k_{2R}%
+k_{1L}\cdot k_{3L}+k_{1R}\cdot k_{3R}+k_{2L}\cdot k_{3L}+k_{2R}\cdot k_{3R}%
}\left(  q\sin\phi\right)  ^{N_{L}+N_{R}}\nonumber\\
&  \cdot\left(  -\frac{\sqrt{q^{2}+M_{3}^{2}}\sqrt{p^{2}+M_{2}^{2}}-pq\cos
\phi}{M_{2}q^{2}\sin^{2}\phi}\right)  ^{q_{L}+q_{R}}\nonumber\\
&  \cdot\sum_{i=0}^{q_{L}}\sum_{j=0}^{q_{R}}\binom{q_{L}}{i}\binom{q_{R}}%
{j}\left(  \frac{\sqrt{p^{2}+M_{1}^{2}}\sqrt{p^{2}+M_{2}^{2}}+p^{2}}%
{-\sqrt{q^{2}+M_{3}^{2}}\sqrt{p^{2}+M_{2}^{2}}+pq\cos\phi}\right)
^{i+j}\nonumber\\
&  \cdot\frac{\sin\left[  -\pi k_{1R}\cdot k_{2R}\right]  \sin\left[  -\pi
k_{2R}\cdot k_{3R}\right]  }{\sin\left[  -\pi\left(  1+k_{1R}\cdot
k_{2R}+k_{2R}\cdot k_{3R}\right)  \right]  }\nonumber\\
&  \cdot\frac{\Gamma\left(  1+k_{1R}\cdot k_{2R}\right)  \Gamma\left(
1+k_{2R}\cdot k_{3R}\right)  \Gamma\left(  1+k_{1L}\cdot k_{2L}\right)
\Gamma\left(  1+k_{2L}\cdot k_{3L}\right)  }{\Gamma\left(  2+k_{1R}\cdot
k_{2R}+k_{2R}\cdot k_{3R}\right)  \Gamma\left(  2+k_{1L}\cdot k_{2L}%
+k_{2L}\cdot k_{3L}\right)  }\nonumber\\
&  \cdot\dfrac{\left(  k_{1R}\cdot k_{2R}\right)  ^{-2j}\left(  k_{2R}\cdot
k_{3R}\right)  ^{-N_{R}+2j}}{\left(  k_{1R}\cdot k_{2R}+k_{2R}\cdot
k_{3R}\right)  ^{-N_{R}}}\dfrac{\left(  k_{1L}\cdot k_{2L}\right)
^{-2i}\left(  k_{2L}\cdot k_{3L}\right)  ^{-N_{L}+2i}}{\left(  k_{1L}\cdot
k_{2L}+k_{2L}\cdot k_{3L}\right)  ^{-N_{L}}}.
\end{align}
One can now do the double summation\ and drop out the $M_{iL,R}$ terms to get%
\begin{align}
A  &  \simeq\left(  -\dfrac{q\sin\phi\left(  s_{L}+t_{L}\right)  }{t_{L}%
}\right)  ^{N_{L}}\left(  -\dfrac{q\sin\phi\left(  s_{R}+t_{R}\right)  }%
{t_{R}}\right)  ^{N_{R}}\left(  \frac{1}{2M_{2}q^{2}\sin^{2}\phi}\right)
^{q_{L}+q_{R}}\nonumber\\
&  \cdot\left(  \left(  t_{R}-2\vec{K}_{2R}\cdot\vec{K}_{3R}\right)
+\dfrac{t_{R}^{2}\left(  s_{R}-2\vec{K}_{1R}\cdot\vec{K}_{2R}\right)  }%
{s_{R}^{2}}\right)  ^{q_{R}}\nonumber\\
&  \cdot\left(  \left(  t_{L}-2\vec{K}_{2L}\cdot\vec{K}_{3L}\right)
+\dfrac{t_{L}^{2}\left(  s_{L}-2\vec{K}_{1L}\cdot\vec{K}_{2L}\right)  }%
{s_{L}^{2}}\right)  ^{q_{L}}\nonumber\\
&  \cdot\frac{\sin\left(  \pi s_{L}/2\right)  \sin\left(  \pi t_{R}/2\right)
}{\sin\left(  \pi u_{L}/2\right)  }B\left(  -1-\dfrac{t_{R}}{2},-1-\dfrac
{u_{R}}{2}\right)  B\left(  -1-\dfrac{t_{L}}{2},-1-\dfrac{u_{L}}{2}\right)  .
\label{amplitude}%
\end{align}
Eq.(\ref{amplitude}) is valid for $E^{2}\gg N_{R}+N_{L},$ $M_{2}^{2}\gg
N_{R}+N_{L}.$

\subsubsection{The infinite linear relations in the GR}

For the special case of GR with $E^{2}\gg M_{2}^{2}$, one can identify $q$
with $p$, and the amplitude in Eq.(\ref{amplitude}) further reduces to%
\begin{align}
\lim_{E^{2}\gg M_{2}^{2}}A  &  =\left(  \frac{2p\cos^{3}\frac{\phi}{2}}%
{\sin\frac{\phi}{2}}\right)  ^{N_{L}+N_{R}}\left(  -\frac{1}{2M_{2}}\right)
^{q_{L}+q_{R}}\frac{\sin\left(  \pi s_{L}/2\right)  \sin\left(  \pi
t_{R}/2\right)  }{\sin\left(  \pi u_{L}/2\right)  }\nonumber\\
&  \cdot B\left(  -1-\dfrac{t_{R}}{2},-1-\dfrac{u_{R}}{2}\right)  B\left(
-1-\dfrac{t_{L}}{2},-1-\dfrac{u_{L}}{2}\right)  . \label{beta.}%
\end{align}
It is crucial to note that the high energy limit of the beta function with
$s+t+u=2n-8$ is \cite{ChanLee,ChanLee1}%
\begin{align}
B\left(  -1-\dfrac{t}{2},-1-\dfrac{u}{2}\right)   &  =\frac{\Gamma(-\frac
{t}{2}-1)\Gamma(-\frac{u}{2}-1)}{\Gamma(\frac{s}{2}+2)}\nonumber\\
&  \simeq E^{-1-2n}\left(  \sin\frac{\phi}{2}\right)  ^{-3}\left(  \cos
\frac{\phi}{2}\right)  ^{5-2n}\nonumber\\
&  \cdot\exp\left(  -\frac{t\ln t+u\ln u-(t+u)\ln(t+u)}{2}\right)
\end{align}
where we have calculated the approximation up to the next leading order in
energy $E$. Note the appearance of the prepower factors in front of the
exponential fall-off factor. For our purpose here, with Eq.(\ref{sum}), we
have%
\begin{equation}
s_{L,R}+t_{L,R}+u_{L,R}=\sum_{i}M_{iL,R}^{2}=2N_{L,R}-8,
\end{equation}
and the high energy limit of the beta functions in Eq.(\ref{beta}) can be
further calculated to be%
\begin{align}
&  B\left(  -1-\dfrac{t_{R}}{2},-1-\dfrac{u_{R}}{2}\right)  B\left(
-1-\dfrac{t_{L}}{2},-1-\dfrac{u_{L}}{2}\right) \nonumber\\
&  \simeq E^{-1-2(N_{L}+N_{R})}\left(  \sin\frac{\phi}{2}\right)  ^{-3}\left(
\cos\frac{\phi}{2}\right)  ^{5-2(N_{L}+N_{R})}\nonumber\\
&  \cdot\exp\left(  -\frac{t_{L}\ln t_{L}+u_{L}\ln u_{L}-(t_{L}+u_{L}%
)\ln(t_{L}+u_{L})}{2}\right) \nonumber\\
&  \cdot\exp\left(  -\frac{t_{R}\ln t_{R}+u_{R}\ln u_{R}-(t_{R}+u_{R}%
)\ln(t_{R}+u_{R})}{2}\right) \nonumber\\
&  \simeq E^{-1-2(N_{L}+N_{R})}\left(  \sin\frac{\phi}{2}\right)  ^{-3}\left(
\cos\frac{\phi}{2}\right)  ^{5-2(N_{L}+N_{R})}\exp\left(  -\frac{t\ln t+u\ln
u-(t+u)\ln(t+u)}{4}\right)  \label{final..}%
\end{align}
where we have implicitly used the relation $\alpha_{\text{closed}}^{\prime
}=4\alpha_{\text{open}}^{\prime}=2.$ By combining Eq.(\ref{beta.}) and
Eq.(\ref{final..}), we end up with%
\begin{align}
\lim_{E^{2}\gg M_{2}^{2}}A  &  \simeq\left(  -\frac{2\cot\frac{\phi}{2}}%
{E}\right)  ^{N_{L}+N_{R}}\left(  -\frac{1}{2M_{2}}\right)  ^{q_{L}+q_{R}%
}E^{-1}\left(  \sin\frac{\phi}{2}\right)  ^{-3}\left(  \cos\frac{\phi}%
{2}\right)  ^{5}\nonumber\\
&  \cdot\frac{\sin\left(  \pi s_{L}/2\right)  \sin\left(  \pi t_{R}/2\right)
}{\sin\left(  \pi u_{L}/2\right)  }\exp\left(  -\frac{t\ln t+u\ln
u-(t+u)\ln(t+u)}{4}\right)  .
\end{align}
We see that there is a $\left(  \frac{m}{R},\frac{1}{2}nR\right)  $ dependence
in the $\frac{\sin\left(  \pi s_{L}/2\right)  \sin\left(  \pi t_{R}/2\right)
}{\sin\left(  \pi u_{L}/2\right)  }$ factor in our final result. This is
physically consistent as one expects a $\left(  \frac{m}{R},\frac{1}%
{2}nR\right)  $ dependent Regge-pole and zero structures in the high energy
string scattering amplitudes.

In conclusion, in the GR, for each fixed mass level with given quantized and
winding momenta $\left(  \frac{m}{R},\frac{1}{2}nR\right)  $ (thus fixed
$N_{L}$ and $N_{R}$ by Eq.(\ref{mass})), we have obtained infinite linear
relations among high energy scattering amplitudes of different string states
with various $(q_{L},q_{R})$. Note also that this result reproduces the
correct ratios $\left(  -\frac{1}{2M_{2}}\right)  ^{q_{L}+q_{R}}$ obtained in
the previous works \cite{Dscatt, Wall, Decay}. However, the mass parameter
$M_{2}$ here depends on $\left(  \frac{m}{R},\frac{1}{2}nR\right)  $. It is
also interesting to see that, if not for the $\left(  \frac{m}{R},\frac{1}%
{2}nR\right)  $ dependence in the $\frac{\sin\left(  \pi s_{L}/2\right)
\sin\left(  \pi t_{R}/2\right)  }{\sin\left(  \pi u_{L}/2\right)  }$ factor in
the high energy scattering amplitudes in the GR, we would have had a linear
relation among scattering amplitudes of different string states in different
mass levels with fixed $\left(  N_{R}+N_{L}\right)  $.

Presumably, the infinite linear relations obtained above can be reproduced by
using the method of high energy ZNS, or high energy Ward identities, adopted
in the previous chapters \cite{ChanLee,ChanLee1,ChanLee2,
CHL,CHLTY1,CHLTY2,CHLTY3,susy,Closed,Dscatt,Decay}. The existence of Soliton
ZNS at arbitrary mass levels was constructed in chapter IV \cite{Lee}. A
closer look in this direction seems worthwhile. In the chapter, however, we
are more interested in understanding the power-law behavior of the high energy
string scattering amplitudes and breakdown of the infinite linear relations as
we will discuss in the next section.

\subsubsection{Power-law and breakdown of the infinite linear relations in the
MR}

In this section we discuss the power-law behavior of high energy string
scattering amplitudes in a compact space. We will see that, in the MR, the
infinite linear relations derived in section B.4 \textit{break down} and,
simultaneously, the UV exponential fall-off behavior of high energy string
scattering amplitudes \textit{enhances} to power-law behavior. The power-law
behavior of high energy string scatterings in a compact space was first
suggested by Mende \cite{Mende}. Here we give a mathematically more concrete
description. It is easy to see that the "power law" condition, i.e. Eq.(3.7)
in Mende's paper \cite{Mende},%
\begin{equation}
k_{1L}\cdot k_{2L}+k_{1R}\cdot k_{2R}=\text{constant,}
\label{mandy's condition}%
\end{equation}
turns out to be%
\begin{align}
&  -\left(  k_{1L}\cdot k_{2L}+k_{1R}\cdot k_{2R}\right) \nonumber\\
&  =\sqrt{p^{2}+M_{1}^{2}}\cdot\sqrt{p^{2}+M_{2}^{2}}+p^{2}+\left(  \vec
{K}_{1L}\cdot\vec{K}_{2L}+\vec{K}_{1R}\cdot\vec{K}_{2R}\right) \nonumber\\
&  =\sqrt{p^{2}+M_{1}^{2}}\cdot\sqrt{p^{2}+M_{2}^{2}}+p^{2}+2\left(  \vec
{K}_{1}\cdot\vec{K}_{2}+\vec{L}_{1}\cdot\vec{L}_{2}\right) \nonumber\\
&  =\text{constant}.
\end{align}
The condition to achieve power-law behavior for the compactified open string
scatterings, Eq.(\ref{const}), is replaced by
\begin{equation}
s_{L}=\text{ constant, }s_{R}=\text{ constant } \label{closedpower}%
\end{equation}
It is easy to see that the condition, Eq.(\ref{closedpower}), leads to the
power-law behavior of the compactified closed string scattering amplitudes. To
satisfy the condition in Eq.(\ref{closedpower}), we define the following
"super-highly" winding kinematic regime%
\begin{equation}
n_{i}^{2}\gg p^{2}\simeq q^{2}\gg N_{R}+N_{L}. \label{closedregime}%
\end{equation}
Note that all $m_{i}$ were chosen to vanish in order to satisfy the
conservations of compactified momentum and winding number respectively
\cite{Compact}. For the choice of the kinematic regime in
Eq.(\ref{closedregime}), Eq.(\ref{k1*k2}) and Eq.(\ref{closedpower}) imply%
\begin{equation}
\lim_{p\rightarrow\infty}\frac{\sqrt{p^{2}+M_{1}^{2}}\cdot\sqrt{p^{2}%
+M_{2}^{2}}+p^{2}}{2(K_{1}K_{2}+L_{1}L_{2})}=\lim_{p\rightarrow\infty}%
\frac{\sqrt{p^{2}+M_{1}^{2}}\cdot\sqrt{p^{2}+M_{2}^{2}}+p^{2}}{2\left(
\dfrac{m_{1}m_{2}}{R^{2}}+\dfrac{1}{4}n_{1}n_{2}R^{2}\right)  }=-1.
\label{exist}%
\end{equation}
Note that since we have set $m_{i}=0,$ Eq.(\ref{exist}) is similar to
Eq.(\ref{condition}) for the compactified open string case, and one can get
nontrivial solution for Eq.(\ref{lamda}) with signs of $\lambda_{1}=\frac
{2p}{n_{1}R}$ and $\lambda_{2}=-\frac{2p}{n_{2}R}$ the same. This completes
the discussion of power-law regime at fixed angle for high energy compactified
closed string scatterings. The "super-highly" winding regime derived in this
section is to correct the "Mende regime"
\begin{equation}
E^{2}\simeq M^{2}\gg N_{R}+N_{L} \label{wrong}%
\end{equation}
discussed in \cite{Compact}. The regime defined in Eq.(\ref{wrong}) is indeed
exponential fall-off behaved rather than power-law claimed in \cite{Compact}.

\part{Stringy symmetries of Regge string scattering amplitudes}

In this part of the review, we are going to discuss string scatterings in the
high energy, fixed momentum transfer regime or Regge regime (RR)
\cite{RR1,RR2,RR3,RR4,RR5,RR6}. See also \cite{OA,DL,KP}. We will see that
Regge string scattering amplitudes contain information of the theory in
complementary to string scattering amplitudes in the high energy, fixed angle
or Gross regime(GR). The UV behavior of high energy string scatterings in the
GR (hard string scatterings) is well known to be very soft exponential
fall-off, while that of RR is power-law. There are some other fundamental
differences\ and links between the calculations of Regge string scatterings
and hard string scatterings, which we list below :

A. The number of high energy scattering amplitudes for each fixed mass level
in the RR is much more numerous than that of GR calculated previously
\cite{bosonic}.

B. The saddle-point method is not applicable in the calculation of Regge
string scattering amplitudes. However, a direct calculation is manageable and
one finds that all string amplitudes in the RR can be expressed in terms of a
finite sum of Kummer functions of the second kind \cite{bosonic,KLY1,KLY2,LY}.

C. There is an interesting link between scattering amplitudes of the RR and
GR. By proving a set of new Stirling number identities \cite{LYAM}, one can
reproduce from amplitudes in the RR the ratios calculated in the GR
\cite{bosonic,RRsusy}.

D. For the high energy scattering amplitudes in the fixed angle regime, one
has the identification $e^{P}=e^{L}$, while in the Regge regime $e^{P}\neq
e^{L}$ \cite{bosonic}$.$

E. The decoupling of ZNS applies to all kinematic regimes including the RR and
the GR. However, the linear relations obtained from decoupling of ZNS in the
RR are not good enough to solve all the amplitudes in the RR at each fixed
mass level. Instead, the recurrence relations among RR amplitudes obtained
from Kummer functions can be used to fix all RR amplitudes in terms of one
amplitude \cite{LY}.

F. All the RR amplitudes can be expressed in terms of one single Appell
function $F_{1}$ \cite{AppellLY}. This result enables us to derive infinite
number of recurrence relations among RR amplitudes at arbitrary mass levels,
which are conjectured to be related to the known $SL(5,C)$ dynamical symmetry
of $F_{1}$.

The part III of the paper is organized as following. In chapter XI we
calculate Regge string scattering amplitudes in terms of Kummer functions
\cite{bosonic}. We then prove a set of Stirling number identities \cite{LYAM}
and use them to reproduce ratios among hard string scattering amplitudes in
the GR discussed in chapter V. Finally we calculate recurrence relations among
Regge string scattering amplitudes, and use them to prove Regge stringy Ward
identities or decoupling of ZNS in the RR for the first few mass levels
\cite{LY}. In Chapter XII, we generalize the calculations to four classes of
Regge superstring scattering amplitudes and reproduce the ratios calculated in
chapter VIII. In addition, discover new high energy superstring scattering
amplitudes with polarizations \textit{orthogonal} to the scattering plane
\cite{RRsusy}.

In Chapter XIII we generalize the calculation of four tachyon BPST vertex
operator \cite{RR6} to the general high spin cases \cite{Tan}, and derive the
recurrence relations among these higher spin BPST vertex operators. In Chapter
XIV we discuss higher spin Regge string states scattered from D-particle
\cite{LMY}. We will obtain the complete GR ratios, which include a subset
calculated in section IX.A.3, from Regge string states scattered from
D-particle. In addition, we will see that although there is no factorization
for closed string D-particle scattering amplitudes into two channels of open
string D-particle scattering amplitudes, the complete ratios are
\textit{factorized} which came as a surprise.

In chapter XV we discover that all RR amplitudes can be expressed in terms of
one single Appell function $F_{1}$\cite{AppellLY}. More general recurrence
relations among RR amplitudes will be derived. We will also discuss the
$SL(5,C)$ dynamical symmetry of $F_{1}$ which is argued to be closely related
to high energy spacetime symmetry of $26D$ bosonic string theory.%

\setcounter{equation}{0}
\renewcommand{\theequation}{\arabic{section}.\arabic{equation}}%

\section{Kummer functions $U$ and patterns of Regge string scattering
amplitudes (RSSA)}

In this chapter, we first calculate a subclass of Regge string scattering
amplitudes (RSSA) and expressed them in terms of Kummer functions of the
second kind \cite{KLY1,KLY2}. We then prove in section B a set of Stirling
number identities \cite{LYAM} and use them to reproduce ratios among hard
string scattering amplitudes calculated in chapter V. In section C, we show
that all RSSA are power law behaved, and the exponents of the power law are
universal and are independent of the mass levels \cite{bosonic}. In section D,
we calculate the most general RSSA and derive recurrence relations among them
\cite{LY}. We show that, for the first few mass levels, the decoupling of ZNS
in the RR or Regge stringy Ward identities can be derived from these
recurrence relations \cite{LY}. This shows that, in contrast to the GR
considered in chapter V, recurrence relations are more fundamental than Regge
stringy Ward identities. Finally we prove that all RSSA can be solved by these
recurrence relations and expressed in terms of one single RSSA \cite{LY}.

\subsection{Kummer functions and RSSA}

We now begin to discuss high energy string scatterings in the RR. That is in
the kinematic regime%
\begin{equation}
s\rightarrow\infty,\sqrt{-t}=\text{fixed (but }\sqrt{-t}\neq\infty).
\end{equation}
As in the case of GR, we only need to consider the polarizations on the
scattering plane, which is defined in Appendix \ref{RR Kinematic}. Appendix
\ref{RR Kinematic} also includes the kinematic set up. Instead of using
$(E,\theta)$ as the two independent kinematic variables in the GR, we choose
to use $(s,t)$ in the RR. One of the reason has been, in the RR, $t\sim
E\theta$ is fixed, and it is more convenient to use $(s,t)$ rather than
$(E,\theta).$ In the RR, to the lowest order, Eqs.(\ref{A13}) to (\ref{A18})
reduce to%
\begin{subequations}
\begin{align}
e^{P}\cdot k_{1}  &  =-\frac{1}{M_{2}}\left(  \sqrt{p^{2}+M_{1}^{2}}%
\sqrt{p^{2}+M_{2}^{2}}+p^{2}\right)  \simeq-\frac{s}{2M_{2}},\\
e^{L}\cdot k_{1}  &  =-\frac{p}{M_{2}}\left(  \sqrt{p^{2}+M_{1}^{2}}%
+\sqrt{p^{2}+M_{2}^{2}}\right)  \simeq-\frac{s}{2M_{2}},\\
e^{T}\cdot k_{1}  &  =0
\end{align}
and%
\end{subequations}
\begin{subequations}
\begin{align}
e^{P}\cdot k_{3}  &  =\frac{1}{M_{2}}\left(  \sqrt{q^{2}+M_{3}^{2}}\sqrt
{p^{2}+M_{2}^{2}}-pq\cos\theta\right)  \simeq-\frac{\tilde{t}}{2M_{2}}%
\equiv-\frac{t-M_{2}^{2}-M_{3}^{2}}{2M_{2}},\label{ttt}\\
e^{L}\cdot k_{3}  &  =\frac{1}{M_{2}}\left(  p\sqrt{q^{2}+M_{3}^{2}}%
-q\sqrt{p^{2}+M_{2}^{2}}\cos\theta\right)  \simeq-\frac{\tilde{t}^{\prime}%
}{2M_{2}}\equiv-\frac{t+M_{2}^{2}-M_{3}^{2}}{2M_{2}},\\
e^{T}\cdot k_{3}  &  =-q\sin\phi\simeq-\sqrt{-{t}}.
\end{align}
Note that $e^{P}$ \textit{does not} approach to $e^{L}$ in the RR. This is
very different from the case of GR. In the following discussion, we will
calculate the amplitudes for the longitudinal polarization $e^{L}.$ For the
$e^{P}$ amplitudes, the results can be trivially modified. We will find that
the number of high energy scattering amplitudes for each fixed mass level in
the RR is much more numerous than that of GR calculated previously. On the
other hand, it seems that the saddle-point method used in the GR is not
applicable in the RR. We will first calculate the string scattering amplitudes
on the scattering plane $\left(  e^{L},e^{T}\right)  $ for the mass level
$M_{2}^{2}=4$. In the mass level $M_{2}^{2}=4$ $\left(  M_{1}^{2}=M_{3}%
^{2}=M_{4}^{2}=-2\right)  $, it turns out that there are eight high energy
amplitudes in the RR%
\end{subequations}
\begin{align}
&  \alpha_{-1}^{T}\alpha_{-1}^{T}\alpha_{-1}^{T}|0\rangle,\alpha_{-1}%
^{L}\alpha_{-1}^{T}\alpha_{-1}^{T}|0\rangle,\alpha_{-1}^{L}\alpha_{-1}%
^{L}\alpha_{-1}^{T}|0\rangle,\alpha_{-1}^{L}\alpha_{-1}^{L}\alpha_{-1}%
^{L}|0\rangle,\nonumber\\
&  \alpha_{-1}^{T}\alpha_{-2}^{T}|0\rangle,\alpha_{-1}^{T}\alpha_{-2}%
^{L}|0\rangle,\alpha_{-1}^{L}\alpha_{-2}^{T}|0\rangle,\alpha_{-1}^{L}%
\alpha_{-2}^{L}|0\rangle.
\end{align}

The $s-t$ channel of these amplitudes can be calculated to be%
\begin{align}
A^{TTT}  &  =\int_{0}^{1}dx\cdot x^{k_{1}\cdot k_{2}}\left(  1-x\right)
^{k_{2}\cdot k_{3}}\cdot\left(  \frac{ie^{T}\cdot k_{1}}{x}-\frac{ie^{T}\cdot
k_{3}}{1-x}\right)  ^{3}\nonumber\\
&  \simeq-i\left(  \sqrt{-t}\right)  ^{3}\frac{\Gamma\left(  -\frac{s}%
{2}-1\right)  \Gamma\left(  -\frac{\tilde{t}}{2}-1\right)  }{\Gamma\left(
\frac{u}{2}+3\right)  }\cdot\left(  -\frac{1}{8}s^{3}+\frac{1}{2}s\right)  ,
\end{align}%
\begin{align}
A^{LTT}  &  =\int_{0}^{1}dx\cdot x^{k_{1}\cdot k_{2}}\left(  1-x\right)
^{k_{2}\cdot k_{3}}\cdot\left(  \frac{ie^{T}\cdot k_{1}}{x}-\frac{ie^{T}\cdot
k_{3}}{1-x}\right)  ^{2}\left(  \frac{ie^{L}\cdot k_{1}}{x}-\frac{ie^{L}\cdot
k_{3}}{1-x}\right) \nonumber\\
&  \simeq-i\left(  \sqrt{-t}\right)  ^{2}\left(  -\frac{1}{2M_{2}}\right)
\frac{\Gamma\left(  -\frac{s}{2}-1\right)  \Gamma\left(  -\frac{\tilde{t}}%
{2}-1\right)  }{\Gamma\left(  \frac{u}{2}+3\right)  }\cdot\left[  \frac{3}%
{4}s^{3}-\frac{t}{4}s^{2}-\left(  \frac{t}{2}+3\right)  s\right]  ,
\end{align}%
\begin{align}
A^{LLT}  &  =\int_{0}^{1}dx\cdot x^{k_{1}\cdot k_{2}}\left(  1-x\right)
^{k_{2}\cdot k_{3}}\cdot\left(  \frac{ie^{T}\cdot k_{1}}{x}-\frac{ie^{T}\cdot
k_{3}}{1-x}\right)  \left(  \frac{ie^{L}\cdot k_{1}}{x}-\frac{ie^{L}\cdot
k_{3}}{1-x}\right)  ^{2}\nonumber\\
&  \simeq-i\left(  \sqrt{-t}\right)  \left(  -\frac{1}{2M_{2}}\right)
^{2}\frac{\Gamma\left(  -\frac{s}{2}-1\right)  \Gamma\left(  -\frac{\tilde{t}%
}{2}-1\right)  }{\Gamma\left(  \frac{u}{2}+3\right)  }\nonumber\\
&  \cdot\left[  \left(  \frac{1}{4}t-\frac{9}{2}\right)  s^{3}+\left(
\frac{1}{4}t^{2}+\frac{7}{2}t\right)  s^{2}+\frac{\left(  t+6\right)  ^{2}}%
{2}s\right]  ,
\end{align}%
\begin{align}
A^{LLL}  &  =\int_{0}^{1}dx\cdot x^{k_{1}\cdot k_{2}}\left(  1-x\right)
^{k_{2}\cdot k_{3}}\cdot\left(  \frac{ie^{L}\cdot k_{1}}{x}-\frac{ie^{L}\cdot
k_{3}}{1-x}\right)  ^{3}\nonumber\\
&  \simeq-i\left(  -\frac{1}{2M_{2}}\right)  ^{3}\frac{\Gamma\left(  -\frac
{s}{2}-1\right)  \Gamma\left(  -\frac{\tilde{t}}{2}-1\right)  }{\Gamma\left(
\frac{u}{2}+3\right)  }\nonumber\\
&  \cdot\left[  -\left(  \frac{11}{2}t-27\right)  s^{3}-6\left(
t^{2}+6t\right)  s^{2}-\frac{\left(  t+6\right)  ^{3}}{2}s\right]  ,
\end{align}%
\begin{align}
A^{TT}  &  =\int_{0}^{1}dx\cdot x^{k_{1}\cdot k_{2}}\left(  1-x\right)
^{k_{2}\cdot k_{3}}\cdot\left(  \frac{ie^{T}\cdot k_{1}}{x}-\frac{ie^{T}\cdot
k_{3}}{1-x}\right)  \left[  \frac{e^{T}\cdot k_{1}}{x^{2}}+\frac{e^{T}\cdot
k_{3}}{\left(  1-x\right)  ^{2}}\right] \nonumber\\
&  \simeq-i\left(  \sqrt{-t}\right)  ^{2}\frac{\Gamma\left(  -\frac{s}%
{2}-1\right)  \Gamma\left(  -\frac{\tilde{t}}{2}-1\right)  }{\Gamma\left(
\frac{u}{2}+3\right)  }\left(  -\frac{1}{8}s^{3}+\frac{1}{2}s\right)  ,
\end{align}%
\begin{align}
A^{TL}  &  =\int_{0}^{1}dx\cdot x^{k_{1}\cdot k_{2}}\left(  1-x\right)
^{k_{2}\cdot k_{3}}\cdot\left(  \frac{ie^{T}\cdot k_{1}}{x}-\frac{ie^{T}\cdot
k_{3}}{1-x}\right)  \left[  \frac{e^{L}\cdot k_{1}}{x^{2}}+\frac{e^{L}\cdot
k_{3}}{\left(  1-x\right)  ^{2}}\right] \nonumber\\
&  \simeq i\left(  \sqrt{-t}\right)  \left(  -\frac{1}{2M_{2}}\right)
\frac{\Gamma\left(  -\frac{s}{2}-1\right)  \Gamma\left(  -\frac{\tilde{t}}%
{2}-1\right)  }{\Gamma\left(  \frac{u}{2}+3\right)  }\nonumber\\
&  \cdot\left[  -\left(  \frac{1}{8}t+\frac{3}{4}\right)  s^{3}-\frac{1}%
{8}\left(  t^{2}-2t\right)  s^{2}-\left(  \frac{1}{4}t^{2}-t-3\right)
s\right]  ,
\end{align}%
\begin{align}
A^{LT}  &  =\int_{0}^{1}dx\cdot x^{k_{1}\cdot k_{2}}\left(  1-x\right)
^{k_{2}\cdot k_{3}}\cdot\left(  \frac{ie^{L}\cdot k_{1}}{x}-\frac{ie^{L}\cdot
k_{3}}{1-x}\right)  \left[  \frac{e^{T}\cdot k_{1}}{x^{2}}+\frac{e^{T}\cdot
k_{3}}{\left(  1-x\right)  ^{2}}\right] \nonumber\\
&  \simeq i\left(  \sqrt{-t}\right)  \left(  -\frac{1}{2M_{2}}\right)
\frac{\Gamma\left(  -\frac{s}{2}-1\right)  \Gamma\left(  -\frac{\tilde{t}}%
{2}-1\right)  }{\Gamma\left(  \frac{u}{2}+3\right)  }\cdot\left[  \frac{3}%
{4}s^{3}-\frac{t}{4}s^{2}-\left(  \frac{t}{2}+3\right)  s\right]  ,
\end{align}
and%
\begin{align}
A^{LL}  &  =\int_{0}^{1}dx\cdot x^{k_{1}\cdot k_{2}}\left(  1-x\right)
^{k_{2}\cdot k_{3}}\cdot\left(  \frac{ie^{L}\cdot k_{1}}{x}-\frac{ie^{L}\cdot
k_{3}}{1-x}\right)  \left[  \frac{e^{L}\cdot k_{1}}{x^{2}}+\frac{e^{L}\cdot
k_{3}}{\left(  1-x\right)  ^{2}}\right] \nonumber\\
&  \simeq i\left(  -\frac{1}{2M_{2}}\right)  ^{2}\frac{\Gamma\left(  -\frac
{s}{2}-1\right)  \Gamma\left(  -\frac{\tilde{t}}{2}-1\right)  }{\Gamma\left(
\frac{u}{2}+3\right)  }\nonumber\\
&  \cdot\left[  \left(  \frac{3}{4}t+\frac{9}{2}\right)  s^{3}+\left(
t^{2}-4t\right)  s^{2}+\left(  \frac{1}{4}t^{3}+\frac{1}{2}t^{2}-9t-18\right)
s\right]  .
\end{align}
From the above calculation, one can easily see that all the amplitudes are in
the same leading order $\left(  \sim s^{3}\right)  $\ in the RR, while in the
GR only $A^{TTT}$, $A^{LLT}$ and $A^{TL}$ are in the leading order $\left(
\sim t^{3/2}s^{3}\text{ or }t^{5/2}s^{2}\right)  $, all other amplitudes are
in the subleading orders. On the other hand, one notes that, for example, the
term $\sim\sqrt{-t}t^{2}s^{2}$ in $A^{LLT}$ and $A^{TL}$ are in the leading
order in the GR, but are in the subleading order in the RR. On the contrary,
the terms $\sqrt{-t}s^{3}$ in $A^{LLT}$ and $A^{TL}$ are in the subleading
order in the GR, but are in the leading order in the RR. These observations
suggest that the high energy string scattering amplitudes in the GR and RR
contain information complementary to each other.

One important observation for high energy amplitudes in the RR is for those
amplitudes with the same structure as those of the GR in Eq.(\ref{relevant}).
For these amplitudes, the relative ratios of the coefficients of the highest
power of $t$ in the leading order amplitudes in the RR can be calculated to
be
\begin{align}
A^{TTT}  &  =-i\left(  \sqrt{-t}\right)  \frac{\Gamma\left(  -\frac{s}%
{2}-1\right)  \Gamma\left(  -\frac{\tilde{t}}{2}-1\right)  }{\Gamma\left(
\frac{u}{2}+3\right)  }\cdot\left(  \frac{1}{8}ts^{3}\right)  \sim\frac{1}%
{8},\\
A^{LLT}  &  =-i\left(  \sqrt{-t}\right)  \left(  -\frac{1}{2M_{2}}\right)
^{2}\frac{\Gamma\left(  -\frac{s}{2}-1\right)  \Gamma\left(  -\frac{\tilde{t}%
}{2}-1\right)  }{\Gamma\left(  \frac{u}{2}+3\right)  }\left(  \frac{1}%
{4}ts^{3}\right)  \sim\frac{1}{64},\label{TLL}\\
A^{TL}  &  =i\left(  \sqrt{-t}\right)  \left(  -\frac{1}{2M_{2}}\right)
\frac{\Gamma\left(  -\frac{s}{2}-1\right)  \Gamma\left(  -\frac{\tilde{t}}%
{2}-1\right)  }{\Gamma\left(  \frac{u}{2}+3\right)  }\cdot\left(  -\frac{1}%
{8}ts^{3}\right)  \sim-\frac{1}{32}, \label{TL}%
\end{align}
which reproduces the ratios in the GR in Eq.(\ref{2.19..}). Note that the
symmetrized and anti-symmetrized amplitudes are defined as%
\begin{align}
T^{\left(  TL\right)  }  &  =\frac{1}{2}\left(  T^{TL}+T^{LT}\right)  ,\\
T^{\left[  TL\right]  }  &  =\frac{1}{2}\left(  T^{TL}-T^{LT}\right)  ;
\end{align}
and similarly for the amplitudes $A^{\left(  TL\right)  }$ and $A^{\left[
TL\right]  }$ in the RR. Note that $T^{LT}$ $\sim(\alpha_{-1}^{L})(\alpha
_{-2}^{T})|0\rangle$ in the GR is of subleading order in energy, while
$A^{LT}$ in the RR is of leading order in energy. However, the contribution of
the amplitude $A^{LT}$ to $A^{\left(  TL\right)  }$ and $A^{\left[  TL\right]
}$ in the RR will not affect the ratios calculated above. As we will see next,
this interesting result can be generalized to all mass levels in the string spectrum.

We first calculate high energy string scattering amplitudes in the RR for the
arbitrary mass levels. Instead of states in Eq.(\ref{relevant}) for the GR,
one can argue that the most general string states (ignore the \ $e^{P}$
amplitudes) one needs to consider at each fixed mass level $N=\sum_{n,m}%
nk_{n}+mq_{m}$ for the RR are%
\begin{equation}
\left\vert k_{n},q_{m}\right\rangle =\prod_{n>0}(\alpha_{-n}^{T})^{k_{n}}%
\prod_{m>0}(\alpha_{-m}^{L})^{q_{m}}|0\rangle.
\end{equation}
These RR amplitudes are good enough to reproduce the GR ratios calculated
previously. By the simple kinematics $e^{T}\cdot k_{1}=0$, and the energy
power counting of the string amplitudes, we end up with the following rules to
simplify the calculation for the leading order amplitudes in the RR:
\begin{align}
&  \alpha_{-n}^{T}:\quad\text{1 term (contraction of $ik_{3}\cdot X$ with
$\varepsilon_{T}\cdot\partial^{n}X$),}\\
&  \alpha_{-n}^{L}:%
\begin{cases}
n>1,\quad\text{1 term}\\
n=1\quad\text{2 terms}\text{ (contraction of $ik_{1}\cdot X$ and $ik_{3}\cdot
X$ with $\varepsilon_{L}\cdot\partial^{n}X$).}%
\end{cases}
\end{align}
The $s-t$ channel scattering amplitudes of this state with three other
tachyonic states can be calculated to be
\begin{align}
A^{(k_{n},q_{m})}  &  =\int_{0}^{1}dx\,x^{k_{1}\cdot k_{2}}(1-x)^{k_{2}\cdot
k_{3}}\left[  \frac{ie^{L}\cdot k_{1}}{-x}+\frac{ie^{L}\cdot k_{3}}%
{1-x}\right]  ^{q_{1}}\nonumber\\
&  \cdot\prod_{n=1}\left[  \frac{ie^{T}\cdot k_{3}\,(n-1)!}{(1-x)^{n}}\right]
^{k_{n}}\prod_{m=2}\left[  \frac{ie^{L}\cdot k_{3}\,(m-1)!}{(1-x)^{m}}\right]
^{q_{m}}\nonumber\\
&  =\left(  \frac{-i\tilde{t}^{\prime}}{2M_{2}}\right)  ^{q_{1}}\sum
_{j=0}^{q_{1}}{\binom{q_{1}}{j}}\left(  \frac{s}{-\tilde{t}}\right)  ^{j}%
\int_{0}^{1}dxx^{k_{1}\cdot k_{2}-j}(1-x)^{k_{2}\cdot k_{3}+j-\sum
_{n,m}(nk_{n}+mq_{m})}\nonumber\\
&  \cdot\prod_{n=1}\left[  i\sqrt{-t}(n-1)!\right]  ^{k_{n}}\prod_{m=2}\left[
i\tilde{t}^{\prime}(m-1)!\left(  -\frac{1}{2M_{2}}\right)  \right]  ^{q_{m}%
}\nonumber\\
&  =\left(  \frac{-i\tilde{t}^{\prime}}{2M_{2}}\right)  ^{q_{1}}\sum
_{j=0}^{q_{1}}{\binom{q_{1}}{j}}\left(  \frac{s}{-\tilde{t}}\right)
^{j}B\left(  k_{1}\cdot k_{2}-j+1\,,\,k_{2}\cdot k_{3}+j-N+1\right)
\nonumber\\
&  \cdot\prod_{n=1}\left[  i\sqrt{-t}(n-1)!\right]  ^{k_{n}}\prod_{m=2}\left[
i\tilde{t}^{\prime}(m-1)!\left(  -\frac{1}{2M_{2}}\right)  \right]  ^{q_{m}}.
\end{align}

The Beta function above can be approximated in the large $s$, but fixed $t$
limit as follows
\begin{align}
&  B\left(  k_{1}\cdot k_{2}-j+1,k_{2}\cdot k_{3}+j-N+1\right) \nonumber\\
&  =B\left(  -1-\frac{s}{2}+N-j,-1-\frac{t}{2}+j\right) \nonumber\\
&  =\frac{\Gamma(-1-\frac{s}{2}+N-j)\Gamma(-1-\frac{t}{2}+j)}{\Gamma(\frac
{u}{2}+2)}\nonumber\\
&  \approx B\left(  -1-\frac{1}{2}s,-1-\frac{t}{2}\right)  \left(  -1-\frac
{s}{2}\right)  ^{N-j}\left(  \frac{u}{2}+2\right)  ^{-N}\left(  -1-\frac{t}%
{2}\right)  _{j}\nonumber\\
&  \approx B\left(  -1-\frac{1}{2}s,-1-\frac{t}{2}\right)  \left(  -\frac
{s}{2}\right)  ^{-j}\left(  -1-\frac{t}{2}\right)  _{j}.
\end{align}
where%
\begin{equation}
(a)_{j}=a(a+1)(a+2)...(a+j-1)
\end{equation}
is the Pochhammer symbol. The leading order amplitude in the RR can then be
written as%
\begin{align}
A^{(k_{n},q_{m})}  &  =\left(  \frac{-i\tilde{t}^{\prime}}{2M_{2}}\right)
^{q_{1}}B\left(  -1-\frac{1}{2}s,-1-\frac{t}{2}\right)  \sum_{j=0}^{q_{1}%
}{\binom{q_{1}}{j}}\left(  \frac{2}{\tilde{t}^{\prime}}\right)  ^{j}\left(
-1-\frac{t}{2}\right)  _{j}\nonumber\\
&  \cdot\prod_{n=1}\left[  i\sqrt{-t}(n-1)!\right]  ^{k_{n}}\prod_{m=2}\left[
i\tilde{t}^{\prime}(m-1)!\left(  -\frac{1}{2M_{2}}\right)  \right]  ^{q_{m}},
\label{A}%
\end{align}
which is UV power-law behaved as expected. The summation in Eq.(\ref{A}) can
be represented by the Kummer function of the second kind $U$ as follows,
\begin{equation}
\sum_{j=0}^{p}{\binom{p}{j}}\left(  \frac{2}{\tilde{t}^{\prime}}\right)
^{j}\left(  -1-\frac{t}{2}\right)  _{j}=2^{p}(\tilde{t}^{\prime}%
)^{-p}\ U\left(  -p,\frac{t}{2}+2-p,\frac{\tilde{t}^{\prime}}{2}\right)
\label{equality}%
\end{equation}
Finally, the amplitudes can be written as
\begin{align}
A^{(k_{n},q_{m})}  &  =\left(  -\frac{i}{M_{2}}\right)  ^{q_{1}}U\left(
-q_{1},\frac{t}{2}+2-q_{1},\frac{\tilde{t}^{\prime}}{2}\right)  B\left(
-1-\frac{s}{2},-1-\frac{t}{2}\right) \nonumber\\
&  \cdot\prod_{n=1}\left[  i\sqrt{-t}(n-1)!\right]  ^{k_{n}}\prod_{m=2}\left[
i\tilde{t}^{\prime}(m-1)!\left(  -\frac{1}{2M_{2}}\right)  \right]  ^{q_{m}}.
\label{general amplitude}%
\end{align}
In the above, $U$ is the Kummer function of the second kind and is defined to
be%
\begin{equation}
U(a,c,x)=\frac{\pi}{\sin\pi c}\left[  \frac{M(a,c,x)}{(a-c)!(c-1)!}%
-\frac{x^{1-c}M(a+1-c,2-c,x)}{(a-1)!(1-c)!}\right]  \text{ \ }(c\neq2,3,4...)
\end{equation}
where $M(a,c,x)=\sum_{j=0}^{\infty}\frac{(a)_{j}}{(c)_{j}}\frac{x^{j}}{j!}$ is
the Kummer function of the first kind. $U$ and $M$ are the two solutions of
the Kummer Equation%
\begin{equation}
xy^{^{\prime\prime}}(x)+(c-x)y^{\prime}(x)-ay(x)=0.
\end{equation}
It is crucial to note that $c=\frac{t}{2}+2-q_{1},$ and is not a constant as
in the usual case, so $U$ in Eq.(\ref{general amplitude}) is not a solution of
the Kummer equation. This will make our analysis in the next section more
complicated as we will see soon. On the contrary, since $a=-q_{1}$ an integer,
the Kummer function in Eq.(\ref{equality}) terminated to be a finite sum. This
will simplify the manipulation of Kummer function used in this chapter.

\subsection{Reproducing ratios among hard scattering amplitudes}

\bigskip It can be seen from Eq.(\ref{A}) that the Regge scattering amplitudes
at each f\/ixed mass level are no longer proportional to each other. The
ratios are $t$ dependent functions and can be calculated to be \cite{bosonic}
\begin{gather}
\frac{A^{(N,2m,q)}(s,t)}{A^{(N,0,0)}(s,t)}=(-1)^{m}\left(  -\frac{1}{2M_{2}%
}\right)  ^{2m+q}(\tilde{t}^{\prime}-2N)^{-m-q}(\tilde{t}^{\prime}%
)^{2m+q}\nonumber\\
\hphantom{\frac{A^{(N,2m,q)}(s,t)}{A^{(N,0,0)}(s,t)} =}{}\times\sum_{j=0}%
^{2m}(-2m)_{j}\left(  -1+N-\frac{\tilde{t}^{\prime}}{2}\right)  _{j}%
\frac{(-2/\tilde{t}^{\prime})^{j}}{j!}+\mathit{O}\left\{  \left(  \frac
{1}{\tilde{t}^{\prime}}\right)  ^{m+1}\right\}  , \label{Ratio}%
\end{gather}
where $(x)_{j}=x(x+1)(x+2)\cdots(x+j-1)$ is the Pochhammer symbol which can be
expressed in terms of the signed Stirling number of the f\/irst kind $s\left(
n,k\right)  $ as following%
\[
\left(  x\right)  _{n}=\sum_{k=0}^{n}(-)^{n-k}s\left(  n,k\right)  x^{k}.
\]
To ensure the identif\/ication for the general mass levels%
\[
\lim_{\tilde{t}^{\prime}\rightarrow\infty}\frac{A^{(N,2m,q)}}{A^{(N,0,0,)}%
}=\frac{T^{(N,2m,q)}}{T^{(N,0,0)}}=\left(  -\frac{1}{M_{2}}\right)
^{2m+q}\left(  \frac{1}{2}\right)  ^{m+q}(2m-1)!!
\]
suggested by the calculation for the mass level $M_{2}^{2}=4$, one needs the
following identity
\begin{gather}
\sum_{j=0}^{2m}(-2m)_{j}\left(  -L-\frac{\tilde{t}^{\prime}}{2}\right)
_{j}\frac{(-2/\tilde{t}^{\prime})^{j}}{j!}\nonumber\\
\qquad{}=0\cdot(-\tilde{t}^{\prime})^{0}\!+0\cdot(-\tilde{t}^{\prime}%
)^{-1}\!+\dots+0\cdot(-\tilde{t}^{\prime})^{-m+1}\!+\frac{(2m)!}{m!}%
(-\tilde{t}^{\prime})^{-m}+\mathit{O}\left\{  \left(  \frac{1}{\tilde
{t}^{\prime}}\right)  ^{m+1}\right\}  \! \label{master.}%
\end{gather}
where $L=1-N$ and is an integer. Similar identif\/ication can be extended to
the case of closed string as well \cite{Closed}. For all four classes of high
energy superstring scattering amplitudes, $L$ is an integer too~\cite{RRsusy}.
A recent work on string D-particle scattering amplitudes \cite{LMY} also gives
an integer value of~$L$. Note that $L$ af\/fects only the sub-leading terms in
$\mathit{O}\Big\{\left(  \frac{1}{\tilde{t}^{\prime}}\right)  ^{m+1}\Big\}$.
Here we give a simple example for $m=3$ \cite{RRsusy}
\begin{gather*}
\sum_{j=0}^{6}(-2m)_{j}\left(  -L-\frac{\tilde{t}^{\prime}}{2}\right)
_{j}\frac{(-2/\tilde{t}^{\prime})^{j}}{j!}\\
\qquad{}=\frac{120}{(-\tilde{t}^{\prime})^{3}}+\frac{720L^{2}-2640L+2080}%
{(-\tilde{t}^{\prime})^{4}}+\frac{480L^{4}-4160L^{3}+12000L^{2}-12928L+3840}%
{(-\tilde{t}^{\prime})^{5}}\\
\quad\qquad{}+\frac{64L^{6}-960L^{5}+5440L^{4}-14400L^{3}+17536L^{2}%
-7680L}{(-\tilde{t}^{\prime})^{6}}.
\end{gather*}
Mathematically, Eq.(\ref{master.}) was exactly proved \cite{RRsusy, bosonic}
for $L=0,1$ by a calculation based on a set of signed Stirling number
identities developed recently in combinatorial theory in~\cite{MK}. For
general integer $L$ cases, only the identity corresponding to the nontrivial
leading term $\frac{(2m)!}{m!}(-\tilde{t}^{\prime})^{-m}$ was rigorously
proved in \cite{RRsusy}, but not for other \textquotedblleft%
0~identities\textquotedblright. A numerical proof of Eq.(\ref{master.}) was
given in~\cite{RRsusy} for arbitrary real values~$L$ and for non-negative
integer~$m$ up to $m=10.$ It was then conjectured that~\cite{RRsusy}
Eq.(\ref{master.}) is valid for any \textit{real} number~$L$ and any
non-negative integer~$m$. It is important to prove Eq.(\ref{master.}) for any
non-negative integer~$m$ and arbitrary real values~$L$, since these values can
be realized in the high energy scattering of \textit{compactified} string
states, as was shown recently in~\cite{HLY}. Real values of $L$ appear in
string compactif\/ications due to the dependence on the generalized KK
internal momenta $K_{i}^{25}$ \cite{HLY}%
\[
L=1-N-\big(K_{2}^{25}\big)^{2}+K_{2}^{25}K_{3}^{25}.
\]

\subsubsection{Proof of the new Stirling number identity}

We now proceed to prove Eq.(\ref{master.}) \cite{LYAM}. We first rewrite the
left-hand side of Eq.(\ref{master.}) in the following form%
\begin{align}
&  \sum_{j=0}^{2m}(-2m)_{j}\left(  -L-\frac{\tilde{t}^{\prime}}{2}\right)
_{j}\frac{(-2/\tilde{t}^{\prime})^{j}}{j!}\nonumber\\
=  &  \sum_{j=0}^{2m}\left(  -1\right)  ^{j}\binom{2m}{j}\sum_{l=0}^{j}%
\binom{j}{l}(-L)_{j-l}\sum_{s=0}^{l}c\left(  l,s\right)  \left(  -\dfrac
{2}{\tilde{t}^{\prime}}\right)  ^{j-s} \label{Stirling expansion}%
\end{align}
where we have used the signless Stirling number of the first kind $c\left(
l,s\right)  $ to expand the Pochhammer symbol%
\begin{equation}
\left(  x\right)  _{n}=\sum_{k=0}^{n}c\left(  n,k\right)  x^{k}
\label{Stirling}%
\end{equation}
The coefficient of $(-2/\tilde{t}^{\prime})^{i}$ in
Eq.(\ref{Stirling expansion}), which will be defined as $G\left(  m,i\right)
$, can be read off from the equation as%
\begin{equation}
G\left(  m,i\right)  =\sum_{j=0}^{2m}\sum_{l=0}^{j}\left(  -1\right)
^{j+i}\binom{2m}{j}\binom{j}{l}(-L)_{j-l}c\left(  l,j-i\right)  .
\end{equation}
One needs to prove that%
\begin{align}
1.G\left(  m,m\right)   &  =\left(  2m-1\right)  !!,\text{ for all }L\in%
\mathbb{R}
;\label{A11}\\
2.G\left(  m,i\right)   &  =0,\text{ for all }L\in%
\mathbb{R}
\text{ and }0\leq i<m. \label{A12.}%
\end{align}
From the definition of $c\left(  n,k\right)  $ in Eq.(\ref{Stirling}), we note
that $c\left(  n,k\right)  \neq0$ only if $0\leq k\leq n$. Thus $c\left(
l,j-i\right)  \neq0$ only if $j\geq i$ and $l\geq j-i$. We can rewrite
$G\left(  m,i\right)  $ as%
\begin{align}
G\left(  m,i\right)   &  =\sum_{j=i}^{2m}\sum_{l=j-i}^{j}\left(  -1\right)
^{j}\binom{2m}{j}\binom{j}{l}(-L)_{j-l}c\left(  l,j-i\right) \nonumber\\
&  =\sum_{k=0}^{2m-i}\sum_{l=k}^{k+i}\left(  -1\right)  ^{k+i}\binom{2m}%
{i+k}\binom{i+k}{l}(-L)_{k+i-l}c\left(  l,k\right) \nonumber\\
&  =\sum_{k=0}^{2m-i}\sum_{p=0}^{i}\left(  -1\right)  ^{k+i}\binom{2m}%
{i+k}\binom{i+k}{p+k}(-L)_{i-p}c\left(  k+p,k\right) \nonumber\\
&  =\sum_{p=0}^{i}(-L)_{i-p}\sum_{k=0}^{2m-i}\left(  -1\right)  ^{k+i}%
\binom{2m}{i+k}\binom{i+k}{p+k}c\left(  k+p,k\right) \nonumber\\
&  =\left(  -1\right)  ^{i}\sum_{p=0}^{i}(-L)_{i-p}\binom{2m}{i-p}\sum
_{k=0}^{2m-i}\left(  -1\right)  ^{k}\binom{2m-i+p}{k+p}c\left(  k+p,k\right)
\nonumber\\
&  \equiv\left(  -1\right)  ^{i}\sum_{p=0}^{i}(-L)_{i-p}\binom{2m}%
{i-p}S_{2m-i}\left(  p\right)  \label{G(m,i)}%
\end{align}
where we have defined%
\begin{equation}
S_{N}\left(  p\right)  =\sum_{k=0}^{N}\left(  -1\right)  ^{k}\binom{N+p}%
{k+p}c\left(  k+p,k\right)  .
\end{equation}
It is easy to see that for fixed $m$ and $0\leq i<m$, $G\left(  m,i\right)  $
is a polynomial of $L$ of degree $i$, expanded with the basis $1$, $\left(
-L\right)  _{1}$, $\left(  -L\right)  _{2}$,\ldots. Note that $p\leq i<m$, so
$2m-i\geq p+1.$ For Eq.(\ref{A12.}), we want to show that $S_{N}\left(
p\right)  =0$ for $N\geq p+1$. For this purpose, we define the functions%
\begin{equation}
C_{n}\left(  x\right)  =\sum_{k\geq0}c\left(  k+n,k\right)  x^{k+n}.
\end{equation}
The recurrence of the signless Stirling number identity%
\begin{equation}
c\left(  k+n,k\right)  =\left(  n+k-1\right)  c\left(  n+k-1,k\right)
+c\left(  n+k-1,k-1\right)
\end{equation}
leads to the equation%
\begin{equation}
C_{n}\left(  x\right)  =\dfrac{x^{2}}{1-x}\dfrac{d}{dx}C_{n-1}\left(
x\right)  ,
\end{equation}
with the initial value%
\begin{equation}
C_{0}\left(  x\right)  =\dfrac{1}{1-x}.
\end{equation}
The first couple of $C_{n}\left(  x\right)  $\ can be calculated to be%
\begin{equation}
C_{1}\left(  x\right)  =\dfrac{x^{2}}{\left(  1-x\right)  ^{3}}\text{,
\ }C_{2}\left(  x\right)  =\dfrac{x^{4}+2x^{3}}{\left(  1-x\right)  ^{5}%
}\text{, \ }C_{3}\left(  x\right)  =\dfrac{x^{6}+8x^{5}+6x^{4}}{\left(
1-x\right)  ^{7}}.
\end{equation}
Now by induction, it is easy to show that%
\begin{equation}
C_{n}\left(  x\right)  =\dfrac{f_{n}\left(  x\right)  }{\left(  1-x\right)
^{2n+1}}\text{, \ where }f_{n}\left(  x\right)  =x^{2n}+\mathcal{O}\left(
x^{2n-1}\right)  ,
\end{equation}
and%
\begin{equation}
f_{n}\left(  1\right)  =\left(  2n-1\right)  !!. \label{fn}%
\end{equation}
In order to prove Eq.(\ref{A12.}), we note that $(-1)^{N}S_{N}\left(
p\right)  $\ is the coefficient of $x^{N+p}$\ in the function%
\begin{equation}
\left(  1-x\right)  ^{N+p}C_{p}\left(  x\right)  =f_{p}\left(  x\right)
\left(  1-x\right)  ^{N-p-1}=x^{N+p-1}+\mathcal{O}\left(  \cdots\right)  ,
\end{equation}
which is obviously zero for $N\geq p+1.$ This proves $S_{N}\left(  p\right)
=0$ for $N\geq p+1$ and thus Eq.(\ref{A12.}).

In order to prove the first identity in Eq.(\ref{A11}), we first note that the
above argument remains true for $i=m$ and $0\leq p<i$. So Eq.(\ref{A11})
corresponds to the case $p=i=m.$ By using Eq.(\ref{G(m,i)}), we can evaluate
\begin{equation}
G\left(  m,m\right)  =\sum_{k=0}^{m}\left(  -1\right)  ^{k+m}\binom{2m}%
{k+m}\binom{k+m}{k+m}c\left(  k+p,k\right)  =\sum_{k=0}^{m}\left(  -1\right)
^{k+m}\binom{2m}{k+m}c\left(  k+p,k\right)  . \label{GMM}%
\end{equation}
Eq.(\ref{GMM}) corresponds to the coefficient of $x^{2m}$\ in the function%
\begin{equation}
(1-x)^{2m}C_{m}(x)=\dfrac{f_{m}\left(  x\right)  }{1-x}=f_{m}\left(  x\right)
(1+x+x^{2}+....).
\end{equation}
By Eq.(\ref{fn}), this coefficient is%
\begin{equation}
f_{m}\left(  1\right)  =\left(  2m-1\right)  !!.
\end{equation}
This proves Eq.(\ref{A11}). We thus have completed the proof of
Eq.(\ref{master.}) for any non-negative integer $m$ and any real value $L.$

It was remarkable to first predict \cite{bosonic} the mathematical identities
in Eq.(\ref{master.}) provided by string theory, and then a rigorous
mathematical proof followed \cite{LYAM}. It was interesting to see that the
validity of Eq.(\ref{master.}) includes non-integer values of $L$ which were
later realized by Regge string scatterings in compact space \cite{HLY}.

\subsubsection{Subleading orders}

In this section, we calculate the next few subleading order amplitudes in the
RR for the mass level $M_{2}^{2}=4,6$ \cite{bosonic}. We will see that the
ratios in Eq.(\ref{2.19..}) and Eq.(\ref{1111}) persist to subleading order
amplitudes in the RR. For the even mass levels with $(N-1)=\frac{M_{2}^{2}}%
{2}$= even, we conjecture and give evidences that the existence of these
ratios in the RR persists to all orders in the Regge expansion of all high
energy string scattering amplitudes . For the odd mass levels with
$(N-1)=\frac{M_{2}^{2}}{2}$= odd, the existence of these ratios will show up
only in the first $[N/2]+1$ terms in the Regge expansion of the amplitudes.

We will extend the kinematic relations in the RR to the subleading orders. We
first express all kinematic variables in terms of $s$ and $t$, and then expand
all relevant quantities in $s:$
\begin{align}
E_{1}  &  =\frac{s-(m_{2}^{2}+2)}{2\sqrt{2}},\\
E_{2}  &  =\frac{s+(m_{2}^{2}+2)}{2\sqrt{2}},\\
|\mathbf{k_{2}}|  &  =\sqrt{E_{1}^{2}+2},\quad|\mathbf{K_{3}}|=\sqrt{\frac
{s}{4}+2};
\end{align}%
\begin{equation}
e_{P}\cdot k_{1}=-\frac{1}{2m_{2}}s+\left(  -\frac{1}{m_{2}}+\frac{m_{2}}%
{2}\right)  ,\quad(\text{exact})
\end{equation}%
\begin{align}
e_{L}\cdot k_{1}  &  =-\frac{1}{2m_{2}}s+\left(  -\frac{1}{m_{2}}+\frac{m_{2}%
}{2}\right)  -{2m_{2}}s^{-1}-2m_{2}(m_{2}^{2}-2)s^{-2}\nonumber\\
&  -2m_{2}(m_{2}^{4}-6m_{2}^{2}+4)s^{-3}-2m_{2}(m_{2}^{6}-12m_{2}^{4}%
+24m_{2}^{2}-8)s^{-4}+O(s^{-5}),
\end{align}%
\begin{equation}
e_{T}\cdot k_{1}=0.
\end{equation}
A key step is to express the scattering angle $\theta$ in terms of $s$ and
$t$. This can be achieved by solving
\begin{equation}
t=-\left(  -(E_{2}-\frac{\sqrt{s}}{2})^{2}+(|\mathbf{k_{2}}|-|\mathbf{k_{3}%
}|\cos\theta)^{2}+|\mathbf{k}_{3}|^{2}\sin^{2}\theta\right)
\end{equation}
to obtain%
\begin{equation}
\theta=\arccos\left(  \frac{s+2t-m_{2}^{2}+6}{\sqrt{s+8}\sqrt{\frac
{(s+2)^{2}-2(s-2)m_{2}^{2}+m_{2}^{4}}{s}}}\right)  .\text{ (exact)}
\label{t-angle}%
\end{equation}
One can then calculate the following expansions
\begin{equation}
e_{P}\cdot k_{3}=\frac{1}{m_{2}}(E_{2}\frac{\sqrt{s}}{2}-|\mathbf{k_{2}%
}||\mathbf{k_{3}}|\cos\theta)=-\frac{t+2-m_{2}^{2}}{2m_{2}},
\end{equation}%
\begin{align}
e_{L}\cdot k_{3}  &  =\frac{1}{m_{2}}(k_{2}\frac{\sqrt{2}}{2}-E_{2}k_{3}%
\cos\theta)\nonumber\\
&  =-\frac{t+2+m_{2}^{2}}{2m_{2}}-m_{2}ts^{-1}-m_{2}[-4(t+1)+m_{2}%
^{2}(t-2)]s^{-2}\nonumber\\
&  -m_{2}[4(4+3t)-12tm_{2}^{2}+(t-4)m_{2}^{4}]s^{-3}\nonumber\\
&  \quad-m_{2}[-16(3+2t)+24(2+3t)m_{2}^{2}\nonumber\\
&  -24(-1+t)m_{2}^{4}+(-6+t)m_{2}^{6}]s^{-4}+O(s^{-5}), \label{tpower1}%
\end{align}%
\begin{align}
e_{T}\cdot k_{3}  &  =-|\mathbf{k_{3}}|\sin\theta\nonumber\\
&  =-\sqrt{-t}-\frac{1}{2}\sqrt{-t}(2+t+m_{2}^{2})s^{-1}\nonumber\\
&  \quad-\frac{1}{8\sqrt{-t}}[32+52t+20t^{2}+t^{3}+(32+20t-6t^{2})m_{2}%
^{2}+(8-3t)m_{2}^{4}]s^{-2}\nonumber\\
&  \quad+\frac{1}{16\sqrt{-t}}[320+456t+188t^{2}+22t^{3}+t^{4}%
-(-224+36t+132t^{2}+5t^{3})m_{2}^{2}.\nonumber\\
&  \quad\quad\quad\quad\quad\quad+(-16-122t+15t^{2})m_{2}^{4}+(-24+5t)m_{2}%
^{6}]s^{-3}\nonumber\\
&  \quad+\frac{1}{128(-t)^{3/2}}[1024+12032t+16080t^{2}+7520t^{3}%
+1432t^{4}+136t^{5}+5t^{6}\nonumber\\
&  -4(-512-896t+2232t^{2}+1844t^{3}+170t^{4}+7t^{5})m_{2}^{2}\nonumber\\
&  +2(768-2240t-2372t^{2}+1172t^{3}+35t^{4})m_{2}^{4}\nonumber\\
&  -4(-128+288t-450t^{2}+35t^{3})m_{2}^{6}+(64+240t-35t^{2})m_{2}^{8}%
]s^{-4}+O(s^{-5}). \label{tpower2}%
\end{align}
We are now ready to calculate the expansions of the four amplitudes
$A_{TTT},A_{LLT},A_{[LT]},A_{(LT)}$ for the mass level $M_{2}^{2}=4$ to
subleading orders in $s$ in the RR. These are
\begin{equation}
A^{TTT}=\frac{1}{8}\sqrt{-t}ts^{3}+\frac{3}{16}\sqrt{-t}t(t+6)s^{2}%
+\frac{3t^{3}+84t^{2}-68t-864}{64}\sqrt{-t}\,s+O(1),
\end{equation}%
\begin{align}
A^{LLT}  &  =\frac{1}{64}\sqrt{-t}(t-6)s^{3}+\frac{3}{128}\sqrt{-t}%
(t^{2}-20t-12)s^{2}\nonumber\\
&  \quad\quad+\frac{3t^{3}-342t^{2}-92t+5016+1728(-t)^{-1/2}}{512}\sqrt
{-t}\,s+O(1),
\end{align}%
\begin{align}
A^{[LT]}  &  =-\frac{1}{64}\sqrt{-t}(t+2)s^{3}-\frac{3}{128}\sqrt{-t}%
(t+2)^{2}s^{2}\nonumber\\
&  \quad\quad-\frac{(3t-8)(t+6)^{2}[1-2(-t)^{-1/2}]}{512}\sqrt{-t}\,s+O(1),
\end{align}%
\begin{align}
A^{(LT)}  &  =-\frac{1}{64}\sqrt{-t}(t+10)s^{3}-\frac{1}{128}\sqrt{-t}%
(3t^{2}+52t+60)s^{2}\nonumber\\
&  \quad\quad-\frac{3[t^{3}+30t^{2}+76t-1080-960(-t)^{-1/2}]}{512}\sqrt
{-t}\,s+O(1).
\end{align}

One can now easily see that the ratios of the coefficients of the highest
power of $t$ in the leading order coefficient functions $\frac{1}{8}:\frac
{1}{64}:-\frac{1}{64}:-\frac{1}{64}$ agree with the ratios in the GR
$8:1:-1:-1$ calculated in Eq.(\ref{2.19..}) as expected. Moreover, one further
observation is that these ratios remain the same for the coefficients of the
highest power of $t$ in the subleading orders $(s^{2})$ $\frac{3}{16}:\frac
{3}{128}:-\frac{3}{128}:-\frac{3}{128}$ and $(s)$ $\frac{3}{64}:\frac{3}%
{512}:-\frac{3}{512}:-\frac{3}{512}$. We conjecture that these ratios persist
to all energy orders in the Regge expansion of the amplitudes. This is
consistent with the results of GR by taking both $s,-t\rightarrow\infty.$ For
the mass level $M_{2}^{2}=6$ \cite{ChanLee2}, the amplitudes can be calculated
to be
\begin{align}
A^{TTTT}  &  =\frac{t^{2}}{16}s^{4}+\frac{t^{2}(t+6)}{8}s^{3}+\frac
{t(t^{3}+24t^{2}-4t-256)}{16}s^{2}\nonumber\\
&  +\frac{t(3t^{3}-2t^{2}-396t-768)}{4}s-\left(  \frac{t^{4}}{4}%
+166t^{3}+960t^{2}-64t-1024\right)  s^{0}\nonumber\\
&  +(-83t^{4}-1536t^{3}+384t^{2}+21248t+12288)s^{-1}+O(s^{-2}),
\end{align}%
\begin{align}
A^{TTLL}  &  =\frac{t(t-16)}{192}s^{4}+\frac{t(t^{2}-41t-32)}{96}s^{3}%
+\frac{t^{4}-132t^{3}-328t^{2}+1984t+2048}{192}s^{2}\nonumber\\
&  +\left(  -\frac{11t^{4}}{32}-\frac{11t^{3}}{4}+\frac{163t^{2}}%
{3}+184t+\frac{128}{3}\right)  s^{1}\nonumber\\
&  +\left(  -\frac{11}{8}t^{4}+88t^{3}+744t^{2}+304t-1408\right)
s^{0}\nonumber\\
&  +4\left(  11t^{4}+280t^{3}+204t^{2}-4448t-4480\right)  s^{-1}+O(s^{-2}),
\end{align}%
\begin{align}
A^{LLLL}  &  =\frac{t(t-52)}{768}s^{4}+\frac{t(t^{2}-140t+256)}{384}%
s^{3}+\frac{t^{4}-456t^{3}+2816t^{2}-512t-16384}{768}s^{2}\nonumber\\
&  \left(  -\frac{19t^{4}}{64}+6t^{3}-\frac{17t^{2}}{3}-176t-\frac{256}%
{3}\right)  s^{1}\nonumber\\
&  +(3t^{4}-10t^{3}-528t^{2}-672t+1792)s^{0}+O(s^{-1}),
\end{align}%
\begin{align}
A^{TTL}  &  =-\frac{(t+20)t}{96\sqrt{6}}s^{4}-\frac{t(t^{2}+31t+40)}%
{48\sqrt{6}}s^{3}-\frac{t^{4}+38t^{3}+224t^{2}-1520t-2560}{96\sqrt{6}}%
s^{2}\nonumber\\
&  +\frac{-3t^{4}-72t^{3}+2248t^{2}+12000t+5120}{48\sqrt{6}}s^{1}\nonumber\\
&  +\frac{67t^{3}+1194t^{2}+1344t-3712}{2\sqrt{6}}s^{0}+O(s^{-1}),
\end{align}%
\begin{align}
A^{LLL}  &  =-\frac{t^{2}-8t-128}{384\sqrt{6}}s^{4}-\frac{t^{3}-52t^{2}%
-412t+256}{192\sqrt{6}}s^{3}\nonumber\\
&  -\frac{t^{4}-236t^{3}-1272t^{2}+4832t+15872}{384\sqrt{6}}s^{2}\nonumber\\
&  +\frac{35t^{4}+50t^{3}-3008t^{2}-23728t-14848}{96\sqrt{6}}s^{1}\nonumber\\
&  -\frac{47t^{4}+1432t^{3}+24796t^{2}+40640t-101376}{48\sqrt{6}}%
s^{0}+O(s^{-1}),
\end{align}%
\begin{align}
\tilde{A}^{LT,T}  &  =-\frac{t(t+2)}{64\sqrt{6}}s^{4}-\frac{t(t+2)^{2}%
}{32\sqrt{6}}s^{3}-\frac{t^{4}+12t^{3}+8t^{2}-152t-256}{64\sqrt{6}}%
s^{2}\nonumber\\
&  +\frac{-3t^{4}+196t^{2}+624t+512}{32\sqrt{6}}s^{1}+\sqrt{\frac{3}{8}%
}(5t^{3}+30t^{2}+24t-32)s^{0}+O(s^{-1}),
\end{align}%
\begin{align}
A^{LL}  &  =\frac{(t+8)^{2}}{384}s^{4}+\frac{(t^{3}+20t^{2}+80t-128)}%
{192}s^{3}+\frac{t^{4}+16t^{3}+96t^{2}-880t-3328}{384}s^{2}\nonumber\\
&  +\frac{-t^{4}+8t^{3}-110t^{2}-1648t-1408}{48}s^{1}\nonumber\\
&  +\frac{t^{4}-4t^{3}-202t^{2}-704t+1728}{6}s^{0}+O(s^{-1}).
\end{align}

In the above calculations, as in the case of $M_{2}^{2}=4$, we have ignored a
common overall factor which will be discussed in the next section. Note that
the ratios of the coefficients in the leading order $t$ for the energy orders
$s^{4},s^{3},s^{2}$ reproduced the GR ratios in Eq.(\ref{2.19..}). However,
the subleading terms for orders $s^{1},s^{0}$ contain no GR ratios.
Mathematically, this is because the highest power of $t$ in the coefficient
functions of $s^{1}$ is $4$ rather than $5$, and those of $s^{0}$ is $4$
rather than $6$. This is because the power of $t$ in the kinematic relation
Eq.(\ref{tpower2}) can be as high as one wants if one goes to subleading
orders, while that of Eq.(\ref{tpower1}) is not. The $\sin\theta$ factor in
Eq.(\ref{tpower2}) contributes terms of higher order powers of $t$, while
$\cos\theta$ factor in in Eq.(\ref{tpower1}) does not. This can be seen from
the kinematic relation in Eq.(\ref{t-angle}). In general, one can easily show
that the $\sin\theta$ factor will contribute only for the even mass levels
with $(N-1)=\frac{M_{2}^{2}}{2}$= even.

We thus conjecture that the existence of the GR ratios in the RR persists to
all orders in the Regge expansion of all string amplitudes for the even mass
level. For the odd mass levels with $(N-1)=\frac{M_{2}^{2}}{2}$= odd, the
existence of the GR ratios will show up only in the first $[N/2]+1$ terms in
the Regge expansion of the amplitudes. An interesting question is whether this
phenomena persists for the case of superstring where GSO projection needs to
be imposed.

\subsection{Universal power law behavior}

In the discussion of the last section, we ignored an overall common
factor$\frac{\Gamma(-1-s/2)\Gamma(-1-t/2)}{\Gamma(u/2+2)}$ of the amplitudes
for mass levels $M_{2}^{2}=4,6$. We paid attention only to the ratios among
scattering amplitudes of different string states. In this section, we
calculate the high energy behavior of string scattering amplitudes for string
states at arbitrary mass levels in the RR. The power law behavior $\sim
s^{\alpha(t)}$ of the four-tachyon amplitude in the RR is well known in the
literature. Here we want to generalize this result to string states at
arbitrary mass levels. We can use the saddle point method to calculate the
leading term of gamma functions in the RR
\begin{equation}
\frac{\Gamma(-1-s/2)\Gamma(-1-t/2)}{\Gamma(u/2+2)}=\frac{\Gamma(-1-s/2)\Gamma
(-1-t/2)}{\Gamma(-s/2-t/2+N-2)}\sim s^{t/2-N+1}\text{ \ (in the RR)}.
\end{equation}
Thus, the overall $s$-dependence in the amplitudes is of the form%
\begin{equation}
A^{(k_{n},q_{m})}\sim s^{\alpha(t)}\text{ \ \ (in the RR)} \label{universal}%
\end{equation}
where
\begin{equation}
\alpha(t)=\alpha(0)+\alpha^{\prime}t\text{, \ }\alpha(0)=1\text{ and }%
\alpha^{\prime}=1/2.
\end{equation}

This generalizes the high energy behavior of the four-tachyon amplitude in the
RR to string states at arbitrary mass levels. The new result here is that the
behavior is universal and is mass level independent. In fact, as a simple
application, one can also derive Eq.(\ref{universal}) directly from
Eq.(\ref{general amplitude}) by using%
\begin{equation}
B\left(  -1-\frac{s}{2},-1-\frac{t}{2}\right)  \sim s^{\alpha(t)}.\text{ \ (in
the RR)}%
\end{equation}
We conclude that the well known $\sim s^{\alpha(t)}$ power-law behavior of the
four tachyon string scattering amplitude in the RR can be extended to high
energy string scattering amplitudes of arbitrary string states.

\subsection{Recurrence relations of RSSA}

To discuss relations among RSSA, one need to consider the complete RR string
states \cite{LY}. The complete leading order high energy open string states in
the Regge regime at each fixed mass level $N=\sum_{n,m,l>0}np_{n}%
+mq_{m}+lr_{l}$ are%
\begin{equation}
\left\vert p_{n},q_{m},r_{l}\right\rangle =\prod_{n>0}(\alpha_{-n}^{T}%
)^{p_{n}}\prod_{m>0}(\alpha_{-m}^{P})^{q_{m}}\prod_{l>0}(\alpha_{-l}%
^{L})^{r_{l}}|0,k\rangle. \label{RR}%
\end{equation}
The case for $q_{m}=0$ has been calculated previously in \cite{bosonic,RRsusy}
We stress that the inclusion of both $\alpha_{-m}^{P}$ and $\alpha_{-l}^{L}$
operators in Eq.(\ref{RR}) will be crucial to study Regge string Ward
identities to be discussed in the later part of this chapter. It is also
important to discuss the conformal invariant property of high energy string
scattering amplitudes \cite{RRsusy}. The momenta of the four particles on the
scattering plane are%

\begin{align}
k_{1}  &  =\left(  +\sqrt{p^{2}+M_{1}^{2}},-p,0\right)  ,\\
k_{2}  &  =\left(  +\sqrt{p^{2}+M_{2}^{2}},+p,0\right)  ,\\
k_{3}  &  =\left(  -\sqrt{q^{2}+M_{3}^{2}},-q\cos\phi,-q\sin\phi\right)  ,\\
k_{4}  &  =\left(  -\sqrt{q^{2}+M_{4}^{2}},+q\cos\phi,+q\sin\phi\right)
\end{align}
where $p\equiv\left\vert \mathrm{\vec{p}}\right\vert $, $q\equiv\left\vert
\mathrm{\vec{q}}\right\vert $ and $k_{i}^{2}=-M_{i}^{2}$. The relevant
kinematics are%
\begin{equation}
e^{P}\cdot k_{1}\simeq-\frac{s}{2M_{2}},\text{ \ }e^{P}\cdot k_{3}\simeq
-\frac{\tilde{t}}{2M_{2}}=-\frac{t-M_{2}^{2}-M_{3}^{2}}{2M_{2}};
\end{equation}%
\begin{equation}
e^{L}\cdot k_{1}\simeq-\frac{s}{2M_{2}},\text{ \ }e^{L}\cdot k_{3}\simeq
-\frac{\tilde{t}^{\prime}}{2M_{2}}=-\frac{t+M_{2}^{2}-M_{3}^{2}}{2M_{2}};
\end{equation}
and%
\begin{equation}
e^{T}\cdot k_{1}=0\text{, \ \ }e^{T}\cdot k_{3}\simeq-\sqrt{-{t}}%
\end{equation}
where $\tilde{t}$ and $\tilde{t}^{\prime}$ are related to $t$ by finite mass
square terms%
\begin{equation}
\tilde{t}=t-M_{2}^{2}-M_{3}^{2}\text{ , \ }\tilde{t}^{\prime}=t+M_{2}%
^{2}-M_{3}^{2}.
\end{equation}
\qquad

Note that, unlike the case of GR, here $e^{P}$ does not approach to $e^{L}$ in
the RR. The Regge string scattering amplitudes can then be explicitly
calculated to be%
\begin{align}
A\left(  s,t\right)   &  \simeq\int_{0}^{1}dy\ y^{k_{1}k_{2}}(1-y)^{k_{2}%
k_{3}}\cdot\prod_{n}\left[  -\frac{\left(  n-1\right)  !e^{T}\cdot k_{1}%
}{\left(  -y\right)  {}^{n}}-\frac{\left(  n-1\right)  !e^{T}\cdot k_{3}%
}{\left(  1-y\right)  ^{n}}\right]  ^{p_{n}}\nonumber\\
&  \quad\cdot\prod_{m}\left[  \frac{\left(  m-1\right)  !e^{P}\cdot k_{1}%
}{\left(  -y\right)  ^{m}}+\frac{\left(  m-1\right)  !e^{P}\cdot k_{3}%
}{\left(  1-y\right)  ^{m}}\right]  ^{q_{m}}\nonumber\\
&  \cdot\prod_{l}\left[  -\frac{\left(  l-1\right)  !e^{L}\cdot k_{1}}{\left(
-y\right)  ^{l}}-\frac{\left(  l-1\right)  !e^{L}\cdot k_{3}}{\left(
1-y\right)  ^{l}}\right]  ^{r_{l}}\nonumber\\
&  \approx\int_{0}^{1}dy\ y^{-\frac{s}{2}+N-2}\left(  1-y\right)  ^{-\frac
{t}{2}+N-2}\nonumber\\
&  \quad\cdot\prod_{n>0}\left[  \frac{\left(  n-1\right)  !\sqrt{-t}}{\left(
1-y\right)  ^{n}}\right]  ^{p_{n}}\cdot\prod_{m>1}\left[  -\frac{\left(
m-1\right)  !\frac{\tilde{t}}{2M_{2}}}{\left(  1-y\right)  ^{m}}\right]
^{q_{m}}\cdot\prod_{l>1}\left[  \frac{\left(  l-1\right)  !\frac{\tilde
{t}^{\prime}}{2M_{2}}}{\left(  1-y\right)  ^{l}}\right]  ^{r_{l}}\nonumber\\
&  \quad\cdot\left[  \frac{\frac{s}{2M_{2}}}{y}-\frac{\frac{\tilde{t}}{2M_{2}%
}}{\left(  1-y\right)  }\right]  ^{q_{1}}\left[  -\frac{\frac{s}{2M_{2}}}%
{y}+\frac{\frac{\tilde{t}^{\prime}}{2M_{2}}}{\left(  1-y\right)  }\right]
^{r_{1}}\nonumber\\
&  =\prod_{n>0}\left[  \left(  n-1\right)  !\sqrt{-t}\right]  ^{p_{n}}%
\cdot\prod_{m>1}\left[  -\left(  m-1\right)  !\frac{\tilde{t}}{2M_{2}}\right]
^{q_{m}}\cdot\prod_{l>1}\left[  \left(  l-1\right)  !\frac{\tilde{t}^{\prime}%
}{2M_{2}}\right]  ^{r_{l}}\nonumber\\
&  \quad\cdot\sum_{i,j}\binom{q_{1}}{i}\binom{r_{1}}{j}\left(  -\frac
{s}{\tilde{t}}\right)  ^{i}\left(  -\frac{s}{\tilde{t}^{\prime}}\right)
^{j}B\left(  -\frac{s}{2}+N-1-i-j,-\frac{t}{2}-1+i+j\right)  . \label{beta..}%
\end{align}
In the second equality of the above equation, we have dropped the first term
in the bracket with power of $p_{n}$, and the first terms in the brackets with
powers of $q_{m}$ and $r_{l}$ for $m,l>1.$ These terms lead to subleading
order terms in energy in the Regge limit \cite{bosonic,RRsusy}. Now the beta
function in Eq.(\ref{beta..}) can be approximated in the RR by
\cite{bosonic,RRsusy}
\begin{equation}
B\left(  -\frac{s}{2}+N-1-i-j,-\frac{t}{2}-1+i+j\right)  =B\left(  -\frac
{s}{2}-1,-\frac{t}{2}-1\right)  \left(  -\frac{s}{2}\right)  ^{-i-j}\left(
-\frac{t}{2}-1\right)  _{i+j}%
\end{equation}
where $(a)_{j}=a(a+1)(a+2)...(a+j-1)$ is the Pochhammer symbol. Finally we
arrive at the amplitude with two equivalent expressions
\begin{align}
A\left(  s,t\right)   &  =\prod_{n>0}\left[  \left(  n-1\right)  !\sqrt
{-t}\right]  ^{p_{n}}\cdot\prod_{m>0}\left[  -\left(  m-1\right)
!\frac{\tilde{t}}{2M}\right]  ^{q_{m}}\cdot\prod_{l>1}\left[  \left(
l-1\right)  !\frac{\tilde{t}^{\prime}}{2M}\right]  ^{r_{l}}\label{factor1}\\
&  \quad\cdot B\left(  -\frac{s}{2}-1,-\frac{t}{2}+1\right)  \left(  \frac
{1}{M}\right)  ^{r_{1}}\nonumber\\
&  \cdot\sum_{i=0}^{q_{1}}\binom{q_{1}}{i}\left(  \frac{2}{\tilde{t}}\right)
^{i}\left(  -\frac{t}{2}-1\right)  _{i}U\left(  -r_{1},\frac{t}{2}%
+2-i-r_{1},\frac{\tilde{t}^{\prime}}{2}\right) \nonumber\\
&  =\prod_{n>0}\left[  \left(  n-1\right)  !\sqrt{-t}\right]  ^{p_{n}}%
\cdot\prod_{m>1}\left[  -\left(  m-1\right)  !\frac{\tilde{t}}{2M}\right]
^{q_{m}}\cdot\prod_{l>0}\left[  \left(  l-1\right)  !\frac{\tilde{t}^{\prime}%
}{2M}\right]  ^{r_{l}}\label{factor2}\\
&  \cdot B\left(  -\frac{s}{2}-1,-\frac{t}{2}+1\right)  \left(  -\frac{1}%
{M}\right)  ^{q_{1}}\nonumber\\
&  \cdot\sum_{j=0}^{r_{1}}\binom{r_{1}}{j}\left(  \frac{2}{\tilde{t}^{\prime}%
}\right)  ^{j}\left(  -\frac{t}{2}-1\right)  _{j}U\left(  -q_{1},\frac{t}%
{2}+2-j-q_{1},\frac{\tilde{t}}{2}\right)  .\nonumber
\end{align}
\qquad\qquad

It is interesting to note that the Regge behavior is again universal and is
mass level independent as in the case of previous section \cite{bosonic}%
\begin{equation}
B\left(  -1-\frac{s}{2},-1-\frac{t}{2}\right)  \sim s^{\alpha(t)}\text{ \ (in
the RR)}%
\end{equation}
where $\alpha(t)=\alpha(0)+\alpha^{\prime}t$, \ $\alpha(0)=1$ and
$\alpha^{\prime}=1/2.$ That is, the well known $\sim s^{\alpha(t)}$ power-law
behavior of the four tachyon string scattering amplitude in the RR can be
extended to arbitrary higher string states. This result will be used to
construct an inter-mass level recurrence relation for Regge string scattering
amplitudes later in Eq.(\ref{RRT3}).

\subsubsection{Recurrence relations and RR stringy Ward identities}

In this section, we first discuss Regge stringy Ward identities derived from
Regge string ZNS (RZNS) for mass level $M^{2}=2$ and$\ 4$ \cite{LY}$.$ We will
see that, unlike the case for the GR stringy Ward identities, Regge string
Wayrd identities are not good enough to solve all Regge string scattering
amplitudes algebraically. On the other hand, we found that the recurrence
relations of Kummer functions Eq.(\ref{RC1}) to Eq.(\ref{RC6}) discussed in
the appendix \ref{recurrence of Kummer} can be used to prove all Regge stringy
Ward identities. Presumably the calculation can be generalized to arbitrary
mass levels. Another reason to work on recurrence relations of Kummer
functions instead of \ Regge stringy Ward identities is that the former is
very easy to generalize to arbitrary higher mass levels while the latter is not.

Most importantly, for Kummer functions $U(a,c,x)$ in Regge string amplitudes
in Eq.(\ref{factor1}) and Eq.(\ref{factor2}) with $a=-q_{1}($or $-r_{1})$ a
non-positive integer, one can use recurrence relations to solve all
$U(-q_{1},c,x)$ functions algebraically and thus determine all Regge string
scattering amplitudes at arbitrary mass levels algebraically up to
multiplicative factors \cite{LY}. We stress that for general values of $a$,
the best one can obtain from recurrence relations is to express any Kummer
function in terms of any two of its associated function (see the appendix
\ref{recurrence of Kummer}).

There are $9$ Regge string amplitudes for the mass level $M^{2}=2$,
$A^{PP}(\alpha_{-1}^{P}\alpha_{-1}^{P})$,$\ A^{PL}(\alpha_{-1}^{P}\alpha
_{-1}^{L})$, $A^{PT}(\alpha_{-1}^{P}\alpha_{-1}^{T})$, $A^{LL}(\alpha_{-1}%
^{T}\alpha_{-1}^{T})$, $A^{LT}(\alpha_{-1}^{L}\alpha_{-1}^{T})$,
$A^{TT}(\alpha_{-1}^{T}\alpha_{-1}^{T})$, $A^{P}(\alpha_{-2}^{P})$,
$A^{L}(\alpha_{-2}^{L})$, $A^{T}(\alpha_{-2}^{T}).$ For this mass level
$\tilde{t}=t,$ $\tilde{t}^{\prime}=t+4.$ The Regge string ZNS (RZNS) in
Eq.(\ref{R2.1}) and Eq.(\ref{R2.2}) gives two Regge stringy Ward identities%
\begin{equation}
A^{T}-\sqrt{2}A^{PT}=0, \label{W2.1}%
\end{equation}%
\begin{equation}
A^{L}-\sqrt{2}A^{PL}=0. \label{W2.2}%
\end{equation}
The RZNS in Eq.(\ref{R2.3}) gives%
\begin{equation}
\sqrt{2}A^{P}-A^{PP}-\frac{1}{5}A^{LL}-\frac{1}{5}A^{TT}=0. \label{W2.3}%
\end{equation}
It's obvious to see that these three Regge stringy Ward identities
Eq.(\ref{R2.1}) to Eq.(\ref{R2.3}) are not good enough to solve all the $9$
Regge string scattering amplitudes algebraically. Indeed, the amplitude
$A^{LT}$ does not even show up in any of these three Ward identities.

Instead of Regge stringy Ward identities, in the following we will do the
calculation based on recurrence relations of Kummer functions. We want to
prove these three Regge stringy Ward identities by using recurrence relations%
\begin{align}
U(a-1,c,x)-(2a-c+x)U(a,c,x)+a(1+a-c)U(a+1,c,x)  &  =0,\label{A1}\\
U(a,c,x)-aU(a+1,c,x)-U(a,c-1,x)  &  =0. \label{A2}%
\end{align}
First, by taking some special values of arguments of Kummer function in
Eq.(\ref{A1}) and Eq.(\ref{A2}), one easily obtain
\begin{equation}
U\left(  -1,x,x\right)  =0, \label{R1}%
\end{equation}%
\begin{equation}
U\left(  -2,x,x\right)  +xU(0,x,x)=0 \label{R2}%
\end{equation}
and%

\begin{equation}
U\left(  0,c,x\right)  -U(0,c-1,x)=0. \label{R3}%
\end{equation}
By using Eq.(\ref{factor2}), one easily see that the Ward identity
Eq.(\ref{W2.1}) implies
\begin{equation}
U\left(  0,\frac{t}{2}+2,\frac{\tilde{t}}{2}\right)  +U\left(  -1,\frac{t}%
{2}+1,\frac{\tilde{t}}{2}\right)  =0\ \label{3.1}%
\end{equation}

\noindent To prove Eq.(\ref{3.1}) by recurrence relations, we note that for
the case of $a=0,\ c=\frac{t}{2}+1,\ x=\frac{\tilde{t}}{2}$, Eq.(\ref{A1})
says
\begin{equation}
U\left(  -1,\frac{t}{2}+1,\frac{\tilde{t}}{2}\right)  +U\left(  0,\frac{t}%
{2}+1,\frac{\tilde{t}}{2}\right)  =0\ . \label{3.2}%
\end{equation}
We then apply Eq.(\ref{R3}) for the second term of Eq.(\ref{3.2}) to obtain
Eq.(\ref{3.1}). This completes the proof of Regge stringy Ward identity
Eq.(\ref{W2.1}) based on recurrence relations Eq.(\ref{A1}) and Eq.(\ref{A2}).
The Ward identity in Eq.(\ref{W2.2}) implies
\begin{equation}
\frac{1}{\sqrt{2}}\frac{\tilde{t}^{\prime}}{2}\left[  U\left(  0,\frac{t}%
{2}+2,\frac{\tilde{t}}{2}\right)  +U\left(  -1,\frac{t}{2}+1,\frac{\tilde{t}%
}{2}\right)  \right]  +\left(  -\frac{t}{2}-1\right)  U\left(  -1,\frac{t}%
{2},\frac{\tilde{t}}{2}\right)  =0. \label{3.3}%
\end{equation}

\noindent To prove Eq.(\ref{3.3}) by using recurrence relations, we note that
Eq.(\ref{3.1}) implies the first and the second terms of Eq.(\ref{3.3}) cancel
out. Eq.(\ref{R3}) and $t=\tilde{t}$ say that the last term of Eq.(\ref{3.3})
vanishes. Finally, to prove Eq.(\ref{W2.3}) by using recurrence relations, one
needs to prove%
\begin{align}
&  \left[  \frac{1}{10}\left(  \frac{\tilde{t}^{\prime}}{2}\right)  ^{2}%
+\frac{\tilde{t}}{2}-\frac{t}{5}\right]  U\left(  0,\frac{t}{2}+2,\frac
{\tilde{t}}{2}\right)  +\frac{1}{2}U\left(  -2,\frac{t}{2},\frac{\tilde{t}}%
{2}\right) \nonumber\\
&  \quad+\frac{1}{5}\left(  \frac{\tilde{t}^{\prime}}{2}\right)  \left(
-\frac{t}{2}-1\right)  U\left(  0,\frac{t}{2}+1,\frac{\tilde{t}}{2}\right)
+\frac{1}{10}\left(  -\frac{t}{2}-1\right)  \left(  -\frac{t}{2}\right)
U\left(  0,\frac{t}{2},\frac{\tilde{t}}{2}\right)  =0. \label{3.4}%
\end{align}

\noindent Now Eq.(\ref{R2}) implies
\begin{equation}
U\left(  0,\frac{t}{2}+2,\frac{\tilde{t}}{2}\right)  =U\left(  0,\frac{t}%
{2}+1,\frac{\tilde{t}}{2}\right)  =U\left(  0,\frac{t}{2},\frac{\tilde{t}}%
{2}\right)  . \label{3.5}%
\end{equation}
Therefore Eq.(\ref{3.4}) is equivalent to
\begin{equation}
\frac{t}{2}U\left(  0,\frac{t}{2},\frac{\tilde{t}}{2}\right)  +U\left(
-2,\frac{t}{2},\frac{\tilde{t}}{2}\right)  =0. \label{3.6}%
\end{equation}
Finally one can use Eq.(\ref{R2}) and $\tilde{t}=t$ to prove Eq.(\ref{3.6}).
This completes the proof of Regge stringy Ward identities for mass level
$M^{2}=2$ by using recurrence relations of Kummer functions.

We will give a brief description for the case of mass level $M^{2}=4.$ There
are $22$ Regge string amplitudes for the mass level $M^{2}=4$, $A^{PPP}%
$,$\ A^{PPL}$, $A^{PPT}$, $A^{PLL}$, $A^{PLT}$, $A^{PTT}$, $A^{PP}$, $A^{LP}$,
$A^{TP}$, $A^{LLL}$, $A^{LLT}$, $A^{LTT}$, $A^{TTT}$,$A^{PL}$, $A^{LL}$,
$A^{TL}$, $A^{PT}$, $A^{LT}$, $A^{TT}$, $A^{P}$, $A^{L}$, $A^{T}.$ To fix the
notation, we adopt the convention of mass ordered in the $\alpha_{-n}^{\alpha
}$ operators, for example, $A^{LT}(\alpha_{-2}^{L}\alpha_{-1}^{T})$ and
$A^{TL}(\alpha_{-2}^{T}\alpha_{-1}^{L})$ etc. For this mass level $\tilde
{t}=t-2,$ $\tilde{t}^{\prime}=t+6.$ The $8$ RZNS Eqs.(\ref{R4.1}),
(\ref{R4.2}), (\ref{R4.3}), (\ref{R4.4}), (\ref{R4.5}), (\ref{R4.6}),
(\ref{R4.7}) and (\ref{R4.8}) calculated in the appendix \ref{Regge ZNS} give
$8$ Regge stringy Ward identities%
\begin{equation}
25A^{PPP}+9A^{PLL}+9A^{PTT}-9A^{LL}-9A^{TT}-75A^{PP}+50A^{P}=0, \label{R4}%
\end{equation}%
\begin{equation}
A^{PLL}-A^{LL}=0, \label{R5}%
\end{equation}%
\begin{equation}
A^{PTT}-A^{TT}=0, \label{R6}%
\end{equation}%
\begin{equation}
A^{PLT}-A^{(LT)}=0, \label{R7}%
\end{equation}%
\begin{equation}
9A^{PPT}+A^{LLT}+A^{TTT}-18A^{(PT)}+6A^{T}=0, \label{R8}%
\end{equation}%
\begin{equation}
9A^{PPL}+A^{LLL}+A^{LTT}-18A^{(PL)}+6A^{L}=0, \label{R9}%
\end{equation}
\
\begin{equation}
A^{LLT}+A^{TTT}-9A^{[PT]}-3A^{T}=0, \label{R10}%
\end{equation}%
\begin{equation}
A^{LLL}+A^{LTT}-9A^{[PL]}-3A^{L}=0. \label{R11}%
\end{equation}
It is obvious to see that these eight Regge stringy Ward identities are not
good enough to solve the $22$ Regge string scattering amplitudes
algebraically. Indeed, for example, the amplitude $A^{[LT]}$ does not even
show up in any of these eight Ward identities. However, in the GR, one can
identify $e^{P}$ and $e^{L}$ components \cite{ChanLee,ChanLee1,ChanLee2}
(Correspondingly the creation operators $\alpha_{-n}^{P}$ and $-\alpha
_{-n}^{L}$ are identified, where the sign comes from the difference between
the timelike and spacelike directions specified by the metric of the
scattering plane $\eta_{\mu\nu}=diag(-1,1,1)$ .), and take high energy fixed
angle limit to get three Ward identities in leading order energy
\cite{ChanLee,ChanLee1,ChanLee2}
\begin{align}
T^{LLT}+T^{(LT)}  &  =0,\\
10T^{LLT}+T^{TTT}+18T^{(LT)}  &  =0,\\
T^{LLT}+T^{TTT}+9T^{[LT]}  &  =0,
\end{align}
which can be easily solved to get \cite{ChanLee,ChanLee1,ChanLee2}
\begin{equation}
T^{TTT}:T^{LLT}:T^{(LT)}:T^{[LT]}=8:1:-1:-1.
\end{equation}
The ratios above are consistent with Eq.(\ref{2.19..}).

\bigskip For illustration, we now proceed to prove Regge stringy Ward
identities Eq.(\ref{R4}) to Eq.(\ref{R7}) by using recurrence relations
Eq.(\ref{A1}), Eq.(\ref{A2}) and%
\begin{equation}
\left(  c-a-1\right)  U(a,c-1,x)-\left(  x+c-1\right)  U\left(  a,c,x\right)
+xU\left(  a,c+1,x\right)  =0. \label{A4}%
\end{equation}
Other Regge stringy Ward identities Eq.(\ref{R8}) to Eq.(\ref{R11}) can be
similarly proved by using recurrence relations. For the case of $a=-1$ ,
$c=x+1$, Eq.(\ref{A4}) reduces to%

\begin{equation}
\left(  x+1\right)  U\left(  -1,x,x\right)  -2xU\left(  -1,x+1,x\right)
+xU\left(  -1,x+2,x\right)  =0. \label{R4-1}%
\end{equation}
For the case of $a=-1$, $c=x+2$, Eq.(\ref{A2}) reduces to
\begin{equation}
U\left(  -1,x+2,x\right)  +U\left(  0,x+2,x\right)  -U\left(  -1,x+1,x\right)
=0. \label{R4-2}%
\end{equation}
Finally Eq.(\ref{R4-1}), Eq.(\ref{R4-2}), and Eq.(\ref{R1}) say
\begin{align}
U(-1,x+2,x)  &  =-2U\left(  0,x+2,x\right)  ,\label{R4-3}\\
U\left(  -1,x+1,x\right)   &  =-U\left(  0,x+2,x\right)  . \label{R4-4}%
\end{align}

\bigskip We are now ready to prove Regge stringy Ward identities. We first
prove Regge stringy Ward identity Eq.(\ref{R5}). The two terms in
Eq.(\ref{R5}) divided by the beta function can be calculated to be
\begin{align}
\frac{1}{B}A^{PLL}  &  =-\frac{1}{M}\left(  \frac{\tilde{t}^{\prime}}%
{2M}\right)  ^{2}\left[  U\left(  -1,\frac{t}{2}+1,\frac{t}{2}-1\right)
+2\left(  \frac{2}{\tilde{t^{\prime}}}\right)  \left(  -\frac{t}{2}-1\right)
U\left(  -1,\frac{t}{2},\frac{t}{2}-1\right)  \right. \nonumber\\
&  \quad\qquad\qquad\qquad\left.  +\left(  \frac{2}{\tilde{t^{\prime}}%
}\right)  ^{2}\left(  -\frac{t}{2}-1\right)  \left(  -\frac{t}{2}\right)
U\left(  -1,\frac{t}{2}-1,\frac{t}{2}-1\right)  \right] \nonumber\\
&  =-\frac{1}{M}\left(  \frac{t+6}{2M}\right)  ^{2}\left[
\begin{array}
[c]{c}%
U\left(  -1,\frac{t}{2}+1,\frac{t}{2}-1\right)  -2\frac{t+2}{t+6}U\left(
-1,\frac{t}{2},\frac{t}{2}-1\right) \\
+\frac{t\left(  t+2\right)  }{\left(  t+6\right)  ^{2}}U\left(  -1,\frac{t}%
{2}-1,\frac{t}{2}-1\right)
\end{array}
\right]  , \label{3.7}%
\end{align}%
\begin{align}
\frac{1}{B}A^{LL}  &  =\left(  \frac{\tilde{t}^{\prime}}{2M}\right)
^{2}\left[  U\left(  0,\frac{t}{2}+2,\frac{t}{2}-1\right)  +\left(  \frac
{2}{\tilde{t^{\prime}}}\right)  \left(  -\frac{t}{2}-1\right)  U\left(
0,\frac{t}{2}+1,\frac{t}{2}-1\right)  \right] \nonumber\\
&  =\left(  \frac{t+6}{2M}\right)  ^{2}\left[  U\left(  0,\frac{t}{2}%
+2,\frac{t}{2}-1\right)  -\frac{t+2}{t+6}U\left(  0,\frac{t}{2}+1,\frac{t}%
{2}-1\right)  \right]  \ . \label{3.8}%
\end{align}
Therefore we want to show
\begin{align}
&  -\frac{1}{M}\left[  U\left(  -1,\frac{t}{2}+1,\frac{t}{2}-1\right)
-2\frac{t+2}{t+6}U\left(  -1,\frac{t}{2},\frac{t}{2}-1\right)  +\frac{t\left(
t+2\right)  }{\left(  t+6\right)  ^{2}}U\left(  -1,\frac{t}{2}-1,\frac{t}%
{2}-1\right)  \right] \nonumber\\
&  \qquad\qquad-U\left(  0,\frac{t}{2}+2,\frac{t}{2}-1\right)  +\frac
{t+2}{t+6}U\left(  0,\frac{t}{2}+1,\frac{t}{2}-1\right)  \overset{?}{=}0
\label{3.9}%
\end{align}
Eq.(\ref{R1}) implies the third term of Eq.(\ref{3.9}) vanishes and therefore
Eq.(\ref{R3}) implies that Eq.(\ref{3.9}) is equivalent to
\begin{align}
&  -\frac{1}{M}U\left(  -1,\frac{t}{2}+1,\frac{t}{2}-1\right)  +\frac{2}%
{M}\frac{t+2}{t+6}U\left(  -1,\frac{t}{2},\frac{t}{2}-1\right)  -\frac{4}%
{t+6}U\left(  0,\frac{t}{2}+1,\frac{t}{2}-1\right) \nonumber\\
&  =\frac{1}{M}\left[  -U\left(  -1,\frac{t}{2}+1,\frac{t}{2}-1\right)
+2\frac{t+2}{t+6}U\left(  -1,\frac{t}{2},\frac{t}{2}-1\right)  -\frac{8}%
{t+6}U\left(  0,\frac{t}{2}+1,\frac{t}{2}-1\right)  \right] \nonumber\\
&  =0 \label{3.10}%
\end{align}
For the case of $x=\frac{t}{2}-1$, Eq.(\ref{R4-3}) and Eq.(\ref{R4-4})
implies
\begin{equation}
U\left(  -1,\frac{t}{2}+1,\frac{t}{2}-1\right)  =2U\left(  -1,\frac{t}%
{2},\frac{t}{2}-1\right)  =-2U\left(  0,\frac{t}{2}+1,\frac{t}{2}-1\right)
\ . \label{3.11}%
\end{equation}
Hence Eq.(\ref{3.10}) is easily proved.

We now prove Regge stringy Ward identity Eq.(\ref{R6}). The two terms in
Eq.(\ref{R6}) divided by the beta function can be calculated to be
\[
\frac{1}{B}A^{PTT}=\left(  -t\right)  \left(  -\frac{1}{M}\right)  U\left(
-1,\frac{t}{2}+1,\frac{t}{2}-1\right)  ,
\]%
\[
\frac{1}{B}A^{TT}=\left(  -t\right)  U\left(  0,\frac{t}{2}+2,\frac{t}%
{2}-1\right)  .
\]
Therefore we want to show
\begin{equation}
\frac{t}{M}U\left(  -1,\frac{t}{2}+1,\frac{t}{2}-1\right)  +tU\left(
0,\frac{t}{2}+2,\frac{t}{2}-1\right)  \overset{?}{=}0 \label{3.12}%
\end{equation}
For the case of $x=\frac{t}{2}-1$, Eq.(\ref{R4-3}) means
\begin{equation}
U\left(  -1,\frac{t}{2}+1,\frac{t}{2}-1\right)  =-2U\left(  0,\frac{t}%
{2}+1,\frac{t}{2}-1\right)  \ . \label{3.13}%
\end{equation}
Eq.(\ref{3.13}) and Eq.(\ref{R3}) prove Eq.(\ref{3.12}).

We can now turn to prove Regge stringy Ward identity Eq.(\ref{R4}). We first
note that Eq.(\ref{R5}) and Eq.(\ref{R6}) implies that Eq.(\ref{R4}) is
equivalent to
\begin{equation}
25A^{PPP}-75A^{PP}+50A^{P}=0. \label{3.14}%
\end{equation}
The three terms in Eq.(\ref{3.14}) divided by the beta function are
\begin{align}
\frac{1}{B}A^{P}  &  =\left(  -\frac{t-2}{2M}\right)  U\left(  0,\frac{t}%
{2}+2,\frac{t}{2}-1\right)  ,\label{3.15}\\
\frac{1}{B}A^{PP}  &  =-\frac{1}{M}\left(  -\frac{t-2}{2M}\right)  U\left(
-1,\frac{t}{2}+1,\frac{t}{2}-1\right)  ,\label{3.16}\\
\frac{1}{B}A^{PPP}  &  =\left(  -\frac{1}{M}\right)  ^{3}U\left(  -3,\frac
{t}{2}-1,\frac{t}{2}-1\right)  \ . \label{3.17}%
\end{align}
Therefore we want to show
\begin{align}
&  2\left(  -\frac{t-2}{2M}\right)  U\left(  0,\frac{t}{2}+2,\frac{t}%
{2}-1\right)  -3\left(  -\frac{1}{M}\right)  \left(  -\frac{t-2}{2M}\right)
U\left(  -1,\frac{t}{2}+1,\frac{t}{2}-1\right) \nonumber\\
&  +\left(  -\frac{1}{M}\right)  ^{3}U\left(  -3,\frac{t}{2}-1,\frac{t}%
{2}-1\right)  \overset{?}{=}0 \label{3.18}%
\end{align}
For the case of $a=-2$, $c=x$, Eq.(\ref{A1}) gives%
\begin{equation}
U\left(  -3,x,x\right)  +4U\left(  -2,x,x\right)  +2\left(  1+x\right)
U\left(  -1,x,x\right)  =0. \label{3.19}%
\end{equation}
Using Eq.(\ref{R1}), we obtain
\begin{equation}
U\left(  -3,x,x\right)  +4U\left(  -2,x,x\right)  =0\ . \label{3.20}%
\end{equation}
From Eq.(\ref{3.20}) and Eq.(\ref{R2}), we obtain
\begin{equation}
U\left(  -3,x,x\right)  -4xU\left(  0,x,x\right)  =0\ . \label{3.21}%
\end{equation}
From Eq.(\ref{3.21}) and Eq.(\ref{R3}), we obtain%
\begin{align}
&  2\left(  -\frac{t-2}{2M}\right)  U\left(  0,\frac{t}{2}+2,\frac{t}%
{2}-1\right)  -3\left(  -\frac{1}{M}\right)  \left(  -\frac{t-2}{2M}\right)
U\left(  -1,\frac{t}{2}+1,\frac{t}{2}-1\right) \nonumber\\
&  +\left(  -\frac{1}{M}\right)  ^{3}U\left(  -3,\frac{t}{2}-1,\frac{t}%
{2}-1\right) \nonumber\\
=  &  \left(  2\left(  -\frac{t-2}{2M}\right)  +4\left(  \frac{t-2}{2}\right)
\left(  -\frac{1}{M}\right)  ^{3}\right)  U\left(  0,\frac{t}{2}+2,\frac{t}%
{2}-1\right) \nonumber\\
&  +3\frac{1}{M}\left(  -\frac{t-2}{2M}\right)  U\left(  -1,\frac{t}%
{2}+1,\frac{t}{2}-1\right) \nonumber\\
=  &  \left(  2\left(  -\frac{t-2}{4}\right)  +\left(  \frac{t-2}{2}\right)
\left(  -\frac{1}{2}\right)  \right)  U\left(  0,\frac{t}{2}+2,\frac{t}%
{2}-1\right) \nonumber\\
&  +\frac{3}{2}\left(  -\frac{t-2}{4}\right)  U\left(  -1,\frac{t}{2}%
+1,\frac{t}{2}-1\right) \nonumber\\
=  &  \frac{t-2}{2}\left[  -\frac{3}{2}U\left(  0,\frac{t}{2}+2,\frac{t}%
{2}-1\right)  -\frac{3}{2}\frac{1}{2}U\left(  -1,\frac{t}{2}+1,\frac{t}%
{2}-1\right)  \right]  \ . \label{3.22}%
\end{align}
Finally Eq.(\ref{R3}) and Eq.(\ref{R4-3}) implies that Eq.(\ref{3.22})
vanishes. This proves Eq.(\ref{3.18}).

For the fourth stringy Ward identity at mass level $M^{2}=4$, the two terms in
Eq.(\ref{R7}) divided by the beta function are
\begin{align}
\frac{1}{B}A^{PLT}=  &  -\frac{\sqrt{-t}\tilde{t}^{\prime}}{2M^{2}}\left[
U\left(  -1,\frac{t}{2}+1,\frac{\tilde{t}}{2}\right)  +\left(  \frac{2}%
{\tilde{t}^{\prime}}\right)  \left(  -\frac{t}{2}-1\right)  U\left(
-1,\frac{t}{2},\frac{\tilde{t}}{2}\right)  \right] \nonumber\\
=  &  -\frac{\sqrt{-t}\left(  t+6\right)  }{2M^{2}}\left[
\begin{array}
[c]{c}%
U\left(  -1,\frac{t}{2}+1,\frac{t}{2}-1\right) \\
+\left(  \frac{2}{t+6}\right)  \left(  -\frac{t}{2}-1\right)  U\left(
-1,\frac{t}{2},\frac{t}{2}-1\right)
\end{array}
\right]  ,\label{3.23}\\
\frac{1}{B}A^{LT}=  &  \sqrt{-t}\frac{\tilde{t}^{\prime}}{2M}U\left(
0,\frac{t}{2}+2,\frac{\tilde{t}}{2}\right)  =\sqrt{-t}\frac{t+6}{2M}U\left(
0,\frac{t}{2}+2,\frac{t}{2}-1\right)  ,\label{3.24}\\
\frac{1}{B}A^{TL}=  &  \sqrt{-t}\frac{\tilde{t}^{\prime}}{2M}\left[  U\left(
0,\frac{t}{2}+2,\frac{\tilde{t}}{2}\right)  +\left(  \frac{2}{\tilde
{t}^{\prime}}\right)  \left(  -\frac{t}{2}-1\right)  U\left(  0,\frac{t}%
{2}+1,\frac{\tilde{t}}{2}\right)  \right] \nonumber\\
=  &  \sqrt{-t}\frac{t+6}{2M}\left[  U\left(  0,\frac{t}{2}+2,\frac{t}%
{2}-1\right)  +\left(  \frac{2}{t+6}\right)  \left(  -\frac{t}{2}-1\right)
U\left(  0,\frac{t}{2}+1,\frac{t}{2}-1\right)  \right]  \ . \label{3.25}%
\end{align}
Therefore we want to show%
\begin{align}
&  2A^{PLT}-A^{LT}-A^{TL}\nonumber\\
&  =B\sqrt{t}\left[  -\frac{t+6}{M^{2}}U\left(  -1,\frac{t}{2}+1,\frac{t}%
{2}-1\right)  -\frac{2}{M^{2}}\left(  -\frac{t}{2}-1\right)  U\left(
-1,\frac{t}{2},\frac{t}{2}-1\right)  \right. \nonumber\\
&  \quad\left.  -\frac{t+6}{M}U\left(  0,\frac{t}{2}+2,\frac{t}{2}-1\right)
-\frac{1}{M}\left(  -\frac{t}{2}-1\right)  U\left(  0,\frac{t}{2}+1,\frac
{t}{2}-1\right)  \right]  \overset{?}{=}0\ . \label{3.26}%
\end{align}
Using Eq.(\ref{R3}), we obtain
\begin{align}
&  \frac{1}{B}(2A^{PLT}-A^{LT}-A^{TL})\nonumber\\
&  =\sqrt{t}\left[  -\frac{t+6}{M^{2}}U\left(  -1,\frac{t}{2}+1,\frac{t}%
{2}-1\right)  -\frac{2}{M^{2}}\left(  -\frac{t}{2}-1\right)  U\left(
-1,\frac{t}{2},\frac{t}{2}-1\right)  \right. \nonumber\\
&  \qquad\left.  +\frac{1}{M}\left(  -t-6+\frac{t}{2}+1\right)  U\left(
0,\frac{t}{2},\frac{t}{2}-1\right)  \right] \nonumber\\
&  =\sqrt{t}\left[
\begin{array}
[c]{c}%
-\frac{t+6}{M^{2}}U\left(  -1,\frac{t}{2}+1,\frac{t}{2}-1\right)  -\frac
{2}{M^{2}}\left(  -\frac{t}{2}-1\right)  U\left(  -1,\frac{t}{2},\frac{t}%
{2}-1\right) \\
+\frac{1}{M}\left(  -\frac{t}{2}-5\right)  U\left(  0,\frac{t}{2},\frac{t}%
{2}-1\right)
\end{array}
\right] \nonumber\\
&  =\sqrt{t}\left[
\begin{array}
[c]{c}%
-\frac{t+6}{4}U\left(  -1,\frac{t}{2}+1,\frac{t}{2}-1\right)  +\frac{t+2}%
{4}U\left(  -1,\frac{t}{2},\frac{t}{2}-1\right) \\
-\frac{t+10}{4}U\left(  0,\frac{t}{2},\frac{t}{2}-1\right)
\end{array}
\right]  \label{3.27}%
\end{align}
One can now use Eq.(\ref{R4-3}) and Eq.(\ref{R4-4}) to prove that
Eq.(\ref{3.27}) vanishes. This completes the explicit proof of four Regge
stringy Ward identities for mass level $M^{2}=4$ by using recurrence relations
of Kummer functions. Other four Regge stringy Ward identities can be similarly proved.

\subsubsection{Solving all RSSA by Kummer recurrence relations}

We observe that the recurrence relations of Kummer functions are more powerful
than Regge stringy Ward identities in relating Regge string scattering
amplitudes. This is indeed the case as we will show \cite{LY} now in the
following that all Regge string scattering amplitudes can be algebraically
solved by using recurrence relations up to multiplicative factors in the first
line of Eq.(\ref{factor1}) (or Eq.(\ref{factor2})).

To be more precise, we will first show that the ratio%
\begin{equation}
\frac{U(a,c,x)}{U(0,x,x)}=f(a,c,x),a=0,-1,-2,-3,... \label{Lemma}%
\end{equation}
is fixed and can be calculated by using recurrence relations Eq.(\ref{A1}),
Eq.(\ref{A2}) and%
\begin{equation}
(c-a)U(a,c,x)+U(a-1,c,x)-xU(a,c+1,x)=0. \label{A3}%
\end{equation}
We stress that Eq.(\ref{Lemma}) is nontrivial in the sense that, for general
values of $a$, the best one can obtain from recurrence relations is to express
any Kummer function in terms of any two of its associated function (see
Appendix \ref{recurrence of Kummer}). However, Eq.(\ref{Lemma}) states that
for non-positive integer values of $a,$ $U(a,c,x)$ can be fixed up to an
overall factor by using recurrence relations.

To prove Eq.(\ref{Lemma}), we first note that, for $a=0,c=x,$ recurrence
relation Eq.(\ref{A1}) implies Eq.(\ref{R1}). \ This determines $\frac
{U(a,x,x)}{U(0,x,x)}$ for $a$ is a non-positive integer$.$ For illustration,
we list examples of relations%
\begin{align*}
a  &  =-1,U(-2,x,x)+0+xU(0,x,x)=0,\\
a  &  =-2,U(-3,x,x)+4U(-2,x,x)+0=0,\\
a  &  =-3,U(-4,x,x)+6U(-3,x,x)+3(2+x)U(-2,x,x)=0,\\
&  ...............
\end{align*}
which determines $\frac{U(-2,x,x)}{U(0,x,x)},\frac{U(-3,x,x)}{U(0,x,x)}%
,\frac{U(-4,x,x)}{U(0,x,x)},....$ recursively.

Next we extend the result to $\frac{U(a,c,x)}{U(0,x,x)}$ for $c=x+Z,Z=$
integer. We first consider the simple case with $a=0$. From Eq.(\ref{A2}), we
obtain for $a=0,c=x+i,i\in Z$%
\begin{equation}
U(0,x+i,x)-U(0,x+i-1,x)=0,
\end{equation}
which gives $\frac{U(0,x+i,x)}{U(0,x,x)}=1.$ This proves Eq.(\ref{Lemma}) for
$a=0.$ For $a\in Z_{-},c=x+Z_{-}$, we obtain from Eq.(\ref{A2}) with $c=x-i$%
\begin{equation}
U(a,x-i,x)-aU(a+1,x-i,x)-U(a,x-i-1,x)=0.
\end{equation}
Since $\frac{U(a,x,x)}{U(0,x,x)},\frac{U(a+1,x,x)}{U(0,x,x)}$ have been
determined for $a\in Z_{-},$ this determines $\frac{U(a,x-i,x)}{U(0,x,x)}$ for
$a\in Z_{-},i=1,2,3...$ recursively. For $a\in Z_{-},c=x+Z_{+},$ we obtain
from Eq.(\ref{A3}) with $c=x+i$%
\begin{equation}
(x-a+i)U(a,x+i,x)+U(a-1,x+i,x)-xU(a,x+i+1,x)=0.
\end{equation}
Since $\frac{U(a-1,x,x)}{U(0,x,x)},\frac{U(a,x,x)}{U(0,x,x)}$ have been
determined for $a\in Z_{-},$ this determines $\frac{U(a,x+i,x)}{U(0,x,x)}$ for
$a\in Z_{-},i=1,2,3...$ recursively. This completes the proof of
Eq.(\ref{Lemma}) by using recurrence relations of Kummer functions.

Secondly, we want to show that each Kummer function in the summation of
Eq.(\ref{factor2}) can be expressed in terms of Regge string scattering
amplitudes. To show this, we first consider $r_{1}=0$ amplitudes in a fixed
mass level and a fixed $q_{1}$ with no summation over Kummer functions. These
amplitudes contain only one Kummer function. Then let us take the amplitude
with the maximum $p_{1}$. By decreasing $p_{1}$ and increasing $r_{1}$ by $1$,
we can create an amplitude with two Kummer functions in the same mass level
and the same $q_{1}$. The first one of the two Kummer functions is the one
appeared in the previous amplitude with $r_{1}=0$, so we can write the second
Kummer function in terms of the two amplitudes, one with $r_{1}=0$ and the
other with $r_{1}=1$.

By decreasing $p_{1}$ and increasing $r_{1}$ by $1$ again, we can create an
amplitude with three Kummer functions in the same mass level and the same
$q_{1}$. The first two of the three Kummer functions is the ones appeared in
the previous two amplitudes, so we can write the third Kummer functions in
terms of the three amplitudes. We can repeat this process until $p_{1}=0$. In
this way, we can express all the Kummer functions in Eq.(\ref{factor2}) in
terms of the RR amplitudes.

In the following, as an example, let us illustrate the above process for the
mass level $4$ amplitudes. There are $22$ Regge string amplitudes for the mass
level $M^{2}=4.$ We first consider the group of amplitudes with $q_{1}=0,$
$(T^{TTT},T^{LTT},T^{LLT},T^{LLL})$. The corresponding $r_{1}$ for each
amplitude are $(0,1,2,3)$. By using Eq.(\ref{factor2}), one can easily see
that $U\left(  0,\frac{t}{2}+2,\frac{t}{2}-1\right)  $ can be expressed in
terms of $T^{TTT}$, $U\left(  0,\frac{t}{2}+1,\frac{t}{2}-1\right)  $ can be
expressed in terms of $(T^{TTT},T^{LTT})$, $U\left(  0,\frac{t}{2},\frac{t}%
{2}-1\right)  $ can be expressed in terms of $(T^{TTT},T^{LTT},T^{LLT})$, and
finally $U\left(  0,\frac{t}{2}-1,\frac{t}{2}-1\right)  $ can be expressed in
terms of $(T^{TTT},T^{LTT},T^{LLT},T^{LLL})$.

Similarly, we can consider groups of amplitudes $(T^{PT},T^{PL})$,
$(T^{LT},T^{LL})$ and $(T^{TT},T^{TL})$ with $q_{1}=0$; group of amplitude
$(T^{PTT},T^{PLT},T^{PLL})$ with $q_{1}=1$ and group of amplitude
$(T^{PPT},T^{PPL})$ with $q_{1}=2$. All the remaining $7$ amplitudes are with
$r_{1}=0$, and each amplitude contains only one Kummer function. Due to the
multiplicative factors, there are much more RR amplitudes than the number of
Kummer functions involved at each fixed mass level. At mass level $4$, for
example, there are $22$ RR amplitudes and only $10$ Kummer functions involved.
So there is an onto correspondence between RR amplitudes and Kummer functions.
We have done the analysis by using Eq.(\ref{factor2}). Similar analysis can be
performed by using Eq.(\ref{factor1}) to get the same results.

An important application of the above prescription is the construction of an
infinite number of recurrence relations among Regge string scattering
amplitudes. One can use the recurrence relations of Kummer functions
Eq.(\ref{RC1}) to Eq.(\ref{RC6}) to systematically construct recurrence
relations among Regge string scattering amplitudes.

Note that a simple calculation by using the explicit form of Kummer function
in Eq.(\ref{finite}) gives $U(0,x,x)=1.$ However, when applying to the case of
Regge string scattering amplitudes, it will bring back a multiplicative factor
in the first line of Eq.(\ref{factor1}), (Eq.(\ref{factor2})) for each
amplitude. We thus conclude that all Regge string scattering amplitudes can be
algebraically solved by recurrence relations of Kummer functions up to
multiplicative factors.

Finally we calculate some examples of recurrence relations among Regge string
scattering amplitudes. At mass level $M^{2}=2,$ by using Eq.(\ref{factor2})
and the recurrence relation
\begin{equation}
U\left(  -2,\frac{t}{2},\frac{t}{2}\right)  +\left(  \frac{t}{2}+1\right)
U(-1,\frac{t}{2},\frac{t}{2})-\frac{t}{2}U\left(  -1,\frac{t}{2}+1,\frac{t}%
{2}\right)  =0,
\end{equation}
one can obtain the following recurrence relation among Regge string scattering
amplitudes \cite{LY}%
\begin{equation}
M\sqrt{-t}A^{PP}-\frac{t}{2}A^{PT}=0. \label{RRI1}%
\end{equation}
In contrast to the Regge stringy Ward identities Eq.(\ref{W2.1}),
Eq.(\ref{W2.2}) and Eq.(\ref{W2.3}) which contain only constant coefficients,
the recurrence relation in Eq.(\ref{RRI1}) contains kinematic variable $t$ in
its coefficients. Note that Eq.(\ref{RRI1}) is independent of all three Regge
stringy Ward identities at mass level $M^{2}=2$.

At mass level $M^{2}=4,$ by using Eq.(\ref{factor2}), one can calculate%
\begin{align}
\frac{1}{B}A^{PPP}=  &  \left(  -\frac{1}{M}\right)  ^{3}U\left(  -3,\frac
{t}{2}-1,\frac{t}{2}-1\right)  ,\\
\frac{1}{B}A^{PPT}=  &  \left(  -\frac{1}{M}\right)  ^{2}\sqrt{-t}U\left(
-2,\frac{t}{2},\frac{t}{2}-1\right)  ,\\
\frac{1}{B}A^{PPL}=  &  \frac{t+6}{2M^{3}}U\left(  -2,\frac{t}{2},\frac{t}%
{2}-1\right)  +\frac{1}{M^{3}}\left(  -\frac{t}{2}-1\right)  U\left(
-2,\frac{t}{2}-1,\frac{t}{2}-1\right)  .
\end{align}
The recurrence relation%
\begin{equation}
U\left(  -3,\frac{t}{2}-1,\frac{t}{2}-1\right)  +\left(  \frac{t}{2}+1\right)
U(-2,\frac{t}{2}-1,\frac{t}{2}-1)-(\frac{t}{2}-1)U\left(  -2,\frac{t}{2}%
,\frac{t}{2}-1\right)  =0 \label{RRTT1}%
\end{equation}
leads to the following recurrence relation among Regge string scattering
amplitudes \cite{LY}%
\begin{equation}
M\sqrt{-t}A^{PPP}-4A^{PPT}+M\sqrt{-t}A^{PPL}=0. \label{RRI2}%
\end{equation}
We have explicitly verified Eq.(\ref{RRI1}) and Eq.(\ref{RRI2}). The
generalization of Eq.(\ref{RRI2}) to arbitrary mass levels will be derived in
Eq.(\ref{aaa}) in chapter XV. It will be difficult to identify identity like
Eq.(\ref{RRI2}) without using the recurrence relation Eq.(\ref{RRTT1}). One
can similarly construct infinite number of them for amplitudes at arbitrary
higher mass levels based on the recurrence relations of Kummer functions and
their associated functions (see Appendix \ref{recurrence of Kummer}).

For the third example, we construct an inter-mass level recurrence relation
for Regge string scattering amplitudes at mass level $M^{2}=2,4.$ We begin
with the addition theorem of Kummer function \cite{Slater}%
\begin{equation}
U(a,c,x+y)=\sum_{k=0}^{\infty}\frac{1}{k!}\left(  a\right)  _{k}(-1)^{k}%
y^{k}U(a+k,c+k,x)
\end{equation}
which terminates to a finite sum for a non-positive integer $a.$ By taking,
for example, $a=-1,c=\frac{t}{2}+1,x=\frac{t}{2}-1$ and $y=1,$ the theorem
gives%
\begin{equation}
U\left(  -1,\frac{t}{2}+1,\frac{t}{2}\right)  -U(-1,\frac{t}{2}+1,\frac{t}%
{2}-1)-U\left(  0,\frac{t}{2}+2,\frac{t}{2}-1\right)  =0. \label{inter1}%
\end{equation}
Note that, unlike all previous cases, the last arguments of Kummer functions
in Eq.(\ref{inter1}) can be different. Eq.(\ref{inter1}) leads to an
inter-mass level recurrence relation \cite{LY}%
\begin{equation}
M(2)(t+6)A_{2}^{TP}-2M(4)^{2}\sqrt{-t}A_{4}^{LP}+2M(4)A_{4}^{LT}=0
\label{RRI3}%
\end{equation}
where \ masses $M(2)=\sqrt{2},M(4)=\sqrt{4}=2,$ and $A_{2},A_{4}$ are Regge
string scattering amplitudes for mass levels $M^{2}=2,4$ respectively. In
deriving Eq.(\ref{RRI3}), it is important to use the fact that the Regge power
law behavior in Eq.(\ref{power}) is universal and is mass level independent
\cite{bosonic}. Following the same procedure, one can construct infinite
number of recurrence relations among Regge string scattering amplitudes at
arbitrary mass levels which, in general, are independent of Regge stringy Ward identities.

In this chapter, we calculate the complete set of high energy string
scattering amplitudes in the Regge regime. We derive Regge stringy Ward
identities for the first few mass levels based on the decoupling of ZNS. These
results are valid even for higher point functions and higher point loops as
well by unitarity. We found that, unlike the case for the fixed angle regime,
the Regge stringy Ward identities were not good enough to solve all the Regge
string scattering amplitudes algebraically. On the other hand, we found that
all the Regge stringy Ward identities can be explicitly proved by the
recurrence relations of Kummer functions of the second kind. We then show
that, instead of Regge stringy Ward identities, one can use these recurrence
relations to solve all Regge string scattering amplitudes algebraically up to
multiplicative factors.

Finally, for illustration, we calculate some examples of recurrence relations
among Regge string scattering amplitudes of different string states based on
recurrence relations and addition theorem of Kummer functions. In contrast to
the Regge stringy Ward identities which contain only constant coefficients,
these recurrence relations contains kinematic variable $t$ in its coefficients
and are in general independent of Regge stringy Ward identities. The dynamical
origin of these recurrence relations remain to be studied. These recurrence
relations among Regge string scattering amplitudes are dual to linear
relations or symmetries among high energy fixed angle string scattering
amplitudes discovered previously
\cite{ChanLee,ChanLee1,ChanLee2,CHLTY1,CHLTY2,CHLTY3}.

Recently, five-point tachyon amplitude was considered in the context of BCFW
application of string theory in \cite{BCFW}. It will be interesting to
consider both RR and GR of higher spin five-point scattering amplitudes.%

\setcounter{equation}{0}
\renewcommand{\theequation}{\arabic{section}.\arabic{equation}}%

\section{Four classes of Regge superstring scattering amplitudes}

In this chapter we will calculate \cite{RRsusy} four classes of scattering
amplitudes considered in chapter VIII corresponding to states in
Eq.(\ref{T/2}) to Eq.(\ref{TLL/3}) in the RR. Moreover, we will extract from
the RR superstring amplitudes the ratios among GR superstring amplitudes
calculated in chapter VIII \cite{susy}%
\begin{align}
\left\vert N,2m,q\right\rangle \otimes\left\vert b_{-\frac{3}{2}}%
^{P}\right\rangle  &  =\left(  -\frac{1}{2M_{2}}\right)  ^{q+m}\frac{\left(
2m-1\right)  !!}{\left(  -M_{2}\right)  ^{m}}\left\vert N,0,0\right\rangle
\otimes\left\vert b_{-\frac{3}{2}}^{P}\right\rangle ,\label{SS1}\\
\left\vert N+1,2m+1,q\right\rangle \otimes\left\vert b_{-\frac{1}{2}}%
^{P}\right\rangle  &  =\left(  -\frac{1}{2M_{2}}\right)  ^{q+m}\frac{\left(
2m+1\right)  !!}{\left(  -M_{2}\right)  ^{m+1}}\left\vert N,0,0\right\rangle
\otimes\left\vert b_{-\frac{3}{2}}^{P}\right\rangle ,\label{SS2}\\
\left\vert N+1,2m,q\right\rangle \otimes\left\vert b_{-\frac{1}{2}}%
^{T}\right\rangle  &  =\left(  -\frac{1}{2M_{2}}\right)  ^{q+m}\frac{\left(
2m-1\right)  !!}{\left(  -M_{2}\right)  ^{m-1}}\left\vert N,0,0\right\rangle
\otimes\left\vert b_{-\frac{3}{2}}^{P}\right\rangle ,\label{SS3}\\
\left\vert N-1,2m,q-1\right\rangle \otimes\left\vert b_{-\frac{1}{2}}%
^{T}b_{-\frac{1}{2}}^{P}b_{-\frac{3}{2}}^{P}\right\rangle  &  =\left(
-\frac{1}{2M_{2}}\right)  ^{q+m}\frac{\left(  2m-1\right)  !!}{\left(
-M_{2}\right)  ^{m}}\left\vert N,0,0\right\rangle \otimes\left\vert
b_{-\frac{3}{2}}^{P}\right\rangle . \label{SS4}%
\end{align}

Note that, in order to simplify the notation, we have only shown the second
state of the four point functions to represent the scattering amplitudes on
both sides of each equation above. This notation will be used throughout the
paper whenever is necessary. Eqs.(\ref{SS1}) to (\ref{SS4}) are thus the SUSY
generalization of Eq.(\ref{mainA}) for the bosonic string. There are much more
high energy fermionic string scattering amplitudes other than states we will
consider in this chapter.

We stress that, in addition to high energy scatterings of string states with
polarizations orthogonal to the scattering plane considered previously in the
GR \cite{susy}, there are more high energy string scattering amplitudes with
more worldsheet fermionic operators $b_{-\frac{n}{2}}^{P,T}$ in the string vertex.

\subsection{Amplitude $\left\vert N,2m,q\right\rangle \otimes\left\vert
b_{-\frac{3}{2}}^{P}\right\rangle $}

The first RR scattering amplitude we want to calculate corresponding to state
in Eq.(\ref{L/3}) is%
\begin{align}
A_{1}^{(N,2m,q)}  &  =\langle\psi_{1}^{T^{1}}e^{-\phi_{1}}e^{ik_{1}X_{1}}%
\cdot(\partial X_{2}^{T})^{N-2m-2q}(\partial X_{2}^{L})^{2m}(\partial^{2}%
X_{2}^{L})^{q}\partial\psi_{2}^{P}e^{-\phi_{2}}e^{ik_{2}X_{2}}\nonumber\\
&  \text{ \ \ \ }\cdot k_{\lambda3}\psi_{3}^{\lambda}e^{ik_{3}X_{3}}\cdot
k_{\sigma4}\psi_{4}^{\sigma}e^{ik_{4}X_{4}}\rangle\label{3.1.}%
\end{align}
where we have dropped out an overall factor. In Eq.(\ref{3.1.}), the first
vertex is a vector state in the $(-)$ ghost picture, and the last two states
are tachyons in the $(0)$ ghost picture. The second state is a tensor in the
$(-)$ ghost picture, so that the total superconformal ghost charges sum up to
$-2$. The $s-t$ channel of the amplitude can be calculated to be
\begin{align}
A_{1}^{(N,2m,q)}  &  =\int_{0}^{1}dx\,x^{k_{1}\cdot k_{2}}(1-x)^{k_{2}\cdot
k_{3}}\left[  \frac{e^{T}\cdot k_{3}}{1-x}\right]  ^{N-2m-2q}\\
&  \cdot\left[  \frac{e^{P}\cdot k_{1}}{-x}+\frac{e^{P}\cdot k_{3}}%
{1-x}\right]  ^{2m}\left[  \frac{e^{P}\cdot k_{1}}{x^{2}}+\frac{e^{P}\cdot
k_{3}}{(1-x)^{2}}\right]  ^{q}\cdot\frac{1}{x}\label{3.3..}\\
&  \cdot\left\{  \langle\psi_{1}^{T^{1}}\partial\psi_{2}^{P}\rangle\langle
\psi_{3}^{\lambda}\psi_{4}^{\sigma}\rangle-\langle\psi_{1}^{T^{1}}\psi
_{3}^{\lambda}\rangle\langle\partial\psi_{2}^{P}\psi_{4}^{\sigma}%
\rangle+\langle\psi_{1}^{T^{1}}\psi_{4}^{\sigma}\rangle\langle\partial\psi
_{2}^{P}\psi_{3}^{\lambda}\rangle\right\}  k_{\lambda3}k_{\sigma
4}\label{3.4..}\\
&  \simeq\int_{0}^{1}dx\,x^{k_{1}\cdot k_{2}}(1-x)^{k_{2}\cdot k_{3}}\left[
\frac{e^{T}\cdot k_{3}}{1-x}\right]  ^{N-2m-2q}\\
&  \cdot\left[  \frac{e^{P}\cdot k_{1}}{-x}+\frac{e^{P}\cdot k_{3}}%
{1-x}\right]  ^{2m}\left[  \frac{e^{P}\cdot k_{3}}{(1-x)^{2}}\right]
^{q}\cdot\frac{1}{x}\frac{1}{M_{2}}\left[  -\frac{(e^{T}\cdot k_{4}%
)(k_{2}\cdot k_{3})}{(1-x)^{2}}\right]  .
\end{align}
In Eq.(\ref{3.3..}),$\frac{e^{P}\cdot k_{1}}{x^{2}}$ is of subleading order in
the RR and $\frac{1}{x}$ is the ghost contribution. The second term of
Eq.(\ref{3.4..}) vanishes due to the $SL(2,R)$ gauge fixing $x_{1}%
=0,x_{2}=x,x_{3}=1$ and $x_{4}=\infty.$ The first term of Eq.(\ref{3.4..})
vanishes due to $e^{T^{1}}\cdot e^{P^{2}}=0.$ The amplitude then reduces to%
\begin{align}
A_{1}^{(N,2m,q)}  &  \simeq\frac{\tilde{t}}{2M_{2}}(\sqrt{-{t}})^{N-2m-2q+1}%
\left(  \frac{\tilde{t}}{2M_{2}}\right)  ^{q}\int_{0}^{1}dx\,x^{k_{1}\cdot
k_{2}-1}(1-x)^{k_{2}\cdot k_{3}-N+2m-2}\nonumber\\
\cdot &  \sum_{j=0}^{2m}{\binom{2m}{j}}\left(  \frac{s}{2M_{2}x}\right)
^{j}\left(  \frac{-\tilde{t}}{2M_{2}(1-x)}\right)  ^{2m-j}\nonumber\\
&  =\frac{\tilde{t}}{2M_{2}}(\sqrt{-{t}})^{N-2m-2q+1}\left(  \frac{\tilde{t}%
}{2M_{2}}\right)  ^{2m+q}\nonumber\\
\cdot &  \sum_{j=0}^{2m}{\binom{2m}{j}}(-1)^{j}\left(  \frac{s}{\tilde{t}%
}\right)  ^{j}B\left(  k_{1}\cdot k_{2}-j,k_{2}\cdot k_{3}-N+j-1\right)  .
\end{align}
The Beta function above can be approximated in the large $s$, but fixed $t$
limit as follows
\begin{align}
&  B\left(  k_{1}\cdot k_{2}-j,k_{2}\cdot k_{3}+j-N-1\right) \nonumber\\
&  =B\left(  1-\frac{s}{2}+N-j,-\frac{1}{2}-\frac{t}{2}+j\right) \nonumber\\
&  =\frac{\Gamma(1-\frac{s}{2}+N-j)\Gamma(-\frac{1}{2}-\frac{t}{2}+j)}%
{\Gamma(\frac{u}{2}-1)}\nonumber\\
&  \approx B\left(  1-\frac{s}{2},-\frac{1}{2}-\frac{t}{2}\right)  \left(
1-\frac{s}{2}\right)  ^{N-j}\left(  \frac{u}{2}-1\right)  ^{-N}\left(
-\frac{1}{2}-\frac{t}{2}\right)  _{j}\nonumber\\
&  \approx B\left(  1-\frac{s}{2},-\frac{1}{2}-\frac{t}{2}\right)  \left(
-\frac{s}{2}\right)  ^{-j}\left(  -\frac{1}{2}-\frac{t}{2}\right)  _{j}%
\end{align}
where%
\begin{equation}
(a)_{j}=a(a+1)(a+2)...(a+j-1)
\end{equation}
is the Pochhammer symbol. The leading order amplitude in the RR can then be
written as%
\begin{align}
A_{1}^{(N,2m,q)}  &  \simeq\frac{\tilde{t}}{2M_{2}}B\left(  1-\frac{s}%
{2},-\frac{1}{2}-\frac{t}{2}\right)  \sqrt{-t}^{N-2m-2q+1}\left(  \frac
{1}{2M_{2}}\right)  ^{2m+q}\nonumber\\
\cdot &  (\tilde{t})^{2m+q}\sum_{j=0}^{2m}{\binom{2m}{j}}\left(  \frac
{2}{\tilde{t}}\right)  ^{j}\left(  -\frac{1}{2}-\frac{t}{2}\right)  _{j},
\label{A.}%
\end{align}
which is UV power-law behaved as expected. The summation in Eq. (\ref{A.}) can
be represented by the Kummer function of the second kind $U$ as follows,
\begin{equation}
\sum_{j=0}^{p}{\binom{p}{j}}\left(  \frac{2}{\tilde{t}}\right)  ^{j}\left(
-\frac{1}{2}-\frac{t}{2}\right)  _{j}=2^{p}(\tilde{t})^{-p}\ U\left(
-p,\frac{t}{2}-p+\frac{3}{2},\frac{\tilde{t}}{2}\right)  .
\end{equation}
Finally, the amplitudes can be written as%
\begin{align}
A_{1}^{(N,2m,q)}  &  \simeq B\left(  1-\frac{s}{2},-\frac{1}{2}-\frac{t}%
{2}\right)  \sqrt{-t}^{N-2m-2q+1}\left(  \frac{1}{2M_{2}}\right)
^{2m+q+1}\nonumber\\
\cdot &  2^{2m}(\tilde{t})^{q+1}U\left(  -2m\,,\,\frac{t}{2}-2m+\frac{3}%
{2}\,,\,\frac{\tilde{t}}{2}\right)  . \label{A1.}%
\end{align}

There are some important observations for the high energy amplitude in
Eq.(\ref{A1.}). First, the amplitude gives the universal power-law behavior
for string states at \textit{all} mass levels%
\begin{equation}
A_{1}^{(N,2m,q)}\sim s^{\alpha(t)}\text{ \ (in the RR)}%
\end{equation}
where
\begin{equation}
\alpha(t)=a_{0}+\alpha^{\prime}t\text{, \ }a_{0}=\frac{1}{2}\text{ and }%
\alpha^{\prime}=1/2.
\end{equation}
This generalizes the high energy behavior of the four massless vector
amplitude in the RR to string states at arbitrary mass levels. Second, the
amplitude gives the correct intercept $a_{0}=\frac{1}{2}$ of fermionic string.
Finally, the amplitude can be used to reproduce the ratios calculated in the
GR as we will see in section E.

\subsection{Amplitude $\left\vert N+1,2m+1,q\right\rangle \otimes\left\vert
b_{-\frac{1}{2}}^{P}\right\rangle $}

Note that this is the only case with odd integer $2m+1.$ The RR scattering
amplitude corresponding to state in Eq.(\ref{L/2}) can be written as%
\begin{align}
A_{2}^{(N+1,2m+1,q)}  &  =\langle\psi_{1}^{T^{1}}e^{-\phi_{1}}e^{ik_{1}X_{1}%
}\cdot(\partial X_{2}^{T})^{N-2m-2q}(\partial X_{2}^{L})^{2m+1}(\partial
^{2}X_{2}^{L})^{q}\psi_{2}^{P}e^{-\phi_{2}}e^{ik_{2}X_{2}}\nonumber\\
&  \text{ \ \ \ }\cdot k_{\lambda3}\psi_{3}^{\lambda}e^{ik_{3}X_{3}}\cdot
k_{\sigma4}\psi_{4}^{\sigma}e^{ik_{4}X_{4}}\rangle
\end{align}
where we have dropped out an overall factor. The amplitude can be calculated
to be%
\begin{align}
A_{2}^{(N+1,2m+1,q)}  &  =\int_{0}^{1}dx\,x^{k_{1}\cdot k_{2}}(1-x)^{k_{2}%
\cdot k_{3}}\left[  \frac{e^{T}\cdot k_{3}}{1-x}\right]  ^{N-2m-2q}\nonumber\\
&  \cdot\left[  \frac{e^{P}\cdot k_{1}}{-x}+\frac{e^{P}\cdot k_{3}}%
{1-x}\right]  ^{2m+1}\left[  \frac{e^{P}\cdot k_{1}}{x^{2}}+\frac{e^{P}\cdot
k_{3}}{(1-x)^{2}}\right]  ^{q}\cdot\frac{1}{x}\nonumber\\
&  \cdot\left\{  \langle\psi_{1}^{T^{1}}\psi_{2}^{P}\rangle\langle\psi
_{3}^{\lambda}\psi_{4}^{\sigma}\rangle-\langle\psi_{1}^{T^{1}}\psi
_{3}^{\lambda}\rangle\langle\psi_{2}^{P}\psi_{4}^{\sigma}\rangle+\langle
\psi_{1}^{T^{1}}\psi_{4}^{\sigma}\rangle\langle\psi_{2}^{P}\psi_{3}^{\lambda
}\rangle\right\}  k_{\lambda3}k_{\sigma4}\nonumber\\
&  =\int_{0}^{1}dx\,x^{k_{1}\cdot k_{2}}(1-x)^{k_{2}\cdot k_{3}}\left[
\frac{e^{T}\cdot k_{3}}{1-x}\right]  ^{N-2m-2q}\left[  \frac{e^{P}\cdot k_{1}%
}{-x}+\frac{e^{P}\cdot k_{3}}{1-x}\right]  ^{2m+1}\nonumber\\
&  \cdot\left[  \frac{e^{P}\cdot k_{3}}{(1-x)^{2}}\right]  ^{q}\frac{1}%
{x}\frac{1}{M_{2}}\left[  (e^{T^{1}}\cdot k_{3})(k_{2}\cdot k_{4}%
)-\frac{(e^{T^{1}}\cdot k_{4})(k_{2}\cdot k_{3})}{1-x}\right] \nonumber\\
&  \simeq\left(  -1\right)  ^{N}\left[  \sqrt{-{t}}\right]  ^{N-2m-2q+1}%
\left(  -\frac{1}{2M_{2}}\right)  ^{2m+q+2}\tilde{t}^{2m+q+1}\sum_{j=0}%
^{2m+1}\binom{2m+1}{j}\left(  -\frac{s}{\tilde{t}}\right)  ^{j}\nonumber\\
\cdot &  \left[
\begin{array}
[c]{c}%
-\left(  s+t+1\right)  \int_{0}^{1}dx\,x^{k_{1}\cdot k_{2}-j-1}(1-x)^{k_{2}%
\cdot k_{3}-N+j-1}\\
+\tilde{t}\int_{0}^{1}dx\,x^{k_{1}\cdot k_{2}-j-1}(1-x)^{k_{2}\cdot
k_{3}-N+j-2}%
\end{array}
\right] \nonumber\\
&  \simeq\left[  \sqrt{-{t}}\right]  ^{N-2m-2q+1}\left(  \frac{1}{2M_{2}%
}\right)  ^{2m+q+2}\tilde{t}^{2m+q+1}\sum_{j=0}^{2m+1}\binom{2m+1}{j}\left(
-\frac{s}{\tilde{t}}\right)  ^{j}\nonumber\\
\cdot &  \left[
\begin{array}
[c]{c}%
-\left(  s+t+1\right)  B\left(  k_{1}\cdot k_{2}-j,k_{2}\cdot k_{3}-N+j\right)
\\
+\tilde{t}B\left(  k_{1}\cdot k_{2}-j,k_{2}\cdot k_{3}-N+j-1\right)
\end{array}
\right]  .
\end{align}
We then do an approximation for beta function similar to the calculation for
$A_{1}^{(N,2m,q)}$ and end up with
\begin{align}
A_{2}^{(N+1,2m+1,q)}  &  \simeq B\left(  1-\frac{s}{2},-\dfrac{1}{2}-\frac
{t}{2}\right)  \left[  \sqrt{-{t}}\right]  ^{N-2m-2q+1}\left(  \frac{1}%
{2M_{2}}\right)  ^{2m+q+2}\tilde{t}^{2m+q+1}\nonumber\\
&  \cdot\sum_{j=0}^{2m+1}\binom{2m+1}{j}\left[  \left(  1+t\right)  \left(
\frac{2}{\tilde{t}}\right)  ^{j}\left(  \dfrac{1}{2}-\frac{t}{2}\right)
_{j}-\tilde{t}\left(  \frac{2}{\tilde{t}}\right)  ^{j}\left(  -\dfrac{1}%
{2}-\frac{t}{2}\right)  _{j}\right] \nonumber\\
&  \simeq B\left(  1-\frac{s}{2},-\dfrac{1}{2}-\frac{t}{2}\right)  \left[
\sqrt{-{t}}\right]  ^{N-2m-2q+1}\left(  \frac{1}{2M_{2}}\right)
^{2m+q+2}2^{2m+1}\left(  \tilde{t}\right)  ^{q}\nonumber\\
&  \cdot\left[  (1+t)U\left(  -1-2m,\frac{t}{2}-2m-\frac{1}{2},\frac{\tilde
{t}}{2}\right)  -\tilde{t}U\left(  -1-2m,\frac{t}{2}-2m+\frac{1}{2}%
,\frac{\tilde{t}}{2}\right)  \right]  . \label{Odd}%
\end{align}
Note that there are two terms in Eq.(\ref{Odd}), and the first argument of the
$U$ function $a=-1-2m$ is odd. These differences will make the calculation of
the ratios in the next section more complicated. Finally, the amplitude gives
the universal power-law behavior for string states at \textit{all} mass levels
with the correct intercept $a_{0}=\frac{1}{2}$ of fermionic string.

\subsection{Amplitude $\left\vert N+1,2m,q\right\rangle \otimes\left\vert
b_{-\frac{1}{2}}^{T}\right\rangle $}

The third RR scattering amplitude corresponding to state in Eq.(\ref{T/2}) is%
\begin{align}
A_{3}^{\left(  N+1,2m,q\right)  }  &  =\langle\psi_{1}^{T^{1}}e^{-\phi_{1}%
}e^{ik_{1}X_{1}}\cdot(\partial X_{2}^{T})^{N-2m-2q+1}(\partial X_{2}^{L}%
)^{2m}(\partial^{2}X_{2}^{L})^{q}\psi_{2}^{T}e^{-\phi_{2}}e^{ik_{2}X_{2}%
}\nonumber\\
&  \text{ \ \ \ }\cdot k_{\lambda3}\psi_{3}^{\lambda}e^{ik_{3}X_{3}}\cdot
k_{\sigma4}\psi_{4}^{\sigma}e^{ik_{4}X_{4}}\rangle
\end{align}
where we have dropped out an overall factor. The scattering amplitude can be
calculated to be%
\begin{align}
A_{3}^{\left(  N+1,2m,q\right)  }  &  =\int_{0}^{1}dx\,x^{k_{1}\cdot k_{2}%
}(1-x)^{k_{2}\cdot k_{3}}\left[  \frac{e^{T}\cdot k_{3}}{1-x}\right]
^{N-2m-2q+1}\nonumber\\
&  \cdot\left[  \frac{e^{P}\cdot k_{1}}{-x}+\frac{e^{P}\cdot k_{3}}%
{1-x}\right]  ^{2m}\left[  \frac{e^{P}\cdot k_{1}}{x^{2}}+\frac{e^{P}\cdot
k_{3}}{(1-x)^{2}}\right]  ^{q}\cdot\frac{1}{x}\nonumber\\
&  \cdot\left\{  \langle\psi_{1}^{T^{1}}\psi_{2}^{T}\rangle\langle\psi
_{3}^{\lambda}\psi_{4}^{\sigma}\rangle-\langle\psi_{1}^{T^{1}}\psi
_{3}^{\lambda}\rangle\langle\psi_{2}^{T}\psi_{4}^{\sigma}\rangle+\langle
\psi_{1}^{T^{1}}\psi_{4}^{\sigma}\rangle\langle\psi_{2}^{T}\psi_{3}^{\lambda
}\rangle\right\}  k_{\lambda3}k_{\sigma4}\nonumber\\
&  =\int_{0}^{1}dx\,x^{k_{1}\cdot k_{2}}(1-x)^{k_{2}\cdot k_{3}}\left[
\frac{e^{T}\cdot k_{3}}{1-x}\right]  ^{N-2m-2q+1}\left[  \frac{e^{P}\cdot
k_{1}}{-x}+\frac{e^{P}\cdot k_{3}}{1-x}\right]  ^{2m}\left[  \frac{e^{P}\cdot
k_{3}}{(1-x)^{2}}\right]  ^{q}\nonumber\\
&  \cdot\frac{1}{x}\left[  \frac{(e^{T^{1}}\cdot e^{T})(k_{3}\cdot k_{4})}%
{-x}+(e^{T^{1}}\cdot k_{3})(e^{T}\cdot k_{4})-\frac{(e^{T^{1}}\cdot
k_{4})(e^{T}\cdot k_{3})}{1-x}\right] \nonumber\\
&  \simeq\left[  \sqrt{-{t}}\right]  ^{N-2m-2q+1}\left(  \frac{1}{2M_{2}%
}\right)  ^{2m+q}\tilde{t}^{2m+q}\sum_{j=0}^{2m}\binom{2m}{j}\left(  -\frac
{s}{\tilde{t}}\right)  ^{j}\nonumber\\
&  \cdot\left[  -\frac{s}{2}\int_{0}^{1}dx\,x^{k_{1}\cdot k_{2}-j-2}%
(1-x)^{k_{2}\cdot k_{3}-N+j-1}+t\int_{0}^{1}dx\,x^{k_{1}\cdot k_{2}%
-j}(1-x)^{k_{2}\cdot k_{3}-N+j-2}\right] \nonumber\\
&  \simeq\left[  \sqrt{-{t}}\right]  ^{N-2m-2q+1}\left(  \frac{1}{2M_{2}%
}\right)  ^{2m+q}\tilde{t}^{2m+q}\sum_{j=0}^{2m}\binom{2m}{j}\left(  -\frac
{s}{\tilde{t}}\right)  ^{j}\nonumber\\
&  \cdot\left[
\begin{array}
[c]{c}%
-\frac{s}{2}B\left(  k_{1}\cdot k_{2}-j-1,k_{2}\cdot k_{3}-N+j\right) \\
+tB\left(  k_{1}\cdot k_{2}-j+1,k_{2}\cdot k_{3}-N+j-1\right)
\end{array}
\right]  .
\end{align}
We then do an approximation for beta function similar to the calculation for
$A_{1}^{(N,2m,q)}$ and end up with%

\begin{align}
&  \simeq-B\left(  1-\frac{s}{2},-\dfrac{1}{2}-\frac{t}{2}\right)  \left[
\sqrt{-{t}}\right]  ^{N-2m-2q+1}\left(  \frac{1}{2M_{2}}\right)  ^{2m+q}%
\tilde{t}^{2m+q}\nonumber\\
&  \cdot\sum_{j=0}^{2m}\binom{2m}{j}\left[  \dfrac{\left(  1+t\right)  }%
{2}\left(  \frac{2}{\tilde{t}}\right)  ^{j}\left(  \dfrac{1}{2}-\frac{t}%
{2}\right)  _{j}-t\left(  \frac{2}{\tilde{t}}\right)  ^{j}\left(  -\dfrac
{1}{2}-\frac{t}{2}\right)  _{j}\right] \nonumber\\
&  \simeq-B\left(  1-\frac{s}{2},-\dfrac{1}{2}-\frac{t}{2}\right)  \left[
\sqrt{-{t}}\right]  ^{N-2m-2q+1}\left(  \frac{1}{2M_{2}}\right)
^{2m+q}2^{2m-1}\tilde{t}^{q}\nonumber\\
&  \cdot\left[  \left(  1+t\right)  U\left(  -2m,\frac{t}{2}-2m+\frac{1}%
{2},\frac{\tilde{t}}{2}\right)  -2tU\left(  -2m,\frac{t}{2}-2m+\frac{3}%
{2},\frac{\tilde{t}}{2}\right)  \right]  .
\end{align}
In this case there are again two terms as in the amplitude $A_{2}$ but with an
even argument $a=-2m$. Finally, the amplitude gives the universal power-law
behavior for string states at \textit{all} mass levels with the correct
intercept $a_{0}=\frac{1}{2}$ of fermionic string.

\subsection{Amplitude $\left\vert N-1,2m,q-1\right\rangle \otimes\left\vert
b_{-\frac{1}{2}}^{T}b_{-\frac{1}{2}}^{P}b_{-\frac{3}{2}}^{P}\right\rangle $}

The fourth RR scattering amplitude corresponding to state in Eq.(\ref{TLL/3})
is%
\begin{align}
A_{4}^{\left(  N-1,2m,q-1\right)  }  &  =\langle\psi_{1}^{T^{1}}e^{-\phi_{1}%
}e^{ik_{1}X_{1}}\cdot(\partial X_{2}^{T})^{N-2m-2q}(\partial X_{2}^{L}%
)^{2m}(\partial^{2}X_{2}^{L})^{q-1}\psi_{2}^{T}\psi_{2}^{P}\partial\psi
_{2}^{P}e^{-\phi_{2}}e^{ik_{2}X_{2}}\nonumber\\
&  \text{ \ \ \ }\cdot k_{\lambda3}\psi_{3}^{\lambda}e^{ik_{3}X_{3}}\cdot
k_{\sigma4}\psi_{4}^{\sigma}e^{ik_{4}X_{4}}\rangle
\end{align}
where we have dropped out an overall factor. The scattering amplitude can be
calculated to be%

\begin{align}
A_{4}^{\left(  N-1,2m,q-1\right)  }  &  =\int_{0}^{1}dx\,x^{k_{1}\cdot k_{2}%
}(1-x)^{k_{2}\cdot k_{3}}\left[  \frac{e^{T}\cdot k_{3}}{1-x}\right]
^{N-2m-2q}\nonumber\\
&  \cdot\left[  \frac{e^{P}\cdot k_{1}}{-x}+\frac{e^{P}\cdot k_{3}}%
{1-x}\right]  ^{2m}\left[  \frac{e^{P}\cdot k_{1}}{x^{2}}+\frac{e^{P}\cdot
k_{3}}{(1-x)^{2}}\right]  ^{q-1}\frac{1}{x}\nonumber\\
&  \cdot\langle\psi_{1}^{T^{1}}\psi_{2}^{T}\rangle\langle\psi_{2}^{P}\psi
_{4}^{\sigma}\rangle\langle\partial\psi_{2}^{P}\psi_{3}^{\lambda}\rangle
k_{\lambda3}k_{\sigma4}\nonumber\\
&  \simeq\int_{0}^{1}dx\,x^{k_{1}\cdot k_{2}}(1-x)^{k_{2}\cdot k_{3}}\left[
\frac{e^{T}\cdot k_{3}}{1-x}\right]  ^{N-2m-2q}\left[  \frac{e^{P}\cdot k_{1}%
}{-x}+\frac{e^{P}\cdot k_{3}}{1-x}\right]  ^{2m}\nonumber\\
&  \cdot\left[  \frac{e^{P}\cdot k_{3}}{(1-x)^{2}}\right]  ^{q-1}\frac{1}%
{x}\frac{1}{M_{2}^{2}}\left[  \frac{(e^{T^{1}}\cdot e^{T})(k_{2}\cdot
k_{4})(k_{2}\cdot k_{3})}{(1-x)^{2}}\right] \nonumber\\
&  \simeq\left[  \sqrt{-{t}}\right]  ^{N-2m-2q}\left(  \frac{1}{2M_{2}%
}\right)  ^{2m+q+1}\tilde{t}^{2m+q}s\nonumber\\
&  \cdot\sum_{j=0}^{2m}\binom{2m}{j}\left(  -\frac{s}{\tilde{t}}\right)
^{j}\int_{0}^{1}dx\,x^{k_{1}\cdot k_{2}-j}(1-x)^{k_{2}\cdot k_{3}%
-N+j-1}\nonumber\\
&  \simeq\left[  \sqrt{-{t}}\right]  ^{N-2m-2q}\left(  \frac{1}{2M_{2}%
}\right)  ^{2m+q+1}\tilde{t}^{2m+q}s\nonumber\\
&  \cdot\sum_{j=0}^{2m}\binom{2m}{j}\left(  -\frac{s}{\tilde{t}}\right)
^{j}B\left(  k_{1}\cdot k_{2}-j+1,k_{2}\cdot k_{3}-N+j\right)  .
\end{align}
With a similar approximation for the beta function, we get%
\begin{align}
A_{4}^{\left(  N-1,2m,q-1\right)  }  &  \simeq B\left(  1-\frac{s}{2}%
,-\dfrac{1}{2}-\frac{t}{2}\right)  \left[  \sqrt{-{t}}\right]  ^{N-2m-2q}%
\left(  \frac{1}{2M_{2}}\right)  ^{2m+q+1}\tilde{t}^{2m+q}\nonumber\\
&  \cdot\left(  1+t\right)  \sum_{j=0}^{2m}\binom{2m}{j}\left(  -\frac
{2}{\tilde{t}}\right)  ^{j}\left(  \dfrac{1}{2}-\frac{t}{2}\right)
_{j}\nonumber\\
&  =B\left(  1-\frac{s}{2},-\dfrac{1}{2}-\frac{t}{2}\right)  \left[
\sqrt{-{t}}\right]  ^{N-2m-2q}\left(  \frac{1}{2M_{2}}\right)  ^{2m+q+1}%
\nonumber\\
&  \cdot2^{2m}(\tilde{t})^{q}\left(  1+t\right)  U\left(  -2m,\frac{t}%
{2}-2m+\frac{1}{2},\frac{\tilde{t}}{2}\right)  .
\end{align}
Again the amplitude gives the universal power-law behavior for string states
at \textit{all} mass levels with the correct intercept $a_{0}=\frac{1}{2}$ of
fermionic string. In the next section we are going to use the four amplitudes
calculated in this section to extract ratios calculated in the fixed angle regime.

\subsection{Reproducing ratios among hard SUSY scattering amplitudes}

In the bosonic string calculation discussed in chapter XI\cite{bosonic}, we
learned that the relative coefficients of the highest power $t$ terms in the
leading order amplitudes in the RR can be used to reproduce the ratios of the
amplitudes in the GR for each fixed mass level. In this section, we are going
to generalize the calculation \cite{RRsusy} to four classes of fermionic
string states for arbitrary mass levels. We begin with the first amplitude of
Eq.(\ref{A1.}).

\subsubsection{Ratios for $\left\vert N,2m,q\right\rangle \otimes\left\vert
b_{-\frac{3}{2}}^{P}\right\rangle $}

It is important to note that there are no linear relations among high energy
string scattering amplitudes, Eq.(\ref{A1.}), of different string states for
each fixed mass level in the RR. In other words, the ratios $A_{1}%
^{(N,2m,q)}/A_{1}^{(N,0,0)}$ are $t$-dependent functions and can be calculated
to be
\begin{align}
\frac{A_{1}^{(N,2m,q)}}{A_{1}^{(N,0,0)}}  &  =\left(  -\frac{1}{2M_{2}%
}\right)  ^{2m+q}(-)^{m}(\tilde{t}+2N+1)^{-m-q}(\tilde{t})^{2m+q}\nonumber\\
&  \cdot\sum_{j=0}^{2m}(-2m)_{j}\left(  -N-1-\frac{\tilde{t}}{2}\right)
_{j}\frac{(-2/\tilde{t})^{j}}{j!} \label{4.1}%
\end{align}
where we have used Eq.(\ref{ttt}) to replace $t$ by $\tilde{t}$. If the
leading order coefficients in Eq.(\ref{4.1}) extracted from the amplitudes in
the RR are to be identified with the ratios calculated in the GR in
Eq.(\ref{SS1}), we need the following identity
\begin{align}
&  \sum_{j=0}^{2m}(-2m)_{j}\left(  -L-\frac{\tilde{t}}{2}\right)  _{j}%
\frac{(-2/\tilde{t})^{j}}{j!}\label{4.2}\\
&  =0(-\tilde{t})^{0}+0(-\tilde{t})^{-1}+...+0(-\tilde{t})^{-m+1}+\frac
{(2m)!}{m!}(-\tilde{t})^{-m}+\mathit{O}\left\{  \left(  \frac{1}{\tilde{t}%
}\right)  ^{m+1}\right\}  \label{4.3}%
\end{align}
where $L=N+1$ and is an integer. This identity was proved in \cite{LYAM}. The
coefficients of the terms $\mathit{O}\left\{  \left(  1/\tilde{t}\right)
^{m+1}\right\}  $ in Eq.(\ref{4.3}) is irrelevant for string amplitudes. We
thus have shown that high energy superstring scattering amplitudes
$A_{1}^{(N,2m,q)}$ of Eq.(\ref{A1.}) in the RR can be used to extract the
ratios $T_{1}^{(N,2m,q)}/T_{1}^{(N,0,0)}$ of Eq.(\ref{SS1}) in the GR by using
the Stirling number identities. That is%
\begin{align}
\lim_{t\rightarrow\infty}\frac{A_{1}^{(N,2m,q)}}{A_{1}^{(N,0,0)}}  &
=\lim_{t\rightarrow\infty}\left(  -\frac{1}{2M_{2}}\right)  ^{2m+q}%
2^{2m}(-t)^{m+2q}U\left(  -2m\,,\,\frac{t}{2}-2m+\frac{3}{2}\,,\,\frac{t}%
{2}\right) \nonumber\\
&  =\left(  -\frac{1}{2M_{2}}\right)  ^{q+m}\frac{\left(  2m-1\right)
!!}{\left(  -M_{2}\right)  ^{m}}=\frac{T_{1}^{(N,2m,q)}}{T_{1}^{(N,0,0)}}.
\end{align}

\subsubsection{Ratios for $\left\vert N+1,2m+1,q\right\rangle \otimes
\left\vert b_{-\frac{1}{2}}^{P}\right\rangle $}

The ratios $A_{2}^{(N+1,2m+1,q)}/A_{1}^{(N,0,0)}$ can be calculated to be%
\begin{align}
\frac{A_{2}^{(N+1,2m+1,q)}}{A_{1}^{(N,0,0)}}  &  =\left(  -\frac{1}{2M_{2}%
}\right)  ^{2m+q+1}\left(  -\tilde{t}\right)  ^{m}\cdot\left[  (1+t)\sum
_{j=0}^{2m+1}{\binom{1+2m}{j}}\left(  \frac{2}{\tilde{t}}\right)  ^{j}\left(
\frac{1}{2}-\frac{t}{2}\right)  _{j}\right. \nonumber\\
&  \left.  -\tilde{t}\sum_{j=0}^{2m+1}{\binom{1+2m}{j}}\left(  \frac{2}%
{\tilde{t}}\right)  ^{j}\left(  -\frac{1}{2}-\frac{t}{2}\right)  _{j}\right]
.
\end{align}
The bracket in the above equation can be simplified by dropping out the
subleading order terms in the calculation, and one obtains%
\begin{align}
&  (1+t)\sum_{j=0}^{2m+1}{\binom{2m+1}{j}}\left(  \frac{2}{\tilde{t}}\right)
^{j}\left(  \frac{1}{2}-\frac{t}{2}\right)  _{j}-\tilde{t}\sum_{j=0}%
^{2m+1}{\binom{2m+1}{j}}\left(  \frac{2}{\tilde{t}}\right)  ^{j}\left(
-\frac{1}{2}-\frac{t}{2}\right)  _{j}\nonumber\\
&  =(1+t)\sum_{j=0}^{2m+1}(-2m-1)_{j}\left(  -N-\frac{\tilde{t}}{2}\right)
_{j}\frac{(-2/\tilde{t})^{j}}{j!}\nonumber\\
&  -\tilde{t}\sum_{j=0}^{2m+1}(-2m-1)_{j}\left(  -N-1-\frac{\tilde{t}}%
{2}\right)  _{j}\frac{(-2/\tilde{t})^{j}}{j!}\nonumber\\
&  \approx\tilde{t}\sum_{j=0}^{2m+1}(-2m-1)_{j}\left(  -N-\frac{\tilde{t}}%
{2}\right)  _{j}\frac{(-2/\tilde{t})^{j}}{j!}-\tilde{t}\sum_{j=0}%
^{2m+1}(-2m-1)_{j}\left(  -N-1-\frac{\tilde{t}}{2}\right)  _{j}\frac
{(-2/\tilde{t})^{j}}{j!}\nonumber\\
&  =2\left(  2m+1\right)  \cdot\sum_{j=1}^{2m+1}(-2m)_{j-1}\left(
-N-\frac{\tilde{t}}{2}\right)  _{j-1}\frac{(-2/\tilde{t})^{j-1}}{\left(
j-1\right)  !}\nonumber\\
&  =2\left(  2m+1\right)  \cdot\sum_{j=0}^{2m}(-2m)_{j}\left(  -N-\frac
{\tilde{t}}{2}\right)  _{j}\frac{(-2/\tilde{t})^{j}}{j!}%
\end{align}
where we have dropped out the subleading order terms in the second equality of
the calculation. Finally, the ratios can be calculated to be
\begin{align}
\frac{A_{2}^{(N+1,2m+1,q)}}{A_{1}^{(N,0,0)}}  &  =\left(  -\frac{1}{2M_{2}%
}\right)  ^{2m+q+1}\left(  -\tilde{t}\right)  ^{m}\cdot\left[  (1+t)\sum
_{j=0}^{2m+1}{\binom{1+2m}{j}}\left(  \frac{2}{\tilde{t}}\right)  ^{j}\left(
\frac{1}{2}-\frac{t}{2}\right)  _{j}\right. \nonumber\\
&  \left.  -\tilde{t}\sum_{j=0}^{2m+1}{\binom{1+2m}{j}}\left(  \frac{2}%
{\tilde{t}}\right)  ^{j}\left(  -\frac{1}{2}-\frac{t}{2}\right)  _{j}\right]
\nonumber\\
&  \simeq\left(  -\frac{1}{2M_{2}}\right)  ^{2m+q+1}\left(  -\tilde{t}\right)
^{m}2\left(  2m+1\right)  \sum_{j=0}^{2m}(-2m)_{j}\left(  -N-1-\frac{\tilde
{t}}{2}\right)  _{j}\frac{(-2/\tilde{t})^{j}}{j!}. \label{RatioB}%
\end{align}
By using the identity Eq.(\ref{4.3}), one can show that the leading order
coefficients in Eq.(\ref{RatioB}) can be identified with the ratios calculated
in the GR in Eq.(\ref{SS2}). That is%
\begin{equation}
\lim_{t\rightarrow\infty}\frac{A_{2}^{(N+1,2m+1,q)}}{A_{1}^{(N,0,0)}}%
=\frac{T_{2}^{(N+1,2m+1,q)}}{T_{1}^{(N,0,0)}}.
\end{equation}
In the calculation for this case, it is crucial to reduce the upper limit of
the summation $2m+1$ to $2m$ in Eq.(\ref{RatioB}). Otherwise, the identity
Eq.(\ref{4.3}) will not be applicable. It is remarkable to see that the
leading order coefficients of Eq.(\ref{RatioB}) can be identified with ratios
of Eq.(\ref{SS2}) in the GR.

\subsubsection{Ratios for $\left\vert N+1,2m,q\right\rangle \otimes\left\vert
b_{-\frac{1}{2}}^{T}\right\rangle $}

The ratios $A_{3}^{(N+1,2m,q)}/A_{1}^{(N,0,0)}$ can be calculated to be%
\begin{equation}
\frac{A_{3}^{(N+1,2m,q)}}{A_{1}^{(N,0,0)}}=\frac{1}{2}\left(  -\frac{1}%
{2M_{2}}\right)  ^{2m+q-1}(-\tilde{t})^{m}\sum_{j=0}^{2m}(-2m)_{j}\left(
-N-1-\frac{\tilde{t}}{2}\right)  _{j}\frac{(-2/\tilde{t})^{j}}{j!}.
\label{RatioC1}%
\end{equation}
By using the identity Eq.(\ref{4.3}), one can show that the leading order
coefficients in Eq.(\ref{RatioC1}) can be identified with the ratios
calculated in the GR in Eq.(\ref{SS3}). That is%
\begin{equation}
\lim_{t\rightarrow\infty}\frac{A_{3}^{(N+1,2m,q)}}{A_{1}^{(N,0,0)}}%
=\frac{T_{3}^{(N+1,2m,q)}}{T_{1}^{(N,0,0)}}.
\end{equation}

\subsubsection{Ratios for $\left\vert N-1,2m,q-1\right\rangle \otimes
\left\vert b_{-\frac{1}{2}}^{T}b_{-\frac{1}{2}}^{P}b_{-\frac{3}{2}}%
^{P}\right\rangle $}

The ratios $A_{4}^{(N-1,2m,q-1)}/A_{1}^{(N,0,0)}$ can be calculated to be%
\begin{equation}
\frac{A_{4}^{(N+1,2m+1,q)}}{A_{1}^{(N,0,0)}}=\left(  -\frac{1}{2M_{2}}\right)
^{2m+q}(-\tilde{t})^{m}\sum_{j=0}^{2m}(-2m)_{j}\left(  -N-1-\frac{\tilde{t}%
}{2}\right)  _{j}\frac{(-2/\tilde{t})^{j}}{j!}. \label{RatioD1}%
\end{equation}
By using the identity Eq.(\ref{4.3}), one can show that the leading order
coefficients in Eq.(\ref{RatioD1}) can be identified with the ratios
calculated in the GR in Eq.(\ref{SS4}). That is%
\begin{equation}
\lim_{t\rightarrow\infty}\frac{A_{4}^{(N+1,2m+1,q)}}{A_{1}^{(N,0,0)}}%
=\frac{T_{4}^{(N+1,2m+1,q)}}{T_{1}^{(N,0,0)}}.
\end{equation}

We thus have succeeded in extracting the Ratios of high energy superstring
scattering amplitudes in the GR from the high energy superstring scattering
amplitudes in the RR. In the next section, we will study the subleading order amplitudes.

\subsubsection{Subleading order amplitudes}

In this section, we calculate the next few subleading order amplitudes in the
RR for the mass levels $M_{2}^{2}=2(N+1)=4,6.$ The calculation for $M_{2}%
^{2}=8$ can be found in \cite{RRsusy}. The relevant kinematic can be found in
the Appendix \ref{RR Kinematic}. We will see that the ratios derived in the
previous section persist to subleading order amplitudes in the RR. For the
even mass levels with $(N+1)=\frac{M_{2}^{2}}{2}$= odd, we conjecture and give
evidences that the existence of these ratios in the RR persists to all orders
in the Regge expansion of all high energy string scattering amplitudes . For
the odd mass levels with $(N+1)=\frac{M_{2}^{2}}{2}$= even, the existence of
these ratios will show up only in the first $\frac{N+1}{2}+1$ terms in the
Regge expansion of the amplitudes. For the mass level $M_{2}^{2}=4$, there are
three states for Eq.(\ref{T/2}), and we obtain the subleading order expansions
as follows.
\begin{align}
|2,0,0\rangle|b_{-\frac{1}{2}}^{T}\rangle &  \rightarrow(\frac{1}{4}%
t^{2}-\frac{1}{4}t)s+(\frac{1}{4}t^{3}+\frac{9}{4}t^{2}+\frac{7}{4}t-\frac
{5}{4})s^{0}\nonumber\\
&  +(\frac{5}{2}t^{3}+18t^{2}+\frac{39}{2}t+4)s^{-1}+O[s^{-2}],
\end{align}%
\begin{align}
|2,2,0\rangle|b_{-\frac{1}{2}}^{T}\rangle &  \rightarrow(\frac{1}{32}%
t^{2}+\frac{1}{8}t+\frac{19}{32})s+(\frac{1}{32}t^{3}+\frac{23}{32}t^{2}%
+\frac{35}{32}t-\frac{19}{32})s^{0}\nonumber\\
&  +(\frac{3}{4}t^{3}-\frac{13}{4}t^{2}-\frac{39}{4}t-\frac{23}{4}%
)s^{-1}+O[s^{-2}],\\
|2,0,1\rangle|b_{-\frac{1}{2}}^{T}\rangle &  \rightarrow(-\frac{1}{16}%
t^{2}-\frac{1}{4}t+\frac{5}{16})s+(-\frac{1}{16}t^{3}-\frac{15}{16}t^{2}%
-\frac{27}{16}t-\frac{29}{16})s^{0}\nonumber\\
&  +(-\frac{3}{4}t^{3}-\frac{17}{4}t^{2}-\frac{45}{4}t-\frac{31}{4}%
)s^{-1}+O[s^{-2}].
\end{align}
In order to simply the notation in the above equations, we have only shown the
second state of the four-point functions in the correction functions to
represent the scattering amplitudes on the left hand side of each equation. We
find that the ratios of the leading order coefficients of $st^{2}$ are
$\frac{1}{4}:\frac{1}{32}:-\frac{1}{16}$, and it is easy to check that these
are the same as the ratios in the fixed angle limit. Moreover, the ratios
persist in the second subleading order terms $s^{0}t^{3}$ as $\frac{1}%
{4}:\frac{1}{32}:-\frac{1}{16}$. The ratios terminate to this order. We can
also compare the ratios among different worldsheet fermionic states but with
the same mass level $M_{2}^{2}=4$. We have the expansions:
\begin{align}
|2,1,0\rangle|b_{-\frac{1}{2}}^{L}\rangle &  \rightarrow(\frac{1}{16}%
t^{2}-\frac{7}{16}t)s+(\frac{1}{16}t^{3}-\frac{29}{16}t^{2}-\frac{49}%
{16}t-\frac{35}{16})s^{0}\nonumber\\
&  +(-\frac{7}{4}t^{3}-\frac{67}{4}t^{2}-\frac{117}{4}t-\frac{57}{4}%
)s^{-1}+O[s^{-2}],\\
|1,0,0\rangle|b_{-\frac{3}{2}}^{L}\rangle &  \rightarrow(-\frac{1}{8}%
t^{2}-\frac{5}{8}t)s+(-\frac{1}{8}t^{3}-\frac{17}{8}t^{2}-\frac{33}{8}%
t-\frac{25}{8})s^{0}\nonumber\\
&  +(-\frac{7}{4}t^{3}-\frac{61}{4}t^{2}-\frac{109}{4}t-\frac{55}{4}%
)s^{-1}+O[s^{-2}],\\
|0,0,0\rangle|b_{-\frac{1}{2}}^{T}b_{-\frac{1}{2}}^{L}b_{-\frac{3}{2}}%
^{L}\rangle &  \rightarrow(\frac{1}{32}t^{2}+\frac{3}{16}t+\frac{5}%
{32})s+(\frac{1}{32}t^{3}+\frac{15}{32}t^{2}+\frac{27}{32}t+\frac{13}%
{32})s^{0}\nonumber\\
&  +(\frac{1}{2}t^{3}+\frac{7}{2}t^{2}+\frac{11}{2}t+\frac{5}{2}%
)s^{-1}+O[s^{-2}].
\end{align}
The ratios of the leading order coefficients are proportional to that of state
$|2,0,0\rangle|b_{-\frac{1}{2}}^{T}\rangle,$ and can be calculated to be
\begin{equation}
|2,0,0\rangle|b_{-\frac{1}{2}}^{T}\rangle:|2,1,0\rangle|b_{-\frac{1}{2}}%
^{L}\rangle:|1,0,0\rangle|b_{-\frac{3}{2}}^{L}\rangle:|0,0,0\rangle
|b_{-\frac{1}{2}}^{T}b_{-\frac{1}{2}}^{L}b_{-\frac{3}{2}}^{L}\rangle=\frac
{1}{4}:\frac{1}{16}:-\frac{1}{8}:\frac{1}{32}.
\end{equation}
They again match with the ratios in the fixed angle limit. One can also find
that the second subleading order ratios are the same $\frac{1}{4}:\frac{1}%
{16}:-\frac{1}{8}:\frac{1}{32}$. Again the ratios terminate to this order.

For the mass level $M_{2}^{2}=6$, there are three states in Eq.(\ref{T/2}). We
again calculate the subleading order expansions. Interestingly, in this case
the ratios of the coefficients seem to be the same in all orders as can be
seen in the following:
\begin{align}
|3,0,0\rangle|b_{-\frac{1}{2}}^{T}\rangle &  \rightarrow\sqrt{-t}(\frac{1}%
{8}t^{2}-\frac{1}{8}t)s^{2}+\sqrt{-t}(\frac{3}{16}t^{3}+\frac{25}{16}%
t^{2}+\frac{25}{16}t-\frac{21}{16})s\nonumber\\
&  +\sqrt{-t}(\frac{3}{64}t^{4}+\frac{197}{64}t^{3}+\frac{625}{32}t^{2}%
+\frac{743}{32}t+\frac{411}{64})s^{0}+O[s^{-1}],\\
|3,2,0\rangle|b_{-\frac{1}{2}}^{T}\rangle &  \rightarrow\sqrt{-t}(\frac{1}%
{96}t^{2}-\frac{1}{48}t+\frac{11}{32})s^{2}+\sqrt{-t}(\frac{1}{64}t^{3}%
+\frac{13}{32}t-\frac{5}{8})s\nonumber\\
&  +\sqrt{-t}(\frac{1}{256}t^{4}+\frac{9}{128}t^{3}-\frac{925}{256}t^{2}%
-\frac{729}{64}t-\frac{1481}{256})s^{0}+O[s^{-1}],
\end{align}%
\begin{align}
|3,0,1\rangle|b_{-\frac{1}{2}}^{T}\rangle &  \rightarrow\sqrt{-t}(-\frac
{1}{16\sqrt{6}}t^{2}-\frac{3}{8\sqrt{6}}t+\frac{7}{16\sqrt{6}})s^{2}%
\nonumber\\
&  +\sqrt{-t}(-\frac{3}{32\sqrt{6}}t^{3}-\frac{3}{2\sqrt{6}}t^{2}-\frac
{51}{16\sqrt{6}}t-\frac{19}{4\sqrt{6}})s\nonumber\\
&  +\sqrt{-t}(-\frac{3}{128\sqrt{6}}t^{4}-\frac{111}{64\sqrt{6}}t^{3}%
-\frac{1841}{128\sqrt{6}}t^{2}-\frac{1209}{32\sqrt{6}}t-\frac{3573}%
{128\sqrt{6}})s^{0}+O[s^{-1}].
\end{align}
We find that the ratios of the leading order coefficients of $s^{2}t^{5/2}$
are $\frac{1}{8}:\frac{1}{96}:-\frac{1}{16\sqrt{6}}$, and they agree with the
ratios in the fixed angle limit. The ratios of the second and the third order
coefficients of $st^{7/2}$ and $s^{0}t^{9/2}$ are $\frac{3}{16}:\frac{1}%
{64}:-\frac{3}{32\sqrt{6}}$ and $\frac{3}{64}:\frac{1}{256}:-\frac{3}%
{128\sqrt{6}}$, respectively. We find that these two set of ratios are the
same with one another. We predict that the ratios persist to all orders in the expansions.

The expansions among different worldsheet fermionic states but with same mass
level $M_{2}^{2}=6$ are
\begin{align}
|3,0,0\rangle|b_{-\frac{1}{2}}^{L}\rangle &  \rightarrow\sqrt{-t}(\frac{1}%
{48}t^{2}-\frac{17}{48}t)s^{2}+\sqrt{-t}(\frac{1}{32}t^{3}-\frac{151}{96}%
t^{2}-\frac{295}{96}t-\frac{119}{32})s\nonumber\\
&  +\sqrt{-t}(\frac{1}{128}t^{4}-\frac{249}{128}t^{3}-\frac{1317}{64}%
t^{2}-\frac{2883}{64}t-\frac{3831}{128})s^{0}+O[s^{-1}],\\
|2,0,0\rangle|b_{-\frac{3}{2}}^{L}\rangle &  \rightarrow\sqrt{-t}(-\frac
{1}{8\sqrt{6}}t^{2}-\frac{7}{8\sqrt{6}}t)s^{2}\nonumber\\
&  +\sqrt{-t}(-\frac{3}{16\sqrt{6}}t^{3}-\frac{57}{16\sqrt{6}}t^{2}-\frac
{129}{16\sqrt{6}}t-\frac{147}{16\sqrt{6}})s\nonumber\\
&  +\sqrt{-t}(-\frac{3}{64\sqrt{6}}t^{4}-\frac{285}{64\sqrt{6}}t^{3}%
-\frac{1289}{32\sqrt{6}}t^{2}-\frac{2831}{32\sqrt{6}}t-\frac{4011}{64\sqrt{6}%
})s^{0}+O[s^{-1}],\\
|1,0,0\rangle|b_{-\frac{1}{2}}^{T}b_{-\frac{1}{2}}^{L}b_{-\frac{3}{2}}%
^{L}\rangle &  \rightarrow\sqrt{-t}(\frac{1}{96}t^{2}+\frac{1}{12}t+\frac
{7}{96})s^{2}+\sqrt{-t}(\frac{1}{64}t^{3}+\frac{9}{32}t^{2}+\frac{31}%
{48}t+\frac{61}{96})s\nonumber\\
&  +\sqrt{-t}(\frac{1}{256}t^{4}+\frac{77}{192}t^{3}+\frac{2531}{768}%
t^{2}+\frac{643}{96}t+\frac{3569}{768})s^{0}+O[s^{-1}].
\end{align}
The ratios of the leading order coefficients are given by
\begin{align}
|3,0,0\rangle|b_{-\frac{1}{2}}^{T}\rangle &  :|3,1,0\rangle|b_{-\frac{1}{2}%
}^{L}\rangle:|2,0,0\rangle|b_{-\frac{3}{2}}^{L}\rangle:|1,0,0\rangle
|b_{-\frac{1}{2}}^{T}b_{-\frac{1}{2}}^{L}b_{-\frac{3}{2}}^{L}\rangle
\nonumber\\
&  =\frac{1}{8}:\frac{1}{48}:-\frac{1}{8\sqrt{6}}:\frac{1}{96}.
\end{align}
We have checked that they agree with the ratios in the fixed angle limit. The
second and the third subleading order ratios are $\frac{3}{16}:\frac{1}%
{32}:-\frac{3}{16\sqrt{6}}:\frac{1}{64}$ and $\frac{3}{64}:\frac{1}%
{128}:-\frac{3}{64\sqrt{6}}:\frac{1}{256},$ respectively. Again they agree
with the ratios in the fixed angle limit. We expect that the ratios persist to
all orders in the expansions.%

\setcounter{equation}{0}
\renewcommand{\theequation}{\arabic{section}.\arabic{equation}}%

\section{Recurrence relations of higher spin BPST vertex operators}

In this chapter, we study higher spin Regge string scattering amplitudes from
BPST vertex operator approach \cite{Tan}. Note that in the original BPST paper
\cite{RR6}, the authors calculated the case of four tachyon closed string and
thus Pomeron vertex operators. Here, for simplicity, we will calculate higher
spin BPST vertex operators at arbitrary mass levels of open bosonic string.
The calculation can be easily generalized to closed string case.

We find that all BPST vertex operators can be expressed in terms of Kummer
functions of the second kind. We can then derive infinite number of recurrence
relations among BPST vertex operators of different string states. These
recurrence relations among BPST vertex operators lead to the recurrence
relations among Regge string scattering amplitudes discovered in chapters XI
and XV. \cite{LY,AppellLY}.

\subsection{Four tachyon scattering}

We will calculate high energy open string scatterings in the Regge Regime
\begin{equation}
s\rightarrow\infty,\sqrt{-t}=\text{fixed (but }\sqrt{-t}\neq\infty)
\end{equation}
where%
\begin{equation}
s=-(k_{1}+k_{2})^{2}\text{ and }t=-(k_{2}+k_{3})^{2}.
\end{equation}
Note that the convention for $s$ and $t$ adopted here is different from the
original BPST paper in \cite{RR6}.

We first review the calculation of tachyon BPST vertex operator \cite{RR6}.
The $s-t$ channel of open string four tachyon amplitude can be written as%
\begin{equation}
A=\int_{0}^{1}d\omega\cdot\omega^{k_{1}\cdot k_{2}}\left(  1-\omega\right)
^{k_{2}\cdot k_{3}}=\int_{0}^{1}d\omega\cdot\omega^{-2-\frac{s}{2}}\left(
1-\omega\right)  ^{-2-\frac{t}{2}}.
\end{equation}
Since $s\rightarrow\infty$, the integral is dominated around $\omega=1$.
Making the variable transformation $\omega=1-x$, the integral is dominated
around $x=0$, we obtain%
\begin{equation}
A=\int_{0}^{1}dx\cdot\left(  1-x\right)  ^{-2-\frac{s}{2}}x^{-2-\frac{t}{2}%
}\simeq\int dx\cdot x^{-2-\frac{t}{2}}e^{\frac{s}{2}x}=\Gamma\left(
-1-\frac{t}{2}\right)  \left(  -\frac{s}{2}\right)  ^{1+\frac{t}{2}}.
\label{Regge}%
\end{equation}
Alternatively, the integral in $A$ can be expressed as%
\begin{equation}
A=\int d\omega\left\langle e^{ik_{1}X(0)}e^{ik_{2}X(\omega)}e^{ik_{3}%
X(1)}e^{ik_{4}X(\infty)}\right\rangle . \label{AMP}%
\end{equation}
One can calculate the operator product expansion (OPE) in the Regge limit%
\begin{equation}
e^{ik_{2}X(w)}e^{ik_{3}X(z)}\sim\left\vert w-z\right\vert ^{k_{2}\cdot k_{3}%
}e^{i(k_{2}+k_{3})X(z)+ik_{2}(w-z)\partial X(z)+\cdots}.\nonumber
\end{equation}
This means%
\begin{equation}
e^{ik_{2}X(\omega)}e^{ik_{3}X(1)}\sim\left(  1-\omega\right)  ^{k_{2}\cdot
k_{3}}e^{ikX(1)-ik_{2}\left(  1-\omega\right)  \partial X(1)+higher\text{
}power\text{ }of\text{ }\left(  1-\omega\right)  },k=k_{2}+k_{3}
\label{OPE-tachyon}%
\end{equation}
In evaluating Eq.(\ref{AMP}), one can instead carry out the $\omega$
integration first in Eq.(\ref{OPE-tachyon}) at the operator level to obtain
the BPST vertex operator \cite{RR6}%
\begin{align}
V_{BPST}  &  =\int d\omega e^{ik_{2}X(\omega)}e^{ik_{3}X(1)}\nonumber\\
&  \sim\int d\omega\left(  1-\omega\right)  ^{k_{2}\cdot k_{3}}%
e^{ikX(1)-ik_{2}\left(  1-\omega\right)  \partial X(1)}\nonumber\\
&  =\int dxx^{k_{2}\cdot k_{3}}e^{ikX(1)-ik_{2}x\partial X(1)}\nonumber\\
&  =\Gamma\left(  -1-\frac{t}{2}\right)  \left[  ik_{2}\partial X(1)\right]
^{1+\frac{t}{2}}e^{ikX(1)},
\end{align}
which leads to the same amplitude as in Eq.(\ref{Regge})%
\begin{align}
A  &  =\left\langle e^{ik_{1}X(0)}V_{P}e^{ik_{4}X(\infty)}\right\rangle
\nonumber\\
&  =\Gamma\left(  -1-\frac{t}{2}\right)  \left\langle e^{ik_{1}X(0)}\left[
ik_{2}\partial X(1)\right]  ^{1+\frac{t}{2}}e^{ikX(1)}e^{ik_{4}X(\infty
)}\right\rangle \nonumber\\
&  =\Gamma\left(  -1-\frac{t}{2}\right)  \left(  k_{1}k_{2}\right)
^{1+\frac{t}{2}}\nonumber\\
&  \sim\Gamma\left(  -1-\frac{t}{2}\right)  \left(  -\frac{s}{2}\right)
^{1+\frac{t}{2}}.
\end{align}

\subsection{Higher spin BPST vertex}

\subsubsection{A spin two state}

It was shown \cite{bosonic,RRsusy,LY} that for the $26D$ open bosonic string
states of leading order in energy in the Regge limit at mass level $M_{2}%
^{2}=2(N-1)$, $N=\sum_{n,m,l>0}np_{n}+mq_{m}+lr_{l}$ are of the form
Eq.(\ref{RR}). In this section, we first consider a simple case of a spin two
state $\alpha_{-1}^{P}\alpha_{-1}^{P}|0\rangle$ corresponding to the vertex
$\left(  \partial X^{P}\right)  ^{2}e^{ik_{2}X}\left(  \omega\right)  $. The
four-point amplitude of the spin two state with three tachyons can be
calculated by using the conventional method%
\begin{align}
A^{(q_{1}=2)}  &  =\int d\omega\left\langle e^{ik_{1}X(0)}\left(  \partial
X^{P}\right)  ^{2}e^{ik_{2}X}\left(  \omega\right)  e^{ik_{3}X(1)}%
e^{ik_{4}X(\infty)}\right\rangle \nonumber\\
&  =\int d\omega\omega^{k_{1}\cdot k_{2}}(1-\omega)^{k_{2}\cdot k_{3}}\left[
\frac{ie^{P}\cdot k_{1}}{-\omega}+\frac{ie^{P}\cdot k_{3}}{1-\omega}\right]
^{2}\nonumber\\
&  =-(e^{P}\cdot k_{1})^{2}\Gamma\left(  -1-\frac{t}{2}\right)  \left(
-\frac{s}{2}\right)  ^{\frac{t}{2}-1}+2(e^{P}\cdot k_{1})(e^{P}\cdot
k_{3})\Gamma\left(  -2-\frac{t}{2}\right)  \left(  -\frac{s}{2}\right)
^{\frac{t}{2}}\nonumber\\
&  -(e^{P}\cdot k_{3})^{2}\Gamma\left(  -3-\frac{t}{2}\right)  \left(
-\frac{s}{2}\right)  ^{\frac{t}{2}+1}. \label{amplitude.}%
\end{align}
The momenta of the four particles on the scattering plane are%

\begin{align}
k_{1}  &  =\left(  +\sqrt{p^{2}+M_{1}^{2}},-p,0\right)  ,\\
k_{2}  &  =\left(  +\sqrt{p^{2}+M_{2}^{2}},+p,0\right)  ,\\
k_{3}  &  =\left(  -\sqrt{q^{2}+M_{3}^{2}},-q\cos\phi,-q\sin\phi\right)  ,\\
k_{4}  &  =\left(  -\sqrt{q^{2}+M_{4}^{2}},+q\cos\phi,+q\sin\phi\right)
\end{align}
where $p\equiv\left\vert \mathrm{\vec{p}}\right\vert $, $q\equiv\left\vert
\mathrm{\vec{q}}\right\vert $ and $k_{i}^{2}=-M_{i}^{2}$. The relevant
kinematics in the Regge limit are \cite{bosonic,RRsusy,LY}%
\begin{equation}
e^{P}\cdot k_{1}\simeq-\frac{s}{2M_{2}},\text{ \ }e^{P}\cdot k_{3}\simeq
-\frac{\tilde{t}}{2M_{2}}=-\frac{t-M_{2}^{2}-M_{3}^{2}}{2M_{2}}; \label{st..}%
\end{equation}%
\begin{equation}
e^{L}\cdot k_{1}\simeq-\frac{s}{2M_{2}},\text{ \ }e^{L}\cdot k_{3}\simeq
-\frac{\tilde{t}^{\prime}}{2M_{2}}=-\frac{t+M_{2}^{2}-M_{3}^{2}}{2M_{2}};
\label{st2}%
\end{equation}
and%
\begin{equation}
e^{T}\cdot k_{1}=0\text{, \ \ }e^{T}\cdot k_{3}\simeq-\sqrt{-{t}}%
\end{equation}
where $\tilde{t}$ and $\tilde{t}^{\prime}$ are related to $t$ by finite mass
square terms%
\begin{equation}
\tilde{t}=t-M_{2}^{2}-M_{3}^{2}\text{ , \ }\tilde{t}^{\prime}=t+M_{2}%
^{2}-M_{3}^{2}.
\end{equation}
By using Eq.(\ref{st..}), one easily see that the three terms in
Eq.(\ref{amplitude.}) share the same order of energy in the Regge limit. We
stress that this key observation on the polarizations for higher spin states
was not discussed in \cite{RR6,CCW}.

One can calculate the OPE in the Regge limit%
\begin{equation}
\partial X^{P}\partial X^{P}e^{ik_{2}X}\left(  w\right)  e^{ik_{3}X}\left(
z\right)  \sim\left\vert w-z\right\vert ^{k_{2}\cdot k_{3}}\left[  \partial
X\left(  z\right)  ^{P}+\frac{ie^{P}\cdot k_{3}}{w-z}\right]  ^{2}%
e^{ikX(z)+ik_{2}\left(  w-z\right)  \partial X(z)}.\nonumber
\end{equation}
This means%
\begin{equation}
\partial X^{P}\partial X^{P}e^{ik_{2}X}\left(  \omega\right)  e^{ik_{3}%
X}\left(  1\right)  \sim\left(  1-\omega\right)  ^{k_{2}\cdot k_{3}}\left[
\partial X\left(  1\right)  ^{P}-\frac{ie^{P}\cdot k_{3}}{1-\omega}\right]
^{2}e^{ikX(1)-ik_{2}\left(  1-\omega\right)  \partial X(1)},k=k_{2}+k_{3}.
\label{OPE-vector}%
\end{equation}
One can carry out the $\omega$ integration in Eq.(\ref{OPE-vector}) at the
operator level to obtain the BPST vertex operator%
\begin{align}
V_{BPST}^{(q_{1}=2)}  &  =\int d\omega(\partial X^{P})^{2}e^{ik_{2}X}\left(
\omega\right)  e^{ik_{3}X}\left(  1\right) \nonumber\\
&  \sim\int d\omega\left(  1-\omega\right)  ^{k_{2}\cdot k_{3}}\left[
\partial X\left(  1\right)  ^{P}-\frac{ie^{P}\cdot k_{3}}{1-\omega}\right]
^{2}e^{ikX(1)-ik_{2}\left(  1-\omega\right)  \partial X(1)}\nonumber\\
&  =\partial X\left(  1\right)  ^{P}\partial X\left(  1\right)  ^{P}\int
dxx^{k_{2}\cdot k_{3}}e^{ikX(1)-ik_{2}x\partial X(1)}\nonumber\\
&  -2ie^{P}\cdot k_{3}\partial X\left(  1\right)  ^{P}\int dxx^{k_{2}\cdot
k_{3}-1}e^{ikX(1)-ik_{2}x\partial X(1)}\nonumber\\
&  -(e^{P}\cdot k_{3})^{2}\int dxx^{k_{2}\cdot k_{3}-2}e^{ikX(1)-ik_{2}%
x\partial X(1)}\nonumber\\
&  =\Gamma\left(  -1-\frac{t}{2}\right)  \left[  ik_{2}\partial X(1)\right]
^{\frac{t}{2}-1}\partial X\left(  1\right)  ^{P}\partial X\left(  1\right)
^{P}e^{ikX(1)}\nonumber\\
&  -2ie^{P}\cdot k_{3}\Gamma\left(  -2-\frac{t}{2}\right)  \left[
ik_{2}\partial X(1)\right]  ^{\frac{t}{2}}\partial X\left(  1\right)
^{P}e^{ikX(1)}\nonumber\\
&  -(e^{P}\cdot k_{3})^{2}\Gamma\left(  -3-\frac{t}{2}\right)  \left[
ik_{2}\partial X(1)\right]  ^{\frac{t}{2}+1}e^{ikX(1)} \label{three}%
\end{align}
which leads to the same amplitude%
\begin{align}
A^{(q_{1}=2)}  &  =\left\langle e^{ik_{1}X(0)}V_{BPST}^{(q_{1}=2)}%
e^{ik_{4}X(\infty)}\right\rangle \nonumber\\
&  =\Gamma\left(  -1-\frac{t}{2}\right)  \left\langle e^{ik_{1}X(0)}\left[
ik_{2}\partial X(1)\right]  ^{\frac{t}{2}-1}\partial X\left(  1\right)
^{P}\partial X\left(  1\right)  ^{P}e^{ikX(1)}e^{ik_{4}X(\infty)}\right\rangle
\nonumber\\
&  -2ie^{P}\cdot k_{3}\Gamma\left(  -2-\frac{t}{2}\right)  \left\langle
e^{ik_{1}X(0)}\left[  ik_{2}\partial X(1)\right]  ^{\frac{t}{2}}\partial
X\left(  1\right)  ^{P}e^{ikX(1)}e^{ik_{4}X(\infty)}\right\rangle \nonumber\\
&  -(e^{P}\cdot k_{3})^{2}\Gamma\left(  -3-\frac{t}{2}\right)  \left\langle
e^{ik_{1}X(0)}\left[  ik_{2}\partial X(1)\right]  ^{\frac{t}{2}+1}%
e^{ikX(1)}e^{ik_{4}X(\infty)}\right\rangle \nonumber\\
&  \sim-(e^{P}\cdot k_{1})^{2}\Gamma\left(  -1-\frac{t}{2}\right)  \left(
-\frac{s}{2}\right)  ^{\frac{t}{2}-1}+2(e^{P}\cdot k_{1})(e^{P}\cdot
k_{3})\Gamma\left(  -2-\frac{t}{2}\right)  \left(  -\frac{s}{2}\right)
^{\frac{t}{2}}\nonumber\\
&  -(e^{P}\cdot k_{3})^{2}\Gamma\left(  -3-\frac{t}{2}\right)  \left(
-\frac{s}{2}\right)  ^{\frac{t}{2}+1}. \label{same}%
\end{align}
Note that the three terms in Eq.(\ref{three}) lead to the three terms
respectively in Eq.(\ref{same}) with the same order of energy in the Regge limit.

\subsubsection{Higher spin states}

We now consider the higher spin state%
\begin{equation}
\left\vert p_{n},q_{m}\right\rangle =\prod_{n=1}(\alpha_{-n}^{T})^{p_{n}}%
\prod_{m=1}(\alpha_{-m}^{P})^{q_{m}}|0\rangle, \label{state1}%
\end{equation}
which corresponds to the vertex%
\begin{equation}
V_{2}\left(  \omega\right)  =\left[  \prod_{n=1}\left(  \partial^{n}%
X^{T}\right)  ^{p_{n}}\prod_{m=1}\left(  \partial^{m}X^{P}\right)  ^{q_{m}%
}\right]  e^{ik_{2}X}\left(  \omega\right)  .
\end{equation}
The four-point amplitude of the above state with three tachyons was calculated
to be (from now on we set $M_{2}=M$) \cite{bosonic,RRsusy,LY}
\begin{align}
A^{(p_{n},q_{m})}  &  =\int d\omega\left\langle e^{ik_{1}X(0)}V_{2}\left(
\omega\right)  e^{ik_{3}X(1)}e^{ik_{4}X(\infty)}\right\rangle \nonumber\\
&  =\left(  -\frac{1}{M}\right)  ^{q_{1}}U\left(  -q_{1},\frac{t}{2}%
+2-q_{1},\frac{\tilde{t}}{2}\right)  B\left(  -1-\frac{s}{2},-1-\frac{t}%
{2}\right) \nonumber\\
&  \cdot\prod_{n=1}\left[  \sqrt{-t}(n-1)!\right]  ^{p_{n}}\prod_{m=2}\left[
\tilde{t}(m-1)!\left(  -\frac{1}{2M}\right)  \right]  ^{q_{m}}\\
&  \sim\left(  -\frac{1}{M}\right)  ^{q_{1}}U\left(  -q_{1},\frac{t}%
{2}+2-q_{1},\frac{\tilde{t}}{2}\right)  \Gamma\left(  -1-\frac{t}{2}\right)
\left(  -\frac{s}{2}\right)  ^{1+\frac{t}{2}}\label{power.}\\
&  \cdot\prod_{n=1}\left[  \sqrt{-t}(n-1)!\right]  ^{p_{n}}\prod_{m=2}\left[
\tilde{t}(m-1)!\left(  -\frac{1}{2M}\right)  \right]  ^{q_{m}}
\label{amplitude-tensor}%
\end{align}
where $U$ is the Kummer function of the second kind. One can calculate the OPE
in the Regge limit%
\begin{align}
&  V_{2}\left(  \omega\right)  e^{ik_{3}X(1)}\nonumber\\
&  =\left[  \prod_{n=1}\left(  \partial^{n}X^{T}\right)  ^{p_{n}}\prod
_{m=1}\left(  \partial^{m}X^{P}\right)  ^{q_{m}}\right]  e^{ik_{2}X}\left(
\omega\right)  e^{ik_{3}X(1)}\nonumber\\
&  \sim\prod_{n=1}\left[  \frac{\left(  n-1\right)  !k_{3}\cdot e^{T}}{\left(
1-\omega\right)  ^{n}}\right]  ^{p_{n}}\prod_{m=2}\left[  \frac{\left(
m-1\right)  !k_{3}\cdot e^{P}}{\left(  1-\omega\right)  ^{m}}\right]  ^{q_{m}%
}\nonumber\\
&  \cdot\left[  \partial X\left(  1\right)  \cdot e^{P}-\frac{ik_{3}\cdot
e^{P}}{1-\omega}\right]  ^{q_{1}}\left(  1-\omega\right)  ^{k_{2}\cdot k_{3}%
}e^{ikX(1)-ik_{2}\left(  1-\omega\right)  \partial X(1)}\\
&  =\left(  \frac{-\tilde{t}}{2M}\right)  ^{q_{1}}\prod_{n=1}\left[  \sqrt
{-t}(n-1)!\right]  ^{p_{n}}\prod_{m=2}\left[  \tilde{t}(m-1)!\left(  -\frac
{1}{2M}\right)  \right]  ^{q_{m}}\nonumber\\
&  \cdot\sum_{j=0}^{q_{1}}{\binom{q_{1}}{j}}\left(  \frac{2iM_{2}\partial
X\left(  1\right)  \cdot e^{P}}{\tilde{t}}\right)  ^{j}\left(  1-\omega
\right)  ^{k_{2}\cdot k_{3}-N+j}e^{ikX(1)-ik_{2}\left(  1-\omega\right)
\partial X(1)} \label{OPE-tensor}%
\end{align}
where $N=\sum_{n,m}\left(  np_{n}+mq_{m}\right)  $. We can carry out the
$\omega$ integration in Eq.(\ref{OPE-tensor}) to obtain the BPST vertex
operator%
\begin{align}
V_{BPST}^{(p_{n};q_{m})}  &  =\int d\omega V_{2}\left(  \omega\right)
e^{ik_{3}X}\left(  1\right) \nonumber\\
&  \sim\left(  \frac{-\tilde{t}}{2M}\right)  ^{q_{1}}\prod_{n=1}\left[
\sqrt{-t}(n-1)!\right]  ^{p_{n}}\prod_{m=2}\left[  \tilde{t}(m-1)!\left(
-\frac{1}{2M}\right)  \right]  ^{q_{m}}\nonumber\\
&  \cdot\sum_{j=0}^{q_{1}}{\binom{q_{1}}{j}}\left(  \frac{2iM_{2}\partial
X\left(  1\right)  \cdot e^{P}}{\tilde{t}}\right)  ^{j}\int d\omega\left(
1-\omega\right)  ^{k_{2}\cdot k_{3}-N+j}e^{ikX(1)-ik_{2}\left(  1-\omega
\right)  \partial X(1)}\nonumber\\
&  =\left(  \frac{-\tilde{t}}{2M}\right)  ^{q_{1}}\prod_{n=1}\left[  \sqrt
{-t}(n-1)!\right]  ^{p_{n}}\prod_{m=2}\left[  \tilde{t}(m-1)!\left(  -\frac
{1}{2M}\right)  \right]  ^{q_{m}}\nonumber\\
&  \cdot\sum_{j=0}^{q_{1}}{\binom{q_{1}}{j}}\left(  \frac{2iM\partial X\left(
1\right)  \cdot e^{P}}{\tilde{t}}\right)  ^{j}\int dxx^{k_{2}\cdot k_{3}%
-N+j}e^{ikX(1)-ik_{2}x\partial X(1)}\nonumber\\
&  =\left(  \frac{-\tilde{t}}{2M}\right)  ^{q_{1}}\prod_{n=1}\left[  \sqrt
{-t}(n-1)!\right]  ^{p_{n}}\prod_{m=2}\left[  \tilde{t}(m-1)!\left(  -\frac
{1}{2M}\right)  \right]  ^{q_{m}}\nonumber\\
&  \cdot\sum_{j=0}^{q_{1}}{\binom{q_{1}}{j}}\left(  \frac{2iM\partial X\left(
1\right)  \cdot e^{P}}{\tilde{t}}\right)  ^{j}\Gamma\left(  -1-\frac{t}%
{2}+j\right)  \left[  ik_{2}\cdot\partial X(1)\right]  ^{1+\frac{t}{2}%
-j}e^{ikX(1)}. \label{dot}%
\end{align}
One notes that, in Eq.(\ref{dot}), $M\partial X\left(  1\right)  \cdot
e^{P}=k_{2}\cdot\partial X(1)$ and the summation over $j$ can be simplified.
The BPST vertex operator can be further reduced to
\begin{align}
V_{BPST}^{(p_{n};q_{m})}  &  =\left(  \frac{-\tilde{t}}{2M_{2}}\right)
^{q_{1}}\prod_{n=1}\left[  \sqrt{-t}(n-1)!\right]  ^{p_{n}}\prod_{m=2}\left[
\tilde{t}(m-1)!\left(  -\frac{1}{2M}\right)  \right]  ^{q_{m}}\nonumber\\
&  \cdot\sum_{j=0}^{q_{1}}{\binom{q_{1}}{j}}\left(  \frac{2}{\tilde{t}%
}\right)  ^{j}\left(  -1-\frac{t}{2}\right)  _{j}\Gamma\left(  -1-\frac{t}%
{2}\right)  \left[  ik_{2}\cdot\partial X(1)\right]  ^{1+\frac{t}{2}%
}e^{ikX(1)}\nonumber\\
&  =\left(  \frac{-1}{M}\right)  ^{q_{1}}\prod_{n=1}\left[  \sqrt
{-t}(n-1)!\right]  ^{p_{n}}\prod_{m=2}\left[  \tilde{t}(m-1)!\left(  -\frac
{1}{2M}\right)  \right]  ^{q_{m}}\nonumber\\
&  \cdot U\left(  -q_{1},\frac{t}{2}+2-q_{1},\frac{\tilde{t}}{2}\right)
\Gamma\left(  -1-\frac{t}{2}\right)  \left[  ik_{2}\cdot\partial X(1)\right]
^{1+\frac{t}{2}}e^{ikX(1)} \label{pomeron}%
\end{align}
where we have used%
\begin{equation}
\sum_{j=0}^{l}{\binom{l}{j}}\left(  \frac{2}{\tilde{t}}\right)  ^{j}\left(
-1-\frac{t}{2}\right)  _{j}=2^{l}(\tilde{t})^{-l}\ U\left(  -l,\frac{t}%
{2}+2-l,\frac{\tilde{t}}{2}\right)  . \label{after}%
\end{equation}

One notes that the exponent of $\left[  ik_{2}\cdot\partial X(1)\right]
^{1+\frac{t}{2}}$ in Eq.(\ref{pomeron}) is mass level $N$ independent. This is
related to the fact that the well known $\sim s^{\alpha(t)}$ power-law
behavior of the four tachyon string scattering amplitude in the RR can be
extended to arbitrary higher string states and is mass level independent as
can be seen from Eq.(\ref{power.}). This interesting result was first pointed
out in section XI.C \cite{bosonic} and will be crucial to derive inter-mass
level recurrence relations among BPST vertex operators to be discussed later.

The BPST vertex operator in Eq.(\ref{pomeron}) leads to exactly the same
amplitude as in Eq.(\ref{amplitude-tensor}).

\subsection{Recurrence relations}

For any confluent hypergeometric function $U(a,c,x)$ with parameters $(a,c)$
the four functions with parameters $(a-1,c),(a+1,c),(a,c-1)$ and $(a,c+1)$ are
called the contiguous functions. Recurrence relation exists between any such
function and any two of its contiguous functions. There are six recurrence
relations \cite{Slater}%
\begin{align}
U(a-1,c,x)-(2a-c+x)U(a,c,x)+a(1+a-c)U(a+1,c,x)  &  =0,\label{RR1}\\
(c-a-1)U(a,c-1,x)-(x+c-1))U(a,c,x)+xU(a,c+1,x)  &  =0,\label{RR2}\\
U(a,c,x)-aU(a+1,c,x)-U(a,c-1,x)  &  =0,\label{RR3}\\
(c-a)U(a,c,x)+U(a-1,c,x)-xU(a,c+1,x)  &  =0,\label{RR4}\\
(a+x)U(a,c,x)-xU(a,c+1,x)+a(c-a-1)U(a+1,c,x)  &  =0,\label{RR5}\\
(a+x-1)U(a,c,x)-U(a-1,c,x)+(1+a-c)U(a,c-1,x)  &  =0. \label{RR6}%
\end{align}
From any two of these six relations the remaining four recurrence relations
can be deduced.

The confluent hypergeometric function $U(a,c,x)$ with parameters $(a\pm m,c\pm
n)$ for $m,n=0,1,2...$are called associated functions. Again it can be shown
that there exist relations between any three associated functions, so that any
confluent hypergeometric function can be expressed in terms of any two of its
associated functions.

Recently it was shown \cite{LY} that Recurrence relations exist among higher
spin Regge string scattering amplitudes of different string states. This was
discussed in section XI.D. The key to derive these relations was to use
recurrence relations and addition theorem of Kummer functions. In view of the
form of higher spin BPST vertex operators in Eq.(\ref{pomeron}), one can
easily calculate recurrence relations among higher spin BPST vertex operators.
By using the recurrence relation of Kummer functions \cite{LY}, for example,
\begin{equation}
U\left(  -2,\frac{t}{2},\frac{t}{2}\right)  +\left(  \frac{t}{2}+1\right)
U(-1,\frac{t}{2},\frac{t}{2})-\frac{t}{2}U\left(  -1,\frac{t}{2}+1,\frac{t}%
{2}\right)  =0,
\end{equation}
one can obtain the following recurrence relation among BPST vertex operators
at mass level $M^{2}=2$ \cite{Tan}
\begin{equation}
M\sqrt{-t}V_{BPST}^{(q_{1}=2)}-\frac{t}{2}V_{BPST}^{(p_{1}=1,q_{1}=1)}=0.
\label{Recurrence}%
\end{equation}
Rather than constant coefficients in the RR Regge stringy Ward identities
derived in \cite{LY}, the coefficients of this recurrence relation
Eq.(\ref{Recurrence}) among BPST vertex operators are kinematic variable
dependent, similar to BCJ relations among field theory amplitudes
\cite{BCJ1,BCJ2,BCJ3,BCJ4,BCJ5}. The recurrence relation among BPST vertex
operators in Eq.(\ref{Recurrence}) leads to the recurrence relation among
Regge string scattering amplitudes \cite{LY}
\begin{equation}
M\sqrt{-t}A^{(q_{1}=2)}-\frac{t}{2}A^{(p_{1}=1,q_{1}=1)}=0, \label{RRT1}%
\end{equation}
which is the same with Eq.(\ref{RRI1}).

\subsection{More general recurrence relations}

To derive more general recurrence relations, we need to calculate BPST vertex
operators corresponding to the general higher spin states in Eq.(\ref{RR}). We
first calculate the BPST vertex operator correspond to the state%
\begin{equation}
\left\vert p_{n},r_{l}\right\rangle =\prod_{n=1}(\alpha_{-n}^{T})^{p_{n}}%
\prod_{m=1}(\alpha_{-l}^{L})^{r_{l}}|0\rangle. \label{state2}%
\end{equation}
The calculation is very similar to that of Eq.(\ref{state1}) up to some
modification. One can easily get that Eq.(\ref{dot}) is now replaced by
\begin{align}
V_{BPST}^{(p_{n};r_{l})}  &  =\left(  \frac{-\tilde{t}^{\prime}}{2M}\right)
^{r_{1}}\prod_{n=1}\left[  \sqrt{-t}(n-1)!\right]  ^{p_{n}}\prod_{l=2}\left[
\tilde{t}^{\prime}(l-1)!\left(  -\frac{1}{2M}\right)  \right]  ^{r_{l}%
}\nonumber\\
&  \cdot\sum_{j=0}^{r_{1}}{\binom{r_{1}}{j}}\left(  \frac{2iM\partial X\left(
1\right)  \cdot e^{L}}{\tilde{t}^{\prime}}\right)  ^{j}\Gamma\left(
-1-\frac{t}{2}+j\right)  \left[  ik_{2}\cdot\partial X(1)\right]  ^{1+\frac
{t}{2}-j}e^{ikX(1)}. \label{dot2}%
\end{align}
One notes that, in Eq.(\ref{dot2}), $M\partial X\left(  1\right)  \cdot
e^{L}\neq k_{2}\cdot\partial X(1)$ and, in contrast to Eq.(\ref{dot}), the two
factors with exponents $j$ and $-j$ do not cancel out. The BPST vertex
operator for this case thus reduces to
\begin{align}
V_{BPST}^{(p_{n};r_{l})}  &  =\left(  \frac{-1}{M}\right)  ^{r_{1}}\prod
_{n=1}\left[  \sqrt{-t}(n-1)!\right]  ^{p_{n}}\prod_{l=2}\left[  \tilde
{t}^{\prime}(l-1)!\left(  -\frac{1}{2M}\right)  \right]  ^{r_{l}}\nonumber\\
&  \cdot U\left(  -r_{1},\frac{t}{2}+2-r_{1},\frac{\tilde{t}^{\prime}}{2}%
\frac{e^{P}\cdot\partial X(1)}{e^{L}\cdot\partial X\left(  1\right)  }\right)
\Gamma\left(  -1-\frac{t}{2}\right)  \left[  ik_{2}\cdot\partial X(1)\right]
^{1+\frac{t}{2}}e^{ikX(1)}. \label{pomeron2}%
\end{align}
The BPST vertex operator in Eq.(\ref{pomeron2}) leads to the amplitude
\begin{align}
A^{(p_{n},r_{l})}  &  =\left(  -\frac{1}{M}\right)  ^{r_{1}}U\left(
-r_{1},\frac{t}{2}+2-r_{1},\frac{\tilde{t}^{\prime}}{2}\right)  \Gamma\left(
-1-\frac{t}{2}\right)  \left(  -\frac{s}{2}\right)  ^{1+\frac{t}{2}%
}\nonumber\\
&  \cdot\prod_{n=1}\left[  \sqrt{-t}(n-1)!\right]  ^{p_{n}}\prod_{l=2}\left[
\tilde{t}^{\prime}(l-1)!\left(  -\frac{1}{2M}\right)  \right]  ^{r_{l}},
\label{amplitude-tensor2}%
\end{align}
which is consistent with the one calculated in \cite{bosonic,RRsusy,LY}. Note
that the contribution of $\frac{e^{P}\cdot\partial X(1)}{e^{L}\cdot\partial
X\left(  1\right)  }$ in the correlation function reduces to $1$ in the Regge
limit by using first equations of Eq.(\ref{st..}) and Eq.(\ref{st2}). One sees
that Eq.(\ref{amplitude-tensor2}) can be obtained from
Eq.(\ref{amplitude-tensor}) by doing the replacement $\tilde{t}\rightarrow$
$\tilde{t}^{\prime}$.

We are now ready to calculate the BPST vertex operator corresponding to the
most general Regge state in Eq.(\ref{RR}). Similar to the RR amplitude
calculated in Eq.(\ref{factor1}) and Eq.(\ref{factor2}) \cite{LY}, the BPST
vertex operator can be expressed in two equivalent forms%
\begin{align}
V_{BPST}^{(p_{n};q_{m;}r_{l})}  &  =\prod_{n=1}\left[  \left(  n-1\right)
!\sqrt{-t}\right]  ^{p_{n}}\cdot\prod_{m=1}\left[  -\left(  m-1\right)
!\frac{\tilde{t}}{2M}\right]  ^{q_{m}}\cdot\prod_{l=2}\left[  \left(
l-1\right)  !\frac{\tilde{t}^{\prime}}{2M}\right]  ^{r_{l}}\nonumber\\
&  \quad\cdot\left(  \frac{1}{M}\right)  ^{r_{1}}\Gamma\left(  -1-\frac{t}%
{2}\right)  \left[  ik_{2}\cdot\partial X(1)\right]  ^{1+\frac{t}{2}%
}e^{ikX(1)}\nonumber\\
&  \cdot\sum_{i=0}^{q_{1}}\binom{q_{1}}{i}\left(  \frac{2}{\tilde{t}}\right)
^{i}\left(  -\frac{t}{2}-1\right)  _{i}U\left(  -r_{1},\frac{t}{2}%
+2-i-r_{1},\frac{\tilde{t}^{\prime}}{2}\frac{e^{P}\cdot\partial X(1)}%
{e^{L}\cdot\partial X\left(  1\right)  }\right) \label{factor1.}\\
&  =\prod_{n=1}\left[  \left(  n-1\right)  !\sqrt{-t}\right]  ^{p_{n}}%
\cdot\prod_{m=2}\left[  -\left(  m-1\right)  !\frac{\tilde{t}}{2M}\right]
^{q_{m}}\cdot\prod_{l=1}\left[  \left(  l-1\right)  !\frac{\tilde{t}^{\prime}%
}{2M}\right]  ^{r_{l}}\nonumber\\
&  \cdot\left(  -\frac{1}{M}\right)  ^{q_{1}}\Gamma\left(  -1-\frac{t}%
{2}\right)  \left[  ik_{2}\cdot\partial X(1)\right]  ^{1+\frac{t}{2}%
}e^{ikX(1)}\nonumber\\
&  \cdot\sum_{j=0}^{r_{1}}\binom{r_{1}}{j}\left(  \frac{2}{\tilde{t}^{\prime}%
}\frac{e^{L}\cdot\partial X(1)}{e^{P}\cdot\partial X\left(  1\right)
}\right)  ^{j}\left(  -\frac{t}{2}-1\right)  _{j}U\left(  -q_{1},\frac{t}%
{2}+2-j-q_{1},\frac{\tilde{t}}{2}\right)  . \label{factor2.}%
\end{align}

\bigskip

\bigskip

\bigskip

\bigskip

Either form Eq.(\ref{factor1.}) or Eq.(\ref{factor2.}) of the above BPST
vertex operator leads consistently to the amplitude calculated previously in
Eq.(\ref{factor1}) and Eq.(\ref{factor2}) and is re-listed here \cite{LY}%
\begin{align}
A^{(p_{n},q_{m}:r_{l})}  &  =\prod_{n=1}\left[  \left(  n-1\right)  !\sqrt
{-t}\right]  ^{p_{n}}\cdot\cdot\prod_{m=1}\left[  -\left(  m-1\right)
!\frac{\tilde{t}}{2M}\right]  ^{q_{m}}\cdot\prod_{l=2}\left[  \left(
l-1\right)  !\frac{\tilde{t}^{\prime}}{2M}\right]  ^{r_{l}}\nonumber\\
&  \cdot\left(  \frac{1}{M}\right)  ^{r_{1}}\Gamma\left(  -1-\frac{t}%
{2}\right)  \left(  -\frac{s}{2}\right)  ^{1+\frac{t}{2}}\nonumber\\
&  \cdot\sum_{i=0}^{q_{1}}\binom{q_{1}}{i}\left(  \frac{2}{\tilde{t}}\right)
^{i}\left(  -\frac{t}{2}-1\right)  _{i}U\left(  -r_{1},\frac{t}{2}%
+2-i-r_{1},\frac{\tilde{t}^{\prime}}{2}\right) \label{RRamp1}\\
&  =\prod_{n=1}\left[  \left(  n-1\right)  !\sqrt{-t}\right]  ^{p_{n}}%
\cdot\prod_{m=2}\left[  -\left(  m-1\right)  !\frac{\tilde{t}}{2M}\right]
^{q_{m}}\cdot\prod_{l=1}\left[  \left(  l-1\right)  !\frac{\tilde{t}^{\prime}%
}{2M}\right]  ^{r_{l}}\nonumber\\
&  \cdot\left(  -\frac{1}{M}\right)  ^{q_{1}}\Gamma\left(  -1-\frac{t}%
{2}\right)  \left(  -\frac{s}{2}\right)  ^{1+\frac{t}{2}}\nonumber\\
&  \cdot\sum_{j=0}^{r_{1}}\binom{r_{1}}{j}\left(  \frac{2}{\tilde{t}^{\prime}%
}\right)  ^{j}\left(  -\frac{t}{2}-1\right)  _{j}U\left(  -q_{1},\frac{t}%
{2}+2-j-q_{1},\frac{\tilde{t}}{2}\right)  . \label{RRamp2}%
\end{align}

One can now derive more general recurrence relations among BPST vertex
operators. As an example, the three BPST vertex operators $V_{BPST}^{q_{1}=3}%
$, $V_{BPST}^{p_{1}=1,q_{1}=2}$ and $V_{BPST}^{q_{1}=2,r_{1}=1}$ can be
calculated by using Eq.(\ref{factor2.}) to be%
\begin{align}
V_{BPST}^{(q_{1}=3)}  &  =\left(  -\frac{1}{M}\right)  ^{3}\Gamma\left(
-1-\frac{t}{2}\right)  \left[  ik_{2}\cdot\partial X(1)\right]  ^{1+\frac
{t}{2}}e^{ikX(1)}U\left(  -3,\frac{t}{2}-1,\frac{t}{2}-1\right)  ,\\
V_{BPST}^{(p_{1}=1,q_{1}=2)}  &  =\left(  -\frac{1}{M}\right)  ^{2}\sqrt
{-t}\Gamma\left(  -1-\frac{t}{2}\right)  \left[  ik_{2}\cdot\partial
X(1)\right]  ^{1+\frac{t}{2}}e^{ikX(1)}U\left(  -2,\frac{t}{2},\frac{t}%
{2}-1\right)  ,\\
V_{BPST}^{(q_{1}=2,r_{1}=1)}  &  =\frac{t+6}{2M}\left(  -\frac{1}{M}\right)
^{2}\Gamma\left(  -1-\frac{t}{2}\right)  \left[  ik_{2}\cdot\partial
X(1)\right]  ^{1+\frac{t}{2}}e^{ikX(1)}\nonumber\\
&  \left[  U\left(  -2,\frac{t}{2},\frac{t}{2}-1\right)  +\frac{2}{t+6}\left(
-\frac{t}{2}-1\right)  U\left(  -2,\frac{t}{2}-1,\frac{t}{2}-1\right)
\frac{e^{L}\cdot\partial X(1)}{e^{P}\cdot\partial X\left(  1\right)  }\right]
.
\end{align}
The recurrence relation among Kummer functions derived from Eq.(\ref{RR4})
\cite{LY}%
\begin{equation}
U\left(  -3,\frac{t}{2}-1,\frac{t}{2}-1\right)  +\left(  \frac{t}{2}+1\right)
U(-2,\frac{t}{2}-1,\frac{t}{2}-1)-\left(  \frac{t}{2}-1\right)  U\left(
-2,\frac{t}{2},\frac{t}{2}-1\right)  =0 \label{RRTT}%
\end{equation}
leads to the following recurrence relation among BPST vertex operators at mass
level $M^{2}=4$ \cite{Tan}%
\begin{align}
M\sqrt{-t}e^{L}\cdot\partial X\left(  1\right)  V_{BPST}^{q_{1}=3}+M\sqrt
{-t}e^{P}\cdot\partial X(1)V_{BPST}^{q_{1}=2,r_{1}=1}  & \nonumber\\
-\left[  \left(  \frac{t}{2}+3\right)  e^{P}\cdot\partial X(1)-\left(
\frac{t}{2}-1\right)  e^{L}\cdot\partial X\left(  1\right)  \right]
V_{BPST}^{p_{1}=1,q_{1}=2}  &  =0. \label{RRT2}%
\end{align}
In addition to the $t$ dependence, the coefficients of the recurrence relation
in Eq.(\ref{RRT2}) are operator dependent. The recurrence relation among BPST
vertex operators in Eq.(\ref{RRT2}) leads to the recurrence relation among
Regge string scattering amplitudes \cite{LY}%
\begin{equation}
M\sqrt{-t}A^{(q_{1}=3)}-4A^{(p_{1}=1,q_{1}=2)}+M\sqrt{-t}A^{(q_{1}=2,r_{1}%
=1)}=0,
\end{equation}
which is the same with Eq.(\ref{RRI2}).\qquad

For the next example, we construct an inter-mass level recurrence relation for
BPST vertex operators at mass level $M^{2}=2,4.$ We begin with the addition
theorem of Kummer function \cite{Slater}%
\begin{equation}
U(a,c,x+y)=\sum_{k=0}^{\infty}\frac{1}{k!}\left(  a\right)  _{k}(-1)^{k}%
y^{k}U(a+k,c+k,x)
\end{equation}
which terminates to a finite sum for a non-positive integer $a.$ By taking,
for example, $a=-1,c=\frac{t}{2}+1,x=\frac{t}{2}-1$ and $y=1,$ the theorem
gives \cite{LY}%
\begin{equation}
U\left(  -1,\frac{t}{2}+1,\frac{t}{2}\right)  -U\left(  -1,\frac{t}{2}%
+1,\frac{t}{2}-1\right)  -U\left(  0,\frac{t}{2}+2,\frac{t}{2}-1\right)  =0.
\label{inter}%
\end{equation}
Eq.(\ref{inter}) leads to an inter-mass level recurrence relation among BPST
vertex operators \cite{Tan}
\begin{equation}
M(2)(t+6)V_{BPST}^{(p_{1}=1,q_{1}=1)}-2M(4)^{2}\sqrt{-t}V_{BPST}%
^{(q_{1}=1,r_{2}=1)}+2M(4)V_{BPST}^{(p_{1}=1,r_{2}=1)}=0 \label{RRT3}%
\end{equation}
where \ masses $M(2)=\sqrt{2},M(4)=\sqrt{4}=2,$ and $V_{BPST}^{p_{1}%
=1,q_{1}=1}$ is a BPST vertex operator at mass level $M^{2}=2$, and
$V_{BPST}^{q_{1}=1,r_{2}=1}$, $V_{BPST}^{p_{1}=1,r_{2}=1}$are BPST vertex
operators at mass levels $M^{2}=4$ respectively. In deriving Eq.(\ref{RRT3}),
it is important to use the fact that the exponent of $\left[  ik_{2}%
\cdot\partial X(1)\right]  ^{1+\frac{t}{2}}$ in the BPST vertex operator in
Eq.(\ref{factor2.}) is mass level $N$ independent as mentioned in the
paragraph after Eq.(\ref{after}). The recurrence relation among BPST vertex
operators in Eq.(\ref{RRT3}) leads to the recurrence relation among Regge
string scattering amplitudes \cite{LY}%
\begin{equation}
M(2)(t+6)A^{(p_{1}=1,q_{1}=1)}-2M(4)^{2}\sqrt{-t}A^{(q_{1}=1,r_{2}%
=1)}+2M(4)A^{(p_{1}=1,r_{2}=1)}=0,
\end{equation}
which is the same with Eq.(\ref{RRI3}).

In section XI.D.2 \cite{LY}, it was shown that, at each fixed mass level, each
Kummer function in the summation of Eq.(\ref{RRamp2}) can be expressed in
terms of Regge string scattering amplitudes $A^{(p_{n},q_{m}:r_{l})}$ at the
same mass level. Moreover, although for general values of $a$, the best one
can obtain from recurrence relations of Kummer function $U(a,c,x)$ is to
express any Kummer function in terms of any two of its associated function,
for non-positive integer values of $a$ in the RR string amplitude case
however, $U(a,c,x)$ can be fixed up to an overall factor by using Kummer
function recurrence relations \cite{LY}. As a result, all Regge string
scattering amplitudes can be algebraically solved by Kummer function
recurrence relations up to multiplicative factors. An important application of
the above properties is the construction of an infinite number of recurrence
relations among Regge string scattering amplitudes. One can use the recurrence
relations of Kummer functions Eq.(\ref{RR1}) to Eq.(\ref{RR6}) to
systematically construct recurrence relations among Regge string scattering amplitudes.

In view of the form of BPST vertex operators calculated in Eq.(\ref{factor2.}%
), one can similarly solve \cite{LY} all Kummer functions $U(a,c,x)$ in
Eq.(\ref{factor2.}) in terms of BPST vertex operators and use the recurrence
relations of Kummer functions Eq.(\ref{RR1}) to Eq.(\ref{RR6}) to
systematically construct an infinite number of recurrence relations among BPST
vertex operators. Moreover, the forms of all BPST vertex operators can be
fixed by these recurrence relations up to multiplicative factors. These
recurrence relations among BPST vertex operators are dual to linear relations
or symmetries among high energy fixed angle string scattering amplitudes
discovered previously \cite{ChanLee,ChanLee1,ChanLee2,CHLTY1,CHLTY2,CHLTY3}.

We illustrate the prescription here to construct other examples of recurrence
relations among BPST vertex operators at mass level $M^{2}=4.$ Generalization
to arbitrary mass levels will be given in the next section. There are $22$
BPST vertex operators for the mass level $M^{2}=4.$ We first consider the
group of BPST vertex operators with $q_{1}=0,$ $(V_{BPST}^{TTT},V_{BPST}%
^{LTT},V_{BPST}^{LLT},V_{BPST}^{LLL})$ \cite{LY}. The corresponding $r_{1}$
for each BPST vertex operator are $(0,1,2,3)$. Here we use a new notation for
BPST vertex operator, for example, $V_{BPST}^{LLT}\equiv V_{BPST}%
^{(p_{1}=1,r_{1}=2)}$ ,$V_{BPST}^{LT}=V_{BPST}^{(p_{1}=1,r_{2}=1)}$and
$V_{BPST}^{TL}=V_{BPST}^{(p_{2}=1,r_{1}=1)}$etc. By using Eq.(\ref{factor2.}),
one can easily calculate that%
\begin{align}
V_{BPST}^{TTT}  &  =\left(  \sqrt{-t}\right)  ^{3}\Gamma\left(  -1-\frac{t}%
{2}\right)  \left[  ik_{2}\cdot\partial X(1)\right]  ^{1+\frac{t}{2}%
}e^{ikX(1)}U\left(  0,\frac{t}{2}+2,\frac{t}{2}-1\right)  ,\\
V_{BPST}^{LTT}  &  =\frac{t+6}{2M}\left(  \sqrt{-t}\right)  ^{2}\Gamma\left(
-1-\frac{t}{2}\right)  \left[  ik_{2}\cdot\partial X(1)\right]  ^{1+\frac
{t}{2}}e^{ikX(1)}\nonumber\\
&  \cdot\left[  U\left(  0,\frac{t}{2}+2,\frac{t}{2}-1\right)  +\frac{2}%
{t+6}\left(  -\frac{t}{2}-1\right)  U\left(  0,\frac{t}{2}+1,\frac{t}%
{2}-1\right)  \frac{e^{L}\cdot\partial X(1)}{e^{P}\cdot\partial X\left(
1\right)  }\right]  ,\\
V_{BPST}^{LLT}  &  =(\frac{t+6}{2M})^{2}\left(  \sqrt{-t}\right)
\Gamma\left(  -1-\frac{t}{2}\right)  \left[  ik_{2}\cdot\partial X(1)\right]
^{1+\frac{t}{2}}e^{ikX(1)}\nonumber\\
&  \cdot\left[
\begin{array}
[c]{c}%
U\left(  0,\frac{t}{2}+2,\frac{t}{2}-1\right)  +\frac{4}{t+6}\left(  -\frac
{t}{2}-1\right)  U\left(  0,\frac{t}{2}+1,\frac{t}{2}-1\right)  \frac
{e^{L}\cdot\partial X(1)}{e^{P}\cdot\partial X\left(  1\right)  }\\
+(\frac{2}{t+6})^{2}\left(  -\frac{t}{2}-1\right)  (-\frac{t}{2})U\left(
0,\frac{t}{2},\frac{t}{2}-1\right)  \left[  \frac{e^{L}\cdot\partial
X(1)}{e^{P}\cdot\partial X\left(  1\right)  }\right]  ^{2}%
\end{array}
\right]  ,\\
V_{BPST}^{LLL}  &  =(\frac{t+6}{2M})^{3}\Gamma\left(  -1-\frac{t}{2}\right)
\left[  ik_{2}\cdot\partial X(1)\right]  ^{1+\frac{t}{2}}e^{ikX(1)}\nonumber\\
&  \cdot\left[
\begin{array}
[c]{c}%
U\left(  0,\frac{t}{2}+2,\frac{t}{2}-1\right)  +\frac{6}{t+6}\left(  -\frac
{t}{2}-1\right)  U\left(  0,\frac{t}{2}+1,\frac{t}{2}-1\right)  \frac
{e^{L}\cdot\partial X(1)}{e^{P}\cdot\partial X\left(  1\right)  }\\
+3(\frac{2}{t+6})^{2}\left(  -\frac{t}{2}-1\right)  (-\frac{t}{2})U\left(
0,\frac{t}{2},\frac{t}{2}-1\right)  \left[  \frac{e^{L}\cdot\partial
X(1)}{e^{P}\cdot\partial X\left(  1\right)  }\right]  ^{2}\\
+(\frac{2}{t+6})^{3}\left(  -\frac{t}{2}-1\right)  (-\frac{t}{2})\left(
-\frac{t}{2}+1\right)  U\left(  0,\frac{t}{2}-1,\frac{t}{2}-1\right)  \left[
\frac{e^{L}\cdot\partial X(1)}{e^{P}\cdot\partial X\left(  1\right)  }\right]
^{3}%
\end{array}
\right]  .
\end{align}
From the above equations, one can easily see that $U\left(  0,\frac{t}%
{2}+2,\frac{t}{2}-1\right)  $ can be expressed in terms of $V_{BPST}^{TTT}$,
$U\left(  0,\frac{t}{2}+1,\frac{t}{2}-1\right)  $ can be expressed in terms of
$(V_{BPST}^{TTT},V_{BPST}^{LTT})$, $U\left(  0,\frac{t}{2},\frac{t}%
{2}-1\right)  $ can be expressed in terms of $(V_{BPST}^{TTT},V_{BPST}%
^{LTT},V_{BPST}^{LLT})$, and finally $U\left(  0,\frac{t}{2}-1,\frac{t}%
{2}-1\right)  $ can be expressed in terms of $(V_{BPST}^{TTT},V_{BPST}%
^{LTT},V_{BPST}^{LLT},V_{BPST}^{LLL})$. We have%
\begin{align}
U\left(  0,\frac{t}{2}+2,\frac{t}{2}-1\right)   &  =\Omega^{-1}\left(
\sqrt{-t}\right)  ^{-3}V_{BPST}^{TTT},\\
U\left(  0,\frac{t}{2}+1,\frac{t}{2}-1\right)   &  =\Omega^{-1}\left(
\sqrt{-t}\right)  ^{-3}\frac{t+6}{t+2}\left[  \frac{e^{P}\cdot\partial
X(1)}{e^{L}\cdot\partial X\left(  1\right)  }\right] \nonumber\\
&  \cdot\left[  V_{BPST}^{TTT}-\frac{2M}{t+6}\sqrt{-t}V_{BPST}^{LTT}\right]
,\\
U\left(  0,\frac{t}{2},\frac{t}{2}-1\right)   &  =\Omega^{-1}\left(  \sqrt
{-t}\right)  ^{-3}\frac{(t+6)^{2}}{t(t+2)}\left[  \frac{e^{P}\cdot\partial
X(1)}{e^{L}\cdot\partial X\left(  1\right)  }\right]  ^{2}\nonumber\\
&  \cdot\left[  V_{BPST}^{TTT}-2\frac{2M}{t+6}\sqrt{-t}V_{BPST}^{LTT}+\left(
\frac{2M}{t+6}\sqrt{-t}\right)  ^{2}V_{BPST}^{LLT}\right]  ,\\
U\left(  0,\frac{t}{2}-1,\frac{t}{2}-1\right)   &  =\Omega^{-1}\left(
\sqrt{-t}\right)  ^{-3}\frac{(t+6)^{3}}{t(t^{2}-4)}\left[  \frac{e^{P}%
\cdot\partial X(1)}{e^{L}\cdot\partial X\left(  1\right)  }\right]
^{3}\nonumber\\
&  \cdot\left[
\begin{array}
[c]{c}%
V_{BPST}^{TTT}-3\frac{2M}{t+6}\sqrt{-t}V_{BPST}^{LTT}\\
+3\left(  \frac{2M}{t+6}\sqrt{-t}\right)  ^{2}V_{BPST}^{LLT}-\left(  \frac
{2M}{t+6}\sqrt{-t}\right)  ^{3}V_{BPST}^{LLL}%
\end{array}
\right]  \label{higher}%
\end{align}
where $\Omega\equiv\Gamma\left(  -1-\frac{t}{2}\right)  \left[  ik_{2}%
\cdot\partial X(1)\right]  ^{1+\frac{t}{2}}e^{ikX(1)}$. To derive an example
of recurrence relation, one notes that Eq.(\ref{RR2}) gives%
\begin{equation}
\frac{t}{2}U\left(  0,\frac{t}{2},\frac{t}{2}-1\right)  -(t-1)U\left(
0,\frac{t}{2}+1,\frac{t}{2}-1\right)  +(\frac{t}{2}-1)U\left(  0,\frac{t}%
{2}+2,\frac{t}{2}-1\right)  =0,
\end{equation}
which leads to the recurrence relation among BPST vertex operators%
\begin{align}
\left[  \left(  \frac{t}{2}-1\right)  -\frac{(t-1)(t+6)}{t+2}\frac{e^{P}%
\cdot\partial X(1)}{e^{L}\cdot\partial X\left(  1\right)  }+\frac{(t+6)^{2}%
}{2(t+2)}\left[  \frac{e^{P}\cdot\partial X(1)}{e^{L}\cdot\partial X\left(
1\right)  }\right]  ^{2}\right]  V_{BPST}^{TTT}  & \nonumber\\
+\left[  \frac{(t-1)}{t+2}\frac{e^{P}\cdot\partial X(1)}{e^{L}\cdot\partial
X\left(  1\right)  }-\frac{(t+6)}{(t+2)}\left[  \frac{e^{P}\cdot\partial
X(1)}{e^{L}\cdot\partial X\left(  1\right)  }\right]  ^{2}\right]
(2M\sqrt{-t})V_{BPST}^{LTT}  & \nonumber\\
+\left[  \frac{1}{2(t+2)}\left[  \frac{e^{P}\cdot\partial X(1)}{e^{L}%
\cdot\partial X\left(  1\right)  }\right]  ^{2}\right]  (2M\sqrt{-t}%
)^{2}V_{BPST}^{LLT}  &  =0. \label{BPST}%
\end{align}
Again one can use Eq.(\ref{BPST}) to deduce recurrence relation among Regge
string scattering amplitudes \cite{Tan}%
\begin{equation}
(t+22)A^{(p_{1}=3)}-14M\sqrt{-t}A^{(p_{1}=2,r_{1}=1)}+2M^{2}(\sqrt{-t}%
)^{2}A^{(p_{1}=1,r_{1}=2)}=0. \label{LY}%
\end{equation}
Other recurrence relations of Kummer functions can be used to derive more
recurrence relations among BPST vertex operators. For example, Eq.(\ref{RR2})
gives a recurrence relation of $U\left(  0,\frac{t}{2}+1,\frac{t}{2}-1\right)
$ and its associated functions $U\left(  0,\frac{t}{2}-1,\frac{t}{2}-1\right)
$ and $U\left(  0,\frac{t}{2}+2,\frac{t}{2}-1\right)  $%
\begin{equation}
tU\left(  0,\frac{t}{2}-1,\frac{t}{2}-1\right)  -(3t-4)U\left(  0,\frac{t}%
{2}+1,\frac{t}{2}-1\right)  +2(t-2)U\left(  0,\frac{t}{2}+2,\frac{t}%
{2}-1\right)  =0,
\end{equation}
which leads to the recurrence relation among BPST vertex operators%
\begin{align}
\left[  2(t-2)-\frac{(3t-4)(t+6)}{t+2}\frac{e^{P}\cdot\partial X(1)}%
{e^{L}\cdot\partial X\left(  1\right)  }+\frac{(t+6)^{3}}{(t^{2}-4)}\left[
\frac{e^{P}\cdot\partial X(1)}{e^{L}\cdot\partial X\left(  1\right)  }\right]
^{3}\right]  V_{BPST}^{TTT}  & \nonumber\\
+\left[  \frac{(3t-4)}{t+2}\frac{e^{P}\cdot\partial X(1)}{e^{L}\cdot\partial
X\left(  1\right)  }-3\frac{(t+6)^{2}}{(t^{2}-4)}\left[  \frac{e^{P}%
\cdot\partial X(1)}{e^{L}\cdot\partial X\left(  1\right)  }\right]
^{3}\right]  (2M\sqrt{-t})V_{BPST}^{LTT}  & \nonumber\\
+\left[  \frac{3(t+6)}{(t^{2}-4)}\left[  \frac{e^{P}\cdot\partial X(1)}%
{e^{L}\cdot\partial X\left(  1\right)  }\right]  ^{3}\right]  (2M\sqrt
{-t})^{2}V_{BPST}^{LLT}  & \nonumber\\
-\left[  \frac{1}{(t^{2}-4)}\left[  \frac{e^{P}\cdot\partial X(1)}{e^{L}%
\cdot\partial X\left(  1\right)  }\right]  ^{3}\right]  (2M\sqrt{-t}%
)^{3}V_{BPST}^{LLL}  &  =0. \label{BPST2}%
\end{align}
one can use Eq.(\ref{BPST2}) to deduce recurrence relation among Regge string
scattering amplitudes \cite{Tan}%
\begin{align}
(3t^{2}+76t+92)A^{(p_{1}=3)}-2(23t+50)M\sqrt{-t}A^{(p_{1}=2,r_{1}=1)}  &
\nonumber\\
+6M^{2}(t+6)(\sqrt{-t})^{2}A^{(p_{1}=1,r_{1}=2)}-4M^{3}(\sqrt{-t}%
)^{3}A^{(r_{1}=3)}  &  =0.
\end{align}
Similarly, we can consider groups of BPST vertex operators $(V_{BPST}%
^{PT},V_{BPST}^{PL})$, $(V_{BPST}^{LT},V_{BPST}^{LL})$ and $(V_{BPST}%
^{TT},V_{BPST}^{TL})$ with $q_{1}=0$; group of BPST vertex operators
$(V_{BPST}^{PTT},V_{BPST}^{PLT},V_{BPST}^{PLL})$ with $q_{1}=1$ and group of
BPST vertex operators $(V_{BPST}^{PPT},V_{BPST}^{PPL})$ with $q_{1}=2$. All
the remaining $7$ BPST vertex operators are with $r_{1}=0$, and each BPST
vertex operators contains only one Kummer function. Thus all Kummer functions
involved at mass level $M^{2}=4$ can be algebraically solved and expressed in
terms of BPST vertex operators. One can then use recurrence relations of
Kummer functions to derive more recurrence relations among BPST vertex operators.

\subsection{Arbitrary mass levels}

In this section, we solve the Kummer functions in terms of the highest spin
string states scattering amplitudes for arbitrary mass levels. The highest
spin string states\ at the mass level $M^{2}=2\left(  N-1\right)  $ are
defined as%
\begin{equation}
\left\vert N-q_{1}-r_{1},q_{1},r_{1}\right\rangle =\left(  \alpha_{-1}%
^{T}\right)  ^{N-q_{1}-r_{1}}\left(  \alpha_{-1}^{P}\right)  ^{q_{1}}\left(
\alpha_{-1}^{L}\right)  ^{r_{1}}|0,k\rangle
\end{equation}
where only $\alpha_{-1}$ operator appears. The highest spin string states BPST
vertex operators can be easily obtained from Eq.(\ref{factor2.}) as%
\begin{align}
&  \left(  V^{T}\right)  ^{N-q_{1}-r_{1}}\left(  V^{P}\right)  ^{q_{1}}\left(
V^{L}\right)  ^{r_{1}}\equiv V_{BPST}^{(N-q_{1}-r_{1},q_{1},r_{1})}\nonumber\\
&  =\Gamma\left(  -\frac{t}{2}-1\right)  \left[  ik_{2}\cdot\partial
X(1)\right]  ^{1+\frac{t}{2}}e^{ikX(1)}\left(  \sqrt{-t}\right)
^{N-q_{1}-r_{1}}\left(  -\frac{1}{M}\right)  ^{q_{1}}\left(  \frac{\tilde
{t}^{\prime}}{2M}\right)  ^{r_{1}}\nonumber\\
&  \cdot\sum_{j=0}^{r_{1}}\binom{r_{1}}{j}\left(  \frac{2}{\tilde{t}^{\prime}%
}\frac{e^{L}\cdot\partial X(1)}{e^{P}\cdot\partial X\left(  1\right)
}\right)  ^{j}\left(  -\frac{t}{2}-1\right)  _{j}U\left(  -q_{1},\frac{t}%
{2}+2-j-q_{1},\frac{\tilde{t}}{2}\right)  . \label{highest spin vertex}%
\end{align}
In view of the form of Eq.(\ref{higher}), we can solve the Kummer function
from Eq.(\ref{highest spin vertex}) and express it in terms of the highest
spin BPST vertex operators as%
\begin{align}
U\left(  -q_{1},\frac{t}{2}+2-q_{1}-r_{1},\frac{\tilde{t}}{2}\right)   &
=\frac{\Gamma\left(  -\frac{t}{2}-1\right)  }{\left(  -\frac{t}{2}-1\right)
_{r_{1}}}\left[  ik_{2}\cdot\partial X(1)\right]  ^{1+\frac{t}{2}}%
e^{ikX(1)}\nonumber\\
&  \cdot\left(  -MV^{P}\right)  ^{q_{1}}\left(  \frac{V^{T}}{\sqrt{-t}%
}\right)  ^{N-q_{1}}\left[  \frac{e^{P}\cdot\partial X(1)}{e^{L}\cdot\partial
X\left(  1\right)  }\left(  \sqrt{-t}M\frac{V^{L}}{V^{T}}-\frac{\tilde
{t}^{\prime}}{2}\right)  \right]  ^{r_{1}}. \label{Kummer functions}%
\end{align}
Putting the Kummer functions Eq.(\ref{Kummer functions}) into the recurrence
relations Eqs.(\ref{RR1}-\ref{RR6}), we can then obtain recurrence relations
among BPST vertex operators.

Let us consider, for example, the recurrence relation%
\begin{equation}
\left(  c-a-1\right)  U(a,c-1,x)-(x+c-1))U(a,c,x)+xU(a,c+1,x)=0.
\end{equation}
With%
\begin{equation}
a=-q_{1},c=\frac{t}{2}+1-q_{1}-r_{1},x=\frac{\tilde{t}}{2}=\frac{t-M^{2}+2}%
{2},
\end{equation}
the above recurrence relation becomes%
\begin{align}
\left(  \frac{t}{2}-r_{1}\right)  U\left(  -q_{1},\frac{t}{2}-q_{1}%
-r_{1},\frac{\tilde{t}}{2}\right)   & \nonumber\\
-\left(  \frac{\tilde{t}}{2}+\frac{t}{2}-q_{1}-r_{1}\right)  U\left(
-q_{1},\frac{t}{2}+1-q_{1}-r_{1},\frac{\tilde{t}}{2}\right)   & \nonumber\\
+\frac{\tilde{t}}{2}U\left(  -q_{1},\frac{t}{2}+2-q_{1}-r_{1},\frac{\tilde{t}%
}{2}\right)   &  =0.
\end{align}
Plug the Kummer functions Eq.(\ref{Kummer functions}) into the above
recurrence relation, we obtain the recurrence relation among BPST vertex
operators at general mass level $N$%
\begin{equation}
\left(  V^{P}\right)  ^{q_{1}}\left(  V^{T}\right)  ^{N-q_{1}}\left(
X\right)  ^{r_{1}}\left[  X^{2}+\left(  \frac{\tilde{t}}{2}+\frac{t}{2}%
-q_{1}-r_{1}\right)  X+\frac{\tilde{t}}{2}\left(  \frac{t}{2}+1-r_{1}\right)
\right]  =0
\end{equation}
where we have defined%
\begin{equation}
X\equiv\frac{e^{P}\cdot\partial X(1)}{e^{L}\cdot\partial X\left(  1\right)
}\left(  \sqrt{-t}M\frac{V^{L}}{V^{T}}-\frac{\tilde{t}^{\prime}}{2}\right)
=\frac{e^{P}\cdot\partial X(1)}{e^{L}\cdot\partial X\left(  1\right)  }\left(
\sqrt{-t}M\frac{V^{L}}{V^{T}}-\frac{t+M^{2}+2}{2}\right)  .
\end{equation}
As an example, at the mass level $M^{2}=4$ with $q_{1}=r_{1}=0$, we get%
\begin{equation}
\left(  V^{T}\right)  ^{3}\left[  X^{2}+\left(  t-1\right)  X+\left(
\frac{t^{2}}{4}-1\right)  \right]  =0 \label{example}%
\end{equation}
where%
\begin{equation}
X=\frac{e^{P}\cdot\partial X(1)}{e^{L}\cdot\partial X\left(  1\right)
}\left(  \sqrt{-t}M\frac{V^{L}}{V^{T}}-\frac{t+6}{2}\right)  .
\end{equation}
A simple calculation shows that Eq.(\ref{example}) is exactly the same as
Eq.(\ref{BPST}), and the same recurrence relation among Regge string
scattering amplitudes Eq.(\ref{LY}) follows.%

\setcounter{equation}{0}
\renewcommand{\theequation}{\arabic{section}.\arabic{equation}}%

\section{Regge string scattered from D-particle}

In this chapter we study \cite{LMY} scattering of higher spin closed string
states at arbitrary mass levels from D-particle in the RR. The scattering of
massless string states from D-brane was well studied in the literature and can
be found in
\cite{Klebanov,Myers,Klebanov3,barbon1996d,bachas1999high,hirano1997scattering}
Since the mass of D-brane scales as the inverse of the string coupling
constant $1/g$, it was assumed that it was infinitely heavy to leading order
in $g$ and did not recoil.

We will extract the \textit{complete} infinite ratios in Eq.(\ref{mainA})
among high energy amplitudes of different string states in the fixed angle
regime from these Regge string scattering amplitudes. The complete ratios
calculated by this indirect method include a subset of ratios in
Eq.(\ref{D-particle}) calculated previously by direct fixed angle calculation
\cite{Dscatt}.

More importantly, we discover that the RR amplitudes calculated in this
chapter for closed string D-particle scatterings can NOT be factorized and
thus are different from amplitudes for the high-energy closed string-string
scattering calculated previously \cite{Closed,bosonic2}. GR Amplitudes for the
high-energy closed string-string scattering calculated in chapter VII can be
factorized into two open string scattering amplitudes by using a calculation
\cite{bosonic2} based on the KLT formula \cite{KLT}. Similarly the RR closed
string-string amplitudes \cite{Closed} can be factorized too. Presumably, this
non-factorization is due to the non-existence of a KLT-like formula for the
string D-brane scattering amplitudes. There is no physical picture for open
string D-particle tree scattering amplitudes and thus no factorization for
closed string D-particle scatterings into two channels of open string
D-particle scatterings.

However, surprisingly, we will find \cite{LMY} that in spite of the
non-factorizability of the closed string D-particle scattering amplitudes, the
complete ratios derived for the fixed angle regime are found to be
\textit{factorized}. These ratios are consistent with the decoupling of
high-energy ZNS calculated in Eq.(\ref{mainA}) of chapter V.
\cite{ChanLee,ChanLee1,ChanLee2, CHL,CHLTY1,CHLTY2,CHLTY3,susy}.

\subsection{Kinematics Set-up}

In this chapter, we consider an incoming string state with momentum $k_{2}$
scattered from an infinitely heavy D-particle and end up with string state
with momentum $k_{1}$in the RR. The high energy scattering plane will be
assumed to be the $X-Y$ plane, and the momenta are arranged to be%
\begin{align}
k_{1}  &  =\left(  E,\mathrm{k}_{1}\cos\phi,-\mathrm{k}_{1}\sin\phi\right)
,\\
k_{2}  &  =\left(  -E,-\mathrm{k}_{2},0\right)
\end{align}
where%
\begin{equation}
E=\sqrt{\mathrm{k}_{2}^{2}+M_{2}^{2}}=\sqrt{\mathrm{k}_{1}^{2}+M_{1}^{2}},
\label{II3}%
\end{equation}
and $\phi$ is the scattering angle. For simplicity, we will calculate the disk
amplitude in this paper. The relevant propagators for the left-moving string
coordinate $X^{\mu}\left(  z\right)  $ and the right-moving one $\tilde
{X}^{\nu}\left(  \bar{w}\right)  $ are%
\begin{align}
\left\langle X^{\mu}\left(  z\right)  ,X^{\nu}\left(  w\right)  \right\rangle
&  =-\eta^{\mu\nu}\left\langle X\left(  z\right)  ,X\left(  w\right)
\right\rangle =-\eta^{\mu\nu}\ln\left(  z-w\right)  ,\label{DD1}\\
\left\langle \tilde{X}^{\mu}\left(  \bar{z}\right)  ,\tilde{X}^{\nu}\left(
\bar{w}\right)  \right\rangle  &  =-\eta^{\mu\nu}\left\langle \tilde{X}\left(
\bar{z}\right)  ,\tilde{X}\left(  \bar{w}\right)  \right\rangle =-\eta^{\mu
\nu}\ln\left(  \bar{z}-\bar{w}\right)  ,\label{DD2}\\
\left\langle X^{\mu}\left(  z\right)  ,\tilde{X}^{\nu}\left(  \bar{w}\right)
\right\rangle  &  =-D^{\mu\nu}\left\langle X\left(  z\right)  ,\tilde
{X}\left(  \bar{w}\right)  \right\rangle =-D^{\mu\nu}\ln\left(  1-z\bar
{w}\right)  \text{\ \ (for Disk)} \label{DDD}%
\end{align}
where matrix $D$ has the standard form for the fields satisfying Neumann
boundary condition, while $D$ reverses the sign for the fields satisfying
Dirichlet boundary condition. Instead of the Mandelstam variables used in the
string-string scatterings, we define%
\begin{align}
a_{0}  &  \equiv k_{1}\cdot D\cdot k_{1}=-E^{2}-\mathrm{k}_{1}^{2}\sim
-2E^{2},\\
a_{0}^{\prime}  &  \equiv k_{2}\cdot D\cdot k_{2}=-E^{2}-\mathrm{k}_{2}%
^{2}\sim-2E^{2},\\
b_{0}  &  \equiv2k_{1}\cdot k_{2}+1=2\left(  E^{2}-\mathrm{k}_{1}%
\mathrm{k}_{2}\cos\phi\right)  +1=fixed,\label{II4}\\
c_{0}  &  \equiv2k_{1}\cdot D\cdot k_{2}+1=2\left(  E^{2}+\mathrm{k}%
_{1}\mathrm{k}_{2}\cos\phi\right)  +1,
\end{align}
so that%
\begin{equation}
2a_{0}+b_{0}+c_{0}=2M_{1}^{2}+2.
\end{equation}
Since we are going to calculate Regge scattering amplitudes, $b_{0}=fixed$. We
can use Eq.(\ref{II3}) and Eq.(\ref{II4}) to calculate%
\begin{align}
\cos\phi\sim &  1-\frac{b_{0}-M_{1}^{2}-M_{2}^{2}-1}{2\mathrm{k}_{1}^{2}}\\
\sin\phi\sim &  \frac{\sqrt{b_{0}-M_{1}^{2}-M_{2}^{2}-1}}{\mathrm{k}_{1}%
}\equiv\frac{\sqrt{\tilde{b}_{0}}}{\mathrm{k}_{1}}%
\end{align}
The normalized polarization vectors on the high energy scattering plane of the
$k_{2}$ string state are defined to be \cite{ChanLee,ChanLee1,ChanLee2}
\begin{equation}
e_{P}=\frac{1}{M_{2}}(-E,-\mathrm{k}_{2},0)=\frac{k_{2}}{M_{2}},
\end{equation}%
\begin{equation}
e_{L}=\frac{1}{M_{2}}(-\mathrm{k}_{2},-E,0),
\end{equation}%
\begin{equation}
e_{T}=(0,0,1).
\end{equation}
One can then easily calculate the following kinematics%
\begin{align}
e^{T}\cdot k_{2}  &  =0,\nonumber\\
e^{T}\cdot k_{1}  &  =-\mathrm{k}_{1}\sin\phi\sim-\sqrt{\tilde{b}_{0}%
},\nonumber\\
e^{T}\cdot D\cdot k_{1}  &  =\mathrm{k}_{1}\sin\phi\sim\sqrt{\tilde{b}_{0}%
},\nonumber\\
e^{T}\cdot D\cdot k_{2}  &  =0,\nonumber\\
e^{P}\cdot k_{2}  &  =-M_{2},\nonumber\\
e^{P}\cdot k_{1}  &  =\frac{1}{M_{2}}\left[  E^{2}-\mathrm{k}_{1}%
\mathrm{k}_{2}\cos\phi\right]  =\frac{b_{0}-1}{2M_{2}},\nonumber\\
e^{P}\cdot D\cdot k_{1}  &  =\frac{1}{M_{2}}\left[  E^{2}+\mathrm{k}%
_{1}\mathrm{k}_{2}\cos\phi\right]  =\frac{c_{0}-1}{2M_{2}},\nonumber\\
e^{P}\cdot D\cdot k_{2}  &  =\frac{1}{M_{2}}\left[  -E^{2}-\mathrm{k}_{2}%
^{2}\right]  =\frac{a_{0}^{\prime}}{M_{2}}\sim\frac{a_{0}}{M_{2}},\nonumber\\
e^{T}\cdot D\cdot e^{T}  &  =-1,\nonumber\\
e^{T}\cdot D\cdot e^{P}  &  =e^{P}\cdot D\cdot e^{T}=0,\nonumber\\
e^{P}\cdot D\cdot e^{P}  &  =\frac{1}{M_{2}^{2}}\left[  -E^{2}-\mathrm{k}%
_{2}^{2}\right]  =\frac{a_{0}^{\prime}}{M_{2}^{2}}\sim\frac{a_{0}}{M_{2}^{2}},
\label{Kine}%
\end{align}
which will be useful in the amplitude calculation in the next section.

\subsection{Regge String D-particle scatterings}

\bigskip We now begin to calculate the scattering amplitudes. For simplicity,
we will take $k_{1}$ to be the tachyon and $k_{2}$ to be the tensor states.
One can easily argue that a class of high energy string states for $k_{2}$ in
the RR are \cite{bosonic,RRsusy}%
\begin{equation}
|p_{n},p_{n}^{\prime},q_{m},q_{m}^{\prime}\rangle=\left[  \prod_{n>0}\left(
\alpha_{-n}^{T}\right)  ^{p_{n}}\prod_{m>0}\left(  \alpha_{-m}^{P}\right)
^{q_{m}}\right]  \left[  \prod_{n>0}\left(  \tilde{\alpha}_{-n}^{T}\right)
^{p_{n}^{\prime}}\prod_{m>0}\left(  \tilde{\alpha}_{-m}^{P}\right)
^{q_{m}^{\prime}}\right]  |0,k\rangle\label{general states}%
\end{equation}
with%
\begin{align}
\sum_{n}n\left(  p_{n}-p_{n}^{\prime}\right)  +\sum_{m}m\left(  q_{m}%
-q_{m}^{\prime}\right)   &  =0,\\
\sum_{n}n\left(  p_{n}+p_{n}^{\prime}\right)  +\sum_{m}m\left(  q_{m}%
+q_{m}^{\prime}\right)   &  =N=\text{const}%
\end{align}
where $M_{2}^{2}=(N-2).$

\subsubsection{An example}

Before calculating the string D-particle scattering amplitudes for general
cases, we take an example and illustrate the method of calculation. We
consider the case
\begin{equation}
p_{1}=p_{1}^{\prime}=q_{1}=q_{1}^{\prime}=q_{2}=q_{2}^{\prime}=1,\quad
\mathrm{others}=0. \label{example-state}%
\end{equation}
As we will see in the next section, the string D-particle scattering
amplitudes with the general states (\ref{general states}) are reduced to
simple forms in the Regge limit, in which most of the ways of contracting the
operators are discarded as subleading. For a fixed number of the contractions
between $\partial X^{P}$ and $\bar{\partial}\tilde{X}^{P}$, the ways of
contracting the other factors are determined by the following rules.
\begin{align}
\alpha_{-n}^{T}  &  \quad\text{1 term (contraction of $ik_{1}X$ with
$\partial_{n}X^{T}$)}\label{rule1}\\
\tilde{\alpha}_{-n}^{T}  &  \quad\text{1 term (contraction of $ik_{1}\tilde
{X}$ with $\bar{\partial}_{n}\tilde{X}^{T}$)}\label{rule2}\\
\alpha_{-n}^{P}  &  \quad%
\begin{cases}
\left(  n>1\right)  \quad\text{1 term (contraction of $ik_{1}X$ with
$\partial_{n}X^{P}$)}\\
\left(  n=1\right)  \quad\text{2 terms (contraction of $ik_{1}X$ and $ik_{2}X$
with $\partial X^{P}$)}%
\end{cases}
\label{rule3}\\
\tilde{\alpha}_{-n}^{P}  &  \quad%
\begin{cases}
\left(  n>1\right)  \quad\text{1 term (contraction of $ik_{1}\tilde{X}$ with
$\bar{\partial}_{n}\tilde{X}^{P}$ )}\\
\left(  n=1\right)  \quad\text{2 terms (contraction of $ik_{1}\tilde{X}$ and
$ik_{2}\tilde{X}$ with $\bar{\partial}\tilde{X}^{P}$ )}%
\end{cases}
\label{rule4}%
\end{align}
Therefore we take the state Eq.(\ref{example-state}) as the simplest example
for the purpose of this section.

We start with the procedure in \cite{KLT} to treat the vertex operator
corresponding to the state (\ref{example-state}).
\begin{align}
V  &  =i^{6}\varepsilon_{\mu_{1}\cdots\mu_{6}}:\partial X^{\mu_{1}}\partial
X^{\mu_{2}} \partial^{2} X^{\mu_{3}}e^{ik_{2} X}\left(  z\right)
:\ :\bar{\partial} \tilde{X}^{\mu_{4}}\bar{\partial} \tilde{X}^{\mu_{5}}
\bar{\partial}^{2} \tilde{X}^{\mu_{6}}e^{ik_{2} \tilde{X}}\left(  \bar
{z}\right)  :\nonumber\\
&  =i^{6}:\partial X^{T}\partial X^{P} \partial^{2} X^{P}e^{ik_{2} X}\left(
z\right)  :\ :\bar{\partial} \tilde{X}^{T}\bar{\partial} \tilde{X}^{P}
\bar{\partial}^{2} \tilde{X}^{P}e^{ik_{2} \tilde{X}}\left(  \bar{z}\right)
:\nonumber\\
&  = i^{6} \left[  :\exp\left\{  ik_{2} X(z) + \varepsilon_{T}^{(1)}\partial
X^{T} (z) + \varepsilon_{P}^{(1)}\partial X^{P} (z) + \varepsilon_{P}%
^{(2)}\partial^{2} X ^{P} (z) \right\}  : \right. \nonumber\\
&  \quad\quad\times\left.  :\exp\left\{  ik_{2} \tilde{X}(\bar{z}) +
\varepsilon_{T}^{\prime(1)}\partial\tilde{X}^{T} (\bar{z}) + \varepsilon
_{P}^{\prime(1)}\partial\tilde{X}^{P} (\bar{z}) + \varepsilon_{P}^{\prime
(2)}\partial^{2} \tilde{X}^{P} (\bar{z}) \right\}  : \right]
_{\mathrm{linear\ terms}} \label{exponentiation-ex}%
\end{align}
In the last equation, we have introduced the dummy variables $\varepsilon
_{T}^{(1)}, \varepsilon_{P}^{(1)}, \varepsilon_{P}^{(2)}, \varepsilon
_{T}^{\prime(1)},\varepsilon_{P}^{\prime(1)},\varepsilon_{P}^{\prime(2)}$
associated with the non-vanishing component $\varepsilon_{TPPTPP}$ of the
polarization tensor and written the operator in the exponential form. ``linear
terms'' indicate that we take the sum of the terms linear in all of
$\varepsilon_{T}^{(1)},\varepsilon_{P}^{(1)}, \varepsilon_{P}^{(2)},
\varepsilon_{T}^{\prime(1)},\varepsilon_{P}^{\prime(1)}$, and $\varepsilon
_{P}^{\prime(2)}$. This sum can be rephrased as the coefficient of the product
$\varepsilon_{T}^{(1)} \varepsilon_{P}^{(1)} \varepsilon_{P}^{(2)}
\varepsilon_{T}^{\prime(1)}\varepsilon_{P}^{\prime(1)} \varepsilon_{P}%
^{\prime(2)}$ because we set the dummy variables to be 1 at the end of calculation.

The string D-particle scattering amplitudes can be calculated to be
\begin{align}
A  &  =\int d^{2}z_{1}d^{2}z_{2}\left(  1-z_{1}\bar{z}_{1}\right)  ^{a_{0}%
}\left(  1-z_{2}\bar{z}_{2}\right)  ^{a_{0}^{\prime}}\left\vert z_{1}%
-z_{2}\right\vert ^{b_{0}-1}\left\vert 1-z_{1}\bar{z}_{2}\right\vert
^{c_{0}-1}\nonumber\\
&  \cdot\Big[\exp\Big\{\nonumber\\
&  \varepsilon_{T}^{(1)}\left[  \dfrac{ie^{T}k_{1}}{\left(  z_{1}%
-z_{2}\right)  }+\dfrac{ie^{T}Dk_{1}\bar{z}_{1}}{\left(  1-\bar{z}_{1}%
z_{2}\right)  }+\dfrac{ie^{T}Dk_{2}\bar{z}_{2}}{\left(  1-\bar{z}_{2}%
z_{2}\right)  }\right]  +\varepsilon_{T}^{\prime(1)}\left[  \dfrac
{ie^{T}Dk_{1}z_{1}}{\left(  1-z_{1}\bar{z}_{2}\right)  }+\dfrac{ie^{T}k_{1}%
}{\left(  \bar{z}_{1}-\bar{z}_{2}\right)  }+\dfrac{ie^{T}Dk_{2}z_{2}}{\left(
1-z_{2}\bar{z}_{2}\right)  }\right] \nonumber\\
&  +\varepsilon_{P}^{(1)}\left[  \dfrac{ie^{P}k_{1}}{\left(  z_{1}%
-z_{2}\right)  }+\dfrac{ie^{P}Dk_{1}\bar{z}_{1}}{\left(  1-\bar{z}_{1}%
z_{2}\right)  }+\dfrac{ie^{P}Dk_{2}\bar{z}_{2}}{\left(  1-\bar{z}_{2}%
z_{2}\right)  }\right]  +\varepsilon_{P}^{(2)}\left[  \dfrac{ie^{P}k_{1}%
}{\left(  z_{1}-z_{2}\right)  ^{2}}+\dfrac{ie^{P}Dk_{1}\bar{z}_{1}^{2}%
}{\left(  1-\bar{z}_{1}z_{2}\right)  ^{2}}+\dfrac{ie^{P}Dk_{2}\bar{z}_{2}^{2}%
}{\left(  1-\bar{z}_{2}z_{2}\right)  ^{2}}\right] \nonumber\\
&  +\varepsilon_{P}^{\prime(1)}\left[  \dfrac{ie^{P}Dk_{1}z_{1}}{\left(
1-z_{1}\bar{z}_{2}\right)  }+\dfrac{ie^{P}k_{1}}{\left(  \bar{z}_{1}-\bar
{z}_{2}\right)  }+\dfrac{ie^{P}Dk_{2}z_{2}}{\left(  1-z_{2}\bar{z}_{2}\right)
}\right]  +\varepsilon_{P}^{\prime(2)}\left[  \dfrac{ie^{P}Dk_{1}z_{1}^{2}%
}{\left(  1-z_{1}\bar{z}_{2}\right)  ^{2}}+\dfrac{ie^{P}k_{1}}{\left(  \bar
{z}_{1}-\bar{z}_{2}\right)  ^{2}}+\dfrac{ie^{P}Dk_{2}z_{2}^{2}}{\left(
1-z_{2}\bar{z}_{2}\right)  ^{2}}\right] \nonumber\\
&  +\varepsilon_{T}^{(1)}\varepsilon_{T}^{\prime(1)}\dfrac{e^{T}De^{T}%
}{\left(  1-z_{2}\bar{z}_{2}\right)  ^{2}}\nonumber\\
&  +\varepsilon_{P}^{(1)}\varepsilon_{P}^{\prime(1)}\dfrac{e^{P}De^{P}%
}{\left(  1-z_{2}\bar{z}_{2}\right)  ^{2}}+2\varepsilon_{P}^{(1)}%
\varepsilon_{P}^{\prime(2)}\dfrac{e^{P}De^{P}z_{2}}{\left(  1-z_{2}\bar{z}%
_{2}\right)  ^{3}}+2\varepsilon_{P}^{(2)}\varepsilon_{P}^{\prime(1)}%
\dfrac{e^{P}De^{P}\bar{z}_{2}}{\left(  1-z_{2}\bar{z}_{2}\right)  ^{3}%
}+2\varepsilon_{P}^{(2)}\varepsilon_{P}^{\prime(2)}\dfrac{e^{P}De^{P}\left(
1+2z_{2}\bar{z}_{2}\right)  }{\left(  1-z_{2}\bar{z}_{2}\right)  ^{4}%
}\nonumber\\
&  +\varepsilon_{T}^{(1)}\varepsilon_{P}^{\prime(1)}\dfrac{e^{T}De^{P}%
}{\left(  1-z_{2}\bar{z}_{2}\right)  ^{2}}+2\varepsilon_{T}^{(1)}%
\varepsilon_{P}^{\prime(2)}\dfrac{e^{T}De^{P}z_{2}}{\left(  1-z_{2}\bar{z}%
_{2}\right)  ^{3}}+\varepsilon_{P}^{(1)}\varepsilon_{T}^{\prime(1)}%
\dfrac{e^{P}De^{T}}{\left(  1-z_{2}\bar{z}_{2}\right)  ^{2}}+2\varepsilon
_{P}^{(2)}\varepsilon_{T}^{\prime(1)}\dfrac{e^{P}De^{T}\bar{z}_{2}}{\left(
1-z_{2}\bar{z}_{2}\right)  ^{3}}\nonumber\\
&  \Big\}\Big]_{\text{linear terms}} \label{def-amp}%
\end{align}
To fix the $SL(2,R)$ modulus group on the disk, we set $z_{1}=0$ and $z_{2}%
=r$, then $d^{2}z_{1}d^{2}z_{2}=d\left(  r^{2}\right)  .$ By using
Eq.(\ref{Kine}), the amplitude can then be reduced to%
\begin{align}
A  &  =\int_{0}^{1}d\left(  r^{2}\right)  \left(  1-r^{2}\right)
^{a_{0}^{\prime}}r^{b_{0}-1}\nonumber\\
&  \cdot\Big[\nonumber\\
&  \exp\left\{
\begin{array}
[c]{l}%
\varepsilon_{T}^{(1)}\left[  -\dfrac{i\sqrt{\tilde{b}_{0}}}{-r}\right]
+\varepsilon_{T}^{\prime(1)}\left[  -\dfrac{i\sqrt{\tilde{b}_{0}}}{-r}\right]
\\
+\varepsilon_{P}^{(1)}\left[  \dfrac{i\frac{b_{0}-1}{2M_{2}}}{-r}%
+\dfrac{i\frac{a_{0}}{M_{2}}}{\left(  1-r^{2}\right)  /r}\right]
+\varepsilon_{P}^{(2)}\left[  \dfrac{i\frac{b_{0}-1}{2M_{2}}}{\left(
-r\right)  ^{2}}+\dfrac{i\frac{a_{0}}{M_{2}}}{\left[  \left(  1-r^{2}\right)
/r\right]  ^{2}}\right] \\
+\varepsilon_{P}^{\prime(1)}\left[  \dfrac{i\frac{b_{0}-1}{2M_{2}}}{-r}%
+\dfrac{i\frac{a_{0}}{M_{2}}}{\left(  1-r^{2}\right)  /r}\right]
+\varepsilon_{P}^{\prime(2)}\left[  \dfrac{i\frac{b_{0}-1}{2M_{2}}}{\left(
-r\right)  ^{2}}+\dfrac{i\frac{a_{0}}{M_{2}}}{\left[  1-r^{2}/r\right]  ^{2}%
}\right] \\
-\varepsilon_{T}^{(1)}\varepsilon_{T}^{\prime(1)}\dfrac{1}{\left(
1-r^{2}\right)  ^{2}}\\
+\varepsilon_{P}^{(1)}\varepsilon_{P}^{\prime(1)}\dfrac{\frac{a_{0}}{M_{2}%
^{2}}}{\left(  1-r^{2}\right)  ^{2}}+2\varepsilon_{P}^{(1)}\varepsilon
_{P}^{\prime(2)}\dfrac{\frac{a_{0}}{M_{2}^{2}}r}{\left(  1-r^{2}\right)  ^{3}%
}+2\varepsilon_{P}^{(2)}\varepsilon_{P}^{\prime(1)}\dfrac{\frac{a_{0}}%
{M_{2}^{2}}r}{\left(  1-r^{2}\right)  ^{3}}+2\varepsilon_{P}^{(2)}%
\varepsilon_{P}^{\prime(2)}\dfrac{\frac{a_{0}}{M_{2}^{2}}\left(
1+2r^{2}\right)  }{\left(  1-r^{2}\right)  ^{4}}%
\end{array}
\right\} \nonumber\\
&  \Big]_{\text{linear terms}} \label{ex1}%
\end{align}

Although in Eq.(\ref{ex1}) we have dropped several subleading terms by using
the kinematic relations Eq.(\ref{Kine}), Eq.(\ref{ex1}) still has subleading
terms. We can see that by performing the integration of a generic term in
Eq.(\ref{ex1}) and looking at its behavior in the Regge limit explicitly.
\begin{align}
\int_{0}^{1} d\left(  r^{2}\right)  \left(  1-r^{2}\right)  ^{a_{0}^{\prime
}+n_{a}}r^{b_{0}-1-N+n_{b}}  &  =B\left(  a^{\prime}_{0} +1 +n_{a},
\frac{b_{0} -N+1}{2}+\frac{n_{b}}{2}\right) \nonumber\\
&  =B\left(  a^{\prime}_{0} +1 , \frac{b_{0} -N+1}{2} \right)  \frac{\left(
a^{\prime}_{0} +1\right)  _{n_{a}} \left(  \frac{b_{0} -N+1}{2}\right)
_{\frac{n_{b}}{2}}} {\left(  a^{\prime}_{0} +1 + \frac{b_{0} -N+1}{2}\right)
_{n_{a} +\frac{n_{b}}{2}}}\nonumber\\
&  \sim B\left(  a_{0} +1 , \frac{b_{0} -N+1}{2} \right)  \left(  \frac{b_{0}
-N+1}{2}\right)  _{\frac{n_{b}}{2}} \left(  a_{0} \right)  ^{-\frac{n_{b}}{2}}
\label{Regge behavior}%
\end{align}
Here the Pochhammer symbol is defined by $\left(  x\right)  _{y}=\frac
{\Gamma\left(  x+y\right)  }{\Gamma\left(  x\right)  }$ , which, if $y$ is a
positive integer, is reduced to $\left(  x\right)  _{y} = x (x+1)(x+2)\cdots
(x+y-1). $ From the Regge behavior Eq.(\ref{Regge behavior}), we see that
increasing one power of $1/r$ in the integrand results in increasing one-half
power of $a_{0}$. Thus we obtain the following rules to determine which terms
in the exponent of Eq.(\ref{ex1}) contribute to the leading behavior of the
amplitude:
\begin{equation}
1/r\rightarrow E,\quad a_{0}\rightarrow E^{2}. \label{rule5}%
\end{equation}

We can now drop the subleading terms in energy to get%
\begin{align}
A  &  =\int_{0}^{1}d\left(  r^{2}\right)  \left(  1-r^{2}\right)
^{a_{0}^{\prime}}r^{b_{0}-1}\nonumber\\
&  \cdot\left[  \exp\left\{  \varepsilon_{T}^{(1)}\left[  -\dfrac
{i\sqrt{\tilde{b}_{0}}}{-r}\right]  +\varepsilon_{T}^{\prime(1)}\left[
-\dfrac{i\sqrt{\tilde{b}_{0}}}{-r}\right]  +\varepsilon_{P}^{(2)}\left[
\dfrac{i\frac{b_{0}-1}{2M_{2}}}{\left(  -r\right)  ^{2}}\right]
+\varepsilon_{P}^{\prime(2)}\left[  \dfrac{i\frac{b_{0}-1}{2M_{2}}}{\left(
-r\right)  ^{2}}\right]  \right\}  \right]  _{\epsilon_{TPTP}}\nonumber\\
&  \cdot\left[  \exp\left\{  \varepsilon_{P}^{(1)}\left[  \dfrac{i\frac
{b_{0}-1}{2M_{2}}}{-r}+\dfrac{i\frac{a_{0}}{M_{2}}}{\left(  1-r^{2}\right)
/r}\right]  +\varepsilon_{P}^{\prime(1)}\left[  \dfrac{i\frac{b_{0}-1}{2M_{2}%
}}{-r}+\dfrac{i\frac{a_{0}}{M_{2}}}{\left(  1-r^{2}\right)  /r}\right]
+\varepsilon_{P}^{(1)}\varepsilon_{P}^{\prime(1)}\dfrac{\frac{a_{0}}{M_{2}%
^{2}}}{\left(  1-r^{2}\right)  ^{2}}\right\}  \right]  _{\epsilon_{PP}}
\label{ex2}%
\end{align}
where $[\cdots]_{\epsilon_{TPTP}}$ in the second line and $[\cdots
]_{\epsilon_{PP}}$ in the third line indicate that we take the coefficients of
$\varepsilon_{T}^{(1)}\varepsilon_{T}^{\prime(1)}\varepsilon_{P}%
^{(2)}\varepsilon_{P}^{\prime(2)}$ and $\varepsilon_{P}^{(1)}\varepsilon
_{P}^{\prime(1)}$ respectively. Because of the difference in the powers of
$1/r$ and $a_{0}$ in the exponent of Eq.(\ref{ex1}), Eq.(\ref{ex2}) has much
more structure for $\varepsilon_{P}^{(1)}$ and $\varepsilon_{P}^{\prime(1)}$
than for $\varepsilon_{T}^{(1)}$, $\varepsilon_{T}^{\prime(1)}$,
$\varepsilon_{P}^{(2)}$, and $\varepsilon_{P}^{\prime(2)}$, and fits into the
rules Eqs.(\ref{rule1}),(\ref{rule2}),(\ref{rule3}) and (\ref{rule4}). It is
also worth noting that the appearance of the last term in the second exponent
of Eq.(\ref{ex2}) originates from the contraction between $\partial X\left(
z_{2}\right)  $ and $\bar{\partial}\tilde{X}\left(  \bar{z}_{2}\right)  $ in
Eq.(\ref{def-amp}), which is a characteristic of string D-brane scattering.

The explicit form of the amplitude for the current example is
\begin{align}
A  &  =\int_{0}^{1} d\left(  r^{2}\right)  \left(  1-r^{2}\right)
^{a_{0}^{\prime}}r^{b_{0}-1} \left(  -\dfrac{i\sqrt{\tilde{b}_{0}}}%
{-r}\right)  \left(  -\dfrac{i\sqrt{\tilde{b}_{0}}}{-r}\right)  \left(
\dfrac{i\frac{b_{0}-1}{2M_{2}}}{\left(  -r\right)  ^{2}} \right)  \left(
\dfrac{i\frac{b_{0}-1}{2M_{2}}}{\left(  -r \right)  ^{2}} \right) \nonumber\\
&  \cdot\left[  \left(  \dfrac{i\frac{b_{0}-1}{2M_{2}}}{-r} +\dfrac
{i\frac{a_{0}}{M_{2}}}{ \left(  1-r^{2}\right)  /r } \right)  \left(
\dfrac{i\frac{b_{0}-1}{2M_{2}}}{-r} +\dfrac{i\frac{a_{0}}{M_{2}}}{ \left(
1-r^{2}\right)  /r} \right)  +\dfrac{\frac{a_{0}}{M_{2}^{2}}}{\left(
1-r^{2}\right)  ^{2}} \right] \label{ex-reduction1}\\
&  = -\left(  \sqrt{\tilde{b}_{0}}\right)  ^{2} \left(  \frac{b_{0} -1}%
{2M_{2}}\right)  ^{4} \int_{0}^{1} d\left(  r^{2}\right)  \left(
1-r^{2}\right)  ^{a_{0}^{\prime}}r^{b_{0}-9}\nonumber\\
&  \cdot\left[  \left(  \sum_{l=0}^{2} \binom{2}{l}\left(  \dfrac{-r^{2}}{
\left(  1-r^{2}\right)  }\frac{2 a_{0}}{b_{0} -1}\right)  ^{l} \right)
-\dfrac{r^{2}}{\left(  1-r^{2}\right)  ^{2}}\frac{ 4 a_{0}}{\left(  b_{0} -1
\right)  ^{2}} \right] \label{ex-reduction2}\\
&  \sim-\left(  \sqrt{\tilde{b}_{0}}\right)  ^{2} \left(  \frac{b_{0}
-1}{2M_{2}}\right)  ^{4} B\left(  a_{0} +1,\frac{b_{0} -7}{2}\right)
\nonumber\\
&  \cdot\left[  \left(  \sum_{l=0}^{2} \binom{2}{l}\left(  -\frac{2 }{b_{0}
-1}\right)  ^{l} \left(  \frac{b_{0} -7}{2}\right)  _{l} \right)  -\frac{ 4
}{\left(  b_{0} -1 \right)  ^{2}} \left(  \frac{b_{0} -7}{2}\right)  \right]
\label{ex-reduction3}\\
&  = -\left(  \sqrt{\tilde{b}_{0}}\right)  ^{2} \left(  \frac{b_{0} -1}%
{2M_{2}}\right)  ^{4} B\left(  a_{0} +1,\frac{b_{0} -7}{2}\right) \nonumber\\
&  \cdot\left[  \ _{2} F_{0} \left(  -2, \frac{b_{0} -7}{2},\frac{2 }{b_{0}
-1}\right)  -\frac{ 4 }{\left(  b_{0} -1 \right)  ^{2}} \left(  \frac{b_{0}
-7}{2}\right)  \right]  \label{ex-reduction4}%
\end{align}
where we have used Eq.(\ref{Regge behavior}).

\subsubsection{General cases}

Now we move on to general cases. The vertex operator corresponding to a
general massive state with $d$ left-modes and $d^{\prime}$ right-modes is of
the following form.
\begin{align}
V=i^{d+d^{\prime}}\varepsilon_{\mu_{1}\cdots\mu_{d+d^{\prime}}}:\partial
^{n_{1}} X^{\mu_{1}}\cdots\partial^{n_{d}} X^{\mu_{d}}e^{ik_{2} X}\left(
z\right)  :\ :\bar{\partial}^{n_{d+1}} \tilde{X}^{\mu_{d+1}}\cdots
\bar{\partial}^{n_{d+d^{\prime}}} \tilde{X}^{\mu_{d+d^{\prime}}}e^{ik_{2}
\tilde{X}}\left(  \bar{z}\right)  : \label{op}%
\end{align}
The vertex operators corresponding to the states Eq.(\ref{general states}) are
expressed in this covariant form by
\begin{align*}
&  d =\sum_{n>0} p_{n}+q_{n},\quad d^{\prime}=\sum_{n>0} p^{\prime}_{n} +
q^{\prime}_{n}\\
&  \left(  n_{1},n_{2},\cdots,n_{d+d^{\prime}} \right)  =\left(
\cdots,\underbrace{m,\cdots,m}_{p_{m}},\cdots,\underbrace{n,\cdots,n}_{q_{n}%
},\cdots,\underbrace{m^{\prime},\cdots,m^{\prime}}_{p^{\prime}_{m^{\prime}}%
},\cdots,\underbrace{n^{\prime},\cdots,n^{\prime}}_{q^{\prime}_{n^{\prime}}%
},\cdots\right) \\
&  \varepsilon_{\cdots\underbrace{T\cdots T}_{p_{m}}\cdots\underbrace{P\cdots
P}_{q_{n}}\cdots\underbrace{T\cdots T}_{p^{\prime}_{m^{\prime}}}%
\cdots\underbrace{P\cdots P}_{q^{\prime}_{n^{\prime}}}\cdots}=1.
\end{align*}
For the calculation of the correlator involving the operator Eq.(\ref{op}), we
introduce parameters associated with the polarization tensor and exponentiate
the kinematic factors.
\begin{align*}
\varepsilon_{TTT\cdots PPP\cdots TTT\cdots PPP\cdots}  &  \rightarrow
\prod_{n>0} \prod_{i=1}^{p_{n}} \prod_{j=1}^{q_{n}}\prod_{i^{\prime}%
=1}^{p^{\prime}_{n}}\prod_{j^{\prime}=1}^{q^{\prime}_{n}} \varepsilon_{T_{i}%
}^{(n)} \varepsilon_{P_{j}}^{(n)} \varepsilon_{T_{i^{\prime}}}^{\prime(n)}
\varepsilon_{P_{j^{\prime}}}^{\prime(n)}%
\end{align*}
\begin{align}
V =  &  \left(  i\right)  ^{\sum_{n>0}p_{n} + p^{\prime}_{n} + q_{n}
+q^{\prime}_{n}} \left[  :\exp\left\{  ik_{2} X(z) + \sum_{n>0} \sum
_{i=1}^{p_{n}} \varepsilon_{T_{i}}^{(n)}\partial^{n} X^{T} (z) + \sum_{m>0}
\sum_{j=1}^{q_{m}} \varepsilon_{P_{j}}^{(m)}\partial^{m} X^{P} (z) \right\}  :
\right. \nonumber\\
&  \quad\quad\times\left.  :\exp\left\{  ik_{2} \tilde{X}(\bar{z}) +
\sum_{n>0} \sum_{i=1}^{p^{\prime}_{n}} \varepsilon_{T_{i}}^{\prime(n)}%
\partial^{n} \tilde{X}^{T} (\bar{z}) + \sum_{m>0} \sum_{j=1}^{q^{\prime}_{m}}
\varepsilon_{P_{j}}^{\prime(m)}\partial^{m} \tilde{X}^{P} (\bar{z}) \right\}
: \right]  _{\mathrm{linear\ terms}} \label{exponentiation}%
\end{align}
where ``linear terms'' means the terms linear in all of $\varepsilon_{T_{i}%
}^{(n)}, \varepsilon_{P_{j}}^{(m)}, \varepsilon_{T_{i}}^{\prime(n)}$, and
$\varepsilon_{P_{j}}^{\prime(m)}$. Below we use symbols like
\begin{align*}
\varepsilon_{T^{3} P^{2} T P^{3}} \equiv\varepsilon_{T_{1}}^{(1)}%
\varepsilon_{T_{1}}^{(3)}\varepsilon_{T_{2}}^{(3)} \varepsilon_{P_{1}}%
^{(2)}\varepsilon_{P_{1}}^{(5)}\varepsilon_{T_{1}}^{\prime(1)} \varepsilon
_{P_{1}}^{\prime(1)}\varepsilon_{P_{2}}^{\prime(1)}\varepsilon_{P_{1}}%
^{\prime(2)}, \qquad\varepsilon_{T}\sum_{n} p_{n}\equiv\sum_{n>0} \sum
_{i=1}^{p_{n}}\varepsilon_{T_{i}}^{(n)}%
\end{align*}
(the meanings of these symbols are not unique.) and do not write the normal
ordering symbol : : to avoid messy expressions.

The string D-particle scattering amplitudes of these string states can be
calculated to be%
\begin{align}
A  &  =\int d^{2}z_{1}d^{2}z_{2}\cdot\varepsilon_{T^{\sum p_{n}}P^{\sum q_{n}%
}T^{\sum p_{n}^{\prime}}P^{\sum q_{n}^{\prime}}}\\
&  \quad\cdot\left\langle
\begin{array}
[c]{c}%
e^{ik_{1}X}\left(  z_{1}\right)  e^{ik_{1}\tilde{X}}\left(  \bar{z}%
_{1}\right)  \cdot\prod\limits_{n>0}\left(  i\partial^{n}X^{T}\right)
^{p_{n}}\prod\limits_{m>0}\left(  i\partial^{m}X^{P}\right)  ^{q_{m}}%
e^{ik_{2}X}\left(  z_{2}\right) \\
\cdot\prod\limits_{n>0}\left(  i\bar{\partial}^{n}\tilde{X}^{T}\right)
^{p_{n}^{\prime}}\prod\limits_{m>0}\left(  i\bar{\partial}^{m}\tilde{X}%
^{P}\right)  ^{q_{m}^{\prime}}e^{ik_{2}\tilde{X}}\left(  \bar{z}_{2}\right)
\end{array}
\right\rangle \nonumber\\
&  \equiv(i)^{\sum\limits_{n>0}p_{n}+p_{n}^{\prime}+q_{n}+q_{n}^{\prime}%
}A^{\prime}\label{phase}\\
&  =(i)^{\sum\limits_{n>0}p_{n}+p_{n}^{\prime}+q_{n}+q_{n}^{\prime}}\int
d^{2}z_{1}d^{2}z_{2}\nonumber\\
&  \cdot\exp\left\{
\begin{array}
[c]{c}%
\left\langle \left(  ik_{1}X\right)  \left(  z_{1}\right)  \left(
ik_{1}\tilde{X}\right)  \left(  \bar{z}_{1}\right)  \right\rangle \\
+\left\langle
\begin{array}
[c]{c}%
\left(  \varepsilon_{T}\sum\limits_{n>0}p_{n}\partial^{n}X^{T}+\varepsilon
_{P}\sum\limits_{m>0}q_{m}\partial^{m}X^{P}+ik_{2}X\right)  \left(
z_{2}\right) \\
\left(  \varepsilon_{T}^{\prime}\sum\limits_{n>0}p_{n}^{\prime}\bar{\partial
}^{n}\tilde{X}^{T}+\varepsilon_{P}^{\prime}\sum\limits_{m>0}q_{m}^{\prime}%
\bar{\partial}^{m}\tilde{X}^{P}+ik_{2}\tilde{X}\right)  \left(  \bar{z}%
_{2}\right)
\end{array}
\right\rangle \\
+\left\langle \left(  ik_{1}X\right)  \left(  z_{1}\right)  \left(
\varepsilon_{T}\sum\limits_{n>0}p_{n}\partial^{n}X^{T}+\varepsilon_{P}%
\sum\limits_{m>0}q_{m}\partial^{m}X^{P}+ik_{2}X\right)  \left(  z_{2}\right)
\right\rangle \\
+\left\langle \left(  ik_{1}\tilde{X}\right)  \left(  \bar{z}_{1}\right)
\left(  \varepsilon_{T}^{\prime}\sum\limits_{n>0}p_{n}^{\prime}\bar{\partial
}^{n}\tilde{X}^{T}+\varepsilon_{P}^{\prime}\sum\limits_{m>0}q_{m}^{\prime}%
\bar{\partial}^{m}\tilde{X}^{P}+ik_{2}\tilde{X}\right)  \left(  \bar{z}%
_{2}\right)  \right\rangle \\
+\left\langle \left(  ik_{1}X\right)  \left(  z_{1}\right)  \left(
\varepsilon_{T}^{\prime}\sum\limits_{n>0}p_{n}^{\prime}\bar{\partial}%
^{n}\tilde{X}^{T}+\varepsilon_{P}^{\prime}\sum\limits_{m>0}q_{m}^{\prime}%
\bar{\partial}^{m}\tilde{X}^{P}+ik_{2}\tilde{X}\right)  \left(  \bar{z}%
_{2}\right)  \right\rangle \\
+\left\langle \left(  ik_{1}\tilde{X}\right)  \left(  \bar{z}_{1}\right)
\left(  \varepsilon_{T}\sum\limits_{n>0}p_{n}\partial^{n}X^{T}+\varepsilon
_{P}\sum\limits_{m>0}q_{m}\partial^{m}X^{P}+ik_{2}X\right)  \left(
z_{2}\right)  \right\rangle
\end{array}
\right\}  \label{GG}%
\end{align}
where only linear terms are taken in the expansion of the exponential (in the
sense of Eq.(\ref{exponentiation})). In Eq.(\ref{GG}), we have used the
simplified notation $\varepsilon_{T_{j}}^{(n)}\equiv\varepsilon_{T},$
$j=1,2,...p_{n}$, $n\in Z_{+}$ for the spin polarizations, and similarly for
the other polarizations. Note that there will be terms corresponding to
quadratic in the spin polarization. After fixing the $SL(2,R)$ modulus group
on the disk, we set $z_{1}=0$ and $z_{2}=r$, then $d^{2}z_{1}d^{2}%
z_{2}=d\left(  r^{2}\right)  .$ By using Eq.(\ref{Kine}), the amplitude can
then be reduced to%
\begin{align}
A^{\prime}  &  =\int d\left(  r^{2}\right)  \left(  1-r^{2}\right)
^{a_{0}^{\prime}}r^{b_{0}-1}\nonumber\\
&  \exp\left\{
\begin{array}
[c]{l}%
\varepsilon_{T}\sum\limits_{n>0}p_{n}\left[  -\dfrac{i\left(  n-1\right)
!\sqrt{\tilde{b}_{0}}}{(-r)^{n}}\right]  +\varepsilon_{T}^{\prime}%
\sum\limits_{n^{\prime}>0}p_{n^{\prime}}^{\prime}\left[  -\dfrac{i\left(
n^{\prime}-1\right)  !\sqrt{\tilde{b}_{0}}}{(-r)^{n^{\prime}}}\right] \\
+\varepsilon_{P}\sum\limits_{m>0}q_{m}\left[  \dfrac{i\left(  m-1\right)
!\frac{b_{0}-1}{2M_{2}}}{(-r)^{m}}+\dfrac{i\left(  m-1\right)  !\frac{a_{0}%
}{M_{2}}}{\left[  \left(  1-r^{2}\right)  /r\right]  ^{m}}\right] \\
+\varepsilon_{P}^{\prime}\sum\limits_{m^{\prime}>0}q_{m^{\prime}}^{\prime
}\left[  \dfrac{i\left(  m^{\prime}-1\right)  !\frac{b_{0}-1}{2M_{2}}%
}{(-r)^{m^{\prime}}}+\dfrac{i\left(  m^{\prime}-1\right)  !\frac{a_{0}}{M_{2}%
}}{\left[  \left(  1-r^{2}\right)  /r\right]  ^{m^{\prime}}}\right] \\
-\varepsilon_{T}\varepsilon_{T}^{\prime}\sum\limits_{n,n^{\prime}>0}%
p_{n}p_{n^{\prime}}^{\prime}\partial^{n}\bar{\partial}^{n^{\prime}}\ln\left(
1-z_{2}\bar{z}_{2}\right)  \big|_{z_{2}=\bar{z}_{2}=r}\\
-\varepsilon_{P}\varepsilon_{P}^{\prime}\sum\limits_{m,m^{\prime}>0}%
q_{m}q_{m^{\prime}}^{\prime}\partial^{m}\bar{\partial}^{m^{\prime}}\ln\left(
1-z_{2}\bar{z}_{2}\right)  \big|_{z_{2}=\bar{z}_{2}=r}\frac{a_{0}}{M_{2}^{2}}%
\end{array}
\right\}  \label{reduction1}%
\end{align}
where only linear terms are taken in the expansion of the exponential.

Now we use the energy counting Eq.(\ref{rule5}) and show how we reach the
rules Eqs.(\ref{rule1}),(\ref{rule2}),(\ref{rule3}) and (\ref{rule4}). We can
see immediately that in the exponent of Eq.(\ref{reduction1}), the terms
linear in $\varepsilon_{P_{i}}^{(n)}$ or $\varepsilon_{P_{i}}^{\prime(n)}$ are
dominated by their first terms if $m\geq2$ or $m^{\prime}\geq2$. We can see
also that most of the terms in the forth and fifth lines of the exponent are
discarded as subleading. If we start with the terms consisting of only the
factors coming from the first three lines, the other terms are obtained by
series of replacements of two factors in them with one factors coming from the
forth and fifth lines, and for each of the replacements we can see how it
changes the power of energy. We do not need to calculate the infinite number
of derivatives. For each differentiation the increase of the power of $1/r$ is
less than or equal to 1, while the powers of $1/r$ in the first three lines
increase with $n,n^{\prime},m$ or $m^{\prime}$, which implies that if one term
in the forth or fifth line is discarded, the terms with higher $n,n^{\prime
},m,m^{\prime}$ in the same line are also discarded. The sequences of those
discarded terms start at $\left(  n,n^{\prime}\right)  =\left(  1,1\right)  $,
$\left(  m,m^{\prime}\right)  =\left(  1,2\right)  $, and $\left(
m,m^{\prime}\right)  =\left(  2,1\right)  $. In this way, we can see that only
the terms with $m=m^{\prime}=1$ in the fifth line contribute to the leading
behavior. Thus we obtain the generalization of Eq.(\ref{ex2})
\begin{align}
&  A^{\prime}=\int d\left(  r^{2}\right)  \left(  1-r^{2}\right)
^{a_{0}^{\prime}}r^{b_{0}-1}\nonumber\\
&  \exp\left\{
\begin{array}
[c]{l}%
\varepsilon_{T}\sum\limits_{n>0}p_{n}\left[  -\dfrac{i\left(  n-1\right)
!\sqrt{\tilde{b}_{0}}}{(-r)^{n}}\right]  +\varepsilon_{T}^{\prime}%
\sum\limits_{n^{\prime}>0}p_{n^{\prime}}^{\prime}\left[  -\dfrac{i\left(
n^{\prime}-1\right)  !\sqrt{\tilde{b}_{0}}}{(-r)^{n^{\prime}}}\right] \\
+\varepsilon_{P}\sum\limits_{m>1}q_{m}\left[  \dfrac{i\left(  m-1\right)
!\frac{b_{0}-1}{2M_{2}}}{(-r)^{m}}\right]  +\varepsilon_{P}^{\prime}%
\sum\limits_{m^{\prime}>1}q_{m^{\prime}}^{\prime}\left[  \dfrac{i\left(
m^{\prime}-1\right)  !\frac{b_{0}-1}{2M_{2}}}{(-r)^{m^{\prime}}}\right]
\end{array}
\right\}  _{\varepsilon_{T^{\sum p_{n}}P^{\sum^{\prime}q_{n}}T^{\sum
p_{n}^{\prime}}P^{\sum^{\prime}q_{n}^{\prime}}}}\nonumber\\
&  \exp\left\{  \varepsilon_{P}q_{1}\left[  \dfrac{i\frac{b_{0}-1}{2M_{2}}%
}{-r}+\dfrac{i\frac{a_{0}}{M_{2}}r}{1-r^{2}}\right]  +\varepsilon_{P}^{\prime
}q_{1}^{\prime}\left[  \dfrac{i\frac{b_{0}-1}{2M_{2}}}{-r}+\dfrac{i\frac
{a_{0}}{M_{2}}r}{1-r^{2}}\right]  +\varepsilon_{P}\varepsilon_{P}^{\prime
}q_{1}q_{1}^{\prime}\dfrac{\frac{a_{0}}{M_{2}^{2}}}{\left(  1-r^{2}\right)
^{2}}\right\}  _{\varepsilon_{P^{q_{1}}P^{q_{1}^{\prime}}}} \label{drop}%
\end{align}
where the symbols $\varepsilon_{\cdots}$ are similar to the ones in
Eq.(\ref{ex2}) and indicate that we take the coefficients of the products of
the dummy variables in the exponents. ( $\varepsilon_{P_{i}}^{(1)}$ and
$\varepsilon_{P_{i}}^{\prime(1)}$ are excluded in the \textquotedblleft
sums\textquotedblright\ $\sum^{\prime}$.) Note that the last term in the last
line of Eq.(\ref{drop}) is quadratic in the polarization. This term is a
characteristic of string D-brane scattering and has no analog in any of the
previous works. It will play a crucial role in the following calculation in
this paper.

For further calculation, we first note that%
\begin{align}
&  \exp\left\{  \varepsilon_{P}q_{1}\left[  \dfrac{i\frac{b_{0}-1}{2M_{2}}%
}{-r}+\dfrac{i\frac{a_{0}}{M_{2}}r}{1-r^{2}}\right]  +\varepsilon_{P}^{\prime
}q_{1}^{\prime}\left[  \dfrac{i\frac{b_{0}-1}{2M_{2}}}{-r}+\dfrac{i\frac
{a_{0}}{M_{2}}r}{1-r^{2}}\right]  +\varepsilon_{P}\varepsilon_{P}^{\prime
}q_{1}q_{1}^{\prime}\dfrac{\frac{a_{0}}{M_{2}^{2}}}{\left(  1-r^{2}\right)
^{2}}\right\}  _{\varepsilon_{P^{q_{1}}P^{q_{1}^{\prime}}}}\nonumber\\
&  =\varepsilon_{P^{q_{1}}P^{q_{1}^{\prime}}}\sum_{j=0}^{\min\left\{
q_{1},q_{1}^{\prime}\right\}  }\binom{q_{1}}{j}\binom{q_{1}^{\prime}}%
{j}j!\left(  \dfrac{i\frac{b_{0}-1}{2M_{2}}}{-r}+\dfrac{i\frac{a_{0}}{M_{2}}%
r}{1-r^{2}}\right)  ^{q_{1}+q_{1}^{\prime}-2j}\left(  \dfrac{\frac{a_{0}%
}{M_{2}^{2}}}{\left(  1-r^{2}\right)  ^{2}}\right)  ^{j}.
\end{align}
Thus the amplitude can be further reduced to%
\begin{align}
A^{\prime}  &  =\int d\left(  r^{2}\right)  \left(  1-r^{2}\right)
^{a_{0}^{\prime}}r^{b_{0}-1}\nonumber\\
&  \cdot\prod\limits_{n>0}\left[  -\dfrac{i\left(  n-1\right)  !\sqrt
{\tilde{b}_{0}}}{(-r)^{n}}\right]  ^{p_{n}}\prod\limits_{n^{\prime}>0}\left[
-\dfrac{i\left(  n^{\prime}-1\right)  !\sqrt{\tilde{b}_{0}}}{(-r)^{n^{\prime}%
}}\right]  ^{p_{n^{\prime}}^{\prime}}\nonumber\\
&  \cdot\prod\limits_{m>1}\left[  \dfrac{i\left(  m-1\right)  !\frac{b_{0}%
-1}{2M_{2}}}{(-r)^{m}}\right]  ^{q_{m}}\prod\limits_{m^{\prime}>1}\left[
\dfrac{i\left(  m^{\prime}-1\right)  !\frac{b_{0}-1}{2M_{2}}}{(-r)^{m^{\prime
}}}\right]  ^{q_{m^{\prime}}}\nonumber\\
&  \cdot\sum_{j=0}^{\min\left\{  q_{1},q_{1}^{\prime}\right\}  }\sum
_{l=0}^{q_{1}+q_{1}^{\prime}-2j}j!\binom{q_{1}}{j}\binom{q_{1}^{\prime}}%
{j}\binom{q_{1}+q_{1}^{\prime}-2j}{l}\nonumber\\
&  \cdot\left(  \dfrac{i\frac{b_{0}-1}{2M_{2}}}{-r}\right)  ^{q_{1}%
+q_{1}^{\prime}-2j-l}\left(  \dfrac{i\frac{a_{0}}{M_{2}}r}{1-r^{2}}\right)
^{l}\left(  \dfrac{\frac{a_{0}}{M_{2}^{2}}}{\left(  1-r^{2}\right)  ^{2}%
}\right)  ^{j},
\end{align}
which, in the case of the state (\ref{example-state}), is reduced to
Eq.(\ref{ex-reduction2}). We can now do the integration to get%
\begin{align}
A^{\prime}  &  =\left(  i\frac{b_{0}-1}{2M_{2}}\right)  ^{q_{1}+q_{1}^{\prime
}}\cdot\prod\limits_{n>0}\left(  \left[  -i\left(  n-1\right)  !\sqrt
{\tilde{b}_{0}}\right]  ^{p_{n}}\left[  -i\left(  n-1\right)  !\sqrt{\tilde
{b}_{0}}\right]  ^{p_{n}^{\prime}}\right) \nonumber\\
&  \cdot\prod\limits_{m>1}\left(  \left[  i\left(  m-1\right)  !\frac{b_{0}%
-1}{2M_{2}}\right]  ^{q_{m}}\left[  i\left(  m-1\right)  !\frac{b_{0}%
-1}{2M_{2}}\right]  ^{q_{m}}\right) \nonumber\\
&  \cdot\sum_{j=0}^{\min\left\{  q_{1},q_{1}^{\prime}\right\}  }\sum
_{l=0}^{q_{1}+q_{1}^{\prime}-2j}j!\binom{q_{1}}{j}\binom{q_{1}^{\prime}}%
{j}\binom{q_{1}+q_{1}^{\prime}-2j}{l}\left(  \frac{-2}{b_{0}-1}\right)
^{l}\left(  \frac{-4}{\left(  b_{0}-1\right)  ^{2}}\right)  ^{j}\nonumber\\
&  \qquad\cdot B\left(  a_{0}+1,\frac{b_{0}+1-N}{2}\right)  \left(
\frac{b_{0}+1-N}{2}\right)  _{j}\left(  \frac{b_{0}+1-N}{2}+j\right)  _{l}
\label{min}%
\end{align}
where we have done the expansion of the beta function in the RR as following
\begin{align}
&  B\left(  a_{0}^{\prime}+1-l-2j,\dfrac{b_{0}+1-N}{2}+l+j\right) \nonumber\\
&  \approx B\left(  a_{0}+1,\frac{b_{0}+1-N}{2}\right)  \frac{\left(
\frac{b_{0}+1-N}{2}\right)  _{l+j}}{a_{0}^{l+j}}\nonumber\\
&  =B\left(  a_{0}+1,\frac{b_{0}+1-N}{2}\right)  \frac{\left(  \frac
{b_{0}+1-N}{2}\right)  _{j}\left(  \frac{b_{0}+1-N}{2}+j\right)  _{l}}%
{a_{0}^{l+j}}.
\end{align}
Note that in the case of the state Eq.(\ref{example-state}), Eq.(\ref{min}) is
reduced to Eq.(\ref{ex-reduction3}). Performing the summation over $n$, we
obtain%
\begin{align}
A^{\prime}  &  =\left(  i\frac{b_{0}-1}{2M_{2}}\right)  ^{q_{1}+q_{1}}%
\cdot\prod\limits_{n>0}\left(  \left[  -i\left(  n-1\right)  !\sqrt{\tilde
{b}_{0}}\right]  ^{p_{n}+p_{n}^{\prime}}\right)  \prod\limits_{m>1}\left(
\left[  i\left(  m-1\right)  !\frac{b_{0}-1}{2M_{2}}\right]  ^{q_{m}%
+q_{m}^{\prime}}\right) \nonumber\\
&  \cdot B\left(  a_{0}+1,\frac{b_{0}+1-N}{2}\right)  \sum_{j=0}^{\min\left\{
q_{1},q_{1}^{\prime}\right\}  }(-1)^{j}j!\binom{q_{1}}{j}\binom{q_{1}^{\prime
}}{j}\left(  \frac{b_{0}+1-N}{2}\right)  _{j}\left(  \frac{2}{b_{0}-1}\right)
^{2j}\nonumber\\
&  \cdot_{2}F_{0}\left(  -q_{1}-q_{1}^{\prime}+2j,\frac{b_{0}+1-N}{2}%
+j,\frac{2}{b_{0}-1}\right)  ,
\end{align}
which, in the case of the state Eq.(\ref{example-state}), is reduced to
Eq.(\ref{ex-reduction4}). Finally we can use the identity of the Kummer
function%
\begin{align}
&  2^{2m}\ \tilde{t}^{-2m}U\left(  -2m,\frac{t}{2}+2-2m,\frac{\tilde{t}}%
{2}\right) \nonumber\\
&  =\,_{2}F_{0}\left(  -2m,-1-\frac{t}{2},-\frac{2}{\tilde{t}}\right)
\nonumber\\
&  \equiv\sum_{j=0}^{2m}\left(  -2m\right)  _{j}\left(  -1-\frac{t}{2}\right)
_{j}\frac{\left(  -\frac{2}{\tilde{t}}\right)  ^{j}}{j!}\nonumber\\
&  =\sum_{j=0}^{2m}{\binom{2m}{j}}\left(  -1-\frac{t}{2}\right)  _{j}\left(
\frac{2}{\tilde{t}}\right)  ^{j} \label{Kummer}%
\end{align}
to get the final form of the amplitude%
\begin{align}
A^{\prime}  &  =\prod\limits_{n>0}\left(  \left[  -i\left(  n-1\right)
!\sqrt{\tilde{b}_{0}}\right]  ^{p_{n}+p_{n}^{\prime}}\right)  \prod
\limits_{m>1}\left(  \left[  i\left(  m-1\right)  !\frac{b_{0}-1}{2M_{2}%
}\right]  ^{q_{m}+q_{m}^{\prime}}\right)  \left(  -\frac{i}{M_{2}}\right)
^{q_{1}+q_{1}^{\prime}}\nonumber\\
&  \cdot B\left(  a_{0}+1,\frac{b_{0}+1-N}{2}\right)  \sum_{j=0}^{\min\left\{
q_{1},q_{1}^{\prime}\right\}  }(-1)^{j}j!\binom{q_{1}}{j}\binom{q_{1}^{\prime
}}{j}\left(  \frac{b_{0}+1-N}{2}\right)  _{j}\nonumber\\
&  \cdot U\left(  -q_{1}-q_{1}^{\prime}+2j,\frac{-b_{0}+N+1}{2}-q_{1}%
-q_{1}^{\prime}+j,-\frac{b_{0}-1}{2}\right)  . \label{RRamp}%
\end{align}
Note that the amplitudes in Eq.(\ref{RRamp}) can \textit{NOT} be factorized
into two open string D-particle scattering amplitudes as in the case of closed
string-string scattering amplitudes \cite{Closed,bosonic2}.

An interesting application of Eq.(\ref{RRamp}) is the universal power law
behavior of the amplitudes. We first define the Mandelstam variables as
$s=2E^{2}$ and $t=-(k_{1}+k_{2})^{2}.$ The second argument of the beta
function in Eq.(\ref{RRamp}) can be calculated to be%
\begin{equation}
\frac{b_{0}+1-N}{2}=\frac{2k_{1}\cdot k_{2}+1+1-N}{2}=\frac{(k_{1}+k_{2}%
)^{2}-k_{1}^{2}-k_{2}^{2}+2-N}{2}=\frac{-t-2}{2}%
\end{equation}
where we have used Eq.(\ref{II4}) and $M_{2}^{2}=(N-2).$ The amplitudes thus
give the universal power-law behavior for string states at \textit{all} mass
levels%
\begin{equation}
A\sim s^{\alpha(t)}\text{ \ (in the RR)}%
\end{equation}
where
\begin{equation}
\alpha(t)=a(0)+\alpha^{\prime}t\text{, \ }a(0)=1\text{ and }\alpha^{\prime
}=\frac{1}{2}.
\end{equation}

\subsection{Reproducing ratios at the fixed angle regime\qquad}

To compare the RR amplitudes Eq.(\ref{RRamp}) with the fixed angle amplitudes
corresponding to states in Eq.(\ref{relevant}), we consider the RR amplitudes
of the following closed string states%
\begin{align}
&  |N;2m,2m^{\prime};q,q^{\prime}\rangle\nonumber\\
&  =\left(  \alpha_{-1}^{T}\right)  ^{N/2-2m-2q}\left(  \alpha_{-1}%
^{P}\right)  ^{2m}\left(  \alpha_{-2}^{P}\right)  ^{q}\otimes\left(
\tilde{\alpha}_{-1}^{T}\right)  ^{N/2-2m^{\prime}-2q^{\prime}}\left(
\tilde{\alpha}_{-1}^{P}\right)  ^{2m^{\prime}}\left(  \tilde{\alpha}_{-2}%
^{P}\right)  ^{q^{\prime}}|0,k\rangle.
\end{align}
where $m,m^{\prime},q$ and $q^{\prime}$ are non-negative integers. We can take
the following values%
\begin{align}
p_{1}  &  =N/2-2m-2q,p_{1}^{\prime}=N/2-2m^{\prime}-2q^{\prime},\\
q_{1}  &  =2m,q_{1}^{\prime}=2m^{\prime},\\
q_{2}  &  =q,q_{2}^{\prime}=q^{\prime}%
\end{align}
in Eq.(\ref{RRamp}), and include the phase factor in Eq.(\ref{phase}) to get%
\begin{align}
&  A^{(N;2m,2m^{\prime};q,q^{\prime})}=(i)^{N-q-q^{\prime}}\left(
-i\sqrt{\tilde{b}_{0}}\right)  ^{N-2\left(  m+m^{\prime}\right)  -2\left(
q+q^{\prime}\right)  }\left(  i\frac{b_{0}-1}{2M_{2}}\right)  ^{q+q^{\prime}%
}\left(  -\frac{i}{M_{2}}\right)  ^{2m+2m^{\prime}{}^{\prime}}\nonumber\\
&  \cdot B\left(  a_{0}+1,\frac{b_{0}+1-N}{2}\right)  \sum_{j=0}^{\min\left\{
2m,2m^{\prime}\right\}  }(-1)^{j}j!\binom{2m}{j}\binom{2m^{\prime}}{j}\left(
\frac{b_{0}+1-N}{2}\right)  _{j}\nonumber\\
&  \cdot U\left(  -2m-2m^{\prime}+2j,\frac{-b_{0}+N+1}{2}-2m-2m^{\prime
}+j,-\frac{b_{0}-1}{2}\right)  .
\end{align}
It is now easy to calculate the RR ratios for each fixed mass level
\begin{align}
\frac{A^{(N;2m,2m^{\prime};q,q^{\prime})}}{A^{(N,0,0,0,0)}}  &
=(i)^{-q-q^{\prime}}\left(  -i\frac{b_{0}-1}{2\tilde{b}_{0}M_{2}}\right)
^{q+q^{\prime}}\left(  \frac{1}{\tilde{b}_{0}M_{2}^{2}}\right)  ^{m+m^{\prime
}{}}\nonumber\\
&  \cdot\sum_{j=0}^{\min\left\{  2m,2m^{\prime}\right\}  }(-1)^{j}j!\binom
{2m}{j}\binom{2m^{\prime}}{j}\left(  \frac{b_{0}+1-N}{2}\right)
_{j}\nonumber\\
&  \cdot U\left(  -2m-2m^{\prime}+2j,\frac{-b_{0}+N+1}{2}-2m-2m^{\prime
}+j,-\frac{b_{0}-1}{2}\right)  \label{DRatio}%
\end{align}
which is a $b_{0}$-dependent function.

Before studying the fixed angle ratios for string D-particle scatterings, we
first make a pause to review previous results on \textit{string-string} scatterings.

\subsubsection{String-string scatterings}

\paragraph{Open string}

For open string-string scatterings, either the saddle-point method ($t-u$
channel only) or the decoupling of high energy ZNS (ZNS) discussed in chapter
V can be used to calculate the fixed angle ratios
\cite{ChanLee,ChanLee1,ChanLee2, CHL,CHLTY1,CHLTY2,CHLTY3}. It was discovered
that there was an interesting link between high energy fixed angle amplitudes
$T$ and RR amplitudes $A.$ To the leading order in energy, the ratios among
fixed angle amplitudes are $\phi$-independent numbers, whereas the ratios
among RR amplitudes are $t$-dependent functions. However, It was discovered
\cite{bosonic} in chapter XI.B that the coefficients of the high energy RR
ratios in the leading power of $t$ can be identified with the fixed angle
ratios, namely \cite{bosonic}%
\begin{equation}
\lim_{\tilde{t}^{\prime}\rightarrow\infty}\frac{A^{(N,2m,q)}}{A^{(N,0,0,)}%
}=\left(  -\frac{1}{M_{2}}\right)  ^{2m+q}\left(  \frac{1}{2}\right)
^{m+q}(2m-1)!!=\frac{T^{(N,2m,q)}}{T^{(N,0,0)}}. \label{Ratios2}%
\end{equation}
To ensure this identification, one needs the following identity
\cite{bosonic,bosonic2,RRsusy,LYAM}
\begin{align}
&  \sum_{j=0}^{2m}(-2m)_{j}\left(  -L-\frac{\tilde{t}^{\prime}}{2}\right)
_{j}\frac{(-2/\tilde{t}^{\prime})^{j}}{j!}\nonumber\\
&  =0(-\tilde{t}^{\prime})^{0}+0(-\tilde{t}^{\prime})^{-1}+...+0(-\tilde
{t}^{\prime})^{-m+1}+\frac{(2m)!}{m!}(-\tilde{t}^{\prime})^{-m}+\mathit{O}%
\left\{  \left(  \frac{1}{\tilde{t}^{\prime}}\right)  ^{m+1}\right\}
\label{master..}%
\end{align}
where $L=1-N$ and is an integer. Note that $L$ effects only the subleading
terms in $\mathit{O}\left\{  \left(  \frac{1}{\tilde{t}^{\prime}}\right)
^{m+1}\right\}  .$ Mathematically, the complete proof of Eq.(\ref{master..})
for \textit{arbitrary real values} $L$ was worked out in \cite{LYAM} by using
an identity of signless Stirling number of the first kind in combinatorial theory.

\paragraph{Open superstring}

For all four classes \cite{susy} of high energy fixed angle open superstring
scattering amplitudes considered in chapter VIII, both the corresponding RR
amplitudes and the complete ratios of the leading (in $t$) RR amplitudes
considered in chapter XII can be calculated \cite{RRsusy}. For the fixed angle
regime \cite{susy}, the complete ratios can be calculated by the decoupling of
high energy ZNS. It turns out that the identification in Eq.(\ref{Ratios2})
continues to work, and $L$ is an integer again for this case \cite{RRsusy}.

\paragraph{Compactified open string}

For compactified open string scatterings, both the amplitudes and the complete
ratios of leading (in $t$) RR can be calculated \cite{HLY}. For the fixed
angle regime or GR, the complete ratios can be calculated by the decoupling of
high energy ZNS. The identification in Eq.(\ref{Ratios2}) continues to work.
However, only a subset of scattering amplitudes corresponding to the case
$m=0$ was calculated. The difficulties has been as following. First, it seems
that the saddle-point method is not applicable here. On the other hand, it was
shown that \cite{ChanLee,ChanLee1,ChanLee2,CHL} the leading order amplitudes
containing $(\alpha_{-1}^{L})^{2m}$ component will drop from energy order
$E^{4m}$ to $E^{2m}$, and one needs to calculate the complicated naive
subleading order terms in order to get the real leading order amplitude. One
encounters this difficulty even for some cases in the non-compactified string
calculation. In these cases, the method of decoupling of high energy ZNS was adopted.

It was important to discover \cite{HLY} that the identity in
Eq.(\ref{master..}) for arbitrary real values $L$ can only be realized in high
energy \textit{compactified} string scatterings. This is due to the dependence
of the value $L$ on winding momenta $K_{i}^{25}$ \cite{HLY}%
\begin{equation}
L=1-N-(K_{2}^{25})^{2}+K_{2}^{25}K_{3}^{25}. \label{L}%
\end{equation}
All other high energy string scatterings calculated previously
\cite{bosonic,bosonic2,RRsusy} correspond to integer value of $L$ only.

\paragraph{ Closed string}

For closed string scatterings \cite{bosonic2}, one can use the KLT formula
\cite{KLT}, which expresses the relation between tree amplitudes of closed and
two channels of open string $(\alpha_{\text{closed}}^{\prime}=4\alpha
_{\text{open}}^{\prime}=2),$ to simplify the calculations. Both ratios of
leading (in $t$) RR amplitudes and GR amplitudes were found to be the tensor
product of two ratios in Eq.(\ref{Ratios2}), namely \cite{bosonic2}%
\begin{align}
\lim_{\tilde{t}^{\prime}\rightarrow\infty}\frac{A_{\text{closed}}^{\left(
N;2m,2m^{^{\prime}};q,q^{^{\prime}}\right)  }}{A_{\text{closed}}^{\left(
N;0,0;0,0\right)  }}  &  =\left(  -\frac{1}{M_{2}}\right)  ^{2(m+m^{^{\prime}%
})+q+q^{^{\prime}}}\left(  \frac{1}{2}\right)  ^{m+m^{^{\prime}}%
+q+q^{^{\prime}}}(2m-1)!!(2m^{^{\prime}}-1)!!\nonumber\\
&  =\frac{T_{\text{closed}}^{\left(  N;2m,2m^{^{\prime}};q,q^{^{\prime}%
}\right)  }}{T_{\text{closed}}^{\left(  N;0,0;0,0\right)  }}. \label{RRclosed}%
\end{align}
\qquad

We now begin to discuss the RR \textit{closed string, D-particle} scatterings
considered in this chapter.

\subsubsection{Closed string D-particle scatterings}

\paragraph{$m=m^{^{\prime}}=0$ Case}

In chapter IX \cite{Dscatt}, the high energy scattering amplitudes and ratios
of fixed angle closed string D-particle scatterings were calculated only for
the case $m=m^{^{\prime}}=0$. For nonzero $m$ or $m^{^{\prime}}$ cases, one
encounters similar difficulties stated in the paragraph before Eq.(\ref{L}) to
calculate the complete fixed angle amplitudes. A subset of ratios can be
extracted from Eq.(\ref{D-particle}) and was found to be \cite{Dscatt}%
\begin{equation}
\frac{T_{SD}^{(N,0,0,q,q^{^{\prime}})}}{T_{SD}^{(N,0,0,0,0)}}=\left(
-\frac{1}{2M_{2}}\right)  ^{q+q^{^{\prime}}}. \label{subset}%
\end{equation}
In view of the non-factorizability of Regge string D-particle scattering
amplitudes calculated in Eq.(\ref{RRamp}), one is tempted to conjecture that
the complete ratios of fixed angle closed string D-particle scatterings may
not be factorized. But on the other hand, the decoupling of high energy ZNS
seems to imply the factorizability of the fixed angle ratios.

\paragraph{General case}

We can show explicitly that the leading behaviors of the inner products in
Eq.(\ref{reduction1}) involving $k_{1},k_{2},e^{T},e^{P}$ and $D$ are not
affected by the replacement of $e^{P}$ with $e^{L}$ if we take the limit
$b_{0}\rightarrow\infty$ after taking the Regge limit. Therefore we proceed as
in the previous works on Regge scattering. The calculation for the complete
ratios of leading (in $b_{0}$) RR \textit{closed string, D-particle}
scatterings from Eq.(\ref{DRatio}) gives%
\begin{align}
&  \lim_{b_{0}\rightarrow\infty}\frac{A_{SD}^{(N;2m,2m^{\prime};q,q^{\prime}%
)}}{A_{SD}^{(N,0,0,0,0)}}\nonumber\\
&  =(i)^{-q-q^{\prime}}\left(  -i\frac{b_{0}}{2b_{0}M_{2}}\right)
^{q+q^{\prime}}\left(  \frac{1}{{b}_{0}M_{2}^{2}}\right)  ^{m+m^{\prime}%
}\nonumber\\
&  \cdot\sum_{j=0}^{\min\left\{  2m,2m^{\prime}\right\}  }(-1)^{j}j!\binom
{2m}{j}\binom{2m^{\prime}}{j}\left(  \frac{b_{0}}{2}\right)  ^{j}\frac{\left(
2m+2m^{\prime}-2j\right)  !}{\left(  m+m^{\prime}-j\right)  !}%
2^{-2m-2m^{\prime}+2j}b_{0}^{m+m^{\prime}-j}\nonumber\\
&  =(i)^{-q-q^{\prime}}\left(  -i\frac{1}{2M_{2}}\right)  ^{q+q^{\prime}%
}\left(  \frac{1}{2M_{2}}\right)  ^{2m+2m^{\prime}}\nonumber\\
&  \cdot\sum_{j=0}^{\min\left\{  2m,2m^{\prime}\right\}  }j!\binom{2m}%
{j}\binom{2m^{\prime}}{j}\left(  -2\right)  ^{j}\frac{\left(  2m+2m^{\prime
}-2j\right)  !}{\left(  m+m^{\prime}-j\right)  !}. \label{sum.}%
\end{align}
In deriving Eq.(\ref{sum.}), we have made use of Eq.(\ref{Kummer}) and
Eq.(\ref{master..}). Note that each term in the summation of Eq.(\ref{sum.})
is not factorized.

Surprisingly, the summation in Eq.(\ref{sum.}) can be performed, and the
ratios can be calculated to be%
\begin{align}
&  \lim_{b_{0}\rightarrow\infty}\frac{A_{SD}^{(N;2m,2m^{\prime};q,q^{\prime}%
)}}{A_{SD}^{(N,0,0,0,0)}}\nonumber\\
&  =(-)^{q+q^{\prime}}\left(  \frac{1}{2}\right)  ^{q+q^{\prime}%
+2m+2m^{\prime}}\left(  \frac{1}{M_{2}}\right)  ^{2m+2m^{\prime}+q+q^{\prime}%
}\nonumber\\
&  \quad\cdot\frac{2^{2m+2m^{\prime}}\pi\sec\left[  \frac{\pi}{2}\left(
2m+2m^{\prime}\right)  \right]  }{\Gamma\left(  \frac{1-2m}{2}\right)
\Gamma\left(  \frac{1-2m^{\prime}}{2}\right)  }\nonumber\\
&  =\left(  -\frac{1}{M_{2}}\right)  ^{2m+q}\left(  \frac{1}{2}\right)
^{m+q}(2m-1)!!\left(  -\frac{1}{M_{2}}\right)  ^{2m^{\prime}+q^{\prime}%
}\left(  \frac{1}{2}\right)  ^{m^{\prime}+q^{\prime}}(2m^{^{\prime}%
}-1)!!\text{ } \label{final2}%
\end{align}
which are \textit{factorized}. They are exactly the same with the ratios of
the high energy, fixed angle closed string-string scattering amplitudes
calculated in Eq.(\ref{RRclosed}) and again consistent with the decoupling of
high energy ZNS in Eq.(\ref{mainA}) \cite{ChanLee,ChanLee1,ChanLee2,
CHL,CHLTY1,CHLTY2,CHLTY3,susy,Closed}. We thus conclude that the
identification in Eq.(\ref{Ratios2}) continues to work for string D-particle
scatterings. So the complete ratios of fixed angle closed string D-particle
scatterings are%
\begin{align}
\frac{T_{SD}^{\left(  N;2m,2m^{^{\prime}};q,q^{^{\prime}}\right)  }}%
{T_{SD}^{\left(  N;0,0;0,0\right)  }}  &  =\left(  -\frac{1}{M_{2}}\right)
^{2(m+m^{^{\prime}})+q+q^{^{\prime}}}\left(  \frac{1}{2}\right)
^{m+m^{^{\prime}}+q+q^{^{\prime}}}(2m-1)!!(2m^{\prime}-1)!!\nonumber\\
&  =\lim_{b_{0}\rightarrow\infty}\frac{A_{SD}^{(N;2m,2m^{\prime};q,q^{\prime
})}}{A_{SD}^{(N,0,0,0,0)}} \label{old2}%
\end{align}
where the first equality can be deduced from the decoupling of high energy
ZNS. Note that, for $m=m^{^{\prime}}=0,$ Eq.(\ref{old2}) reduces to
Eq.(\ref{subset}) calculated previously in chapter IX \cite{Dscatt}.

It is well known that the closed string-string scattering amplitudes can be
factorized into two open string-string scattering amplitudes due to the
existence of the KLT formula \cite{KLT}. On the contrary, there is no physical
picture for open string D-particle tree scattering amplitudes and thus no
factorization for closed string D-particle scatterings into two channels of
open string D-particle scatterings, and hence no KLT-like formula there. Here
what we really mean is: two string, two D-particle scattering in the limit of
infinite D-particle mass. This can also be seen from the nontrivial string
D-particle propagator in Eq.(\ref{DDD}), which vanishes for the case of closed
string-string scattering.

Thus the factorized ratios in high energy fixed angle regime calculated in the
RR in Eq.(\ref{final2}) and Eq.(\ref{old2}) came as a surprise. However, these
ratios are consistent with the decoupling of high energy ZNS calculated
previously in Eq.(\ref{mainA})\cite{ChanLee,ChanLee1,ChanLee2,
CHL,CHLTY1,CHLTY2,CHLTY3,susy,Closed}. It will be interesting if one can
calculate the complete fixed angle amplitudes directly and see how the
non-factorized amplitudes can give the result of factorized ratios.%

\setcounter{equation}{0}
\renewcommand{\theequation}{\arabic{section}.\arabic{equation}}%

\section{The Appell functions $F_{1}$ and the complete RSSA}

In this chapter, we will show that \cite{AppellLY} each 26D open bosonic Regge
string scattering amplitude (RSSA) can be expressed in terms of one single
Appell function $F_{1}$ in the Regge limit. This result enables us to derive
infinite number of recurrence relations among RSSA at arbitrary mass levels,
which are conjectured to be related to the known $SL(5,C)$ dynamical symmetry
of $F_{1}$. Since there is only one single Appell function in the expression
of the amplitudes in contrast to a sum of Kummer functions discussed in
chapter XI, it is easier to systematically construct recurrence relations
among RSSA by directly using recurrence relations of Appell functions.

In addition, we show that these recurrence relations in the Regge limit can be
systematically solved so that all RSSA can be expressed in terms of one
amplitude. All these results are dual to high energy symmetries of fixed angle
string scattering amplitudes discovered previously in chapter
V\cite{ChanLee,ChanLee1,ChanLee2, CHL,CHLTY1,CHLTY2,CHLTY3,susy,Closed}.

\subsection{Appell functions and RSSA}

The leading order high energy open string states in the Regge regime at each
fixed mass level $N=\sum_{n,m,l>0}np_{n}+mq_{m}+lr_{l}$ are \cite{LY,AppellLY}%
\begin{equation}
\left\vert p_{n},q_{m},r_{l}\right\rangle =\prod_{n>0}(\alpha_{-n}^{T}%
)^{p_{n}}\prod_{m>0}(\alpha_{-m}^{P})^{q_{m}}\prod_{l>0}(\alpha_{-l}%
^{L})^{r_{l}}|0,k\rangle.
\end{equation}
The momenta of the four particles on the scattering plane are%

\begin{align}
k_{1}  &  =\left(  +\sqrt{p^{2}+M_{1}^{2}},-p,0\right)  ,\\
k_{2}  &  =\left(  +\sqrt{p^{2}+M_{2}^{2}},+p,0\right)  ,\\
k_{3}  &  =\left(  -\sqrt{q^{2}+M_{3}^{2}},-q\cos\phi,-q\sin\phi\right)  ,\\
k_{4}  &  =\left(  -\sqrt{q^{2}+M_{4}^{2}},+q\cos\phi,+q\sin\phi\right)
\end{align}
where $p\equiv\left\vert \mathrm{\vec{p}}\right\vert $, $q\equiv\left\vert
\mathrm{\vec{q}}\right\vert $ and $k_{i}^{2}=-M_{i}^{2}$. The relevant
kinematics in the Regge regime are%
\begin{equation}
e^{P}\cdot k_{1}\simeq-\frac{s}{2M_{2}},\text{ \ }e^{P}\cdot k_{3}\simeq
-\frac{\tilde{t}}{2M_{2}}=-\frac{t-M_{2}^{2}-M_{3}^{2}}{2M_{2}};
\end{equation}%
\begin{align}
e^{L}\cdot k_{1}  &  \simeq-\frac{s}{2M_{2}},\text{ \ }e^{L}\cdot k_{3}%
\simeq-\frac{\tilde{t}^{\prime}}{2M_{2}}=-\frac{t+M_{2}^{2}-M_{3}^{2}}{2M_{2}%
};\\
e^{T}\cdot k_{1}  &  =0\text{, \ \ }e^{T}\cdot k_{3}\simeq-\sqrt{-{t}}%
\end{align}
where $\tilde{t}$ $=t-M_{2}^{2}-M_{3}^{2}$ and $\tilde{t}^{\prime}=t+M_{2}%
^{2}-M_{3}^{2}$ . The $s-t$ channel one higher spin and three tachyons string
scattering amplitudes in the Regge limit can be calculated as%
\begin{align}
A^{(p_{n};q_{m};r_{l})}  &  =\int_{0}^{1}dx\,x^{k_{1}\cdot k_{2}}%
(1-x)^{k_{2}\cdot k_{3}}\cdot\left[  \frac{e^{P}\cdot k_{1}}{x}-\frac
{e^{P}\cdot k_{3}}{1-x}\right]  ^{q_{1}}\left[  \frac{e^{L}\cdot k_{1}}%
{x}+\frac{e^{L}\cdot k_{3}}{1-x}\right]  ^{r_{1}}\nonumber\\
&  \cdot\prod_{n=1}\left[  \frac{(n-1)!e^{T}\cdot k_{3}}{(1-x)^{n}}\right]
^{p_{n}}\prod_{m=2}\left[  \frac{(m-1)!e^{P}\cdot k_{3}}{(1-x)^{m}}\right]
^{q_{m}}\prod_{l=2}\left[  \frac{(l-1)!e^{L}\cdot k_{3}}{(1-x)^{l}}\right]
^{r_{l}}\nonumber\\
&  =\prod_{n=1}\left[  (n-1)!\sqrt{-t}\right]  ^{p_{n}}\prod_{m=1}\left[
-(m-1)!\dfrac{\tilde{t}}{2M_{2}}\right]  ^{q_{m}}\prod_{l=1}\left[
(l-1)!\dfrac{\tilde{t}^{\prime}}{2M_{2}}\right]  ^{r_{l}}\nonumber\\
&  \cdot\sum_{j=0}^{r_{1}}\sum_{i=0}^{q_{1}}\binom{r_{1}}{j}{\binom{q_{1}}{i}%
}\left(  -\dfrac{s}{\tilde{t}}\right)  ^{i}\left(  -\dfrac{s}{\tilde
{t}^{\prime}}\right)  ^{j}B\left(  -\frac{s}{2}+N-1-i-j,-\frac{t}%
{2}-1+i+j\right)
\end{align}
where in the Regge limit the beta function $B$ can be further reduced to%
\begin{align}
&  B\left(  -\frac{s}{2}-1+N-i-j,-\frac{t}{2}-1+i+j\right) \nonumber\\
&  \simeq B\left(  -\frac{s}{2}-1,-\frac{t}{2}-1\right)  \dfrac{\left(
-1\right)  ^{i+j}\left(  -\frac{t}{2}-1\right)  _{i+j}}{\left(  \frac{s}%
{2}\right)  _{i+j}}.
\end{align}
Thus%
\begin{align}
A^{(p_{n};q_{m};r_{l})}  &  =B\left(  -\frac{s}{2}-1,-\frac{t}{2}-1\right)
\nonumber\\
&  \cdot\prod_{n=1}\left[  (n-1)!\sqrt{-t}\right]  ^{p_{n}}\prod_{m=1}\left[
-(m-1)!\dfrac{\tilde{t}}{2M_{2}}\right]  ^{q_{m}}\prod_{l=1}\left[
(l-1)!\dfrac{\tilde{t}^{\prime}}{2M_{2}}\right]  ^{r_{l}}\nonumber\\
&  \cdot\sum_{j=0}^{r_{1}}\sum_{i=0}^{q_{1}}\binom{r_{1}}{j}{\binom{q_{1}}{i}%
}\dfrac{\left(  -\frac{t}{2}-1\right)  _{i+j}}{\left(  \frac{s}{2}\right)
_{i+j}}\left(  \dfrac{s}{\tilde{t}}\right)  ^{i}\left(  \dfrac{s}{\tilde
{t}^{\prime}}\right)  ^{j}%
\end{align}
in which the double summation can be expressed in terms of the Appell function
$F_{1}$ as%
\begin{align}
&  \sum_{j=0}^{r_{1}}\sum_{i=0}^{q_{1}}\binom{r_{1}}{j}{\binom{q_{1}}{i}%
}\dfrac{\left(  -\frac{t}{2}-1\right)  _{i+j}}{\left(  \frac{s}{2}\right)
_{i+j}}\left(  \dfrac{s}{\tilde{t}}\right)  ^{i}\left(  \dfrac{s}{\tilde
{t}^{\prime}}\right)  ^{j}\nonumber\\
&  =\sum_{j=0}^{r_{1}}\sum_{i=0}^{q_{1}}\dfrac{\left(  -q_{1}\right)
_{i}\left(  -r_{1}\right)  _{j}}{i!j!}\dfrac{\left(  -\frac{t}{2}-1\right)
_{i+j}}{\left(  \frac{s}{2}\right)  _{i+j}}\left(  -\dfrac{s}{\tilde{t}%
}\right)  ^{i}\left(  -\dfrac{s}{\tilde{t}^{\prime}}\right)  ^{j}\nonumber\\
&  =F_{1}\left(  -\frac{t}{2}-1;-q_{1},-r_{1};\frac{s}{2};-\dfrac{s}{\tilde
{t}},-\dfrac{s}{\tilde{t}^{\prime}}\right)  . \label{finite}%
\end{align}
The Appell function $F_{1}$ is one of the four extensions of the
hypergeometric function $_{2}F_{1}$ to two variables and is defined to be%
\begin{equation}
F_{1}\left(  a;b,b^{\prime};c;x,y\right)  =\sum_{m=0}^{\infty}\sum
_{n=0}^{\infty}\dfrac{\left(  a\right)  _{m+n}\left(  b\right)  _{m}\left(
b^{\prime}\right)  _{n}}{m!n!\left(  c\right)  _{m+n}}x^{m}y^{n}%
\end{equation}
where $(a)_{n}=a\cdot\left(  a+1\right)  \cdots\left(  a+n-1\right)  $ is the
rising Pochhammer symbol. Note that when $a$ or $b(b^{\prime})$ is a
non-positive integer, the Appell function truncates to a polynomial. This is
the case for the Appell function in the RSSA calculated in Eq.(\ref{main.}) in
the following%
\begin{align}
A^{(p_{n};q_{m};r_{l})}  &  =\prod_{n=1}\left[  (n-1)!\sqrt{-t}\right]
^{p_{n}}\prod_{m=1}\left[  -(m-1)!\dfrac{\tilde{t}}{2M_{2}}\right]  ^{q_{m}%
}\prod_{l=1}\left[  (l-1)!\dfrac{\tilde{t}^{\prime}}{2M_{2}}\right]  ^{r_{l}%
}\nonumber\\
&  \cdot F_{1}\left(  -\frac{t}{2}-1;-q_{1},-r_{1};\frac{s}{2};-\dfrac
{s}{\tilde{t}},-\dfrac{s}{\tilde{t}^{\prime}}\right)  \cdot B\left(  -\frac
{s}{2}-1,-\frac{t}{2}-1\right)  . \label{main.}%
\end{align}
Alternatively, it is interesting to note that the result calculated in
Eq.(\ref{main.}) can be directly\ obtained from an integral representation of
$F_{1}$ due to Emile Picard (1881) \cite{Picard}%
\begin{equation}
F_{1}\left(  a;b_{1},b_{2};c;x,y\right)  =\frac{\Gamma(c)}{\Gamma
(a)\Gamma(c-a)}\int_{0}^{1}dt\,t^{a-1}(1-t)^{c-a-1}(1-xt)^{-b_{1}%
}(1-yt)^{-b_{2}}, \label{picard}%
\end{equation}
which was later generalized by Appell and Kampe de Feriet (1926) \cite{Appell}
to $n$ variables%
\begin{align}
F_{1}\left(  a;b_{1},b_{2}...,b_{n};c;x_{1},x_{2}...,x_{n}\right)   &
=\frac{\Gamma(c)}{\Gamma(a)\Gamma(c-a)}\int_{0}^{1}dt\,t^{a-1}(1-t)^{c-a-1}%
\nonumber\\
&  \cdot(1-x_{1}t)^{-b_{1}}(1-x_{2}t)^{-b_{2}}...(1-x_{n}t)^{-b_{n}}.
\label{Kam}%
\end{align}
Eq.(\ref{Kam}) may have application for higher point RSSA. To apply the Picard
formula in Eq.(\ref{picard}), we do the transformation $x\rightarrow\left(
1-x\right)  $, and RSSA can be calculated to be%
\begin{align}
A^{(p_{n};q_{m};r_{l})}  &  =\int_{0}^{1}dx\,\left(  1-x\right)  ^{-\tfrac
{s}{2}+N-2}x^{-\tfrac{t}{2}-2}\cdot\left[  1-\frac{s}{\tilde{t}}\dfrac{x}%
{1-x}\right]  ^{q_{1}}\left[  1-\frac{s}{\tilde{t}^{\prime}}\dfrac{x}%
{1-x}\right]  ^{r_{1}}\nonumber\\
&  \cdot\prod_{n=1}\left[  (n-1)!\sqrt{-t}\right]  ^{p_{n}}\prod_{m=1}\left[
-(m-1)!\dfrac{\tilde{t}}{2M_{2}}\right]  ^{q_{m}}\prod_{l=1}\left[
(l-1)!\dfrac{\tilde{t}^{\prime}}{2M_{2}}\right]  ^{r_{l}}\nonumber\\
&  \simeq B\left(  -\tfrac{t}{2}-1,-\frac{s}{2}-1\right)  \cdot F_{1}\left(
-\tfrac{t}{2}-1,-q_{1},-r_{1},-\tfrac{s}{2};\dfrac{s}{\tilde{t}},\dfrac
{s}{\tilde{t}^{\prime}}\right) \nonumber\\
&  \cdot\prod_{n=1}\left[  (n-1)!\sqrt{-t}\right]  ^{p_{n}}\prod_{m=1}\left[
-(m-1)!\dfrac{\tilde{t}}{2M_{2}}\right]  ^{q_{m}}\prod_{l=1}\left[
(l-1)!\dfrac{\tilde{t}^{\prime}}{2M_{2}}\right]  ^{r_{l}},
\end{align}
which is consistent with the result calculated in Eq.(\ref{main.}). It is
important to note that although $F_{1}$ in Eq.(\ref{main.}) is a polynomial in
$s$, the result in Eq.(\ref{main.}) is valid only for the \textit{leading
order} in $s$ \ in the Regge limit. Note that in contrast to the previous
calculation \cite{LY} in Eq.(\ref{factor1}) and Eq.(\ref{factor2}) where a
finite sum of Kummer functions was obtained, here we get only one single
Appell function in Eq.(\ref{main.}). This simplification will greatly simplify
the calculation of recurrence relations among RSSA to be discussed in the next section.

\subsection{Solving all RSSA by Appell recurrence relations}

The Appell function $F_{1}$ entails four recurrence relations among contiguous
functions%
\begin{align}
\left(  a-b_{1}-b_{2}\right)  F_{1}\left(  a;b_{1},b_{2};c;x,y\right)
-aF_{1}\left(  a+1;b_{1},b_{2};c;x,y\right)   & \nonumber\\
+b_{1}F_{1}\left(  a;b_{1}+1,b_{2};c;x,y\right)  +b_{2}F_{1}\left(
a;b_{1},b_{2}+1;c;x,y\right)   &  =0,\label{Re1}\\
cF_{1}\left(  a;b_{1},b_{2};c;x,y\right)  -\left(  c-a\right)  F_{1}\left(
a;b_{1},b_{2};c+1;x,y\right)   & \nonumber\\
-aF_{1}\left(  a+1;b_{1},b_{2};c+1;x,y\right)   &  =0,\label{Re2}\\
cF_{1}\left(  a;b_{1},b_{2};c;x,y\right)  +c\left(  x-1\right)  F_{1}\left(
a;b_{1}+1,b_{2};c;x,y\right)   & \nonumber\\
-\left(  c-a\right)  xF_{1}\left(  a;b_{1}+1,b_{2};c+1;x,y\right)   &
=0,\label{Re3}\\
cF_{1}\left(  a;b_{1},b_{2};c;x,y\right)  +c\left(  y-1\right)  F_{1}\left(
a;b_{1},b_{2}+1;c;x,y\right)   & \nonumber\\
-\left(  c-a\right)  yF_{1}\left(  a;b_{1},b_{2}+1;c+1;x,y\right)   &  =0.
\label{Re4}%
\end{align}
All other recurrence relations can be deduced from these four relations. We
can easily solve the Appell function in Eq.(\ref{main.}) and express it in
terms of the RSSA%
\begin{align}
&  F_{1}\left(  -\frac{t}{2}-1;-q_{1},-r_{1};\frac{s}{2};-\dfrac{s}{\tilde{t}%
},-\dfrac{s}{\tilde{t}^{\prime}}\right) \nonumber\\
&  =\dfrac{A^{(p_{n};q_{m};r_{l})}}{B\left(  -\frac{s}{2}-1,-\frac{t}%
{2}-1\right)  }\prod_{n=1}\left[  (n-1)!\sqrt{-t}\right]  ^{-p_{n}}\prod
_{m=1}\left[  -(m-1)!\dfrac{\tilde{t}}{2M_{2}}\right]  ^{-q_{m}}\prod
_{l=1}\left[  (l-1)!\dfrac{\tilde{t}^{\prime}}{2M_{2}}\right]  ^{-r_{l}}.
\label{F1}%
\end{align}
Note that among the set of integers $(p_{n},q_{m},r_{l})$ on the right hand
side of Eq.(\ref{F1}), only $(-q_{1},-r_{1})$ dependence shows up on the
Appell function $F_{1}$ on the left hand side of Eq.(\ref{F1}). Indeed, for
those highest spin string states at the mass level $M_{2}^{2}=2\left(
N-1\right)  $%
\begin{equation}
\left\vert N;q_{1},r_{1}\right\rangle \equiv\left(  \alpha_{-1}^{T}\right)
^{N-q_{1}-r_{1}}\left(  \alpha_{-1}^{P}\right)  ^{q_{1}}\left(  \alpha
_{-1}^{L}\right)  ^{r_{1}}|0,k\rangle,
\end{equation}
the string amplitudes reduce to%
\begin{align}
A^{(N;q_{1},r_{1})}  &  =\left(  \sqrt{-t}\right)  ^{N-q_{1}-r_{1}}\left(
-\dfrac{\tilde{t}}{2M_{2}}\right)  ^{q_{1}}\left(  \dfrac{\tilde{t}^{\prime}%
}{2M_{2}}\right)  ^{r_{1}}\nonumber\\
&  \cdot F_{1}\left(  -\frac{t}{2}-1;-q_{1},-r_{1};\frac{s}{2};-\dfrac
{s}{\tilde{t}},-\dfrac{s}{\tilde{t}^{\prime}}\right)  B\left(  -\frac{s}%
{2}-1,-\frac{t}{2}-1\right)  ,
\end{align}
which can be used to solve easily the Appell function $F_{1}$ in terms of the
RSSA $A^{(N;q_{1},r_{1})}.$

We now proceed to show that the recurrence relations of the Appell function
$F_{1}$ \textit{in the Regge limit} in\textit{ }Eq.(\ref{main.}) can be
systematically solved so that all RSSA can be expressed in terms of one
amplitude. As the first step, we note that in \cite{LY} the RSSA was expressed
in terms of finite sum of Kummer functions. There are two equivalent
expressions \cite{LY} which were written in Eq.(\ref{factor1}) and
Eq.(\ref{factor2}) in section XI.D.

It is easy to see that, for $q_{1}=0$ in Eq.(\ref{factor1}) or $r_{1}=0$ in
Eq.(\ref{factor2}), the RSSA can be expressed in terms of only one single
Kummer function $U\left(  -r_{1},\frac{t}{2}+2-r_{1},\frac{\tilde{t}^{\prime}%
}{2}\right)  $ or $U\left(  -q_{1},\frac{t}{2}+2-q_{1},\frac{\tilde{t}}%
{2}\right)  $, which are thus related to the Appell function $F_{1}\left(
-\frac{t}{2}-1;0,-r_{1};\frac{s}{2};-\dfrac{s}{\tilde{t}},-\dfrac{s}{\tilde
{t}^{\prime}}\right)  $ or $F_{1}\left(  -\frac{t}{2}-1;-q_{1},0;\frac{s}%
{2};-\dfrac{s}{\tilde{t}},-\dfrac{s}{\tilde{t}^{\prime}}\right)  $
respectively in the Regge limit in\textit{ }Eq.(\ref{main.}). Indeed, one can
easily calculate%
\begin{align}
\lim_{s\rightarrow\infty}F_{1}\left(  -\frac{t}{2}-1;0,-r_{1};\frac{s}%
{2};-\dfrac{s}{\tilde{t}},-\dfrac{s}{\tilde{t}^{\prime}}\right)   &  =\left(
\frac{2}{\tilde{t}^{\prime}}\right)  ^{r_{1}}U\left(  -r_{1},\frac{t}%
{2}+2-r_{1},\frac{\tilde{t}^{\prime}}{2}\right)  ,\\
\lim_{s\rightarrow\infty}F_{1}\left(  -\frac{t}{2}-1;-q_{1},0;\frac{s}%
{2};-\dfrac{s}{\tilde{t}},-\dfrac{s}{\tilde{t}^{\prime}}\right)   &  =\left(
\frac{2}{\tilde{t}}\right)  ^{q_{1}}U\left(  -q_{1},\frac{t}{2}+2-q_{1}%
,\frac{\tilde{t}}{2}\right)  .
\end{align}
On the other hand, it was shown in Eq.(\ref{Lemma}) \cite{LY} that the ratio%
\begin{equation}
\frac{U(\alpha,\gamma,z)}{U(0,z,z)}=f(\alpha,\gamma,z),\alpha=0,-1,-2,-3,...
\end{equation}
is determined and $f(\alpha,\gamma,z)$ can be calculated by using recurrence
relations of $U(\alpha,\gamma,z)$. Note that $U(0,z,z)=1$ by explicit
calculation. We thus conclude that in the Regge limit%
\begin{equation}
c=\dfrac{s}{2}\rightarrow\infty;x,y\rightarrow\infty;a,b_{1},b_{2}\text{
fixed,}%
\end{equation}
the Appell functions $F_{1}\left(  a;0,b_{2};c;x,y\right)  $ and $F_{1}\left(
a;b_{1},0;c;x,y\right)  $ are determined up to an overall factor by recurrence
relations. The next step is to derive the recurrence relation%
\begin{equation}
yF_{1}\left(  a;b_{1},b_{2};c;x,y\right)  -xF_{1}\left(  a;b_{1}%
+1,b_{2}-1;c;x,y\right)  +\left(  x-y\right)  F_{1}\left(  a;b_{1}%
+1,b_{2};c;x,y\right)  =0, \label{1re}%
\end{equation}
which can be obtained from Eq.(\ref{Re3}) and Eq.(\ref{Re4}). We are now ready
to show that the recurrence relations of the Appell function $F_{1}$
\textit{in the Regge limit} in\textit{ }Eq.(\ref{main.}) can be systematically
solved so that all RSSA can be expressed in terms of one amplitude. We will
use the short notation $F_{1}\left(  a;b_{1},b_{2};c;x,y\right)  =F_{1}\left(
b_{1},b_{2}\right)  $ in the following. For $b_{2}=-1$, by using
Eq.(\ref{1re}) and the known $F_{1}\left(  b_{1},0\right)  $ and $F_{1}\left(
0,b_{2}\right)  $, one can easily show that $F_{1}\left(  b_{1},-1\right)  $
are determined for all $b_{1}=-1,-2,-3...$. Similarly, $F_{1}\left(
b_{1},-2\right)  $ are determined for all $b_{1}=-1,-2,-3...$.if one uses the
result of $F_{1}\left(  b_{1},-1\right)  $ in addition to Eq.(\ref{1re}) and
the known $F_{1}\left(  b_{1},0\right)  $ and $F_{1}\left(  0,b_{2}\right)  $.
This process can be continued and one ends up with the result that
$F_{1}\left(  b_{1},b_{2}\right)  $ are determined for all $b_{1}%
,b_{2}=-1,-2,-3...$. This completes the proof that the recurrence relations of
the Appell function $F_{1}$ in the Regge limit in Eq.(\ref{main.}) can be
systematically solved so that all RSSA can be expressed in terms of one amplitude.

\subsection{Higher recurrence relations}

With the result calculated in Eq.(\ref{main.}), one can easily derive many
recurrence relations among RSSA at arbitrary mass levels. For example, the
identity in Eq.(\ref{1re}) leads to%
\begin{equation}
\sqrt{-t}\left[  A^{(N;q_{1},r_{1})}+A^{(N;q_{1}-1,r_{1}+1)}\right]
-M_{2}A^{(N;q_{1}-1,r_{1})}=0, \label{aaa}%
\end{equation}
which is the generalization of Eq.(\ref{RRI2}) discussed in chapter XI
\cite{LY} for mass level $M_{2}^{2}=4$ to arbitrary mass levels $M_{2}%
^{2}=2(N-1)$. Incidentally, one should keep in mind that the recurrence
relations among RSSA are valid only in the Regge limit. We give one example to
illustrate the calculation. By using Eq.(\ref{Re1}) and Eq.(\ref{Re2}), we
have
\begin{align}
\left(  c-b_{1}-b_{2}\right)  F_{1}\left(  a;b_{1},b_{2};c+1;x,y\right)
-cF_{1}\left(  a;b_{1},b_{2};c;x,y\right)   & \nonumber\\
+b_{1}F_{1}\left(  a;b_{1}+1,b_{2};c+1;x,y\right)  +b_{2}F_{1}\left(
a;b_{1},b_{2}+1;c+1;x,y\right)   &  =0.
\end{align}
Then with Eq.(\ref{Re3}) and Eq.(\ref{Re4}), we obtain%
\begin{align}
\left(  c-b_{1}-b_{2}\right)  yF_{1}\left(  a;b_{1}-1,b_{2};c;x,y\right)   &
\nonumber\\
+\left[  \left(  a-b_{1}-b_{2}\right)  xy-\left(  c-2b_{1}-b_{2}\right)
y+b_{2}x\right]  F_{1}\left(  a;b_{1},b_{2};c;x,y\right)   & \nonumber\\
+b_{1}\left(  x-1\right)  yF_{1}\left(  a;b_{1}+1,b_{2};c;x,y\right)
+b_{2}x\left(  y-1\right)  F_{1}\left(  a;b_{1},b_{2}+1;c;x,y\right)   &
=0,\\
\left(  c-b_{1}-b_{2}\right)  xF_{1}\left(  a;b_{1},b_{2}-1;c;x,y\right)   &
\nonumber\\
+\left[  \left(  a-b_{1}-b_{2}\right)  xy-\left(  c-b_{1}-2b_{2}\right)
x+b_{1}y\right]  F_{1}\left(  a;b_{1},b_{2};c;x,y\right)   & \nonumber\\
+b_{1}\left(  x-1\right)  yF_{1}\left(  a;b_{1}+1,b_{2};c;x,y\right)
+b_{2}x\left(  y-1\right)  F_{1}\left(  a;b_{1},b_{2}+1;c;x,y\right)   &  =0.
\label{bb}%
\end{align}
Finally by Combining Eq.(\ref{1re}) and Eq.(\ref{bb}), and taking the leading
term of $s$ in the Regge limit, we end up with the recurrence relation for
$b_{2}$%
\begin{align}
cx^{2}F_{1}\left(  a;b_{1},b_{2};c;x,y\right)   & \nonumber\\
+\left[  \left(  a-b_{1}-b_{2}-1\right)  xy^{2}+cx^{2}-2cxy\right]
F_{1}\left(  a;b_{1},b_{2}+1;c;x,y\right)   & \nonumber\\
-\left[  \left(  a+1\right)  x^{2}y-\left(  a-b_{2}-1\right)  xy^{2}%
-cx^{2}+cxy\right]  F_{1}\left(  a;b_{1},b_{2}+2;c;x,y\right)   & \nonumber\\
-\left(  b_{2}+2\right)  x\left(  x-y\right)  yF_{1}\left(  a;b_{1}%
,b_{2}+3;c;x,y\right)   &  =0,
\end{align}
which leads to a recurrence relation for RSSA at arbitrary mass levels%
\begin{align}
\tilde{t}^{\prime2}A^{(N;q_{1},r_{1})}  & \nonumber\\
+\left[  \tilde{t}^{\prime2}+\tilde{t}\left(  t-2\tilde{t}^{\prime}%
-2q_{1}-2r_{1}+4\right)  \right]  \left(  \frac{\frac{\tilde{t}^{\prime}%
}{2M_{2}}}{\sqrt{-t}}\right)  A^{(N;q_{1},r_{1}+1)}  & \nonumber\\
+\left[  \tilde{t}^{\prime2}-\tilde{t}^{\prime}\left(  \tilde{t}+t\right)
+\tilde{t}\left(  t-2r_{1}+4\right)  \right]  \left(  \frac{\frac{\tilde
{t}^{\prime}}{2M_{2}}}{\sqrt{-t}}\right)  ^{2}A^{(N;q_{1},r_{1}+2)}  &
\nonumber\\
-2\left(  r_{1}-2\right)  \left(  \tilde{t}^{\prime}-\tilde{t}\right)  \left(
\frac{\frac{\tilde{t}^{\prime}}{2M_{2}}}{\sqrt{-t}}\right)  ^{3}%
A^{(N;q_{1},r_{1}+3)}  &  =0.
\end{align}
More higher recurrence relations which contain general number of $l\geq3$
Appell functions can be found in \cite{Wang}.

Since it was shown that \cite{sl5c} the Appell function $F_{1}$ are basis
vectors for models of irreducible representations of $sl(5,C)$ algebra, it is
reasonable to believe that the spacetime symmetry of Regge string theory is
closely related to $SL(5,C)$ non-compact group. In particular, the recurrence
relations of RSSA studied in this chapter are related to the $SL(5,C)$ group
as well. Further investigation remains to be done and more evidences need to
be uncovered.

\begin{acknowledgments}
We thank all of our collaborators to jointly work on this interesting subject.
We acknowledge financial supports from NSC, MoST and NCTS of Taiwan, and ST
Yau Center of NCTU.
\end{acknowledgments}

\appendix%

\setcounter{equation}{0}
\renewcommand{\theequation}{\thesection.\arabic{equation}}%

\section{Linear relations at mass levels $M^{2}=6$ and $M^{2}=8$ in
GR\label{ratios in GR}}

At mass level $M^{2}=6$, the most general form of physical states at mass
level $M^{2}=6$ are given by
\begin{align}
&  [\epsilon_{\mu\nu\lambda\sigma}\alpha_{-1}^{\mu}\alpha_{-1}^{\nu}%
\alpha_{-1}^{\lambda}\alpha_{-1}^{\sigma}+\epsilon_{(\mu\nu\lambda)}%
\alpha_{-1}^{\mu}\alpha_{-1}^{\nu}\alpha_{-2}^{\lambda}+\epsilon_{\mu
\nu,\lambda}\alpha_{-1}^{\mu}\alpha_{-1}^{\nu}\alpha_{-2}^{\lambda}\nonumber\\
&  +\epsilon_{(\mu\nu)}^{(1)}\alpha_{-1}^{\mu}\alpha_{-3}^{\nu}+\epsilon
_{\lbrack\mu\nu]}^{(1)}\alpha_{-1}^{\mu}\alpha_{-3}^{\nu}+\epsilon_{(\mu\nu
)}^{(2)}\alpha_{-2}^{\mu}\alpha_{-2}^{\nu}+\epsilon_{\mu}\alpha_{-4}^{\mu
}]|0,k\rangle,
\end{align}
where $\epsilon_{\mu\nu,\lambda}$ represents the mixed symmetric spin three
states, that is, one first symmetrizes $\mu\nu$ and then anti-symmetrizes
$\mu\lambda.$ The Virasoro constraints are calculated to be
\begin{align}
2k^{\sigma}\epsilon_{(\mu\nu\lambda\sigma)}+\epsilon_{(\mu\nu\lambda)}  &
=0,\\
2k^{\lambda}\epsilon_{(\mu\nu\lambda)}+k^{\lambda}(\epsilon_{\lambda\mu,\nu
}+\epsilon_{\mu\lambda,\nu})+3(\epsilon_{(\mu\nu)}^{(1)}+\epsilon_{\lbrack
\mu\nu]}^{(1)})+4\epsilon_{(\mu\nu)}^{(2)}  &  =0,\\
k^{\mu}\epsilon_{(\mu\nu)}^{(1)}+k^{\mu}\epsilon_{\lbrack\mu\nu]}%
^{(1)}+4\epsilon_{\nu}  &  =0,\\
6\eta^{\lambda\sigma}\epsilon_{(\mu\nu\lambda\sigma)}+2k^{\lambda}%
\epsilon_{(\mu\nu\lambda)}+\frac{1}{2}k^{\lambda}(\epsilon_{\mu\nu,\lambda
}+\epsilon_{\nu\mu,\lambda})+3\epsilon_{(\mu\nu)}^{(1)}  &  =0,\\
\eta^{\mu\nu}\epsilon_{(\mu\nu\lambda)}+\eta^{\mu\nu}\epsilon_{(\mu\nu
,\lambda)}+4k^{\mu}\varepsilon_{(\mu\nu)}^{(2)}+4\epsilon_{\lambda}  &  =0.
\end{align}
In the high energy limit, by replacing $P$ by $L$ and ignoring irrelevant
states, one gets
\begin{equation}
\mathcal{\epsilon}_{(TTTT)}:\mathcal{\epsilon}_{(TTLL)}:\mathcal{\epsilon
}_{(LLLL)}:\mathcal{\epsilon}_{TT,L}:\mathcal{\epsilon}_{(TTL)}%
:\mathcal{\epsilon}_{(LLL)}:\mathcal{\epsilon}_{(LL)}^{(2)}=48:4:1:12\sqrt
{6}:8\sqrt{6}:2\sqrt{6}:6.\nonumber
\end{equation}
After including the normalization factor of the field variables and the
appropriate symmetry factors, one ends up with%
\begin{align}
&  \mathcal{T}_{(TTTT)}:\mathcal{T}_{(TTLL)}:\mathcal{T}_{(LLLL)}%
:\mathcal{T}_{TT,L}:\mathcal{T}_{(TTL)}:\mathcal{T}_{(LLL)}:\mathcal{T}%
_{(LL)}\nonumber\\
&  =4!\mathcal{\epsilon}_{(TTTT)}:4!\mathcal{\epsilon}_{(TTLL)}%
:4!\mathcal{\epsilon}_{(LLLL)}:-4\mathcal{\epsilon}_{TT,L}:-4\mathcal{\epsilon
}_{(TTL)}:-4\mathcal{\epsilon}_{(LLL)}:8\mathcal{\epsilon}_{(LL)}%
^{(2)}\nonumber\\
&  =16:\frac{4}{3}:\frac{1}{3}:-\frac{2\sqrt{6}}{3}:-\frac{4\sqrt{6}}%
{9}:-\frac{\sqrt{6}}{9}:\frac{2}{3}.
\end{align}
At mass level $M^{2}=8$, the most general form of physical states at mass
level $M^{2}=8$ are given by (for simplicity, we neglect terms containing
$\alpha_{-n}^{\mu}$ with $n\geq3$)
\begin{align}
&  [\epsilon_{\mu\nu\lambda\sigma\rho}\alpha_{-1}^{\mu}\alpha_{-1}^{\nu}%
\alpha_{-1}^{\lambda}\alpha_{-1}^{\sigma}\alpha_{-1}^{\rho}+\epsilon_{(\mu
\nu\lambda\sigma)}\alpha_{-1}^{\mu}\alpha_{-1}^{\nu}\alpha_{-1}^{\lambda
}\alpha_{-2}^{\rho}+\epsilon_{(\mu\nu\lambda)}\alpha_{-1}^{\mu}\alpha
_{-2}^{\nu}\alpha_{-2}^{\lambda}\nonumber\\
&  +\epsilon_{\mu\nu\lambda,\sigma}\alpha_{-1}^{\mu}\alpha_{-1}^{\nu}%
\alpha_{-1}^{\lambda}\alpha_{-2}^{\rho}+\epsilon_{\mu,\nu\lambda}\alpha
_{-1}^{\mu}\alpha_{-2}^{\nu}\alpha_{-2}^{\lambda}]|0,k\rangle,
\end{align}
where $\epsilon_{\mu\nu\lambda,\sigma}$ represents the mixed symmetric spin
four states, that is, first symmetrizes $\mu\nu\lambda$ and then
anti-symmetrizes $\mu\sigma.$ Similar definition for the mixed symmetric spin
three states $\epsilon_{\mu,\nu\lambda}$. The Virasoro constraints are
calculated to be
\begin{align}
5k^{\sigma}\epsilon_{(\mu\nu\lambda\sigma\rho)}+2\epsilon_{(\mu\nu
\lambda\sigma)}  &  =0,\\
3k^{\lambda}\epsilon_{(\mu\nu\lambda\sigma)}+\frac{1}{2}k^{\lambda}%
[(\epsilon_{\mu\nu\lambda,\sigma}+\epsilon_{\lambda\mu\nu,\sigma}%
+\epsilon_{\mu\lambda\nu,\sigma})+(\mu &  \leftrightarrow\nu)]\nonumber\\
+4\epsilon_{(\mu\nu\sigma)}+\epsilon_{\mu,\nu\sigma}+\epsilon_{\nu,\mu\sigma}
&  =0,\\
k^{\mu}\epsilon_{(\mu\nu\lambda)}+\frac{1}{2}k^{\mu}(\epsilon_{\mu,\nu\lambda
}+\epsilon_{\mu,\lambda\nu})  &  =0,\\
5\eta^{\rho\sigma}\epsilon_{(\mu\nu\lambda\sigma\rho)}+k^{\sigma}%
\epsilon_{(\mu\nu\lambda\sigma)}+\frac{1}{3}k^{\sigma}(\epsilon_{\mu\nu
\lambda,\sigma}+\epsilon_{\nu\lambda\mu,\sigma}+\epsilon_{\lambda\mu\nu
,\sigma})  &  =0,\\
3\eta^{\nu\lambda}\epsilon_{(\mu\nu\lambda\sigma)}+\eta^{\nu\lambda}%
(\epsilon_{\mu\nu\lambda,\sigma}+\epsilon_{\lambda\mu\nu,\sigma}+\epsilon
_{\nu\lambda\mu,\sigma})+4k^{\lambda}\varepsilon_{(\mu\sigma\lambda
)}+2k^{\lambda}(\varepsilon_{\mu,\sigma\lambda}+\varepsilon_{\mu,\lambda
\sigma})  &  =0.
\end{align}
In the high energy limit, by replacing $P$ by $L$ and ignoring irrelevant
states, one gets
\begin{align}
&  \mathcal{\epsilon}_{(TTTTT)}:\mathcal{\epsilon}_{(TTTL)}:\mathcal{\epsilon
}_{(TTTLL)}:\mathcal{\epsilon}_{(TLLL)}:\mathcal{\epsilon}_{(TLLLL)}%
:\mathcal{\epsilon}_{(TLL)}:\mathcal{\epsilon}_{T,LL}:\mathcal{\epsilon
}_{TLL,L}:\mathcal{\epsilon}_{TTT,L}\nonumber\\
=  &  \frac{4}{15}:\frac{\sqrt{2}}{12}:\frac{2}{120}:\frac{\sqrt{2}}{64}%
:\frac{1}{320}:\frac{1}{24}:\frac{1}{12}:\frac{\sqrt{2}}{192}:\frac{\sqrt{2}%
}{4}.
\end{align}
After including the normalization factor of the field variables and the
appropriate symmetry factors, one ends up with%
\begin{align}
&  \mathcal{T}_{(TTTTT)}:\mathcal{T}_{(TTTL)}:\mathcal{T}_{(TTTLL)}%
:\mathcal{T}_{(TLLL)}:\mathcal{T}_{(TLLLL)}:\mathcal{T}_{(TLL)}:\mathcal{T}%
_{T,LL}:\mathcal{T}_{TLL,L}:\mathcal{T}_{TTT,L}\nonumber\\
&  =5!\mathcal{\epsilon}_{(TTTTT)}:3!\times2\mathcal{\epsilon}_{(TTTL)}%
:5!\mathcal{\epsilon}_{(TTTLL)}:3!\times2\mathcal{\epsilon}_{(TLLL)}%
:5!\mathcal{\epsilon}_{(TLLLL)}\nonumber\\
&  :8\mathcal{\epsilon}_{(TLL)}:8\mathcal{\epsilon}_{T,LL}:3!\times
2\mathcal{\epsilon}_{TLL,L}:3!\times2\mathcal{\epsilon}_{TTT,L}\nonumber\\
&  =32:\sqrt{2}:2:\frac{3\sqrt{2}}{16}:\frac{3}{8}:\frac{1}{3}:\frac{2}%
{3}:\frac{\sqrt{2}}{16}:3\sqrt{2}. \label{M8}%
\end{align}

\section{High energy limit of Virasoro constraints\label{Virasoro}}

\subsection{Bosonic String}

To take the high energy limit for the Virasoro constraints, we replace the
indices $\left(  \mu_{i},\nu_{i}\right)  $ by $L$ or $T$, and%
\begin{equation}
k^{\mu_{i}}\rightarrow Me^{L}\text{, }\eta^{\mu_{1}\mu_{2}}\rightarrow
e^{T}e^{T}.
\end{equation}
where $M$ is the mass operator. Equations (\ref{L1}) and (\ref{L2}) become%
\begin{subequations}%
\begin{align}
0  &  =M%
\begin{tabular}
[c]{|c|c|c|c|}\hline
$L$ & $\mu_{2}^{1}$ & $\cdots$ & $\mu_{m_{1}}^{1}$\\\hline
\end{tabular}
\overset{N}{\underset{j\neq1}{\otimes}}%
\begin{tabular}
[c]{|c|c|c|}\hline
$\mu_{1}^{j}$ & $\cdots$ & $\mu_{m_{j}}^{j}$\\\hline
\end{tabular}
\nonumber\\
&  +\sum_{i=2}^{m_{1}}%
\begin{tabular}
[c]{|c|c|c|c|c|}\hline
$\mu_{2}^{1}$ & $\cdots$ & $\hat{\mu}_{i}^{1}$ & $\cdots$ & $\mu_{m_{1}}^{1}%
$\\\hline
\end{tabular}
\otimes%
\begin{tabular}
[c]{|c|c|c|c|}\hline
$\mu_{i}^{1}$ & $\mu_{1}^{2}$ & $\cdots$ & $\mu_{m_{2}}^{2}$\\\hline
\end{tabular}
\overset{N}{\underset{j\neq1,2}{\otimes}}%
\begin{tabular}
[c]{|c|c|c|}\hline
$\mu_{1}^{j}$ & $\cdots$ & $\mu_{m_{j}}^{j}$\\\hline
\end{tabular}
\nonumber\\
&  +\sum_{l=3}^{N}\left(  l-1\right)
\begin{tabular}
[c]{|c|c|c|}\hline
$\mu_{2}^{1}$ & $\cdots$ & $\mu_{m_{1}}^{1}$\\\hline
\end{tabular}
\nonumber\\
&  \otimes\sum_{i=1}^{m_{l-1}}%
\begin{tabular}
[c]{|c|c|c|c|c|}\hline
$\mu_{1}^{l-1}$ & $\cdots$ & $\hat{\mu}_{i}^{l-1}$ & $\cdots$ & $\mu_{m_{l-1}%
}^{l-1}$\\\hline
\end{tabular}
\otimes%
\begin{tabular}
[c]{|c|c|c|c|}\hline
$\mu_{i}^{l-1}$ & $\mu_{1}^{l}$ & $\cdots$ & $\mu_{m_{l}}^{l}$\\\hline
\end{tabular}
\overset{N}{\underset{j\neq1,l,l-1}{\otimes}}%
\begin{tabular}
[c]{|c|c|c|}\hline
$\mu_{1}^{j}$ & $\cdots$ & $\mu_{m_{j}}^{j}$\\\hline
\end{tabular}
,
\end{align}
and%
\begin{align}
0  &  =\frac{1}{2}%
\begin{tabular}
[c]{|c|c|c|c|c|}\hline
$T$ & $T$ & $\mu_{3}^{1}$ & $\cdots$ & $\mu_{m_{1}}^{1}$\\\hline
\end{tabular}
\overset{N}{\underset{j\neq1}{\otimes}}%
\begin{tabular}
[c]{|c|c|c|}\hline
$\mu_{1}^{j}$ & $\cdots$ & $\mu_{m_{j}}^{j}$\\\hline
\end{tabular}
\nonumber\\
&  +M%
\begin{tabular}
[c]{|c|c|c|}\hline
$\mu_{3}^{1}$ & $\cdots$ & $\mu_{m_{1}}^{1}$\\\hline
\end{tabular}
\otimes%
\begin{tabular}
[c]{|c|c|c|c|}\hline
$\mu_{1}^{2}$ & $\cdots$ & $\mu_{m_{2}}^{2}$ & $L$\\\hline
\end{tabular}
\overset{N}{\underset{j\neq1,2}{\otimes}}%
\begin{tabular}
[c]{|c|c|c|}\hline
$\mu_{1}^{j}$ & $\cdots$ & $\mu_{m_{j}}^{j}$\\\hline
\end{tabular}
\nonumber\\
&  +\sum_{i=3}^{m_{1}}%
\begin{tabular}
[c]{|c|c|c|c|c|}\hline
$\mu_{3}^{1}$ & $\cdots$ & $\hat{\mu}_{i}^{1}$ & $\cdots$ & $\mu_{m_{1}}^{1}%
$\\\hline
\end{tabular}
\otimes%
\begin{tabular}
[c]{|c|c|c|c|}\hline
$\mu_{i}^{1}$ & $\mu_{1}^{3}$ & $\cdots$ & $\mu_{m_{3}}^{3}$\\\hline
\end{tabular}
\overset{N}{\underset{j\neq1,3}{\otimes}}%
\begin{tabular}
[c]{|c|c|c|}\hline
$\mu_{1}^{j}$ & $\cdots$ & $\mu_{m_{j}}^{j}$\\\hline
\end{tabular}
\nonumber\\
&  +\sum_{l=4}^{N}\left(  l-2\right)
\begin{tabular}
[c]{|c|c|c|}\hline
$\mu_{3}^{1}$ & $\cdots$ & $\mu_{m_{1}}^{1}$\\\hline
\end{tabular}
\nonumber\\
&  \otimes\sum_{i=1}^{m_{l-2}}%
\begin{tabular}
[c]{|c|c|c|c|c|}\hline
$\mu_{1}^{l-2}$ & $\cdots$ & $\hat{\mu}_{i}^{l-2}$ & $\cdots$ & $\mu_{m_{l}%
}^{l-2}$\\\hline
\end{tabular}
\otimes%
\begin{tabular}
[c]{|c|c|c|c|}\hline
$\mu_{i}^{l-2}$ & $\mu_{1}^{l}$ & $\cdots$ & $\mu_{m_{l}}^{l}$\\\hline
\end{tabular}
\overset{N}{\underset{j\neq1,l,l-2}{\otimes}}%
\begin{tabular}
[c]{|c|c|c|}\hline
$\mu_{1}^{j}$ & $\cdots$ & $\mu_{m_{j}}^{j}$\\\hline
\end{tabular}
.
\end{align}%
\end{subequations}%
The indices and $\left\{  \mu_{i}^{j}\right\}  $ are symmetric and can be
chosen to have $l_{j}$ of $\left\{  L\right\}  $ which $0\leq l_{j}\leq m_{j}$
and $\left\{  T\right\}  $ for the rest. Thus%
\begin{subequations}
\begin{align}
0  &  =M%
\begin{tabular}
[c]{|c|}\hline
$\mu_{2}^{1}$\\\hline
\end{tabular}
\underset{m_{1}-2-l_{1}}{\underbrace{%
\begin{tabular}
[c]{|l|l|l|}\hline
$T$ & $\cdots$ & $T$\\\hline
\end{tabular}
}}\underset{l_{1}+1}{\underbrace{%
\begin{tabular}
[c]{|l|l|l|}\hline
$L$ & $\cdots$ & $L$\\\hline
\end{tabular}
}}\overset{N}{\underset{j\neq1}{\otimes}}%
\begin{tabular}
[c]{|c|}\hline
$\mu_{1}^{j}$\\\hline
\end{tabular}
\underset{m_{j}-1-l_{j}}{\underbrace{%
\begin{tabular}
[c]{|l|l|l|}\hline
$T$ & $\cdots$ & $T$\\\hline
\end{tabular}
}}\underset{l_{j}}{\underbrace{%
\begin{tabular}
[c]{|l|l|l|}\hline
$L$ & $\cdots$ & $L$\\\hline
\end{tabular}
}}\nonumber\\
&  +\underset{m_{1}-2-l_{1}}{\underbrace{%
\begin{tabular}
[c]{|l|l|l|}\hline
$T$ & $\cdots$ & $T$\\\hline
\end{tabular}
}}\underset{l_{1}}{\underbrace{%
\begin{tabular}
[c]{|l|l|l|}\hline
$L$ & $\cdots$ & $L$\\\hline
\end{tabular}
}}\otimes%
\begin{tabular}
[c]{|c|c|}\hline
$\mu_{2}^{1}$ & $\mu_{1}^{2}$\\\hline
\end{tabular}
\underset{m_{2}-1-l_{2}}{\underbrace{%
\begin{tabular}
[c]{|l|l|l|}\hline
$T$ & $\cdots$ & $T$\\\hline
\end{tabular}
}}\underset{l_{2}}{\underbrace{%
\begin{tabular}
[c]{|l|l|l|}\hline
$L$ & $\cdots$ & $L$\\\hline
\end{tabular}
}}\overset{N}{\underset{j\neq1,2}{\otimes}}%
\begin{tabular}
[c]{|c|}\hline
$\mu_{1}^{j}$\\\hline
\end{tabular}
\underset{m_{j}-1-l_{j}}{\underbrace{%
\begin{tabular}
[c]{|l|l|l|}\hline
$T$ & $\cdots$ & $T$\\\hline
\end{tabular}
}}\underset{l_{j}}{\underbrace{%
\begin{tabular}
[c]{|l|l|l|}\hline
$L$ & $\cdots$ & $L$\\\hline
\end{tabular}
}}\nonumber\\
&  +\left(  m_{1}-2-l_{1}\right)
\begin{tabular}
[c]{|c|}\hline
$\mu_{2}^{1}$\\\hline
\end{tabular}
\underset{m_{1}-3-l_{1}}{\underbrace{%
\begin{tabular}
[c]{|l|l|l|}\hline
$T$ & $\cdots$ & $T$\\\hline
\end{tabular}
}}\underset{l_{1}}{\underbrace{%
\begin{tabular}
[c]{|l|l|l|}\hline
$L$ & $\cdots$ & $L$\\\hline
\end{tabular}
}}\otimes%
\begin{tabular}
[c]{|c|}\hline
$\mu_{1}^{2}$\\\hline
\end{tabular}
\underset{m_{2}-l_{2}}{\underbrace{%
\begin{tabular}
[c]{|l|l|l|}\hline
$T$ & $\cdots$ & $T$\\\hline
\end{tabular}
}}\underset{l_{2}}{\underbrace{%
\begin{tabular}
[c]{|l|l|l|}\hline
$L$ & $\cdots$ & $L$\\\hline
\end{tabular}
}}\overset{N}{\underset{j\neq1,2}{\otimes}}%
\begin{tabular}
[c]{|c|}\hline
$\mu_{1}^{j}$\\\hline
\end{tabular}
\underset{m_{j}-1-l_{j}}{\underbrace{%
\begin{tabular}
[c]{|l|l|l|}\hline
$T$ & $\cdots$ & $T$\\\hline
\end{tabular}
}}\underset{l_{j}}{\underbrace{%
\begin{tabular}
[c]{|l|l|l|}\hline
$L$ & $\cdots$ & $L$\\\hline
\end{tabular}
}}\nonumber\\
&  +l_{1}%
\begin{tabular}
[c]{|c|}\hline
$\mu_{2}^{1}$\\\hline
\end{tabular}
\underset{m_{1}-2-l_{1}}{\underbrace{%
\begin{tabular}
[c]{|l|l|l|}\hline
$T$ & $\cdots$ & $T$\\\hline
\end{tabular}
}}\underset{l_{1}-1}{\underbrace{%
\begin{tabular}
[c]{|l|l|l|}\hline
$L$ & $\cdots$ & $L$\\\hline
\end{tabular}
}}\otimes%
\begin{tabular}
[c]{|c|}\hline
$\mu_{1}^{2}$\\\hline
\end{tabular}
\underset{m_{2}-1-l_{2}}{\underbrace{%
\begin{tabular}
[c]{|l|l|l|}\hline
$T$ & $\cdots$ & $T$\\\hline
\end{tabular}
}}\underset{l_{2}+1}{\underbrace{%
\begin{tabular}
[c]{|l|l|l|}\hline
$L$ & $\cdots$ & $L$\\\hline
\end{tabular}
}}\overset{N}{\underset{j\neq1,2}{\otimes}}%
\begin{tabular}
[c]{|c|}\hline
$\mu_{1}^{j}$\\\hline
\end{tabular}
\underset{m_{j}-1-l_{j}}{\underbrace{%
\begin{tabular}
[c]{|l|l|l|}\hline
$T$ & $\cdots$ & $T$\\\hline
\end{tabular}
}}\underset{l_{j}}{\underbrace{%
\begin{tabular}
[c]{|l|l|l|}\hline
$L$ & $\cdots$ & $L$\\\hline
\end{tabular}
}}\nonumber\\
&  +\sum_{l=3}^{N}\left(  l-1\right)
\begin{tabular}
[c]{|c|}\hline
$\mu_{2}^{1}$\\\hline
\end{tabular}
\underset{m_{1}-2-l_{1}}{\underbrace{%
\begin{tabular}
[c]{|l|l|l|}\hline
$T$ & $\cdots$ & $T$\\\hline
\end{tabular}
}}\underset{l_{1}}{\underbrace{%
\begin{tabular}
[c]{|l|l|l|}\hline
$L$ & $\cdots$ & $L$\\\hline
\end{tabular}
}}\otimes\underset{m_{l-1}-1-l_{l-1}}{\underbrace{%
\begin{tabular}
[c]{|l|l|l|}\hline
$T$ & $\cdots$ & $T$\\\hline
\end{tabular}
}}\underset{l_{l-1}}{\underbrace{%
\begin{tabular}
[c]{|l|l|l|}\hline
$L$ & $\cdots$ & $L$\\\hline
\end{tabular}
}}\nonumber\\
&  \otimes%
\begin{tabular}
[c]{|c|c|}\hline
$\mu_{1}^{l-1}$ & $\mu_{1}^{l}$\\\hline
\end{tabular}
\underset{m_{l}-1-l_{l}}{\underbrace{%
\begin{tabular}
[c]{|l|l|l|}\hline
$T$ & $\cdots$ & $T$\\\hline
\end{tabular}
}}\underset{l_{l}}{\underbrace{%
\begin{tabular}
[c]{|l|l|l|}\hline
$L$ & $\cdots$ & $L$\\\hline
\end{tabular}
}}\overset{N}{\underset{j\neq1,l,l-1}{\otimes}}%
\begin{tabular}
[c]{|c|}\hline
$\mu_{1}^{j}$\\\hline
\end{tabular}
\underset{m_{j}-1-l_{j}}{\underbrace{%
\begin{tabular}
[c]{|l|l|l|}\hline
$T$ & $\cdots$ & $T$\\\hline
\end{tabular}
}}\underset{l_{j}}{\underbrace{%
\begin{tabular}
[c]{|l|l|l|}\hline
$L$ & $\cdots$ & $L$\\\hline
\end{tabular}
}}\nonumber\\
&  +\sum_{l=3}^{N}\left(  l-1\right)  \left(  m_{l-1}-1-l_{l-1}\right)
\begin{tabular}
[c]{|c|}\hline
$\mu_{2}^{1}$\\\hline
\end{tabular}
\underset{m_{1}-2-l_{1}}{\underbrace{%
\begin{tabular}
[c]{|l|l|l|}\hline
$T$ & $\cdots$ & $T$\\\hline
\end{tabular}
}}\underset{l_{1}}{\underbrace{%
\begin{tabular}
[c]{|l|l|l|}\hline
$L$ & $\cdots$ & $L$\\\hline
\end{tabular}
}}\otimes%
\begin{tabular}
[c]{|c|}\hline
$\mu_{1}^{l-1}$\\\hline
\end{tabular}
\underset{m_{l-1}-2-l_{l-1}}{\underbrace{%
\begin{tabular}
[c]{|l|l|l|}\hline
$T$ & $\cdots\cdots$ & $T$\\\hline
\end{tabular}
}}\underset{l_{l-1}}{\underbrace{%
\begin{tabular}
[c]{|l|l|l|}\hline
$L$ & $\cdots$ & $L$\\\hline
\end{tabular}
}}\nonumber\\
&  \otimes%
\begin{tabular}
[c]{|c|}\hline
$\mu_{1}^{l}$\\\hline
\end{tabular}
\underset{m_{l}-l_{l}}{\underbrace{%
\begin{tabular}
[c]{|l|l|l|}\hline
$T$ & $\cdots$ & $T$\\\hline
\end{tabular}
}}\underset{l_{l}}{\underbrace{%
\begin{tabular}
[c]{|l|l|l|}\hline
$L$ & $\cdots$ & $L$\\\hline
\end{tabular}
}}\overset{N}{\underset{j\neq1,l,l-1}{\otimes}}%
\begin{tabular}
[c]{|c|}\hline
$\mu_{1}^{j}$\\\hline
\end{tabular}
\underset{m_{j}-1-l_{j}}{\underbrace{%
\begin{tabular}
[c]{|l|l|l|}\hline
$T$ & $\cdots$ & $T$\\\hline
\end{tabular}
}}\underset{l_{j}}{\underbrace{%
\begin{tabular}
[c]{|l|l|l|}\hline
$L$ & $\cdots$ & $L$\\\hline
\end{tabular}
}}\nonumber\\
&  +\sum_{l=3}^{N}l_{l-1}\left(  l-1\right)
\begin{tabular}
[c]{|c|}\hline
$\mu_{2}^{1}$\\\hline
\end{tabular}
\underset{m_{1}-2-l_{1}}{\underbrace{%
\begin{tabular}
[c]{|l|l|l|}\hline
$T$ & $\cdots$ & $T$\\\hline
\end{tabular}
}}\underset{l_{1}}{\underbrace{%
\begin{tabular}
[c]{|l|l|l|}\hline
$L$ & $\cdots$ & $L$\\\hline
\end{tabular}
}}\otimes%
\begin{tabular}
[c]{|c|}\hline
$\mu_{1}^{l-1}$\\\hline
\end{tabular}
\underset{m_{l-1}-1-l_{l-1}}{\underbrace{%
\begin{tabular}
[c]{|l|l|l|}\hline
$T$ & $\cdots\cdots$ & $T$\\\hline
\end{tabular}
}}\underset{l_{l-1}-1}{\underbrace{%
\begin{tabular}
[c]{|l|l|l|}\hline
$L$ & $\cdots$ & $L$\\\hline
\end{tabular}
}}\nonumber\\
&  \otimes%
\begin{tabular}
[c]{|c|}\hline
$\mu_{1}^{l}$\\\hline
\end{tabular}
\underset{m_{l}-1-l_{l}}{\underbrace{%
\begin{tabular}
[c]{|l|l|l|}\hline
$T$ & $\cdots$ & $T$\\\hline
\end{tabular}
}}\underset{l_{l}+1}{\underbrace{%
\begin{tabular}
[c]{|l|l|l|}\hline
$L$ & $\cdots$ & $L$\\\hline
\end{tabular}
}}\overset{N}{\underset{j\neq1,l,l-1}{\otimes}}%
\begin{tabular}
[c]{|c|}\hline
$\mu_{1}^{j}$\\\hline
\end{tabular}
\underset{m_{j}-1-l_{j}}{\underbrace{%
\begin{tabular}
[c]{|l|l|l|}\hline
$T$ & $\cdots$ & $T$\\\hline
\end{tabular}
}}\underset{l_{j}}{\underbrace{%
\begin{tabular}
[c]{|l|l|l|}\hline
$L$ & $\cdots$ & $L$\\\hline
\end{tabular}
}},
\end{align}
and%
\begin{align}
0  &  =\frac{1}{2}%
\begin{tabular}
[c]{|c|}\hline
$\mu_{3}^{1}$\\\hline
\end{tabular}
\underset{m_{1}-1-l_{1}}{\underbrace{%
\begin{tabular}
[c]{|l|l|l|}\hline
$T$ & $\cdots$ & $T$\\\hline
\end{tabular}
}}\underset{l_{1}}{\underbrace{%
\begin{tabular}
[c]{|l|l|l|}\hline
$L$ & $\cdots$ & $L$\\\hline
\end{tabular}
}}\overset{N}{\underset{j\neq1}{\otimes}}%
\begin{tabular}
[c]{|c|}\hline
$\mu_{1}^{j}$\\\hline
\end{tabular}
\underset{m_{j}-1-l_{j}}{\underbrace{%
\begin{tabular}
[c]{|l|l|l|}\hline
$T$ & $\cdots$ & $T$\\\hline
\end{tabular}
}}\underset{l_{j}}{\underbrace{%
\begin{tabular}
[c]{|l|l|l|}\hline
$L$ & $\cdots$ & $L$\\\hline
\end{tabular}
}}\nonumber\\
&  +M%
\begin{tabular}
[c]{|c|}\hline
$\mu_{3}^{1}$\\\hline
\end{tabular}
\underset{m_{1}-3-l_{1}}{\underbrace{%
\begin{tabular}
[c]{|l|l|l|}\hline
$T$ & $\cdots$ & $T$\\\hline
\end{tabular}
}}\underset{l_{1}}{\underbrace{%
\begin{tabular}
[c]{|l|l|l|}\hline
$L$ & $\cdots$ & $L$\\\hline
\end{tabular}
}}\otimes%
\begin{tabular}
[c]{|c|}\hline
$\mu_{1}^{2}$\\\hline
\end{tabular}
\underset{m_{2}-1-l_{2}}{\underbrace{%
\begin{tabular}
[c]{|l|l|l|}\hline
$T$ & $\cdots$ & $T$\\\hline
\end{tabular}
}}\underset{l_{2}+1}{\underbrace{%
\begin{tabular}
[c]{|l|l|l|}\hline
$L$ & $\cdots$ & $L$\\\hline
\end{tabular}
}}\overset{N}{\underset{j\neq1,2}{\otimes}}%
\begin{tabular}
[c]{|c|}\hline
$\mu_{1}^{j}$\\\hline
\end{tabular}
\underset{m_{j}-1-l_{j}}{\underbrace{%
\begin{tabular}
[c]{|l|l|l|}\hline
$T$ & $\cdots$ & $T$\\\hline
\end{tabular}
}}\underset{l_{j}}{\underbrace{%
\begin{tabular}
[c]{|l|l|l|}\hline
$L$ & $\cdots$ & $L$\\\hline
\end{tabular}
}}\nonumber\\
&  +\underset{m_{1}-3-l_{1}}{\underbrace{%
\begin{tabular}
[c]{|l|l|l|}\hline
$T$ & $\cdots$ & $T$\\\hline
\end{tabular}
}}\underset{l_{1}}{\underbrace{%
\begin{tabular}
[c]{|l|l|l|}\hline
$L$ & $\cdots$ & $L$\\\hline
\end{tabular}
}}\otimes%
\begin{tabular}
[c]{|c|c|}\hline
$\mu_{3}^{1}$ & $\mu_{1}^{3}$\\\hline
\end{tabular}
\underset{m_{3}-1-l_{3}}{\underbrace{%
\begin{tabular}
[c]{|l|l|l|}\hline
$T$ & $\cdots$ & $T$\\\hline
\end{tabular}
}}\underset{l_{3}}{\underbrace{%
\begin{tabular}
[c]{|l|l|l|}\hline
$L$ & $\cdots$ & $L$\\\hline
\end{tabular}
}}\overset{N}{\underset{j\neq1,3}{\otimes}}%
\begin{tabular}
[c]{|c|}\hline
$\mu_{1}^{j}$\\\hline
\end{tabular}
\underset{m_{j}-1-l_{j}}{\underbrace{%
\begin{tabular}
[c]{|l|l|l|}\hline
$T$ & $\cdots$ & $T$\\\hline
\end{tabular}
}}\underset{l_{j}}{\underbrace{%
\begin{tabular}
[c]{|l|l|l|}\hline
$L$ & $\cdots$ & $L$\\\hline
\end{tabular}
}}\nonumber\\
&  +\left(  m_{1}-3-l_{1}\right)
\begin{tabular}
[c]{|c|}\hline
$\mu_{3}^{1}$\\\hline
\end{tabular}
\underset{m_{1}-4-l_{1}}{\underbrace{%
\begin{tabular}
[c]{|l|l|l|}\hline
$T$ & $\cdots$ & $T$\\\hline
\end{tabular}
}}\underset{l_{1}}{\underbrace{%
\begin{tabular}
[c]{|l|l|l|}\hline
$L$ & $\cdots$ & $L$\\\hline
\end{tabular}
}}\otimes%
\begin{tabular}
[c]{|c|}\hline
$\mu_{1}^{3}$\\\hline
\end{tabular}
\underset{m_{3}-l_{3}}{\underbrace{%
\begin{tabular}
[c]{|l|l|l|}\hline
$T$ & $\cdots$ & $T$\\\hline
\end{tabular}
}}\underset{l_{3}}{\underbrace{%
\begin{tabular}
[c]{|l|l|l|}\hline
$L$ & $\cdots$ & $L$\\\hline
\end{tabular}
}}\overset{N}{\underset{j\neq1,3}{\otimes}}%
\begin{tabular}
[c]{|c|}\hline
$\mu_{1}^{j}$\\\hline
\end{tabular}
\underset{m_{j}-1-l_{j}}{\underbrace{%
\begin{tabular}
[c]{|l|l|l|}\hline
$T$ & $\cdots$ & $T$\\\hline
\end{tabular}
}}\underset{l_{j}}{\underbrace{%
\begin{tabular}
[c]{|l|l|l|}\hline
$L$ & $\cdots$ & $L$\\\hline
\end{tabular}
}}\nonumber\\
&  +l_{1}%
\begin{tabular}
[c]{|c|}\hline
$\mu_{3}^{1}$\\\hline
\end{tabular}
\underset{m_{1}-3-l_{1}}{\underbrace{%
\begin{tabular}
[c]{|l|l|l|}\hline
$T$ & $\cdots$ & $T$\\\hline
\end{tabular}
}}\underset{l_{1}-1}{\underbrace{%
\begin{tabular}
[c]{|l|l|l|}\hline
$L$ & $\cdots$ & $L$\\\hline
\end{tabular}
}}\otimes%
\begin{tabular}
[c]{|c|}\hline
$\mu_{1}^{3}$\\\hline
\end{tabular}
\underset{m_{3}-1-l_{3}}{\underbrace{%
\begin{tabular}
[c]{|l|l|l|}\hline
$T$ & $\cdots$ & $T$\\\hline
\end{tabular}
}}\underset{l_{3}+1}{\underbrace{%
\begin{tabular}
[c]{|l|l|l|}\hline
$L$ & $\cdots$ & $L$\\\hline
\end{tabular}
}}\overset{N}{\underset{j\neq1,3}{\otimes}}%
\begin{tabular}
[c]{|c|}\hline
$\mu_{1}^{j}$\\\hline
\end{tabular}
\underset{m_{j}-1-l_{j}}{\underbrace{%
\begin{tabular}
[c]{|l|l|l|}\hline
$T$ & $\cdots$ & $T$\\\hline
\end{tabular}
}}\underset{l_{j}}{\underbrace{%
\begin{tabular}
[c]{|l|l|l|}\hline
$L$ & $\cdots$ & $L$\\\hline
\end{tabular}
}}\nonumber\\
&  +\sum_{l=4}^{N}\left(  l-2\right)
\begin{tabular}
[c]{|c|}\hline
$\mu_{3}^{1}$\\\hline
\end{tabular}
\underset{m_{1}-3-l_{1}}{\underbrace{%
\begin{tabular}
[c]{|l|l|l|}\hline
$T$ & $\cdots$ & $T$\\\hline
\end{tabular}
}}\underset{l_{1}}{\underbrace{%
\begin{tabular}
[c]{|l|l|l|}\hline
$L$ & $\cdots$ & $L$\\\hline
\end{tabular}
}}\otimes\underset{m_{l-2}-1-l_{l-2}}{\underbrace{%
\begin{tabular}
[c]{|l|l|l|}\hline
$T$ & $\cdots$ & $T$\\\hline
\end{tabular}
}}\underset{l_{l-2}}{\underbrace{%
\begin{tabular}
[c]{|l|l|l|}\hline
$L$ & $\cdots$ & $L$\\\hline
\end{tabular}
}}\nonumber\\
&  \otimes%
\begin{tabular}
[c]{|c|c|}\hline
$\mu_{1}^{l-2}$ & $\mu_{1}^{l}$\\\hline
\end{tabular}
\underset{m_{l}-1-l_{l}}{\underbrace{%
\begin{tabular}
[c]{|l|l|l|}\hline
$T$ & $\cdots$ & $T$\\\hline
\end{tabular}
}}\underset{l_{l}}{\underbrace{%
\begin{tabular}
[c]{|l|l|l|}\hline
$L$ & $\cdots$ & $L$\\\hline
\end{tabular}
}}\overset{N}{\underset{j\neq1,l,l-2}{\otimes}}%
\begin{tabular}
[c]{|c|}\hline
$\mu_{1}^{j}$\\\hline
\end{tabular}
\underset{m_{j}-1-l_{j}}{\underbrace{%
\begin{tabular}
[c]{|l|l|l|}\hline
$T$ & $\cdots$ & $T$\\\hline
\end{tabular}
}}\underset{l_{j}}{\underbrace{%
\begin{tabular}
[c]{|l|l|l|}\hline
$L$ & $\cdots$ & $L$\\\hline
\end{tabular}
}}\nonumber\\
&  +\sum_{l=4}^{N}\left(  l-2\right)  \left(  m_{l-2}-1-l_{l-2}\right)
\begin{tabular}
[c]{|c|}\hline
$\mu_{3}^{1}$\\\hline
\end{tabular}
\underset{m_{1}-3-l_{1}}{\underbrace{%
\begin{tabular}
[c]{|l|l|l|}\hline
$T$ & $\cdots$ & $T$\\\hline
\end{tabular}
}}\underset{l_{1}}{\underbrace{%
\begin{tabular}
[c]{|l|l|l|}\hline
$L$ & $\cdots$ & $L$\\\hline
\end{tabular}
}}\otimes%
\begin{tabular}
[c]{|c|}\hline
$\mu_{1}^{l-2}$\\\hline
\end{tabular}
\underset{m_{l-2}-2-l_{l-2}}{\underbrace{%
\begin{tabular}
[c]{|l|l|l|}\hline
$T$ & $\cdots$ & $T$\\\hline
\end{tabular}
}}\underset{l_{l-2}}{\underbrace{%
\begin{tabular}
[c]{|l|l|l|}\hline
$L$ & $\cdots$ & $L$\\\hline
\end{tabular}
}}\nonumber\\
&  \otimes%
\begin{tabular}
[c]{|c|}\hline
$\mu_{1}^{l}$\\\hline
\end{tabular}
\underset{m_{l}-l_{l}}{\underbrace{%
\begin{tabular}
[c]{|l|l|l|}\hline
$T$ & $\cdots$ & $T$\\\hline
\end{tabular}
}}\underset{l_{l}}{\underbrace{%
\begin{tabular}
[c]{|l|l|l|}\hline
$L$ & $\cdots$ & $L$\\\hline
\end{tabular}
}}\overset{N}{\underset{j\neq1,l,l-2}{\otimes}}%
\begin{tabular}
[c]{|c|}\hline
$\mu_{1}^{j}$\\\hline
\end{tabular}
\underset{m_{j}-1-l_{j}}{\underbrace{%
\begin{tabular}
[c]{|l|l|l|}\hline
$T$ & $\cdots$ & $T$\\\hline
\end{tabular}
}}\underset{l_{j}}{\underbrace{%
\begin{tabular}
[c]{|l|l|l|}\hline
$L$ & $\cdots$ & $L$\\\hline
\end{tabular}
}}\nonumber\\
&  +\sum_{l=4}^{N}l_{l-2}\left(  l-2\right)
\begin{tabular}
[c]{|c|}\hline
$\mu_{3}^{1}$\\\hline
\end{tabular}
\underset{m_{1}-3-l_{1}}{\underbrace{%
\begin{tabular}
[c]{|l|l|l|}\hline
$T$ & $\cdots$ & $T$\\\hline
\end{tabular}
}}\underset{l_{1}}{\underbrace{%
\begin{tabular}
[c]{|l|l|l|}\hline
$L$ & $\cdots$ & $L$\\\hline
\end{tabular}
}}\otimes\sum_{i=2}^{m_{l-2}}%
\begin{tabular}
[c]{|c|}\hline
$\mu_{1}^{l-2}$\\\hline
\end{tabular}
\underset{m_{l-2}-1-l_{l-2}}{\underbrace{%
\begin{tabular}
[c]{|l|l|l|}\hline
$T$ & $\cdots$ & $T$\\\hline
\end{tabular}
}}\underset{l_{l-2}-1}{\underbrace{%
\begin{tabular}
[c]{|l|l|l|}\hline
$L$ & $\cdots$ & $L$\\\hline
\end{tabular}
}}\nonumber\\
&  \otimes%
\begin{tabular}
[c]{|c|}\hline
$\mu_{1}^{l}$\\\hline
\end{tabular}
\underset{m_{l}-1-l_{l}}{\underbrace{%
\begin{tabular}
[c]{|l|l|l|}\hline
$T$ & $\cdots$ & $T$\\\hline
\end{tabular}
}}\underset{l_{l}+1}{\underbrace{%
\begin{tabular}
[c]{|l|l|l|}\hline
$L$ & $\cdots$ & $L$\\\hline
\end{tabular}
}}\overset{N}{\underset{j\neq1,l,l-2}{\otimes}}%
\begin{tabular}
[c]{|c|}\hline
$\mu_{1}^{j}$\\\hline
\end{tabular}
\underset{m_{j}-1-l_{j}}{\underbrace{%
\begin{tabular}
[c]{|l|l|l|}\hline
$T$ & $\cdots$ & $T$\\\hline
\end{tabular}
}}\underset{l_{j}}{\underbrace{%
\begin{tabular}
[c]{|l|l|l|}\hline
$L$ & $\cdots$ & $L$\\\hline
\end{tabular}
}}.
\end{align}%
\end{subequations}%
There are still some undetermined parameters $\mu_{2}^{1}$, $\mu_{3}^{1}$ and
$\mu_{1}^{j}\left(  j\geq2\right)  $, which can be chosen to be $L$ or $T$, in
the above equations. However, it is easy to see that both choices lead to the
same equations. Therefore, we will set all of them to be $T$ in the following.
The final Virasoro constraints at high energies become%
\begin{subequations}%
\begin{align}
0  &  =M\underset{m_{1}-1-l_{1}}{\underbrace{%
\begin{tabular}
[c]{|l|l|l|}\hline
$T$ & $\cdots$ & $T$\\\hline
\end{tabular}
}}\underset{l_{1}+1}{\underbrace{%
\begin{tabular}
[c]{|l|l|l|}\hline
$L$ & $\cdots$ & $L$\\\hline
\end{tabular}
}}\overset{N}{\underset{j\neq1}{\otimes}}\underset{m_{j}-l_{j}}{\underbrace{%
\begin{tabular}
[c]{|l|l|l|}\hline
$T$ & $\cdots$ & $T$\\\hline
\end{tabular}
}}\underset{l_{j}}{\underbrace{%
\begin{tabular}
[c]{|l|l|l|}\hline
$L$ & $\cdots$ & $L$\\\hline
\end{tabular}
}}\nonumber\\
&  +\left(  m_{1}-1-l_{1}\right)  \underset{m_{1}-2-l_{1}}{\underbrace{%
\begin{tabular}
[c]{|l|l|l|}\hline
$T$ & $\cdots$ & $T$\\\hline
\end{tabular}
}}\underset{l_{1}}{\underbrace{%
\begin{tabular}
[c]{|l|l|l|}\hline
$L$ & $\cdots$ & $L$\\\hline
\end{tabular}
}}\otimes\underset{m_{2}+1-l_{2}}{\underbrace{%
\begin{tabular}
[c]{|l|l|l|}\hline
$T$ & $\cdots$ & $T$\\\hline
\end{tabular}
}}\underset{l_{2}}{\underbrace{%
\begin{tabular}
[c]{|l|l|l|}\hline
$L$ & $\cdots$ & $L$\\\hline
\end{tabular}
}}\overset{N}{\underset{j\neq1,2}{\otimes}}\underset{m_{j}-l_{j}}{\underbrace{%
\begin{tabular}
[c]{|l|l|l|}\hline
$T$ & $\cdots$ & $T$\\\hline
\end{tabular}
}}\underset{l_{j}}{\underbrace{%
\begin{tabular}
[c]{|l|l|l|}\hline
$L$ & $\cdots$ & $L$\\\hline
\end{tabular}
}}\nonumber\\
&  +l_{1}\underset{m_{1}-1-l_{1}}{\underbrace{%
\begin{tabular}
[c]{|l|l|l|}\hline
$T$ & $\cdots$ & $T$\\\hline
\end{tabular}
}}\underset{l_{1}-1}{\underbrace{%
\begin{tabular}
[c]{|l|l|l|}\hline
$L$ & $\cdots$ & $L$\\\hline
\end{tabular}
}}\otimes\underset{m_{2}-l_{2}}{\underbrace{%
\begin{tabular}
[c]{|l|l|l|}\hline
$T$ & $\cdots$ & $T$\\\hline
\end{tabular}
}}\underset{l_{2}+1}{\underbrace{%
\begin{tabular}
[c]{|l|l|l|}\hline
$L$ & $\cdots$ & $L$\\\hline
\end{tabular}
}}\overset{N}{\underset{j\neq1,2}{\otimes}}\underset{m_{j}-l_{j}}{\underbrace{%
\begin{tabular}
[c]{|l|l|l|}\hline
$T$ & $\cdots$ & $T$\\\hline
\end{tabular}
}}\underset{l_{j}}{\underbrace{%
\begin{tabular}
[c]{|l|l|l|}\hline
$L$ & $\cdots$ & $L$\\\hline
\end{tabular}
}}\nonumber\\
&  +\sum_{l=3}^{N}\left(  l-1\right)  \left(  m_{l-1}-l_{l-1}\right)
\underset{m_{1}-1-l_{1}}{\underbrace{%
\begin{tabular}
[c]{|l|l|l|}\hline
$T$ & $\cdots$ & $T$\\\hline
\end{tabular}
}}\underset{l_{1}}{\underbrace{%
\begin{tabular}
[c]{|l|l|l|}\hline
$L$ & $\cdots$ & $L$\\\hline
\end{tabular}
}}\otimes\underset{m_{l-1}-1-l_{l-1}}{\underbrace{%
\begin{tabular}
[c]{|l|l|l|}\hline
$T$ & $\cdots$ & $T$\\\hline
\end{tabular}
}}\underset{l_{l-1}}{\underbrace{%
\begin{tabular}
[c]{|l|l|l|}\hline
$L$ & $\cdots$ & $L$\\\hline
\end{tabular}
}}\nonumber\\
&  \otimes\underset{m_{l}+1-l_{l}}{\underbrace{%
\begin{tabular}
[c]{|l|l|l|}\hline
$T$ & $\cdots$ & $T$\\\hline
\end{tabular}
}}\underset{l_{l}}{\underbrace{%
\begin{tabular}
[c]{|l|l|l|}\hline
$L$ & $\cdots$ & $L$\\\hline
\end{tabular}
}}\overset{N}{\underset{j\neq1,l,l-1}{\otimes}}\underset{m_{j}-l_{j}%
}{\underbrace{%
\begin{tabular}
[c]{|l|l|l|}\hline
$T$ & $\cdots$ & $T$\\\hline
\end{tabular}
}}\underset{l_{j}}{\underbrace{%
\begin{tabular}
[c]{|l|l|l|}\hline
$L$ & $\cdots$ & $L$\\\hline
\end{tabular}
}}\nonumber\\
&  +\sum_{l=3}^{N}l_{l-1}\left(  l-1\right)  \underset{m_{1}-1-l_{1}%
}{\underbrace{%
\begin{tabular}
[c]{|l|l|l|}\hline
$T$ & $\cdots$ & $T$\\\hline
\end{tabular}
}}\underset{l_{1}}{\underbrace{%
\begin{tabular}
[c]{|l|l|l|}\hline
$L$ & $\cdots$ & $L$\\\hline
\end{tabular}
}}\otimes\underset{m_{l-1}-l_{l-1}}{\underbrace{%
\begin{tabular}
[c]{|l|l|l|}\hline
$T$ & $\cdots$ & $T$\\\hline
\end{tabular}
}}\underset{l_{l-1}-1}{\underbrace{%
\begin{tabular}
[c]{|l|l|l|}\hline
$L$ & $\cdots$ & $L$\\\hline
\end{tabular}
}}\nonumber\\
&  \otimes\underset{m_{l}-l_{l}}{\underbrace{%
\begin{tabular}
[c]{|l|l|l|}\hline
$T$ & $\cdots$ & $T$\\\hline
\end{tabular}
}}\underset{l_{l}+1}{\underbrace{%
\begin{tabular}
[c]{|l|l|l|}\hline
$L$ & $\cdots$ & $L$\\\hline
\end{tabular}
}}\overset{N}{\underset{j\neq1,l,l-1}{\otimes}}\underset{m_{j}-l_{j}%
}{\underbrace{%
\begin{tabular}
[c]{|l|l|l|}\hline
$T$ & $\cdots$ & $T$\\\hline
\end{tabular}
}}\underset{l_{j}}{\underbrace{%
\begin{tabular}
[c]{|l|l|l|}\hline
$L$ & $\cdots$ & $L$\\\hline
\end{tabular}
}}, \label{a}%
\end{align}
and%

\begin{align}
0  &  =\frac{1}{2}\underset{m_{1}-l_{1}}{\underbrace{%
\begin{tabular}
[c]{|l|l|l|}\hline
$T$ & $\cdots$ & $T$\\\hline
\end{tabular}
}}\underset{l_{1}}{\underbrace{%
\begin{tabular}
[c]{|l|l|l|}\hline
$L$ & $\cdots$ & $L$\\\hline
\end{tabular}
}}\overset{N}{\underset{j\neq1}{\otimes}}\underset{m_{j}-l_{j}}{\underbrace{%
\begin{tabular}
[c]{|l|l|l|}\hline
$T$ & $\cdots$ & $T$\\\hline
\end{tabular}
}}\underset{l_{j}}{\underbrace{%
\begin{tabular}
[c]{|l|l|l|}\hline
$L$ & $\cdots$ & $L$\\\hline
\end{tabular}
}}\nonumber\\
&  +M\underset{m_{1}-2-l_{1}}{\underbrace{%
\begin{tabular}
[c]{|l|l|l|}\hline
$T$ & $\cdots$ & $T$\\\hline
\end{tabular}
}}\underset{l_{1}}{\underbrace{%
\begin{tabular}
[c]{|l|l|l|}\hline
$L$ & $\cdots$ & $L$\\\hline
\end{tabular}
}}\otimes\underset{m_{2}-l_{2}}{\underbrace{%
\begin{tabular}
[c]{|l|l|l|}\hline
$T$ & $\cdots$ & $T$\\\hline
\end{tabular}
}}\underset{l_{2}+1}{\underbrace{%
\begin{tabular}
[c]{|l|l|l|}\hline
$L$ & $\cdots$ & $L$\\\hline
\end{tabular}
}}\overset{k}{\overset{N}{\underset{j\neq1,2}{\otimes}}}\underset{m_{j}%
-l_{j}}{\underbrace{%
\begin{tabular}
[c]{|l|l|l|}\hline
$T$ & $\cdots$ & $T$\\\hline
\end{tabular}
}}\underset{l_{j}}{\underbrace{%
\begin{tabular}
[c]{|l|l|l|}\hline
$L$ & $\cdots$ & $L$\\\hline
\end{tabular}
}}\nonumber\\
&  +\left(  m_{1}-2-l_{1}\right)  \underset{m_{1}-3-l_{1}}{\underbrace{%
\begin{tabular}
[c]{|l|l|l|}\hline
$T$ & $\cdots$ & $T$\\\hline
\end{tabular}
}}\underset{l_{1}}{\underbrace{%
\begin{tabular}
[c]{|l|l|l|}\hline
$L$ & $\cdots$ & $L$\\\hline
\end{tabular}
}}\otimes\underset{m_{3}+1-l_{3}}{\underbrace{%
\begin{tabular}
[c]{|l|l|l|}\hline
$T$ & $\cdots$ & $T$\\\hline
\end{tabular}
}}\underset{l_{3}}{\underbrace{%
\begin{tabular}
[c]{|l|l|l|}\hline
$L$ & $\cdots$ & $L$\\\hline
\end{tabular}
}}\overset{N}{\underset{j\neq1,3}{\otimes}}\underset{m_{j}-l_{j}}{\underbrace{%
\begin{tabular}
[c]{|l|l|l|}\hline
$T$ & $\cdots$ & $T$\\\hline
\end{tabular}
}}\underset{l_{j}}{\underbrace{%
\begin{tabular}
[c]{|l|l|l|}\hline
$L$ & $\cdots$ & $L$\\\hline
\end{tabular}
}}\nonumber\\
&  +l_{1}\underset{m_{1}-2-l_{1}}{\underbrace{%
\begin{tabular}
[c]{|l|l|l|}\hline
$T$ & $\cdots$ & $T$\\\hline
\end{tabular}
}}\underset{l_{1}-1}{\underbrace{%
\begin{tabular}
[c]{|l|l|l|}\hline
$L$ & $\cdots$ & $L$\\\hline
\end{tabular}
}}\otimes\underset{m_{3}-l_{3}}{\underbrace{%
\begin{tabular}
[c]{|l|l|l|}\hline
$T$ & $\cdots$ & $T$\\\hline
\end{tabular}
}}\underset{l_{3}+1}{\underbrace{%
\begin{tabular}
[c]{|l|l|l|}\hline
$L$ & $\cdots$ & $L$\\\hline
\end{tabular}
}}\overset{N}{\underset{j\neq1,3}{\otimes}}\underset{m_{j}-l_{j}}{\underbrace{%
\begin{tabular}
[c]{|l|l|l|}\hline
$T$ & $\cdots$ & $T$\\\hline
\end{tabular}
}}\underset{l_{j}}{\underbrace{%
\begin{tabular}
[c]{|l|l|l|}\hline
$L$ & $\cdots$ & $L$\\\hline
\end{tabular}
}}\nonumber\\
&  +\sum_{l=4}^{N}\left(  l-2\right)  \left(  m_{l-2}-l_{l-2}\right)
\underset{m_{1}-2-l_{1}}{\underbrace{%
\begin{tabular}
[c]{|l|l|l|}\hline
$T$ & $\cdots$ & $T$\\\hline
\end{tabular}
}}\underset{l_{1}}{\underbrace{%
\begin{tabular}
[c]{|l|l|l|}\hline
$L$ & $\cdots$ & $L$\\\hline
\end{tabular}
}}\otimes\underset{m_{l-2}-1-l_{l-2}}{\underbrace{%
\begin{tabular}
[c]{|l|l|l|}\hline
$T$ & $\cdots$ & $T$\\\hline
\end{tabular}
}}\underset{l_{l-2}}{\underbrace{%
\begin{tabular}
[c]{|l|l|l|}\hline
$L$ & $\cdots$ & $L$\\\hline
\end{tabular}
}}\nonumber\\
&  \otimes\underset{m_{l}+1-l_{l}}{\underbrace{%
\begin{tabular}
[c]{|l|l|l|}\hline
$T$ & $\cdots$ & $T$\\\hline
\end{tabular}
}}\underset{l_{l}}{\underbrace{%
\begin{tabular}
[c]{|l|l|l|}\hline
$L$ & $\cdots$ & $L$\\\hline
\end{tabular}
}}\overset{N}{\underset{j\neq1,l,l-2}{\otimes}}\underset{m_{j}-l_{j}%
}{\underbrace{%
\begin{tabular}
[c]{|l|l|l|}\hline
$T$ & $\cdots$ & $T$\\\hline
\end{tabular}
}}\underset{l_{j}}{\underbrace{%
\begin{tabular}
[c]{|l|l|l|}\hline
$L$ & $\cdots$ & $L$\\\hline
\end{tabular}
}}\nonumber\\
&  +\sum_{l=4}^{N}l_{l-2}\left(  l-2\right)  \underset{m_{1}-2-l_{1}%
}{\underbrace{%
\begin{tabular}
[c]{|l|l|l|}\hline
$T$ & $\cdots$ & $T$\\\hline
\end{tabular}
}}\underset{l_{1}}{\underbrace{%
\begin{tabular}
[c]{|l|l|l|}\hline
$L$ & $\cdots$ & $L$\\\hline
\end{tabular}
}}\otimes\sum_{i=2}^{m_{l-2}}\underset{m_{l-2}-l_{l-2}}{\underbrace{%
\begin{tabular}
[c]{|l|l|l|}\hline
$T$ & $\cdots$ & $T$\\\hline
\end{tabular}
}}\underset{l_{l-2}-1}{\underbrace{%
\begin{tabular}
[c]{|l|l|l|}\hline
$L$ & $\cdots$ & $L$\\\hline
\end{tabular}
}}\nonumber\\
&  \otimes\underset{m_{l}-l_{l}}{\underbrace{%
\begin{tabular}
[c]{|l|l|l|}\hline
$T$ & $\cdots$ & $T$\\\hline
\end{tabular}
}}\underset{l_{l}+1}{\underbrace{%
\begin{tabular}
[c]{|l|l|l|}\hline
$L$ & $\cdots$ & $L$\\\hline
\end{tabular}
}}\overset{N}{\underset{j\neq1,l,l-2}{\otimes}}\underset{m_{j}-l_{j}%
}{\underbrace{%
\begin{tabular}
[c]{|l|l|l|}\hline
$T$ & $\cdots$ & $T$\\\hline
\end{tabular}
}}\underset{l_{j}}{\underbrace{%
\begin{tabular}
[c]{|l|l|l|}\hline
$L$ & $\cdots$ & $L$\\\hline
\end{tabular}
}}. \label{b}%
\end{align}%
\end{subequations}%
To solve the above constraints, we need the following lemma to further
simplify them.

\textbf{Lemma}%

\begin{equation}%
\begin{tabular}
[c]{|l|l|l|}\hline
$T$ & $\cdots$ & $T$\\\hline
\end{tabular}
\underset{l_{1}}{\underbrace{%
\begin{tabular}
[c]{|l|l|l|}\hline
$L$ & $\cdots$ & $L$\\\hline
\end{tabular}
}}\otimes\underset{m_{2}-l_{2}}{\underbrace{%
\begin{tabular}
[c]{|l|l|l|}\hline
$T$ & $\cdots$ & $T$\\\hline
\end{tabular}
}}%
\begin{tabular}
[c]{|l|l|l|}\hline
$L$ & $\cdots$ & $L$\\\hline
\end{tabular}
\otimes\underset{\left\{  m_{j},j\geq3\right\}  }{\underbrace{%
\begin{tabular}
[c]{|lll|}\hline
& $\cdots$ & \\\hline
\end{tabular}
}}\equiv0, \label{lemma}%
\end{equation}
except for (i) $l_{2}=m_{2}$, $m_{j}=0$ for $j\geq3$ and (ii) $l_{1}=2m$.

\textbf{Proof: }In the high energy limit, we only need to consider the leading
energy terms. To count the energy scaling behavior: each $T$ contributes a
factor of energy $E$ and each $L$ contributes $E^{2}$. Any terms with total
energy order level less than $N$ are sub-leading terms and can be ignored.

(i) If $l_{2}\neq m_{2}$ and $m_{j}\neq0,j\geq3$, then in Eq.(\ref{a}),

\begin{enumerate}
\item for $l_{1}=0$, all terms except the first term are sub-leading, then%
\begin{equation}%
\begin{tabular}
[c]{|l|l|l|}\hline
$T$ & $\cdots$ & $T$\\\hline
\end{tabular}
\underset{1}{\underbrace{%
\begin{tabular}
[c]{|l|l|l|}\hline
$L$ & $\cdots$ & $L$\\\hline
\end{tabular}
}}\overset{N}{\underset{j\neq1}{\otimes}}\underset{m_{j}-l_{j}}{\underbrace{%
\begin{tabular}
[c]{|l|l|l|}\hline
$T$ & $\cdots$ & $T$\\\hline
\end{tabular}
}}\underset{l_{j}}{\underbrace{%
\begin{tabular}
[c]{|l|l|l|}\hline
$L$ & $\cdots$ & $L$\\\hline
\end{tabular}
}}=0, \label{l=0}%
\end{equation}

\item for $l_{1}=1$, the third term is sub-leading, and (\ref{l=0}) implies
all other terms except the first term are vanished, then%
\begin{equation}%
\begin{tabular}
[c]{|l|l|l|}\hline
$T$ & $\cdots$ & $T$\\\hline
\end{tabular}
\underset{2}{\underbrace{%
\begin{tabular}
[c]{|l|l|l|}\hline
$L$ & $\cdots$ & $L$\\\hline
\end{tabular}
}}\overset{N}{\underset{j\neq1}{\otimes}}\underset{m_{j}-l_{j}}{\underbrace{%
\begin{tabular}
[c]{|l|l|l|}\hline
$T$ & $\cdots$ & $T$\\\hline
\end{tabular}
}}\underset{l_{j}}{\underbrace{%
\begin{tabular}
[c]{|l|l|l|}\hline
$L$ & $\cdots$ & $L$\\\hline
\end{tabular}
}}=0,
\end{equation}

\item if for $l_{1}=l^{\prime}$,%
\begin{equation}%
\begin{tabular}
[c]{|l|l|l|}\hline
$T$ & $\cdots$ & $T$\\\hline
\end{tabular}
\underset{l^{\prime}-1}{\underbrace{%
\begin{tabular}
[c]{|l|l|l|}\hline
$L$ & $\cdots$ & $L$\\\hline
\end{tabular}
}}\overset{N}{\underset{j\neq1}{\otimes}}\underset{m_{j}-l_{j}}{\underbrace{%
\begin{tabular}
[c]{|l|l|l|}\hline
$T$ & $\cdots$ & $T$\\\hline
\end{tabular}
}}\underset{l_{j}}{\underbrace{%
\begin{tabular}
[c]{|l|l|l|}\hline
$L$ & $\cdots$ & $L$\\\hline
\end{tabular}
}}=0,
\end{equation}
and%
\begin{equation}%
\begin{tabular}
[c]{|l|l|l|}\hline
$T$ & $\cdots$ & $T$\\\hline
\end{tabular}
\underset{l^{\prime}}{\underbrace{%
\begin{tabular}
[c]{|l|l|l|}\hline
$L$ & $\cdots$ & $L$\\\hline
\end{tabular}
}}\overset{N}{\underset{j\neq1}{\otimes}}\underset{m_{j}-l_{j}}{\underbrace{%
\begin{tabular}
[c]{|l|l|l|}\hline
$T$ & $\cdots$ & $T$\\\hline
\end{tabular}
}}\underset{l_{j}}{\underbrace{%
\begin{tabular}
[c]{|l|l|l|}\hline
$L$ & $\cdots$ & $L$\\\hline
\end{tabular}
}}=0,
\end{equation}
(\ref{a}) implies all terms except the first term are vanished, then%
\begin{equation}%
\begin{tabular}
[c]{|l|l|l|}\hline
$T$ & $\cdots$ & $T$\\\hline
\end{tabular}
\underset{l^{\prime}+1}{\underbrace{%
\begin{tabular}
[c]{|l|l|l|}\hline
$L$ & $\cdots$ & $L$\\\hline
\end{tabular}
}}\overset{N}{\underset{j\neq1}{\otimes}}\underset{m_{j}-l_{j}}{\underbrace{%
\begin{tabular}
[c]{|l|l|l|}\hline
$T$ & $\cdots$ & $T$\\\hline
\end{tabular}
}}\underset{l_{j}}{\underbrace{%
\begin{tabular}
[c]{|l|l|l|}\hline
$L$ & $\cdots$ & $L$\\\hline
\end{tabular}
}}=0.
\end{equation}

\end{enumerate}

(ii) If $l_{2}=m_{2}$ and $m_{j}=0$ for $j\geq3$, then Eq.(\ref{a}) reduces to%
\begin{equation}
M\underset{m_{1}-1-l_{1}}{\underbrace{%
\begin{tabular}
[c]{|l|l|l|}\hline
$T$ & $\cdots$ & $T$\\\hline
\end{tabular}
}}\underset{l_{1}+1}{\underbrace{%
\begin{tabular}
[c]{|l|l|l|}\hline
$L$ & $\cdots$ & $L$\\\hline
\end{tabular}
}}\otimes\underset{m_{2}}{\underbrace{%
\begin{tabular}
[c]{|l|l|l|}\hline
$L$ & $\cdots$ & $L$\\\hline
\end{tabular}
}}+l_{1}\underset{m_{1}-1-l_{1}}{\underbrace{%
\begin{tabular}
[c]{|l|l|l|}\hline
$T$ & $\cdots$ & $T$\\\hline
\end{tabular}
}}\underset{l_{1}-1}{\underbrace{%
\begin{tabular}
[c]{|l|l|l|}\hline
$L$ & $\cdots$ & $L$\\\hline
\end{tabular}
}}\otimes\underset{m_{2}+1}{\underbrace{%
\begin{tabular}
[c]{|l|l|l|}\hline
$L$ & $\cdots$ & $L$\\\hline
\end{tabular}
}}=0. \label{aa}%
\end{equation}
Similarly, we have in Eq.(\ref{aa}),

\begin{enumerate}
\item for $l_{1}=0$,%
\begin{equation}%
\begin{tabular}
[c]{|l|l|l|}\hline
$T$ & $\cdots$ & $T$\\\hline
\end{tabular}
\underset{1}{\underbrace{%
\begin{tabular}
[c]{|l|l|l|}\hline
$L$ & $\cdots$ & $L$\\\hline
\end{tabular}
}}\otimes%
\begin{tabular}
[c]{|l|l|l|}\hline
$L$ & $\cdots$ & $L$\\\hline
\end{tabular}
=0,
\end{equation}

\item if for $l_{1}=2m$,%
\begin{equation}%
\begin{tabular}
[c]{|l|l|l|}\hline
$T$ & $\cdots$ & $T$\\\hline
\end{tabular}
\underset{2m-1}{\underbrace{%
\begin{tabular}
[c]{|l|l|l|}\hline
$L$ & $\cdots$ & $L$\\\hline
\end{tabular}
}}\otimes%
\begin{tabular}
[c]{|l|l|l|}\hline
$L$ & $\cdots$ & $L$\\\hline
\end{tabular}
=0,
\end{equation}
then (\ref{aa}) implies%
\begin{equation}%
\begin{tabular}
[c]{|l|l|l|}\hline
$T$ & $\cdots$ & $T$\\\hline
\end{tabular}
\underset{2m+1}{\underbrace{%
\begin{tabular}
[c]{|l|l|l|}\hline
$L$ & $\cdots$ & $L$\\\hline
\end{tabular}
}}\otimes%
\begin{tabular}
[c]{|l|l|l|}\hline
$L$ & $\cdots$ & $L$\\\hline
\end{tabular}
=0.
\end{equation}

\end{enumerate}

Finally, the Virasoro constraints at high energies reduce to%
\begin{subequations}%
\begin{align}
\underset{n-2q-2-2m}{\underbrace{%
\begin{tabular}
[c]{|l|l|l|}\hline
$T$ & $\cdots$ & $T$\\\hline
\end{tabular}
}}\underset{2m+2}{\underbrace{%
\begin{tabular}
[c]{|l|l|l|}\hline
$L$ & $\cdots$ & $L$\\\hline
\end{tabular}
}}\otimes\underset{q}{\underbrace{%
\begin{tabular}
[c]{|l|l|l|}\hline
$L$ & $\cdots$ & $L$\\\hline
\end{tabular}
}}  &  =-\frac{2m+1}{M}\underset{n-2q-2-2m}{\underbrace{%
\begin{tabular}
[c]{|l|l|l|}\hline
$T$ & $\cdots$ & $T$\\\hline
\end{tabular}
}}\underset{2m}{\underbrace{%
\begin{tabular}
[c]{|l|l|l|}\hline
$L$ & $\cdots$ & $L$\\\hline
\end{tabular}
}}\otimes\underset{q+1}{\underbrace{%
\begin{tabular}
[c]{|l|l|l|}\hline
$L$ & $\cdots$ & $L$\\\hline
\end{tabular}
}},\\
\underset{n-2q-2-2m}{\underbrace{%
\begin{tabular}
[c]{|l|l|l|}\hline
$T$ & $\cdots$ & $T$\\\hline
\end{tabular}
}}\underset{2m}{\underbrace{%
\begin{tabular}
[c]{|l|l|l|}\hline
$L$ & $\cdots$ & $L$\\\hline
\end{tabular}
}}\otimes\underset{q+1}{\underbrace{%
\begin{tabular}
[c]{|l|l|l|}\hline
$L$ & $\cdots$ & $L$\\\hline
\end{tabular}
}}  &  =-\frac{1}{2M}\underset{n-2q-2m}{\underbrace{%
\begin{tabular}
[c]{|l|l|l|}\hline
$T$ & $\cdots$ & $T$\\\hline
\end{tabular}
}}\underset{2m}{\underbrace{%
\begin{tabular}
[c]{|l|l|l|}\hline
$L$ & $\cdots$ & $L$\\\hline
\end{tabular}
}}\otimes\underset{q}{\underbrace{%
\begin{tabular}
[c]{|l|l|l|}\hline
$L$ & $\cdots$ & $L$\\\hline
\end{tabular}
}},
\end{align}
where we have renamed $m_{2}\rightarrow q$ and $m_{1}\rightarrow N-2q$.%
\end{subequations}%

\subsection{Superstring}

Applying Virasoro conditions (\ref{G1/2}) and (\ref{G3/2}) on the states
(\ref{|n>}), we obtain%
\begin{subequations}%
\begin{align}
G_{1/2}\left\vert N\right\rangle  &  =\sum_{\left\{  m_{j},m_{r}\right\}
}\left[  k^{\nu_{1}^{1/2}}%
\begin{tabular}
[c]{|c|c|c|}\hline
$\nu_{1}^{1/2}$ & $\cdots$ & $\nu_{m_{1/2}}^{1/2}$\\\hline
\end{tabular}
^{T}\right.  \overset{N}{\underset{j=1}{\otimes}}%
\begin{tabular}
[c]{|c|c|c|}\hline
$\mu_{1}^{j}$ & $\cdots$ & $\mu_{m_{j}}^{j}$\\\hline
\end{tabular}
\overset{N-1/2}{\underset{r\neq1/2}{\otimes}}%
\begin{tabular}
[c]{|c|c|c|}\hline
$\nu_{1}^{r}$ & $\cdots$ & $\nu_{m_{r}}^{r}$\\\hline
\end{tabular}
^{T}\nonumber\\
&  +\sum_{l\geq1}\sum_{i=1}^{m_{l}}l%
\begin{tabular}
[c]{|c|c|c|c|c|}\hline
$\mu_{1}^{l}$ & $\cdots$ & $\hat{\mu}_{i}^{l}$ & $\cdots$ & $\mu_{m_{l}}^{l}%
$\\\hline
\end{tabular}
\otimes%
\begin{tabular}
[c]{|c|c|c|c|}\hline
$\mu_{i}^{l}$ & $\nu_{1}^{l+1/2}$ & $\cdots$ & $\nu_{m_{l+1/2}}^{l+1/2}%
$\\\hline
\end{tabular}
^{T}\nonumber\\
&  \otimes%
\begin{tabular}
[c]{|c|c|c|}\hline
$\nu_{2}^{1/2}$ & $\cdots$ & $\nu_{m_{1/2}}^{1/2}$\\\hline
\end{tabular}
^{T}\overset{N}{\underset{j\neq l}{\otimes}}%
\begin{tabular}
[c]{|c|c|c|}\hline
$\mu_{1}^{j}$ & $\cdots$ & $\mu_{m_{j}}^{j}$\\\hline
\end{tabular}
\overset{N-1/2}{\underset{r\neq1/2,l+1/2}{\otimes}}%
\begin{tabular}
[c]{|c|c|c|}\hline
$\nu_{1}^{r}$ & $\cdots$ & $\nu_{m_{r}}^{r}$\\\hline
\end{tabular}
^{T}\nonumber\\
&  +\sum_{i=2}^{m_{1/2}}%
\begin{tabular}
[c]{|c|c|c|c|}\hline
$\nu_{i}^{1/2}$ & $\mu_{1}^{1}$ & $\cdots$ & $\mu_{m_{1}}^{1}$\\\hline
\end{tabular}
\otimes\left(  -1\right)  ^{i+1}%
\begin{tabular}
[c]{|c|c|c|c|c|}\hline
$\nu_{2}^{1/2}$ & $\cdots$ & $\hat{\nu}_{i}^{1/2}$ & $\cdots$ & $\nu_{m_{1/2}%
}^{1/2}$\\\hline
\end{tabular}
^{T}\nonumber\\
&  \overset{N}{\underset{j\neq l}{\otimes}}%
\begin{tabular}
[c]{|c|c|c|}\hline
$\mu_{1}^{j}$ & $\cdots$ & $\mu_{m_{j}}^{j}$\\\hline
\end{tabular}
\overset{N-1/2}{\underset{r\neq1/2}{\otimes}}%
\begin{tabular}
[c]{|c|c|c|}\hline
$\nu_{1}^{r}$ & $\cdots$ & $\nu_{m_{r}}^{r}$\\\hline
\end{tabular}
^{T}\nonumber\\
&  +\sum_{l\geq2}\sum_{i=1}^{m_{l-1/2}}\left(  -1\right)  ^{i+1}%
\begin{tabular}
[c]{|c|c|c|c|}\hline
$\nu_{i}^{l-1/2}$ & $\mu_{1}^{l}$ & $\cdots$ & $\mu_{m_{l}}^{l}$\\\hline
\end{tabular}
\otimes%
\begin{tabular}
[c]{|c|c|c|c|c|}\hline
$\nu_{1}^{l-1/2}$ & $\cdots$ & $\hat{\nu}_{i}^{l-1/2}$ & $\cdots$ &
$\nu_{m_{l-1/2}}^{l-1/2}$\\\hline
\end{tabular}
^{T}\nonumber\\
&  \otimes%
\begin{tabular}
[c]{|c|c|c|}\hline
$\nu_{2}^{1/2}$ & $\cdots$ & $\nu_{m_{1/2}}^{1/2}$\\\hline
\end{tabular}
^{T}\left.  \overset{N}{\underset{j\neq l}{\otimes}}%
\begin{tabular}
[c]{|c|c|c|}\hline
$\mu_{1}^{j}$ & $\cdots$ & $\mu_{m_{j}}^{j}$\\\hline
\end{tabular}
\overset{N-1/2}{\underset{r\neq1/2,l-1/2}{\otimes}}%
\begin{tabular}
[c]{|c|c|c|}\hline
$\nu_{1}^{r}$ & $\cdots$ & $\nu_{m_{r}}^{r}$\\\hline
\end{tabular}
^{T}\right] \nonumber\\
&  \cdot\frac{1}{\left(  m_{1/2}-1\right)  !}\psi_{-1/2}^{\nu_{2}^{1/2}%
\cdots\nu_{m_{1/2}}^{1/2}}\prod_{j=1}^{N}\frac{1}{j^{m_{j}}m_{j}!}\alpha
_{-j}^{\mu_{1}^{j}\cdots\mu_{m_{j}}^{j}}\prod_{r\neq1/2}^{N-1/2}\frac{1}%
{m_{r}!}\psi_{-r}^{\nu_{1}^{r}\cdots\nu_{m_{r}}^{r}},
\end{align}
and%
\begin{align}
G_{3/2}\left\vert N\right\rangle  &  =\sum_{\left\{  m_{j},m_{r}\right\}
}\left[
\begin{tabular}
[c]{|c|c|c|}\hline
$\nu_{1}^{3/2}$ & $\cdots$ & $\nu_{m_{3/2}}^{3/2}$\\\hline
\end{tabular}
^{T}k^{\nu_{1}^{3/2}}\right.  \overset{N}{\underset{j=1}{\otimes}}%
\begin{tabular}
[c]{|c|c|c|}\hline
$\mu_{1}^{j}$ & $\cdots$ & $\mu_{m_{j}}^{j}$\\\hline
\end{tabular}
\overset{N-1/2}{\underset{r\neq3/2}{\otimes}}%
\begin{tabular}
[c]{|c|c|c|}\hline
$\nu_{1}^{r}$ & $\cdots$ & $\nu_{m_{r}}^{r}$\\\hline
\end{tabular}
^{T}\nonumber\\
&  +\eta^{\mu\nu}%
\begin{tabular}
[c]{|c|c|c|c|}\hline
$\mu$ & $\mu_{1}^{1}$ & $\cdots$ & $\mu_{m_{1}}^{1}$\\\hline
\end{tabular}
\otimes%
\begin{tabular}
[c]{|c|c|c|c|}\hline
$\nu$ & $\nu_{1}^{1/2}$ & $\cdots$ & $\nu_{m_{1/2}}^{1/2}$\\\hline
\end{tabular}
^{T}\otimes%
\begin{tabular}
[c]{|c|c|c|}\hline
$\nu_{2}^{3/2}$ & $\cdots$ & $\nu_{m_{3/2}}^{3/2}$\\\hline
\end{tabular}
^{T}\nonumber\\
&  \overset{N}{\underset{j\neq1}{\otimes}}%
\begin{tabular}
[c]{|c|c|c|}\hline
$\mu_{1}^{j}$ & $\cdots$ & $\mu_{m_{j}}^{j}$\\\hline
\end{tabular}
\overset{N-1/2}{\underset{r\neq1/2,3/2}{\otimes}}%
\begin{tabular}
[c]{|c|c|c|}\hline
$\nu_{1}^{r}$ & $\cdots$ & $\nu_{m_{r}}^{r}$\\\hline
\end{tabular}
^{T}\nonumber\\
&  +\sum_{l\geq1}\sum_{i=1}^{m_{l}}l%
\begin{tabular}
[c]{|c|c|c|c|c|}\hline
$\mu_{1}^{l}$ & $\cdots$ & $\hat{\mu}_{i}^{l}$ & $\cdots$ & $\mu_{m_{l}}^{l}%
$\\\hline
\end{tabular}
\otimes%
\begin{tabular}
[c]{|c|c|c|c|}\hline
$\mu_{i}^{l}$ & $\nu_{1}^{l+3/2}$ & $\cdots$ & $\nu_{m_{l+3/2}}^{l+3/2}%
$\\\hline
\end{tabular}
^{T}\nonumber\\
&  \otimes%
\begin{tabular}
[c]{|c|c|c|}\hline
$\nu_{2}^{3/2}$ & $\cdots$ & $\nu_{m_{3/2}}^{3/2}$\\\hline
\end{tabular}
^{T}\overset{N}{\underset{j\neq l}{\otimes}}%
\begin{tabular}
[c]{|c|c|c|}\hline
$\mu_{1}^{j}$ & $\cdots$ & $\mu_{m_{j}}^{j}$\\\hline
\end{tabular}
\overset{N-1/2}{\underset{r\neq3/2,l+3/2}{\otimes}}%
\begin{tabular}
[c]{|c|c|c|}\hline
$\nu_{1}^{r}$ & $\cdots$ & $\nu_{m_{r}}^{r}$\\\hline
\end{tabular}
^{T}\nonumber\\
&  +\sum_{i=2}^{m_{3/2}}3%
\begin{tabular}
[c]{|c|c|c|c|}\hline
$\nu_{i}^{3/2}$ & $\mu_{1}^{3}$ & $\cdots$ & $\mu_{m_{3}}^{3}$\\\hline
\end{tabular}
\otimes\left(  -1\right)  ^{i+1}%
\begin{tabular}
[c]{|c|c|c|c|c|}\hline
$\nu_{2}^{3/2}$ & $\cdots$ & $\hat{\nu}_{i}^{3/2}$ & $\cdots$ & $\nu_{m_{3/2}%
}^{3/2}$\\\hline
\end{tabular}
^{T}\nonumber\\
&  \overset{N}{\underset{j\neq3}{\otimes}}%
\begin{tabular}
[c]{|c|c|c|}\hline
$\mu_{1}^{j}$ & $\cdots$ & $\mu_{m_{j}}^{j}$\\\hline
\end{tabular}
\overset{N-1/2}{\underset{r\neq3/2}{\otimes}}%
\begin{tabular}
[c]{|c|c|c|}\hline
$\nu_{1}^{r}$ & $\cdots$ & $\nu_{m_{r}}^{r}$\\\hline
\end{tabular}
^{T}\nonumber\\
&  +\sum_{l\geq2,l\neq3}\sum_{i=1}^{m_{l-3/2}}%
\begin{tabular}
[c]{|c|c|c|c|}\hline
$\nu_{i}^{l-3/2}$ & $\mu_{1}^{l}$ & $\cdots$ & $\mu_{m_{l}}^{l}$\\\hline
\end{tabular}
\otimes%
\begin{tabular}
[c]{|c|c|c|c|c|}\hline
$\nu_{1}^{l-3/2}$ & $\cdots$ & $\hat{\nu}_{i}^{l-3/2}$ & $\cdots$ &
$\nu_{m_{l-3/2}}^{l-3/2}$\\\hline
\end{tabular}
^{T}\nonumber\\
&  \otimes%
\begin{tabular}
[c]{|c|c|c|}\hline
$\nu_{2}^{3/2}$ & $\cdots$ & $\nu_{m_{3/2}}^{3/2}$\\\hline
\end{tabular}
^{T}\left.  \overset{N}{\underset{j\neq l}{\otimes}}%
\begin{tabular}
[c]{|c|c|c|}\hline
$\mu_{1}^{j}$ & $\cdots$ & $\mu_{m_{j}}^{j}$\\\hline
\end{tabular}
\overset{N-1/2}{\underset{r\neq3/2,l-3/2}{\otimes}}%
\begin{tabular}
[c]{|c|c|c|}\hline
$\nu_{1}^{r}$ & $\cdots$ & $\nu_{m_{r}}^{r}$\\\hline
\end{tabular}
^{T}\right] \nonumber\\
&  \cdot\frac{1}{\left(  m_{3/2}-1\right)  !}\psi_{-3/2}^{\nu_{2}^{3/2}%
\cdots\nu_{m_{3/2}}^{3/2}}\prod_{j=1}^{N}\frac{1}{j^{m_{j}}m_{j}!}\alpha
_{-j}^{\mu_{1}^{j}\cdots\mu_{m_{j}}^{j}}\prod_{r\neq3/2}^{N-1/2}\frac{1}%
{m_{r}!}\psi_{-r}^{\nu_{1}^{r}\cdots\nu_{m_{r}}^{r}},
\end{align}%
\end{subequations}%
where we have used the identities of the Young tableaux,%
\begin{align}%
\begin{tabular}
[c]{|c|c|c|}\hline
$1$ & $\cdots$ & $p$\\\hline
\end{tabular}
&  =\frac{1}{p}\left[  1+\sigma_{\left(  21\right)  }+\sigma_{\left(
321\right)  }+\cdots+\sigma_{\left(  p\cdots1\right)  }\right]
\begin{tabular}
[c]{|c|}\hline
$1$\\\hline
\end{tabular}
\otimes%
\begin{tabular}
[c]{|c|c|c|}\hline
$2$ & $\cdots$ & $p$\\\hline
\end{tabular}
\nonumber\\
&  =\frac{1}{p}\sum_{i=1}^{p}\sigma_{\left(  i1\right)  }%
\begin{tabular}
[c]{|c|}\hline
$1$\\\hline
\end{tabular}
\otimes%
\begin{tabular}
[c]{|c|c|c|}\hline
$2$ & $\cdots$ & $p$\\\hline
\end{tabular}
,\\%
\begin{tabular}
[c]{|c|c|c|}\hline
$1$ & $\cdots$ & $p$\\\hline
\end{tabular}
^{T}  &  =\frac{1}{p}\left[  1-\sigma_{\left(  21\right)  }+\sigma_{\left(
321\right)  }-\cdots+\left(  -1\right)  ^{p+1}\sigma_{\left(  p\cdots1\right)
}\right]
\begin{tabular}
[c]{|c|}\hline
$1$\\\hline
\end{tabular}
\otimes%
\begin{tabular}
[c]{|c|c|c|}\hline
$2$ & $\cdots$ & $p$\\\hline
\end{tabular}
^{T}\nonumber\\
&  =\frac{1}{p}\sum_{i=1}^{p}\left(  -1\right)  ^{i+1}\sigma_{\left(
i\cdots1\right)  }%
\begin{tabular}
[c]{|c|}\hline
$1$\\\hline
\end{tabular}
\otimes%
\begin{tabular}
[c]{|c|c|c|}\hline
$2$ & $\cdots$ & $p$\\\hline
\end{tabular}
^{T}.
\end{align}
We then obtain the constraint equations%
\begin{subequations}%
\begin{align}
0  &  =k^{\nu_{1}^{1/2}}%
\begin{tabular}
[c]{|c|c|c|}\hline
$\nu_{1}^{1/2}$ & $\cdots$ & $\nu_{m_{1/2}}^{1/2}$\\\hline
\end{tabular}
^{T}\overset{N}{\underset{j=1}{\otimes}}%
\begin{tabular}
[c]{|c|c|c|}\hline
$\mu_{1}^{j}$ & $\cdots$ & $\mu_{m_{j}}^{j}$\\\hline
\end{tabular}
\overset{N-1/2}{\underset{r\neq1/2}{\otimes}}%
\begin{tabular}
[c]{|c|c|c|}\hline
$\nu_{1}^{r}$ & $\cdots$ & $\nu_{m_{r}}^{r}$\\\hline
\end{tabular}
^{T}\nonumber\\
&  +\sum_{l\geq1}\sum_{i=1}^{m_{l}}l%
\begin{tabular}
[c]{|c|c|c|c|c|}\hline
$\mu_{1}^{l}$ & $\cdots$ & $\hat{\mu}_{i}^{l}$ & $\cdots$ & $\mu_{m_{l}}^{l}%
$\\\hline
\end{tabular}
\otimes%
\begin{tabular}
[c]{|c|c|c|c|}\hline
$\mu_{i}^{l}$ & $\nu_{1}^{l+1/2}$ & $\cdots$ & $\nu_{m_{l+1/2}}^{l+1/2}%
$\\\hline
\end{tabular}
^{T}\nonumber\\
&  \otimes%
\begin{tabular}
[c]{|c|c|c|}\hline
$\nu_{2}^{1/2}$ & $\cdots$ & $\nu_{m_{1/2}}^{1/2}$\\\hline
\end{tabular}
^{T}\overset{N}{\underset{j\neq l}{\otimes}}%
\begin{tabular}
[c]{|c|c|c|}\hline
$\mu_{1}^{j}$ & $\cdots$ & $\mu_{m_{j}}^{j}$\\\hline
\end{tabular}
\overset{N-1/2}{\underset{r\neq1/2,l+1/2}{\otimes}}%
\begin{tabular}
[c]{|c|c|c|}\hline
$\nu_{1}^{r}$ & $\cdots$ & $\nu_{m_{r}}^{r}$\\\hline
\end{tabular}
^{T}\nonumber\\
&  +\sum_{i=2}^{m_{1/2}}%
\begin{tabular}
[c]{|c|c|c|c|}\hline
$\nu_{i}^{1/2}$ & $\mu_{1}^{1}$ & $\cdots$ & $\mu_{m_{1}}^{1}$\\\hline
\end{tabular}
\otimes\left(  -1\right)  ^{i+1}%
\begin{tabular}
[c]{|c|c|c|c|c|}\hline
$\nu_{2}^{1/2}$ & $\cdots$ & $\hat{\nu}_{i}^{1/2}$ & $\cdots$ & $\nu_{m_{1/2}%
}^{1/2}$\\\hline
\end{tabular}
^{T}\nonumber\\
&  \overset{N}{\underset{j\neq l}{\otimes}}%
\begin{tabular}
[c]{|c|c|c|}\hline
$\mu_{1}^{j}$ & $\cdots$ & $\mu_{m_{j}}^{j}$\\\hline
\end{tabular}
\overset{N-1/2}{\underset{r\neq1/2}{\otimes}}%
\begin{tabular}
[c]{|c|c|c|}\hline
$\nu_{1}^{r}$ & $\cdots$ & $\nu_{m_{r}}^{r}$\\\hline
\end{tabular}
^{T}\nonumber\\
&  +\sum_{l\geq2}\sum_{i=1}^{m_{l-1/2}}\left(  -1\right)  ^{i+1}%
\begin{tabular}
[c]{|c|c|c|c|}\hline
$\nu_{i}^{l-1/2}$ & $\mu_{1}^{l}$ & $\cdots$ & $\mu_{m_{l}}^{l}$\\\hline
\end{tabular}
\otimes%
\begin{tabular}
[c]{|c|c|c|c|c|}\hline
$\nu_{1}^{l-1/2}$ & $\cdots$ & $\hat{\nu}_{i}^{l-1/2}$ & $\cdots$ &
$\nu_{m_{l-1/2}}^{l-1/2}$\\\hline
\end{tabular}
^{T}\nonumber\\
&  \otimes%
\begin{tabular}
[c]{|c|c|c|}\hline
$\nu_{2}^{1/2}$ & $\cdots$ & $\nu_{m_{1/2}}^{1/2}$\\\hline
\end{tabular}
^{T}\overset{N}{\underset{j\neq l}{\otimes}}%
\begin{tabular}
[c]{|c|c|c|}\hline
$\mu_{1}^{j}$ & $\cdots$ & $\mu_{m_{j}}^{j}$\\\hline
\end{tabular}
\overset{N-1/2}{\underset{r\neq1/2,l-1/2}{\otimes}}%
\begin{tabular}
[c]{|c|c|c|}\hline
$\nu_{1}^{r}$ & $\cdots$ & $\nu_{m_{r}}^{r}$\\\hline
\end{tabular}
^{T},
\end{align}%
\begin{align}
0  &  =%
\begin{tabular}
[c]{|c|c|c|}\hline
$\nu_{1}^{3/2}$ & $\cdots$ & $\nu_{m_{3/2}}^{3/2}$\\\hline
\end{tabular}
^{T}k^{\nu_{1}^{3/2}}\overset{N}{\underset{j=1}{\otimes}}%
\begin{tabular}
[c]{|c|c|c|}\hline
$\mu_{1}^{j}$ & $\cdots$ & $\mu_{m_{j}}^{j}$\\\hline
\end{tabular}
\overset{N-1/2}{\underset{r\neq3/2}{\otimes}}%
\begin{tabular}
[c]{|c|c|c|}\hline
$\nu_{1}^{r}$ & $\cdots$ & $\nu_{m_{r}}^{r}$\\\hline
\end{tabular}
^{T}\nonumber\\
&  +\eta^{\mu\nu}%
\begin{tabular}
[c]{|c|c|c|c|}\hline
$\mu$ & $\mu_{1}^{1}$ & $\cdots$ & $\mu_{m_{1}}^{1}$\\\hline
\end{tabular}
\otimes%
\begin{tabular}
[c]{|c|c|c|c|}\hline
$\nu$ & $\nu_{1}^{1/2}$ & $\cdots$ & $\nu_{m_{1/2}}^{1/2}$\\\hline
\end{tabular}
^{T}\otimes%
\begin{tabular}
[c]{|c|c|c|}\hline
$\nu_{2}^{3/2}$ & $\cdots$ & $\nu_{m_{3/2}}^{3/2}$\\\hline
\end{tabular}
^{T}\nonumber\\
&  \overset{N}{\underset{j\neq1}{\otimes}}%
\begin{tabular}
[c]{|c|c|c|}\hline
$\mu_{1}^{j}$ & $\cdots$ & $\mu_{m_{j}}^{j}$\\\hline
\end{tabular}
\overset{N-1/2}{\underset{r\neq1/2,3/2}{\otimes}}%
\begin{tabular}
[c]{|c|c|c|}\hline
$\nu_{1}^{r}$ & $\cdots$ & $\nu_{m_{r}}^{r}$\\\hline
\end{tabular}
^{T}\nonumber\\
&  +\sum_{l\geq1}\sum_{i=1}^{m_{l}}l%
\begin{tabular}
[c]{|c|c|c|c|c|}\hline
$\mu_{1}^{l}$ & $\cdots$ & $\hat{\mu}_{i}^{l}$ & $\cdots$ & $\mu_{m_{l}}^{l}%
$\\\hline
\end{tabular}
\otimes%
\begin{tabular}
[c]{|c|c|c|c|}\hline
$\mu_{i}^{l}$ & $\nu_{1}^{l+3/2}$ & $\cdots$ & $\nu_{m_{l+3/2}}^{l+3/2}%
$\\\hline
\end{tabular}
^{T}\nonumber\\
&  \otimes%
\begin{tabular}
[c]{|c|c|c|}\hline
$\nu_{2}^{3/2}$ & $\cdots$ & $\nu_{m_{3/2}}^{3/2}$\\\hline
\end{tabular}
^{T}\overset{N}{\underset{j\neq l}{\otimes}}%
\begin{tabular}
[c]{|c|c|c|}\hline
$\mu_{1}^{j}$ & $\cdots$ & $\mu_{m_{j}}^{j}$\\\hline
\end{tabular}
\overset{N-1/2}{\underset{r\neq3/2,l+3/2}{\otimes}}%
\begin{tabular}
[c]{|c|c|c|}\hline
$\nu_{1}^{r}$ & $\cdots$ & $\nu_{m_{r}}^{r}$\\\hline
\end{tabular}
^{T}\nonumber\\
&  +\sum_{i=2}^{m_{3/2}}3%
\begin{tabular}
[c]{|c|c|c|c|}\hline
$\nu_{i}^{3/2}$ & $\mu_{1}^{3}$ & $\cdots$ & $\mu_{m_{3}}^{3}$\\\hline
\end{tabular}
\otimes\left(  -1\right)  ^{i+1}%
\begin{tabular}
[c]{|c|c|c|c|c|}\hline
$\nu_{2}^{3/2}$ & $\cdots$ & $\hat{\nu}_{i}^{3/2}$ & $\cdots$ & $\nu_{m_{3/2}%
}^{3/2}$\\\hline
\end{tabular}
^{T}\nonumber\\
&  \overset{N}{\underset{j\neq3}{\otimes}}%
\begin{tabular}
[c]{|c|c|c|}\hline
$\mu_{1}^{j}$ & $\cdots$ & $\mu_{m_{j}}^{j}$\\\hline
\end{tabular}
\overset{N-1/2}{\underset{r\neq3/2}{\otimes}}%
\begin{tabular}
[c]{|c|c|c|}\hline
$\nu_{1}^{r}$ & $\cdots$ & $\nu_{m_{r}}^{r}$\\\hline
\end{tabular}
^{T}\nonumber\\
&  +\sum_{l\geq2,l\neq3}\sum_{i=1}^{m_{l-3/2}}%
\begin{tabular}
[c]{|c|c|c|c|}\hline
$\nu_{i}^{l-3/2}$ & $\mu_{1}^{l}$ & $\cdots$ & $\mu_{m_{l}}^{l}$\\\hline
\end{tabular}
\otimes\left(  -1\right)  ^{i+1}%
\begin{tabular}
[c]{|c|c|c|c|c|}\hline
$\nu_{1}^{l-3/2}$ & $\cdots$ & $\hat{\nu}_{i}^{l-3/2}$ & $\cdots$ &
$\nu_{m_{l-3/2}}^{l-3/2}$\\\hline
\end{tabular}
^{T}\nonumber\\
&  \otimes%
\begin{tabular}
[c]{|c|c|c|}\hline
$\nu_{2}^{3/2}$ & $\cdots$ & $\nu_{m_{3/2}}^{3/2}$\\\hline
\end{tabular}
^{T}\overset{N}{\underset{j\neq l}{\otimes}}%
\begin{tabular}
[c]{|c|c|c|}\hline
$\mu_{1}^{j}$ & $\cdots$ & $\mu_{m_{j}}^{j}$\\\hline
\end{tabular}
\overset{N-1/2}{\underset{r\neq3/2,l-3/2}{\otimes}}%
\begin{tabular}
[c]{|c|c|c|}\hline
$\nu_{1}^{r}$ & $\cdots$ & $\nu_{m_{r}}^{r}$\\\hline
\end{tabular}
^{T}.
\end{align}%
\end{subequations}%
Taking the high energy limit in the above equations by letting $\left(
\mu_{i},\nu_{i}\right)  \rightarrow\left(  L,T\right)  $, and%
\begin{equation}
k^{\mu_{i}}\rightarrow M\left(  e^{L}\right)  ^{\mu_{i}}\text{, }\eta^{\mu
_{1}\mu_{2}}\rightarrow\left(  e^{T}\right)  ^{\mu_{1}}\left(  e^{T}\right)
^{\mu_{2}},
\end{equation}
we get%
\begin{subequations}%
\begin{align}
0  &  =M%
\begin{tabular}
[c]{|c|c|c|c|}\hline
$L$ & $\nu_{2}^{1/2}$ & $\cdots$ & $\nu_{m_{1/2}}^{1/2}$\\\hline
\end{tabular}
^{T}\overset{N}{\underset{j=1}{\otimes}}%
\begin{tabular}
[c]{|c|c|c|}\hline
$\mu_{1}^{j}$ & $\cdots$ & $\mu_{m_{j}}^{j}$\\\hline
\end{tabular}
\overset{N-1/2}{\underset{r\neq1/2}{\otimes}}%
\begin{tabular}
[c]{|c|c|c|}\hline
$\nu_{1}^{r}$ & $\cdots$ & $\nu_{m_{r}}^{r}$\\\hline
\end{tabular}
^{T}\nonumber\\
&  +\sum_{l\geq1}\sum_{i=1}^{m_{l}}l%
\begin{tabular}
[c]{|c|c|c|c|c|}\hline
$\mu_{1}^{l}$ & $\cdots$ & $\hat{\mu}_{i}^{l}$ & $\cdots$ & $\mu_{m_{l}}^{l}%
$\\\hline
\end{tabular}
\otimes%
\begin{tabular}
[c]{|c|c|c|c|}\hline
$\mu_{i}^{l}$ & $\nu_{1}^{l+1/2}$ & $\cdots$ & $\nu_{m_{l+1/2}}^{l+1/2}%
$\\\hline
\end{tabular}
^{T}\nonumber\\
&  \otimes%
\begin{tabular}
[c]{|c|c|c|}\hline
$\nu_{2}^{1/2}$ & $\cdots$ & $\nu_{m_{1/2}}^{1/2}$\\\hline
\end{tabular}
^{T}\overset{N}{\underset{j\neq l}{\otimes}}%
\begin{tabular}
[c]{|c|c|c|}\hline
$\mu_{1}^{j}$ & $\cdots$ & $\mu_{m_{j}}^{j}$\\\hline
\end{tabular}
\overset{N-1/2}{\underset{r\neq1/2,l+1/2}{\otimes}}%
\begin{tabular}
[c]{|c|c|c|}\hline
$\nu_{1}^{r}$ & $\cdots$ & $\nu_{m_{r}}^{r}$\\\hline
\end{tabular}
^{T}\nonumber\\
&  +\sum_{i=2}^{m_{1/2}}%
\begin{tabular}
[c]{|c|c|c|c|}\hline
$\nu_{i}^{1/2}$ & $\mu_{1}^{1}$ & $\cdots$ & $\mu_{m_{1}}^{1}$\\\hline
\end{tabular}
\otimes\left(  -1\right)  ^{i+1}%
\begin{tabular}
[c]{|c|c|c|c|c|}\hline
$\nu_{2}^{1/2}$ & $\cdots$ & $\hat{\nu}_{i}^{1/2}$ & $\cdots$ & $\nu_{m_{1/2}%
}^{1/2}$\\\hline
\end{tabular}
^{T}\nonumber\\
&  \overset{N}{\underset{j\neq l}{\otimes}}%
\begin{tabular}
[c]{|c|c|c|}\hline
$\mu_{1}^{j}$ & $\cdots$ & $\mu_{m_{j}}^{j}$\\\hline
\end{tabular}
\overset{N-1/2}{\underset{r\neq1/2}{\otimes}}%
\begin{tabular}
[c]{|c|c|c|}\hline
$\nu_{1}^{r}$ & $\cdots$ & $\nu_{m_{r}}^{r}$\\\hline
\end{tabular}
^{T}\nonumber\\
&  +\sum_{l\geq2}\sum_{i=1}^{m_{l-1/2}}\left(  -1\right)  ^{i+1}%
\begin{tabular}
[c]{|c|c|c|c|}\hline
$\nu_{i}^{l-1/2}$ & $\mu_{1}^{l}$ & $\cdots$ & $\mu_{m_{l}}^{l}$\\\hline
\end{tabular}
\otimes%
\begin{tabular}
[c]{|c|c|c|c|c|}\hline
$\nu_{1}^{l-1/2}$ & $\cdots$ & $\hat{\nu}_{i}^{l-1/2}$ & $\cdots$ &
$\nu_{m_{l-1/2}}^{l-1/2}$\\\hline
\end{tabular}
^{T}\nonumber\\
&  \otimes%
\begin{tabular}
[c]{|c|c|c|}\hline
$\nu_{2}^{1/2}$ & $\cdots$ & $\nu_{m_{1/2}}^{1/2}$\\\hline
\end{tabular}
^{T}\overset{N}{\underset{j\neq l}{\otimes}}%
\begin{tabular}
[c]{|c|c|c|}\hline
$\mu_{1}^{j}$ & $\cdots$ & $\mu_{m_{j}}^{j}$\\\hline
\end{tabular}
\overset{N-1/2}{\underset{r\neq1/2,l-1/2}{\otimes}}%
\begin{tabular}
[c]{|c|c|c|}\hline
$\nu_{1}^{r}$ & $\cdots$ & $\nu_{m_{r}}^{r}$\\\hline
\end{tabular}
^{T},
\end{align}%
\begin{align}
0  &  =M%
\begin{tabular}
[c]{|c|c|c|c|}\hline
$L$ & $\nu_{2}^{3/2}$ & $\cdots$ & $\nu_{m_{3/2}}^{3/2}$\\\hline
\end{tabular}
^{T}\overset{N}{\underset{j=1}{\otimes}}%
\begin{tabular}
[c]{|c|c|c|}\hline
$\mu_{1}^{j}$ & $\cdots$ & $\mu_{m_{j}}^{j}$\\\hline
\end{tabular}
\overset{N-1/2}{\underset{r\neq3/2}{\otimes}}%
\begin{tabular}
[c]{|c|c|c|}\hline
$\nu_{1}^{r}$ & $\cdots$ & $\nu_{m_{r}}^{r}$\\\hline
\end{tabular}
^{T}\nonumber\\
&  +%
\begin{tabular}
[c]{|c|c|c|c|}\hline
$T$ & $\mu_{1}^{1}$ & $\cdots$ & $\mu_{m_{1}}^{1}$\\\hline
\end{tabular}
\otimes%
\begin{tabular}
[c]{|c|c|c|c|}\hline
$T$ & $\nu_{1}^{1/2}$ & $\cdots$ & $\nu_{m_{1/2}}^{1/2}$\\\hline
\end{tabular}
^{T}\otimes%
\begin{tabular}
[c]{|c|c|c|}\hline
$\nu_{2}^{3/2}$ & $\cdots$ & $\nu_{m_{3/2}}^{3/2}$\\\hline
\end{tabular}
^{T}\nonumber\\
&  \overset{N}{\underset{j\neq1}{\otimes}}%
\begin{tabular}
[c]{|c|c|c|}\hline
$\mu_{1}^{j}$ & $\cdots$ & $\mu_{m_{j}}^{j}$\\\hline
\end{tabular}
\overset{N-1/2}{\underset{r\neq1/2,3/2}{\otimes}}%
\begin{tabular}
[c]{|c|c|c|}\hline
$\nu_{1}^{r}$ & $\cdots$ & $\nu_{m_{r}}^{r}$\\\hline
\end{tabular}
^{T}\nonumber\\
&  +\sum_{l\geq1}\sum_{i=1}^{m_{l}}l%
\begin{tabular}
[c]{|c|c|c|c|c|}\hline
$\mu_{1}^{l}$ & $\cdots$ & $\hat{\mu}_{i}^{l}$ & $\cdots$ & $\mu_{m_{l}}^{l}%
$\\\hline
\end{tabular}
\otimes%
\begin{tabular}
[c]{|c|c|c|c|}\hline
$\mu_{i}^{l}$ & $\nu_{1}^{l+3/2}$ & $\cdots$ & $\nu_{m_{l+3/2}}^{l+3/2}%
$\\\hline
\end{tabular}
^{T}\nonumber\\
&  \otimes%
\begin{tabular}
[c]{|c|c|c|}\hline
$\nu_{2}^{3/2}$ & $\cdots$ & $\nu_{m_{3/2}}^{3/2}$\\\hline
\end{tabular}
^{T}\overset{N}{\underset{j\neq l}{\otimes}}%
\begin{tabular}
[c]{|c|c|c|}\hline
$\mu_{1}^{j}$ & $\cdots$ & $\mu_{m_{j}}^{j}$\\\hline
\end{tabular}
\overset{N-1/2}{\underset{r\neq3/2,l+3/2}{\otimes}}%
\begin{tabular}
[c]{|c|c|c|}\hline
$\nu_{1}^{r}$ & $\cdots$ & $\nu_{m_{r}}^{r}$\\\hline
\end{tabular}
^{T}\nonumber\\
&  +\sum_{i=2}^{m_{3/2}}3%
\begin{tabular}
[c]{|c|c|c|c|}\hline
$\nu_{i}^{3/2}$ & $\mu_{1}^{3}$ & $\cdots$ & $\mu_{m_{3}}^{3}$\\\hline
\end{tabular}
\otimes\left(  -1\right)  ^{i+1}%
\begin{tabular}
[c]{|c|c|c|c|c|}\hline
$\nu_{2}^{3/2}$ & $\cdots$ & $\hat{\nu}_{i}^{3/2}$ & $\cdots$ & $\nu_{m_{3/2}%
}^{3/2}$\\\hline
\end{tabular}
^{T}\nonumber\\
&  \overset{N}{\underset{j\neq3}{\otimes}}%
\begin{tabular}
[c]{|c|c|c|}\hline
$\mu_{1}^{j}$ & $\cdots$ & $\mu_{m_{j}}^{j}$\\\hline
\end{tabular}
\overset{N-1/2}{\underset{r\neq3/2}{\otimes}}%
\begin{tabular}
[c]{|c|c|c|}\hline
$\nu_{1}^{r}$ & $\cdots$ & $\nu_{m_{r}}^{r}$\\\hline
\end{tabular}
^{T}\nonumber\\
&  +\sum_{l\geq2,l\neq3}\sum_{i=1}^{m_{l-3/2}}%
\begin{tabular}
[c]{|c|c|c|c|}\hline
$\nu_{i}^{l-3/2}$ & $\mu_{1}^{l}$ & $\cdots$ & $\mu_{m_{l}}^{l}$\\\hline
\end{tabular}
\otimes\left(  -1\right)  ^{i+1}%
\begin{tabular}
[c]{|c|c|c|c|c|}\hline
$\nu_{1}^{l-3/2}$ & $\cdots$ & $\hat{\nu}_{i}^{l-3/2}$ & $\cdots$ &
$\nu_{m_{l-3/2}}^{l-3/2}$\\\hline
\end{tabular}
^{T}\nonumber\\
&  \otimes%
\begin{tabular}
[c]{|c|c|c|}\hline
$\nu_{2}^{3/2}$ & $\cdots$ & $\nu_{m_{3/2}}^{3/2}$\\\hline
\end{tabular}
^{T}\overset{N}{\underset{j\neq l}{\otimes}}%
\begin{tabular}
[c]{|c|c|c|}\hline
$\mu_{1}^{j}$ & $\cdots$ & $\mu_{m_{j}}^{j}$\\\hline
\end{tabular}
\overset{N-1/2}{\underset{r\neq3/2,l-3/2}{\otimes}}%
\begin{tabular}
[c]{|c|c|c|}\hline
$\nu_{1}^{r}$ & $\cdots$ & $\nu_{m_{r}}^{r}$\\\hline
\end{tabular}
^{T}.
\end{align}%
\end{subequations}%
The indices$\left\{  \mu_{i}^{j}\right\}  $ are symmetric and can be chosen to
have $l_{j}$ of $\left\{  L\right\}  $ and $\left\{  T\right\}  $, while
$\left\{  \nu_{i}^{r}\right\}  $ are antisymmetric and we keep them as what
they are at this moment. Thus%
\begin{subequations}%
\begin{align}
0  &  =M%
\begin{tabular}
[c]{|c|c|c|c|}\hline
$L$ & $\nu_{2}^{1/2}$ & $\cdots$ & $\nu_{m_{1/2}}^{1/2}$\\\hline
\end{tabular}
^{T}\overset{N}{\underset{j=1}{\otimes}}%
\begin{tabular}
[c]{|c|}\hline
$\mu_{1}^{j}$\\\hline
\end{tabular}
\underset{m_{j}-1-l_{j}}{\underbrace{%
\begin{tabular}
[c]{|l|l|l|}\hline
$T$ & $\cdots$ & $T$\\\hline
\end{tabular}
}}\underset{l_{j}}{\underbrace{%
\begin{tabular}
[c]{|l|l|l|}\hline
$L$ & $\cdots$ & $L$\\\hline
\end{tabular}
}}\overset{N-1/2}{\underset{r\neq1/2}{\otimes}}%
\begin{tabular}
[c]{|c|c|c|}\hline
$\nu_{1}^{r}$ & $\cdots$ & $\nu_{m_{r}}^{r}$\\\hline
\end{tabular}
^{T}\nonumber\\
&  +\sum_{l\geq1}l\underset{m_{l}-1-l_{l}}{\underbrace{%
\begin{tabular}
[c]{|l|l|l|}\hline
$T$ & $\cdots$ & $T$\\\hline
\end{tabular}
}}\underset{l_{l}}{\underbrace{%
\begin{tabular}
[c]{|l|l|l|}\hline
$L$ & $\cdots$ & $L$\\\hline
\end{tabular}
}}\otimes%
\begin{tabular}
[c]{|c|c|c|c|}\hline
$\mu_{1}^{l}$ & $\nu_{1}^{l+1/2}$ & $\cdots$ & $\nu_{m_{l+1/2}}^{l+1/2}%
$\\\hline
\end{tabular}
^{T}\nonumber\\
&  \otimes%
\begin{tabular}
[c]{|c|c|c|}\hline
$\nu_{2}^{1/2}$ & $\cdots$ & $\nu_{m_{1/2}}^{1/2}$\\\hline
\end{tabular}
^{T}\overset{N}{\underset{j\neq l}{\otimes}}%
\begin{tabular}
[c]{|c|}\hline
$\mu_{1}^{j}$\\\hline
\end{tabular}
\underset{m_{j}-1-l_{j}}{\underbrace{%
\begin{tabular}
[c]{|l|l|l|}\hline
$T$ & $\cdots$ & $T$\\\hline
\end{tabular}
}}\underset{l_{j}}{\underbrace{%
\begin{tabular}
[c]{|l|l|l|}\hline
$L$ & $\cdots$ & $L$\\\hline
\end{tabular}
}}\overset{N-1/2}{\underset{r\neq1/2,l+1/2}{\otimes}}%
\begin{tabular}
[c]{|c|c|c|}\hline
$\nu_{1}^{r}$ & $\cdots$ & $\nu_{m_{r}}^{r}$\\\hline
\end{tabular}
^{T}\nonumber\\
&  +\sum_{l\geq1}l\left(  m_{l}-1-l_{l}\right)
\begin{tabular}
[c]{|c|}\hline
$\mu_{1}^{l}$\\\hline
\end{tabular}
\underset{m_{l}-2-l_{l}}{\underbrace{%
\begin{tabular}
[c]{|l|l|l|}\hline
$T$ & $\cdots$ & $T$\\\hline
\end{tabular}
}}\underset{l_{l}}{\underbrace{%
\begin{tabular}
[c]{|l|l|l|}\hline
$L$ & $\cdots$ & $L$\\\hline
\end{tabular}
}}\otimes%
\begin{tabular}
[c]{|c|c|c|c|}\hline
$T$ & $\nu_{1}^{l+1/2}$ & $\cdots$ & $\nu_{m_{l+1/2}}^{l+1/2}$\\\hline
\end{tabular}
^{T}\nonumber\\
&  \otimes%
\begin{tabular}
[c]{|c|c|c|}\hline
$\nu_{2}^{1/2}$ & $\cdots$ & $\nu_{m_{1/2}}^{1/2}$\\\hline
\end{tabular}
^{T}\overset{N}{\underset{j\neq l}{\otimes}}%
\begin{tabular}
[c]{|c|}\hline
$\mu_{1}^{j}$\\\hline
\end{tabular}
\underset{m_{j}-1-l_{j}}{\underbrace{%
\begin{tabular}
[c]{|l|l|l|}\hline
$T$ & $\cdots$ & $T$\\\hline
\end{tabular}
}}\underset{l_{j}}{\underbrace{%
\begin{tabular}
[c]{|l|l|l|}\hline
$L$ & $\cdots$ & $L$\\\hline
\end{tabular}
}}\overset{N-1/2}{\underset{r\neq1/2,l+1/2}{\otimes}}%
\begin{tabular}
[c]{|c|c|c|}\hline
$\nu_{1}^{r}$ & $\cdots$ & $\nu_{m_{r}}^{r}$\\\hline
\end{tabular}
^{T}\nonumber\\
&  +\sum_{l\geq1}ll_{l}%
\begin{tabular}
[c]{|c|}\hline
$\mu_{1}^{l}$\\\hline
\end{tabular}
\underset{m_{l}-1-l_{l}}{\underbrace{%
\begin{tabular}
[c]{|l|l|l|}\hline
$T$ & $\cdots$ & $T$\\\hline
\end{tabular}
}}\underset{l_{l}-1}{\underbrace{%
\begin{tabular}
[c]{|l|l|l|}\hline
$L$ & $\cdots$ & $L$\\\hline
\end{tabular}
}}\otimes%
\begin{tabular}
[c]{|c|c|c|c|}\hline
$L$ & $\nu_{1}^{l+1/2}$ & $\cdots$ & $\nu_{m_{l+1/2}}^{l+1/2}$\\\hline
\end{tabular}
^{T}\nonumber\\
&  \otimes%
\begin{tabular}
[c]{|c|c|c|}\hline
$\nu_{2}^{1/2}$ & $\cdots$ & $\nu_{m_{1/2}}^{1/2}$\\\hline
\end{tabular}
^{T}\overset{N}{\underset{j\neq l}{\otimes}}%
\begin{tabular}
[c]{|c|}\hline
$\mu_{1}^{j}$\\\hline
\end{tabular}
\underset{m_{j}-1-l_{j}}{\underbrace{%
\begin{tabular}
[c]{|l|l|l|}\hline
$T$ & $\cdots$ & $T$\\\hline
\end{tabular}
}}\underset{l_{j}}{\underbrace{%
\begin{tabular}
[c]{|l|l|l|}\hline
$L$ & $\cdots$ & $L$\\\hline
\end{tabular}
}}\overset{N-1/2}{\underset{r\neq1/2,l+1/2}{\otimes}}%
\begin{tabular}
[c]{|c|c|c|}\hline
$\nu_{1}^{r}$ & $\cdots$ & $\nu_{m_{r}}^{r}$\\\hline
\end{tabular}
^{T}\nonumber\\
&  +\sum_{i=2}^{m_{1/2}}%
\begin{tabular}
[c]{|c|c|}\hline
$\nu_{i}^{1/2}$ & $\mu_{1}^{1}$\\\hline
\end{tabular}
\underset{m_{1}-1-l_{1}}{\underbrace{%
\begin{tabular}
[c]{|l|l|l|}\hline
$T$ & $\cdots$ & $T$\\\hline
\end{tabular}
}}\underset{l_{1}}{\underbrace{%
\begin{tabular}
[c]{|l|l|l|}\hline
$L$ & $\cdots$ & $L$\\\hline
\end{tabular}
}}\otimes\left(  -1\right)  ^{i+1}%
\begin{tabular}
[c]{|c|c|c|c|c|}\hline
$\nu_{2}^{1/2}$ & $\cdots$ & $\hat{\nu}_{i}^{1/2}$ & $\cdots$ & $\nu_{m_{1/2}%
}^{1/2}$\\\hline
\end{tabular}
^{T}\nonumber\\
&  \overset{N}{\underset{j\neq1}{\otimes}}%
\begin{tabular}
[c]{|c|}\hline
$\mu_{1}^{j}$\\\hline
\end{tabular}
\underset{m_{j}-1-l_{j}}{\underbrace{%
\begin{tabular}
[c]{|l|l|l|}\hline
$T$ & $\cdots$ & $T$\\\hline
\end{tabular}
}}\underset{l_{j}}{\underbrace{%
\begin{tabular}
[c]{|l|l|l|}\hline
$L$ & $\cdots$ & $L$\\\hline
\end{tabular}
}}\overset{N-1/2}{\underset{r\neq1/2}{\otimes}}%
\begin{tabular}
[c]{|c|c|c|}\hline
$\nu_{1}^{r}$ & $\cdots$ & $\nu_{m_{r}}^{r}$\\\hline
\end{tabular}
^{T}\nonumber\\
&  +\sum_{l\geq2}\sum_{i=1}^{m_{l-1/2}}\left(  -1\right)  ^{i+1}%
\begin{tabular}
[c]{|c|c|}\hline
$\nu_{i}^{l-1/2}$ & $\mu_{1}^{l}$\\\hline
\end{tabular}
\underset{m_{l}-1-l_{l}}{\underbrace{%
\begin{tabular}
[c]{|l|l|l|}\hline
$T$ & $\cdots$ & $T$\\\hline
\end{tabular}
}}\underset{l_{l}}{\underbrace{%
\begin{tabular}
[c]{|l|l|l|}\hline
$L$ & $\cdots$ & $L$\\\hline
\end{tabular}
}}\otimes%
\begin{tabular}
[c]{|c|c|c|c|c|}\hline
$\nu_{1}^{l-1/2}$ & $\cdots$ & $\hat{\nu}_{i}^{l-1/2}$ & $\cdots$ &
$\nu_{m_{l-1/2}}^{l-1/2}$\\\hline
\end{tabular}
^{T}\nonumber\\
&  \otimes%
\begin{tabular}
[c]{|c|c|c|}\hline
$\nu_{2}^{1/2}$ & $\cdots$ & $\nu_{m_{1/2}}^{1/2}$\\\hline
\end{tabular}
^{T}\overset{N}{\underset{j\neq l}{\otimes}}%
\begin{tabular}
[c]{|c|}\hline
$\mu_{1}^{j}$\\\hline
\end{tabular}
\underset{m_{j}-1-l_{j}}{\underbrace{%
\begin{tabular}
[c]{|l|l|l|}\hline
$T$ & $\cdots$ & $T$\\\hline
\end{tabular}
}}\underset{l_{j}}{\underbrace{%
\begin{tabular}
[c]{|l|l|l|}\hline
$L$ & $\cdots$ & $L$\\\hline
\end{tabular}
}}\overset{N-1/2}{\underset{r\neq1/2,l-1/2}{\otimes}}%
\begin{tabular}
[c]{|c|c|c|}\hline
$\nu_{1}^{r}$ & $\cdots$ & $\nu_{m_{r}}^{r}$\\\hline
\end{tabular}
^{T},
\end{align}%
\begin{align}
0  &  =M%
\begin{tabular}
[c]{|c|c|c|c|}\hline
$L$ & $\nu_{2}^{3/2}$ & $\cdots$ & $\nu_{m_{3/2}}^{3/2}$\\\hline
\end{tabular}
^{T}\overset{N}{\underset{j=1}{\otimes}}%
\begin{tabular}
[c]{|c|}\hline
$\mu_{1}^{j}$\\\hline
\end{tabular}
\underset{m_{j}-1-l_{j}}{\underbrace{%
\begin{tabular}
[c]{|l|l|l|}\hline
$T$ & $\cdots$ & $T$\\\hline
\end{tabular}
}}\underset{l_{j}}{\underbrace{%
\begin{tabular}
[c]{|l|l|l|}\hline
$L$ & $\cdots$ & $L$\\\hline
\end{tabular}
}}\overset{N-1/2}{\underset{r\neq3/2}{\otimes}}%
\begin{tabular}
[c]{|c|c|c|}\hline
$\nu_{1}^{r}$ & $\cdots$ & $\nu_{m_{r}}^{r}$\\\hline
\end{tabular}
^{T}\nonumber\\
&  +%
\begin{tabular}
[c]{|c|}\hline
$\mu_{1}^{1}$\\\hline
\end{tabular}
\underset{m_{1}-l_{1}}{\underbrace{%
\begin{tabular}
[c]{|l|l|l|}\hline
$T$ & $\cdots$ & $T$\\\hline
\end{tabular}
}}\underset{l_{1}}{\underbrace{%
\begin{tabular}
[c]{|l|l|l|}\hline
$L$ & $\cdots$ & $L$\\\hline
\end{tabular}
}}\otimes%
\begin{tabular}
[c]{|c|c|c|c|}\hline
$T$ & $\nu_{1}^{1/2}$ & $\cdots$ & $\nu_{m_{1/2}}^{1/2}$\\\hline
\end{tabular}
^{T}\otimes%
\begin{tabular}
[c]{|c|c|c|}\hline
$\nu_{2}^{3/2}$ & $\cdots$ & $\nu_{m_{3/2}}^{3/2}$\\\hline
\end{tabular}
^{T}\nonumber\\
&  \overset{N}{\underset{j\neq1}{\otimes}}%
\begin{tabular}
[c]{|c|}\hline
$\mu_{1}^{j}$\\\hline
\end{tabular}
\underset{m_{j}-1-l_{j}}{\underbrace{%
\begin{tabular}
[c]{|l|l|l|}\hline
$T$ & $\cdots$ & $T$\\\hline
\end{tabular}
}}\underset{l_{j}}{\underbrace{%
\begin{tabular}
[c]{|l|l|l|}\hline
$L$ & $\cdots$ & $L$\\\hline
\end{tabular}
}}\overset{N-1/2}{\underset{r\neq1/2,3/2}{\otimes}}%
\begin{tabular}
[c]{|c|c|c|}\hline
$\nu_{1}^{r}$ & $\cdots$ & $\nu_{m_{r}}^{r}$\\\hline
\end{tabular}
^{T}\nonumber\\
&  +\sum_{l\geq1}l\underset{m_{l}-1-l_{l}}{\underbrace{%
\begin{tabular}
[c]{|l|l|l|}\hline
$T$ & $\cdots$ & $T$\\\hline
\end{tabular}
}}\underset{l_{l}}{\underbrace{%
\begin{tabular}
[c]{|l|l|l|}\hline
$L$ & $\cdots$ & $L$\\\hline
\end{tabular}
}}\otimes%
\begin{tabular}
[c]{|c|c|c|c|}\hline
$\mu_{1}^{l}$ & $\nu_{1}^{l+3/2}$ & $\cdots$ & $\nu_{m_{l+3/2}}^{l+3/2}%
$\\\hline
\end{tabular}
^{T}\nonumber\\
&  \otimes%
\begin{tabular}
[c]{|c|c|c|}\hline
$\nu_{2}^{3/2}$ & $\cdots$ & $\nu_{m_{3/2}}^{3/2}$\\\hline
\end{tabular}
^{T}\overset{N}{\underset{j\neq l}{\otimes}}%
\begin{tabular}
[c]{|c|}\hline
$\mu_{1}^{j}$\\\hline
\end{tabular}
\underset{m_{j}-1-l_{j}}{\underbrace{%
\begin{tabular}
[c]{|l|l|l|}\hline
$T$ & $\cdots$ & $T$\\\hline
\end{tabular}
}}\underset{l_{j}}{\underbrace{%
\begin{tabular}
[c]{|l|l|l|}\hline
$L$ & $\cdots$ & $L$\\\hline
\end{tabular}
}}\overset{N-1/2}{\underset{r\neq3/2,l+3/2}{\otimes}}%
\begin{tabular}
[c]{|c|c|c|}\hline
$\nu_{1}^{r}$ & $\cdots$ & $\nu_{m_{r}}^{r}$\\\hline
\end{tabular}
^{T}\nonumber\\
&  +\sum_{l\geq1}l\left(  m_{l}-1-l_{l}\right)
\begin{tabular}
[c]{|c|}\hline
$\mu_{1}^{l}$\\\hline
\end{tabular}
\underset{m_{l}-2-l_{l}}{\underbrace{%
\begin{tabular}
[c]{|l|l|l|}\hline
$T$ & $\cdots$ & $T$\\\hline
\end{tabular}
}}\underset{l_{l}}{\underbrace{%
\begin{tabular}
[c]{|l|l|l|}\hline
$L$ & $\cdots$ & $L$\\\hline
\end{tabular}
}}\otimes%
\begin{tabular}
[c]{|c|c|c|c|}\hline
$T$ & $\nu_{1}^{l+3/2}$ & $\cdots$ & $\nu_{m_{l+3/2}}^{l+3/2}$\\\hline
\end{tabular}
^{T}\nonumber\\
&  \otimes%
\begin{tabular}
[c]{|c|c|c|}\hline
$\nu_{2}^{3/2}$ & $\cdots$ & $\nu_{m_{3/2}}^{3/2}$\\\hline
\end{tabular}
^{T}\overset{N}{\underset{j\neq l}{\otimes}}%
\begin{tabular}
[c]{|c|}\hline
$\mu_{1}^{j}$\\\hline
\end{tabular}
\underset{m_{j}-1-l_{j}}{\underbrace{%
\begin{tabular}
[c]{|l|l|l|}\hline
$T$ & $\cdots$ & $T$\\\hline
\end{tabular}
}}\underset{l_{j}}{\underbrace{%
\begin{tabular}
[c]{|l|l|l|}\hline
$L$ & $\cdots$ & $L$\\\hline
\end{tabular}
}}\overset{N-1/2}{\underset{r\neq3/2,l+3/2}{\otimes}}%
\begin{tabular}
[c]{|c|c|c|}\hline
$\nu_{1}^{r}$ & $\cdots$ & $\nu_{m_{r}}^{r}$\\\hline
\end{tabular}
^{T}\nonumber\\
&  +\sum_{l\geq1}ll_{l}%
\begin{tabular}
[c]{|c|}\hline
$\mu_{1}^{l}$\\\hline
\end{tabular}
\underset{m_{l}-1-l_{l}}{\underbrace{%
\begin{tabular}
[c]{|l|l|l|}\hline
$T$ & $\cdots$ & $T$\\\hline
\end{tabular}
}}\underset{l_{l}-1}{\underbrace{%
\begin{tabular}
[c]{|l|l|l|}\hline
$L$ & $\cdots$ & $L$\\\hline
\end{tabular}
}}\otimes%
\begin{tabular}
[c]{|c|c|c|c|}\hline
$L$ & $\nu_{1}^{l+3/2}$ & $\cdots$ & $\nu_{m_{l+3/2}}^{l+3/2}$\\\hline
\end{tabular}
^{T}\nonumber\\
&  \otimes%
\begin{tabular}
[c]{|c|c|c|}\hline
$\nu_{2}^{3/2}$ & $\cdots$ & $\nu_{m_{3/2}}^{3/2}$\\\hline
\end{tabular}
^{T}\overset{N}{\underset{j\neq l}{\otimes}}%
\begin{tabular}
[c]{|c|}\hline
$\mu_{1}^{j}$\\\hline
\end{tabular}
\underset{m_{j}-1-l_{j}}{\underbrace{%
\begin{tabular}
[c]{|l|l|l|}\hline
$T$ & $\cdots$ & $T$\\\hline
\end{tabular}
}}\underset{l_{j}}{\underbrace{%
\begin{tabular}
[c]{|l|l|l|}\hline
$L$ & $\cdots$ & $L$\\\hline
\end{tabular}
}}\overset{N-1/2}{\underset{r\neq3/2,l+3/2}{\otimes}}%
\begin{tabular}
[c]{|c|c|c|}\hline
$\nu_{1}^{r}$ & $\cdots$ & $\nu_{m_{r}}^{r}$\\\hline
\end{tabular}
^{T}\nonumber\\
&  +\sum_{i=2}^{m_{3/2}}3%
\begin{tabular}
[c]{|c|c|}\hline
$\nu_{i}^{3/2}$ & $\mu_{1}^{3}$\\\hline
\end{tabular}
\underset{m_{3}-1-l_{3}}{\underbrace{%
\begin{tabular}
[c]{|l|l|l|}\hline
$T$ & $\cdots$ & $T$\\\hline
\end{tabular}
}}\underset{l_{3}}{\underbrace{%
\begin{tabular}
[c]{|l|l|l|}\hline
$L$ & $\cdots$ & $L$\\\hline
\end{tabular}
}}\otimes\left(  -1\right)  ^{i+1}%
\begin{tabular}
[c]{|c|c|c|c|c|}\hline
$\nu_{2}^{3/2}$ & $\cdots$ & $\hat{\nu}_{i}^{3/2}$ & $\cdots$ & $\nu_{m_{3/2}%
}^{3/2}$\\\hline
\end{tabular}
^{T}\nonumber\\
&  \overset{N}{\underset{j\neq3}{\otimes}}%
\begin{tabular}
[c]{|c|}\hline
$\mu_{1}^{j}$\\\hline
\end{tabular}
\underset{m_{j}-1-l_{j}}{\underbrace{%
\begin{tabular}
[c]{|l|l|l|}\hline
$T$ & $\cdots$ & $T$\\\hline
\end{tabular}
}}\underset{l_{j}}{\underbrace{%
\begin{tabular}
[c]{|l|l|l|}\hline
$L$ & $\cdots$ & $L$\\\hline
\end{tabular}
}}\overset{N-1/2}{\underset{r\neq3/2}{\otimes}}%
\begin{tabular}
[c]{|c|c|c|}\hline
$\nu_{1}^{r}$ & $\cdots$ & $\nu_{m_{r}}^{r}$\\\hline
\end{tabular}
^{T}\nonumber\\
&  +\sum_{l\geq2,l\neq3}\sum_{i=1}^{m_{l-3/2}}%
\begin{tabular}
[c]{|c|c|}\hline
$\nu_{i}^{l-3/2}$ & $\mu_{1}^{l}$\\\hline
\end{tabular}
\underset{m_{l}-1-l_{l}}{\underbrace{%
\begin{tabular}
[c]{|l|l|l|}\hline
$T$ & $\cdots$ & $T$\\\hline
\end{tabular}
}}\underset{l_{l}}{\underbrace{%
\begin{tabular}
[c]{|l|l|l|}\hline
$L$ & $\cdots$ & $L$\\\hline
\end{tabular}
}}\otimes\left(  -1\right)  ^{i+1}%
\begin{tabular}
[c]{|c|c|c|c|c|}\hline
$\nu_{1}^{l-3/2}$ & $\cdots$ & $\hat{\nu}_{i}^{l-3/2}$ & $\cdots$ &
$\nu_{m_{l-3/2}}^{l-3/2}$\\\hline
\end{tabular}
^{T}\nonumber\\
&  \otimes%
\begin{tabular}
[c]{|c|c|c|}\hline
$\nu_{2}^{3/2}$ & $\cdots$ & $\nu_{m_{3/2}}^{3/2}$\\\hline
\end{tabular}
^{T}\overset{N}{\underset{j\neq l}{\otimes}}%
\begin{tabular}
[c]{|c|}\hline
$\mu_{1}^{j}$\\\hline
\end{tabular}
\underset{m_{j}-1-l_{j}}{\underbrace{%
\begin{tabular}
[c]{|l|l|l|}\hline
$T$ & $\cdots$ & $T$\\\hline
\end{tabular}
}}\underset{l_{j}}{\underbrace{%
\begin{tabular}
[c]{|l|l|l|}\hline
$L$ & $\cdots$ & $L$\\\hline
\end{tabular}
}}\overset{N-1/2}{\underset{r\neq3/2,l-3/2}{\otimes}}%
\begin{tabular}
[c]{|c|c|c|}\hline
$\nu_{1}^{r}$ & $\cdots$ & $\nu_{m_{r}}^{r}$\\\hline
\end{tabular}
^{T}.
\end{align}%
\end{subequations}%
There are some undetermined parameters, which can be $L$ or $T$, in the above
equations. However, it is easy to see that both choice lead to the same
equations. Therefore, we will set all of them to be $T$ in the following.
Thus, the constrain equations become%
\begin{subequations}%
\begin{align}
0  &  =M%
\begin{tabular}
[c]{|c|c|c|c|}\hline
$L$ & $\nu_{2}^{1/2}$ & $\cdots$ & $\nu_{m_{1/2}}^{1/2}$\\\hline
\end{tabular}
^{T}\overset{N}{\underset{j=1}{\otimes}}\underset{m_{j}-l_{j}}{\underbrace{%
\begin{tabular}
[c]{|l|l|l|}\hline
$T$ & $\cdots$ & $T$\\\hline
\end{tabular}
}}\underset{l_{j}}{\underbrace{%
\begin{tabular}
[c]{|l|l|l|}\hline
$L$ & $\cdots$ & $L$\\\hline
\end{tabular}
}}\overset{N-1/2}{\underset{r\neq1/2}{\otimes}}%
\begin{tabular}
[c]{|c|c|c|}\hline
$\nu_{1}^{r}$ & $\cdots$ & $\nu_{m_{r}}^{r}$\\\hline
\end{tabular}
^{T}\nonumber\\
&  +\sum_{l\geq1}l\left(  m_{l}-l_{l}\right)  \underset{m_{l}-1-l_{l}%
}{\underbrace{%
\begin{tabular}
[c]{|l|l|l|}\hline
$T$ & $\cdots$ & $T$\\\hline
\end{tabular}
}}\underset{l_{l}}{\underbrace{%
\begin{tabular}
[c]{|l|l|l|}\hline
$L$ & $\cdots$ & $L$\\\hline
\end{tabular}
}}\otimes%
\begin{tabular}
[c]{|c|c|c|c|}\hline
$T$ & $\nu_{1}^{l+1/2}$ & $\cdots$ & $\nu_{m_{l+1/2}}^{l+1/2}$\\\hline
\end{tabular}
^{T}\nonumber\\
&  \otimes%
\begin{tabular}
[c]{|c|c|c|}\hline
$\nu_{2}^{1/2}$ & $\cdots$ & $\nu_{m_{1/2}}^{1/2}$\\\hline
\end{tabular}
^{T}\overset{N}{\underset{j\neq l}{\otimes}}\underset{m_{j}-l_{j}%
}{\underbrace{%
\begin{tabular}
[c]{|l|l|l|}\hline
$T$ & $\cdots$ & $T$\\\hline
\end{tabular}
}}\underset{l_{j}}{\underbrace{%
\begin{tabular}
[c]{|l|l|l|}\hline
$L$ & $\cdots$ & $L$\\\hline
\end{tabular}
}}\overset{N-1/2}{\underset{r\neq1/2,l+1/2}{\otimes}}%
\begin{tabular}
[c]{|c|c|c|}\hline
$\nu_{1}^{r}$ & $\cdots$ & $\nu_{m_{r}}^{r}$\\\hline
\end{tabular}
^{T}\nonumber\\
&  +\sum_{l\geq1}ll_{l}\underset{m_{l}-l_{l}}{\underbrace{%
\begin{tabular}
[c]{|l|l|l|}\hline
$T$ & $\cdots$ & $T$\\\hline
\end{tabular}
}}\underset{l_{l}-1}{\underbrace{%
\begin{tabular}
[c]{|l|l|l|}\hline
$L$ & $\cdots$ & $L$\\\hline
\end{tabular}
}}\otimes%
\begin{tabular}
[c]{|c|c|c|c|}\hline
$L$ & $\nu_{1}^{l+1/2}$ & $\cdots$ & $\nu_{m_{l+1/2}}^{l+1/2}$\\\hline
\end{tabular}
^{T}\nonumber\\
&  \otimes%
\begin{tabular}
[c]{|c|c|c|}\hline
$\nu_{2}^{1/2}$ & $\cdots$ & $\nu_{m_{1/2}}^{1/2}$\\\hline
\end{tabular}
^{T}\overset{N}{\underset{j\neq l}{\otimes}}\underset{m_{j}-l_{j}%
}{\underbrace{%
\begin{tabular}
[c]{|l|l|l|}\hline
$T$ & $\cdots$ & $T$\\\hline
\end{tabular}
}}\underset{l_{j}}{\underbrace{%
\begin{tabular}
[c]{|l|l|l|}\hline
$L$ & $\cdots$ & $L$\\\hline
\end{tabular}
}}\overset{N-1/2}{\underset{r\neq1/2,l+1/2}{\otimes}}%
\begin{tabular}
[c]{|c|c|c|}\hline
$\nu_{1}^{r}$ & $\cdots$ & $\nu_{m_{r}}^{r}$\\\hline
\end{tabular}
^{T}\nonumber\\
&  +\sum_{i=2}^{m_{1/2}}%
\begin{tabular}
[c]{|c|}\hline
$\nu_{i}^{1/2}$\\\hline
\end{tabular}
\underset{m_{1}-l_{1}}{\underbrace{%
\begin{tabular}
[c]{|l|l|l|}\hline
$T$ & $\cdots$ & $T$\\\hline
\end{tabular}
}}\underset{l_{1}}{\underbrace{%
\begin{tabular}
[c]{|l|l|l|}\hline
$L$ & $\cdots$ & $L$\\\hline
\end{tabular}
}}\otimes\left(  -1\right)  ^{i+1}%
\begin{tabular}
[c]{|c|c|c|c|c|}\hline
$\nu_{2}^{1/2}$ & $\cdots$ & $\hat{\nu}_{i}^{1/2}$ & $\cdots$ & $\nu_{m_{1/2}%
}^{1/2}$\\\hline
\end{tabular}
^{T}\nonumber\\
&  \overset{N}{\underset{j\neq1}{\otimes}}\underset{m_{j}-l_{j}}{\underbrace{%
\begin{tabular}
[c]{|l|l|l|}\hline
$T$ & $\cdots$ & $T$\\\hline
\end{tabular}
}}\underset{l_{j}}{\underbrace{%
\begin{tabular}
[c]{|l|l|l|}\hline
$L$ & $\cdots$ & $L$\\\hline
\end{tabular}
}}\overset{N-1/2}{\underset{r\neq1/2}{\otimes}}%
\begin{tabular}
[c]{|c|c|c|}\hline
$\nu_{1}^{r}$ & $\cdots$ & $\nu_{m_{r}}^{r}$\\\hline
\end{tabular}
^{T}\nonumber\\
&  +\sum_{l\geq2}\sum_{i=1}^{m_{l-1/2}}\left(  -1\right)  ^{i+1}%
\begin{tabular}
[c]{|c|}\hline
$\nu_{i}^{l-1/2}$\\\hline
\end{tabular}
\underset{m_{l}-l_{l}}{\underbrace{%
\begin{tabular}
[c]{|l|l|l|}\hline
$T$ & $\cdots$ & $T$\\\hline
\end{tabular}
}}\underset{l_{l}}{\underbrace{%
\begin{tabular}
[c]{|l|l|l|}\hline
$L$ & $\cdots$ & $L$\\\hline
\end{tabular}
}}\otimes%
\begin{tabular}
[c]{|c|c|c|c|c|}\hline
$\nu_{1}^{l-1/2}$ & $\cdots$ & $\hat{\nu}_{i}^{l-1/2}$ & $\cdots$ &
$\nu_{m_{l-1/2}}^{l-1/2}$\\\hline
\end{tabular}
^{T}\nonumber\\
&  \otimes%
\begin{tabular}
[c]{|c|c|c|}\hline
$\nu_{2}^{1/2}$ & $\cdots$ & $\nu_{m_{1/2}}^{1/2}$\\\hline
\end{tabular}
^{T}\overset{N}{\underset{j\neq l}{\otimes}}\underset{m_{j}-l_{j}%
}{\underbrace{%
\begin{tabular}
[c]{|l|l|l|}\hline
$T$ & $\cdots$ & $T$\\\hline
\end{tabular}
}}\underset{l_{j}}{\underbrace{%
\begin{tabular}
[c]{|l|l|l|}\hline
$L$ & $\cdots$ & $L$\\\hline
\end{tabular}
}}\overset{N-1/2}{\underset{r\neq1/2,l-1/2}{\otimes}}%
\begin{tabular}
[c]{|c|c|c|}\hline
$\nu_{1}^{r}$ & $\cdots$ & $\nu_{m_{r}}^{r}$\\\hline
\end{tabular}
^{T},
\end{align}

\begin{align}
0  &  =M%
\begin{tabular}
[c]{|c|c|c|c|}\hline
$L$ & $\nu_{2}^{3/2}$ & $\cdots$ & $\nu_{m_{3/2}}^{3/2}$\\\hline
\end{tabular}
^{T}\overset{N}{\underset{j=1}{\otimes}}\underset{m_{j}-l_{j}}{\underbrace{%
\begin{tabular}
[c]{|l|l|l|}\hline
$T$ & $\cdots$ & $T$\\\hline
\end{tabular}
}}\underset{l_{j}}{\underbrace{%
\begin{tabular}
[c]{|l|l|l|}\hline
$L$ & $\cdots$ & $L$\\\hline
\end{tabular}
}}\overset{N-1/2}{\underset{r\neq3/2}{\otimes}}%
\begin{tabular}
[c]{|c|c|c|}\hline
$\nu_{1}^{r}$ & $\cdots$ & $\nu_{m_{r}}^{r}$\\\hline
\end{tabular}
^{T}\nonumber\\
&  +\underset{m_{1}+1-l_{1}}{\underbrace{%
\begin{tabular}
[c]{|l|l|l|}\hline
$T$ & $\cdots$ & $T$\\\hline
\end{tabular}
}}\underset{l_{1}}{\underbrace{%
\begin{tabular}
[c]{|l|l|l|}\hline
$L$ & $\cdots$ & $L$\\\hline
\end{tabular}
}}\otimes%
\begin{tabular}
[c]{|c|c|c|c|}\hline
$T$ & $\nu_{1}^{1/2}$ & $\cdots$ & $\nu_{m_{1/2}}^{1/2}$\\\hline
\end{tabular}
^{T}\otimes%
\begin{tabular}
[c]{|c|c|c|}\hline
$\nu_{2}^{3/2}$ & $\cdots$ & $\nu_{m_{3/2}}^{3/2}$\\\hline
\end{tabular}
^{T}\nonumber\\
&  \overset{N}{\underset{j\neq1}{\otimes}}\underset{m_{j}-l_{j}}{\underbrace{%
\begin{tabular}
[c]{|l|l|l|}\hline
$T$ & $\cdots$ & $T$\\\hline
\end{tabular}
}}\underset{l_{j}}{\underbrace{%
\begin{tabular}
[c]{|l|l|l|}\hline
$L$ & $\cdots$ & $L$\\\hline
\end{tabular}
}}\overset{N-1/2}{\underset{r\neq1/2,3/2}{\otimes}}%
\begin{tabular}
[c]{|c|c|c|}\hline
$\nu_{1}^{r}$ & $\cdots$ & $\nu_{m_{r}}^{r}$\\\hline
\end{tabular}
^{T}\nonumber\\
&  +\sum_{l\geq1}l\left(  m_{l}-l_{l}\right)  \underset{m_{l}-1-l_{l}%
}{\underbrace{%
\begin{tabular}
[c]{|l|l|l|}\hline
$T$ & $\cdots$ & $T$\\\hline
\end{tabular}
}}\underset{l_{l}}{\underbrace{%
\begin{tabular}
[c]{|l|l|l|}\hline
$L$ & $\cdots$ & $L$\\\hline
\end{tabular}
}}\otimes%
\begin{tabular}
[c]{|c|c|c|c|}\hline
$T$ & $\nu_{1}^{l+3/2}$ & $\cdots$ & $\nu_{m_{l+3/2}}^{l+3/2}$\\\hline
\end{tabular}
^{T}\nonumber\\
&  \otimes%
\begin{tabular}
[c]{|c|c|c|}\hline
$\nu_{2}^{3/2}$ & $\cdots$ & $\nu_{m_{3/2}}^{3/2}$\\\hline
\end{tabular}
^{T}\overset{N}{\underset{j\neq l}{\otimes}}\underset{m_{j}-l_{j}%
}{\underbrace{%
\begin{tabular}
[c]{|l|l|l|}\hline
$T$ & $\cdots$ & $T$\\\hline
\end{tabular}
}}\underset{l_{j}}{\underbrace{%
\begin{tabular}
[c]{|l|l|l|}\hline
$L$ & $\cdots$ & $L$\\\hline
\end{tabular}
}}\overset{N-1/2}{\underset{r\neq3/2,l+3/2}{\otimes}}%
\begin{tabular}
[c]{|c|c|c|}\hline
$\nu_{1}^{r}$ & $\cdots$ & $\nu_{m_{r}}^{r}$\\\hline
\end{tabular}
^{T}\nonumber\\
&  +\sum_{l\geq1}ll_{l}\underset{m_{l}-l_{l}}{\underbrace{%
\begin{tabular}
[c]{|l|l|l|}\hline
$T$ & $\cdots$ & $T$\\\hline
\end{tabular}
}}\underset{l_{l}-1}{\underbrace{%
\begin{tabular}
[c]{|l|l|l|}\hline
$L$ & $\cdots$ & $L$\\\hline
\end{tabular}
}}\otimes%
\begin{tabular}
[c]{|c|c|c|c|}\hline
$L$ & $\nu_{1}^{l+3/2}$ & $\cdots$ & $\nu_{m_{l+3/2}}^{l+3/2}$\\\hline
\end{tabular}
^{T}\nonumber\\
&  \otimes%
\begin{tabular}
[c]{|c|c|c|}\hline
$\nu_{2}^{3/2}$ & $\cdots$ & $\nu_{m_{3/2}}^{3/2}$\\\hline
\end{tabular}
^{T}\overset{N}{\underset{j\neq l}{\otimes}}\underset{m_{j}-l_{j}%
}{\underbrace{%
\begin{tabular}
[c]{|l|l|l|}\hline
$T$ & $\cdots$ & $T$\\\hline
\end{tabular}
}}\underset{l_{j}}{\underbrace{%
\begin{tabular}
[c]{|l|l|l|}\hline
$L$ & $\cdots$ & $L$\\\hline
\end{tabular}
}}\overset{N-1/2}{\underset{r\neq3/2,l+3/2}{\otimes}}%
\begin{tabular}
[c]{|c|c|c|}\hline
$\nu_{1}^{r}$ & $\cdots$ & $\nu_{m_{r}}^{r}$\\\hline
\end{tabular}
^{T}\nonumber\\
&  +\sum_{i=2}^{m_{3/2}}3%
\begin{tabular}
[c]{|c|}\hline
$\nu_{i}^{3/2}$\\\hline
\end{tabular}
\underset{m_{3}-l_{3}}{\underbrace{%
\begin{tabular}
[c]{|l|l|l|}\hline
$T$ & $\cdots$ & $T$\\\hline
\end{tabular}
}}\underset{l_{3}}{\underbrace{%
\begin{tabular}
[c]{|l|l|l|}\hline
$L$ & $\cdots$ & $L$\\\hline
\end{tabular}
}}\otimes\left(  -1\right)  ^{i+1}%
\begin{tabular}
[c]{|c|c|c|c|c|}\hline
$\nu_{2}^{3/2}$ & $\cdots$ & $\hat{\nu}_{i}^{3/2}$ & $\cdots$ & $\nu_{m_{3/2}%
}^{3/2}$\\\hline
\end{tabular}
^{T}\nonumber\\
&  \overset{N}{\underset{j\neq3}{\otimes}}\underset{m_{j}-l_{j}}{\underbrace{%
\begin{tabular}
[c]{|l|l|l|}\hline
$T$ & $\cdots$ & $T$\\\hline
\end{tabular}
}}\underset{l_{j}}{\underbrace{%
\begin{tabular}
[c]{|l|l|l|}\hline
$L$ & $\cdots$ & $L$\\\hline
\end{tabular}
}}\overset{N-1/2}{\underset{r\neq3/2}{\otimes}}%
\begin{tabular}
[c]{|c|c|c|}\hline
$\nu_{1}^{r}$ & $\cdots$ & $\nu_{m_{r}}^{r}$\\\hline
\end{tabular}
^{T}\nonumber\\
&  +\sum_{l\geq2,l\neq3}\sum_{i=1}^{m_{l-3/2}}%
\begin{tabular}
[c]{|c|}\hline
$\nu_{i}^{l-3/2}$\\\hline
\end{tabular}
\underset{m_{l}-l_{l}}{\underbrace{%
\begin{tabular}
[c]{|l|l|l|}\hline
$T$ & $\cdots$ & $T$\\\hline
\end{tabular}
}}\underset{l_{l}}{\underbrace{%
\begin{tabular}
[c]{|l|l|l|}\hline
$L$ & $\cdots$ & $L$\\\hline
\end{tabular}
}}\otimes\left(  -1\right)  ^{i+1}%
\begin{tabular}
[c]{|c|c|c|c|c|}\hline
$\nu_{1}^{l-3/2}$ & $\cdots$ & $\hat{\nu}_{i}^{l-3/2}$ & $\cdots$ &
$\nu_{m_{l-3/2}}^{l-3/2}$\\\hline
\end{tabular}
^{T}\nonumber\\
&  \otimes%
\begin{tabular}
[c]{|c|c|c|}\hline
$\nu_{2}^{3/2}$ & $\cdots$ & $\nu_{m_{3/2}}^{3/2}$\\\hline
\end{tabular}
^{T}\overset{N}{\underset{j\neq l}{\otimes}}\underset{m_{j}-l_{j}%
}{\underbrace{%
\begin{tabular}
[c]{|l|l|l|}\hline
$T$ & $\cdots$ & $T$\\\hline
\end{tabular}
}}\underset{l_{j}}{\underbrace{%
\begin{tabular}
[c]{|l|l|l|}\hline
$L$ & $\cdots$ & $L$\\\hline
\end{tabular}
}}\overset{N-1/2}{\underset{r\neq3/2,l-3/2}{\otimes}}%
\begin{tabular}
[c]{|c|c|c|}\hline
$\nu_{1}^{r}$ & $\cdots$ & $\nu_{m_{r}}^{r}$\\\hline
\end{tabular}
^{T}.
\end{align}%
\end{subequations}%
Next, we will deal with those antisymmetric indices$\left\{  \nu_{i}%
^{r}\right\}  $. In this case, there are much fewer possibilities which we can
chosen, i.e.
\begin{equation}%
\begin{tabular}
[c]{|c|c|c|}\hline
$\nu_{1}^{r}$ & $\cdots$ & $\nu_{m_{r}}^{r}$\\\hline
\end{tabular}
\equiv%
\begin{tabular}
[c]{|c|c|}\hline
$\nu_{1}^{r}$ & $\nu_{2}^{r}$\\\hline
\end{tabular}
\left(  =%
\begin{tabular}
[c]{|c|c|}\hline
$T$ & $L$\\\hline
\end{tabular}
\text{, }%
\begin{tabular}
[c]{|l|}\hline
$T$\\\hline
\end{tabular}
\text{, }%
\begin{tabular}
[c]{|l|}\hline
$L$\\\hline
\end{tabular}
\text{ or }%
\begin{tabular}
[c]{|l|}\hline
$0$\\\hline
\end{tabular}
\right)  \text{.}%
\end{equation}
Therefore%
\begin{align}
0  &  =M%
\begin{tabular}
[c]{|c|c|}\hline
$L$ & $\nu_{2}^{1/2}$\\\hline
\end{tabular}
^{T}\overset{N}{\underset{j=1}{\otimes}}\underset{m_{j}-l_{j}}{\underbrace{%
\begin{tabular}
[c]{|l|l|l|}\hline
$T$ & $\cdots$ & $T$\\\hline
\end{tabular}
}}\underset{l_{j}}{\underbrace{%
\begin{tabular}
[c]{|l|l|l|}\hline
$L$ & $\cdots$ & $L$\\\hline
\end{tabular}
}}\overset{N-1/2}{\underset{r\neq1/2}{\otimes}}%
\begin{tabular}
[c]{|c|c|}\hline
$\nu_{1}^{r}$ & $\nu_{2}^{r}$\\\hline
\end{tabular}
^{T}\nonumber\\
&  +\sum_{l\geq1}l\left(  m_{l}-l_{l}\right)  \underset{m_{l}-1-l_{l}%
}{\underbrace{%
\begin{tabular}
[c]{|l|l|l|}\hline
$T$ & $\cdots$ & $T$\\\hline
\end{tabular}
}}\underset{l_{l}}{\underbrace{%
\begin{tabular}
[c]{|l|l|l|}\hline
$L$ & $\cdots$ & $L$\\\hline
\end{tabular}
}}\otimes%
\begin{tabular}
[c]{|c|c|c|}\hline
$T$ & $\nu_{1}^{l+1/2}$ & $\nu_{2}^{l+1/2}$\\\hline
\end{tabular}
^{T}\nonumber\\
&  \otimes%
\begin{tabular}
[c]{|c|}\hline
$\nu_{2}^{1/2}$\\\hline
\end{tabular}
\overset{N}{\underset{j\neq l}{\otimes}}\underset{m_{j}-l_{j}}{\underbrace{%
\begin{tabular}
[c]{|l|l|l|}\hline
$T$ & $\cdots$ & $T$\\\hline
\end{tabular}
}}\underset{l_{j}}{\underbrace{%
\begin{tabular}
[c]{|l|l|l|}\hline
$L$ & $\cdots$ & $L$\\\hline
\end{tabular}
}}\overset{N-1/2}{\underset{r\neq1/2,l+1/2}{\otimes}}%
\begin{tabular}
[c]{|c|c|}\hline
$\nu_{1}^{r}$ & $\nu_{2}^{r}$\\\hline
\end{tabular}
^{T}\nonumber\\
&  +\sum_{l\geq1}ll_{l}\underset{m_{l}-l_{l}}{\underbrace{%
\begin{tabular}
[c]{|l|l|l|}\hline
$T$ & $\cdots$ & $T$\\\hline
\end{tabular}
}}\underset{l_{l}-1}{\underbrace{%
\begin{tabular}
[c]{|l|l|l|}\hline
$L$ & $\cdots$ & $L$\\\hline
\end{tabular}
}}\otimes%
\begin{tabular}
[c]{|c|c|c|}\hline
$L$ & $\nu_{1}^{l+1/2}$ & $\nu_{2}^{l+1/2}$\\\hline
\end{tabular}
^{T}\nonumber\\
&  \otimes%
\begin{tabular}
[c]{|c|}\hline
$\nu_{2}^{1/2}$\\\hline
\end{tabular}
\overset{N}{\underset{j\neq l}{\otimes}}\underset{m_{j}-l_{j}}{\underbrace{%
\begin{tabular}
[c]{|l|l|l|}\hline
$T$ & $\cdots$ & $T$\\\hline
\end{tabular}
}}\underset{l_{j}}{\underbrace{%
\begin{tabular}
[c]{|l|l|l|}\hline
$L$ & $\cdots$ & $L$\\\hline
\end{tabular}
}}\overset{N-1/2}{\underset{r\neq1/2,l+1/2}{\otimes}}%
\begin{tabular}
[c]{|c|c|}\hline
$\nu_{1}^{r}$ & $\nu_{2}^{r}$\\\hline
\end{tabular}
^{T}\nonumber\\
&  +\left(  -1\right)
\begin{tabular}
[c]{|c|}\hline
$\nu_{2}^{1/2}$\\\hline
\end{tabular}
\underset{m_{1}-l_{1}}{\underbrace{%
\begin{tabular}
[c]{|l|l|l|}\hline
$T$ & $\cdots$ & $T$\\\hline
\end{tabular}
}}\underset{l_{1}}{\underbrace{%
\begin{tabular}
[c]{|l|l|l|}\hline
$L$ & $\cdots$ & $L$\\\hline
\end{tabular}
}}\overset{k}{\underset{j\neq1}{\otimes}}\underset{m_{j}-l_{j}}{\underbrace{%
\begin{tabular}
[c]{|l|l|l|}\hline
$T$ & $\cdots$ & $T$\\\hline
\end{tabular}
}}\underset{l_{j}}{\underbrace{%
\begin{tabular}
[c]{|l|l|l|}\hline
$L$ & $\cdots$ & $L$\\\hline
\end{tabular}
}}\overset{s}{\underset{r\neq1/2}{\otimes}}%
\begin{tabular}
[c]{|c|c|}\hline
$\nu_{1}^{r}$ & $\nu_{2}^{r}$\\\hline
\end{tabular}
^{T}\nonumber\\
&  +\sum_{l\geq2}%
\begin{tabular}
[c]{|c|}\hline
$\nu_{1}^{l-1/2}$\\\hline
\end{tabular}
\underset{m_{l}-l_{l}}{\underbrace{%
\begin{tabular}
[c]{|l|l|l|}\hline
$T$ & $\cdots$ & $T$\\\hline
\end{tabular}
}}\underset{l_{l}}{\underbrace{%
\begin{tabular}
[c]{|l|l|l|}\hline
$L$ & $\cdots$ & $L$\\\hline
\end{tabular}
}}\otimes%
\begin{tabular}
[c]{|c|}\hline
$\nu_{2}^{l-1/2}$\\\hline
\end{tabular}
\nonumber\\
&  \otimes%
\begin{tabular}
[c]{|c|}\hline
$\nu_{2}^{1/2}$\\\hline
\end{tabular}
\overset{N}{\underset{j\neq l}{\otimes}}\underset{m_{j}-l_{j}}{\underbrace{%
\begin{tabular}
[c]{|l|l|l|}\hline
$T$ & $\cdots$ & $T$\\\hline
\end{tabular}
}}\underset{l_{j}}{\underbrace{%
\begin{tabular}
[c]{|l|l|l|}\hline
$L$ & $\cdots$ & $L$\\\hline
\end{tabular}
}}\overset{N-1/2}{\underset{r\neq1/2,l-1/2}{\otimes}}%
\begin{tabular}
[c]{|c|c|}\hline
$\nu_{1}^{r}$ & $\nu_{2}^{r}$\\\hline
\end{tabular}
^{T}\nonumber\\
&  +\sum_{l\geq2}\left(  -1\right)
\begin{tabular}
[c]{|c|}\hline
$\nu_{2}^{l-1/2}$\\\hline
\end{tabular}
\underset{m_{l}-l_{l}}{\underbrace{%
\begin{tabular}
[c]{|l|l|l|}\hline
$T$ & $\cdots$ & $T$\\\hline
\end{tabular}
}}\underset{l_{l}}{\underbrace{%
\begin{tabular}
[c]{|l|l|l|}\hline
$L$ & $\cdots$ & $L$\\\hline
\end{tabular}
}}\otimes%
\begin{tabular}
[c]{|c|}\hline
$\nu_{1}^{l-1/2}$\\\hline
\end{tabular}
\nonumber\\
&  \otimes%
\begin{tabular}
[c]{|c|}\hline
$\nu_{2}^{1/2}$\\\hline
\end{tabular}
\overset{N}{\underset{j\neq l}{\otimes}}\underset{m_{j}-l_{j}}{\underbrace{%
\begin{tabular}
[c]{|l|l|l|}\hline
$T$ & $\cdots$ & $T$\\\hline
\end{tabular}
}}\underset{l_{j}}{\underbrace{%
\begin{tabular}
[c]{|l|l|l|}\hline
$L$ & $\cdots$ & $L$\\\hline
\end{tabular}
}}\overset{N-1/2}{\underset{r\neq1/2,l-1/2}{\otimes}}%
\begin{tabular}
[c]{|c|c|}\hline
$\nu_{1}^{r}$ & $\nu_{2}^{r}$\\\hline
\end{tabular}
^{T}, \label{a.}%
\end{align}%
\begin{align}
0  &  =M%
\begin{tabular}
[c]{|c|c|}\hline
$L$ & $\nu_{2}^{3/2}$\\\hline
\end{tabular}
^{T}\overset{N}{\underset{j=1}{\otimes}}\underset{m_{j}-l_{j}}{\underbrace{%
\begin{tabular}
[c]{|l|l|l|}\hline
$T$ & $\cdots$ & $T$\\\hline
\end{tabular}
}}\underset{l_{j}}{\underbrace{%
\begin{tabular}
[c]{|l|l|l|}\hline
$L$ & $\cdots$ & $L$\\\hline
\end{tabular}
}}\overset{N-1/2}{\underset{r\neq3/2}{\otimes}}%
\begin{tabular}
[c]{|c|c|}\hline
$\nu_{1}^{r}$ & $\nu_{2}^{r}$\\\hline
\end{tabular}
^{T}\nonumber\\
&  +\underset{m_{1}+1-l_{1}}{\underbrace{%
\begin{tabular}
[c]{|l|l|l|}\hline
$T$ & $\cdots$ & $T$\\\hline
\end{tabular}
}}\underset{l_{1}}{\underbrace{%
\begin{tabular}
[c]{|l|l|l|}\hline
$L$ & $\cdots$ & $L$\\\hline
\end{tabular}
}}\otimes%
\begin{tabular}
[c]{|c|c|c|}\hline
$T$ & $\nu_{1}^{1/2}$ & $\nu_{2}^{1/2}$\\\hline
\end{tabular}
^{T}\nonumber\\
&  \otimes%
\begin{tabular}
[c]{|c|}\hline
$\nu_{2}^{3/2}$\\\hline
\end{tabular}
\overset{N}{\underset{j\neq1}{\otimes}}\underset{m_{j}-l_{j}}{\underbrace{%
\begin{tabular}
[c]{|l|l|l|}\hline
$T$ & $\cdots$ & $T$\\\hline
\end{tabular}
}}\underset{l_{j}}{\underbrace{%
\begin{tabular}
[c]{|l|l|l|}\hline
$L$ & $\cdots$ & $L$\\\hline
\end{tabular}
}}\overset{N-1/2}{\underset{r\neq1/2,3/2}{\otimes}}%
\begin{tabular}
[c]{|c|c|}\hline
$\nu_{1}^{r}$ & $\nu_{2}^{r}$\\\hline
\end{tabular}
^{T}\nonumber\\
&  +\sum_{l\geq1}l\left(  m_{l}-l_{l}\right)  \underset{m_{l}-1-l_{l}%
}{\underbrace{%
\begin{tabular}
[c]{|l|l|l|}\hline
$T$ & $\cdots$ & $T$\\\hline
\end{tabular}
}}\underset{l_{l}}{\underbrace{%
\begin{tabular}
[c]{|l|l|l|}\hline
$L$ & $\cdots$ & $L$\\\hline
\end{tabular}
}}\otimes%
\begin{tabular}
[c]{|c|c|c|}\hline
$T$ & $\nu_{1}^{l+3/2}$ & $\nu_{2}^{l+3/2}$\\\hline
\end{tabular}
^{T}\nonumber\\
&  \otimes%
\begin{tabular}
[c]{|c|}\hline
$\nu_{2}^{3/2}$\\\hline
\end{tabular}
\overset{N}{\underset{j\neq l}{\otimes}}\underset{m_{j}-l_{j}}{\underbrace{%
\begin{tabular}
[c]{|l|l|l|}\hline
$T$ & $\cdots$ & $T$\\\hline
\end{tabular}
}}\underset{l_{j}}{\underbrace{%
\begin{tabular}
[c]{|l|l|l|}\hline
$L$ & $\cdots$ & $L$\\\hline
\end{tabular}
}}\overset{N-1/2}{\underset{r\neq3/2,l+3/2}{\otimes}}%
\begin{tabular}
[c]{|c|c|}\hline
$\nu_{1}^{r}$ & $\nu_{2}^{r}$\\\hline
\end{tabular}
^{T}\nonumber\\
&  +\sum_{l\geq1}ll_{l}\underset{m_{l}-l_{l}}{\underbrace{%
\begin{tabular}
[c]{|l|l|l|}\hline
$T$ & $\cdots$ & $T$\\\hline
\end{tabular}
}}\underset{l_{l}-1}{\underbrace{%
\begin{tabular}
[c]{|l|l|l|}\hline
$L$ & $\cdots$ & $L$\\\hline
\end{tabular}
}}\otimes%
\begin{tabular}
[c]{|c|c|c|}\hline
$L$ & $\nu_{1}^{l+3/2}$ & $\nu_{2}^{l+3/2}$\\\hline
\end{tabular}
^{T}\nonumber\\
&  \otimes%
\begin{tabular}
[c]{|c|}\hline
$\nu_{2}^{3/2}$\\\hline
\end{tabular}
\overset{N}{\underset{j\neq l}{\otimes}}\underset{m_{j}-l_{j}}{\underbrace{%
\begin{tabular}
[c]{|l|l|l|}\hline
$T$ & $\cdots$ & $T$\\\hline
\end{tabular}
}}\underset{l_{j}}{\underbrace{%
\begin{tabular}
[c]{|l|l|l|}\hline
$L$ & $\cdots$ & $L$\\\hline
\end{tabular}
}}\overset{N-1/2}{\underset{r\neq3/2,l+3/2}{\otimes}}%
\begin{tabular}
[c]{|c|c|}\hline
$\nu_{1}^{r}$ & $\nu_{2}^{r}$\\\hline
\end{tabular}
^{T}\nonumber\\
&  +3\left(  -1\right)
\begin{tabular}
[c]{|c|}\hline
$\nu_{2}^{3/2}$\\\hline
\end{tabular}
\underset{m_{3}-l_{3}}{\underbrace{%
\begin{tabular}
[c]{|l|l|l|}\hline
$T$ & $\cdots$ & $T$\\\hline
\end{tabular}
}}\underset{l_{3}}{\underbrace{%
\begin{tabular}
[c]{|l|l|l|}\hline
$L$ & $\cdots$ & $L$\\\hline
\end{tabular}
}}\overset{N}{\underset{j\neq3}{\otimes}}\underset{m_{j}-l_{j}}{\underbrace{%
\begin{tabular}
[c]{|l|l|l|}\hline
$T$ & $\cdots$ & $T$\\\hline
\end{tabular}
}}\underset{l_{j}}{\underbrace{%
\begin{tabular}
[c]{|l|l|l|}\hline
$L$ & $\cdots$ & $L$\\\hline
\end{tabular}
}}\overset{N-1/2}{\underset{r\neq3/2}{\otimes}}%
\begin{tabular}
[c]{|c|c|}\hline
$\nu_{1}^{r}$ & $\nu_{2}^{r}$\\\hline
\end{tabular}
^{T}\nonumber\\
&  +\sum_{l\geq2,l\neq3}%
\begin{tabular}
[c]{|c|}\hline
$\nu_{1}^{l-3/2}$\\\hline
\end{tabular}
\underset{m_{l}-l_{l}}{\underbrace{%
\begin{tabular}
[c]{|l|l|l|}\hline
$T$ & $\cdots$ & $T$\\\hline
\end{tabular}
}}\underset{l_{l}}{\underbrace{%
\begin{tabular}
[c]{|l|l|l|}\hline
$L$ & $\cdots$ & $L$\\\hline
\end{tabular}
}}\otimes%
\begin{tabular}
[c]{|c|}\hline
$\nu_{2}^{l-3/2}$\\\hline
\end{tabular}
\nonumber\\
&  \otimes%
\begin{tabular}
[c]{|c|}\hline
$\nu_{2}^{3/2}$\\\hline
\end{tabular}
\overset{N}{\underset{j\neq l}{\otimes}}\underset{m_{j}-l_{j}}{\underbrace{%
\begin{tabular}
[c]{|l|l|l|}\hline
$T$ & $\cdots$ & $T$\\\hline
\end{tabular}
}}\underset{l_{j}}{\underbrace{%
\begin{tabular}
[c]{|l|l|l|}\hline
$L$ & $\cdots$ & $L$\\\hline
\end{tabular}
}}\overset{N-1/2}{\underset{r\neq3/2,l-3/2}{\otimes}}%
\begin{tabular}
[c]{|c|c|}\hline
$\nu_{1}^{r}$ & $\nu_{2}^{r}$\\\hline
\end{tabular}
^{T}\nonumber\\
&  +\sum_{l\geq2,l\neq3}%
\begin{tabular}
[c]{|c|}\hline
$\nu_{2}^{l-3/2}$\\\hline
\end{tabular}
\underset{m_{l}-l_{l}}{\underbrace{%
\begin{tabular}
[c]{|l|l|l|}\hline
$T$ & $\cdots$ & $T$\\\hline
\end{tabular}
}}\underset{l_{l}}{\underbrace{%
\begin{tabular}
[c]{|l|l|l|}\hline
$L$ & $\cdots$ & $L$\\\hline
\end{tabular}
}}\otimes\left(  -1\right)
\begin{tabular}
[c]{|c|}\hline
$\nu_{1}^{l-3/2}$\\\hline
\end{tabular}
\nonumber\\
&  \otimes%
\begin{tabular}
[c]{|c|}\hline
$\nu_{2}^{3/2}$\\\hline
\end{tabular}
\overset{N}{\underset{j\neq l}{\otimes}}\underset{m_{j}-l_{j}}{\underbrace{%
\begin{tabular}
[c]{|l|l|l|}\hline
$T$ & $\cdots$ & $T$\\\hline
\end{tabular}
}}\underset{l_{j}}{\underbrace{%
\begin{tabular}
[c]{|l|l|l|}\hline
$L$ & $\cdots$ & $L$\\\hline
\end{tabular}
}}\overset{N-1/2}{\underset{r\neq3/2,l-3/2}{\otimes}}%
\begin{tabular}
[c]{|c|c|}\hline
$\nu_{1}^{r}$ & $\nu_{2}^{r}$\\\hline
\end{tabular}
^{T}. \label{b.}%
\end{align}

Using the following lemma,

\noindent\textbf{Lemma:}%
\begin{equation}%
\begin{tabular}
[c]{|l|l|l|}\hline
$T$ & $\cdots$ & $T$\\\hline
\end{tabular}
\underset{l_{1}}{\underbrace{%
\begin{tabular}
[c]{|l|l|l|}\hline
$L$ & $\cdots$ & $L$\\\hline
\end{tabular}
}}\otimes\underset{m_{2}-l_{2}}{\underbrace{%
\begin{tabular}
[c]{|l|l|l|}\hline
$T$ & $\cdots$ & $T$\\\hline
\end{tabular}
}}%
\begin{tabular}
[c]{|l|l|l|}\hline
$L$ & $\cdots$ & $L$\\\hline
\end{tabular}
\otimes\underset{m_{1/2}}{\underbrace{%
\begin{tabular}
[c]{|c|c|}\hline
$\nu_{1}^{1/2}$ & $\nu_{2}^{1/2}$\\\hline
\end{tabular}
}}\otimes%
\begin{tabular}
[c]{|c|}\hline
$\nu_{2}^{3/2}$\\\hline
\end{tabular}
\otimes%
\begin{tabular}
[c]{|lll|}\hline
& $\cdots$ & \\\hline
\end{tabular}
\equiv0, \label{lamma}%
\end{equation}
\textit{except for (i) }$m_{j\geq3}=m_{r\geq3/2}=0$\textit{, }$l_{2}=m_{2}%
$\textit{, }$l_{3/2}=m_{3/2}=1$\textit{ and (ii) }$l_{1}+l_{1/2}%
=2k$\textit{\U{ff0c}}

The equations (\ref{a.}) and (\ref{b.}) reduce to%
\begin{subequations}%
\begin{align}
0  &  =M\underset{m_{1}-l_{1}}{\underbrace{%
\begin{tabular}
[c]{|l|l|l|}\hline
$T$ & $\cdots$ & $T$\\\hline
\end{tabular}
}}\underset{l_{1}}{\underbrace{%
\begin{tabular}
[c]{|l|l|l|}\hline
$L$ & $\cdots$ & $L$\\\hline
\end{tabular}
}}\otimes\underset{m_{2}}{\underbrace{%
\begin{tabular}
[c]{|l|l|l|}\hline
$L$ & $\cdots$ & $L$\\\hline
\end{tabular}
}}\otimes%
\begin{tabular}
[c]{|c|c|c|}\hline
$L$ & $\nu_{2}^{1/2}$ & $\nu_{3}^{1/2}$\\\hline
\end{tabular}
^{T}\otimes%
\begin{tabular}
[c]{|c|}\hline
$\nu_{1}^{3/2}$\\\hline
\end{tabular}
\nonumber\\
&  +l_{1}\underset{m_{1}-l_{1}}{\underbrace{%
\begin{tabular}
[c]{|l|l|l|}\hline
$T$ & $\cdots$ & $T$\\\hline
\end{tabular}
}}\underset{l_{1}-1}{\underbrace{%
\begin{tabular}
[c]{|l|l|l|}\hline
$L$ & $\cdots$ & $L$\\\hline
\end{tabular}
}}\otimes\underset{m_{2}}{\underbrace{%
\begin{tabular}
[c]{|l|l|l|}\hline
$L$ & $\cdots$ & $L$\\\hline
\end{tabular}
}}\otimes%
\begin{tabular}
[c]{|c|c|}\hline
$\nu_{2}^{1/2}$ & $\nu_{3}^{1/2}$\\\hline
\end{tabular}
^{T}\otimes%
\begin{tabular}
[c]{|c|c|}\hline
$L$ & $\nu_{1}^{3/2}$\\\hline
\end{tabular}
^{T}\nonumber\\
&  -%
\begin{tabular}
[c]{|c|}\hline
$\nu_{2}^{1/2}$\\\hline
\end{tabular}
\underset{m_{1}-l_{1}}{\underbrace{%
\begin{tabular}
[c]{|l|l|l|}\hline
$T$ & $\cdots$ & $T$\\\hline
\end{tabular}
}}\underset{l_{1}}{\underbrace{%
\begin{tabular}
[c]{|l|l|l|}\hline
$L$ & $\cdots$ & $L$\\\hline
\end{tabular}
}}\otimes\underset{m_{2}}{\underbrace{%
\begin{tabular}
[c]{|l|l|l|}\hline
$L$ & $\cdots$ & $L$\\\hline
\end{tabular}
}}\otimes%
\begin{tabular}
[c]{|c|}\hline
$\nu_{3}^{1/2}$\\\hline
\end{tabular}
\otimes%
\begin{tabular}
[c]{|c|}\hline
$\nu_{1}^{3/2}$\\\hline
\end{tabular}
\nonumber\\
&  +%
\begin{tabular}
[c]{|c|}\hline
$\nu_{3}^{1/2}$\\\hline
\end{tabular}
\underset{m_{1}-l_{1}}{\underbrace{%
\begin{tabular}
[c]{|l|l|l|}\hline
$T$ & $\cdots$ & $T$\\\hline
\end{tabular}
}}\underset{l_{1}}{\underbrace{%
\begin{tabular}
[c]{|l|l|l|}\hline
$L$ & $\cdots$ & $L$\\\hline
\end{tabular}
}}\otimes\underset{m_{2}}{\underbrace{%
\begin{tabular}
[c]{|l|l|l|}\hline
$L$ & $\cdots$ & $L$\\\hline
\end{tabular}
}}\otimes%
\begin{tabular}
[c]{|c|}\hline
$\nu_{2}^{1/2}$\\\hline
\end{tabular}
\otimes%
\begin{tabular}
[c]{|c|}\hline
$\nu_{1}^{3/2}$\\\hline
\end{tabular}
\nonumber\\
&  +\underset{m_{1}-l_{1}}{\underbrace{%
\begin{tabular}
[c]{|l|l|l|}\hline
$T$ & $\cdots$ & $T$\\\hline
\end{tabular}
}}\underset{l_{1}}{\underbrace{%
\begin{tabular}
[c]{|l|l|l|}\hline
$L$ & $\cdots$ & $L$\\\hline
\end{tabular}
}}\otimes%
\begin{tabular}
[c]{|c|}\hline
$\nu_{1}^{3/2}$\\\hline
\end{tabular}
\underset{m_{2}}{\underbrace{%
\begin{tabular}
[c]{|l|l|l|}\hline
$L$ & $\cdots$ & $L$\\\hline
\end{tabular}
}}\otimes%
\begin{tabular}
[c]{|c|c|}\hline
$\nu_{2}^{1/2}$ & $\nu_{3}^{1/2}$\\\hline
\end{tabular}
^{T}\otimes%
\begin{tabular}
[c]{|c|}\hline
$0$\\\hline
\end{tabular}
\nonumber\\
&  -\underset{m_{1}-l_{1}}{\underbrace{%
\begin{tabular}
[c]{|l|l|l|}\hline
$T$ & $\cdots$ & $T$\\\hline
\end{tabular}
}}\underset{l_{1}}{\underbrace{%
\begin{tabular}
[c]{|l|l|l|}\hline
$L$ & $\cdots$ & $L$\\\hline
\end{tabular}
}}\otimes\underset{m_{2}}{\underbrace{%
\begin{tabular}
[c]{|l|l|l|}\hline
$L$ & $\cdots$ & $L$\\\hline
\end{tabular}
}}\otimes%
\begin{tabular}
[c]{|c|c|}\hline
$\nu_{2}^{1/2}$ & $\nu_{3}^{1/2}$\\\hline
\end{tabular}
^{T}\otimes%
\begin{tabular}
[c]{|c|}\hline
$\nu_{1}^{3/2}$\\\hline
\end{tabular}
,
\end{align}%
\begin{align}
0  &  =M\underset{m_{1}-1-l_{1}}{\underbrace{%
\begin{tabular}
[c]{|l|l|l|}\hline
$T$ & $\cdots$ & $T$\\\hline
\end{tabular}
}}\underset{l_{1}}{\underbrace{%
\begin{tabular}
[c]{|l|l|l|}\hline
$L$ & $\cdots$ & $L$\\\hline
\end{tabular}
}}\otimes\underset{m_{2}}{\underbrace{%
\begin{tabular}
[c]{|l|l|l|}\hline
$L$ & $\cdots$ & $L$\\\hline
\end{tabular}
}}\otimes%
\begin{tabular}
[c]{|c|c|}\hline
$\nu_{2}^{1/2}$ & $\nu_{3}^{1/2}$\\\hline
\end{tabular}
^{T}\otimes%
\begin{tabular}
[c]{|c|c|}\hline
$L$ & $\nu_{1}^{3/2}$\\\hline
\end{tabular}
^{T}\nonumber\\
&  +\underset{m_{1}-l_{1}}{\underbrace{%
\begin{tabular}
[c]{|l|l|l|}\hline
$T$ & $\cdots$ & $T$\\\hline
\end{tabular}
}}\underset{l_{1}}{\underbrace{%
\begin{tabular}
[c]{|l|l|l|}\hline
$L$ & $\cdots$ & $L$\\\hline
\end{tabular}
}}\otimes\underset{m_{2}}{\underbrace{%
\begin{tabular}
[c]{|l|l|l|}\hline
$L$ & $\cdots$ & $L$\\\hline
\end{tabular}
}}\otimes%
\begin{tabular}
[c]{|c|c|c|}\hline
$T$ & $\nu_{2}^{1/2}$ & $\nu_{3}^{1/2}$\\\hline
\end{tabular}
^{T}\otimes%
\begin{tabular}
[c]{|c|}\hline
$\nu_{1}^{3/2}$\\\hline
\end{tabular}
\nonumber\\
&  -\underset{m_{1}-1-l_{1}}{\underbrace{%
\begin{tabular}
[c]{|l|l|l|}\hline
$T$ & $\cdots$ & $T$\\\hline
\end{tabular}
}}\underset{l_{1}}{\underbrace{%
\begin{tabular}
[c]{|l|l|l|}\hline
$L$ & $\cdots$ & $L$\\\hline
\end{tabular}
}}\otimes%
\begin{tabular}
[c]{|c|}\hline
$\nu_{2}^{1/2}$\\\hline
\end{tabular}
\underset{m_{2}}{\underbrace{%
\begin{tabular}
[c]{|l|l|l|}\hline
$L$ & $\cdots$ & $L$\\\hline
\end{tabular}
}}\otimes%
\begin{tabular}
[c]{|c|}\hline
$\nu_{3}^{1/2}$\\\hline
\end{tabular}
\otimes%
\begin{tabular}
[c]{|c|}\hline
$\nu_{1}^{3/2}$\\\hline
\end{tabular}
\nonumber\\
&  +\underset{m_{1}-1-l_{1}}{\underbrace{%
\begin{tabular}
[c]{|l|l|l|}\hline
$T$ & $\cdots$ & $T$\\\hline
\end{tabular}
}}\underset{l_{1}}{\underbrace{%
\begin{tabular}
[c]{|l|l|l|}\hline
$L$ & $\cdots$ & $L$\\\hline
\end{tabular}
}}\otimes%
\begin{tabular}
[c]{|c|}\hline
$\nu_{3}^{1/2}$\\\hline
\end{tabular}
\underset{m_{2}}{\underbrace{%
\begin{tabular}
[c]{|l|l|l|}\hline
$L$ & $\cdots$ & $L$\\\hline
\end{tabular}
}}\otimes%
\begin{tabular}
[c]{|c|}\hline
$\nu_{2}^{1/2}$\\\hline
\end{tabular}
\otimes%
\begin{tabular}
[c]{|c|}\hline
$\nu_{1}^{3/2}$\\\hline
\end{tabular}
.
\end{align}%
\end{subequations}%
From the first equation we have:

For $\nu_{2}^{1/2}=0$, $\nu_{3}^{1/2}=0$ and $\nu_{1}^{3/2}=0,$%
\begin{align}
0  &  =M\underset{m_{1}-l_{1}}{\underbrace{%
\begin{tabular}
[c]{|l|l|l|}\hline
$T$ & $\cdots$ & $T$\\\hline
\end{tabular}
}}\underset{l_{1}}{\underbrace{%
\begin{tabular}
[c]{|l|l|l|}\hline
$L$ & $\cdots$ & $L$\\\hline
\end{tabular}
}}\otimes\underset{m_{2}}{\underbrace{%
\begin{tabular}
[c]{|l|l|l|}\hline
$L$ & $\cdots$ & $L$\\\hline
\end{tabular}
}}\otimes%
\begin{tabular}
[c]{|c|}\hline
$L$\\\hline
\end{tabular}
\otimes%
\begin{tabular}
[c]{|c|}\hline
$0$\\\hline
\end{tabular}
\nonumber\\
&  +l_{1}\underset{m_{1}-l_{1}}{\underbrace{%
\begin{tabular}
[c]{|l|l|l|}\hline
$T$ & $\cdots$ & $T$\\\hline
\end{tabular}
}}\underset{l_{1}-1}{\underbrace{%
\begin{tabular}
[c]{|l|l|l|}\hline
$L$ & $\cdots$ & $L$\\\hline
\end{tabular}
}}\otimes\underset{m_{2}}{\underbrace{%
\begin{tabular}
[c]{|l|l|l|}\hline
$L$ & $\cdots$ & $L$\\\hline
\end{tabular}
}}\otimes%
\begin{tabular}
[c]{|c|}\hline
$0$\\\hline
\end{tabular}
\otimes%
\begin{tabular}
[c]{|c|}\hline
$L$\\\hline
\end{tabular}
\label{1-000}%
\end{align}

For $\nu_{2}^{1/2}=T$, $\nu_{3}^{1/2}=L$ and $\nu_{1}^{3/2}=0,$%
\begin{align}
0  &  =l_{1}\underset{m_{1}-l_{1}}{\underbrace{%
\begin{tabular}
[c]{|l|l|l|}\hline
$T$ & $\cdots$ & $T$\\\hline
\end{tabular}
}}\underset{l_{1}-1}{\underbrace{%
\begin{tabular}
[c]{|l|l|l|}\hline
$L$ & $\cdots$ & $L$\\\hline
\end{tabular}
}}\otimes\underset{m_{2}}{\underbrace{%
\begin{tabular}
[c]{|l|l|l|}\hline
$L$ & $\cdots$ & $L$\\\hline
\end{tabular}
}}\otimes%
\begin{tabular}
[c]{|c|c|}\hline
$T$ & $L$\\\hline
\end{tabular}
^{T}\otimes%
\begin{tabular}
[c]{|c|}\hline
$L$\\\hline
\end{tabular}
\nonumber\\
&  -\underset{m_{1}+1-l_{1}}{\underbrace{%
\begin{tabular}
[c]{|l|l|l|}\hline
$T$ & $\cdots$ & $T$\\\hline
\end{tabular}
}}\underset{l_{1}}{\underbrace{%
\begin{tabular}
[c]{|l|l|l|}\hline
$L$ & $\cdots$ & $L$\\\hline
\end{tabular}
}}\otimes\underset{m_{2}}{\underbrace{%
\begin{tabular}
[c]{|l|l|l|}\hline
$L$ & $\cdots$ & $L$\\\hline
\end{tabular}
}}\otimes%
\begin{tabular}
[c]{|c|}\hline
$L$\\\hline
\end{tabular}
\otimes%
\begin{tabular}
[c]{|c|}\hline
$0$\\\hline
\end{tabular}
\nonumber\\
&  +\underset{m_{1}-l_{1}}{\underbrace{%
\begin{tabular}
[c]{|l|l|l|}\hline
$T$ & $\cdots$ & $T$\\\hline
\end{tabular}
}}\underset{l_{1}+1}{\underbrace{%
\begin{tabular}
[c]{|l|l|l|}\hline
$L$ & $\cdots$ & $L$\\\hline
\end{tabular}
}}\otimes\underset{m_{2}}{\underbrace{%
\begin{tabular}
[c]{|l|l|l|}\hline
$L$ & $\cdots$ & $L$\\\hline
\end{tabular}
}}\otimes%
\begin{tabular}
[c]{|c|}\hline
$T$\\\hline
\end{tabular}%
\label{1-L0L}%
\end{align}

From the second equation we have:

For $\nu_{2}^{1/2}=0$, $\nu_{3}^{1/2}=0$ and $\nu_{1}^{3/2}=0,$%
\begin{align}
0  &  =M\underset{m_{1}-1-l_{1}}{\underbrace{%
%
. \label{3-L0L}%
\end{align}
Using the equations (\ref{1-000}) and (\ref{1-L0L}), we get%
\begin{align}
&  \underset{m_{1}-l_{1}}{\underbrace{%
%
,
\end{align}
then using equations (\ref{1-T0L}), (\ref{3-000}) and (\ref{3-L0L}), we
obtain
\begin{align}
&  \underset{N-2m_{2}-2k}{\underbrace{%
%
.
\end{align}

We solve the above equations, the ratios between the physical states in the NS
sector in the high energy limit are given as%
\begin{align}
&  \underset{N-2m_{2}-2k}{\underbrace{%
%
,
\end{align}

\section{Kinematic relations in the RR\label{RR Kinematic}}

In this appendix, we list the expressions of the kinematic variables we used
in the evaluation of 4-point functions in this paper. For convenience, we take
the center of momentum frame and choose the momenta of particles 1 and 2 to be
along the $X^{1}$-direction. The high energy scattering plane is defined to be
on the $X^{1}-X^{2}$ plane.

The momenta of the four particles are%

\begin{align}
k_{1}  &  =\left(  +\sqrt{p^{2}+M_{1}^{2}},-p,0\right)  ,\\
k_{2}  &  =\left(  +\sqrt{p^{2}+M_{2}^{2}},+p,0\right)  ,\\
k_{3}  &  =\left(  -\sqrt{q^{2}+M_{3}^{2}},-q\cos\theta,-q\sin\theta\right)
,\\
k_{4}  &  =\left(  -\sqrt{q^{2}+M_{4}^{2}},+q\cos\theta,+q\sin\theta\right)
\end{align}

where $p\equiv\left\vert \mathrm{\vec{p}}\right\vert $, $q\equiv\left\vert
\mathrm{\vec{q}}\right\vert $ and $k_{i}^{2}=-M_{i}^{2}$. In the calculation
of the string scattering amplitudes, we use the following formulas%

\begin{align}
-k_{1}\cdot k_{2}  &  =\sqrt{p^{2}+M_{1}^{2}}\cdot\sqrt{p^{2}+M_{2}^{2}}%
+p^{2}=\dfrac{1}{2}\left(  s-M_{1}^{2}-M_{2}^{2}\right)  ,\\
-k_{2}\cdot k_{3}  &  =-\sqrt{p^{2}+M_{2}^{2}}\cdot\sqrt{q^{2}+M_{3}^{2}%
}+pq\cos\theta=\dfrac{1}{2}\left(  t-M_{2}^{2}-M_{3}^{2}\right)  ,\\
-k_{1}\cdot k_{3}  &  =-\sqrt{p^{2}+M_{1}^{2}}\cdot\sqrt{q^{2}+M_{3}^{2}%
}-pq\cos\theta=\dfrac{1}{2}\left(  u-M_{1}^{2}-M_{3}^{2}\right)
\end{align}
where the Mandelstam variables are defined as usual with%

\begin{equation}
s+t+u=\sum_{i}M_{i}^{2}=2N-1.
\end{equation}
The center of mass energy $E$ is defined as%
\begin{equation}
E=\dfrac{1}{2}\left(  \sqrt{p^{2}+M_{1}^{2}}+\sqrt{p^{2}+M_{2}^{2}}\right)
=\dfrac{1}{2}\left(  \sqrt{q^{2}+M_{3}^{2}}+\sqrt{q^{2}+M_{4}^{2}}\right)  .
\end{equation}
We define the polarizations of the string state on the scattering plane as%

\begin{align}
e^{P}  &  =\frac{1}{M_{2}}\left(  \sqrt{p^{2}+M_{2}^{2}},p,0\right)  ,\\
e^{L}  &  =\frac{1}{M_{2}}\left(  p,\sqrt{p^{2}+M_{2}^{2}},0\right)  ,\\
e^{T}  &  =\left(  0,0,1\right)  .
\end{align}
The projections of the momenta on the scattering plane can be calculated as
(here we only list the ones we need for our calculations)%

\begin{align}
e^{P}\cdot k_{1}  &  =-\frac{1}{M_{2}}\left(  \sqrt{p^{2}+M_{1}^{2}}%
\sqrt{p^{2}+M_{2}^{2}}+p^{2}\right)  ,\label{A13}\\
e^{L}\cdot k_{1}  &  =-\frac{p}{M_{2}}\left(  \sqrt{p^{2}+M_{1}^{2}}%
+\sqrt{p^{2}+M_{2}^{2}}\right)  ,\\
e^{T}\cdot k_{1}  &  =0
\end{align}
and%
\begin{align}
e^{P}\cdot k_{3}  &  =\frac{1}{M_{2}}\left(  \sqrt{q^{2}+M_{3}^{2}}\sqrt
{p^{2}+M_{2}^{2}}-pq\cos\theta\right)  ,\\
e^{L}\cdot k_{3}  &  =\frac{1}{M_{2}}\left(  p\sqrt{q^{2}+M_{3}^{2}}%
-q\sqrt{p^{2}+M_{2}^{2}}\cos\theta\right)  ,\\
e^{T}\cdot k_{3}  &  =-q\sin\theta. \label{A18}%
\end{align}
We now expand the kinematic relations to the subleading orders in the RR. We
first express all kinematic variables in terms of $s$ and $t$, and then expand
all relevant quantities in $s:$
\begin{align}
E_{1}  &  =\frac{s-(M_{2}^{2}+2)}{2\sqrt{2}},\\
E_{2}  &  =\frac{s+(M_{2}^{2}+2)}{2\sqrt{2}},\\
|\mathbf{k_{2}}|  &  =\sqrt{E_{1}^{2}+2},\quad|\mathbf{K_{3}}|=\sqrt{\frac
{s}{4}+2};
\end{align}%
\begin{equation}
e_{P}\cdot k_{1}=-\frac{1}{2M_{2}}s+\left(  -\frac{1}{M_{2}}+\frac{M_{2}}%
{2}\right)  ,\quad(\text{exact})
\end{equation}%
\begin{align}
e_{L}\cdot k_{1}  &  =-\frac{1}{2M_{2}}s+\left(  -\frac{1}{M_{2}}+\frac{M_{2}%
}{2}\right)  -{2}M_{2}s^{-1}-2M_{2}(M_{2}^{2}-2)s^{-2}\nonumber\\
&  -2m_{2}(M_{2}^{4}-6M_{2}^{2}+4)s^{-3}-2M_{2}(M_{2}^{6}-12M_{2}^{4}%
+24M_{2}^{2}-8)s^{-4}+O(s^{-5}),
\end{align}%
\begin{equation}
e_{T}\cdot k_{1}=0.
\end{equation}
A key step is to express the scattering angle $\theta$ in terms of $s$ and
$t$. This can be achieved by solving
\begin{equation}
t=-\left(  -(E_{2}-\frac{\sqrt{s}}{2})^{2}+(|\mathbf{k_{2}}|-|\mathbf{k_{3}%
}|\cos\theta)^{2}+|\mathbf{k}_{3}|^{2}\sin^{2}\theta\right)
\end{equation}
to obtain%
\begin{equation}
\theta=\arccos\left(  \frac{s+2t-M_{2}^{2}+6}{\sqrt{s+8}\sqrt{\frac
{(s+2)^{2}-2(s-2)M_{2}^{2}+M_{2}^{4}}{s}}}\right)  .\text{ (exact)}%
\end{equation}
One can then calculate the following expansions which we used in the
subleading order calculation in section V
\begin{equation}
e_{P}\cdot k_{3}=\frac{1}{M_{2}}(E_{2}\frac{\sqrt{s}}{2}-|\mathbf{k_{2}%
}||\mathbf{k_{3}}|\cos\theta)=-\frac{t+2-M_{2}^{2}}{2M_{2}},
\end{equation}%
\begin{align}
e_{L}\cdot k_{3}  &  =\frac{1}{M_{2}}(k_{2}\frac{\sqrt{2}}{2}-E_{2}k_{3}%
\cos\theta)\nonumber\\
&  =-\frac{t+2+M_{2}^{2}}{2M_{2}}-M_{2}ts^{-1}-M_{2}[-4(t+1)+M_{2}%
^{2}(t-2)]s^{-2}\nonumber\\
&  -M_{2}[4(4+3t)-12tM_{2}^{2}+(t-4)M_{2}^{4}]s^{-3}-M_{2}%
[-16(3+2t)+24(2+3t)M_{2}^{2}\nonumber\\
&  -24(-1+t)M_{2}^{4}+(-6+t)M_{2}^{6}]s^{-4}+O(s^{-5}),
\end{align}%
\begin{align}
e_{T}\cdot k_{3}  &  =-|\mathbf{k_{3}}|\sin\theta\nonumber\\
&  =-\sqrt{-t}-\frac{1}{2}\sqrt{-t}(2+t+M_{2}^{2})s^{-1}\nonumber\\
&  \quad-\frac{1}{8\sqrt{-t}}[32+52t+20t^{2}+t^{3}+(32+20t-6t^{2})M_{2}%
^{2}+(8-3t)M_{2}^{4}]s^{-2}\nonumber\\
&  \quad+\frac{1}{16\sqrt{-t}}[320+456t+188t^{2}+22t^{3}+t^{4}%
-(-224+36t+132t^{2}+5t^{3})M_{2}^{2}.\nonumber\\
&  \quad\quad\quad\quad\quad\quad+(-16-122t+15t^{2})M_{2}^{4}+(-24+5t)M_{2}%
^{6}]s^{-3}\nonumber\\
&  \quad+\frac{1}{128(-t)^{3/2}}[1024+12032t+16080t^{2}+7520t^{3}%
+1432t^{4}+136t^{5}+5t^{6}\nonumber\\
&  -4(-512-896t+2232t^{2}+1844t^{3}+170t^{4}+7t^{5})M_{2}^{2}\nonumber\\
&  +2(768-2240t-2372t^{2}+1172t^{3}+35t^{4})M_{2}^{4}\nonumber\\
&  -4(-128+288t-450t^{2}+35t^{3})M_{2}^{6}+(64+240t-35t^{2})M_{2}^{8}%
]s^{-4}+O(s^{-5}).
\end{align}
%

\setcounter{equation}{0}
\renewcommand{\theequation}{\thesection.\arabic{equation}}%

\section{Recurrence relations of Kummer functions\label{recurrence of Kummer}}

In this appendix, we review the recurrence relations of Kummer functions of
the second kind \cite{Slater}. The Kummer function of the second kind $U$ \ is
defined to be%
\begin{equation}
U(a,c,x)=\frac{\pi}{\sin\pi c}\left[  \frac{M(a,c,x)}{(a-c)!(c-1)!}%
-\frac{x^{1-c}M(a+1-c,2-c,x)}{(a-1)!(1-c)!}\right]  \text{ \ }(c\neq2,3,4...)
\end{equation}
where $M(a,c,x)=\sum_{j=0}^{\infty}\frac{(a)_{j}}{(c)_{j}}\frac{x^{j}}{j!}$ is
the Kummer function of the first kind. Here $(a)_{j}=a(a+1)(a+2)...(a+j-1)$ is
the Pochhammer symbol. $U$ and $M$ are the two solutions of the Kummer
Equation%
\begin{equation}
xy^{^{\prime\prime}}(x)+(c-x)y^{\prime}(x)-ay(x)=0. \label{KE}%
\end{equation}
For any confluent hypergeometric function with parameters $(a,c)$ the four
functions with parameters $(a-1,c),(a+1,c),(a,c-1)$ and $(a,c+1)$ are called
the contiguous functions. It follows, from the Kummer Equation Eq.(\ref{KE})
and derivatives of Kummer functions%
\begin{align}
U(a+1,c+1,x)  &  =\frac{-1}{a}U^{\prime}(a,c,x),\label{DE1}\\
U(a+1,c,x)  &  =\frac{1}{1+a-c}U(a,c,x)+\frac{x}{a(1+a-c)}U^{\prime}(a,c,x),\\
U(a,c-1,x)  &  =\frac{1-c}{1+a-c}U(a,c,x)-\frac{x}{1+a-c}U^{\prime}(a,c,x),\\
U(a,c+1,x)  &  =U^{\prime}(a,c,x)-U^{\prime}(a,c,x),\\
U(a-1,c,x)  &  =(x+a-c)U(a,c,x)-xU^{\prime}(a,c,x),\\
U(a-1,c-1,x)  &  =(1+x-c)U(a,c,x)-xU^{\prime}(a,c,x), \label{DE6}%
\end{align}
that a recurrence relation exists between any such function and any two of its
contiguous functions. There are six recurrence relations%
\begin{align}
U(a-1,c,x)-(2a-c+x)U(a,c,x)+a(1+a-c)U(a+1,c,x)  &  =0,\label{RC1}\\
(c-a-1)U(a,c-1,x)-(x+c-1))U(a,c,x)+xU(a,c+1,x)  &  =0,\label{RC2}\\
U(a,c,x)-aU(a+1,c,x)-U(a,c-1,x)  &  =0,\label{RC3}\\
(c-a)U(a,c,x)+U(a-1,c,x)-xU(a,c+1,x)  &  =0,\label{RC4}\\
(a+x)U(a,c,x)-xU(a,c+1,x)+a(c-a-1)U(a+1,c,x)  &  =0,\label{RC5}\\
(a+x-1)U(a,c,x)-U(a-1,c,x)+(1+a-c)U(a,c-1,x)  &  =0. \label{RC6}%
\end{align}
From any two of these six relations the remaining four recurrence relations
can be deduced. Thus they are not independent. For example, one can deduces
recurrence relation Eq.(\ref{RC1}) from Eq.(\ref{RC3}) and Eq.(\ref{RC4}). We
start with Eq.(\ref{RC3}) with $c\rightarrow c+1$
\begin{equation}
\quad U\left(  a,c+1,x\right)  -aU\left(  a+1,c+1,x\right)  -U(a,c,x)=0\ .
\label{RC7}%
\end{equation}
We consider Eq.(\ref{RC4})+$x\cdot$Eq.(\ref{RC7}) to deduce
\begin{equation}
\quad\left(  c-a-x\right)  U\left(  a,c,x\right)  +U\left(  a-1,c,x\right)
-axU\left(  a+1,c+1,x\right)  =0\ . \label{RC8}%
\end{equation}
Next we replace Eq.(\ref{RC4}) with $a\rightarrow a+1$ to get
\begin{equation}
\quad\left(  c-a-1\right)  U\left(  a+1,c,x\right)  +U\left(  a,c,x\right)
-xU(a+1,c+1,x)=0\ . \label{RC9}%
\end{equation}
Finally we consider Eq.(\ref{RC8})$-a\cdot$Eq.(\ref{RC9}) to deduce
\begin{equation}
\quad\left(  c-2a-x\right)  U\left(  a,c,x\right)  +U\left(  a-1,c,x\right)
-a\left(  c-a-1\right)  U\left(  a+1,c,x\right)  =0\ , \label{RC10}%
\end{equation}
which is nothing but Eq.(\ref{RC1}).

The confluent hypergeometric function with parameters $(a\pm m,c\pm n)$ for
$m,n=0,1,2...$are called associated functions. Again it can be shown that
there exist relations between any three associated functions, so that any
confluent hypergeometric function can be expressed in terms of any two of its
associated functions.%

\setcounter{equation}{0}
\renewcommand{\theequation}{\thesection.\arabic{equation}}%

\section{Regge string ZNS\label{Regge ZNS}}

There are two types of ZNS in the old covariant first quantized string spectrum%

\begin{equation}
\text{Type I}:L_{-1}\left\vert x\right\rangle ,\text{ where }L_{1}\left\vert
x\right\rangle =L_{2}\left\vert x\right\rangle =0,\text{ }L_{0}\left\vert
x\right\rangle =0; \label{ZN1}%
\end{equation}

\begin{equation}
\text{Type II}:(L_{-2}+\frac{3}{2}L_{-1}^{2})\left\vert \widetilde{x}%
\right\rangle ,\text{ where }L_{1}\left\vert \widetilde{x}\right\rangle
=L_{2}\left\vert \widetilde{x}\right\rangle =0,\text{ }(L_{0}+1)\left\vert
\widetilde{x}\right\rangle =0. \label{ZN2}%
\end{equation}
Eq.(\ref{ZN1}) and Eq.(\ref{ZN2}) can be derived from Kac determinant in
conformal field theory. While type I states have zero-norm at any spacetime
dimension, type II states have zero-norm \textit{only} at D=26. The existence
of type II ZNS signals the importance of ZNS in the structure of the theory of
string. In fact, the linear relations obtained by high energy limit of stringy
Ward identities or decoupling of ZNS in the GR were just good enough to solve
all the high energy amplitudes in terms of one amplitude.

In the RR, however, the Regge stringy Ward identities or decoupling of ZNS in
the RR turned out to be not good enough to solve all the Regge scattering
amplitudes algebraically. This is due to the much more numerous Regge string
scattering amplitudes than those in the GR at each fixed mass level. In this
appendix, we list all ZNS for $M^{2}=2$ and $\ 4$ and calculate their Regge
limit which we use in the text to demonstrate the calculation. At the first
massive level $k^{2}=-2,$ there is a type II ZNS%

\begin{equation}
\lbrack\frac{1}{2}\alpha_{-1}\cdot\alpha_{-1}+\frac{5}{2}k\cdot\alpha
_{-2}+\frac{3}{2}(k\cdot\alpha_{-1})^{2}]\left\vert 0,k\right\rangle
\label{2.1}%
\end{equation}
and a type I ZNS%

\begin{equation}
\lbrack\theta\cdot\alpha_{-2}+(k\cdot\alpha_{-1})(\theta\cdot\alpha
_{-1})]\left\vert 0,k\right\rangle ,\theta\cdot k=0. \label{2.2}%
\end{equation}
In the Regge limit, the polarizations of the 2nd particle with momentum
$k_{2}$ on the scattering plane used in the text were defined to be
$e^{P}=\frac{1}{M_{2}}(E_{2},\mathrm{k}_{2},0)=\frac{k_{2}}{M_{2}}$ as the
momentum polarization, $e^{L}=\frac{1}{M_{2}}(\mathrm{k}_{2},E_{2},0)$ the
longitudinal polarization and $e^{T}=(0,0,1)$ the transverse polarization
which lies on the scattering plane. $\eta_{\mu\nu}=diag(-1,1,1).$ The three
vectors $e^{P}$, $e^{L}$ and $e^{T}$ satisfy the completeness relation
$\eta_{\mu\nu}=\sum_{\alpha,\beta}e_{\mu}^{\alpha}e_{\nu}^{\beta}\eta
_{\alpha\beta}$ where $\mu,\nu=0,1,2$ and $\alpha,\beta=P,L,T$ and
$\alpha_{-1}^{T}=\sum_{\mu}e_{\mu}^{T}\alpha_{-1}^{\mu}$, $\alpha_{-1}%
^{T}\alpha_{-2}^{L}=\sum_{\mu,\nu}e_{\mu}^{T}e_{\nu}^{L}\alpha_{-1}^{\mu
}\alpha_{-2}^{\nu}$ etc.

In the Regge limit, the type II ZNS in Eq.(\ref{2.1}) gives the Regge string
zero-norm state (RZNS)%
\begin{equation}
(\sqrt{2}\alpha_{-2}^{P}-\alpha_{-1}^{P}\alpha_{-1}^{P}-\frac{1}{5}\alpha
_{-1}^{L}\alpha_{-1}^{L}-\frac{1}{5}\alpha_{-1}^{T}\alpha_{-1}^{T}%
)|0,k\rangle. \label{R2.3}%
\end{equation}
Type I ZNS in Eq.(\ref{2.2}) gives two RZNS%
\begin{equation}
(\alpha_{-2}^{T}-\sqrt{2}\alpha_{-1}^{P}\alpha_{-1}^{T})|0,k\rangle,
\label{R2.1}%
\end{equation}%
\begin{equation}
(\alpha_{-2}^{L}-\sqrt{2}\alpha_{-1}^{P}\alpha_{-1}^{L})|0,k\rangle
\label{R2.2}%
\end{equation}
RZNS in Eq.(\ref{R2.1}) and Eq.(\ref{R2.2}) correspond to choose $\theta^{\mu
}=e^{T}$ and $\theta^{\mu}=e^{L}$ respectively. Note that the norms of Regge
"zero-norm" states may not be zero. For instance the norm of Eq.(\ref{R2.3})
is not zero. They are just used to produce Regge stringy Ward identities
Eq.(\ref{W2.3}), Eq.(\ref{W2.1}) and Eq.(\ref{W2.2}) in the text.

At the second massive level $k^{2}=-4,$ there is a type I scalar ZNS%
\begin{align}
&  [\frac{17}{4}(k\cdot\alpha_{-1})^{3}+\frac{9}{2}(k\cdot\alpha_{-1}%
)(\alpha_{-1}\cdot\alpha_{-1})+9(\alpha_{-1}\cdot\alpha_{-2})\nonumber\\
&  +21(k\cdot\alpha_{-1})(k\cdot\alpha_{-2})+25(k\cdot\alpha_{-3})]\left\vert
0,k\right\rangle , \label{41}%
\end{align}
a symmetric type I spin two ZNS%

\begin{equation}
\lbrack2\theta_{\mu\nu}\alpha_{-1}^{(\mu}\alpha_{-2}^{\nu)}+k_{\lambda}%
\theta_{\mu\nu}\alpha_{-1}^{\lambda\mu\nu}]\left\vert 0,k\right\rangle
,k\cdot\theta=\eta^{\mu\nu}\theta_{\mu\nu}=0,\theta_{\mu\nu}=\theta_{\nu\mu}
\label{42}%
\end{equation}
where $\alpha_{-1}^{\lambda\mu\nu}\equiv\alpha_{-1}^{\lambda}\alpha_{-1}^{\mu
}\alpha_{-1}^{\nu}$ and two vector ZNS%
\begin{align}
\lbrack(\frac{5}{2}k_{\mu}k_{\nu}\theta_{\lambda}^{\prime}+\eta_{\mu\nu}%
\theta_{\lambda}^{\prime})\mathcal{\alpha}_{-1}^{(\mu\nu\lambda)}+9k_{\mu
}\theta_{\nu}^{\prime}\mathcal{\alpha}_{-1}^{(\mu\nu)}+6\theta_{\mu}^{\prime
}\mathcal{\alpha}_{-1}^{\mu}]\left\vert 0,k\right\rangle ,\theta\cdot k  &
=0,\label{43}\\
\lbrack(\frac{1}{2}k_{\mu}k_{\nu}\theta_{\lambda}+2\eta_{\mu\nu}%
\theta_{\lambda})\mathcal{\alpha}_{-1}^{(\mu\nu\lambda)}+9k_{\mu}\theta_{\nu
}\mathcal{\alpha}_{-1}^{[\mu\nu]}-6\theta_{\mu}\mathcal{\alpha}_{-1}^{\mu
}]\left\vert 0,k\right\rangle ,\theta\cdot k  &  =0. \label{44}%
\end{align}
Note that Eq.(\ref{43}) and Eq.(\ref{44}) are linear combinations of a type I
and a type II ZNS. This completes the four ZNS at the second massive level
$M^{2}=$ $4$.

In the Regge limit, the scalar ZNS in Eq.(\ref{41}) gives the RZNS%
\begin{equation}
\lbrack25(\alpha_{-1}^{P})^{3}+9\alpha_{-1}^{P}(\alpha_{-1}^{L})^{2}%
+9\alpha_{-1}^{P}(\alpha_{-1}^{T})^{2}-9\alpha_{-2}^{L}\alpha_{-1}^{L}%
-9\alpha_{-2}^{T}\alpha_{-1}^{T}-75\alpha_{-2}^{P}\alpha_{-1}^{P}%
+50\alpha_{-3}^{P}]\left\vert 0,k\right\rangle . \label{R4.1}%
\end{equation}
For the type I spin two ZNS in Eq.(\ref{42}), we define $\theta_{\mu\nu}%
=\sum_{\alpha,\beta}e_{\mu}^{\alpha}e_{\nu}^{\beta}u_{\alpha\beta}$, symmetric
and transverse conditions on $\theta_{\mu\nu}$ implies%
\begin{equation}
u_{\alpha\beta}=u_{\beta\alpha};u_{PP}=u_{PL}=u_{PT}=0. \label{MM}%
\end{equation}
Naively, the traceless condition on $\theta_{\mu\nu}$ implies%
\begin{equation}
u_{PP}-u_{LL}-u_{TT}=0. \label{Naive}%
\end{equation}
However, for the reason which will become clear later that one needs to
include at least one component $u_{NN}$ perpendicular to the scattering plane
and modify Eq.(\ref{Naive}) to%
\begin{equation}
u_{PP}-u_{LL}-u_{TT}-u_{NN}=0. \label{NN}%
\end{equation}
Note that, in the Regge limit, Eq.(\ref{NN}) reduces to Eq.(\ref{Naive}).
However, the solutions for Eq.(\ref{MM}) and Eq.(\ref{NN}) give three RZNS%
\begin{equation}
(\alpha_{-1}^{L}\alpha_{-2}^{L}-\alpha_{-1}^{P}\alpha_{-1}^{L}\alpha_{-1}%
^{L})|0,k\rangle, \label{R4.2}%
\end{equation}%
\begin{equation}
(\alpha_{-1}^{T}\alpha_{-2}^{T}-\alpha_{-1}^{P}\alpha_{-1}^{T}\alpha_{-1}%
^{T})|0,k\rangle, \label{R4.3}%
\end{equation}%
\begin{equation}
(\alpha_{-1}^{(L}\alpha_{-2}^{T)}-\alpha_{-1}^{P}\alpha_{-1}^{L}\alpha
_{-1}^{T})|0,k\rangle, \label{R4.4}%
\end{equation}
while Eq.(\ref{MM}) and Eq.(\ref{Naive}) give only two RZNS%
\begin{equation}
(\alpha_{-1}^{L}\alpha_{-2}^{L}-\alpha_{-1}^{P}\alpha_{-1}^{L}\alpha_{-1}%
^{L}-\alpha_{-1}^{T}\alpha_{-2}^{T}+\alpha_{-1}^{P}\alpha_{-1}^{T}\alpha
_{-1}^{T})|0,k\rangle\label{R4.23}%
\end{equation}
and Eq.(\ref{R4.4}). Note that Eq.(\ref{R4.23}) is just a linear combination
of Eq.(\ref{R4.2}) and Eq.(\ref{R4.3}). For the high energy fixed angle
calculation in \cite{ChanLee1,ChanLee2}, the corresponding extra ZNS will not
affect the final result there. The vector ZNS in Eq.(\ref{43}) gives two RZNS%
\begin{equation}
\lbrack6\alpha_{-3}^{T}-18\alpha_{-1}^{(P}\alpha_{-2}^{T)}+9\alpha_{-1}%
^{P}\alpha_{-1}^{P}\alpha_{-1}^{T}+\alpha_{-1}^{L}\alpha_{-1}^{L}\alpha
_{-1}^{T}+\alpha_{-1}^{T}\alpha_{-1}^{T}\alpha_{-1}^{T}]|0,k\rangle,
\label{R4.5}%
\end{equation}%
\begin{equation}
\lbrack6\alpha_{-3}^{L}-18\alpha_{-1}^{(P}\alpha_{-2}^{L)}+9\alpha_{-1}%
^{P}\alpha_{-1}^{P}\alpha_{-1}^{L}+\alpha_{-1}^{L}\alpha_{-1}^{L}\alpha
_{-1}^{L}+\alpha_{-1}^{L}\alpha_{-1}^{T}\alpha_{-1}^{T}]|0,k\rangle.
\label{R4.6}%
\end{equation}
\ The vector ZNS in Eq.(\ref{44}) gives two RZNS%
\begin{equation}
\lbrack3\alpha_{-3}^{T}+9\alpha_{-1}^{[P}\alpha_{-2}^{T]}-\alpha_{-1}%
^{L}\alpha_{-1}^{L}\alpha_{-1}^{T}-\alpha_{-1}^{T}\alpha_{-1}^{T}\alpha
_{-1}^{T}]|0,k\rangle, \label{R4.7}%
\end{equation}%
\begin{equation}
\lbrack3\alpha_{-3}^{L}+9\alpha_{-1}^{[P}\alpha_{-2}^{L]}-\alpha_{-1}%
^{L}\alpha_{-1}^{L}\alpha_{-1}^{L}-\alpha_{-1}^{L}\alpha_{-1}^{T}\alpha
_{-1}^{T}]|0,k\rangle. \label{R4.8}%
\end{equation}
There are totally 8 RZNS at the mass level $M^{2}=$ $4$.


\begin{thebibliography}{999}                                                                                              %


\bibitem {GM}David~J Gross and Paul~F Mende.
\newblock {The high-energy behavior of string scattering amplitudes}.
\newblock {\em Phys. Lett. B}, 197(1):129--134, 1987.

\bibitem {GM1}David~J Gross and Paul~F Mende.
\newblock {String theory beyond the Planck scale}.
\newblock {\em Nucl. Phys. B}, 303(3):407--454, 1988.

\bibitem {Gross}David~J. Gross.
\newblock {High-Energy Symmetries of String Theory}.
\newblock {\em Phys. Rev. Lett.}, 60:1229, 1988.

\bibitem {Gross1}D.~J. Gross.
\newblock {Strings at superPlanckian energies: In search of the string
symmetry}. \newblock In \emph{{In *London 1988, Proceedings, Physics and
mathematics of strings* 83-95. (Philos. Trans. R. Soc. London A329 (1989)
401-413).}}, 1988.

\bibitem {GrossManes}David~J Gross and JL~Manes.
\newblock {The high energy behavior of open string scattering}.
\newblock {\em Nucl. Phys. B}, 326(1):73--107, 1989.

\bibitem {Lee}Jen-Chi Lee.
\newblock {New symmetries of higher spin states in string theory}.
\newblock {\em Phys. Lett. B}, 241(3):336--342, 1990.

\bibitem {Lee-Ov}Jen-Chi Lee and Burt~A Ovrut.
\newblock {Zero-norm states and enlarged gauge symmetries of the closed bosonic
string with massive background fields}. \newblock {\em Nucl. Phys. B},
336(2):222--244, 1990.

\bibitem {LeePRL}Jen-Chi Lee.
\newblock {Decoupling of degenerate positive-norm states in string theory}.
\newblock {\em Phys. Rev. Lett.}, 64(14):1636, 1990.

\bibitem {West9}P. C.West, A Brief Review of the Group Theoretic Approach to
String Theory, in "conformal Field Theories and Related Topics", Proceedings
of Third Annecy Meeting on Theoretical Physics, LAPP, Annecy le Vieux, France,
Nucl. Phys. B (Proc. Suppl) 5B (1988) 217, edited by P. Binutruy, P. Sorba and
R. Stora, North Holland (1988).

\bibitem {GSW}MB~Green, JH~Schwarz, and E~Witten.
\newblock {\em {Superstring theory, v. 1}}. \newblock Cambridge University,
Cambridge, 1987.

\bibitem {Lee3}Jen-Chi Lee.
\newblock {Reduction of stringy scattering amplitudes through massive ward
identities}. \newblock {\em Z Phys. C}, 63(2):351--355, 1994.

\bibitem {Lee4}Jen-Chi Lee.
\newblock {Heterotic massive Einstein-Yang-Mills-type symmetry and Ward
identity}. \newblock {\em Phys. Lett. B}, 337(1):69--73, 1994.

\bibitem {LEO}Jen-Chi Lee.
\newblock {Spontaneously broken symmetry in string theory}.
\newblock {\em Phys. Lett. B}, 326(1):79--83, 1994.

\bibitem {JCLee}Jen-Chi Lee.
\newblock {Generalized on-shell Ward identities in string theory}.
\newblock {\em Prog. of th. phys.}, 91(2):353--360, 1994.

\bibitem {KaoLee}Hsien-Chung Kao and Jen-Chi Lee.
\newblock {Decoupling of degenerate positive-norm states in Witteno<
s string
field theory}. \newblock {\em Phys. Rev. D}, 67(8):086003, 2003.

\bibitem {CLYang}Chuan-Tsung Chan, Jen-Chi Lee, and Yi~Yang.
\newblock {Anatomy of zero-norm states in string theory}.
\newblock {\em Phys. Rev. D}, 71(8):086005, 2005.

\bibitem {Witten}Edward Witten.
\newblock {Non-commutative geometry and string field theory}.
\newblock {\em Nucl. Phys. B}, 268(2):253--294, 1986.

\bibitem {Winfinity}Jean Avan and Antal Jevicki.
\newblock {Classical integrability and higher symmetries of collective string
field theory}. \newblock {\em Phys. Lett. B}, 266(1):35--41, 1991.

\bibitem {Winfinity2}Jean Avan and Antal Jevicki.
\newblock {Quantum integrability and exact eigenstates of the collective string
field theory}. \newblock {\em Physics Letters B}, 272(1):17--24, 1991.

\bibitem {Klebanov1}Igor~R Klebanov and Alexander~M Polyakov.
\newblock {Interaction of discrete states in two-dimensional string theory}.
\newblock {\em Mod. Phys. Lett. A}, 6(35):3273--3281, 1991.

\bibitem {Ring}Edward Witten.
\newblock {Ground ring of two-dimensional string theory}.
\newblock {\em Nucl. Phys. B}, 373(1):187--213, 1992.

\bibitem {Ring1}Edward Witten and Barton Zwiebach.
\newblock {Algebraic structures and differential geometry in two-dimensional
string theory}. \newblock {\em Nucl. Phys. B}, 377(1):55--112, 1992.

\bibitem {ChungLee1}Tze-Dan Chung and Jen-Chi Lee.
\newblock {Discrete gauge states and wo9" charges in c= 1 2D gravity}.
\newblock {\em Phys. Lett. B}, 350(1):22--27, 1995.

\bibitem {ChungLee2}Tze-Dan Chung and Jen-Chi Lee.
\newblock {Superfield form of discrete gauge states in {\^c}}= 1 2d
supergravity. \newblock {\em Z Phys. C}, 75(3):555--558, 1997.

\bibitem {Lee1}J-C Lee.
\newblock {Soliton gauge states and T-duality of closed bosonic string
compatified on torus}. \newblock {\em Euro. Phys. J. C}, 7(4):669--672, 1999.

\bibitem {Lee2}Jen-Chi Lee.
\newblock {Chan--Paton soliton gauge states of the compactified open string}.
\newblock {\em Euro. Phys. J. C}, 13(4):695--697, 2000.

\bibitem {ChanLee}Chuan-Tsung Chan and Jen-Chi Lee.
\newblock {Stringy symmetries and their high-energy limits}.
\newblock {\em Phys. Lett. B}, 611(1):193--198, 2005.

\bibitem {ChanLee1}Jen-Chi Lee.
\newblock {Stringy symmetries and their high-energy limit}.
\newblock {\em arXiv preprint hep-th/0303012}, 2003.

\bibitem {ChanLee2}Chuan-Tsung Chan and Jen-Chi Lee.
\newblock {Zero-norm states and high-energy symmetries of string theory}.
\newblock {\em Nucl. Phys. B}, 690(1):3--20, 2004.

\bibitem {CHL}Chuan-Tsung Chan, Pei-Ming Ho, and Jen-Chi Lee.
\newblock {Ward identities and high energy scattering amplitudes in string
theory}. \newblock {\em Nucl. Phys. B}, 708(1):99--114, 2005.

\bibitem {CHLTY1}Chuan-Tsung Chan, Pei-Ming Ho, Jen-Chi Lee, Shunsuke
Teraguchi, and Yi~Yang.
\newblock {Solving all 4-point correlation functions for bosonic open string
theory in the high-energy limit}. \newblock {\em Nucl. Phys. B},
725(1):352--382, 2005.

\bibitem {CHLTY2}Chuan-Tsung Chan, Pei-Ming Ho, Jen-Chi Lee, Shunsuke
Teraguchi, and Yi~Yang.
\newblock {High-energy zero-norm states and symmetries of string theory}.
\newblock {\em Phys. Rev. Lett.}, 96(17):171601, 2006.

\bibitem {ChanLee3}Chuan-Tsung Chan and Jen-Chi Lee.
\newblock {One-loop massive scattering amplitudes and Ward identities in string
theory}. \newblock {\em Prog. of th. phys.}, 115(1):229--243, 2006.

\bibitem {susy}Chuan-Tsung Chan, Jen-Chi Lee, and Yi~Yang.
\newblock {High energy scattering amplitudes of superstring theory}.
\newblock {\em Nucl. Phys. B}, 738(1):93--123, 2006.

\bibitem {West2}Nicolas Moeller and Peter West.
\newblock {Arbitrary four string scattering at high energy and fixed angle}.
\newblock {\em Nucl. Phys. B}, 729(1):1--48, 2005.

\bibitem {Closed}Chuan-Tsung Chan, Jen-Chi Lee, and Yi~Yang.
\newblock {Notes on high-energy limit of bosonic closed string scattering
amplitudes}. \newblock {\em Nucl. Phys. B}, 749(1):280--290, 2006.

\bibitem {Veneziano}Gabriele Veneziano.
\newblock {Construction of a crossing-simmetric, Regge-behaved amplitude for
linearly rising trajectories}. \newblock {\em Il Nuovo Cimento A},
57(1):190--197, 1968.

\bibitem {BCJ2}N~Emil~J Bjerrum-Bohr, Poul~H Damgaard, and Pierre Vanhove.
\newblock {Minimal basis for gauge theory amplitudes}.
\newblock {\em Phys. Rev. Lett.}, 103(16):161602, 2009.

\bibitem {LLY1}Sheng-Hong Lai, Jen-Chi Lee, and Yi~Yang. \newblock The string
bcj relations revisited and extended recurrence relations of nonrelativistic
string scattering amplitudes. \newblock {\em arXiv preprint arXiv:1601.03813}, 2016.

\bibitem {LLY2}Sheng-Hong Lai, Jen-Chi Lee, and Yi~Yang. \newblock The exact
sl (k+ 3, c) symmetry of string scattering amplitudes.
\newblock {\em arXiv preprint arXiv:1603.00396}, 2016.

\bibitem {BCJ1}Z~Bern, JJM Carrasco, and Henrik Johansson.
\newblock {New relations for gauge-theory amplitudes}.
\newblock {\em Phys. Rev. D}, 78(8):085011, 2008.

\bibitem {Dscatt}Chuan-Tsung Chan, Jen-Chi Lee, and Yi~Yang.
\newblock {Scatterings of massive string states from D-brane and their linear
relations at high energies}. \newblock {\em Nucl. Phys. B}, 764(1):1--14, 2007.

\bibitem {LMY}Jen-Chi Lee, Yoshihiro Mitsuka, and Yi~Yang.
\newblock {Higher spin string states scattered from d-particle in the Regge
regime and factorized ratios of fixed angle scatterings}.
\newblock {\em Progress of Theoretical Physics}, 126(3):397--417, 2011.

\bibitem {KLT}Hideyuki Kawai, David~C Lewellen, and S-HH Tye.
\newblock {A relation between tree amplitudes of closed and open strings}.
\newblock {\em Nucl. Phys. B}, 269(1):1--23, 1986.

\bibitem {CHLTY3}Chuan-Tsung Chan, Pei-Ming Ho, Jen-Chi Lee, Shunsuke
Teraguchi, and Yi~Yang.
\newblock {Comments on the high energy limit of bosonic open string theory}.
\newblock {\em Nucl. Phys. B}, 749(1):266--279, 2006.

\bibitem {Wall}Chuan~Tsung Chan, Jen-Chi Lee, and Yi~Yang.
\newblock {Power-law Behavior of Strings Scattered from Domain-wall at High
Energies and Breakdown of their Linear Relations}.
\newblock {\em arXiv preprint hep-th/0610219}, 2006.

\bibitem {O-plane}Jen-Chi Lee and Yi~Yang.
\newblock {High-energy massive string scatterings from orientifold planes}.
\newblock {\em Nuclear Physics B}, 798(1):198--209, 2008.

\bibitem {Craps}Ben Craps and Frederik Roose.
\newblock {Anomalous D-brane and orientifold couplings from the boundary
state}. \newblock {\em Phys. Lett. B}, 445(1):150--159, 1998.

\bibitem {Craps1}Bogdan Stefa{\'n}ski.
\newblock {Gravitational couplings of D-branes and O-planes}.
\newblock {\em Nucl. Phys. B}, 548(1):275--290, 1999.

\bibitem {Craps2}Jose~F Morales, Claudio~A Scrucca, and Marco Serone.
\newblock {Anomalous couplings for D-branes and O-planes}.
\newblock {\em Nucl. Phys. B}, 552(1):291--315, 1999.

\bibitem {Schnitzer}Howard~J Schnitzer and Niclas Wyllard.
\newblock {An orientifold of AdS5$\times$ T11 with D7-branes, the associated
$\alpha$'2-corrections and their role in the dual Script N= 1 Sp (2N+
2M)$\times$ Sp (2N) gauge theory}. \newblock {\em JHEP}, 2002(08):012, 2002.

\bibitem {Garousi}Mohammad~R Garousi.
\newblock {Superstring scattering from O-planes}.
\newblock {\em Nucl. Phys. B}, 765(1):166--184, 2007.

\bibitem {Myers}Mohammad~R Garousi and Robert~C Myers.
\newblock {Superstring scattering from D-branes}.
\newblock {\em Nucl. Phys. B}, 475(1):193--224, 1996.

\bibitem {Mende}Paul~F Mende.
\newblock {High energy string collisions in a compact space}.
\newblock {\em Phys. Lett. B}, 326(3):216--222, 1994.

\bibitem {Compact}Jen-Chi Lee and Yi~Yang.
\newblock {Linear relations and their breakdown in high energy massive string
scatterings in compact spaces}. \newblock {\em Nucl. Phys. B}, 784(1):22--35, 2007.

\bibitem {Compact2}Jen-Chi Lee, Tomohisa Takimi, and Yi~Yang.
\newblock {High-energy string scatterings of compactified open string}.
\newblock {\em Nucl. Phys. B}, 804(1):250--261, 2008.

\bibitem {RR1}Daniele Amati, M~Ciafaloni, and G~Veneziano.
\newblock {Superstring collisions at Planckian energies}.
\newblock {\em Phys. Lett. B}, 197(1):81--88, 1987.

\bibitem {RR2}Daniele Amati, M~Ciafaloni, and G~Veneziano.
\newblock {Classical and quantum gravity effects from Planckian energy
superstring collisions}. \newblock {\em Int. J. of Mod. Phys. A},
3(07):1615--1661, 1988.

\bibitem {RR3}Daniele Amati, Marcello Ciafaloni, and Gabriele Veneziano.
\newblock {Can spacetime be probed below the string size?}
\newblock {\em Phys. Lett. B}, 216(1):41--47, 1989.

\bibitem {RR4}M~Soldate.
\newblock {Partial-wave unitarity and closed-string amplitudes}.
\newblock {\em Phys. Lett. B}, 186(3):321--327, 1987.

\bibitem {RR5}IJ~Muzinich and M~Soldate.
\newblock {High-energy unitarity of gravitation and strings}.
\newblock {\em Phys. Rev. D}, 37(2):359, 1988.

\bibitem {RR6}Richard~C Brower, Joseph Polchinski, Matthew~J Strassler, and
Chung-I Tan. \newblock {The Pomeron and gauge/string duality}.
\newblock {\em JHEP}, 2007(12):005, 2007.

\bibitem {RR7}Richard~C Brower, Horatiu Nastase, Howard~J Schnitzer, and
Chung-I Tan.
\newblock {Analyticity for multi-Regge limits of the Bern--Dixon--Smirnov
amplitudes}. \newblock {\em Nucl. Phys. B}, 822(1):301--347, 2009.

\bibitem {bosonic}Sheng-Lan Ko, Jen-Chi Lee, and Yi~Yang.
\newblock {Patterns of high energy massive string scatterings in the Regge
regime}. \newblock {\em JHEP}, 2009(06):028, 2009.

\bibitem {RRsusy}Song He, Jen-Chi Lee, Keijiro Takahashi, and Yi~Yang.
\newblock {Massive superstring scatterings in the Regge regime}.
\newblock {\em Phys. Rev. D}, 83(6):066016, 2011.

\bibitem {hep-th/0410131}Oleg Andreev and Warren Siegel.
\newblock {Quantized tension: Stringy amplitudes with Regge poles and parton
behavior}. \newblock {\em Phys. Rev. D}, 71(8):086001, 2005.

\bibitem {LYAM}Jen-Chi Lee, Catherine~H Yan, and Yi~Yang.
\newblock {High-energy string scattering amplitudes and signless Stirling
number identity}.
\newblock {\em SIGMA. Symmetry, Integrability and Geometry: Methods and
Applications}, 8:045, 2012.

\bibitem {HLY}Song He, Jen-Chi Lee, and Yi~Yang.
\newblock {Exponential fall-off behavior of Regge scatterings in compactified
open string theory}. \newblock {\em Prog. of th. phys.}, 128(5):887--901, 2012.

\bibitem {KLY1}Sheng-Lan Ko, Jen-Chi Lee, and Yi~Yang.
\newblock {Kummer function and high energy string scatterings}.
\newblock {\em arXiv preprint arXiv:0811.4502}, 2008.

\bibitem {KLY2}Jen-Chi Lee, Yi~Yang, and Sheng-Lan Ko.
\newblock {Stirling number identities and High energy String Scatterings}.
\newblock {\em arXiv preprint arXiv:0909.3894}, 2009.

\bibitem {LY}Jen-Chi Lee and Yoshihiro Mitsuka.
\newblock {Recurrence relations of Kummer functions and Regge string scattering
amplitudes}. \newblock {\em JHEP}, 2013(4):1--23, 2013.

\bibitem {Slater}Lucy~Joan Slater.
\newblock {\em {Confluent hypergeometric functions}}. \newblock University
Press Cambridge, 1960.

\bibitem {Tan}Chih-Hao Fu, Jen-Chi Lee, Chung-I Tan, and Yi~Yang.
\newblock {Recurrence relations of higher spin BPST vertex operators for open
strings}. \newblock {\em Phys. Rev. D}, 88(4):046004, 2013.

\bibitem {AppellLY}Jen-Chi Lee and Yi~Yang.
\newblock {The Appell function F1 and Regge string scattering amplitudes}.
\newblock {\em Physics Letters B}, 739:370--374, 2014.

\bibitem {sl5c}Willard Miller~Jr.
\newblock {Lie theory and the Lauricella functions FD}.
\newblock {\em Journal of Mathematical Physics}, 13:1393--1399, 1972.

\bibitem {Wang}Xiaoxia Wang.
\newblock {Recursion formulas for Appell functions}.
\newblock {\em Integral Transforms and Special Functions}, 23(6):421--433, 2012.

\bibitem {Sagnotti}A~Sagnotti and M~Taronna.
\newblock {String lessons for higher-spin interactions}.
\newblock {\em Nucl. Phys. B}, 842(3):299--361, 2011.

\bibitem {WS}J~Isberg, U~Lindstr{\"o}m, B~Sundborg, and G~Theodoridis.
\newblock {Classical and quantized tensionless strings}.
\newblock {\em Nucl. Phys. B}, 411(1):122--156, 1994.

\bibitem {WS1}Bo~Sundborg.
\newblock {Stringy gravity, interacting tensionless strings and massless higher
spins}. \newblock {\em Nucl. Phys. B}, 102:113--119, 2001.

\bibitem {WS2}E~Sezgin and P~Sundell.
\newblock {Massless higher spins and holography}. \newblock In \emph{Nucl.
Phys. B644 (2002) 303, erratum Nucl. Phys. B660 (2003) 403 [hep-th/0205131}.
Citeseer, 2003.

\bibitem {WS3}E~Sezgin and P~Sundell.
\newblock {Massless higher spins and holography}.
\newblock {\em Nucl. Phys. B}, 644(1):303--370, 2002.

\bibitem {WS4}Chong-Sun Chu, Pei-Ming Ho, and Feng-Li Lin.
\newblock {Cubic string field theory in pp-wave background and background
independent Moyal structure}. \newblock {\em JHEP}, 2002(09):003, 2002.

\bibitem {WS5}Giulio Bonelli.
\newblock {On the tensionless limit of bosonic strings, infinite symmetries and
higher spins}. \newblock {\em Nucl. Phys. B}, 669(1):159--172, 2003.

\bibitem {0305052}Massimo Bianchi, Jose~F Morales, and Henning Samtleben.
\newblock {On stringy AdS5$\times$ S5 and higher spin holography}.
\newblock {\em Journal of High Energy Physics}, 2003(07):062, 2003.

\bibitem {0311257}A~Sagnotti and M~Tsulaia. \newblock On higher spins and the
tensionless limit of string theory. \newblock {\em Nuclear Physics B},
682(1):83--116, 2004.

\bibitem {1207.4485}Chi-Ming Chang, Shiraz Minwalla, Tarun Sharma, and Xi~Yin.
\newblock Abj triality: from higher spin fields to strings.
\newblock {\em Journal of Physics A: Mathematical and Theoretical},
46(21):214009, 2013.

\bibitem {Hagedorn}Joseph~J Atick and Edward Witten.
\newblock {The Hagedorn transition and the number of degrees of freedom of
string theory}. \newblock {\em Nucl. Phys. B}, 310(2):291--334, 1988.

\bibitem {Hagedorn1}R~Hagedorn. \newblock {Nuovo Cim. Suppl. 3 (1965) 147}.
\newblock {\em Nuovo Cim. A}, 56:1027, 1968.

\bibitem {hep-th/9908001}Bo~Sundborg.
\newblock {The Hagedorn transition, deconfinement and N= 4 SYM theory}.
\newblock {\em Nucl. Phys. B}, 573(1):349--363, 2000.

\bibitem {Moore}Gregory Moore. \newblock {Finite in all directions}.
\newblock {\em arXiv preprint hep-th/9305139}, 1993.

\bibitem {Moore1}Gregory Moore.
\newblock {Symmetries of the bosonic string S-matrix}.
\newblock {\em arXiv preprint hep-th/9310026}, 1993.

\bibitem {CKT}Chuan-Tsung Chan, Shoichi Kawamoto, and Dan Tomino.
\newblock {To see symmetry in a forest of trees}.
\newblock {\em Nucl. Phys. B}, 885:225--265, 2014.

\bibitem {West1}Peter West. \newblock {Physical states and string symmetries}.
\newblock {\em Mod. Phys. Lett. A}, 10(09):761--771, 1995.

\bibitem {0705.1816}Steven~B Giddings, David~J Gross, and Anshuman Maharana.
\newblock Gravitational effects in ultrahigh-energy string scattering.
\newblock {\em Physical Review D}, 77(4):046001, 2008.

\bibitem {0302123}Jen-Chi Lee.
\newblock {Calculation of zero-norm states and reduction of stringy scattering
amplitudes}. \newblock {\em Prog. of th. phys.}, 114(1):259--273, 2005.

\bibitem {EO}Mark Evans and Burt~A Ovrut.
\newblock {Spontaneously broken inter mass level symmetries in string theory}.
\newblock {\em Phys. Lett. B}, 231(1):80--84, 1989.

\bibitem {Heter}David~J Gross, Jeffrey~A Harvey, Emil Martinec, and Ryan Rohm.
\newblock {Heterotic string theory:(II). The interacting heterotic string}.
\newblock {\em Nucl. Phys. B}, 267(1):75--124, 1986.

\bibitem {Das}Sumit~R Das and B~Sathiapalan.
\newblock {String propagation in a tachyon background}.
\newblock {\em Phys. Rev. Lett.}, 56(25):2664, 1986.

\bibitem {Das1}Sumit~R Das and B~Sathiapalan.
\newblock {New infinities in two-dimensional nonlinear sigma models and
consistent string propagation}. \newblock {\em Phys. Rev. Lett.}, 57(13):1511, 1986.

\bibitem {Weinberg}Steven Weinberg.
\newblock {Coupling constants and vertex functions in string theories}.
\newblock {\em Phys. Lett. B}, 156(5):309--314, 1985.

\bibitem {Sasaki}Ryu Sasaki and Itaru Yamanaka.
\newblock {Vertex operators for a bosonic string}.
\newblock {\em Phys. Lett. B}, 165(4):283--288, 1985.

\bibitem {Labas}JMF Labastida and Maria~AH Vozmediano.
\newblock {Bosonic strings in background massive fields}.
\newblock {\em Nucl. Phys. B}, 312(2):308--340, 1989.

\bibitem {Giddings}Steven~B Giddings.
\newblock {The Veneziano amplitude from interacting string field theory}.
\newblock {\em Nucl. Phys. B}, 278(2):242--255, 1986.

\bibitem {Spradlin}Marcus Spradlin and Anastasia Volovich.
\newblock {Superstring interactions in a pp-wave background}.
\newblock {\em Phys. Rev. D}, 66(8):086004, 2002.

\bibitem {Banks}Thomas Banks and Michael~E Peskin.
\newblock {Gauge invariance of string fields}. \newblock {\em Nucl. Phys. B},
264:513--547, 1986.

\bibitem {Sen}Ashoke Sen.
\newblock {Descent relations among bosonic D-branes}.
\newblock {\em Int. J. of Mod. Phys. A}, 14(25):4061--4077, 1999.

\bibitem {Manes}JL~Manes and Maria~AH Vozmediano.
\newblock {A simple construction of string vertex operators}.
\newblock {\em Nucl. Phys. B}, 326(1):271--284, 1989.

\bibitem {Polyakov2}Alexander~M Polyakov and AM~Poljakov.
\newblock {\em {Gauge fields and strings}}, volume 140. \newblock Harwood
academic publishers Chur, 1987.

\bibitem {2Dstring}Igor~R Klebanov and Andrea Pasquinucci.
\newblock {Infinite symmetry and Ward identities in two-dimensional string
theory}. \newblock {\em arXiv preprint hep-th/9210105}, 1992.

\bibitem {GKN}David~J Gross, Igor~R Klebanov, and Michael~J Newman.
\newblock {The two-point correlation function of the one-dimensional matrix
model}. \newblock {\em Nucl. Phys. B}, 350(3):621--634, 1991.

\bibitem {GKN1}David~J Gross and Igor~R Klebanov.
\newblock {Fermionic string field theory of c= 1 two-dimensional quantum
gravity}. \newblock {\em Nucl. Phys. B}, 352(3):671--688, 1991.

\bibitem {GKN2}Kresimir Demeterfi, Antal Jevicki, and Joao~P. Rodrigues.
\newblock {Scattering amplitudes and loop corrections in collective string
field theory. 2.} \newblock {\em Nucl. Phys.}, B365:499--522, 1991.

\bibitem {GKN3}Ulf~H. Danielsson and David~J. Gross.
\newblock {On the correlation functions of the special operators in c = 1
quantum gravity}. \newblock {\em Nucl. Phys.}, B366:3--26, 1991.

\bibitem {Polyakov}Alexander~M Polyakov.
\newblock {Self-tuning fields and resonant correlations in 2d-gravity}.
\newblock {\em Mod. Phys. Lett. A}, 6(07):635--644, 1991.

\bibitem {Polyakov1}AM~Polyakov.
\newblock {Princeton preprint PUPT-1289, Lectures given at the Jerusalem winter
school (1991); IR Klebanov and AM Polyakov}.
\newblock {\em Mod. Phys. Lett. A}, 6:3273, 1991.

\bibitem {Distler}Jacques Distler, Zvonimir Hlousek, and Hikaru Kawai.
\newblock {Super-Liouville theory as a two-dimensional, superconformal
supergravity theory}. \newblock {\em Int. J. of Mod. Phys. A}, 5(02):391--414, 1990.

\bibitem {Bouwknegt}Peter Bouwknegt, Jim McCarthy, and Krzysztof Pilch.
\newblock {Ground ring for the two-dimensional NSR string}.
\newblock {\em Nucl. Phys. B}, 377(3):541--570, 1992.

\bibitem {Itoh}Katsumi Itoh and Nobuyoshi Ohta.
\newblock {BRST cohomology and physical states in two-dimensional supergravity
coupled to $\hat{c}\leq 1$ matter}. \newblock {\em Nuclear Physics B},
377(1):113--142, 1992.

\bibitem {Itoh1}Katsumi Itoh and Nobuyoshi Ohta.
\newblock {Spectrum of two-dimensional (super) gravity}.
\newblock {\em Prog. of Th. Phys. Sup.}, 110:97--116, 1992.

\bibitem {Narain}Kumar~S Narain.
\newblock {New heterotic string theories in uncompactified dimensions< 10}.
\newblock {\em Phys. Lett. B}, 169(1):41--46, 1986.

\bibitem {Narain1}KS~Narain, MH~Sarmadi, and E~Witten.
\newblock {A note on toroidal compactification of heterotic string theory}.
\newblock {\em Nucl. Phys. B}, 279(3):369--379, 1987.

\bibitem {Giveon}Amit Giveon, Massimo Porrati, and Eliezer Rabinovici.
\newblock {Target space duality in string theory}.
\newblock {\em Phys. Rept.}, 244(2):77--202, 1994.

\bibitem {Goddard}Peter Goddard and David Olive.
\newblock {\em {Kac-Moody and Virasoro algebras: a reprint volume for
physicists}}, volume~3. \newblock World scientific, 1988.

\bibitem {BCJ3}Stephan Stieberger.
\newblock {Open \& closed vs. pure open string disk amplitudes}.
\newblock {\em arXiv:0907.2211}, 2009.

\bibitem {BCJ4}Bo~Feng, Rijun Huang, and Yin Jia.
\newblock {Gauge amplitude identities by on-shell recursion relation in
S-matrix program}. \newblock {\em Phys. Lett. B}, 695(1):350--353, 2011.

\bibitem {BCJ5}Yi-Xin Chen, Yi-Jian Du, and Bo~Feng.
\newblock {A proof of the explicit minimal-basis expansion of tree amplitudes
in gauge field theory}. \newblock {\em JHEP}, 2011(2):1--22, 2011.

\bibitem {Decay}Jen-Chi Lee and Yi~Yang.
\newblock {Linear relations of high energy absorption/emission amplitudes of
D-brane}. \newblock {\em Phys. Lett. B}, 646(2):120--124, 2007.

\bibitem {Giudice}E~Del~Giudice, P~Di~Vecchia, and S~Fubini.
\newblock {General properties of the dual resonance model}.
\newblock {\em Annals of Physics}, 70(2):378--398, 1972.

\bibitem {DDF1}Richard~C Brower.
\newblock {Spectrum-generating algebra and no-ghost theorem for the dual
model}. \newblock {\em Phys. Rev. D}, 6(6):1655, 1972.

\bibitem {DDF2}RC~Brower and P~Goddard.
\newblock {Collinear algebra for the dual model}.
\newblock {\em Nucl. Phys. B}, 40(CERN-TH-1392):437--444, 1972.

\bibitem {1106.0033}Thomas S{\o }ndergaard.
\newblock {Perturbative Gravity and Gauge Theory Relations: A Review}.
\newblock {\em Advances in High Energy Physics}, 2012, 2012.

\bibitem {1108.2381}Pawel Caputa and Shinji Hirano.
\newblock {Observations on open and closed string scattering amplitudes at high
energies}. \newblock {\em JHEP}, 2012(2):1--17, 2012.

\bibitem {1205.6369}Pawe{\l } Caputa.
\newblock {Lightlike contours with fermions}. \newblock {\em Phys. Lett. B},
716(3):475--480, 2012.

\bibitem {GSO}Tamiaki Yoneya.
\newblock {Spontaneously broken space--time supersymmetry in open string theory
without GSO projection}. \newblock {\em Nucl. Phys. B}, 576(1):219--240, 2000.

\bibitem {GSO1}Tabito Hara and Tamiaki Yoneya.
\newblock {Nonlinear supersymmetry without the GSO projection and unstable
D9-brane}. \newblock {\em Nucl. Phys. B}, 602(3):499--513, 2001.

\bibitem {Klebanov}Akikazu Hashimoto and Igor~R Klebanov.
\newblock {Scattering of strings from D-branes}.
\newblock {\em Nucl. Phys. B}, 55(2):118--133, 1997.

\bibitem {Klebanov3}Igor~R Klebanov and Larus Thorlacius.
\newblock {The Size of p-branes}. \newblock {\em Phys. Lett. B},
371(1):51--56, 1996.

\bibitem {barbon1996d}Jos{\'e}~LF Barb{\'o}n.
\newblock {D-brane form factors at high energy}.
\newblock {\em Phys. Lett. B}, 382(1):60--64, 1996.

\bibitem {bachas1999high}Constantin Bachas and Boris Pioline.
\newblock {High energy scattering on distant branes}. \newblock {\em JHEP},
1999(12):004, 1999.

\bibitem {hirano1997scattering}Shinji Hirano and Yoichi Kazama.
\newblock {Scattering of closed string states from a quantized D-particle}.
\newblock {\em Nucl. Phys. B}, 499(1):495--515, 1997.

\bibitem {Decay1}Akikazu Hashimoto and Igor~R Klebanov.
\newblock {Decay of excited D-branes}. \newblock {\em Phys. Lett. B},
381(4):437--445, 1996.

\bibitem {OA}Oleg Andreev.
\newblock {Remarks on the high-energy behavior of string scattering amplitudes
in warped spacetimes. II}. \newblock {\em Phys. Rev. D}, 71(6):066006, 2005.

\bibitem {DL}GS~Danilov and Lev~Nikolaevich Lipatov.
\newblock {BFKL pomeron in string models}. \newblock {\em Nucl. Phys. B},
754(1):187--232, 2006.

\bibitem {KP}M~Kachelriess and M~Plumacher.
\newblock {Remarks on the high-energy behaviour of cross-sections in weak-scale
string theories}. \newblock {\em arXiv preprint hep-ph/0109184}, 2001.

\bibitem {MK}Manuel Kauers.
\newblock {Summation algorithms for Stirling number identities}.
\newblock {\em Journal of Symbolic Computation}, 42(10):948--970, 2007.

\bibitem {BCFW}Yung-Yeh Chang, Bo~Feng, Chih-Hao Fu, Jen-Chi Lee, Yihong Wang,
and Yi~Yang.
\newblock {A note on on-shell recursion relation of string amplitudes}.
\newblock {\em Journal of High Energy Physics}, 2013(2):1--31, 2013.

\bibitem {CCW}Clifford Cheung, Donal O'Connell, and Brian Wecht.
\newblock {BCFW recursion relations and string theory}. \newblock {\em JHEP},
2010(9):1--32, 2010.

\bibitem {bosonic2}Jen-Chi Lee and Yi~Yang.
\newblock {Regge closed string scattering and its implication on fixed angle
closed string scattering}. \newblock {\em Phys. Lett. B}, 687(1):84--88, 2010.

\bibitem {Picard}{\'E}mile Picard.
\newblock {Sur une extension aux fonctions de deux variables du probleme de
Riemann relatif aux fonctions hyperg{\'e}}om{\'e}triques. \newblock In
\emph{Annales scientifiques de l'{\'E}cole Normale Sup{\'e}rieure}, volume~10,
pages 305--322, 1881.

\bibitem {Appell}Joseph~Kamp{\'e} de~F{\'e}riet and Paul Appell.
\newblock {\em {Fonctions hyperg{\'e}}om{\'e}triques et hypersph{\'e}riques}. \newblock 1926.
\end{thebibliography}
\end{document}